\documentclass[fleqn,usenatbib]{mnras} 

\usepackage{newtxtext,newtxmath}
\usepackage[T1]{fontenc}
\usepackage{anyfontsize}
\usepackage{tabularx}
\usepackage{supertabular}
\usepackage[math]{cellspace}

\usepackage{xspace}
\usepackage[dvipsnames]{xcolor}

\usepackage{silence}
\usepackage[super]{nth}
\usepackage{multirow}
\usepackage{titlesec}
\usepackage{orcidlink}
\usepackage{adjustbox}
\usepackage{caption}
\usepackage{url}
\usepackage{longtable}
\usepackage{tabularx}
\usepackage{multirow}
\usepackage{boldline}
\def\simlt{\lower.5ex\hbox{$\; \buildrel < \over \sim \;$}}
\def\simgt{\lower.5ex\hbox{$\; \buildrel > \over \sim \;$}}

\usepackage{subfiles}
\usepackage{amsmath}	

\usepackage[english]{babel}
\usepackage{color}
\usepackage{fontawesome}
\usepackage{natbib}
\usepackage{graphics}
\usepackage{graphicx}
\usepackage{hyperref}
\usepackage[capitalise]{cleveref}

\defcitealias{Etherington2022}{E22}

\newcommand{\github}[1]{%
   \href{#1}{\faGithubSquare}%
}

\newcommand{\LUPM}{%
Laboratoire Univers et Particules de Montpellier, CNRS \& Université de Montpellier, Parvis Alexander Grothendieck, Montpellier, France 34090
}

\newcommand{\Newcastle}{%
School of Mathematics, Statistics and Physics, Newcastle University, Herschel Building, Newcastle-upon-Tyne, NE1 7RU, UK
}

\newcommand{\DurhamICC}{%
Department of Physics, Institute for Computational Cosmology, Durham University, South Road, Durham DH1 3LE, UK
}

\newcommand{\DurhamCEA}{%
Department of Physics, Centre for Extragalactic Astronomy, Durham University, South Road, Durham DH1 3LE, UK
}

\newcommand{\Northeastern}{%
Department of Physics, Northeastern University, 360 Huntington Ave, Boston, MA USA
}
\newcommand{\Liege}{%
STAR Institute, Quartier Agora - All\'ee du six Ao\^ut, 19c B-4000 Li\`ege, Belgium}

\newcommand{\Aalto}{%
Department of Computer Science, Aalto University, PO Box 15400, Espoo, FI-00 076, Finland
}
\newcommand{\Helsinki}{%
Department of Physics, Faculty of Science, University of Helsinki, 00014-Helsinki, Finland}

\newcommand{\UTAustin}{%
Department of Astronomy, The University of Texas at Austin, Austin, TX, USA
}

\newcommand{\DAWN}{%
Cosmic Dawn Centre (DAWN), Denmark
}

\newcommand{\NBI}{%
Niels Bohr Institute, University of Copenhagen, Jagtvej 128, 2200 Copenhagen, Denmark}

\newcommand{\JPL}{%
Jet Propulsion Laboratory, California Institute of Technology, 4800, Oak Grove Drive, Pasadena, CA, USA}

\newcommand{\PMO}{Purple Mountain Observatory, Chinese Academy of Sciences, 10
Yuanhua Road, Nanjing 210023, China}

\newcommand{\IAP}{Institut d’Astrophysique de Paris, UMR 7095, CNRS, Sorbonne Universit\'e, 98 bis boulevard Arago, F-75014 Paris, France}

\newcommand{\LAM}{Laboratoire d'astrophysique de Marseille, Aix Marseille University, CNRS, CNES, Marseille, France}

\newcommand{\UCSB}{Department of Phyiscs University of California Santa Barbara, CA,93106, CA}

\newcommand{\Rochester}{Laboratory for Multiwavelength Astrophysics, School of Physics and Astronomy, Rochester Institute of Technology, 84 Lomb Memorial Drive, Rochester, NY 14623, USA}

\newcommand{\STScI}{Space Telescope Science Institute, 3700 San Martin Drive, Baltimore, MD 21218, USA}

\newcommand{\UCSC}{Department of Astronomy and Astrophysics, University of California, Santa Cruz, 1156 High Street, Santa Cruz, CA 95064 USA}

\newcommand{\Hawaii}{Department of Physics and Astronomy, University of Hawaii, Hilo, 200 W Kawili St, Hilo, HI 96720, USA}

\newcommand{\Caltech}{Caltech/IPAC, 1200 E. California Blvd., Pasadena, CA 91125, USA}

\newcommand{\DTU}{DTU-Space, Technical University of Denmark, Elektrovej 327, DK-2800 Kgs. Lyngby, Denmark}

\title[COWLS: The COSMOS-Web Lens Survey]
{The COSMOS-Web Lens Survey (COWLS) I: Discovery of >100 high redshift strong lenses in contiguous \textit{JWST} imaging}
\author[Nightingale et al.]{James W.\ Nightingale\orcidlink{0000-0002-8987-7401}\thanks{e-mail: James.Nightingale@newcastle.ac.uk},
Guillaume Mahler$^{2,3,4}$\orcidlink{0000-0003-3266-2001}, 
Jacqueline McCleary$^{5}$,
Qiuhan He$^{4}$\orcidlink{0000-0003-3672-9365},
Natalie B.\ Hogg$^{6}$\orcidlink{0000-0001-9346-4477},
\newauthor
Aristeidis Amvrosiadis$^{4}$\orcidlink{0000-0002-4465-1564},
Ghassem Gozaliasl$^{7,8}$,
Wilfried Mercier$^{9}$,
Diana Scognamiglio$^{10}$,
\newauthor
Edward Berman$^{5}$,
Gavin Leroy$^{3,4}$,
Daizhong Liu$^{11}$,
Richard J.\ Massey$^{3,4}$\orcidlink{0000-0002-6085-3780}
Marko Shuntov$^{12,13}$,
\newauthor
Maximilian von Wietersheim-Kramsta$^{3,4}$, 
Maximilien Franco,$^{14}$
Louise Paquereau,$^{15}$
Olivier Ilbert$^{9}$, 
\newauthor
Natalie Allen$^{12, 13}$,
Sune Toft$^{12, 13}$, 
Hollis B. Akins$^{14}$,
Caitlin M. Casey,$^{14,16,12}$
Jeyhan S. Kartaltepe,$^{17}$
\newauthor
Anton M. Koekemoer,$^{18}$
Henry Joy McCracken,$^{15}$
Jason D. Rhodes,$^{10}$
Brant E. Robertson,$^{19}$
\newauthor
Nicole E. Drakos,$^{20}$
Andreas L. Faisst,$^{21}$
and Shuowen Jin$^{12,22}$
\\
Affiliations can be found after the references.
}

\begin{document}

\bibpunct{(}{)}{;}{a}{}{;}
\date{\today}
\pagerange{\pageref{firstpage}--\pageref{lastpage}} 
\pubyear{2025}
\maketitle
\label{firstpage}

\begin{abstract}
We present the COSMOS-Web Lens Survey (COWLS), a sample of over 100 strong lens candidates from the $0.54$\,deg$^2$ COSMOS-Web survey, discovered using exquisite \textit{James Webb} Space Telescope (\textit{JWST}) imaging across four wavebands. Following two rounds of visual inspection, over 100 candidates were ranked as `high confidence' or `likely' by at least $50\%$ of inspectors. The COWLS sample has several notable properties: (i) magnified source galaxies spanning redshifts $z \sim 0.1$ to $z \sim 9$, which therefore extend into the epoch of reionisation; (ii) the highest-redshift lens galaxies known, pushing galaxy density profile evolution studies beyond $z \sim 2$; (iii) all lenses are distributed within a contiguous $0.54$\,deg$^2$ region, allowing for joint strong and weak lensing analyses; and (iv) a subset exhibits lensed source emission ray-traced near the lens galaxy centers, enabling studies of supermassive black holes and dust absorption. A key innovation of our approach is the use of lens modelling to aid in identifying lenses that may otherwise be missed. This paper is accompanied by the first COWLS public release, providing \textit{JWST} NIRCam imaging in four bands, lens models, pixelized source reconstructions and lens redshift estimates \github{https://github.com/Jammy2211/COWLS_COSMOS_Web_Lens_Survey}.

\end{abstract}


\section{Introduction}\label{Intro}

Galaxy-scale strong gravitational lensing, where a background source is multiply imaged by a foreground deflector galaxy, has facilitated numerous advancements in astrophysics and cosmology. This phenomenon has provided magnified views of distant galaxies that would otherwise be inaccessible \citep{Vieira2013, Swinbank2015, Rizzo2020, Rizzo2021, Liu2024, Amvrosiadis2025}, offered insights into the mass distribution of galaxies up to a redshift of $z \sim$ 2 \citep{Koopmans2009, Auger2009, Shajib2021, Etherington2023, Tan2024}, and enabled the measurement of cosmological parameters \citep{Collett2014, Vegetti2014, Birrer2020, Hogg:2023khs}. Our ability to leverage strong lensing as a means to understand the Universe is ultimately limited by the number of known lenses, how high in redshift their lens and source galaxies are and the quality of the imaging data.

Finding strong lenses begins by building a sample of candidate lenses through one of three selection techniques: (i) searching for arcs or rings in large wide-field ground-based imaging surveys \citep{Gavazzi2007, Sonnenfeld2013b, Jacobs2019, Tran2022}; (ii) detecting multiple emission lines at different redshifts in the spectra of galaxies \citep{Bolton2006, Bolton2008a, Bolton2012, Shu2016}, or (iii) applying flux-density cuts on point-source catalogues from wide-area submillimetre surveys to isolate galaxies that are so bright they must have been magnified by lensing \citep{Vieira2010, Negrello2014, Harrington2016}. In all cases, the initial sample of galaxies inspected exceeds millions, to overcome the rarity of strong lensing occurring. For all three methods, confirming that candidates are strong lenses requires expensive high-resolution imaging follow-up, historically from the \textit{Hubble} Space Telescope (\textit{HST}). These collective efforts have led to the discovery of hundreds of galaxy-scale strong lenses. 

The \textit{James Webb} Space Telescope (\textit{JWST}) offers a powerful approach to finding strong lenses through high-resolution imaging at near-infrared (NIR) and infrared (IR) wavelengths. With its 6.5-metre mirror, \textit{JWST}'s increased depth observes a significant fraction of distant and faint galaxies in a single exposure, even with modest exposure times. Consequently, within the relatively small $2.2 \times 2.2$ arcminute field of view, there is still a high probability that an image will contain multiple detectable strong lenses \citep{Casey2023, Holloway2023, Pearson2024}. Any strong lens discovered this way comes with exquisite multi-wavelength \textit{JWST} imaging `for free', which would be challenging and expensive to acquire through individual follow-up observations. 

We therefore search for strong lenses in the \textit{JWST} Cosmic Origins Survey (COSMOS-Web) survey \citep{Casey2023}, a $0.54$\,deg$^2$ field observed in 4 NIR and IR bands using \textit{JWST}, which completed observations in May 2024. Forecasts predict that the survey contains between $30 - 100$ strong lenses \citep{Casey2023, Holloway2023}. COWLS Paper III, \citet[][hereafter H25]{Hogg2025b}, uses the properties of the candidates found in this paper to infer the total number of lenses the COSMOS-Web field contains -- inferring a value of $107$, indicating our sample has high completeness. We use 4 wavebands of Near-Infrared Camera (NIRCam) imaging (F115W, F150W, F277W and F444W), prioritising this shorter wavelength data as it provides the highest spatial resolution and is able to resolve the lower Einstein radius strong lenses. Our lens search therefore omits the higher wavelength COSMOS-Web Mid-Infrared Imager (MIRI) data, which likely contains additional strong lenses \citep{Pearson2024}. However, MIRI data is available for many of the lens candidates we present.

We name this the COSMOS-Web Lens Survey (COWLS), which stands out among lens surveys due to several unique characteristics that shape its primary scientific goals. The exceptional survey depth allows the lensed source population to extend beyond reionisation to $z > 6$, surpassing previous lens surveys, as shown in H25. These magnified sources provide an unprecedented view of distant galaxy structure and morphology. Lens modeling enables detailed pixel-level reconstructions of every unlensed source galaxy \citep{Nightingale2018}, with reconstructions across all four NIRCam bands included in the public data release accompanying this paper~\github{https://github.com/Jammy2211/COWLS_COSMOS_Web_Lens_Survey}.

The COWLS lens population is also distinct, featuring the highest-redshift lenses known (above $z \sim 2$), lower-mass systems, and rare morphologies such as disk galaxies. This enables studies of galaxy density profile evolution at higher redshifts \citep{Shajib2021, Etherington2023, Tan2024} and makes strong lens dark matter detection studies more sensitive to line-of-sight halos \citep{Vegetti2014, Ritondale2019a, Despali2018, He2022b, Despali2022, Nightingale2024}. Unlike other lens surveys that span large sky areas, all COWLS candidates reside within a $0.54$ deg$^2$ region with contiguous JWST imaging, providing a unique opportunity to study strong lenses within their broader cosmological environment \citep{Peng2010}. COSMOS-Web’s deep \textit{JWST} imaging enables high-precision weak lensing analysis, and the dense concentration of strong lenses makes this dataset ideal for joint strong and weak lensing studies to measure cosmic shear \citep{Birrer2017, Birrer:2017sge, Fleury2021, Hogg2022, Duboscq:2024asf, Hogg:2025wac}, provided systematic uncertainties in strong lens shear measurements are addressed \citep{Etherington2023a}. Additionally, we identify lenses where source emission passes within $0.2\arcsec$ of the lens center, potentially probing the influence of supermassive black holes, as demonstrated in Abell 1201 where the source-lens separation was $0.38\arcsec$ \citep{Nightingale2023}.

An important aspect of the lens-finding approach used in this study is the use of traditional lens modelling, via the open-source software {\tt PyAutoLens}~\github{https://github.com/Jammy2211/PyAutoLens} \citep{Nightingale2015, Nightingale2018, pyautolens}, to help determine whether candidates are genuinely strong lenses. Only a small fraction of lens-finding studies have applied lens modelling in this manner (e.g., \citealt{Sonnenfeld2018b, Sonnenfeld2020, Rojas2022}), using low-resolution ground-based imaging which limits the information a lens model can extract. \citet{Acevedo2024} used lens modeling to help find lenses with high resolution \textit{Euclid} early release observations, and we compare our approach to theirs. The lens modelling pipeline is fully automated after initial data preprocessing and has been successfully applied to simulated lenses \citep{Cao2021, He2023} and \textit{HST} datasets \citep{Etherington2022, Etherington2023, Nightingale2024, He2024}. In this work, we fit over 400 candidates across all four imaging wavebands and illustrate multiple ways in which it allows us to find high quality lens candidates we would have otherwise missed.

This work was performed in tandem with COWLS Paper II, \citet[][hereafter M25]{Mahler2025}, who report on the 17 most ``spectacular'' strong lenses in COWLS, showing how JWST's depth, resolution and NIR coverage makes a subset of strong lenses easily identifiable through basic visual inspection. Cowls Paper III, H25, compares the lens sample we found to a forecasting analysis, to address whether the lens sample is consistent with that expected based on known galaxy populations and cosmology. 


The $1.64$\,deg$^2$ \textit{HST} COSMOS field, which contains the $0.54$\,deg$^2$ COSMOS-Web field, has previously been searched for strong lenses, both via human visual inspection \citep{Faure2008, Jackson2008} and using machine learning \citep{Pourrahmani2018}. There are 64 candidates presented in \citet{Faure2008}, 112 in \citet{Jackson2008} and 92 by \citet{Pourrahmani2018}. Where COSMOS-Web overlaps with the \textit{HST} COSMOS field we compare candidates. Boosts in submillimetre counts and far IR emission of COSMOS data have also produced strong lens candidates \citep{Aretxaga2011, Jin2018}. The COSMOS field has also had individual strong lenses discovered and discussed, all of which are in our high ranking sample \citep{Guzzo2007, More2012a, VanDerWel2013, Pearson2024, Jin24}. COWLS includes the COSMOS-Web ring \citep{Mercier2024, VanDokkum2024}, recently spectroscopically confirmed to be the highest redshift known galaxy-scale lens galaxy ($z = 2.02$) and have a $z = 5.10$ source galaxy \citep{Shuntov2025}.

With the recent launch of \textit{Euclid}, strong lenses are now being found with high resolution space based imaging \cite{ORiordan2025,Pearce-Casey2024,Nagam2025,Acevedo2024,EuclidCollaboration2025, EuclidCollaboration2025a, EuclidCollaboration2025b} spanning large patches of the Universe. The COWLS samples complements perfectly the tens of thousands of lenses \textit{Euclid} is poised to discover, with the two samples essentially covering different regions of lens and source redshift parameter space.

Data for all lens candidates is publicly available at the following link:~\github{https://github.com/Jammy2211/COWLS_COSMOS_Web_Lens_Survey}, including .fits data files, point spread functions, {\tt PyAutoLens} modeling results, unlensed source reconstructions, redshift estimates and other metadata.

This paper is structured as follows.
In Section~\ref{COSMOS}, we describe the COSMOS-Web survey and \textit{JWST} data reduction.
In Section~\ref{Modeling}, we describe the {\tt PyAutoLens} method and model fits performed in this work.
In Section~\ref{Visual}, we describe the visual inspection process used to find lens candidates.
In Section~\ref{Results}, we present the lens candidates found after visual inspection and lens modelling.
In Section~\ref{Discussion}, we discuss the implications of our measurements, and we give a summary in \S\ref{Summary}.
\section{Data}\label{COSMOS}

\subsection{NIRCam Imaging \& Catalogue}\label{Data}

We use \textit{JWST} data from the cycle 1 GO program COSMOS-Web (PID 1727; PIs: Kartaltepe \& Casey; \citealt{Casey2023}), a 255 hour \textit{JWST} treasury program. The full survey was completed in May 2024 and maps a $0.54$\,deg$^2$ area using NIRCam (we do not use the $0.19$\,deg$^2$ COSMOS-Web MIRI data which partially overlaps the NIRCam footprint). We use the complete COSMOS-Web dataset including the NIRCam F115W, F150W, F277W, and F444W filters over the full area.  It is the largest \textit{JWST} program both in terms of contiguous area covered at this depth and GO time allocated. The depths of the NIRCam data are measured to be 26.6-27.3 AB (F115W), 26.9-27.7 (F150W), 27.5-28.2 (F277W), and 27.5-28.2 (F444W) for 5$\sigma$ point sources calculated within $0.15''$ radius apertures. Data for each waveband were reduced to two pixel scales, 0.03\,arcsec and 0.06\,arcsec. To minimise correlated noise, lens modelling uses data closest to NIRCam's native pixel-scale, therefore 0.03\,arcsec for F115W and F150W, 0.06\,arcsec for F277W and F444W. 

The COSMOS-Web team has constructed a model-based photometric catalogue combining ground and space-based data for broad general use; this catalogue is described in the forthcoming paper by Shuntov et al. in prep \citep[see also][]{Shuntov2024}. To briefly describe the catalogue construction, the four NIRCam imaging filters were PSF-homogenised to F444W’s PSF then combined in a $\sqrt{\chi^2}$ image \citep[][]{szalay_simultaneous_1999, Drlica-Wagner2018} truncated towards positive values only. The detection was performed on this $\sqrt{\chi^2}$ image following a hot/cold methodology \citep{Grogin2011} using SEP \citep{K.Barbary2016}.  The hot/cold catalogue is then used as input to construct 2D S\'ersic models for each galaxy, fit simultaneously to the four NIRCam filters using \texttt{SourceExtractor++} (SE++; \citealt{Kummel2020,Bertin2020}).  These S\'ersic subtracted images are used in the visual inspection campaign. Once the model shapes are derived, SE++ is used once more to extract photometry from 30+ bands across the field (ground and space) to produce the final photometric catalog. Photometric redshifts and physical parameters are derived with \textsc{LePHARE} \citep{Arnouts1999, Ilbert2006}. 
From these stellar masses and photometric redshifts, we then select candidate lenses for visual vetting. COSMOS-Web data was collected as $\sim 304$ fields of view (FOVs) across the whole COSMOS-Web field, taken together in 152 visits. Data reduction and cataloguing reduced and processed these as a single FOV in January 2023, 10 FOVs in April 2023, and 10 FOVs in January 2024, producing 21 FOVs in total.



\subsection{Point Spread Function}

Point spread functions (PSFs) describe the impulse response of an optical system like a telescope to light. Because the distortion produced by the PSF mimics and dilutes the gravitational lensing shear of interest, the PSF must be modelled and mitigated before further analysis. The NIRCam PSF varies with time, bandpass, and across the field of view and moreover is non-trivially affected by the mosaicing procedure. Accordingly, we create PSF models empirically from the observations themselves, using stars as reference points with which to build the model. 

We begin by using SourceExtractor \citep{2002ASPC..281..228B} to generate astronomical source catalogues on the \texttt{i2d}-format resampled mosaics produced by the \textit{JWST} Science Calibration Pipeline, using the WHT extension as a weight image and a $3 \times 3$ Gaussian convolution kernel with a two-pixel standard deviation. 

A preliminary star sample is created from this catalogue using size (specifically, \texttt{FLUX\_RADIUS}) and magnitude cuts ($19 < $ \texttt{MAG\_AUTO} $ < 25$) designed to isolate the stellar locus in a size-magnitude diagram. This initial sample is scrubbed of stars with saturated or otherwise bad pixels indicated by sentinel value of 0 in the ERR extension of the \texttt{i2d} mosaics. We then pass this star catalogue sample to PSFEx to obtain PSF fits \citep{2011ASPC..442..435B}. We use the \texttt{PIXEL} basis, fix the PSFEx sampling step to 0.5, and fit first-order polynomials in $(X, Y)$, which we find sufficient to capture the spatial variation across the FOV of a single resampled mosaic. Once models are obtained, the {\tt psfex}~\github{https://github.com/esheldon/psfex?tab=readme-ov-file} Python package is used to create a rendering of the PSF at the location of lens galaxies, which can then be passed to \texttt{PyAutoLens}~\github{https://github.com/Jammy2211/PyAutoLens} for further analysis.  This procedure is repeated in all four bandpasses for each of the 21 NIRCam tiles. 


\subsection{Redshifts}\label{Redshift}

The redshift of the lens is a crucial piece of information for strong lensing studies. We collected both spectroscopic and photometric redshift measurements for the foreground lensing galaxies in our sample. For spectroscopic redshifts, we cross-matched our candidates with the DR10 catalogue of the DESI Legacy Imaging Surveys \citep{Dey2019}. We found that approximately one-third of our candidates have spectroscopic redshifts from previous surveys. For photometric redshifts, we used the COSMOS-Web survey catalogue, 
where redshifts were derived from SED fitting to multi-band images using \textsc{LePhare} \citep{Arnouts1999, Ilbert2006}. We show and compare these redshifts in the results section after presenting our lens candidate sample.

\section{Lens modelling}\label{Modeling}

\subsection{Overview}

\begin{figure*}
\centering
\includegraphics[width=0.99\textwidth]{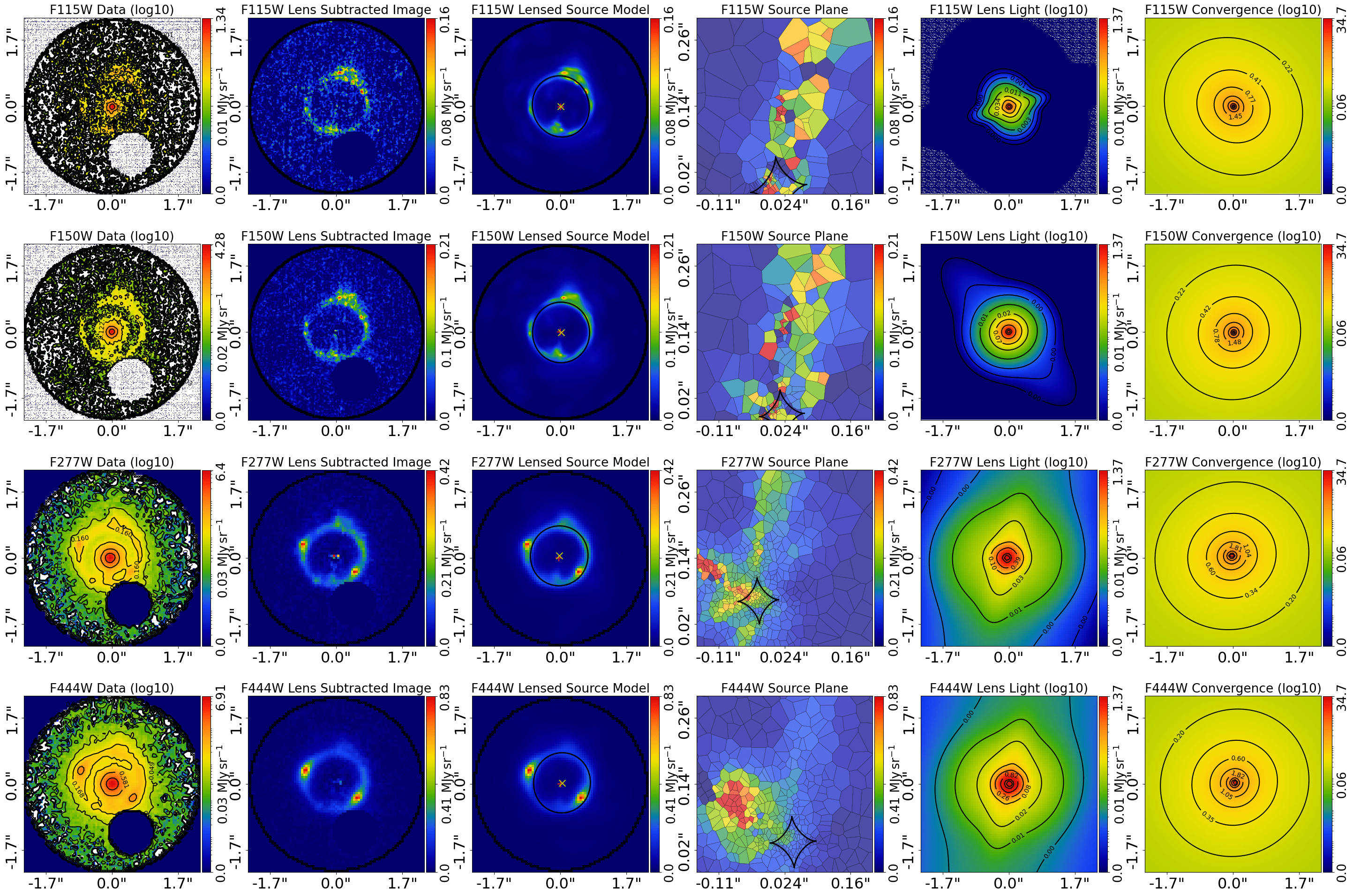}
\caption{
An overview of the lens modelling performed to aid the inspection of strong lens candidates, for the confirmed strong lens COSJ100024+015334, also named the COSMOS-Web ring \citep{Mercier2024, VanDokkum2024, Shuntov2025}. Each row shows the imaging data for the F115W, F150W, F277W and F444W wavebands. From left to right the columns show: (i) the imaging data with a $\log_{10}$ colour scale; (ii) the data where a Multi-Gaussian Expansion (MGE) model subtracts the foreground lens emission; (iii) the image-plane lensed source model; (iv) the source-plane source reconstruction using a Voronoi mesh; (v) the MGE lens light model on a $\log_{10}$ colour scale and; (vi) the inferred lens model convergence. In section \ref{VisualLensModel} we will discuss in detail how this image enables high confidence strong lens candidates to be found. 
}
\label{figure:VisualLensFit}
\end{figure*}

We use version \texttt{2024.11.13.2} of the lens modelling software {\tt PyAutoLens}~\github{https://github.com/Jammy2211/PyAutoLens} \citep{Nightingale2021}. {\tt PyAutoLens} fits the lens galaxy's light and mass and the source galaxy simultaneously. The method assumes a model for the lens's foreground light (a multi-Gaussian expansion \citealt{He2024}), which accounts for blurring by the instrumental PSF and is subtracted from the observed image. A mass model (an isothermal mass distribution) ray-traces image-pixels from the image-plane to the source-plane and a pixelized source reconstruction, using an adaptive Voronoi mesh, is performed. 

Figure \ref{figure:VisualLensFit} shows an overview of a {\tt PyAutoLens} lens model using the COSMOS-Web ring \citep{Mercier2024, VanDokkum2024}, also one of M25's "spectacular" lenses, where models for the image-plane lens galaxy emission, lensed source, source-plane source reconstruction and mass model convergence are shown. This image is used to help inspectors determine if a candidate is a strong lens, as discussed in section \ref{visual:second_round}.

We now describe each step in more detail. The following link contains Jupyter notebooks providing a visual step-by-step guide of the {\tt PyAutoLens} likelihood function used in this work: \url{https://github.com/Jammy2211/autolens_likelihood_function}.

\subsubsection{Lens Light Subtraction}\label{LensLight}

To subtract the foreground emission of the lens galaxy we use a Multi-Gaussian Expansion \citep[][ hereafter MGE]{Cappellari2002}. This decomposes a galaxy's emission into a set of 2D elliptical Gaussian light profiles, unlike many lensing studies which use an elliptical S\'ersic profile. The MGE is implemented within the semi-linear inversion framework for lens modelling \citep{Warren2003}, with the implementation described fully in \citet[][hereafter H24]{He2024}. Its enhanced flexibility for fitting galaxy emission is key to subtracting the complex morphological features seen in lens galaxies when they are observed with deep space based observations (e.g. \textit{HST} or \textit{JWST}). 

The MGE decomposes the light of a galaxy into sets of Gaussians whose overall intensity is given as
\begin{equation}
\label{eqn: MGE}
    I_{\rm set}(x, y) = \sum_{i}^{N} G_i(x, y) \, ,
\end{equation}
with $G_i$ the $i^{\rm th}$ Gaussian profile,
\begin{equation}
\label{eqn: gaussian}
    G_i(x, y) = I_{i}\cdot{\rm exp}\left(-\frac{R_i^2(x, y)}{2\sigma_i^2}\right),
\end{equation}
containing $I_i$ as the intensity normalisation factor and $\sigma_i$ as the full-width at half-maximum of the Gaussian profile. $R_i(x, y)$ is the Gaussian elliptical radius and given is as
\begin{equation}
    \begin{split}
        & R_i(x, y) = \sqrt{x^{\prime2} + \left(\frac{y^\prime}{q_i}\right)^2} \\
        & x^{\prime} = \cos{\phi_i}\cdot\left(x - x^{\rm c}_i\right) + \sin{\phi_i}\cdot\left(y - y^{\rm c}_i\right) \\
        & y^{\prime} = \cos{\phi_i}\cdot\left(y - y^{\rm c}_i\right) -\sin{\phi_i}\cdot\left(x - x^{\rm c}_i\right),
    \end{split}\label{eq: ell_rad}
\end{equation}
where $q_i$ is the axis ratio, $\phi_i$ the position angle, and $\left(x^{\rm c}_i,\ y^{\rm c}_i\right)$ is the centre of the Gaussian. The intensity of each Gaussian is solved for linearly, using the semi-linear inversion formalism for lens analysis \citep{Warren2003}. We enforce positivity on the solution using a modified version of the fast non-negative least-square (fnnls) algorithm \citep{Bro1997}\footnote{The fnnls code we are using is modified from \url{https://github.com/jvendrow/fnnls}.}, because H24 show that allowing for negative intensities produces unphysical solutions whereby the Gaussians alternate between large positive and negative values.

Gaussians are grouped in `sets', where a set of Gaussians share the same centres, position angles, axis ratios, and their $\sigma$ values are fixed to preset values which evenly increase in $\log_{10}$-spaced intervals between one-fifth of the image pixel scale and the circular radius of the mask applied to the lens image. We use two sets of 30 Gaussians for every lens galaxy, where all 60 Gaussians share the same centre. We include an additional set of 10 Gaussians to model central point-like emission, where the $\sigma$ values of these Gaussians are evenly $\log_{10}$-spaced between $0.01\arcsec$ and twice the pixel scale of the data. The centre of these 10 Gaussians is shared with the other 60. The grouping of Gaussians into sets is done to reduce the number of non-linear free parameters in the MGE, which for this study consists of just $N=6$ free parameters for the lens light (see H24).

\begin{figure*}
\centering
\includegraphics[width=0.99\textwidth]{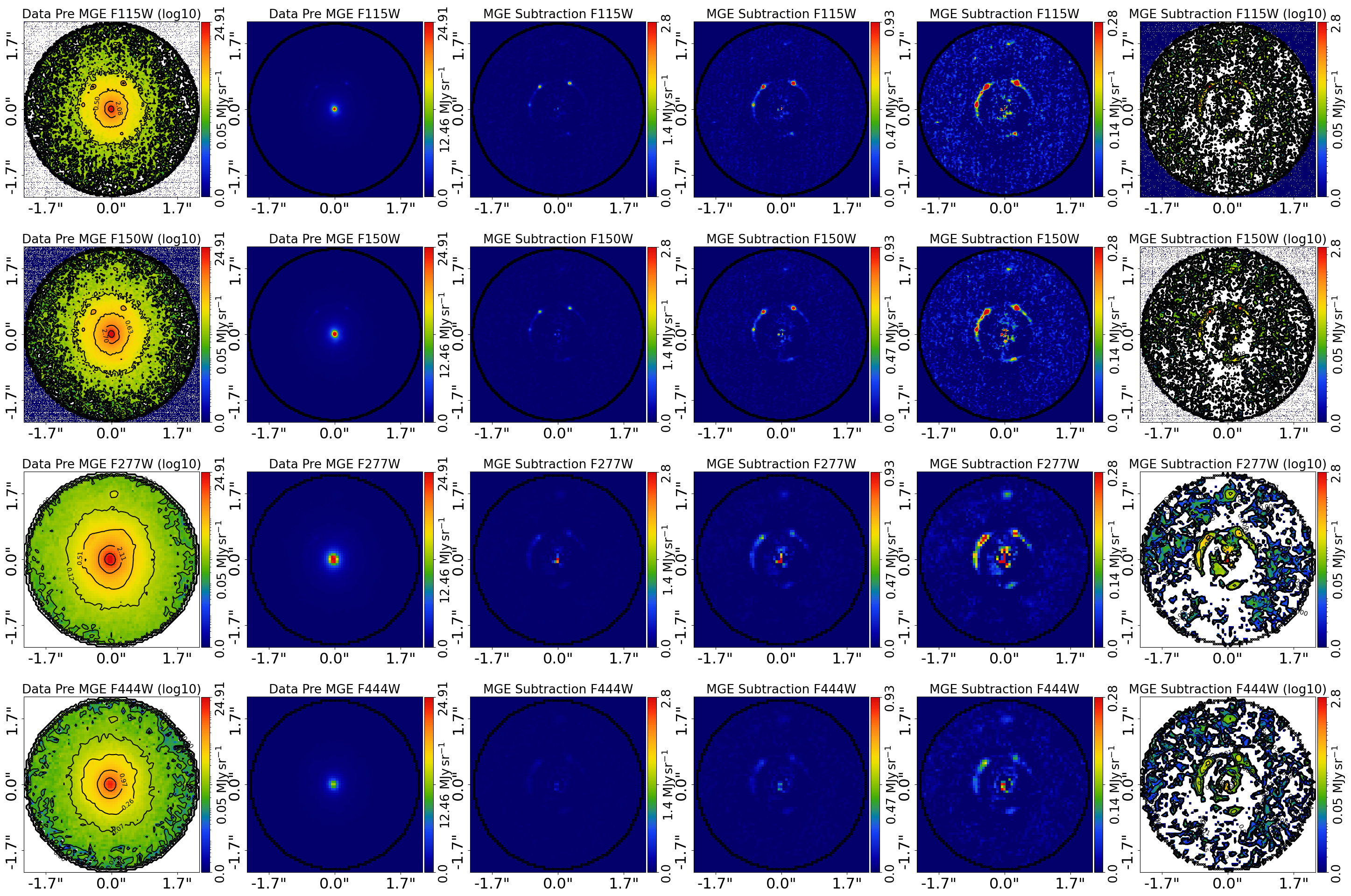}
\caption{
The foreground lens light subtracted images performed using a Multi-Gaussian Expansion (MGE) \citealt{He2024} and no lens mass modelling or source reconstruction, provided to inspectors in the second round of visual inspection. This figure shows an example using the lens COSJ095921+020638. Each row shows the imaging data for the F115W, F150W, F277W and F444W wavebands. The first and second columns show the data before the MGE subtraction with a $\log_{10}$ and linear colourmap, to highlight the lens galaxy in different ways. The remaining four columns show the lensed source with different colour scalings applied, to highlight the lensed source emission in different ways and aid the visual inspection. By performing the MGE subtraction independently of the lens model, an inspector can assert if features in the lensed source are genuinely present in the data as opposed to an artefact of lens modelling. Not including the source in the model can produce residuals in the centre of the lens subtraction, most clearly seen above for the F150W and F277W filters, which the full lens modelling pipeline fits successfully. The right hand column also shows white patches where the $\log_{10}$ colour scale cuts out negative values, this is an intentional choice to provide different information on the candidate source emission.
}
\label{figure:VisualMGELensSub}
\end{figure*}

Each candidate strong lens is fitted twice across all four wavebands using an MGE for the lens light, under two different assumptions. First, an MGE fit is performed without a lens mass model or lensed source model. This produces a lens-light subtracted image used for preprocessing before lens modelling and visual inspection. \Cref{figure:VisualMGELensSub} shows an example of such an image, with different color scales highlighting the lens and source emission. Second, an MGE fit is performed simultaneously with the lens mass model and source model, as shown in \cref{figure:VisualLensFit}.

\subsection{Mass Profiles}\label{MassProfiles}

For the lens mass model (superscript `mass'), we first assume a singular isothermal ellipsoid (SIE) density profile representing the total mass of the lens (e.g., stars and dark matter) of the form
\begin{equation}
\label{eqn:SPLEkap}
\kappa (\xi) = \frac{1}{1 + q^{\rm mass}} \bigg( \frac{\theta^{\rm mass}_{\rm E}}{\xi} \bigg) ,
\end{equation}
where $\theta^{\rm mass}_{\rm E}$ is the model Einstein radius in arc-seconds. Deflection angles are computed via an implementation of the method of \citet{Tessore2015} in {\tt PyAutoLens}.

An external shear (superscript `ext') field is included and parameterised as two elliptical components $(\gamma_{\rm 1}^{\rm ext}, \gamma_{\rm 2}^{\rm ext})$. The shear magnitude, $\gamma^{\rm ext}$, and orientation measured counter-clockwise from north, $\phi^{\rm ext}$, are given by
\begin{equation}
    \label{eq:shear}
    \gamma^{\rm ext} = \sqrt{\gamma_{\rm 1}^{\rm ext^{2}}+\gamma_{\rm 2}^{\rm ext^{2}}}, \, \,
    \tan{2\phi^{\rm ext}} = \frac{\gamma_{\rm 2}^{\rm ext}}{\gamma_{\rm 1}^{\rm ext}}.
\end{equation}
The deflection angles due to the external shear are computed analytically. A recent study by \citealt{Etherington2023a} suggests that this external shear component represents missing complexity in the lens mass distribution.

As an alternative, we also perform fits using an MGE mass model, where the convergence of each Gaussian is related to its intensity as
\begin{equation}
\label{eqn:S\'ersickap}
\kappa_i (\xi) = \Psi G_i(x, y) \, \, ,
\end{equation}
where $\Psi$ is the mass-to-light ratio of the Gaussian and is fixed to the same value across all Gaussians. Every other parameter of each Gaussian in the MGE mass model ($I$, $q$, $\phi$, $\sigma$) comes from the lens light fit. Deflection angles are computed following \citet{Shajib2019}. We do not include a dark matter component. The MGE mass model has been used to perform lens mass modelling in a small number of previous studies (e.g. \citealt{Collett2018, Melo2023}), and in all cases, it was combined with stellar dynamics, which are not used in this study.

The SIE mass model is independent of the MGE lens light model, such that it may infer solutions where the mass and light are geometrically offset and misaligned. In contrast, the MGE mass model is tied to the lens light MGE model, meaning that the stellar mass must be geometrically aligned with the lens light. The two mass models therefore provide different and complementary information, to help the visual inspectors determine if a candidate is a genuine strong lens.

We fit an MGE mass model without an explicit dark matter component (e.g., a Navarro--Frenk--White (NFW) profile; \citealt{Navarro1996}). Our intended use of the MGE mass model is to illustrate whether a mass model which ties lens light to lens mass can still produce a reliable lens model fit. When a dark matter component is included, solutions are often inferred where the dark matter becomes the dominant mass component, meaning that this model can, analogous to the SIE, infer solutions where the lens's mass does not trace its light. We verified that the MGE-only mass model provides physically plausible lens model solutions to the spectacular lenses found in M25, and therefore verified whilst this assumption may not be fully realistic it is sufficient to add information that aids inspectors in judging strong lens candidates.


\subsection{Source Model}\label{Source}

After subtracting the foreground lens emission and ray-tracing coordinates to the source-plane via the mass model, the source is reconstructed in the source-plane. The first stage of our lens modelling pipeline uses an MGE to model the light of the lensed source. As discussed in H24, the MGE source model is an effective tool for automated and efficient lens modelling because it can fit the data in a highly flexible way whilst retaining a relatively low number of non-linear free parameters. For the source, we use only one set of 30 Gaussians, whose $\sigma$ values span $\log_{10}$ increments from 0.001\arcsec to 1.0\arcsec.

However, an MGE source galaxy model is still symmetric around the centre of the Gaussians and thus cannot fit the complex and irregular source morphologies typical of high redshift galaxies. The later stages of our lens modelling pipeline therefore use an adaptive Voronoi mesh which can reconstruct irregular and asymmetric source morphologies. A complete description of the process, including the linear algebra formalism, interpolation scheme and regularisation, is given in \citet{Nightingale2024} and appendix A of H24. An example source reconstruction using this Voronoi mesh is shown in the fourth column of \cref{figure:VisualLensFit}. The Voronoi mesh uses natural neighbour interpolation \citep{Sibson1981} and the cross-like regularisation scheme introduced in H24.

\subsection{Lens modelling pipeline}\label{LensPipeline}

\begin{table}
\renewcommand{\arraystretch}{1.5}
\centering
\begin{tabular}{cccc}
\hline
\textbf{Pipeline} & \textbf{Component} & \textbf{Model} & \textbf{Prior info} \\ \hline
\textbf{Source}                             &  Lens light         & MGE            & -   \\ 
\textbf{Parametric}   & Lens mass          & SIE+Shear      & -                   \\ 
\textbf{(SP)}                                & Source light       & MGE            & -                   \\ \hline
\textbf{Source}                               & Lens light         & MGE            & \textbf{SP (fixed)}         \\
\textbf{Pixelization 1}  & Lens mass          & SIE+Shear      & \textbf{SP}         \\ 
  \textbf{(SPix1)}                               & Source light       & MPR            & -                   \\ \hline
 {\textbf{Source}}                & Lens light         & MGE            & \textbf{SP (fixed)}         \\ 
 {\textbf{Pixelization 2}} & Lens mass          & SIE+Shear      & -        \\
{\textbf{(SPix2)}}                  & Source light       & Voronoi            & -                   \\ \hline
 {\textbf{Mass1}}                & Lens light         & MGE            & \textbf{SP1}         \\ 
 {\textbf{(MP1)}} & Lens mass          & SIE            & \textbf{SP1}        \\ 
               & Source light       & Voronoi            & \textbf{SP2}                   \\ \hline
 {\textbf{Mass2}}                & Lens light         & MGE            & \textbf{SP1}         \\ 
 {\textbf{(MP2)}} & Lens mass          & MGE            & -        \\
               & Source light       & Voronoi            & \textbf{SP2}                   \\ \hline
\end{tabular}
\caption{Pipeline composition used in the analysis built using \texttt{PyAutoLens}.}
\label{tab:pipeline_table}
\end{table}

\cref{tab:pipeline_table} outlines our automated lens modelling pipeline, which iteratively fits various combinations of light, mass, and source models. The pipeline chains together five lens model fits. The pipeline initially fits a simpler model using an MGE source for efficient and robust convergence towards accurate results. Subsequent stages employ the more complex Voronoi source reconstruction. The pipeline performs five chained fits, the first two and final two use the nested sampler \texttt{nautilus}\footnote{\url{https://github.com/johannesulf/nautilus}} \citep{nautilus} and the third stage uses the nested sampler \texttt{dynesty} \citep{dynesty}. The pipeline is based on the \texttt{PyAutoLens} SLaM (Source, Light, and Mass) pipelines used by various other studies (e.g. \citealt{Etherington2022, Cao2021, He2022a, Nightingale2023, Nightingale2024}). A break down of the different fits are as follows: 

{\bf Source Parametric (SP)}: The SP pipeline computes an accurate initial estimate of the lens model parameters. The lens and source light are both modeled using a MGE (see \cref{LensLight}) and the lens mass uses an SIE mass profile with shear (see \cref{MassProfiles}). The non-linear parameter space has $N=17$ free parameters.

{\bf Source Pixelized 1 (SPix1)}: The SPix1 pipeline uses a Voronoi source reconstruction, where the regularization coefficients of the source are free. The lens light uses an MGE, where the centres, axis-ratios and position angles of all Gaussians are fixed to the results of the SP pipeline, but their intensities are linearly solved simultaneously with the source. The mass model is an SIE plus shear where all parameters are free (with priors based on the inferred model from {\bf SP}).

{\bf Source Pixelized 2 (SPix2)}: The lens light MGE and mass model parameters are fixed to the maximum likelihood values of SPix1. The only free parameters are associated with the Voronoi source, which adapt to the source's unlensed morphology (see \citealt{Nightingale2018}) to enable a sharper and more detailed reconstruction of a compact and complex source.

{\bf Mass Pipeline 1 (MP1)}: The lens light MGE parameters are fixed to the maximum likelihood values of SPix1 and Voronoi source parameters fixed to SPix2.  The only free parameters are the SIE mass model and external shear parameters.

{\bf Mass Pipeline 2 (MP2)}: Identical to MP1, but instead of an SIE plus shear the only free parameters are the MGE mass-to-light ratio $\Psi$ and external shear parameters.

\subsection{Multi-wavelength Fits}

For results used during visual inspection, the lens modelling pipeline was run independently on imaging from all four wavebands. As a result, the mass model fitted to each waveband is independent, explaining the slight differences in convergence maps in the right-hand column of \cref{figure:VisualLensFit}. Since lensing is achromatic, mass variations across wavelength are unphysical. This was intentional, as it allows visual inspectors to assess the consistency of mass models across wavebands, providing more information to judge whether a candidate is a lens. The role of this strategy in identifying candidates will be discussed in Section \ref{Results}.

A drawback of fitting each waveband independently is that the source reconstruction also varies across wavelengths, complicating studies of the magnified source. To address this in the COWLS public data release, all candidates were re-fitted after visual inspection using a pipeline that constrained the mass model to a single waveband—the "primary waveband"—selected by JWN based on the clarity of the candidate's source emission in the initial fit. This mass model was then applied to other wavebands, with the source regularisation and MGE lens light refitted as free parameters, alongside two additional parameters accounting for arc-second filter offsets, ensuring alignment across reconstructions. These aligned models are included in the first COWLS public data release~\github{https://github.com/Jammy2211/COWLS_COSMOS_Web_Lens_Survey}. 

When quoting or plotting lens mass model values (e.g., Einstein radius), we refer to those inferred from the primary waveband fit. Magnitudes are computed using models that fit each waveband independently. Lens light magnitudes are derived from the total flux of each MGE lens light model, lensed source magnitudes from the total flux of the source emission in the image plane, and source magnitudes from the sum of all source-plane Voronoi mesh pixels. Magnifications are calculated as the ratio of the total lensed source flux in the image plane to the total source flux in the source plane.

\subsection{User Inputs}

\begin{figure}
\centering
\includegraphics[width=0.235\textwidth]{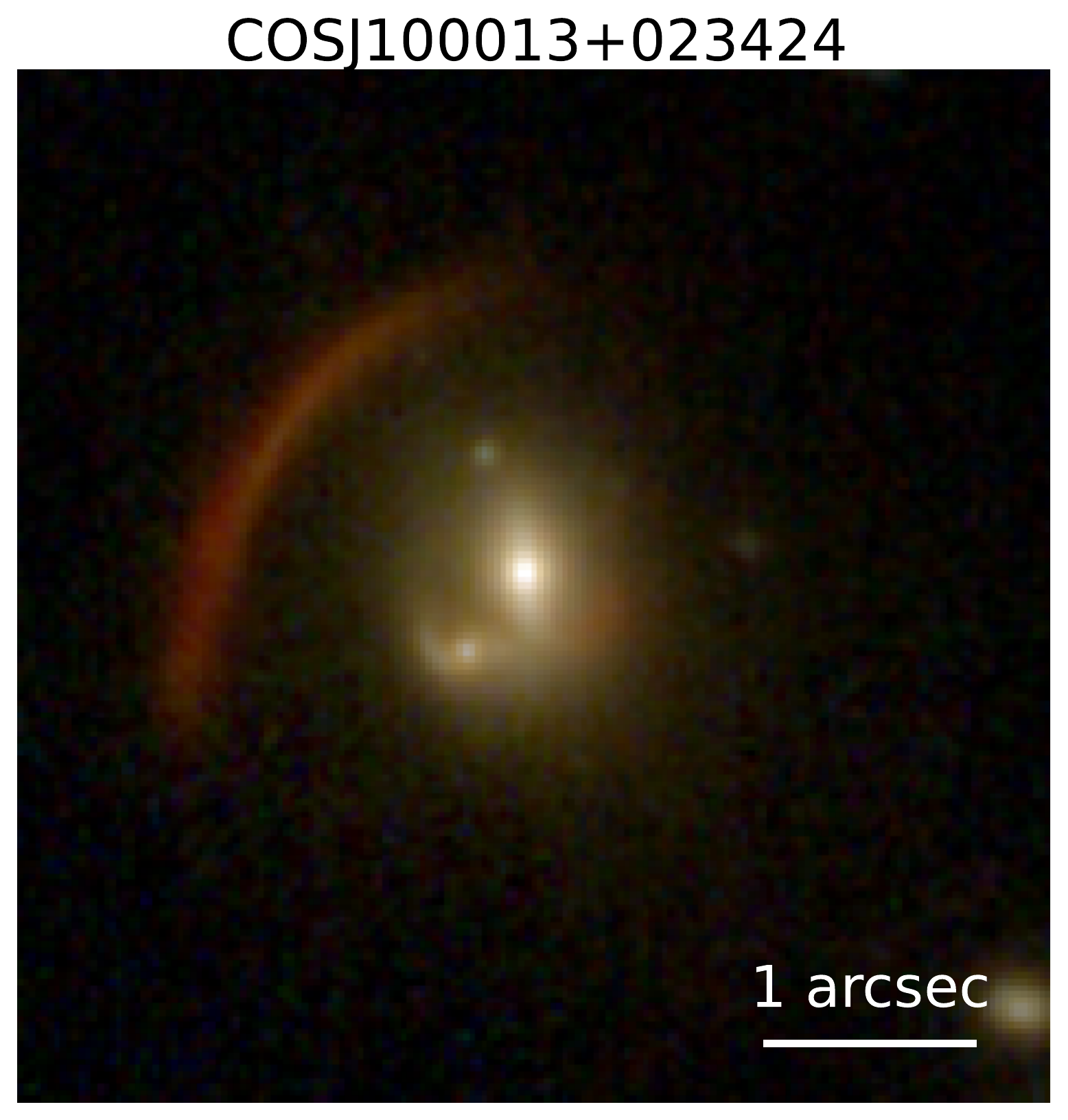}
\includegraphics[width=0.234\textwidth]{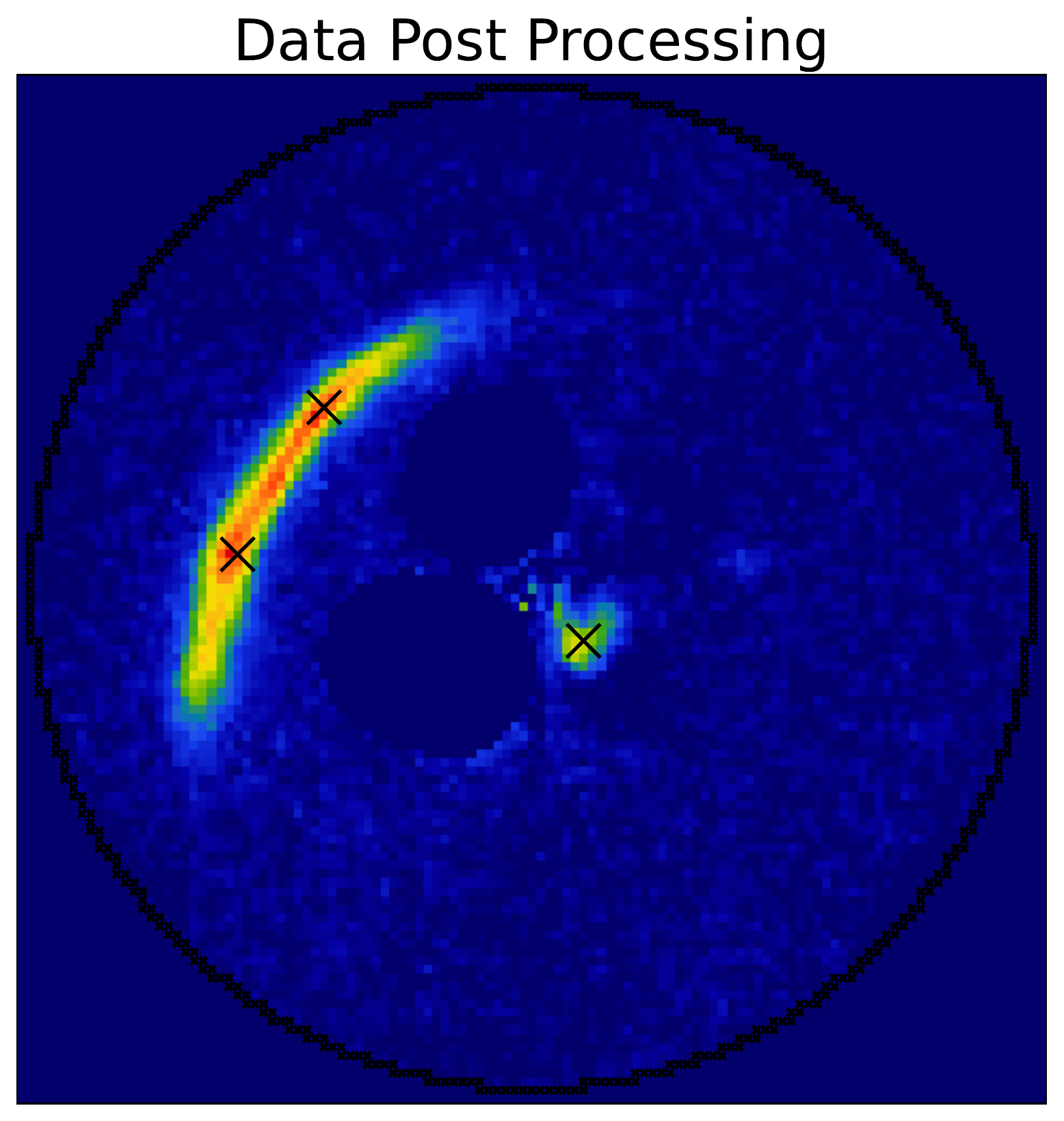}
\caption{
Illustration of preprocessing steps applied to each lens before modelling, using the lens COSJ100013+023424. The left panel shows an RGB image, where colours help distinguish the candidate lensed source (e.g., red arcs) from the foreground lens. The right panel displays the processed F444W image, with the following steps applied: (i) A circular mask (black) is set to encompass the lens and lensed source, in this example of radius $3.5\arcsec$. (ii) A GUI ``spray-paint'' tool removes light from nearby galaxies by setting their values to zero and inflating RMS noise to prevent them from affecting the fit; (iii) The foreground lens light inside this mask is subtracted using an MGE-only model to enhance source visibility independently for all four wavebands and; (iv) a GUI marks the ($x, y$) coordinates of multiple lensed images (black crosses), which during lens modelling must map within $0.1\arcsec$ in the source plane; otherwise, a penalty term is applied to the likelihood, removing unphysical solutions \protect\citep{Maresca2021}.
}
\label{figure:Preprocessing}
\end{figure}

\cref{figure:Preprocessing} illustrates the manual user inputs required before processing by the lens modelling pipeline. All steps are performed manually by JWN, unlike visual inspection tasks which are conducted by multiple inspectors. JWN's subjective determination of what constitutes a candidate's lensed source emission may therefore impact the lens model visuals used later by inspectors to rank lenses. These preprocessing steps are performed using images across all four wavebands in the following order:

\begin{enumerate}

\item \textbf{Mask:} A circular mask is applied to the imaging dataset, such that only image pixels within this mask are fitted by the lens model. \cref{figure:Preprocessing} shows an example $3.5\arcsec$ circular mask with a black circle. The mask's size is chosen to ensure it contains the candidate lens galaxy's emission and all lensed source multiple images.

\item \textbf{Contaminant Removal:} Emission from nearby line-of-sight galaxies expected to not be directly associated with the candidate strong lens may fall within the circular mask. If this emission is not removed, the lens model may incorrectly fit it as part of the lensed source. A GUI is therefore used to ``spray-paint'' any regions of the image with this contaminant emission. The GUI replaces the emission with zeros and increases the RMS noise map to such large values that the lens model fit ignores it. This technique is commonly applied in lens modelling (e.g., \citealt{Etherington2022, Nightingale2024}).

\item \textbf{MGE Lens Subtraction:} The MGE-only lens light fit described in \cref{LensLight} is performed on all four wavebands of data using this mask and with the contaminants removed. This foreground subtracted image is used to make candidate lensed source multiple images more clearly visible.

\item \textbf{Multiple Image Positions:} Using a graphical user interface (GUI), the ($x, y$) coordinates of each multiple image of the candidate lensed source are input, shown by black crosses in \cref{figure:Preprocessing}. Each fit of the lens modelling pipeline requires that the mass model traces all multiple images within a $0.1\arcsec$ threshold of one in the source plane, or else a penalty term is applied to the likelihood, down-weighting that solution. This technique is commonly used in lens modelling to remove unphysical solutions, as described in \citet{Maresca2021}.

\end{enumerate}

\section{Visual Inspection}\label{Visual}

\subsection{First Round}\label{visual:first_round}

\begin{figure*}
\centering
\includegraphics[width=0.9\textwidth]{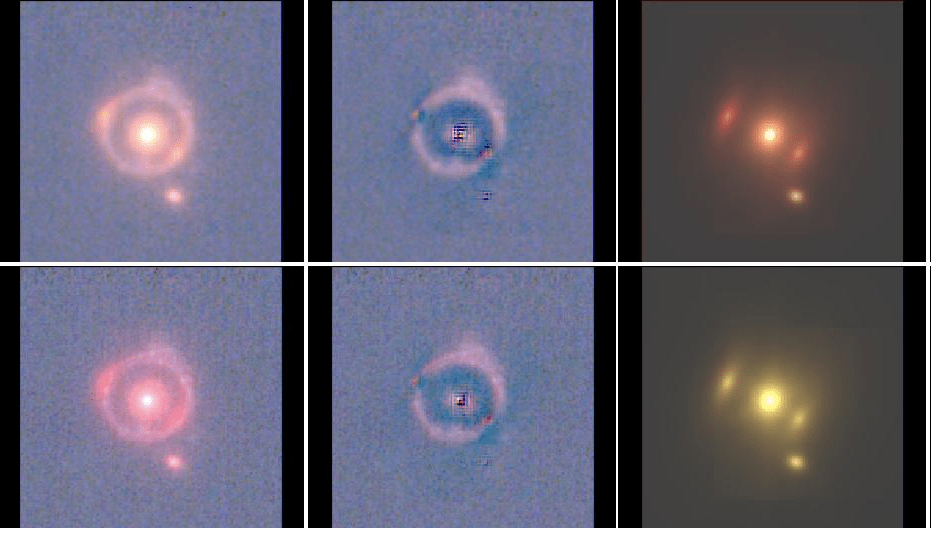}
\caption{
Example image shown to inspectors during the first round of visual inspection, featuring the confirmed strong lens COSJ100024+015334, also named the COSMOS-Web ring \citep{Mercier2024, VanDokkum2024, Shuntov2025}. The left column shows an RGB cutout centred on the candidate lens. The central column shows the same cutout after subtracting one or more Sérsic profiles using the COSMOS-Web photometry pipeline. The right column displays the model Sérsic light profiles used for this subtraction. The top and bottom rows use different RGB weightings to emphasise red and blue emission. Inspectors reviewed similar images for all candidates and classified them as either `Y' (Yes, this is a lens), `M' (Maybe, this is a lens), or `N' (No, this is not a lens).
}
\label{figure:Visual1}
\end{figure*}\label{visual_1}

The COSMOS-Web survey contains millions of galaxies, of which only the most massive are anticipated to act as strong gravitational lenses. The first round of visual inspection therefore isolated the $42\,660$ most massive galaxies in the survey, using luminosity cuts. This used a magnitude cut at $m_{\rm AB} = 23$ in the F277W filter and required a stellarity below 1, the latter removing stars which made up approximately $16\%$ of objects brighter than 23 mag. Based on SED fitting, a magnitude cut of 23 in the F277W band represents a median and \nth{68} percentile stellar mass of $(3.8 \pm 3.3) \times 10^8 \, M_{\odot}$, $(2.9 \pm 2.6) \times 10^9 \, M_{\odot}$, and  $(6 \pm 6) \times 10^9 \, M_{\odot}$ at $z = 0.3 \pm 0.05$, $z = 1.0 \pm 0.1$, and $z = 2.0 \pm 0.1$ respectively.

Visual inspection of the sample was conducted by multiple inspectors using images similar to those in \cref{figure:Visual1}. The first column shows a postage stamp cutout of each galaxy, where RGB colour scaling helps distinguish red and blue emission. The second column presents the same image after subtracting Sérsic light profiles fitted to the central galaxy and nearby sources as part of the COSMOS-Web catalogue \citep{Shuntov2024}. The third column displays the Sérsic model images used for this subtraction. The top and bottom rows apply different RGB scaling using different sets of \textit{JWST} and NIRCam filters, to highlight red and blue emission. In this example, the Sérsic photometry successfully removed the central lens galaxy and two of the lensed source’s multiple images. Inspectors were advised to account for such subtractions when assessing whether a candidate was a strong lens.

Inspectors evaluated all 42\,660 galaxies using this image format, selecting one of three options via a GUI: `Y – Yes, this is a lens', `M – Maybe this is a lens', or `N – No, this is not a lens'. An optional text box allowed for comments, and inspectors could revisit candidates later. Given the large number of galaxies, the GUI and ranking system enabled responses at a cadence of approximately one second per object. For brevity, we refer to these responses as `Yes', `Maybe', and `No' throughout the paper. Visual inspection was paired with the 21 FOVs described in \cref{Data}, with the January 2023 data was inspected from May 2023 to June 2023, the April 2023 data from August 2023 to September 2023, and the January 2024 data from April 2024 to May 2024.

\subsection{Spectacular lenses}\label{visual:obvious}

After the first round of visual inspection, 32 objects were marked as 'Yes' by over 50\% of inspectors. These objects are analysed in M25, where inspectors re-evaluated the cut-outs in \cref{figure:Visual1} to identify those considered 'spectacular' lenses. These objects are included in our lens modelling analysis but are not forwarded to the second round of inspection and are labelled 'M25' throughout this work. The remaining 15 objects, not marked as spectacular in M25, are still potential lenses and proceed to the second round of inspection.

\subsection{Second Round}\label{visual:second_round}


Objects advancing to the second round of visual inspection were selected based on two criteria: (i) those where at least 50\% of inspectors responded `Yes' or `Maybe', and (ii) `edge case' objects that fell below this threshold but received at least one `Yes' or `Maybe', where JWN reviewed all qualifying objects and advanced any deemed strong candidates for further inspection. 


All candidates meeting either selection criterion were fitted with an MGE-only foreground subtraction across all wavebands (\cref{LensLight}) and had independent lens models fitted in each waveband (\cref{LensPipeline}). The resulting images were provided to inspectors in the second round of visual inspection, viewed in the following sequence:

\begin{enumerate}

    \item The initial cut-out image from the first round of visual inspection (see \cref{figure:Visual1}).
    \item Plots of the candidate lens and source in all four wavebands, displayed with linear and base-10 logarithmic colour scales, as well as RGB images. 
    \item The MGE-subtracted image (see \cref{figure:VisualMGELensSub}), where different colour scales highlight the lensed source emission in each waveband.
    \item The \texttt{PyAutoLens} model-fit to all four wavebands (see \cref{figure:VisualLensFit}), including model images of the lens galaxy, lensed source, source-plane reconstruction, and mass model convergence. Separate figures show fits using both SIE and MGE mass models.
\end{enumerate}

Additional images were available in a separate folder, allowing for detailed inspections of specific candidates. These included images where the image-source plane mappings were visualised using coloured polygons, to assess model consistency. Inspectors did not often use these extra images during the review process, with them only required for detailed inspection of select candidates.

Inspectors ranked each lens into one of the following five categories:

\begin{itemize}
    \item \textbf{A}: High confidence this is a strong lens.
    \item \textbf{B}: Likely a strong lens, but there is ambiguity.
    \item \textbf{U}: Unlikely to be a strong lens, but not impossible.
    \item \textbf{S}: A singly imaged strong lens feature/arc (e.g. without an observed counter image).
    \item \textbf{X}: Not a lens. 
    \item \textbf{I (Optional and independent)}: The candidate is interesting and warrants follow up investigation. 
\end{itemize}

Each candidate is assigned a final score based on their ranking, with 2 points for an A, 1 point for a B, and 1 point for an S. For example, a candidate ranked `AAABSX' would score 8 points. All candidates in the second round of visual inspection were assessed by 6 inspectors, giving them a total score out of 12. This score is displayed alongside all candidate visualisations and listings in tables. Plots are colour-coded based on the scores: black for the M25 sample, dark green for scores above 7, yellow for scores of 5 or 6, orange for scores of 3 or 4, and maroon for scores below 3. Some plots are ordered by descending score. The second round of visual inspection took place from August 2nd to August 30th, 2024.



\section{Results}\label{Results}

\subsection{Visual Inspection}

\subsubsection{First Round}

\begin{table*}
\begin{adjustbox}{max width=\textwidth}
\normalsize
\begin{tabular}{ l | l | l | l | l | l | l | l} 
\multicolumn{1}{p{1.2cm}|}{Dataset Group} 
& \multicolumn{1}{p{1.1cm}|}{Total Inspectors} 
& \multicolumn{1}{p{1.1cm}|}{Total Objects} 
& \multicolumn{1}{p{2.00cm}|}{M25 Spectacular Lenses} 
& \multicolumn{1}{p{2.00cm}|}{At Least $50\%$ Yes (Excluding M25)} 
& \multicolumn{1}{p{2.00cm}|}{At Least $50\%$ Maybe (Excluding M25)}  
& \multicolumn{1}{p{2.00cm}|}{Over $1$ Yes or Maybe and Less Than $50\%$ Yes or Maybe} 
& \multicolumn{1}{p{2.00cm}}{Edge Cases}   \\  \hline
& & & & & & \\[-6pt]
Jan 2023 & 4 & 847 & 0 (0.0\%) & 0 (0.0\%) & 8 (0.94\%) & 101 (11.92\%) & 10 (1.18\%)\\[2pt]
April 2023 & 4 & 14688 & 8 (0.05\%) & 6 (0.05\%) & 152 (1.03\%) & 1120 (7.62\%) & 48 (0.33\%)\\[2pt]
Jan 2024 & 5 & 27125 & 9 (0.03\%) & 9 (0.03\%) & 105 (0.39\%) & 2476 (9.13\%) & 96 (0.35\%) \\[2pt]
\hline
& & & & & & \\[-6pt]
Total & 4 or 5 & 42660 & 17 (0.04\%) & 15 (0.04\%) & \textbf{265} (0.62\%) & 3697 (11.53\%) & \textbf{154} (0.40\%) \\[2pt]
\end{tabular}
\end{adjustbox}
\caption{
The results of the first round of visual inspection. The first column lists the time of acquisition of observations the results correspond to, with the bottom providing total values for each group. The second column shows the number of inspectors for that dataset group. The third column shows how many objects are in this dataset group because they met the luminosity selection criteria described in \cref{visual_1}. The fourth column lists the number of objects which are in the M25 spectacular lens sample. Column 5 shows how many objects at least 50\% of visual inspections indicated `Y - Yes, this is a lens' which were not included in M25. The sixth column lists the number of objects for which at least 50\% of visual inspections indicated `Y - Yes, this is a lens' or `M - Maybe this is a lens' (including the $15$ objects in column 5), meaning they were forwarded to the second round of visual inspection. The seventh column lists how many objects at least one inspector identified as a potential lens but less than $50\%$ of inspectors overall did, meaning that JWN reinspected it to determine if it is an edge case that should be forwarded to the second round of visual inspection. The final column lists how many of these objects JWN forwarded as edge cases. Bottom row values in bold are the candidate numbers which made it through to the second round of visual inspection. The percentages in brackets for each row are relative to the total number of objects in that dataset group, (column 3).
}
\label{table:visual_1}
\end{table*}

The results of the first round of visual inspection are provided in \cref{table:visual_1}. In total, $42\,660$ images were inspected by 4 or 5 inspectors. 32 objects had at least $50\%$ of inspectors rank them in the `Yes' category, which were inspected further in M25. Of these 32 objects, M25 presented 17 objects as spectacular lenses, $0.04$\% of the total number of galaxies inspected. A total of $419$ objects made it through to the second round of visual inspection (excluding the 17 M25 lenses). $265$ candidates met the criteria that at least $50\%$ of inspectors ranked them as a `Yes' or `Maybe`. There are $3,697$ objects that at least one inspector said `Yes' or `Maybe` but not $50\%$ of inspectors. JWN inspected all $3,697$ objects and decided to put forward $154$ as edge cases. In \cref{AppVisual} we discuss the first round of visual inspection results in more detail.

\subsubsection{Second Round}

\begin{table}
\begin{adjustbox}{max width=\textwidth}
\begin{tabular}{ l | l | l | l | l | l} 
\multicolumn{1}{p{0.9cm}|}{Score} 
& \multicolumn{1}{p{1.0cm}|}{Total Candidates} 
& \multicolumn{1}{p{1.4cm}|}{Cumulative} 
& \multicolumn{1}{p{1.0cm}|}{Percent} 
& \multicolumn{1}{p{1.2cm}|}{Round 1 Candidate}
& \multicolumn{1}{p{0.7cm}}{Round 1 Edge} 
\\ \hline
& & & & \\[-6pt]
S12 & 2 & 2 & 0.48\% & 1 & 1 \\[2pt]
S11 & 3 & 5 & 0.72\% & 3 & 0 \\[2pt]
S10 & 12 & 17 & 2.86\% & 9 & 3 \\[2pt]
S09 & 11 & 28 & 2.63\% & 9 & 2 \\[2pt]
S08 & 14 & 42 & 3.34\% & 9 & 5 \\[2pt]
S07 & 20 & 62 & 4.77\% & 11 & 9 \\[2pt]
S06 & 37 & 99 & 8.83\% & 19 & 18 \\[2pt]
S05 & 44 & 143 & 10.5\% & 27 & 17 \\[2pt]
S04 & 58 & 201 & 13.84\% & 34 & 24 \\[2pt]
S03 & 62 & 263 & 14.8\% & 36 & 26 \\[2pt]
S02 & 66 & 329 & 15.75\% & 45 & 21 \\[2pt]
S01 & 46 & 375 & 10.98\% & 32 & 14 \\[2pt]
S00 & 44 & 419 & 10.5\% & 30 & 14 \\[2pt]
\end{tabular}
\end{adjustbox}
\caption{
Candidate scores for the second round of visual inspection. The first column lists the score out of 12, as described in \cref{visual:second_round}. The second, third, and fourth columns show the total number of objects for each score, the cumulative sum of candidates with scores equal to or above that score, and the percentage of the total 419 second-round candidates corresponding to each score. The fifth column indicates how many candidates reached the second round as a result of the first round of visual inspection, meaning at least 50\% of inspectors flagged them as `Y - Yes, this is a lens' or `M - Maybe this is a lens'. The sixth column shows the number of edge cases, where less than 50\% of inspectors flagged the object but at least one identified it as a potential lens. These edge cases were reinspected by JWN and forwarded for further evaluation.
}
\label{table:visual_2}
\end{table}

Figure \ref{table:visual_2} presents the scores for the 419 candidates inspected in the second round of visual inspection. Two candidates received the highest possible score of 12, indicating unanimous agreement from all inspectors that they are strong lenses. The third column shows the cumulative total of candidates with scores above certain thresholds, for example, 17 candidates scored 10 or higher, 62 scored 7 or higher, and 143 scored 5 or higher. The final two columns of \cref{table:visual_2} indicate whether candidates came from the first round of visual inspection as a candidate or an edge case. Edge cases are common; for example, of the 143 candidates scoring 5 or higher, 55 (38.4\%) were edge cases, meaning only one inspector initially flagged them as a lens or a possible lens in the first round of visual inspection. A significant portion of high-scoring candidates would have been missed if edge cases had not been carried forward. The high scores of these edge cases suggest that lens modelling provided additional information, allowing inspectors to rank them higher, as discussed in \cref{VisualLensModel}.

\subsection{Candidate Images}

\begin{figure*}
\centering
\includegraphics[width=0.12\textwidth]{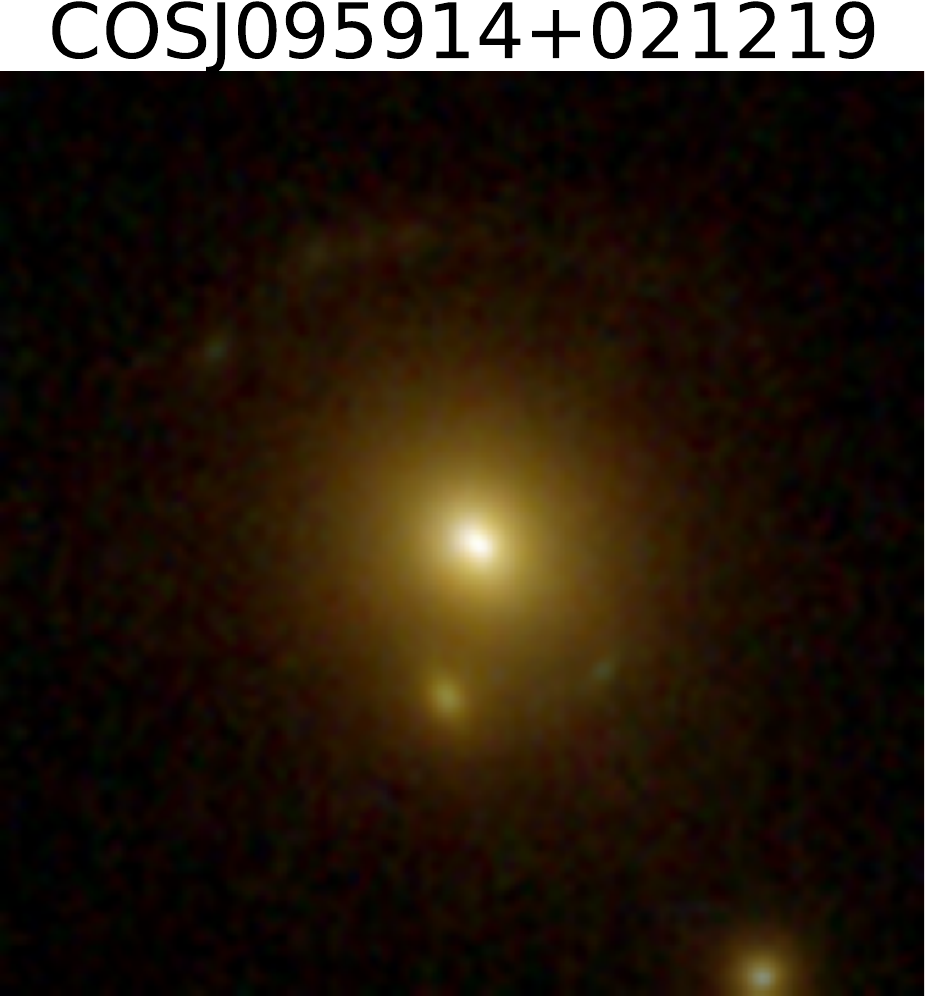}
\includegraphics[width=0.12\textwidth]{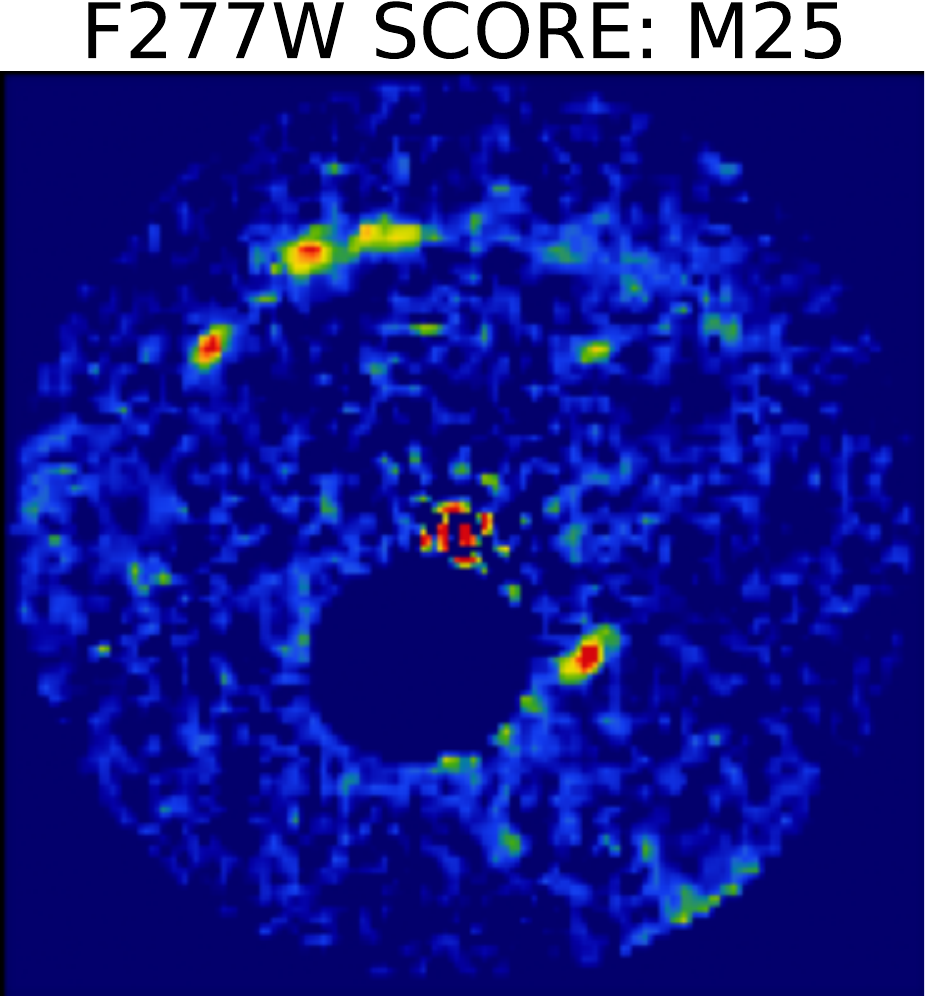}
\includegraphics[width=0.12\textwidth]{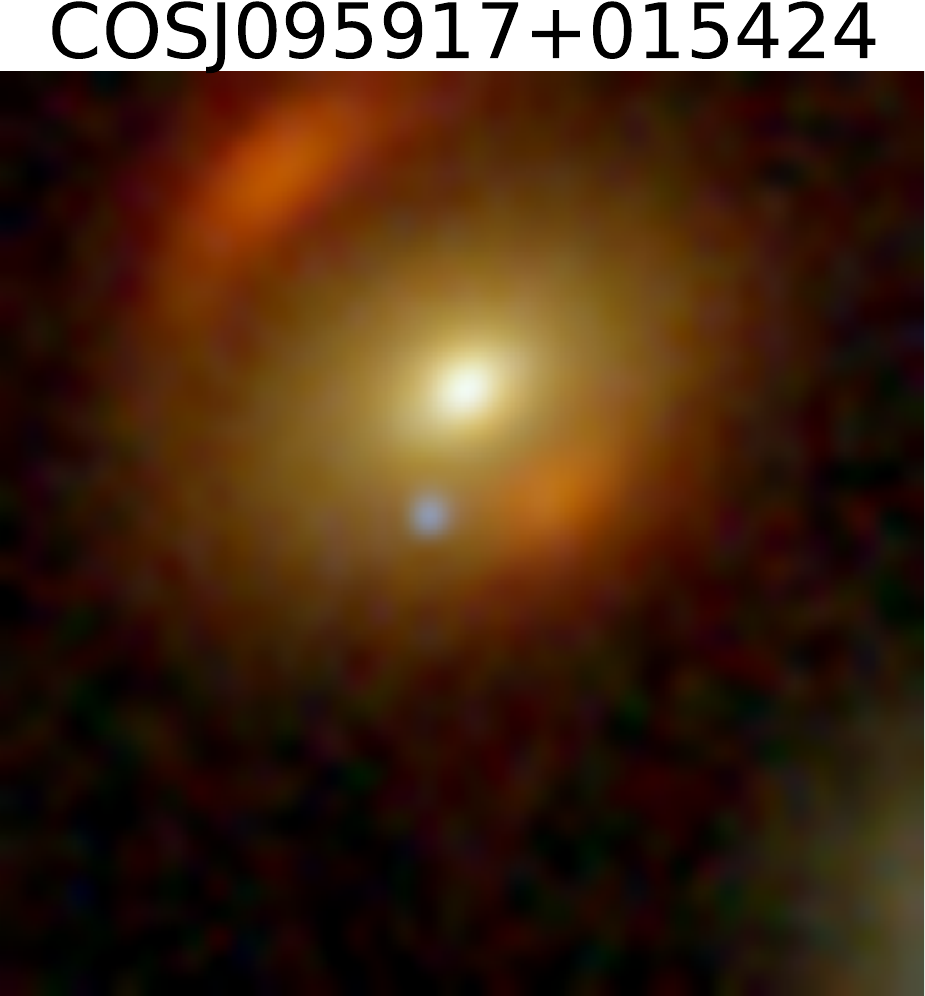}
\includegraphics[width=0.12\textwidth]{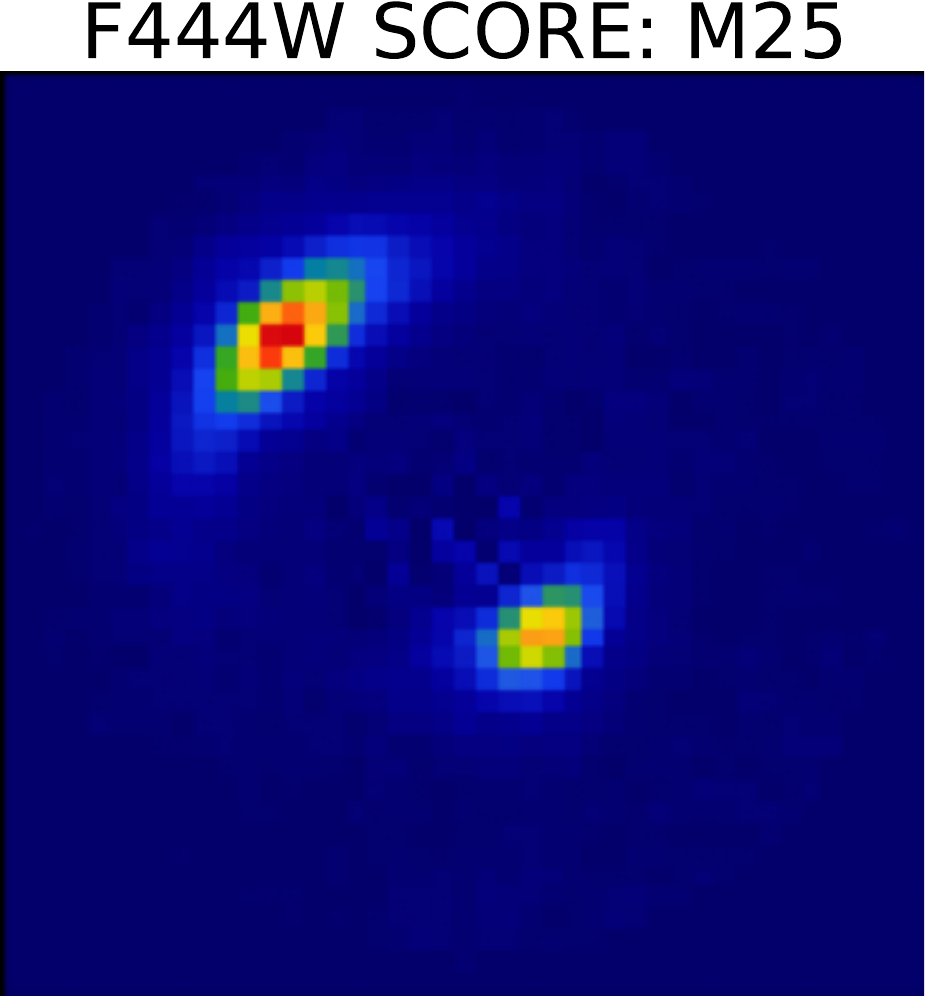}
\includegraphics[width=0.12\textwidth]{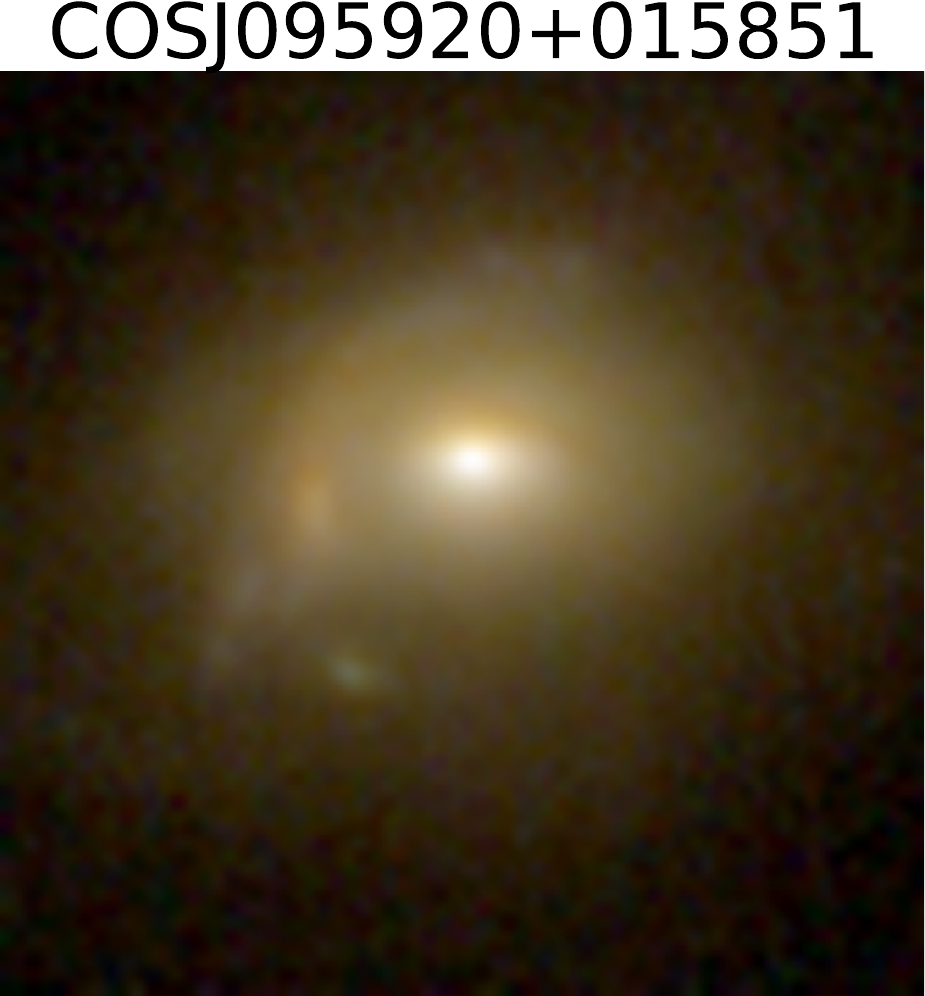}
\includegraphics[width=0.12\textwidth]{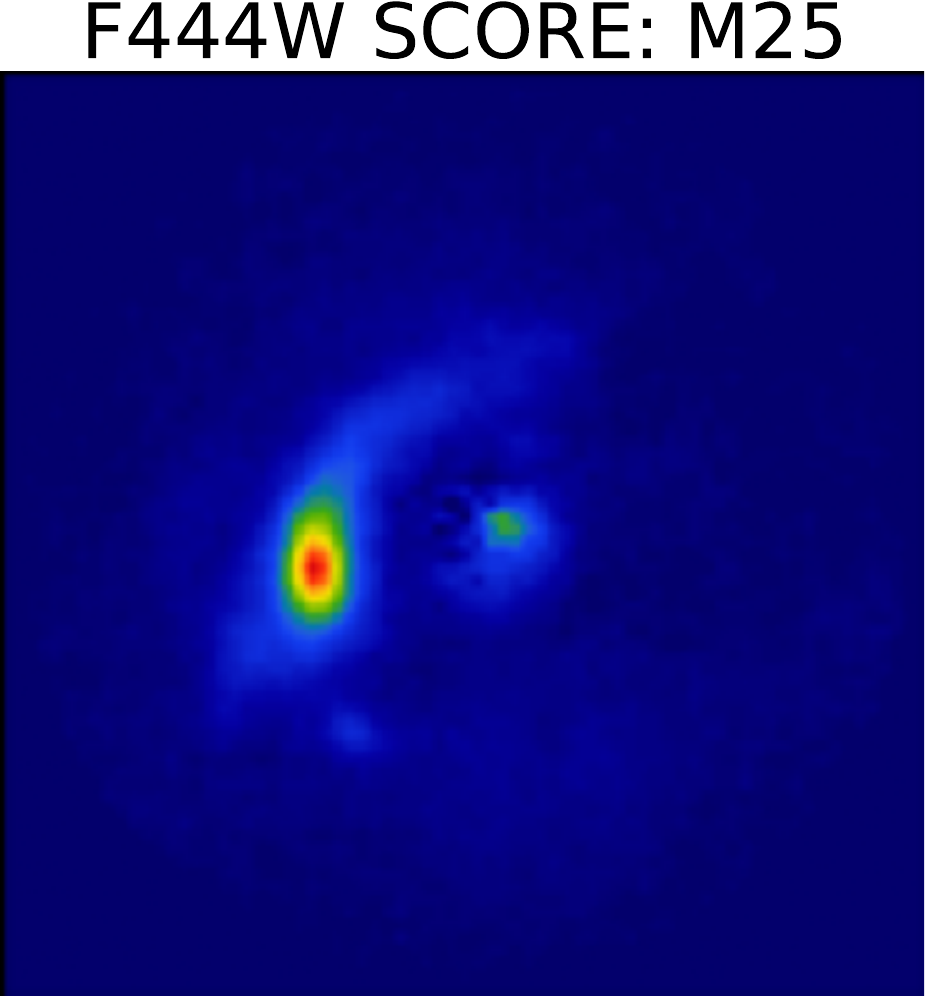}
\includegraphics[width=0.12\textwidth]{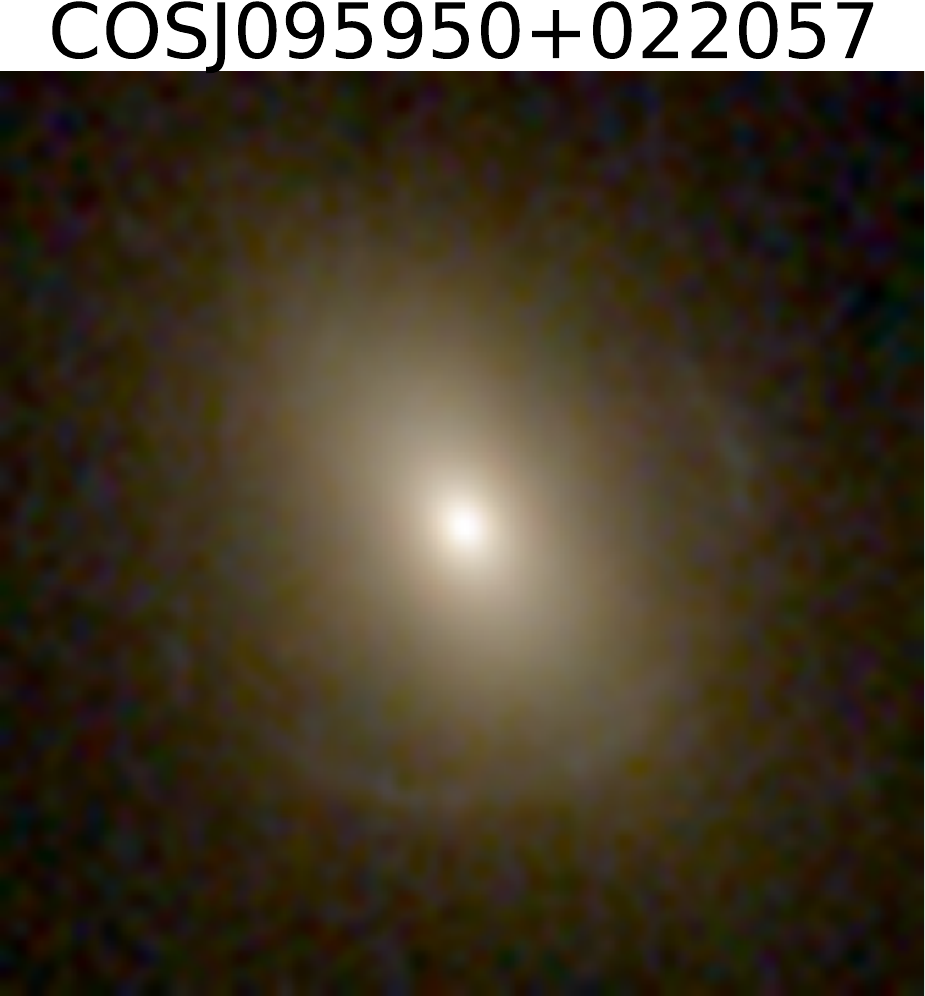}
\includegraphics[width=0.12\textwidth]{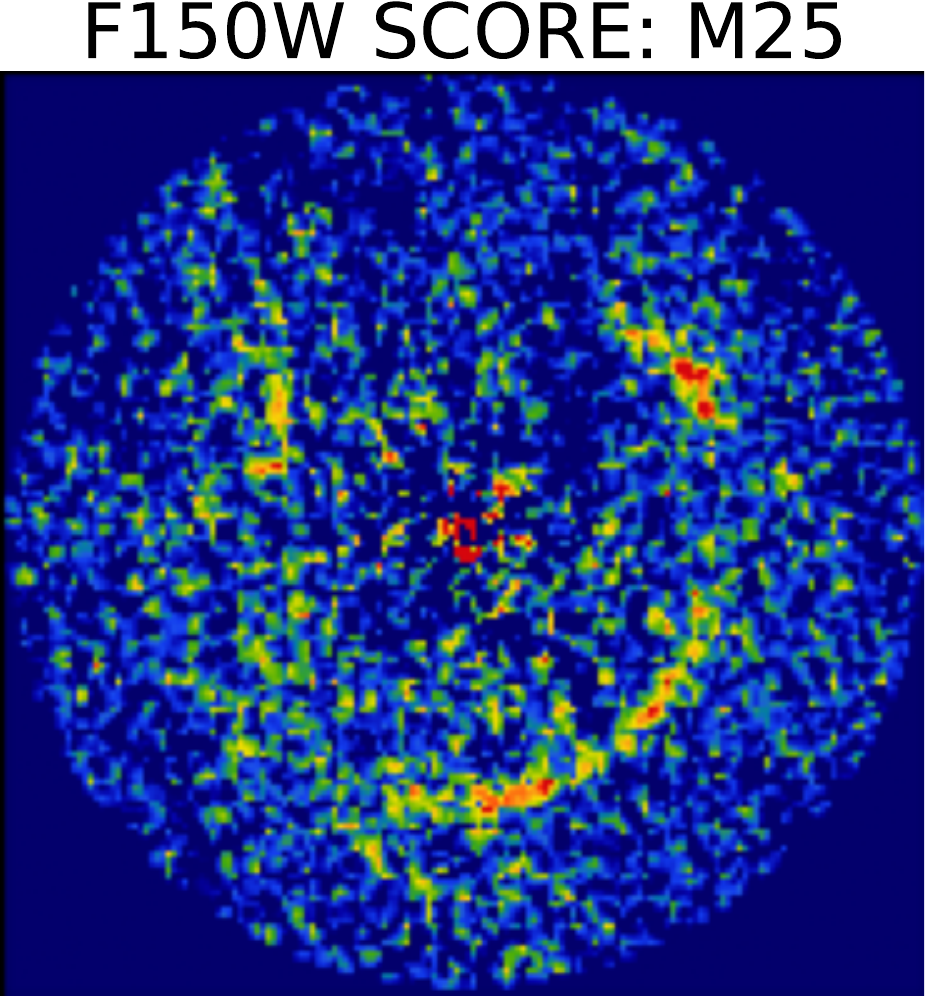}
\includegraphics[width=0.12\textwidth]{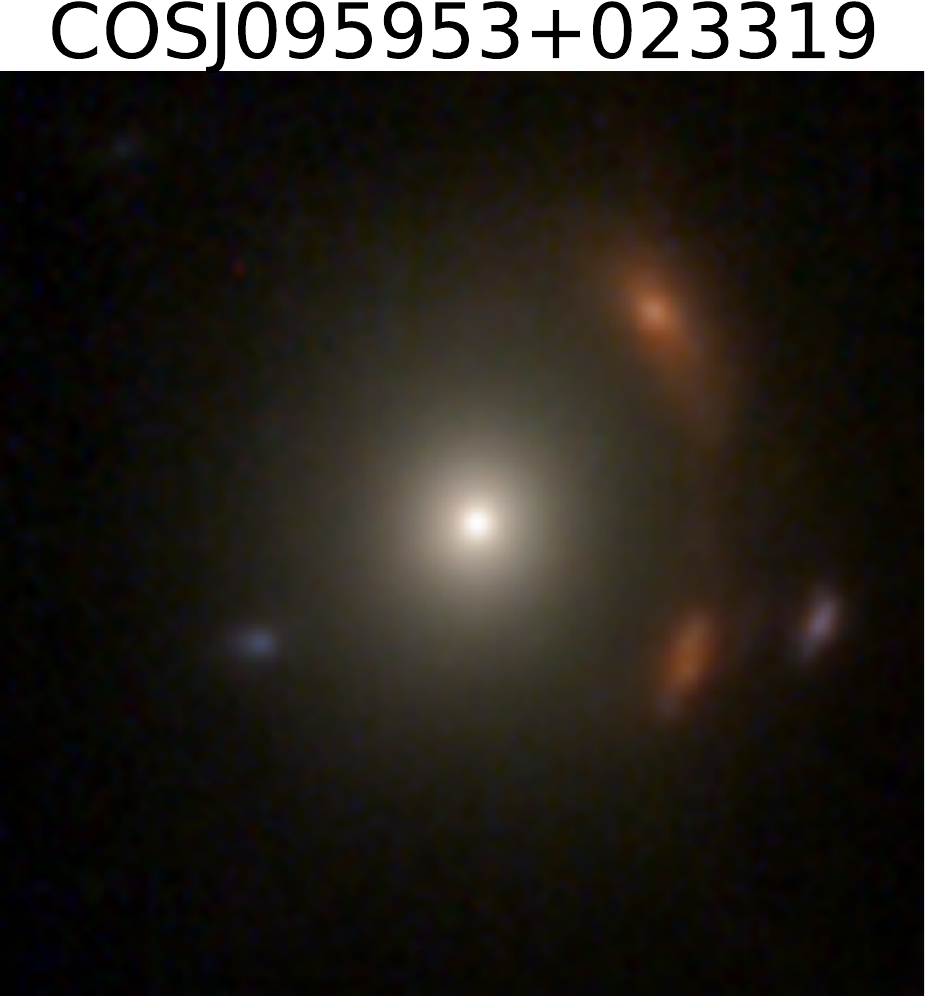}
\includegraphics[width=0.12\textwidth]{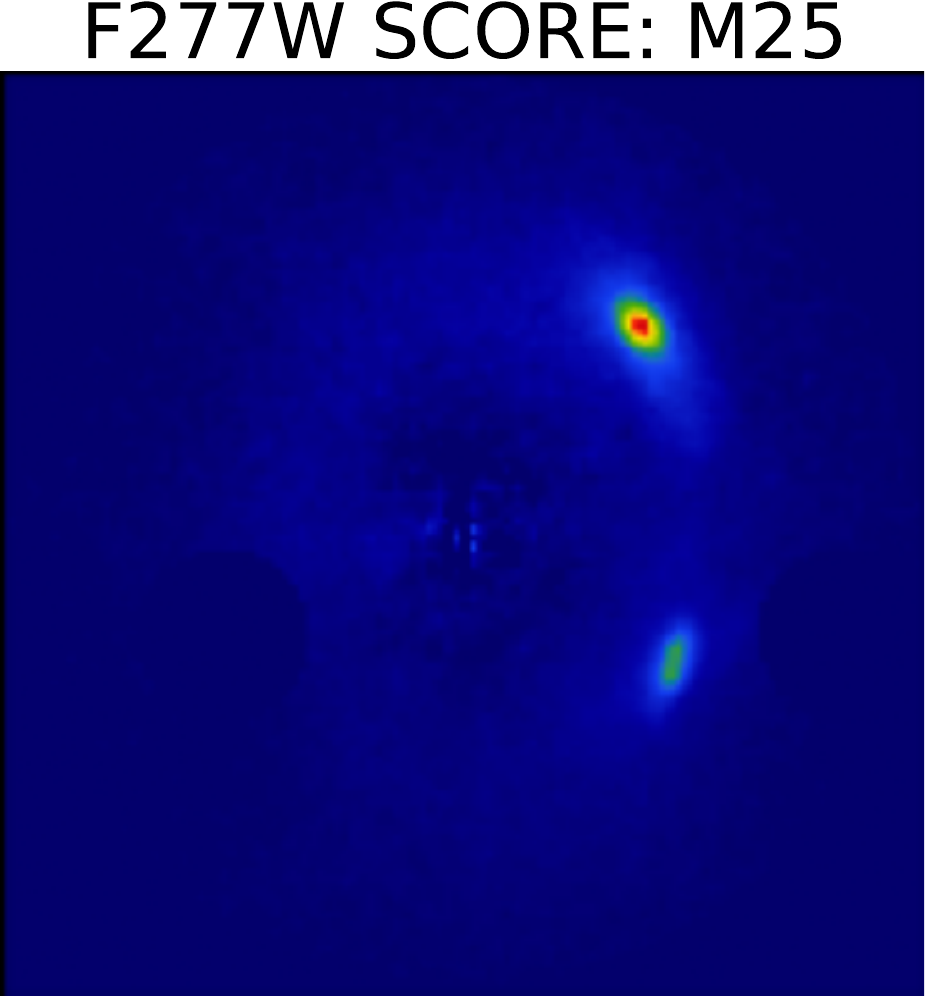}
\includegraphics[width=0.12\textwidth]{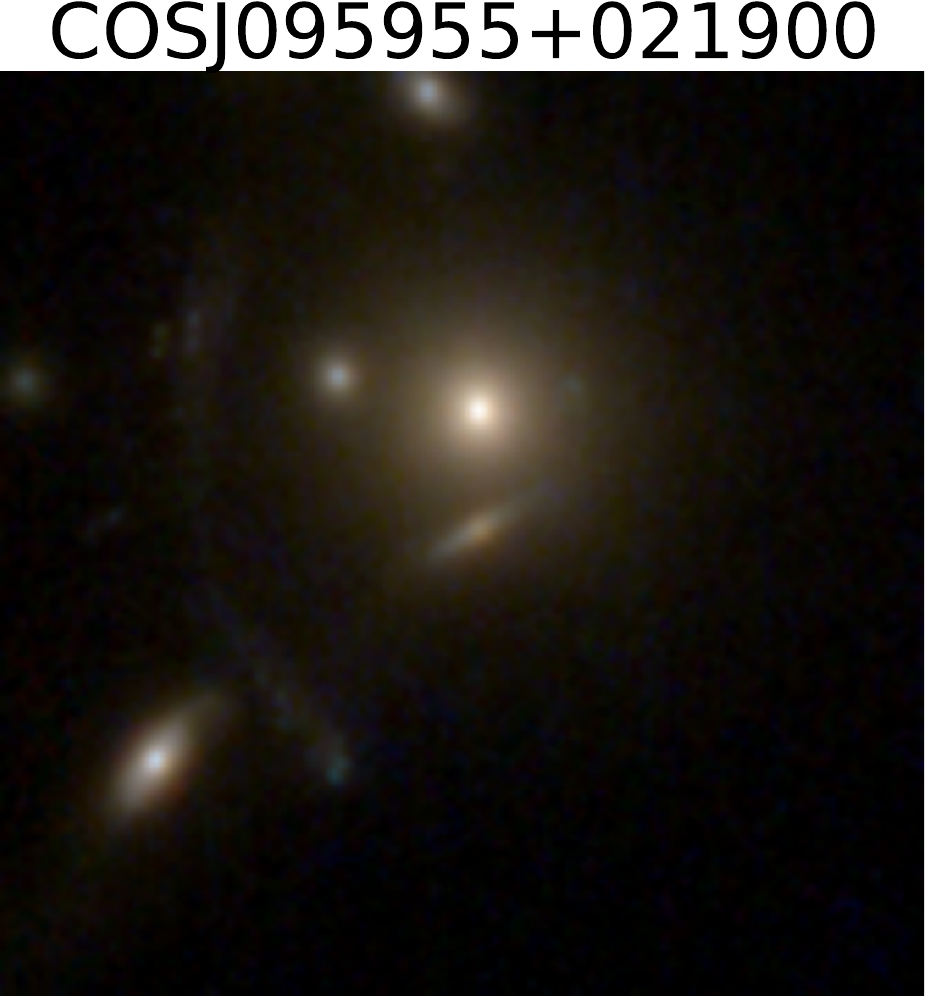}
\includegraphics[width=0.12\textwidth]{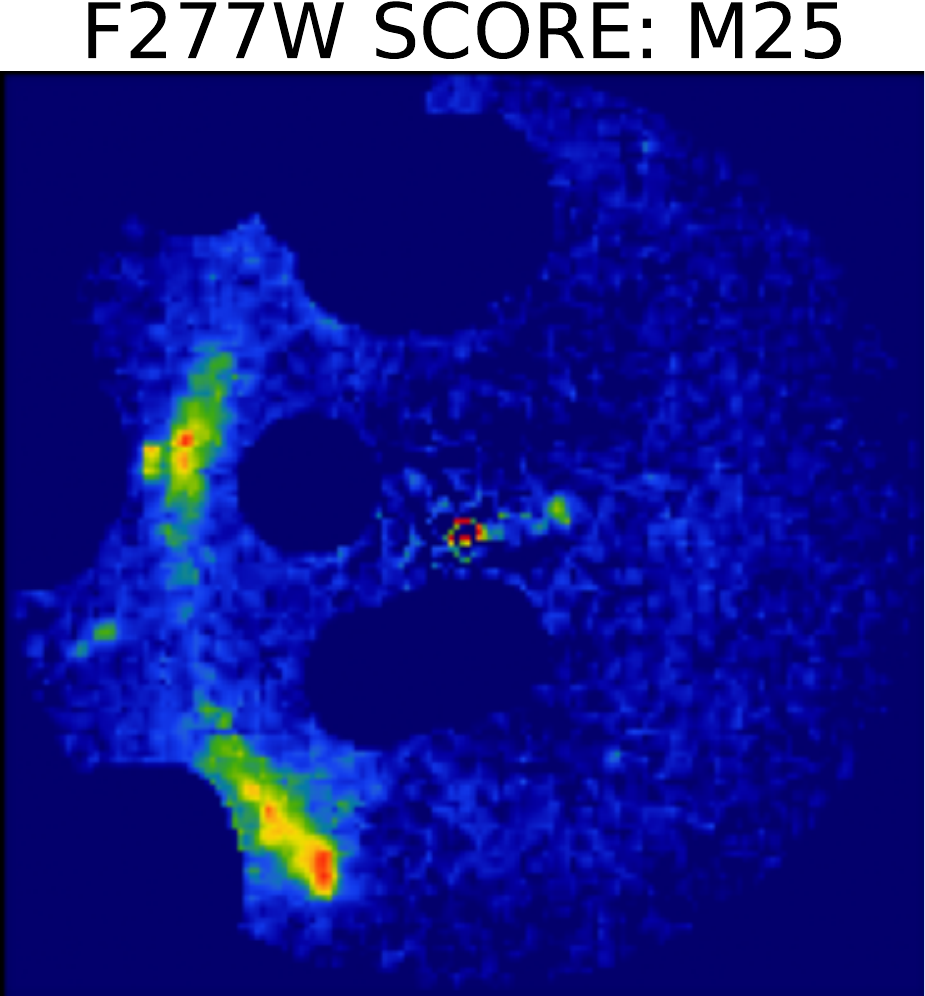}
\includegraphics[width=0.12\textwidth]{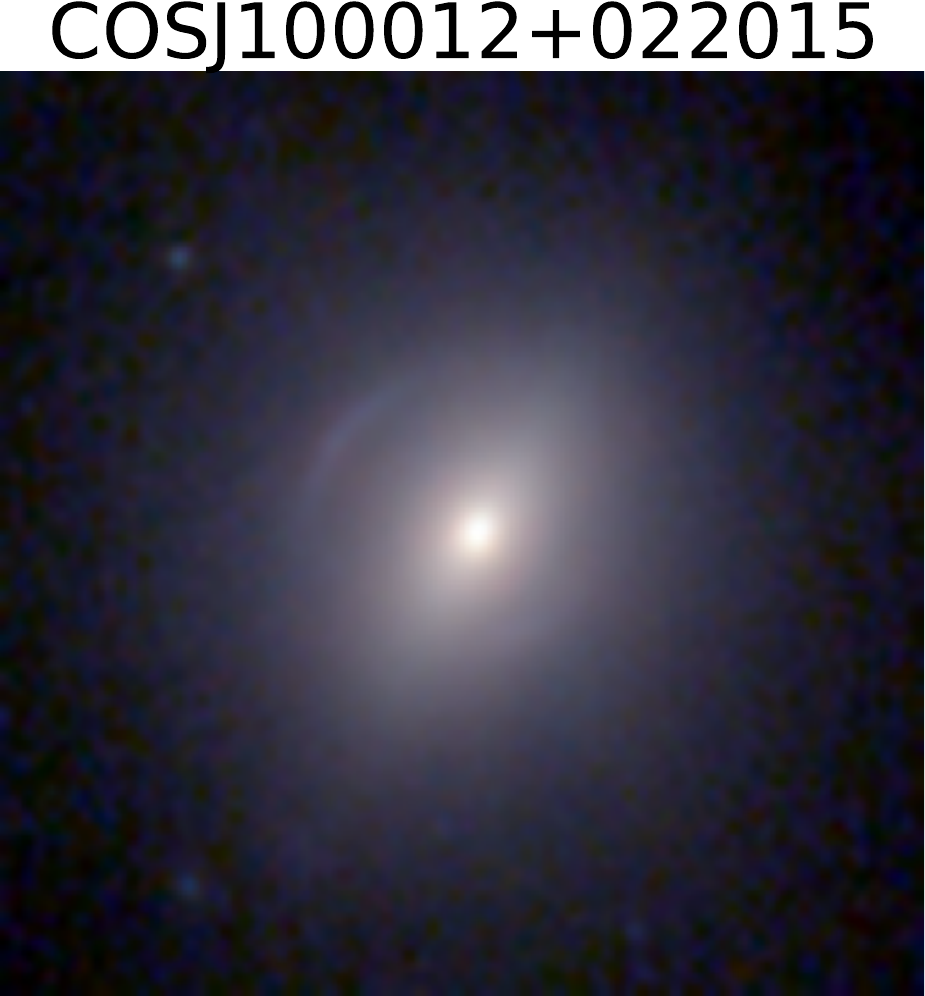}
\includegraphics[width=0.12\textwidth]{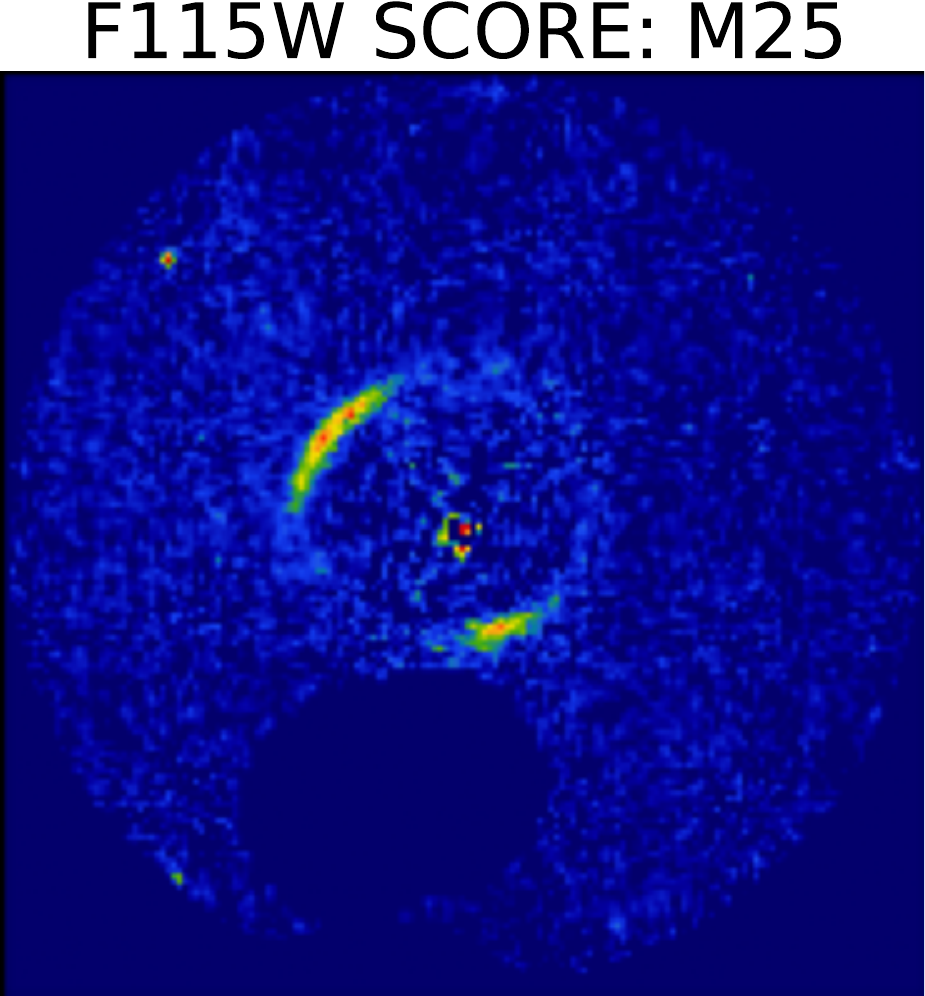}
\includegraphics[width=0.12\textwidth]{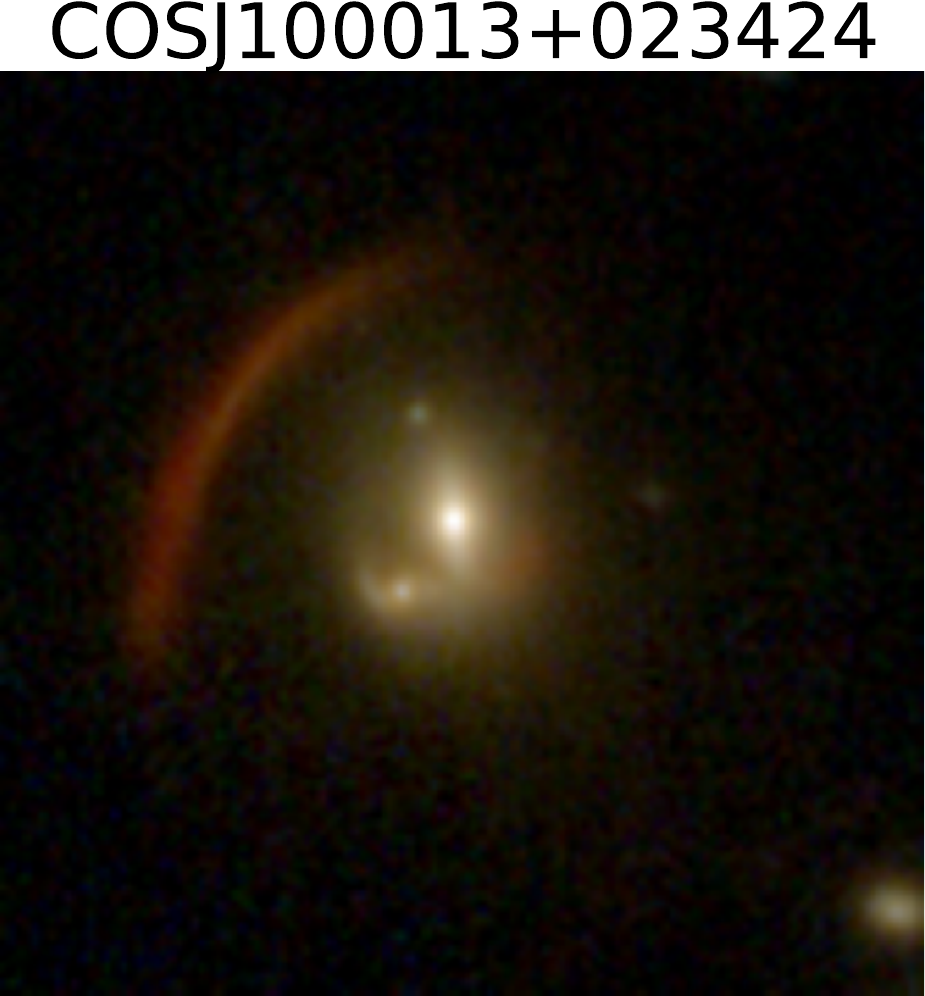}
\includegraphics[width=0.12\textwidth]{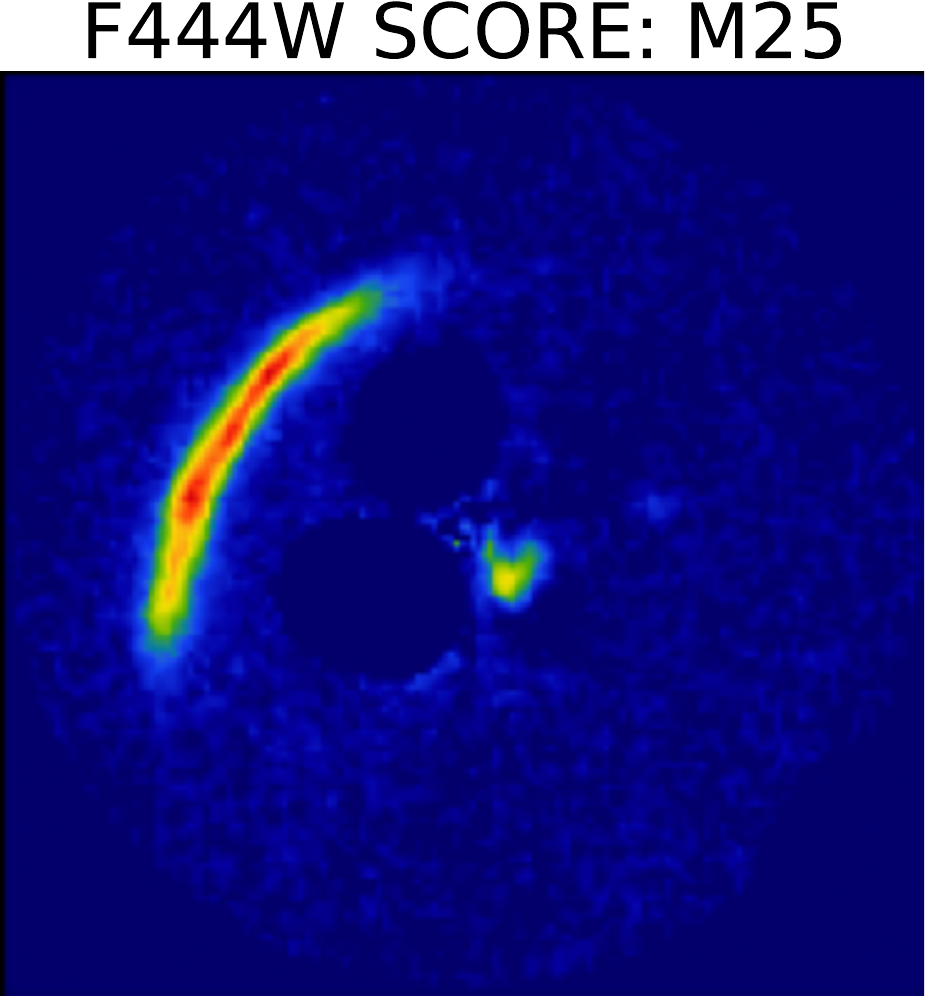}
\includegraphics[width=0.12\textwidth]{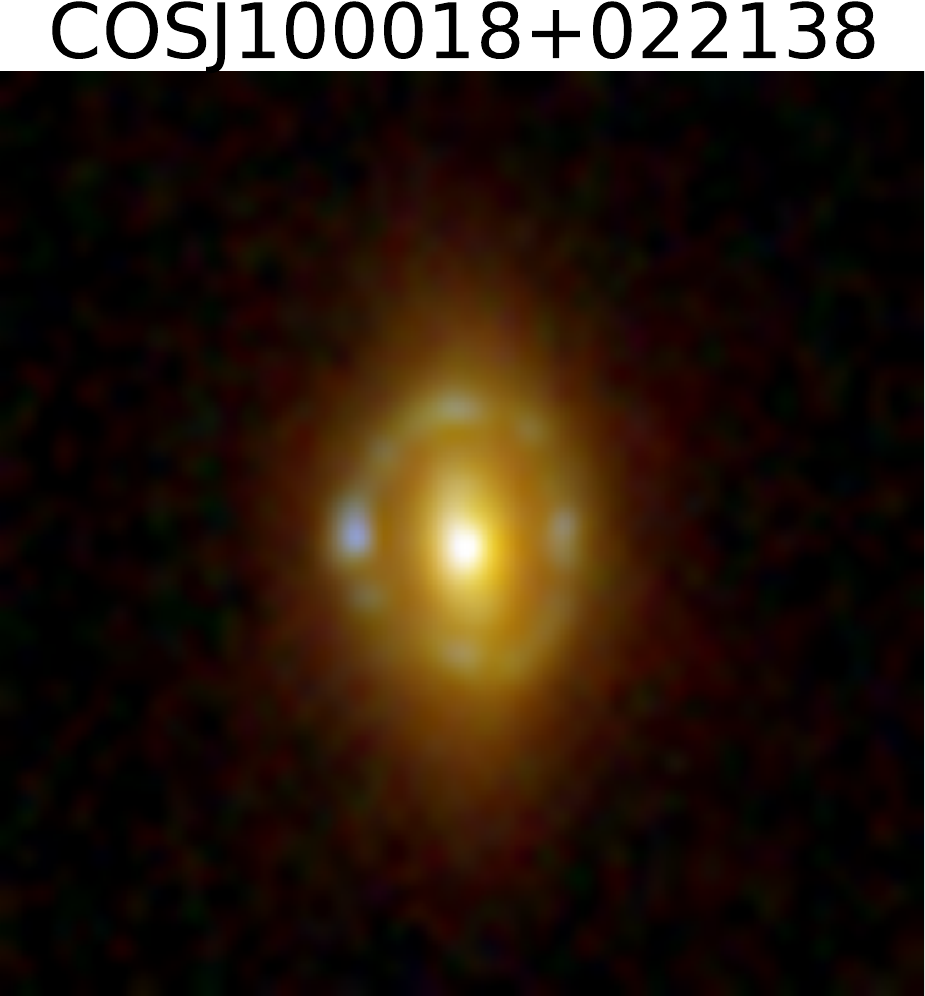}
\includegraphics[width=0.12\textwidth]{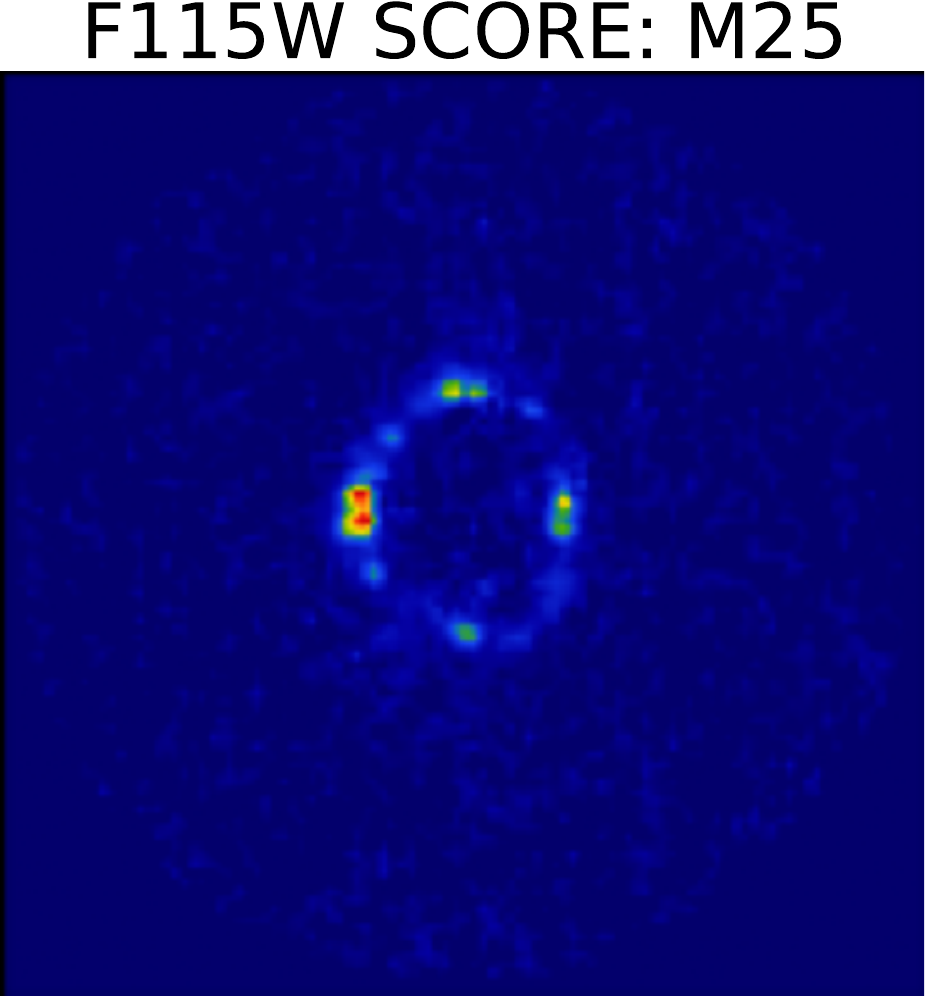}
\includegraphics[width=0.12\textwidth]{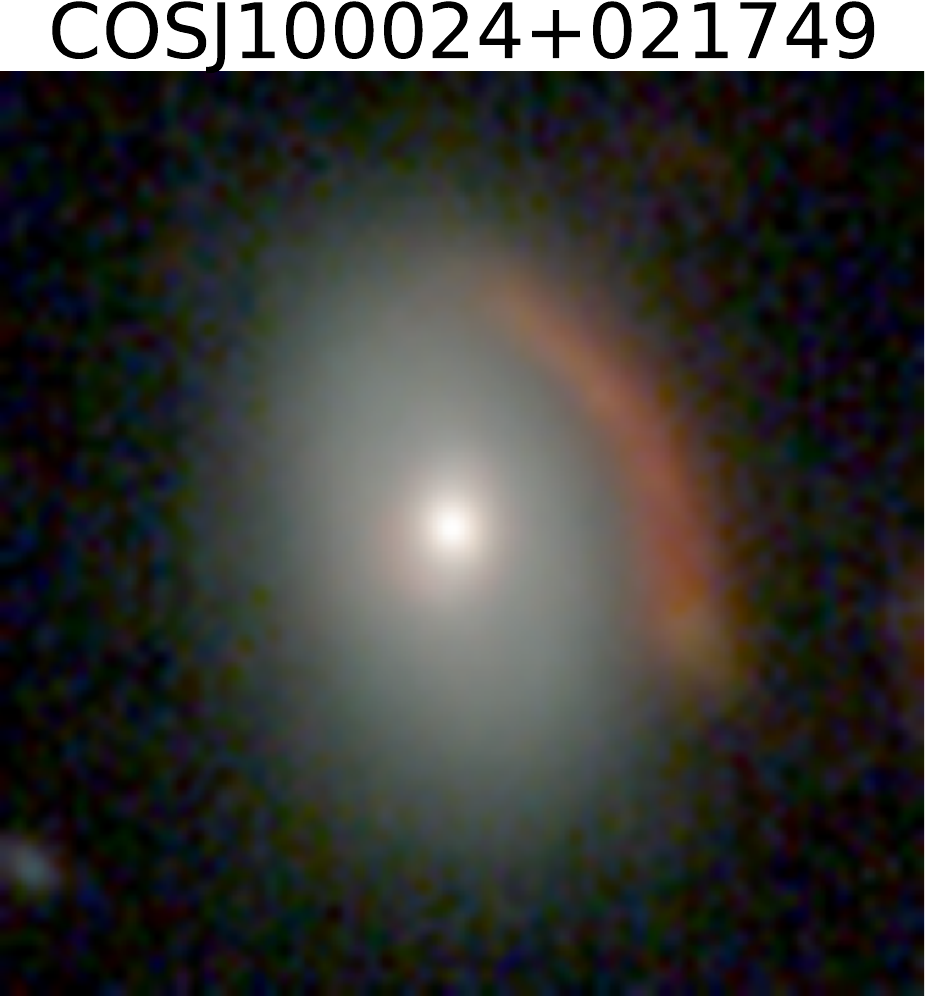}
\includegraphics[width=0.12\textwidth]{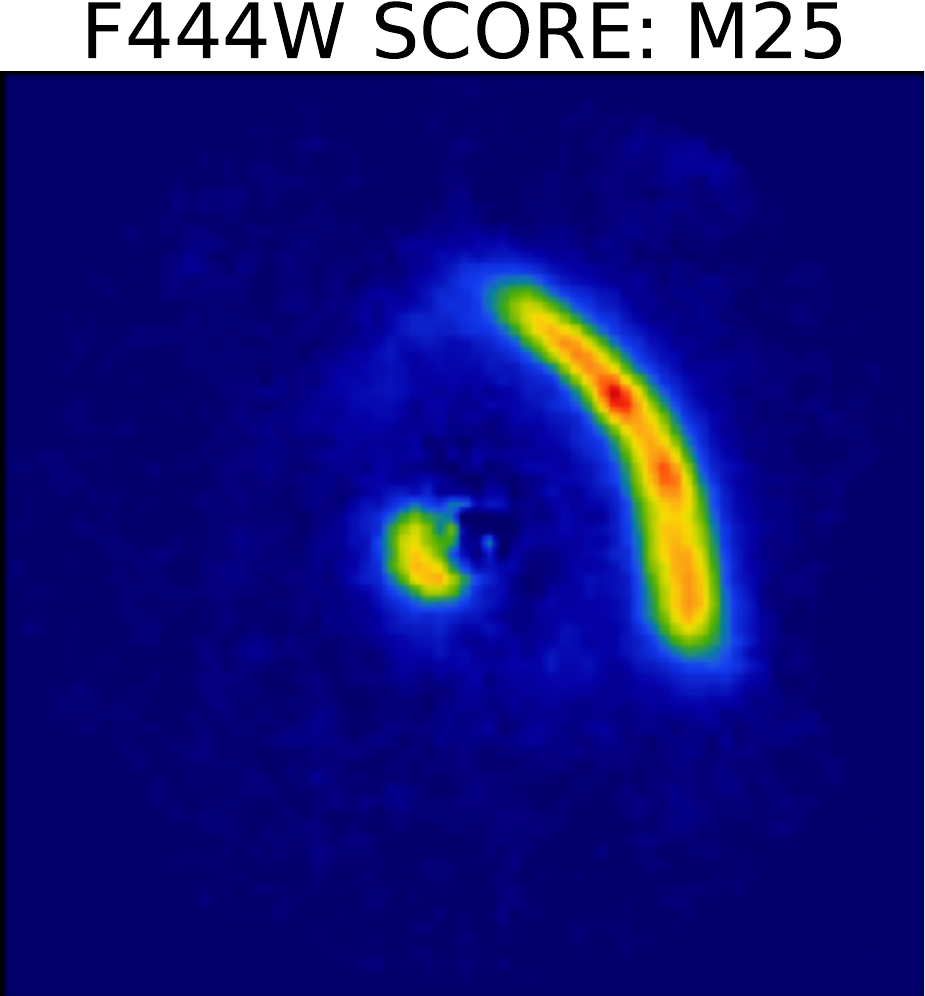}
\includegraphics[width=0.12\textwidth]{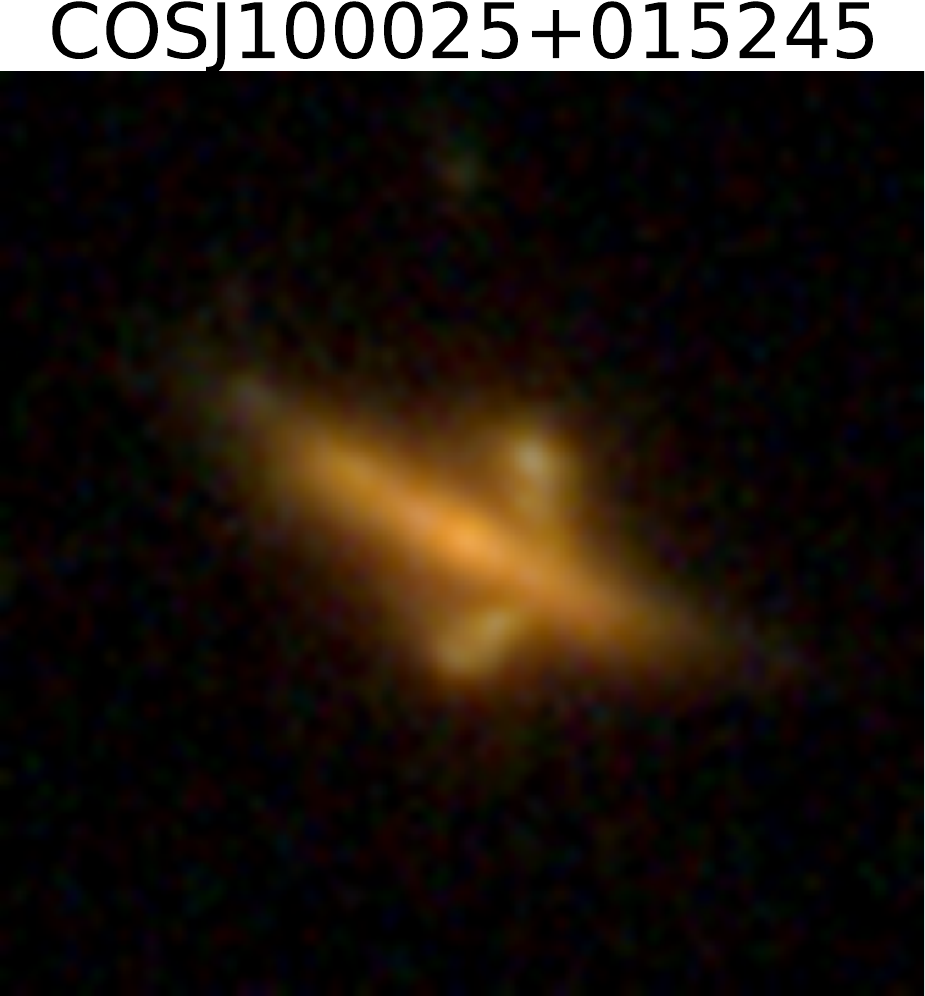}
\includegraphics[width=0.12\textwidth]{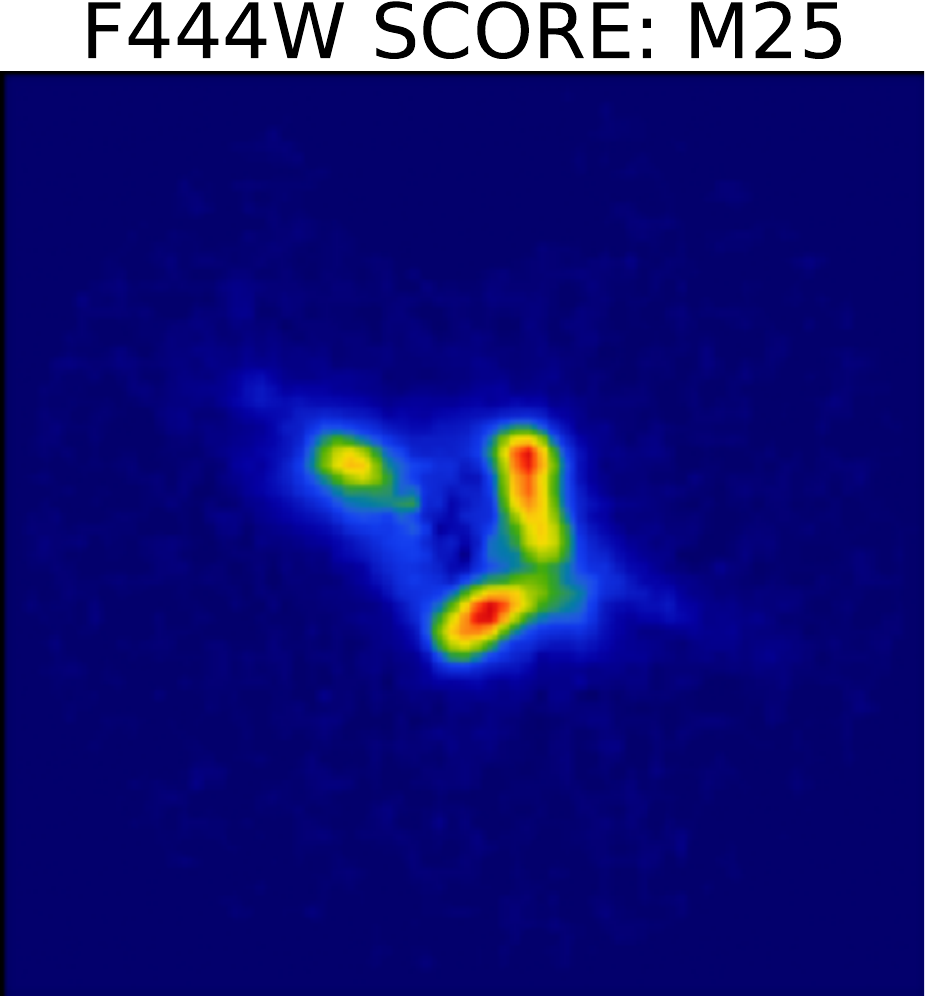}
\includegraphics[width=0.12\textwidth]{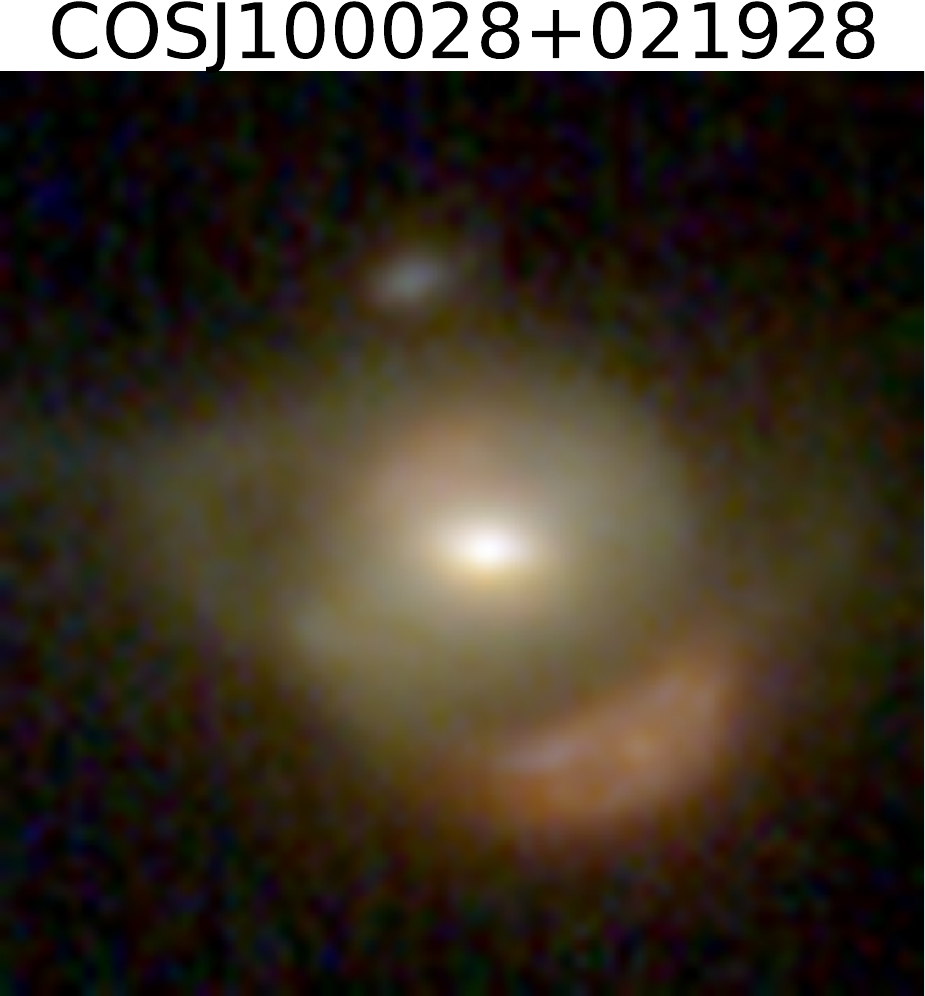}
\includegraphics[width=0.12\textwidth]{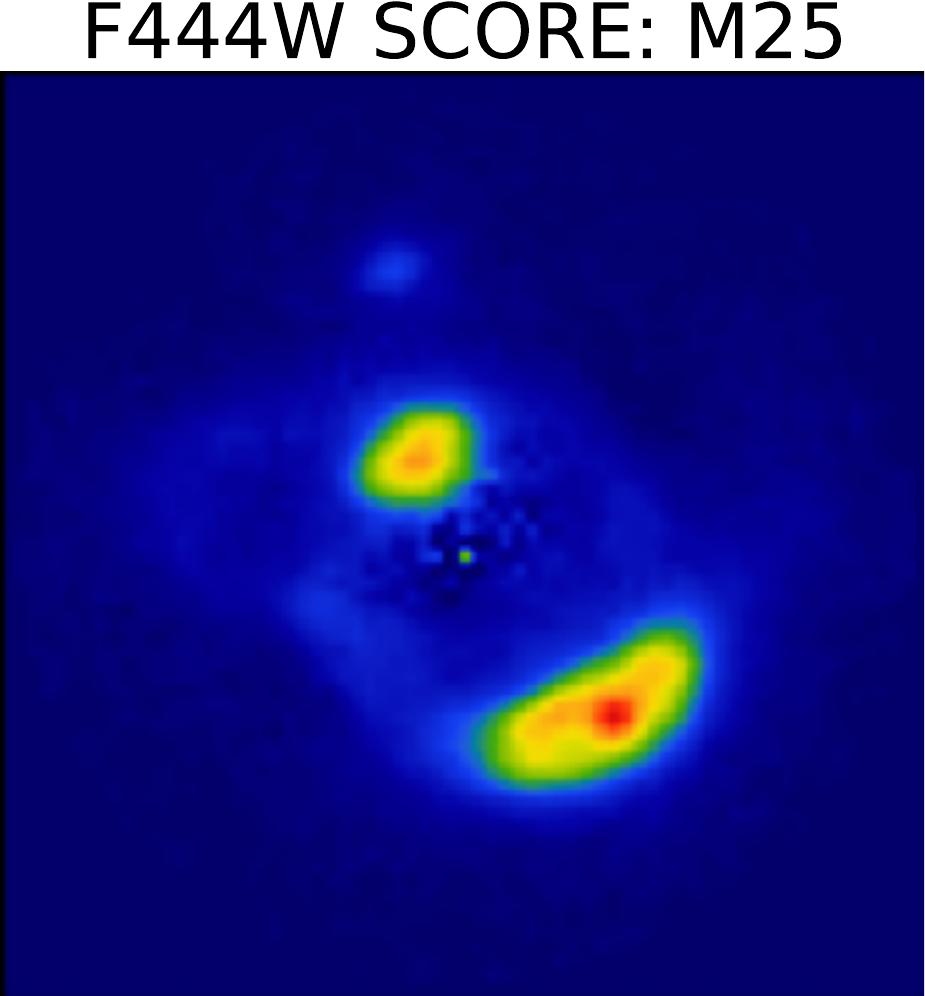}
\includegraphics[width=0.12\textwidth]{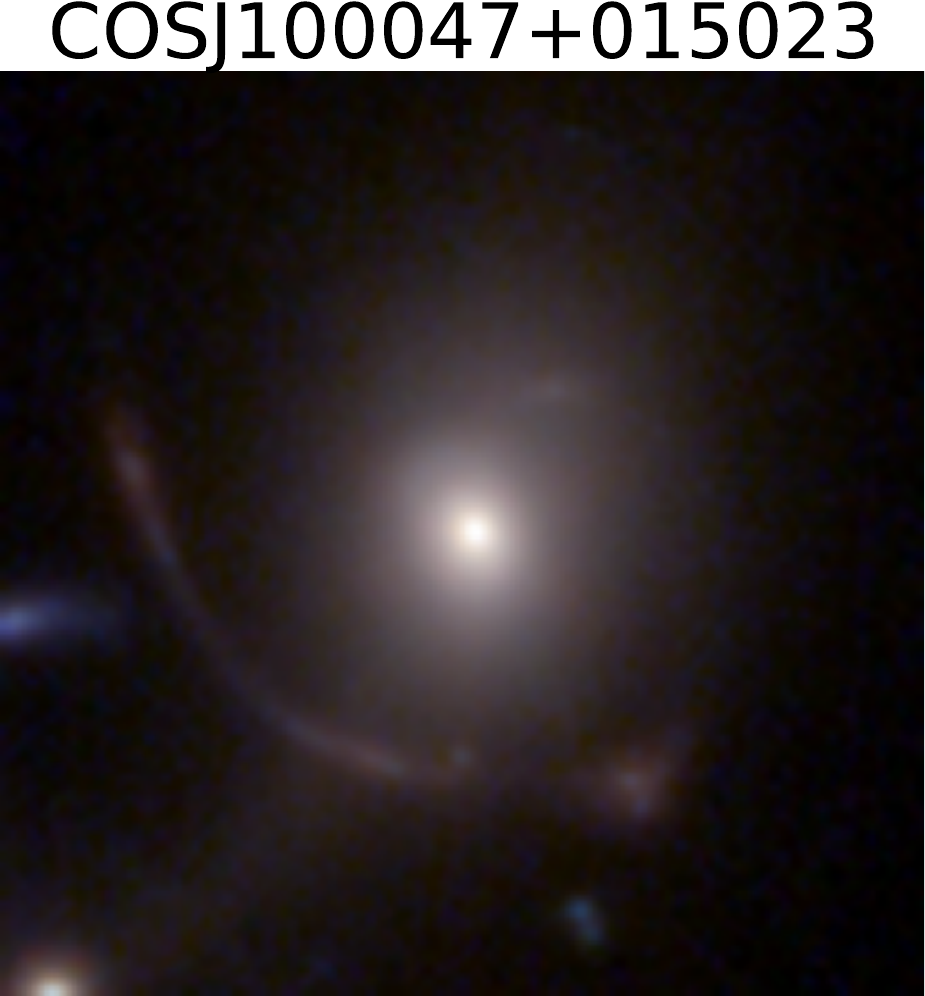}
\includegraphics[width=0.12\textwidth]{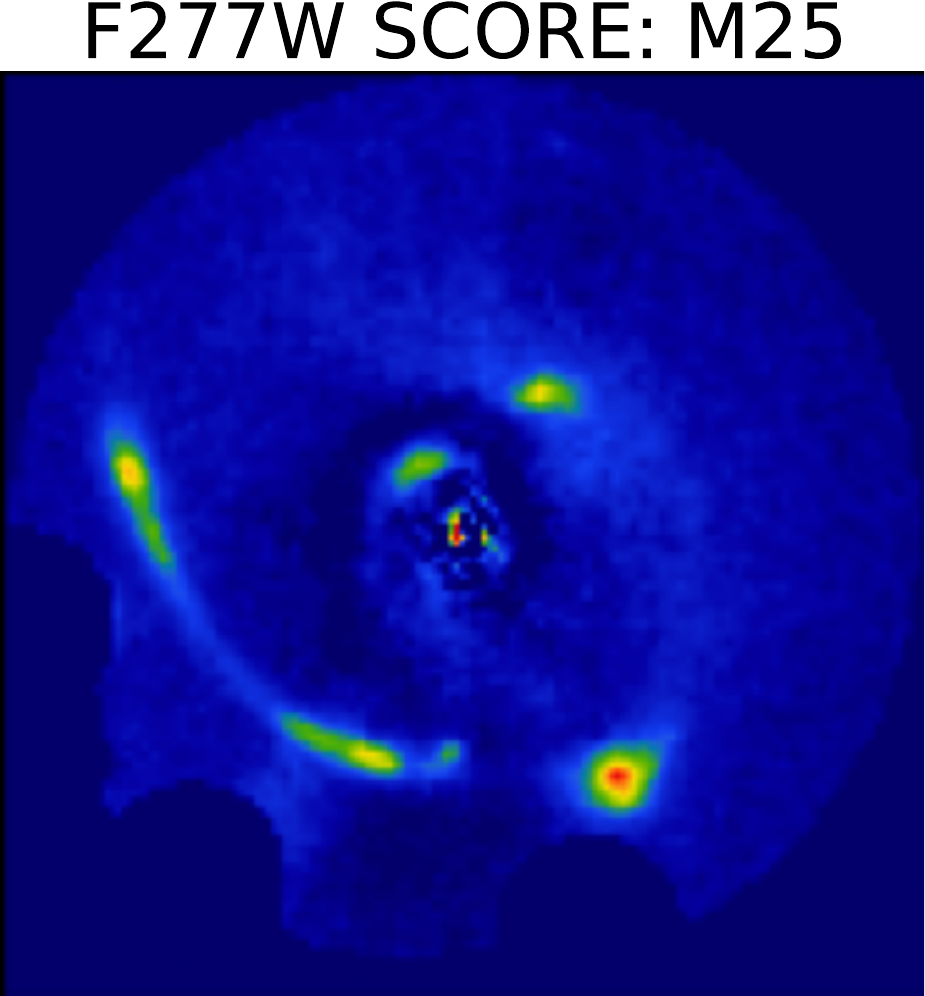}
\includegraphics[width=0.12\textwidth]{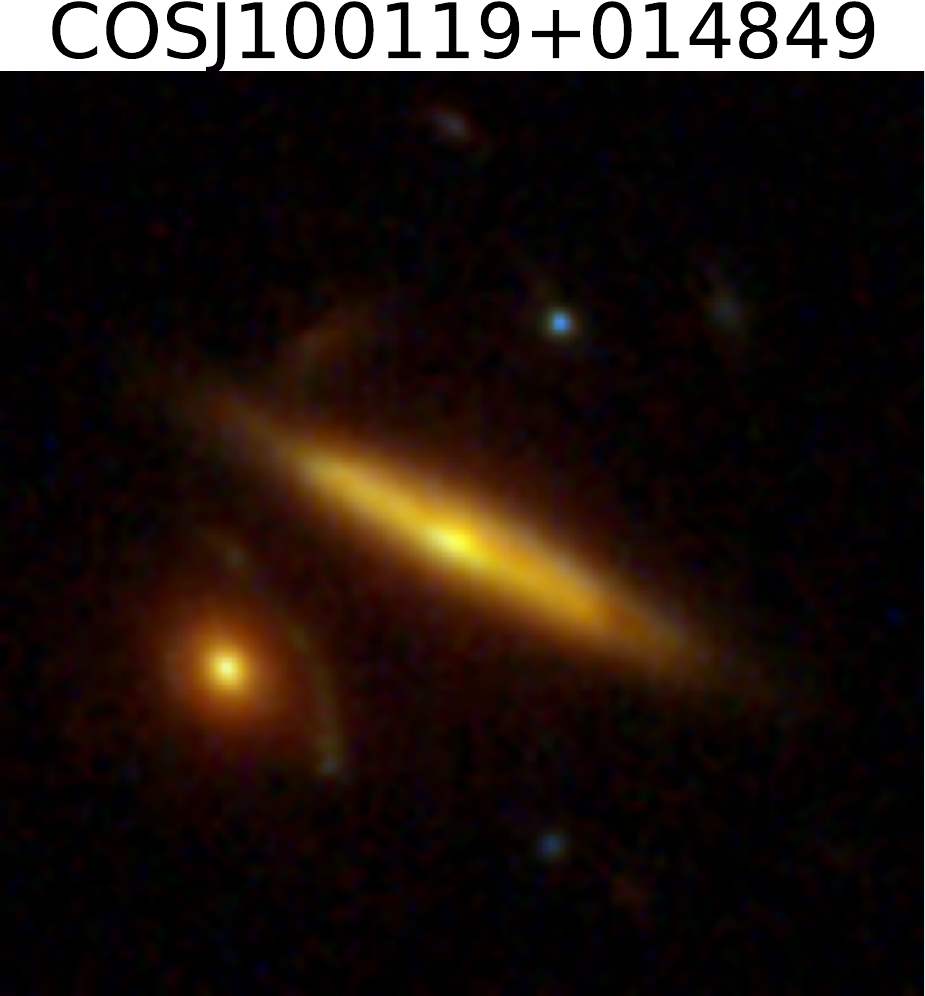}
\includegraphics[width=0.12\textwidth]{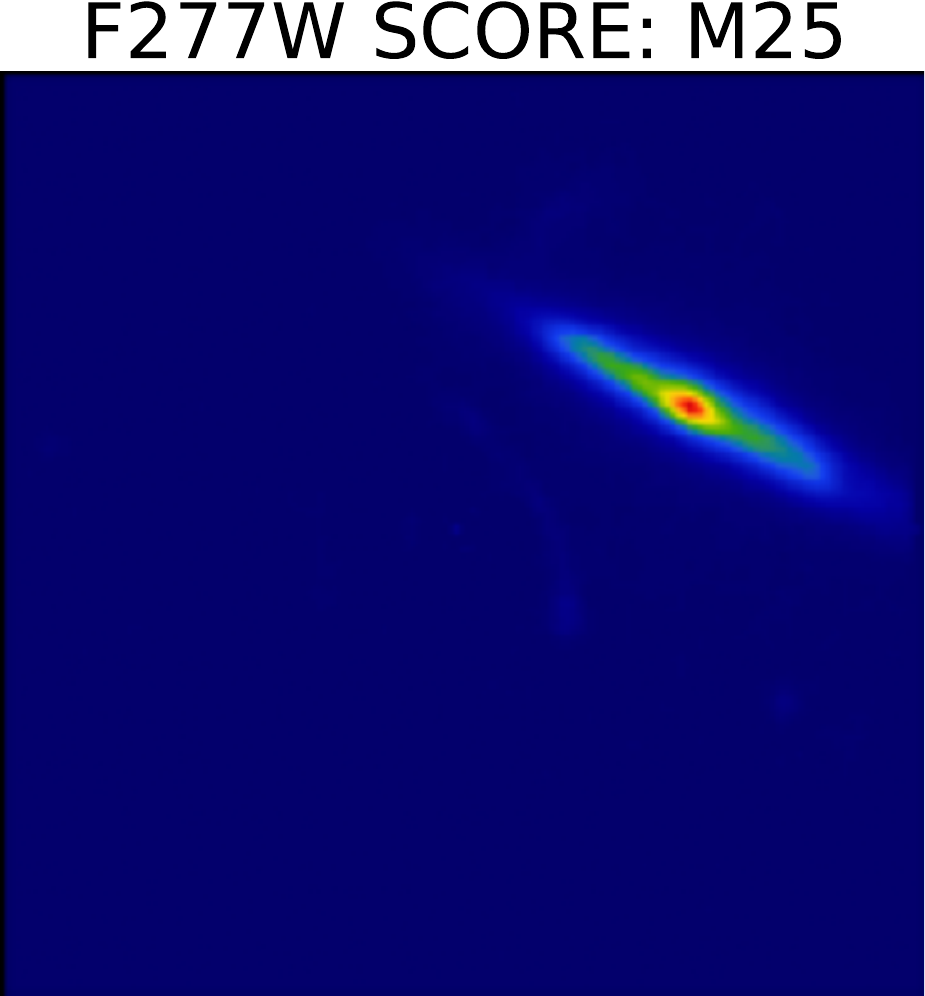}
\includegraphics[width=0.12\textwidth]{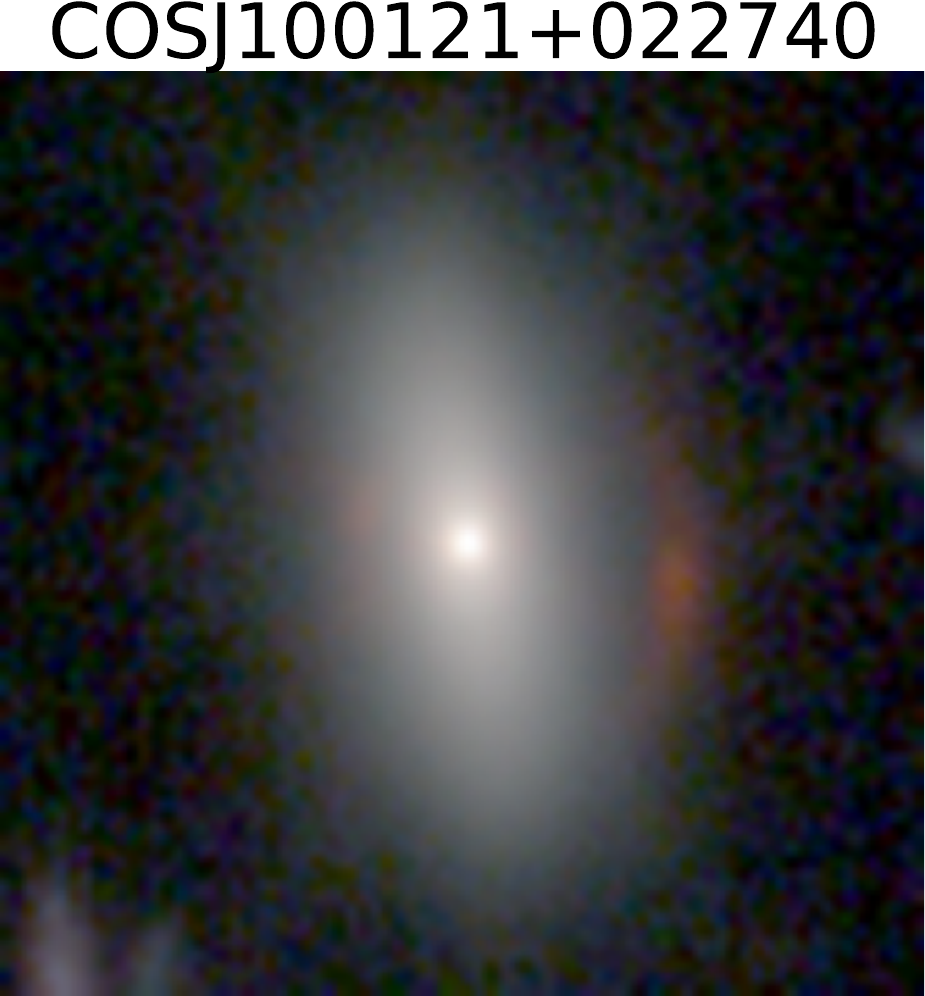}
\includegraphics[width=0.12\textwidth]{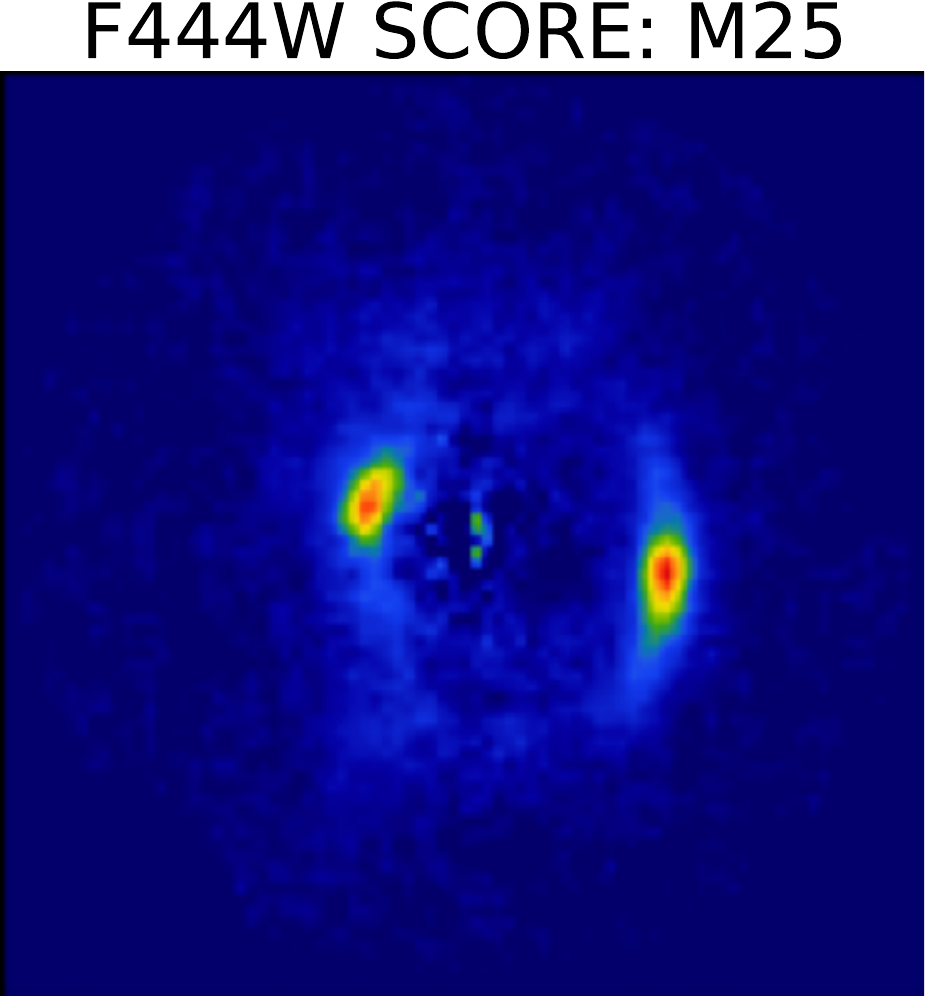}
\includegraphics[width=0.12\textwidth]{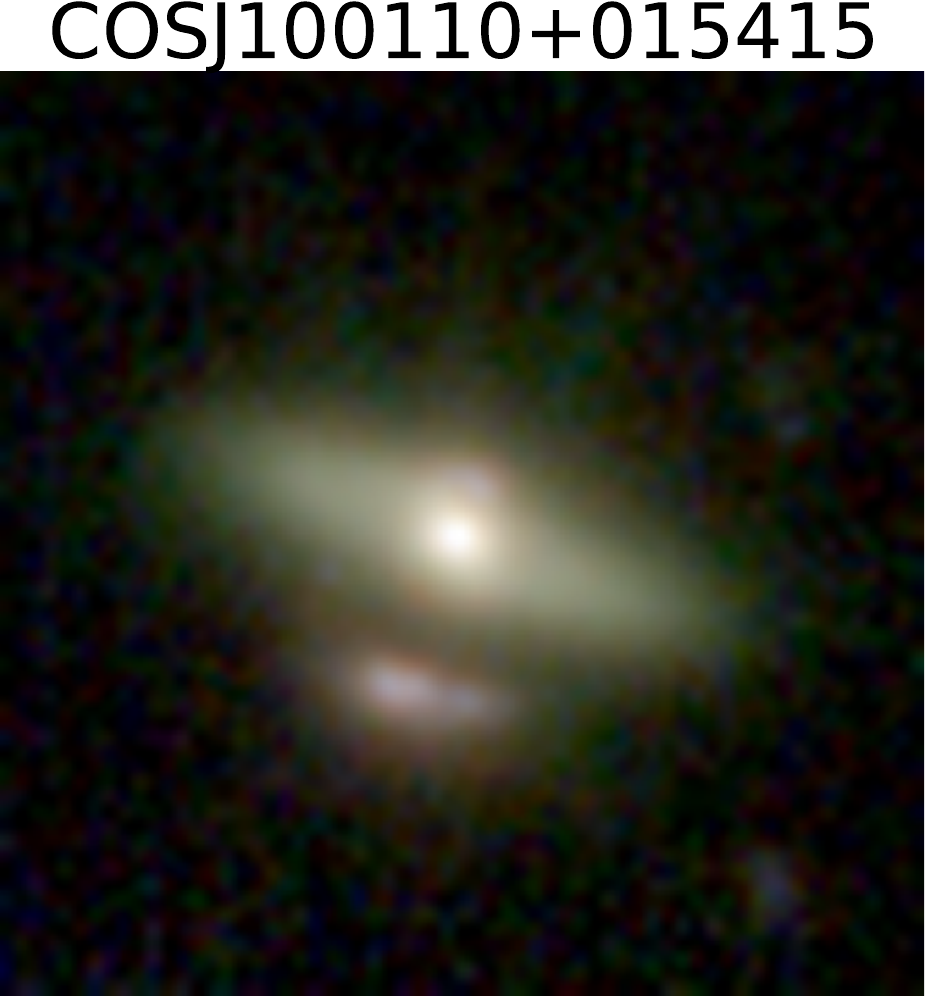}
\includegraphics[width=0.12\textwidth]{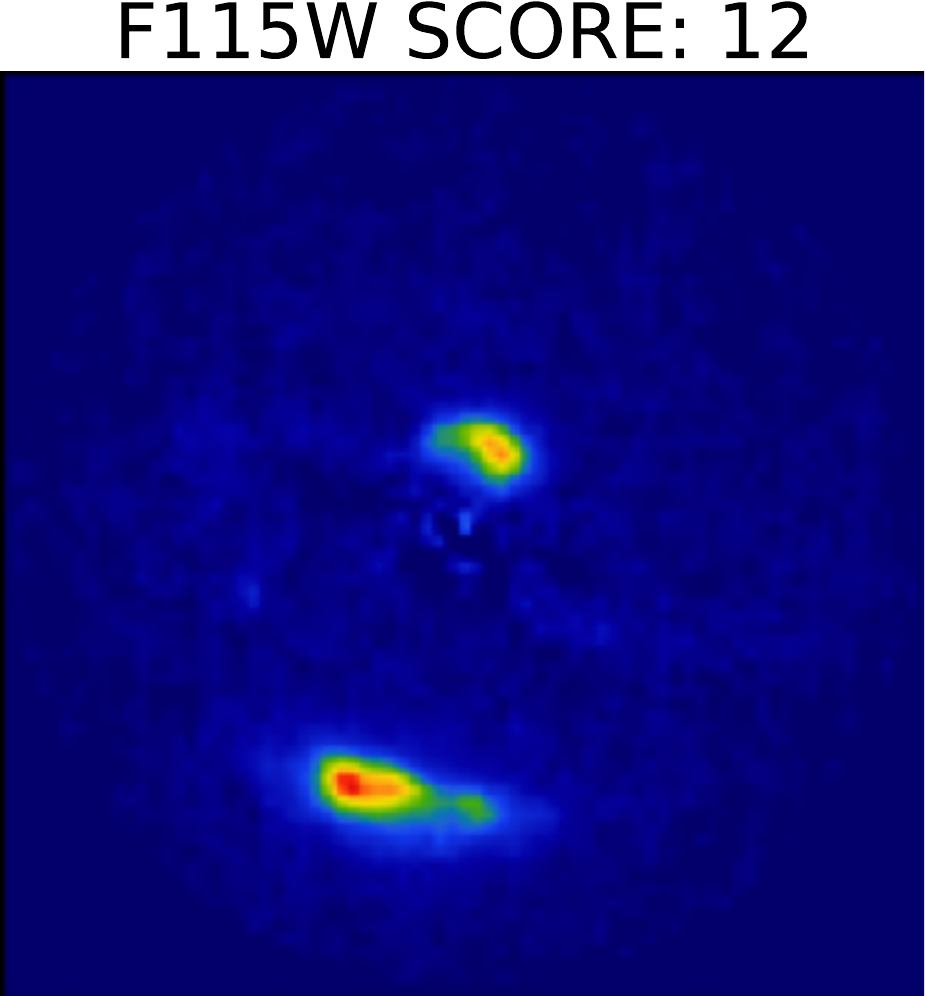}
\includegraphics[width=0.12\textwidth]{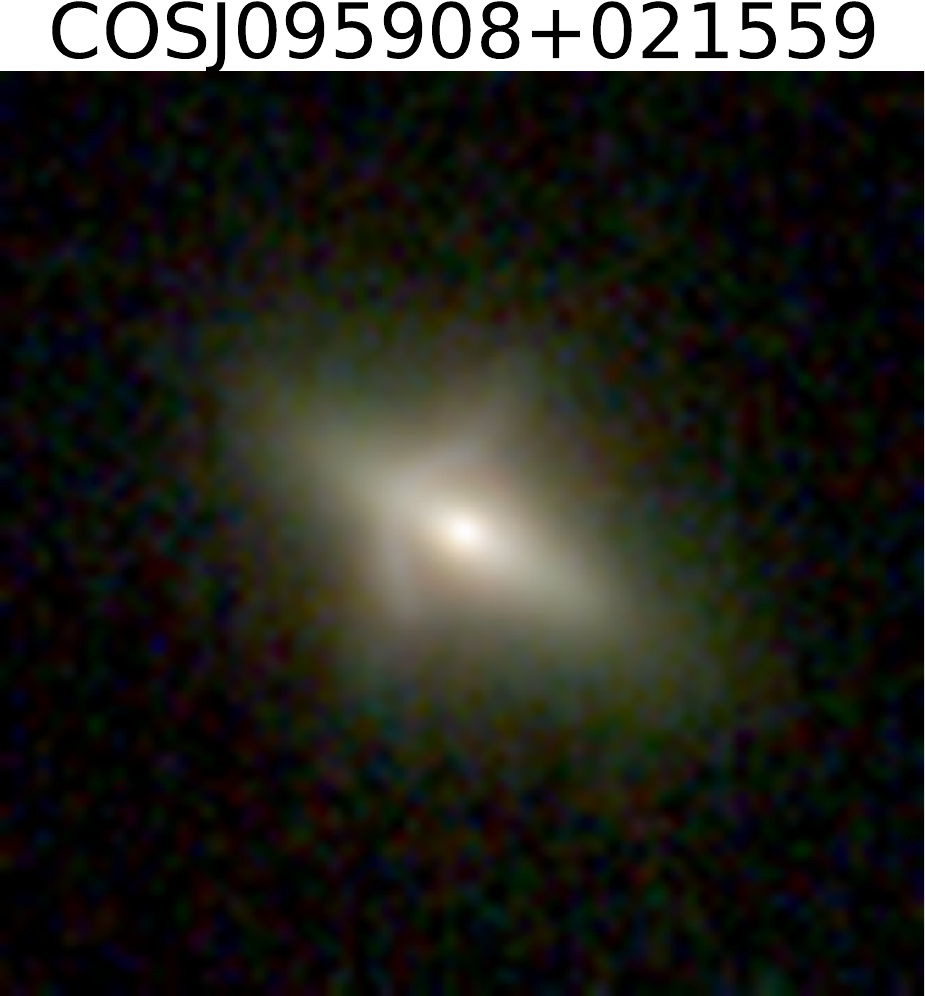}
\includegraphics[width=0.12\textwidth]{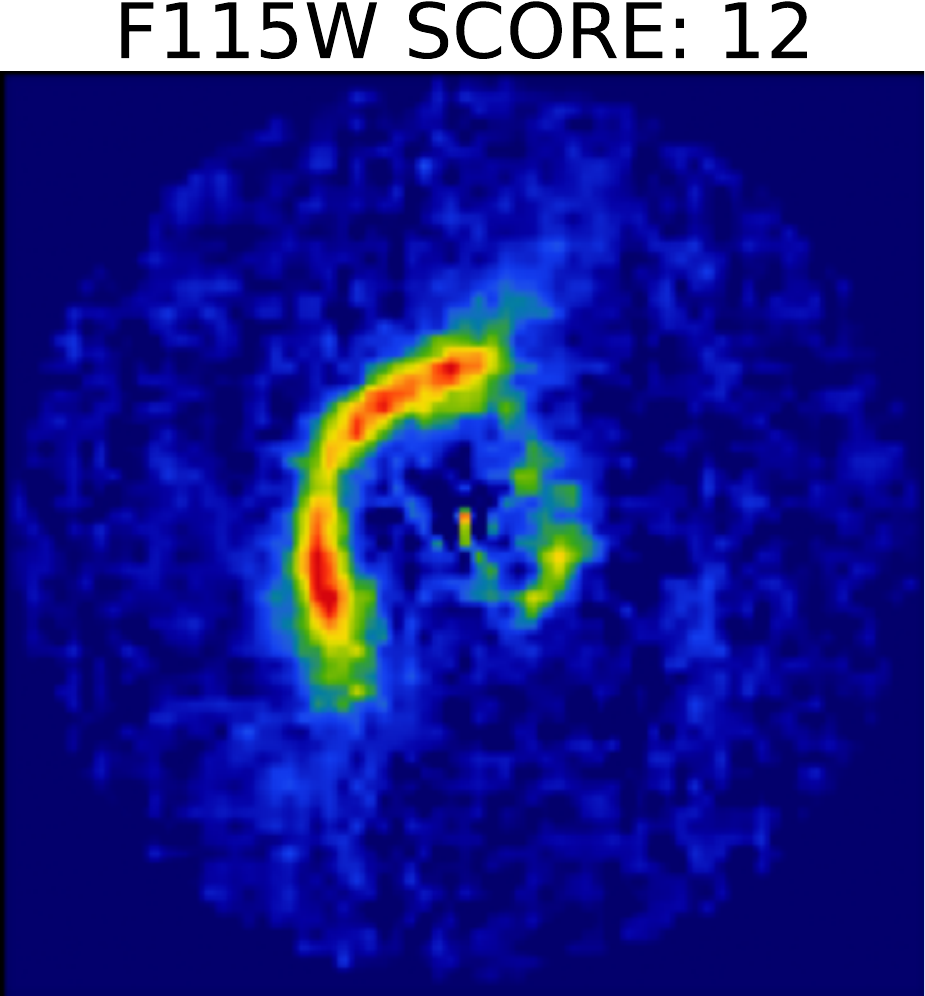}
\includegraphics[width=0.12\textwidth]{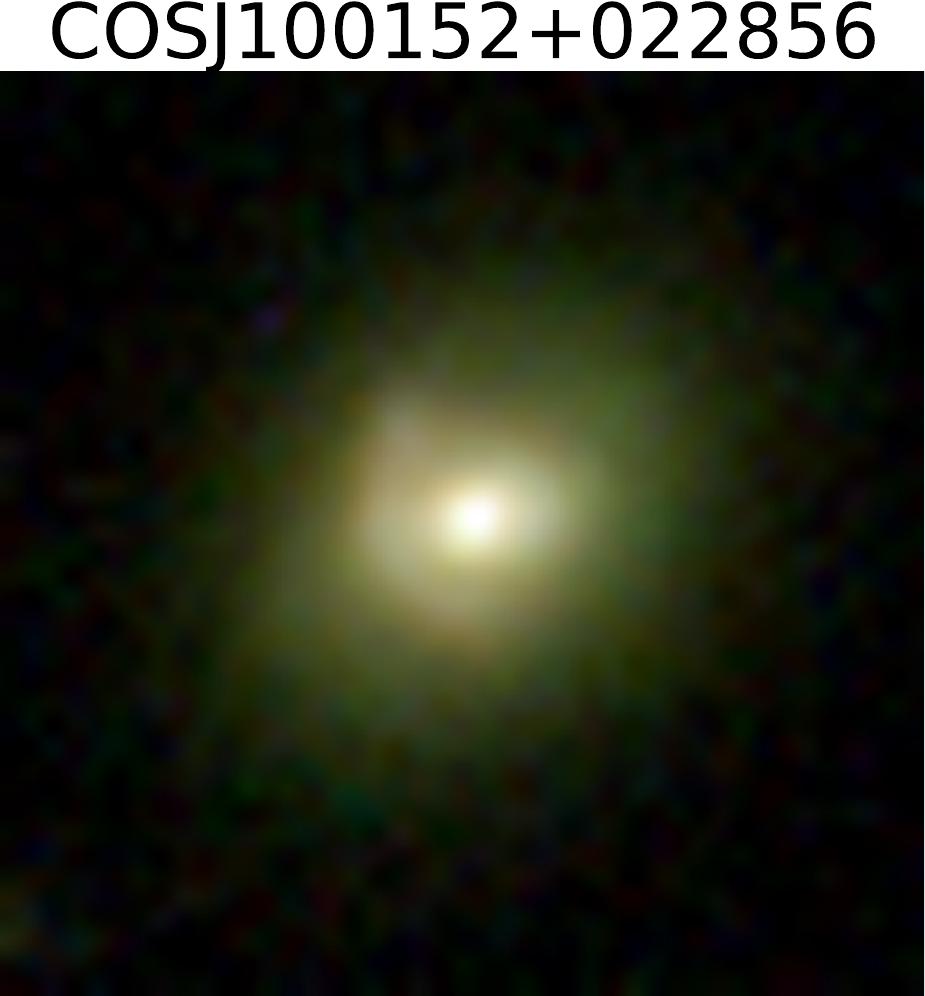}
\includegraphics[width=0.12\textwidth]{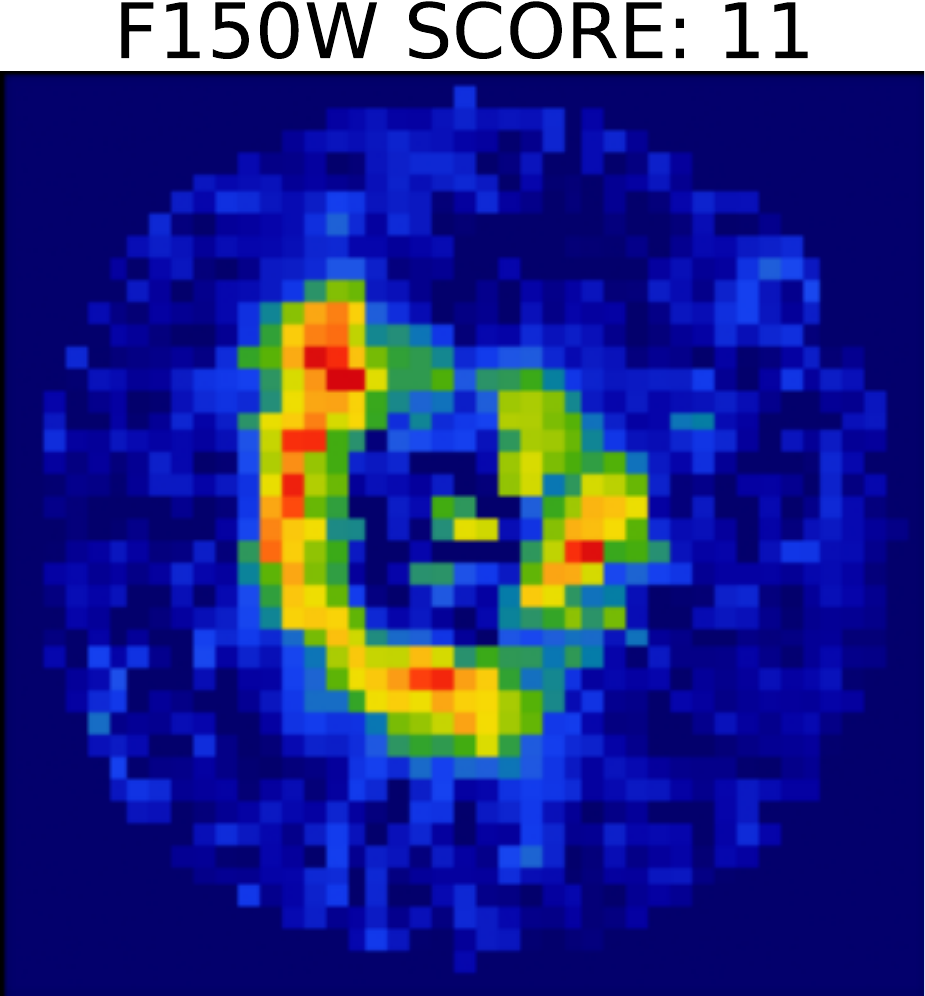}
\includegraphics[width=0.12\textwidth]{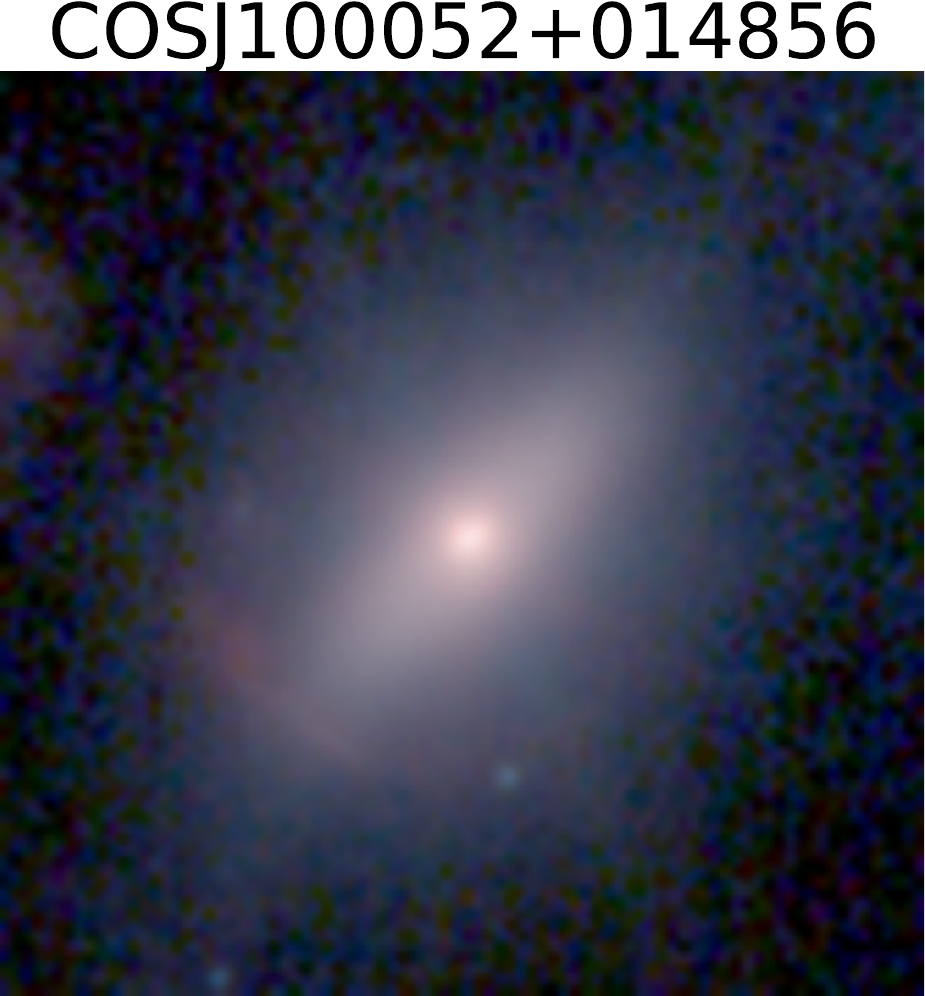}
\includegraphics[width=0.12\textwidth]{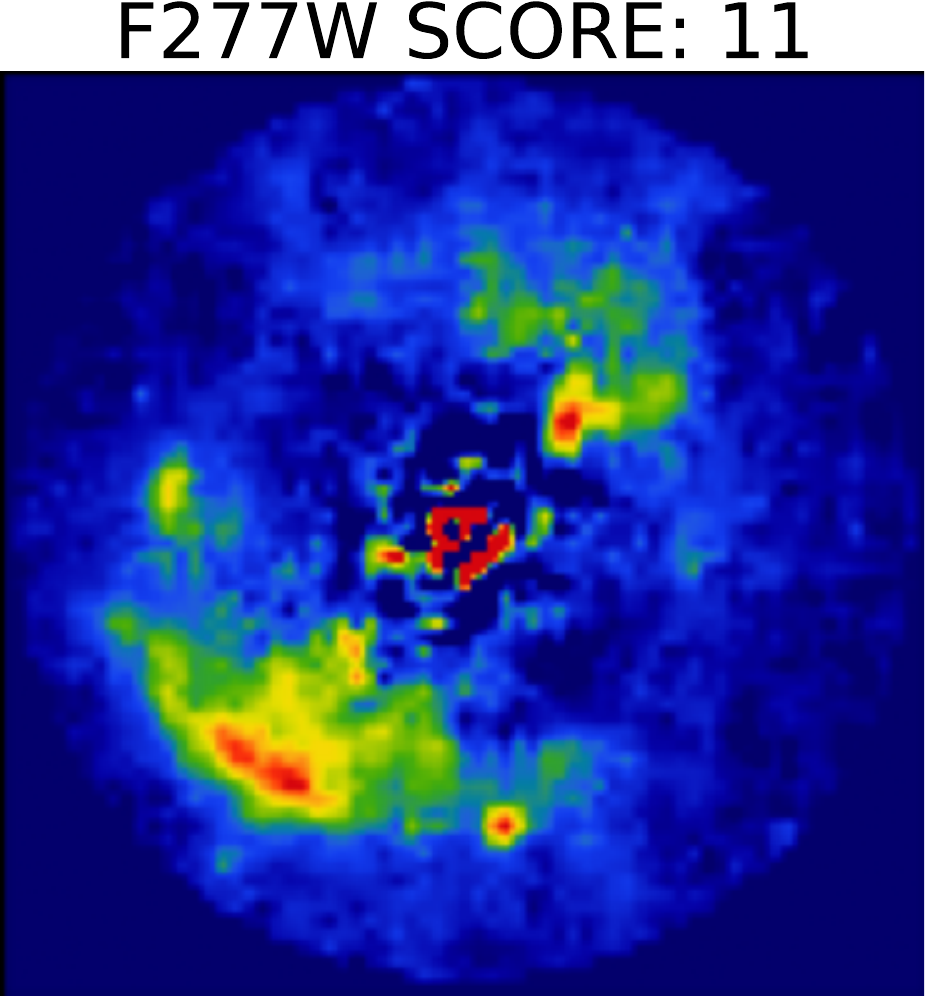}
\includegraphics[width=0.12\textwidth]{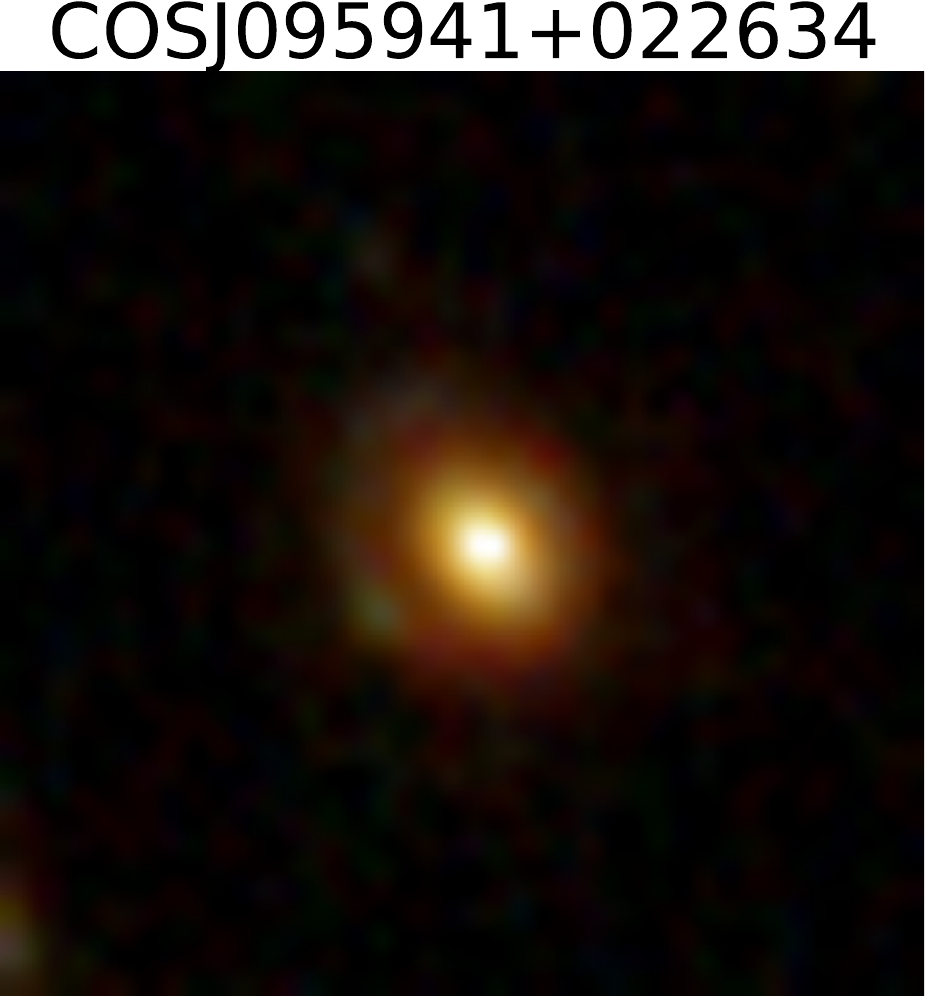}
\includegraphics[width=0.12\textwidth]{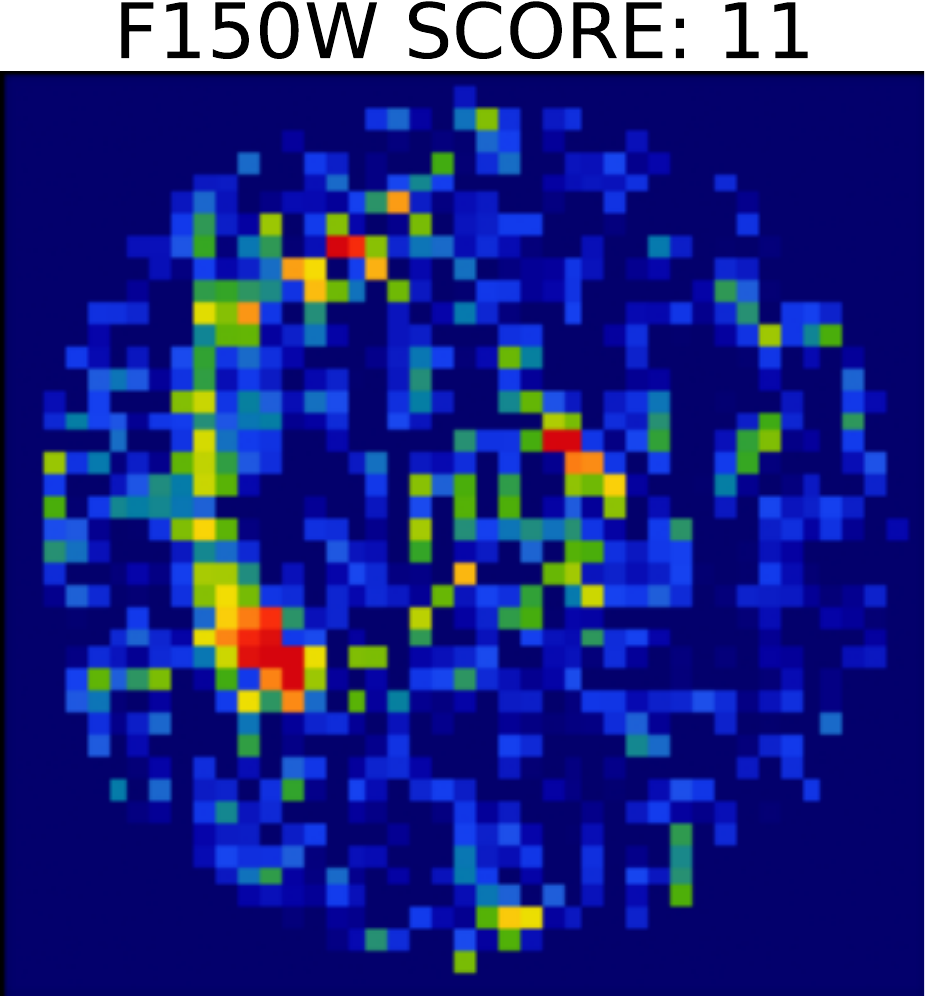}
\includegraphics[width=0.12\textwidth]{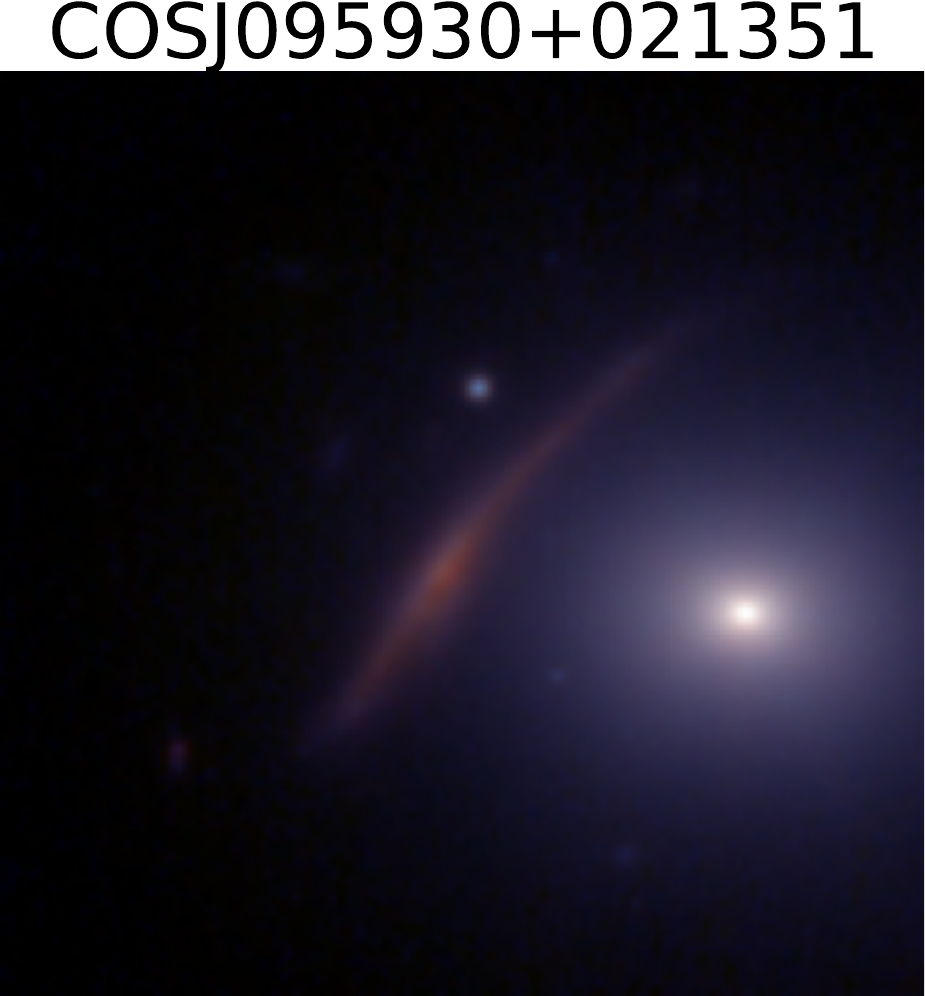}
\includegraphics[width=0.12\textwidth]{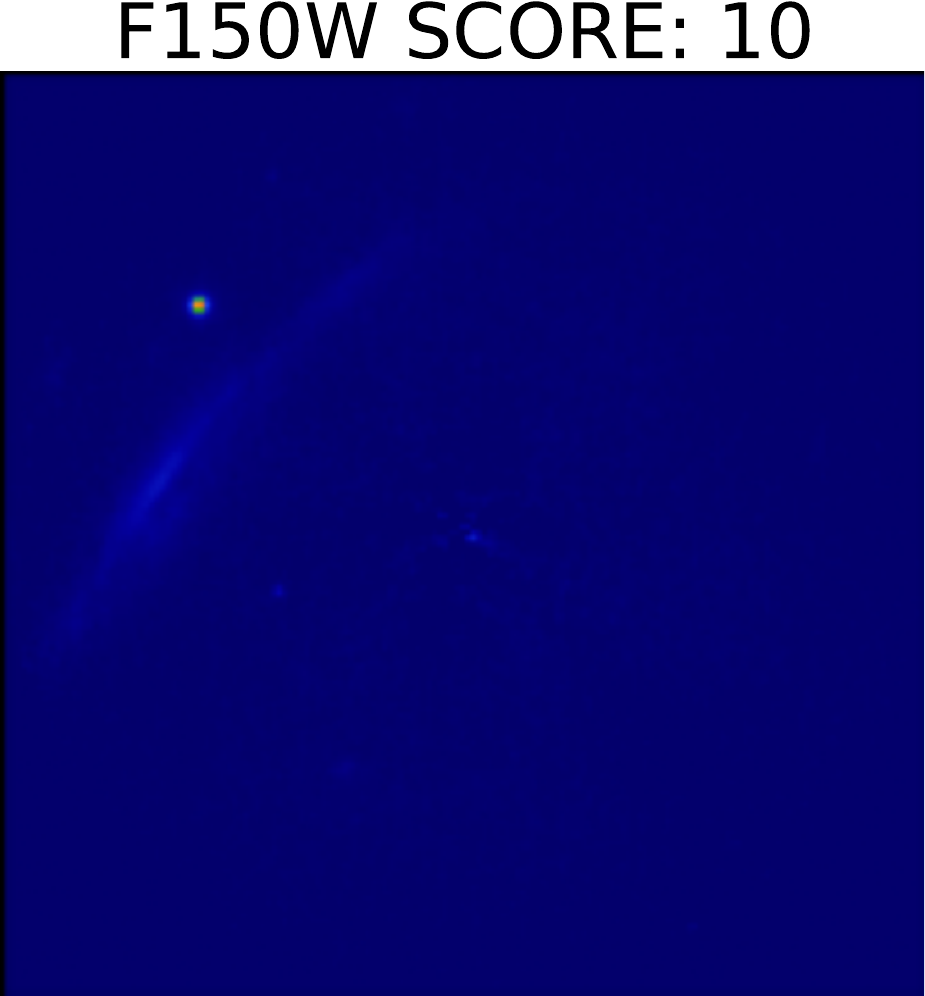}
\includegraphics[width=0.12\textwidth]{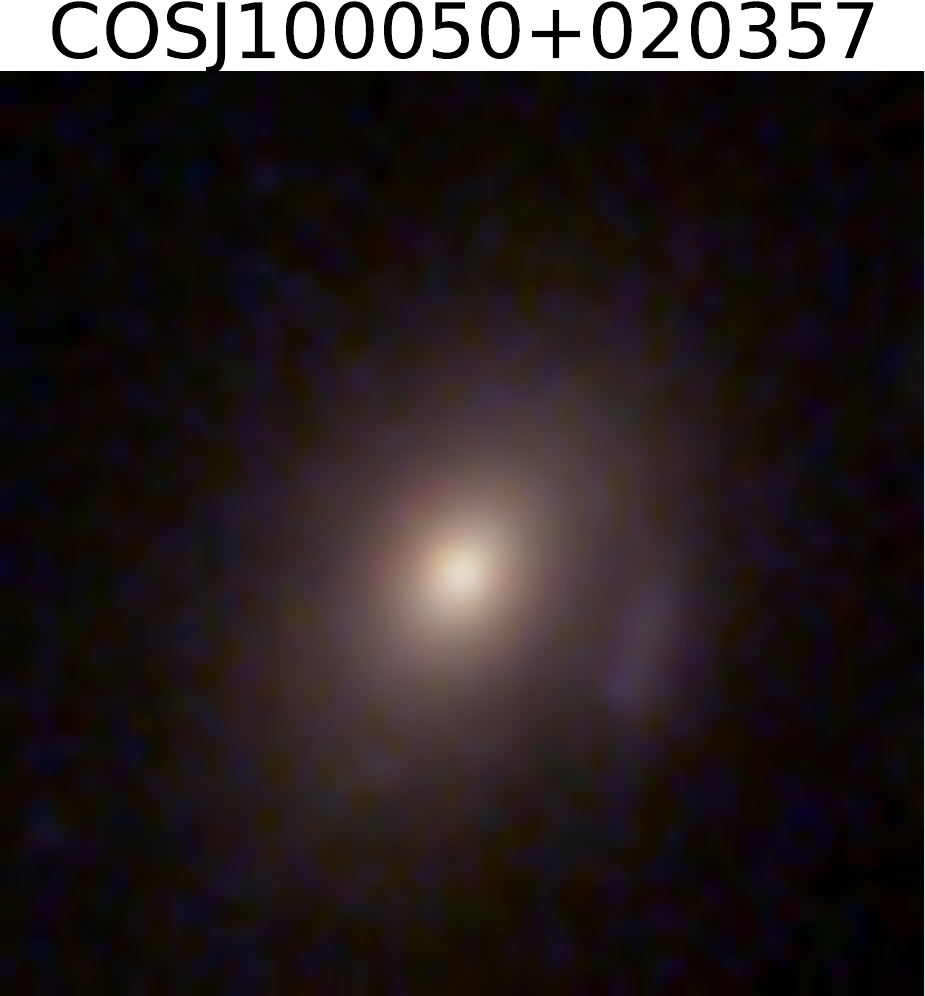}
\includegraphics[width=0.12\textwidth]{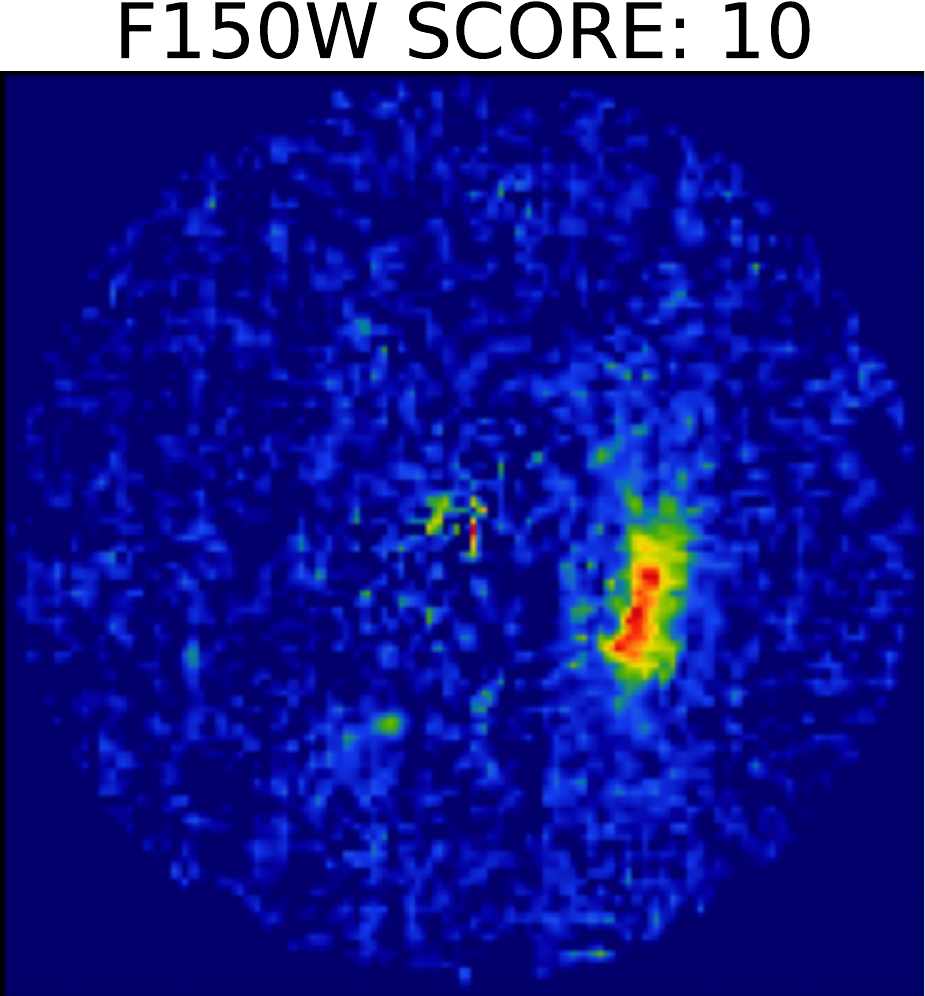}
\includegraphics[width=0.12\textwidth]{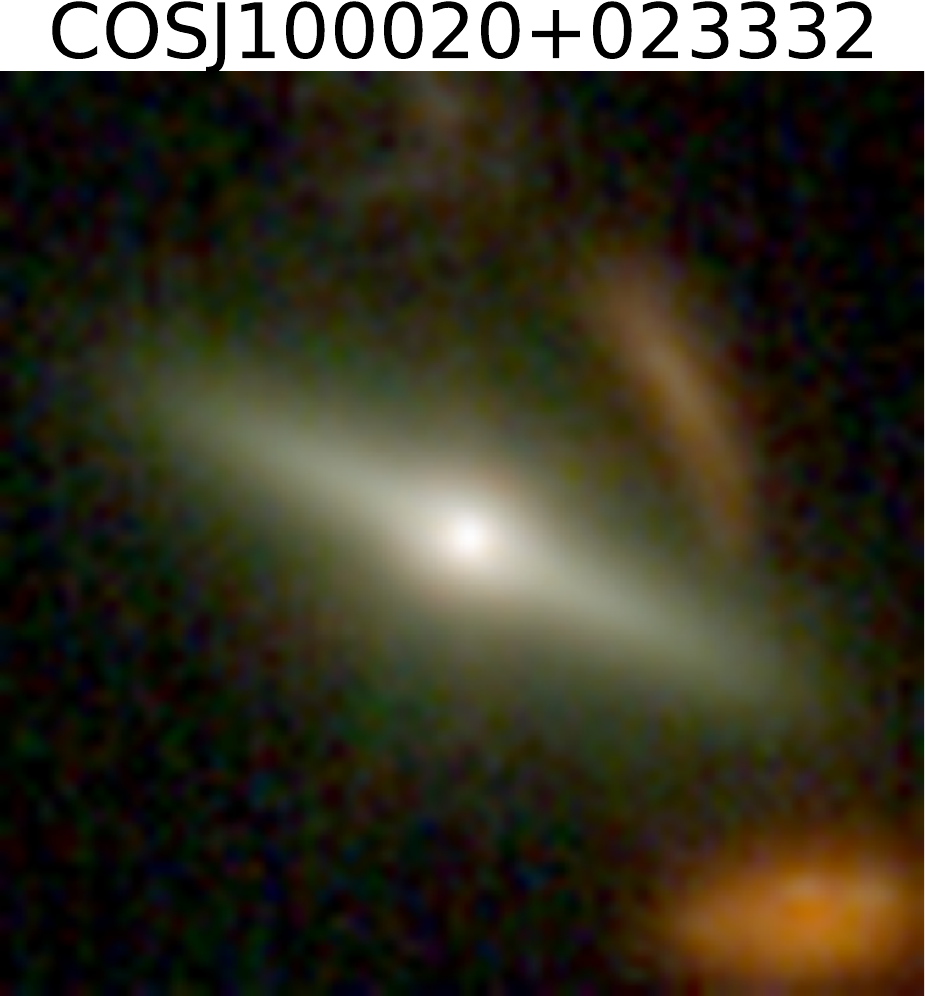}
\includegraphics[width=0.12\textwidth]{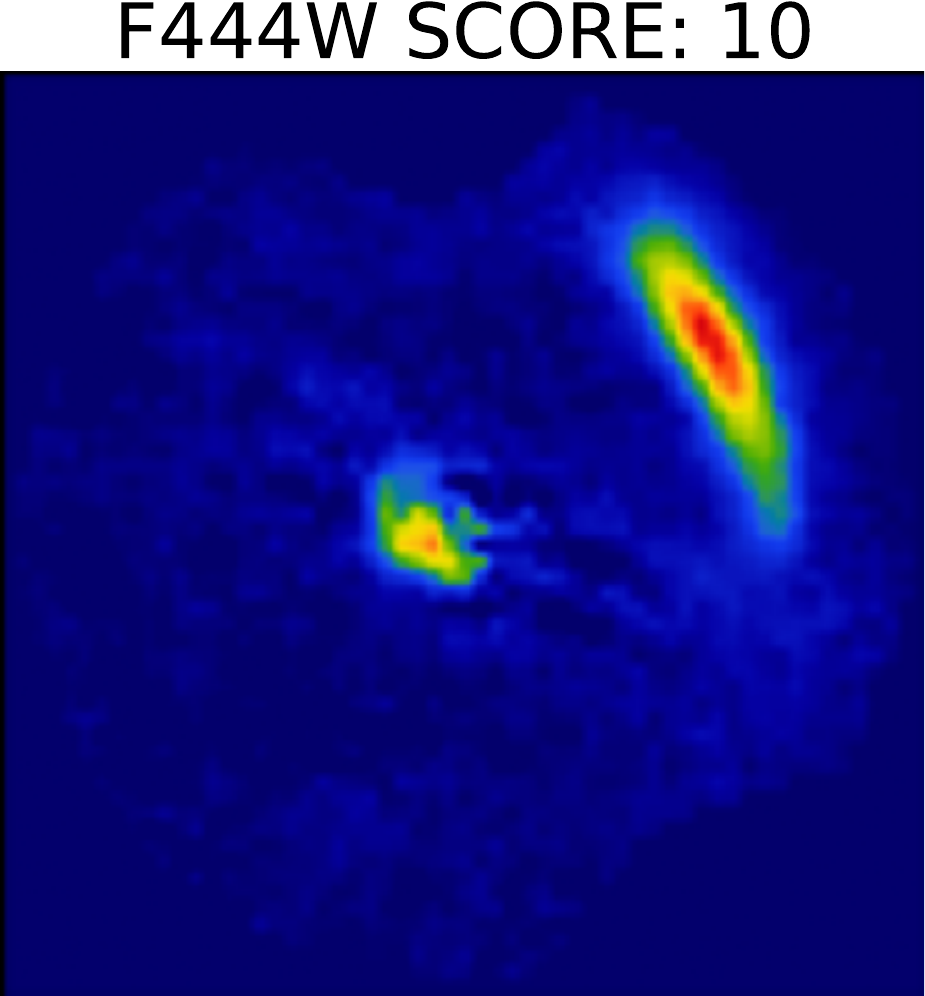}
\includegraphics[width=0.12\textwidth]{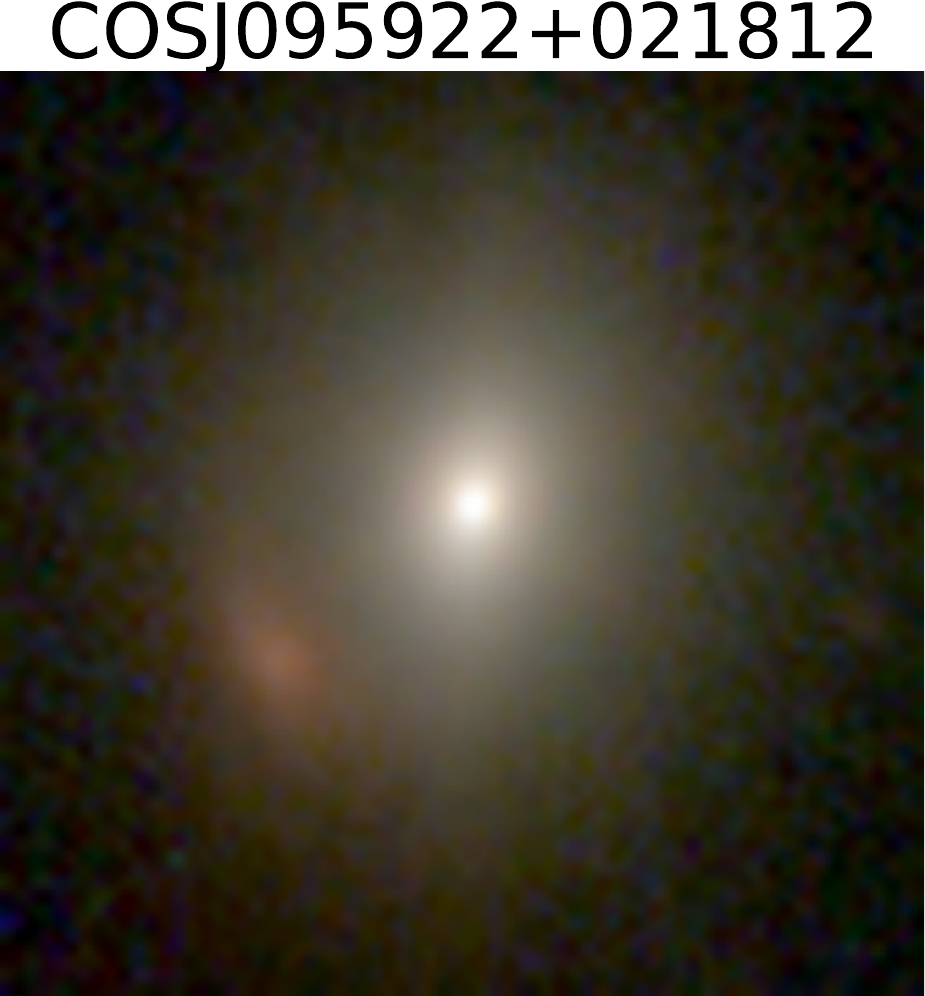}
\includegraphics[width=0.12\textwidth]{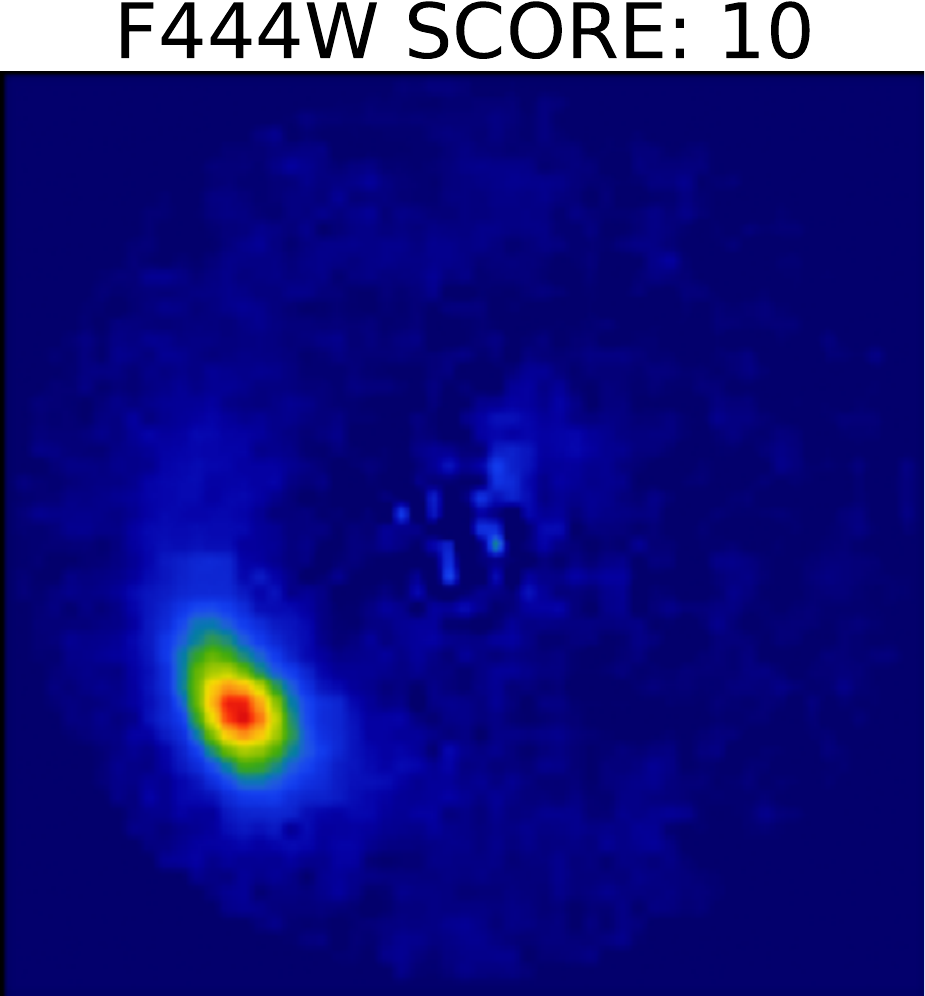}
\includegraphics[width=0.12\textwidth]{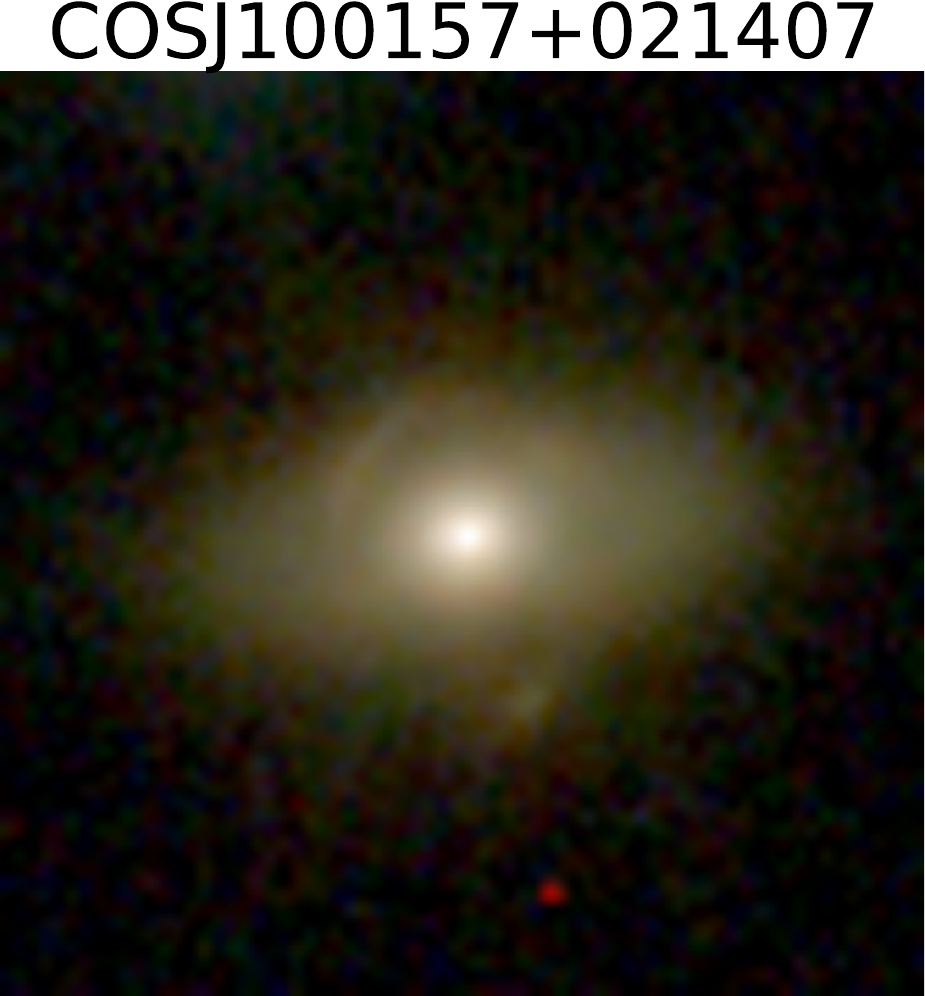}
\includegraphics[width=0.12\textwidth]{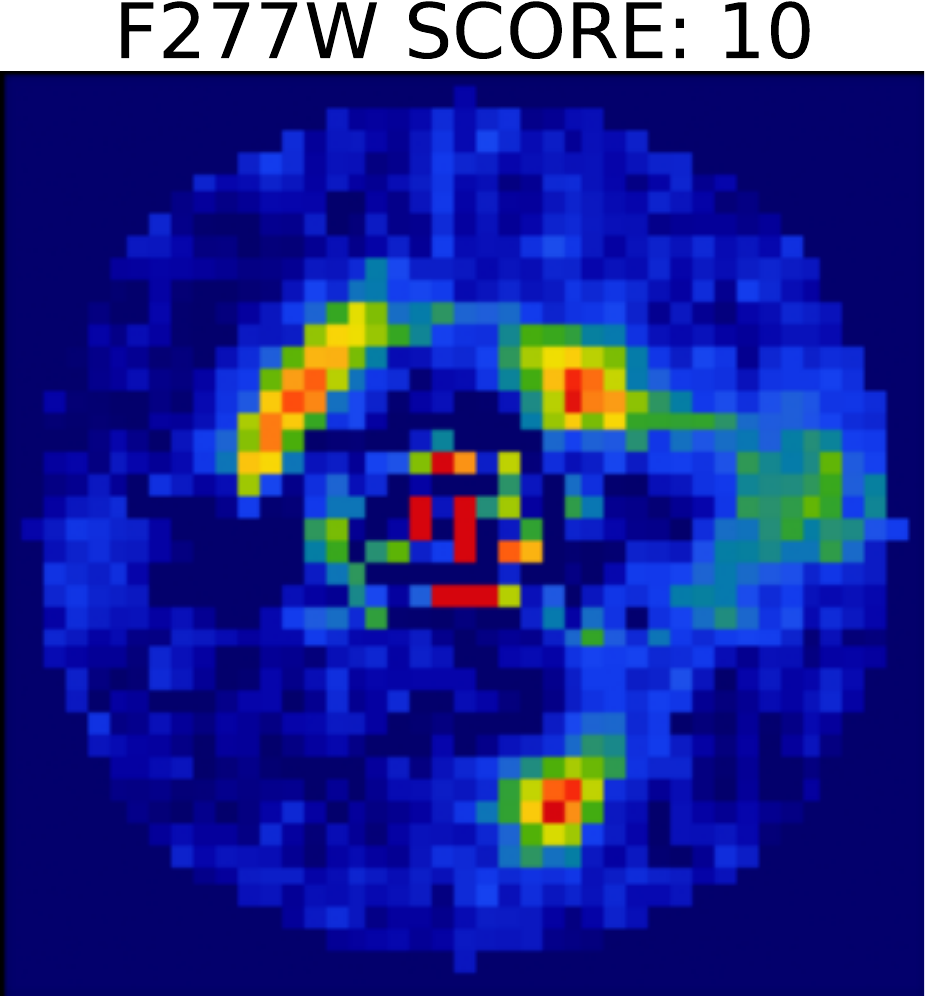}
\includegraphics[width=0.12\textwidth]{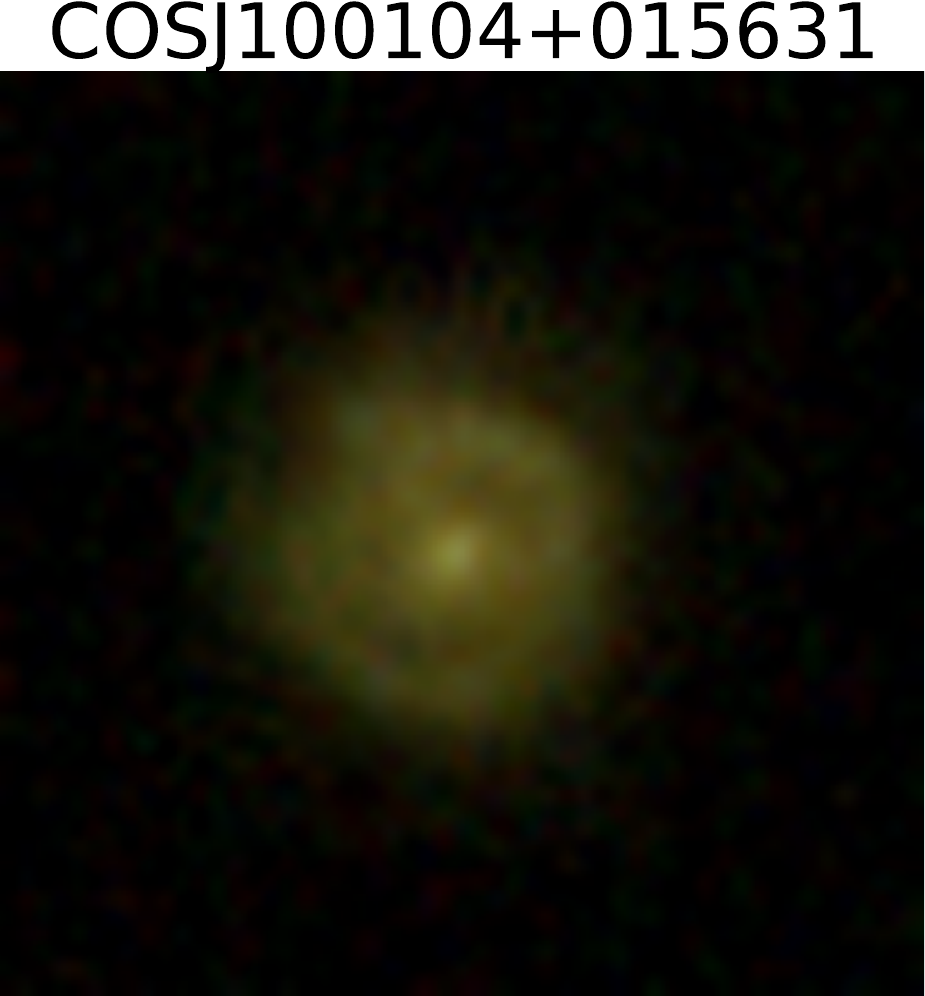}
\includegraphics[width=0.12\textwidth]{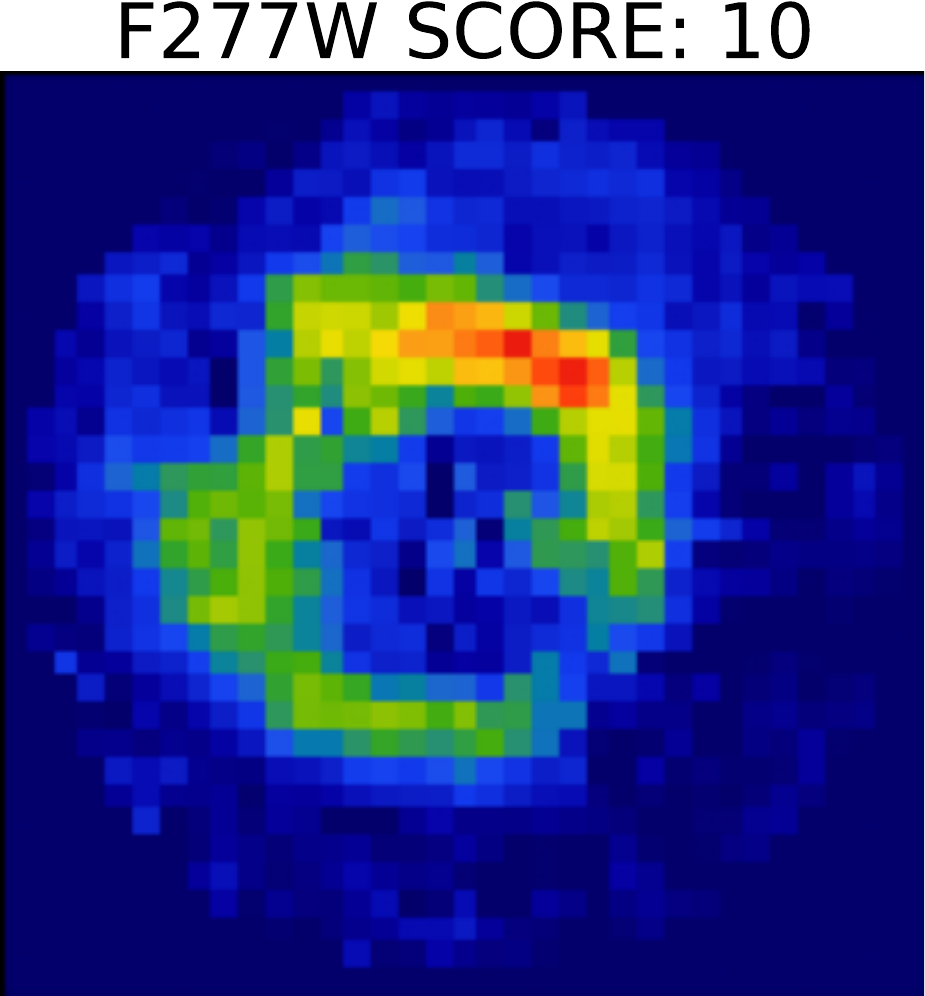}
\includegraphics[width=0.12\textwidth]{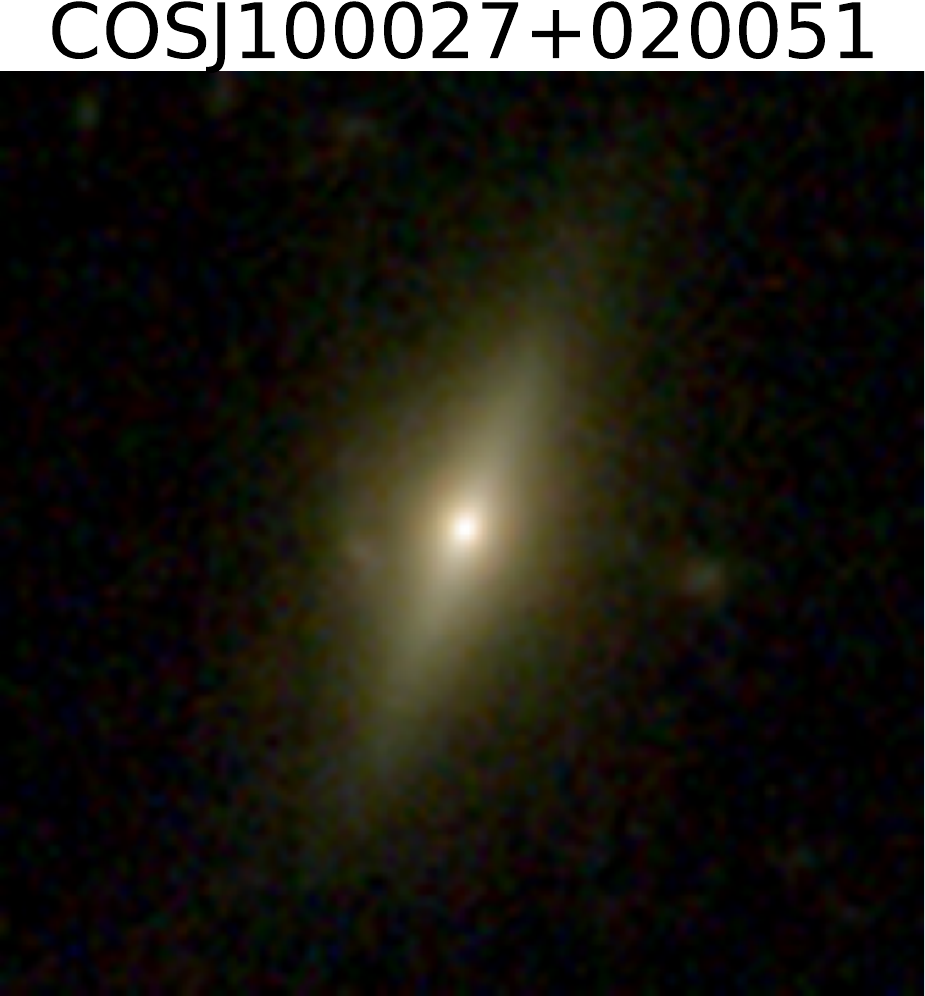}
\includegraphics[width=0.12\textwidth]{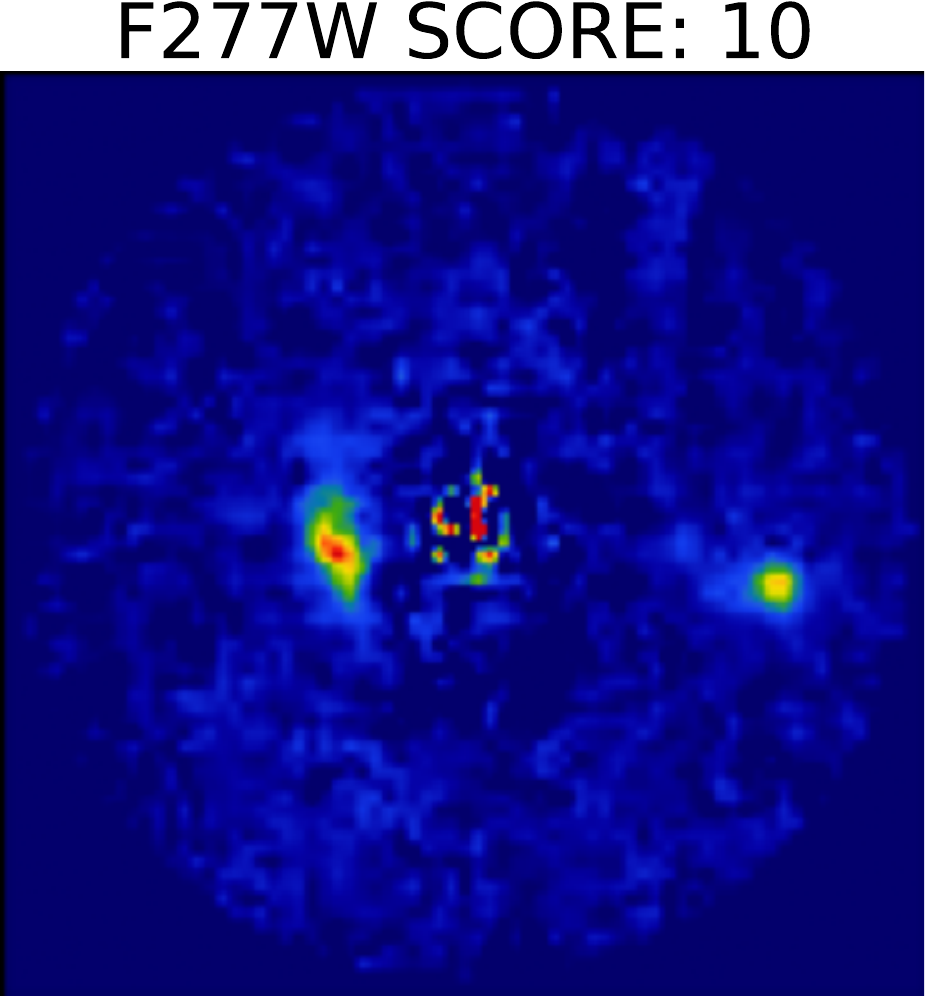}
\includegraphics[width=0.12\textwidth]{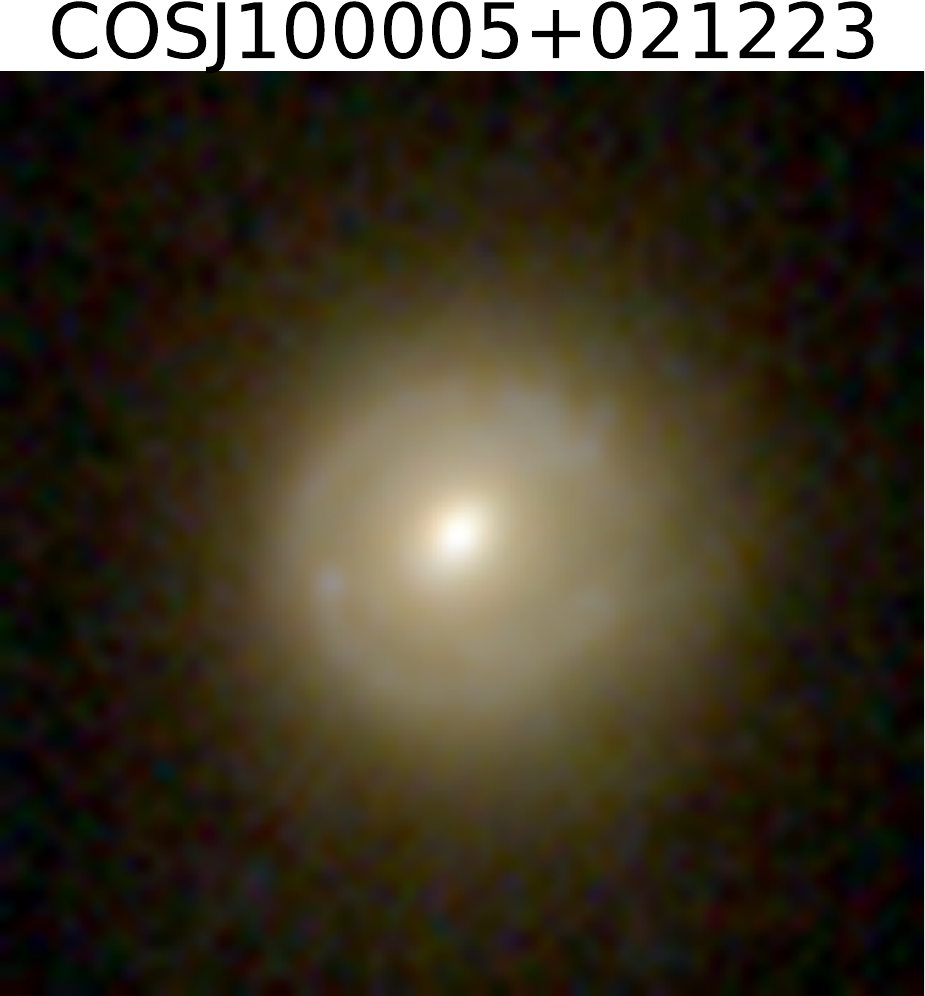}
\includegraphics[width=0.12\textwidth]{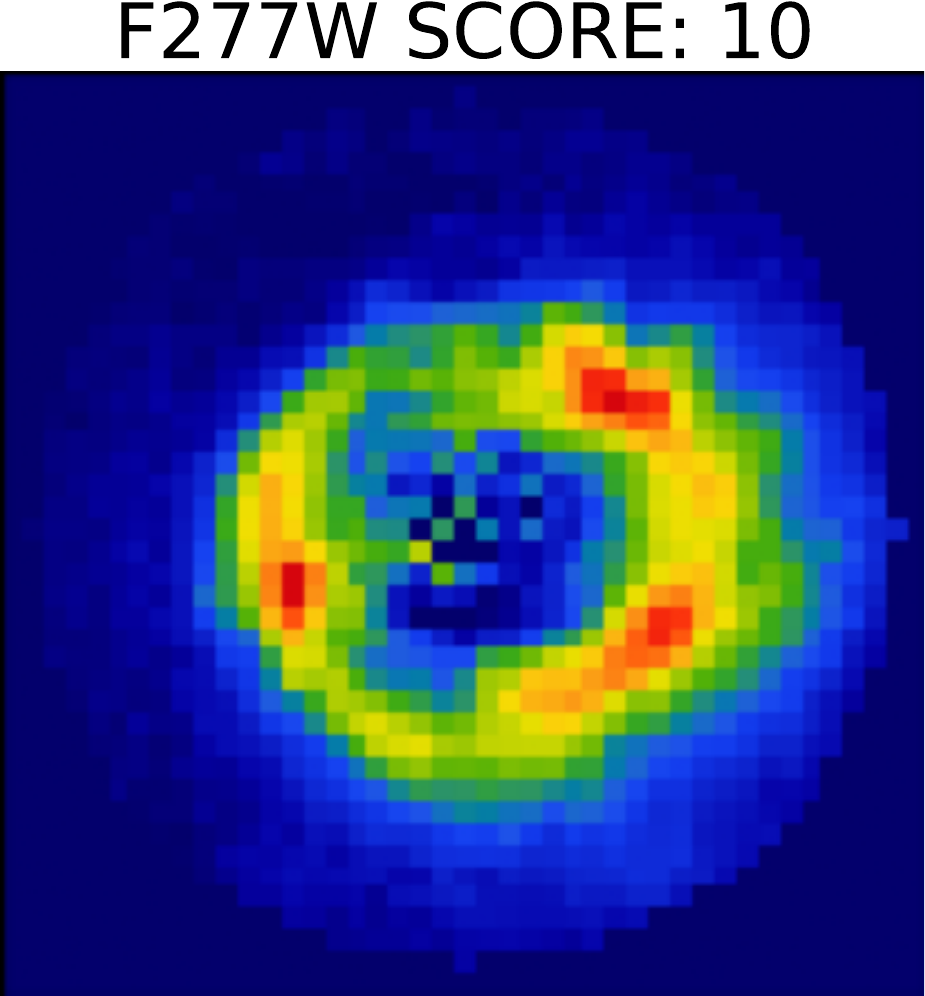}
\includegraphics[width=0.12\textwidth]{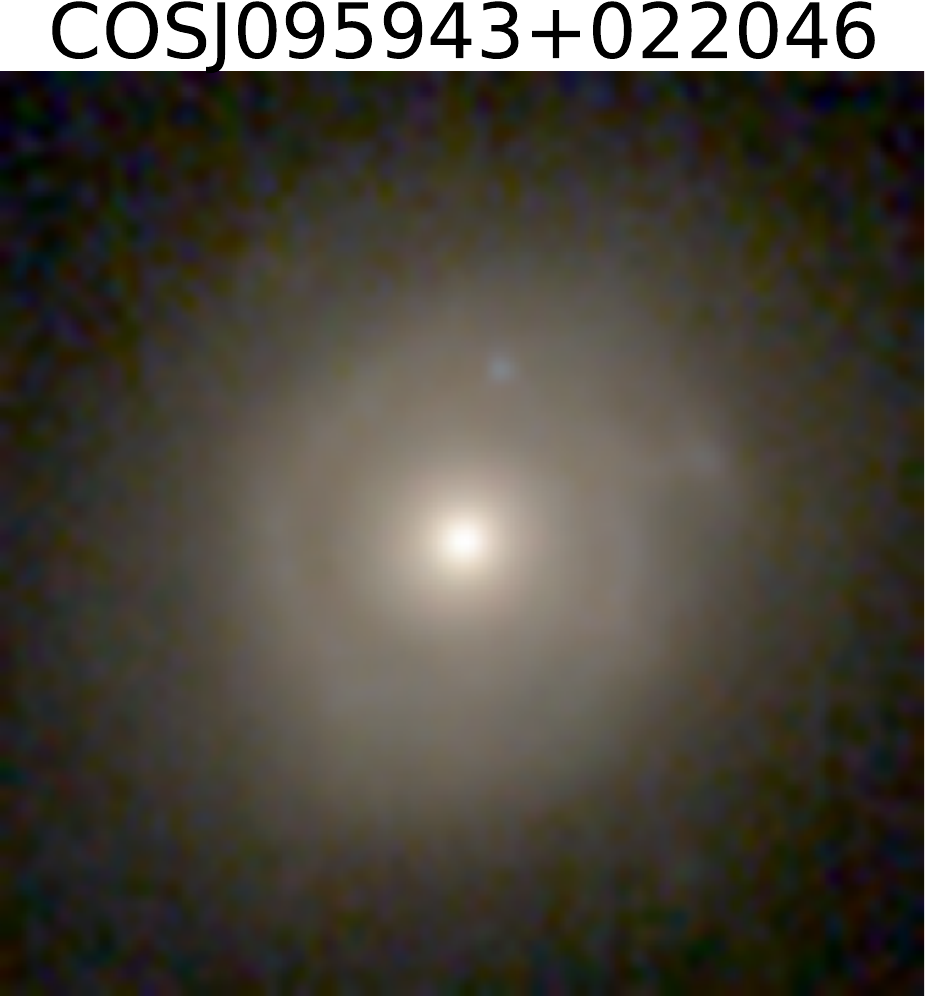}
\includegraphics[width=0.12\textwidth]{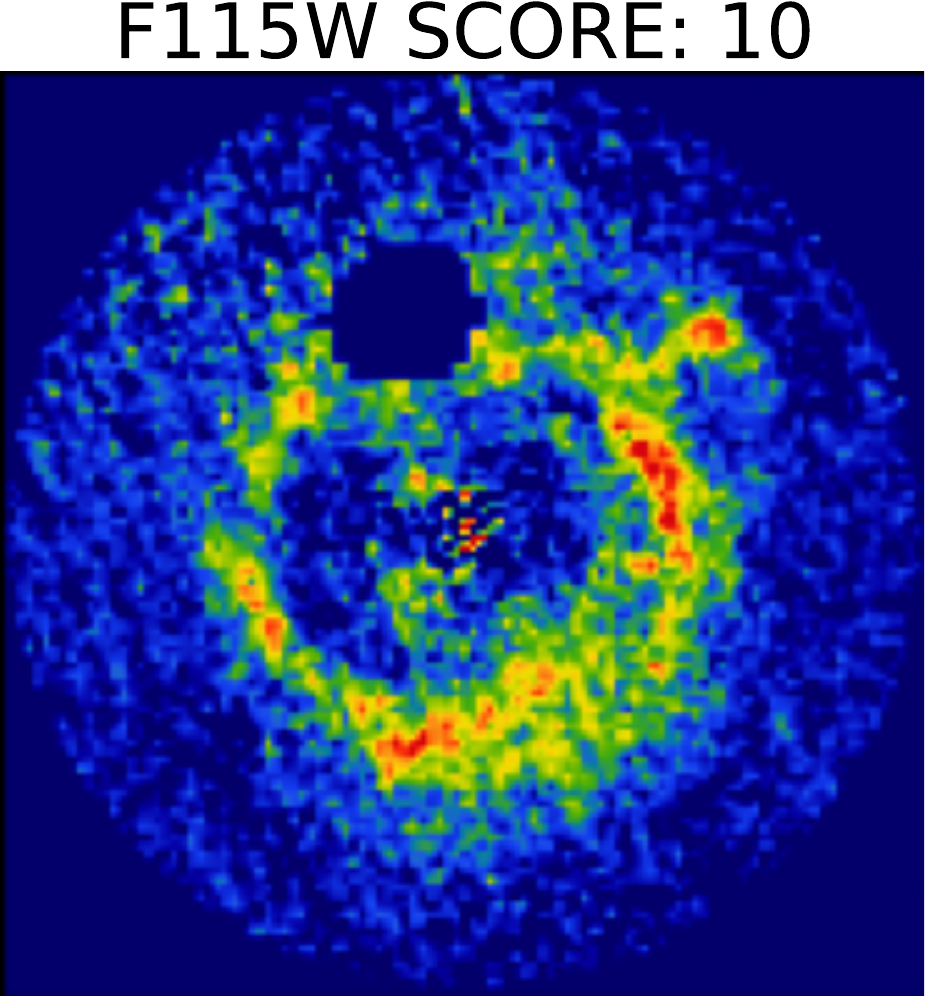}
\includegraphics[width=0.12\textwidth]{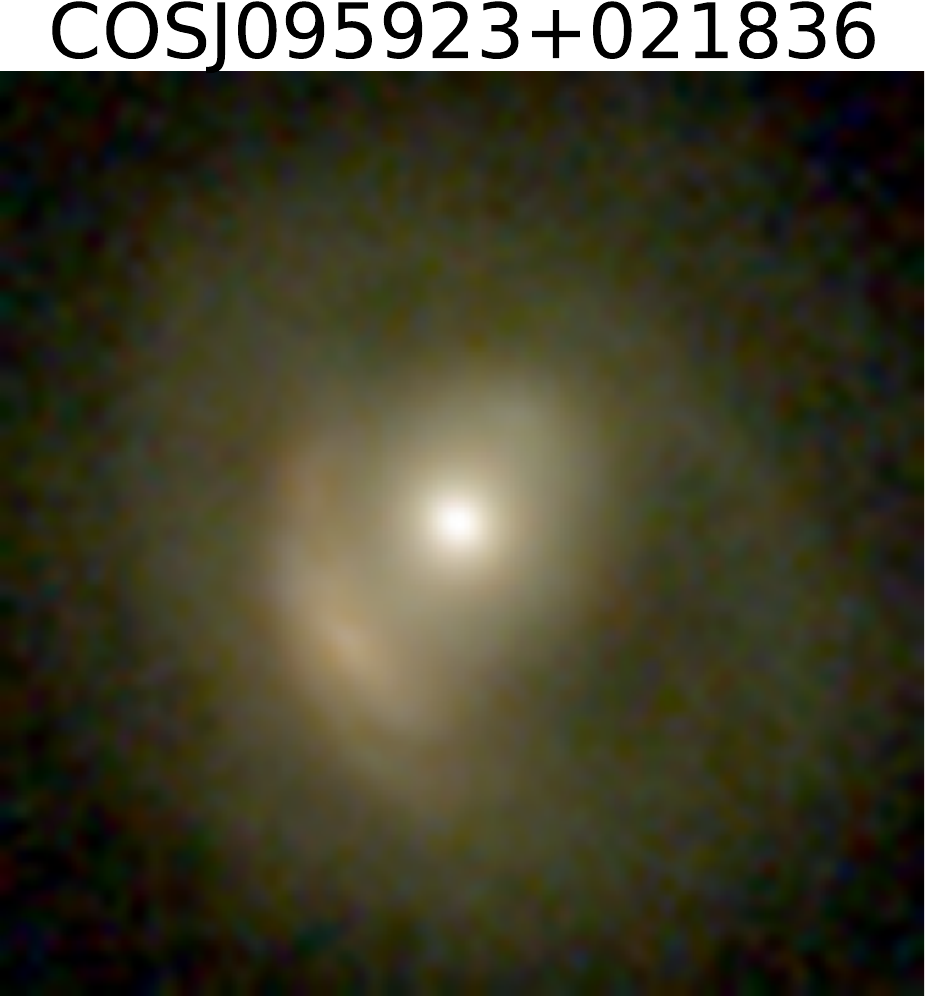}
\includegraphics[width=0.12\textwidth]{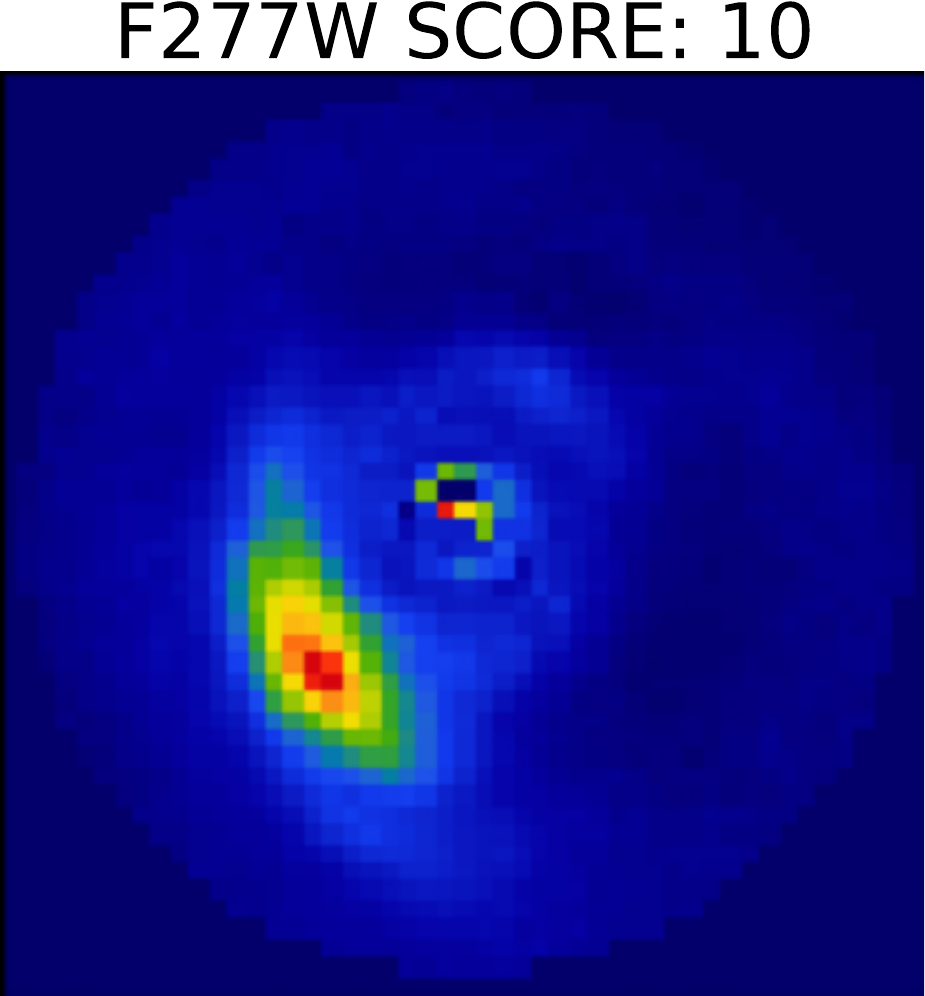}
\includegraphics[width=0.12\textwidth]{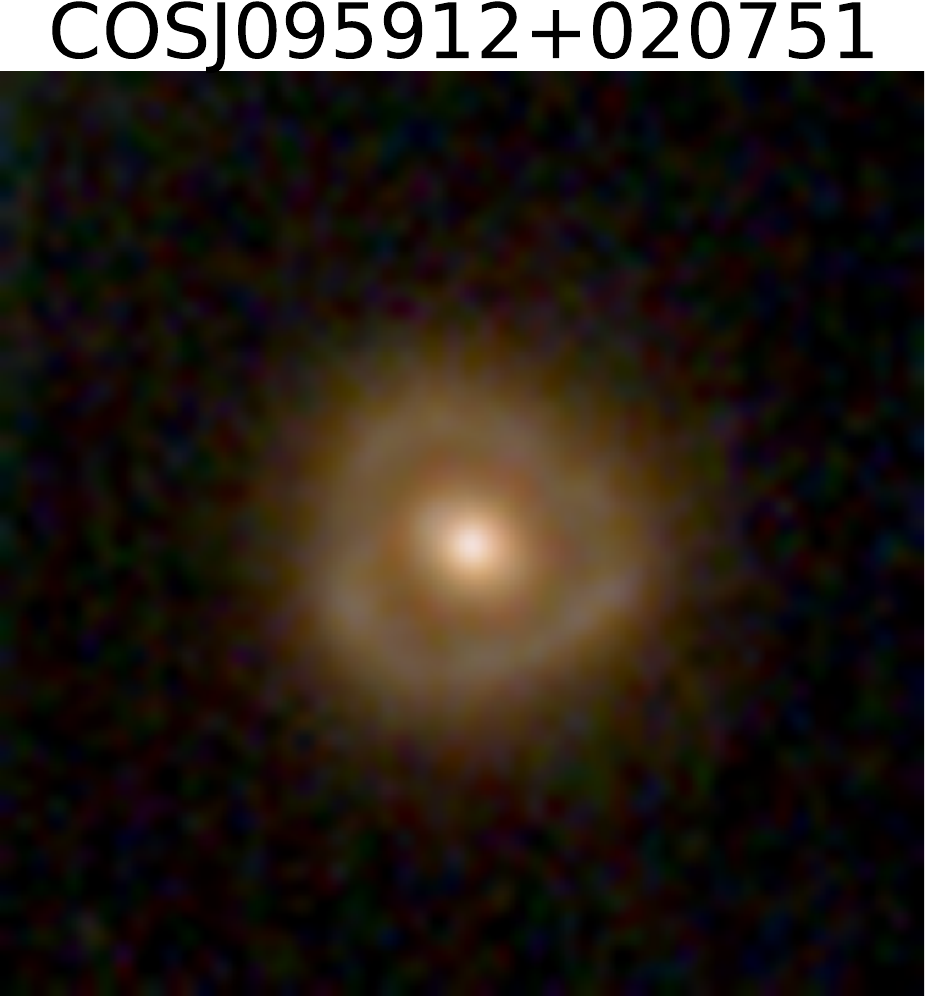}
\includegraphics[width=0.12\textwidth]{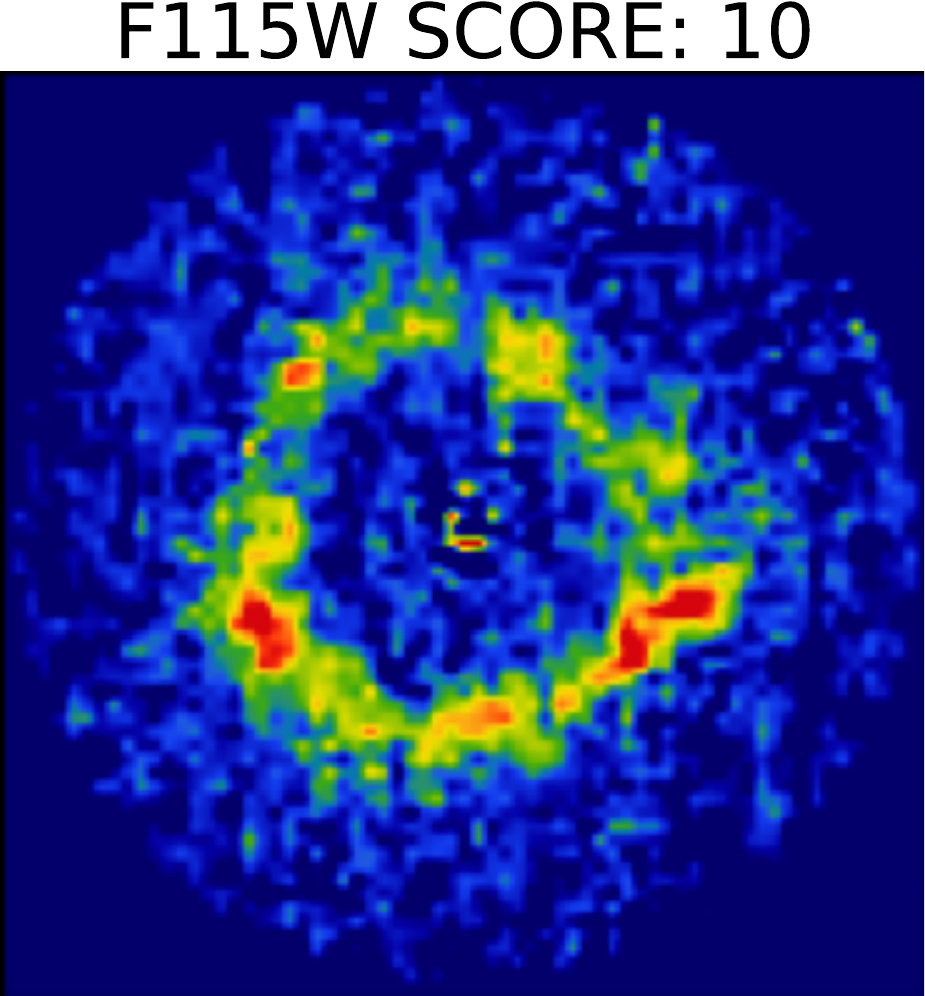}
\includegraphics[width=0.12\textwidth]{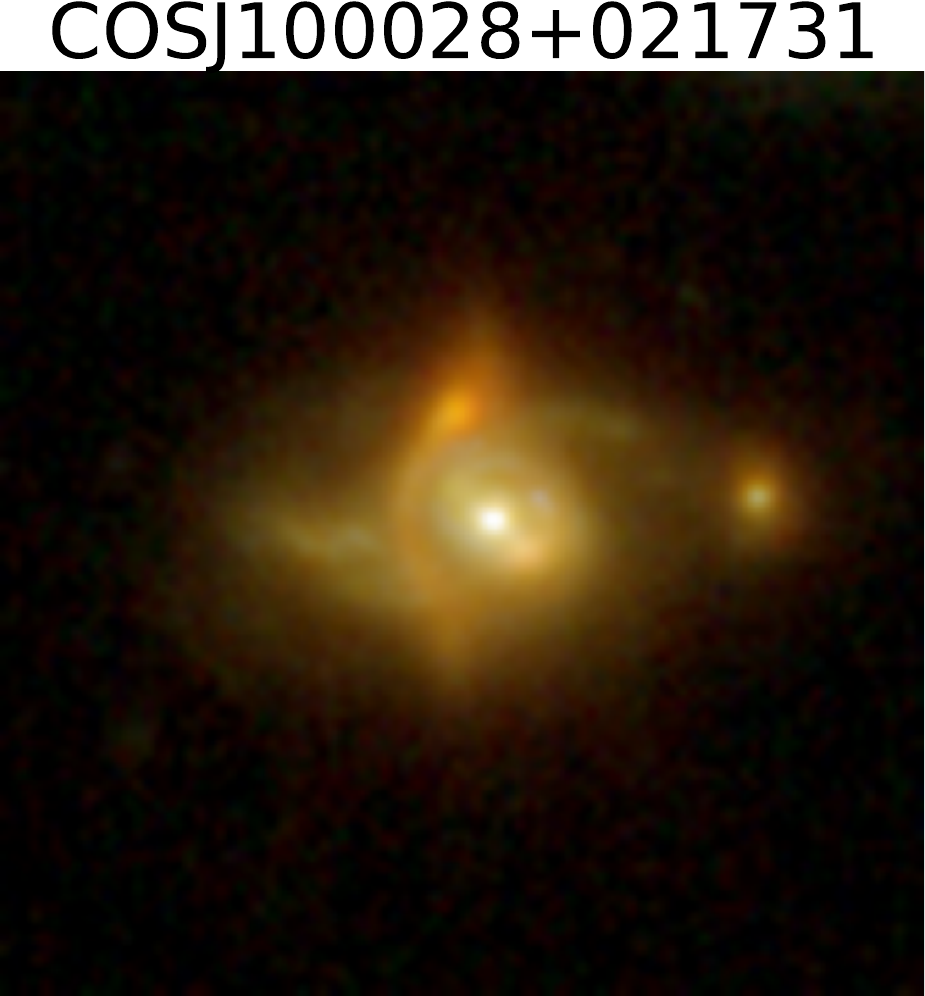}
\includegraphics[width=0.12\textwidth]{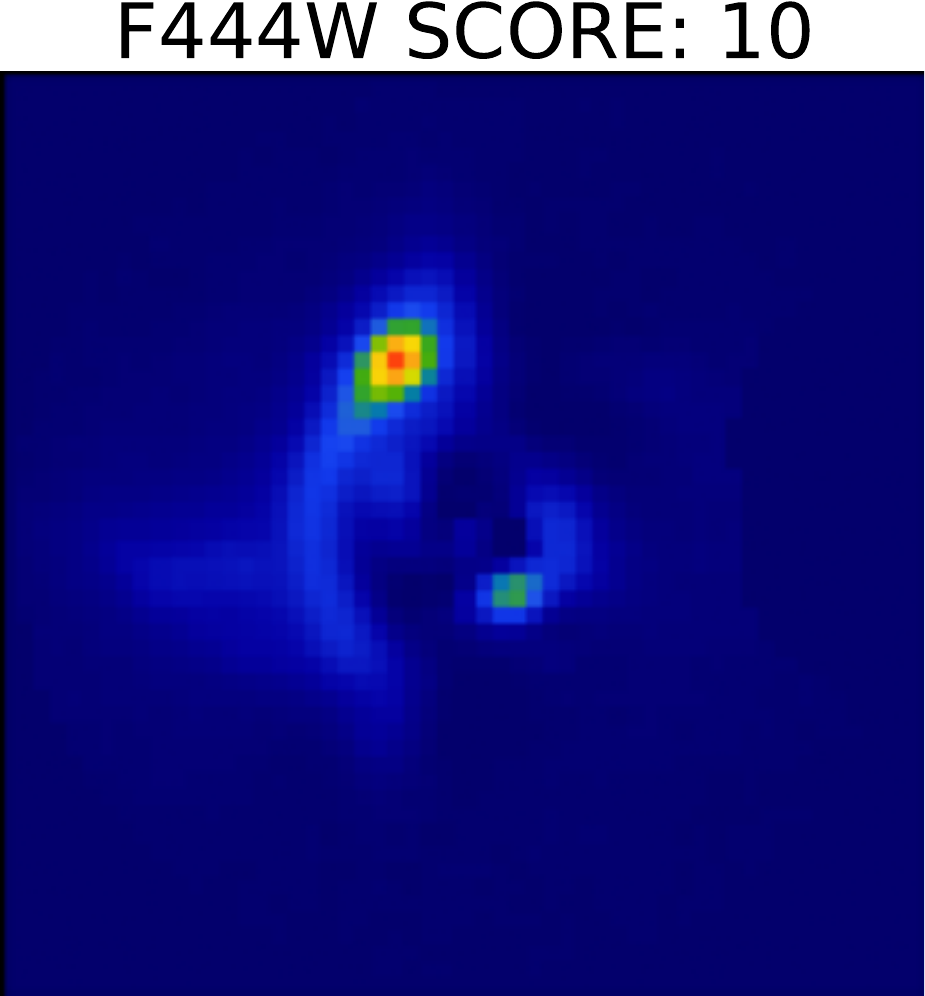}
\includegraphics[width=0.12\textwidth]{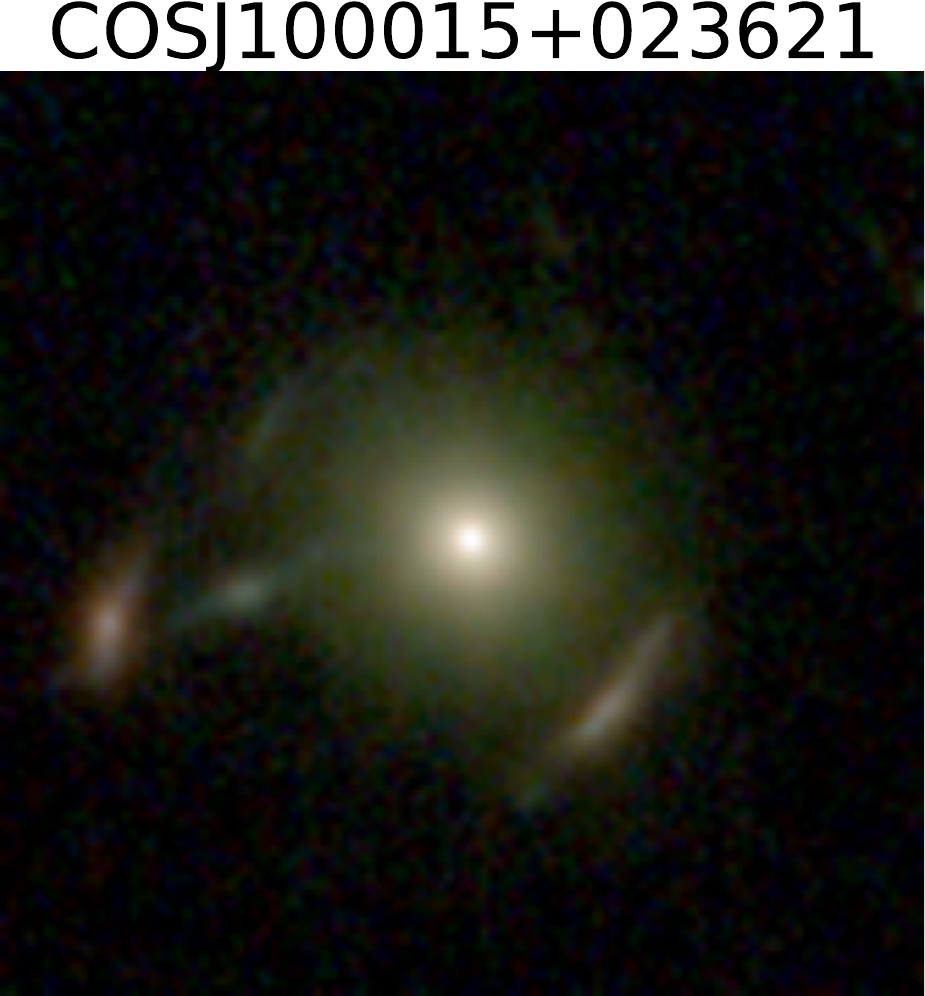}
\includegraphics[width=0.12\textwidth]{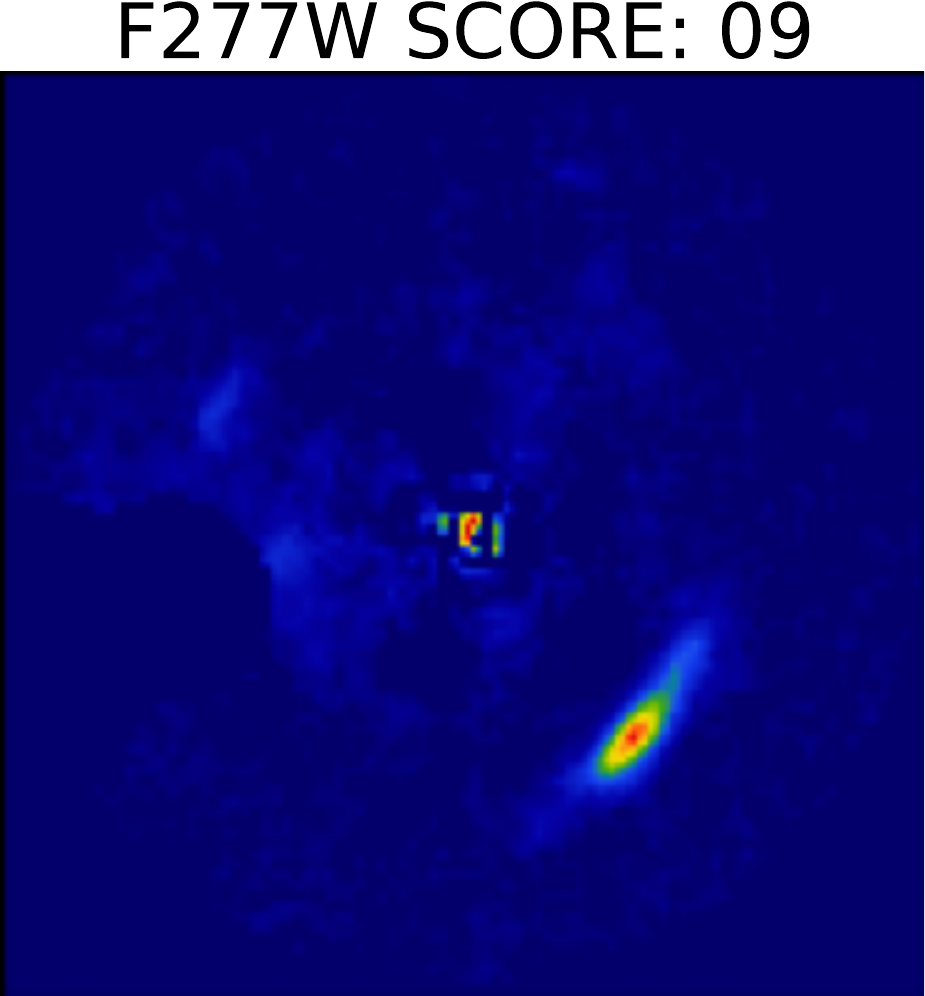}
\includegraphics[width=0.12\textwidth]{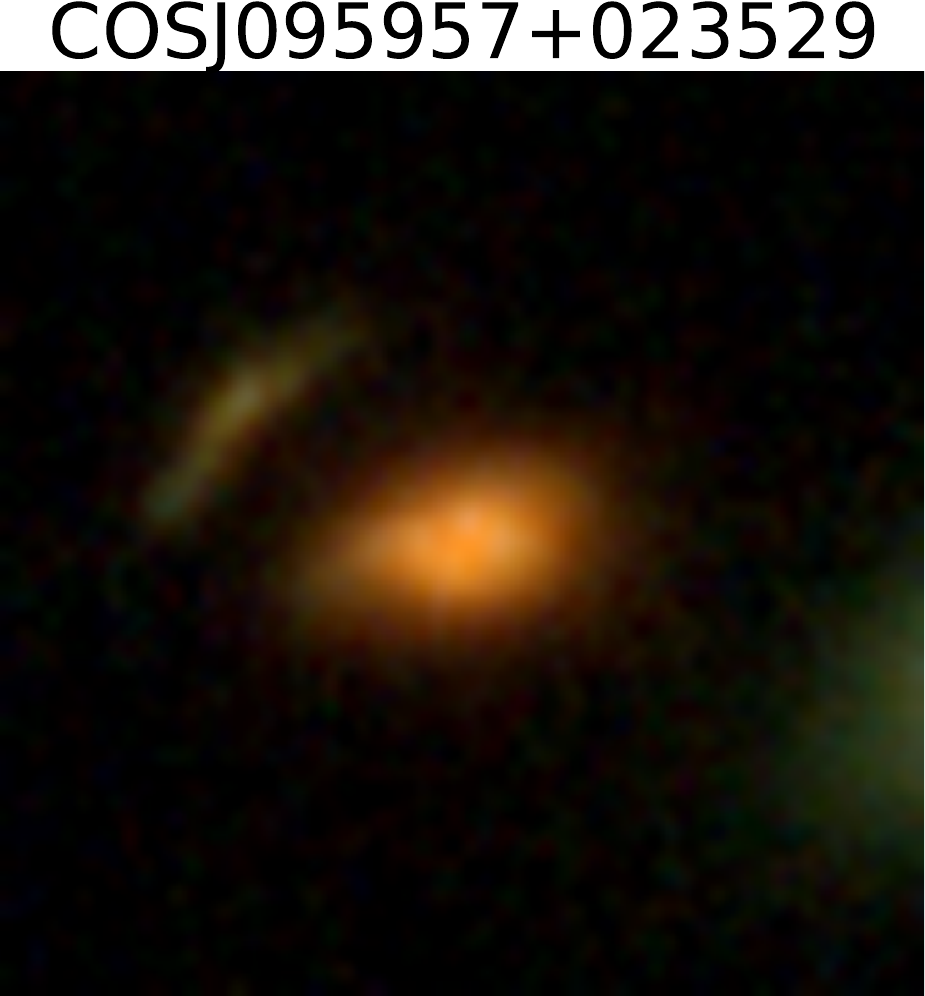}
\includegraphics[width=0.12\textwidth]{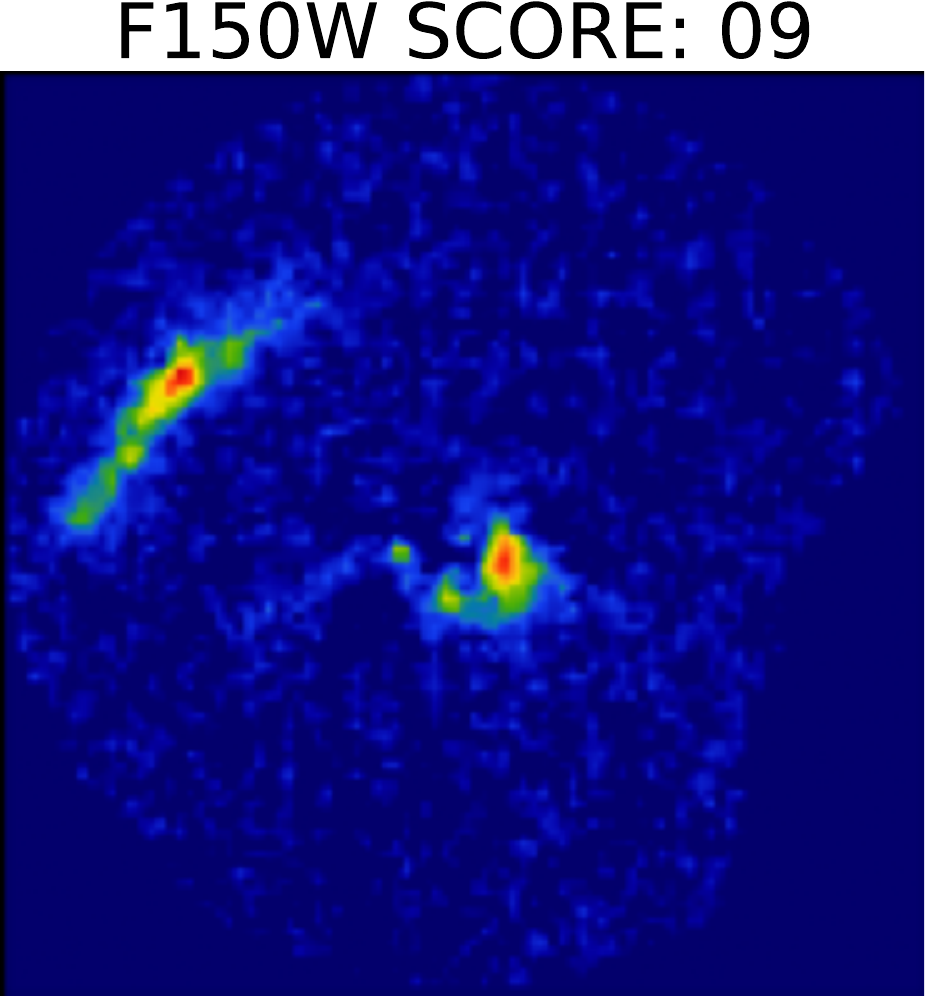}
\includegraphics[width=0.12\textwidth]{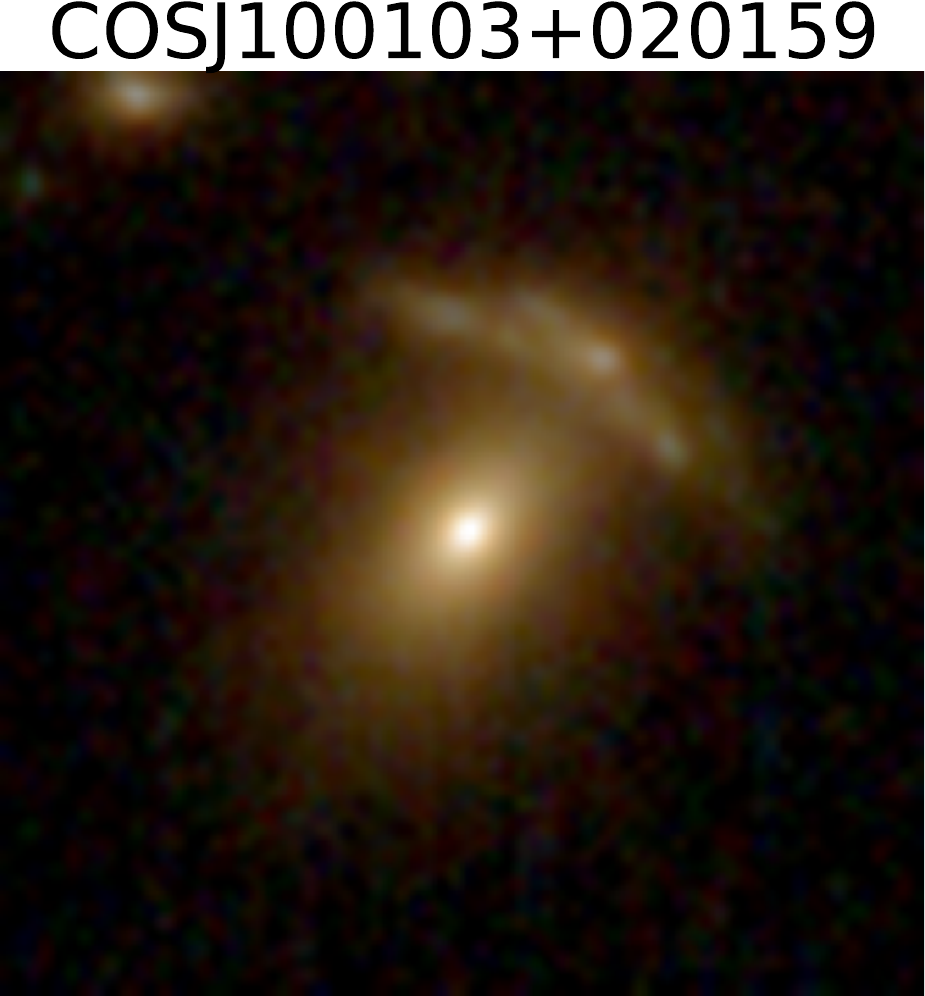}
\includegraphics[width=0.12\textwidth]{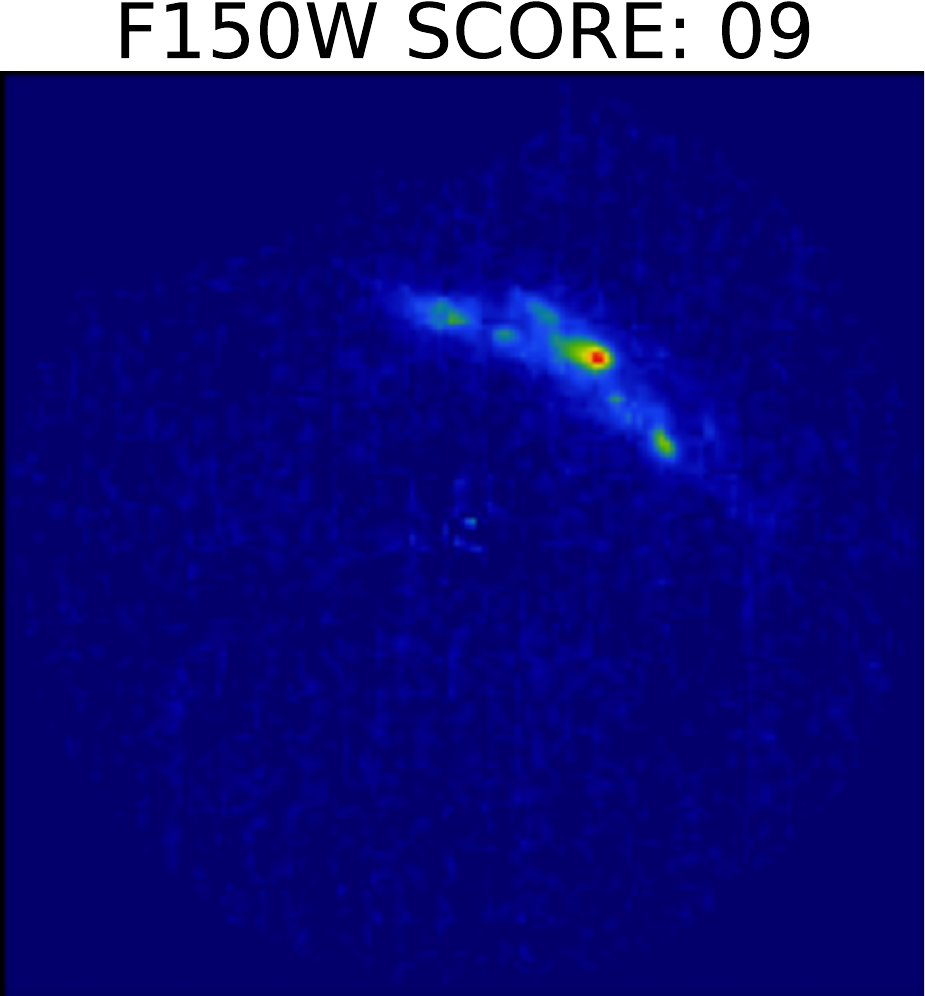}
\includegraphics[width=0.12\textwidth]{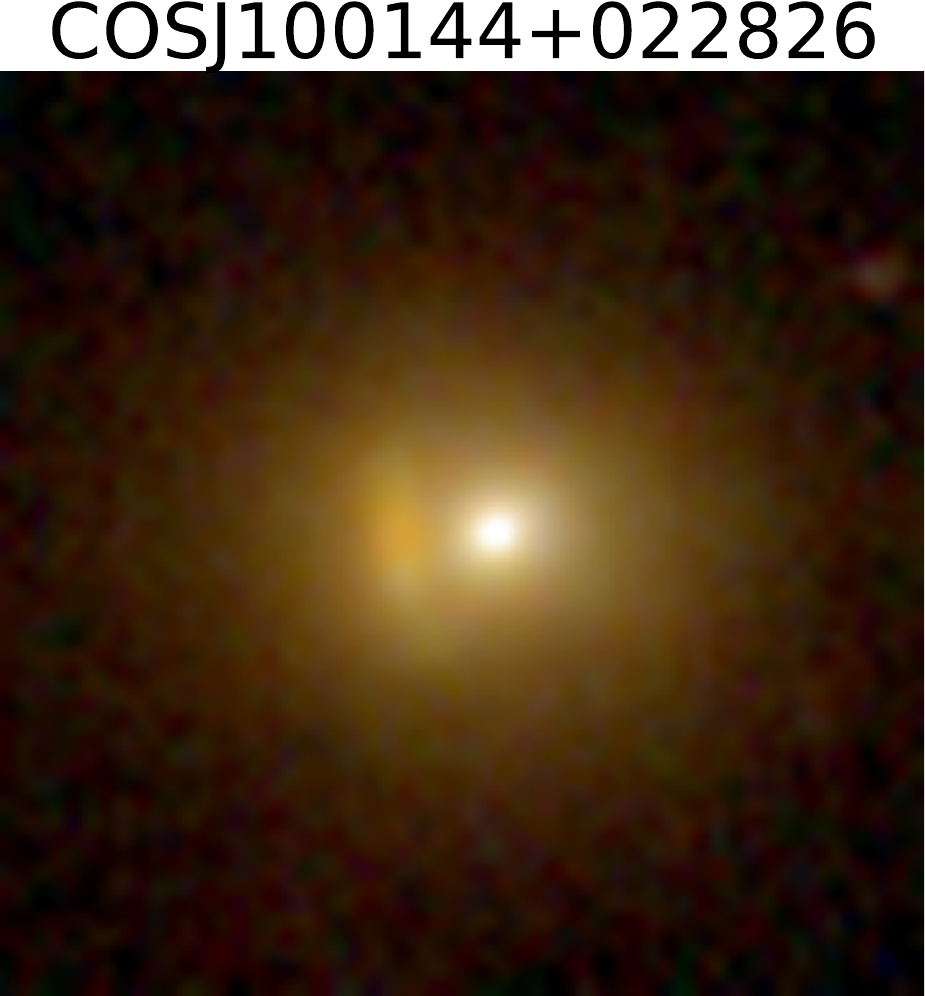}
\includegraphics[width=0.12\textwidth]{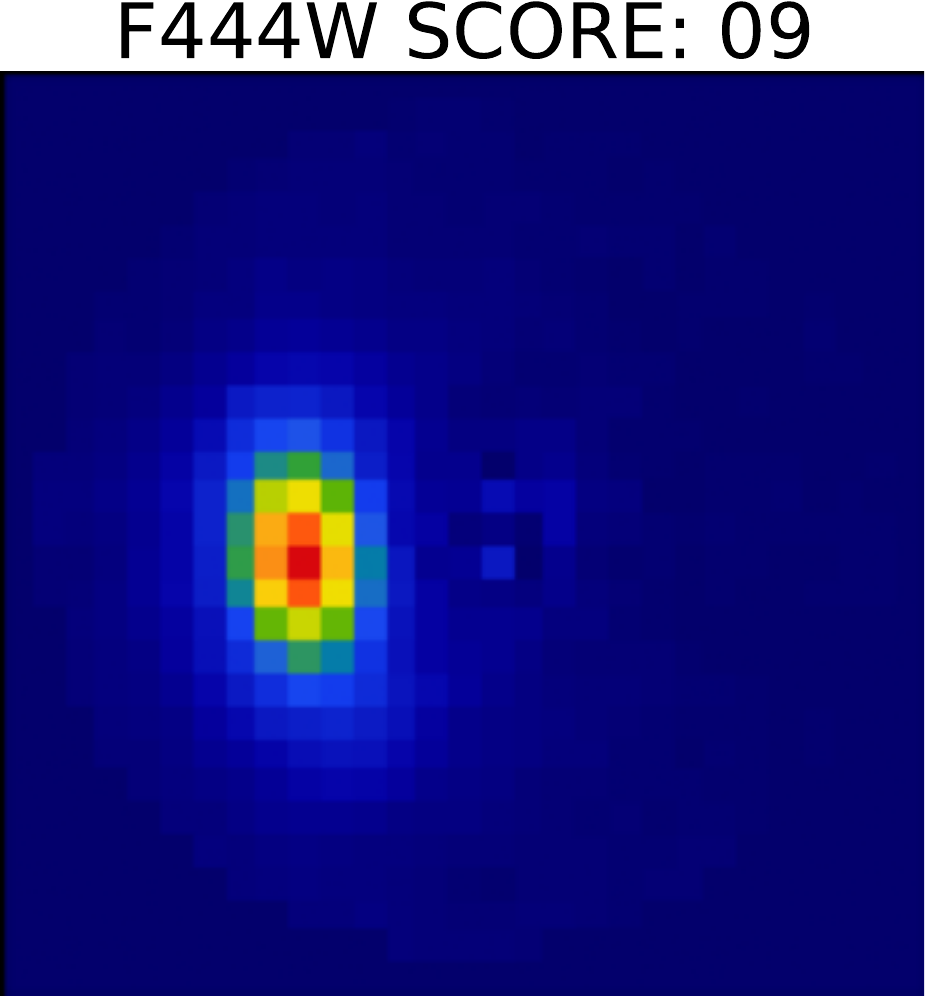}
\caption{
RGB composite images showing the lens and source emission, and a foreground-lens-subtracted image highlighting the lensed source, of the highest ranking COWLS candidates. The first 15 lenses belong to the COWLS Paper II spectacular lens sample of M25. Two M25 lenses, COSJ100024+015334 and COSJ095921+020638, are excluded as they appear in \cref{figure:VisualLensFit} and \cref{figure:VisualMGELensSub}, respectively. The 143 highest-ranked candidates from the second round of visual inspection (all scoring 5 or above) are then shown. This indicates most inspectors selected ``A: High confidence this is a strong lens.'' or ``B: Likely a strong lens, but there is ambiguity.'' Images are arranged in descending order of rank, starting from the top left, with a maximum score of 12 if all six inspectors gave an `A' grade. Inspectors had access to all four wavelengths and lens models during grading, making some candidates more convincing with full data. The complete visualisations are available at the following URL:~\github{https://github.com/Jammy2211/COWLS_COSMOS_Web_Lens_Survey}.
}
\label{figure:CutoutA}
\vspace{-11pt} 
\end{figure*}

\begin{figure*}\ContinuedFloat
\centering
\includegraphics[width=0.12\textwidth]{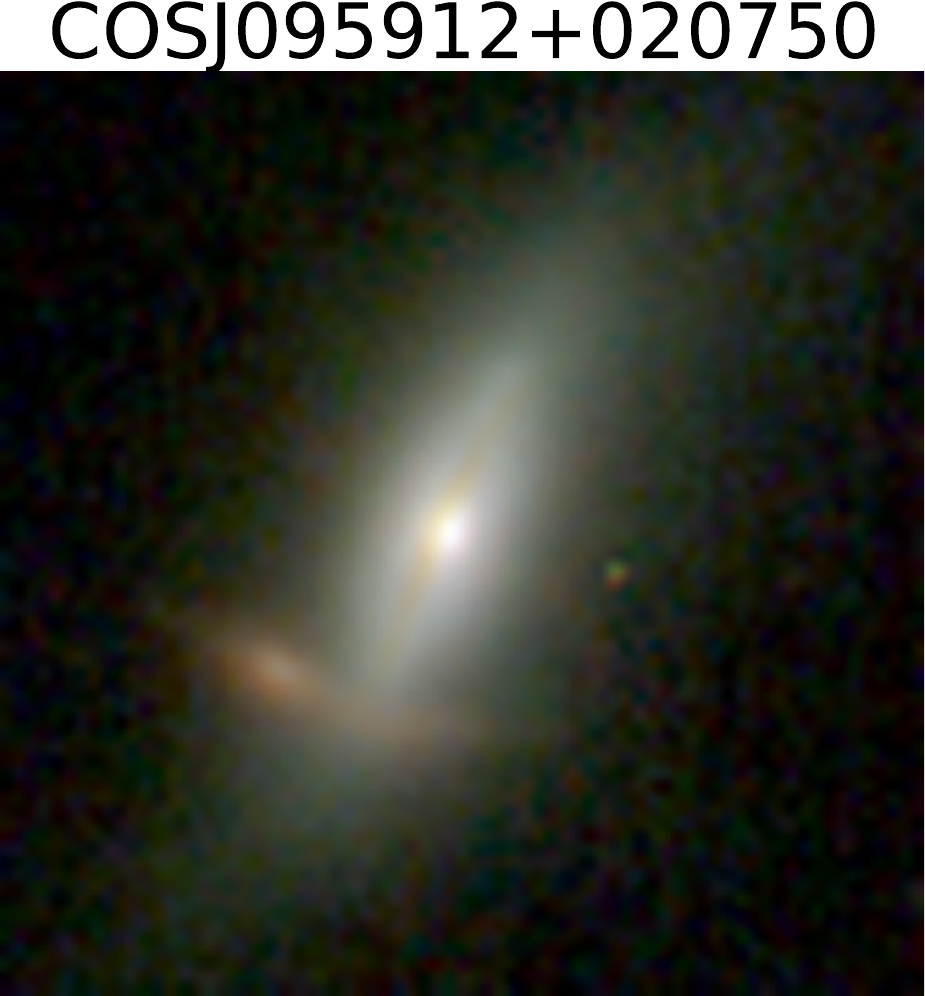}
\includegraphics[width=0.12\textwidth]{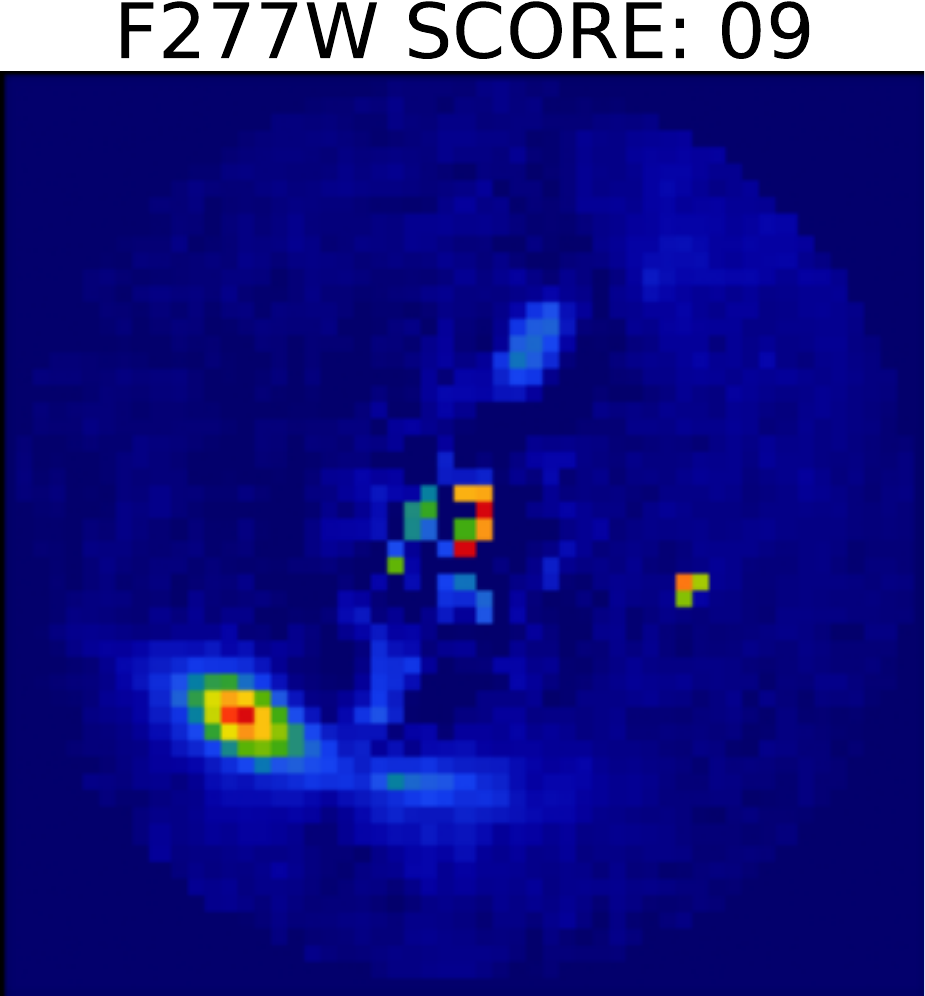}
\includegraphics[width=0.12\textwidth]{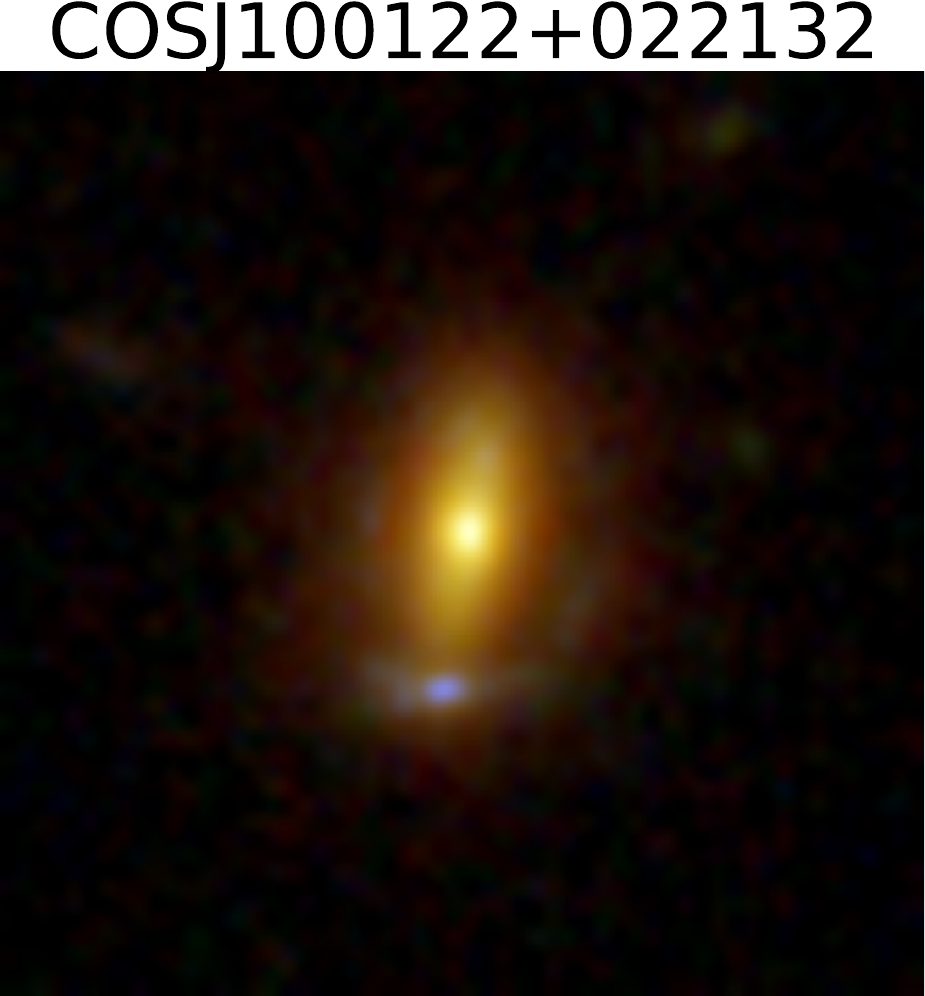}
\includegraphics[width=0.12\textwidth]{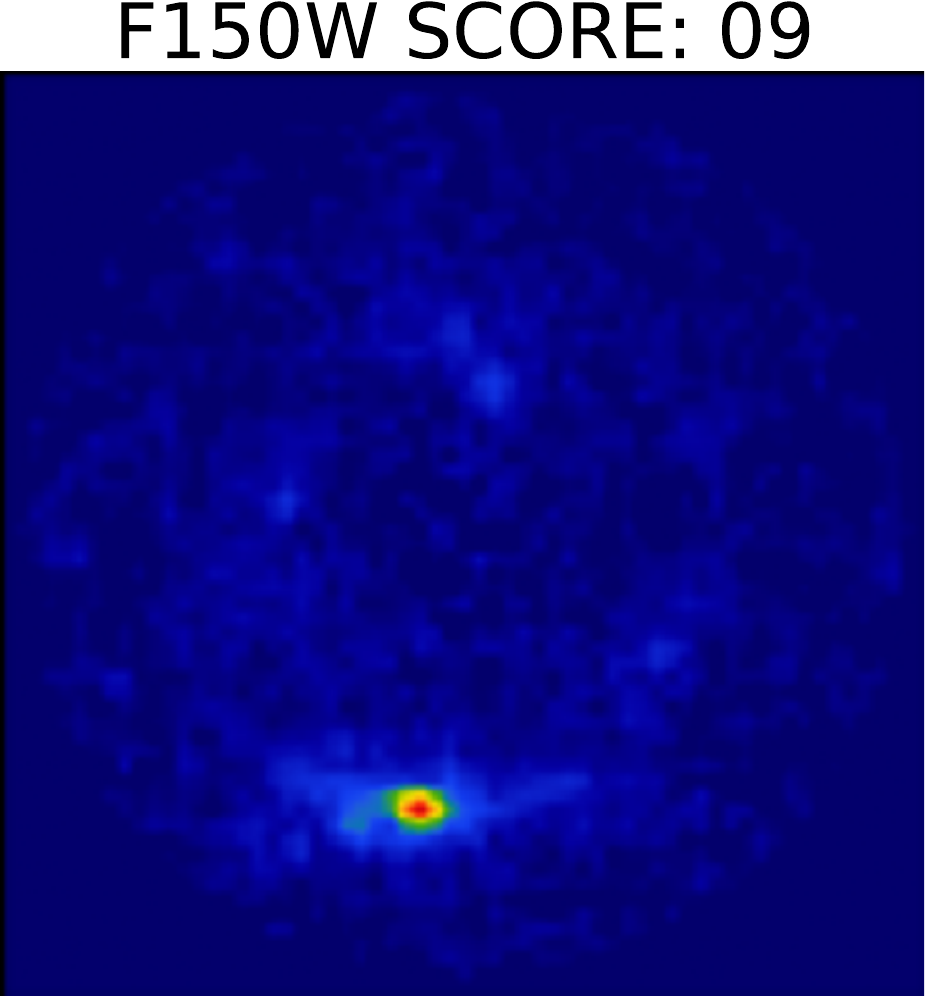}
\includegraphics[width=0.12\textwidth]{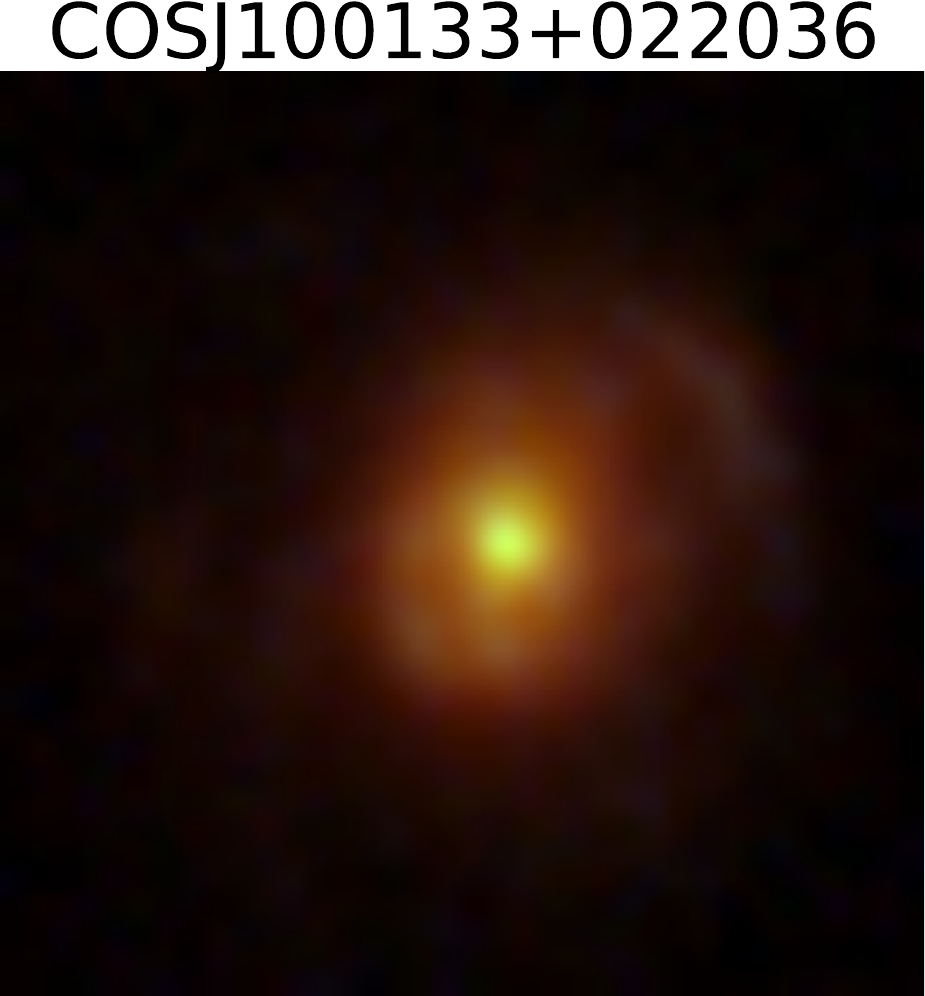}
\includegraphics[width=0.12\textwidth]{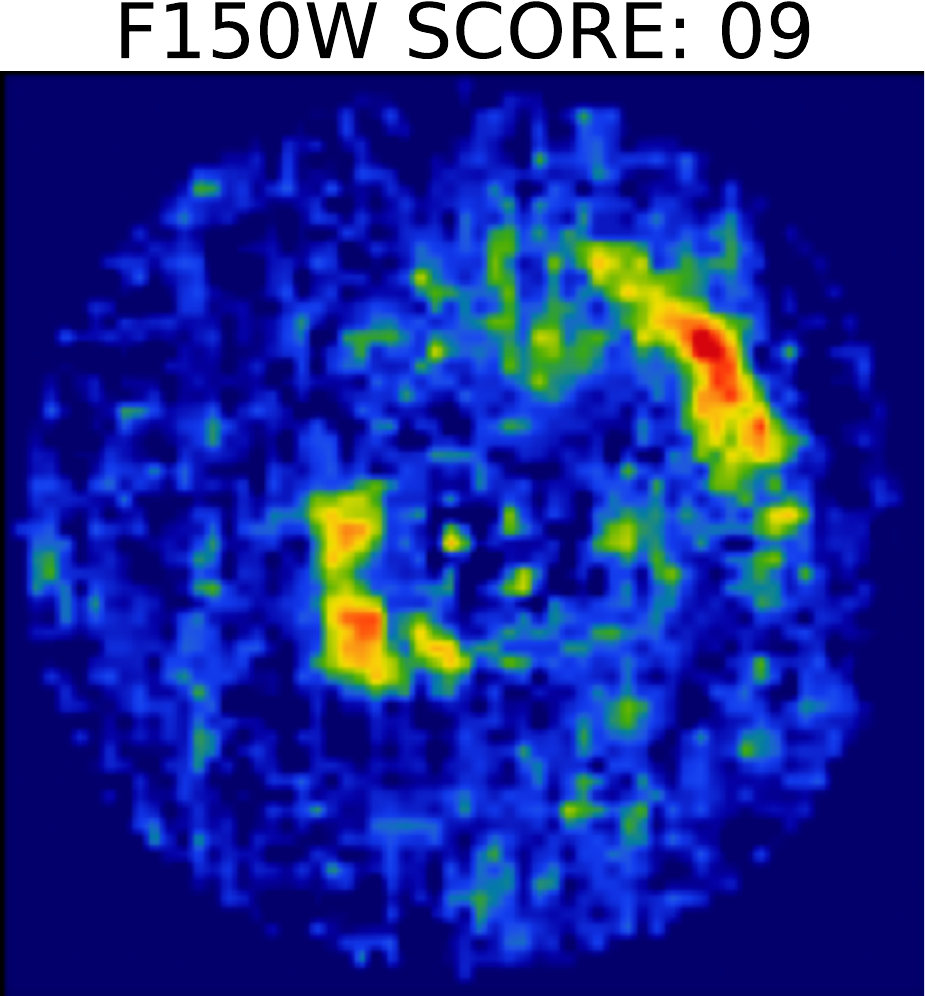}
\includegraphics[width=0.12\textwidth]{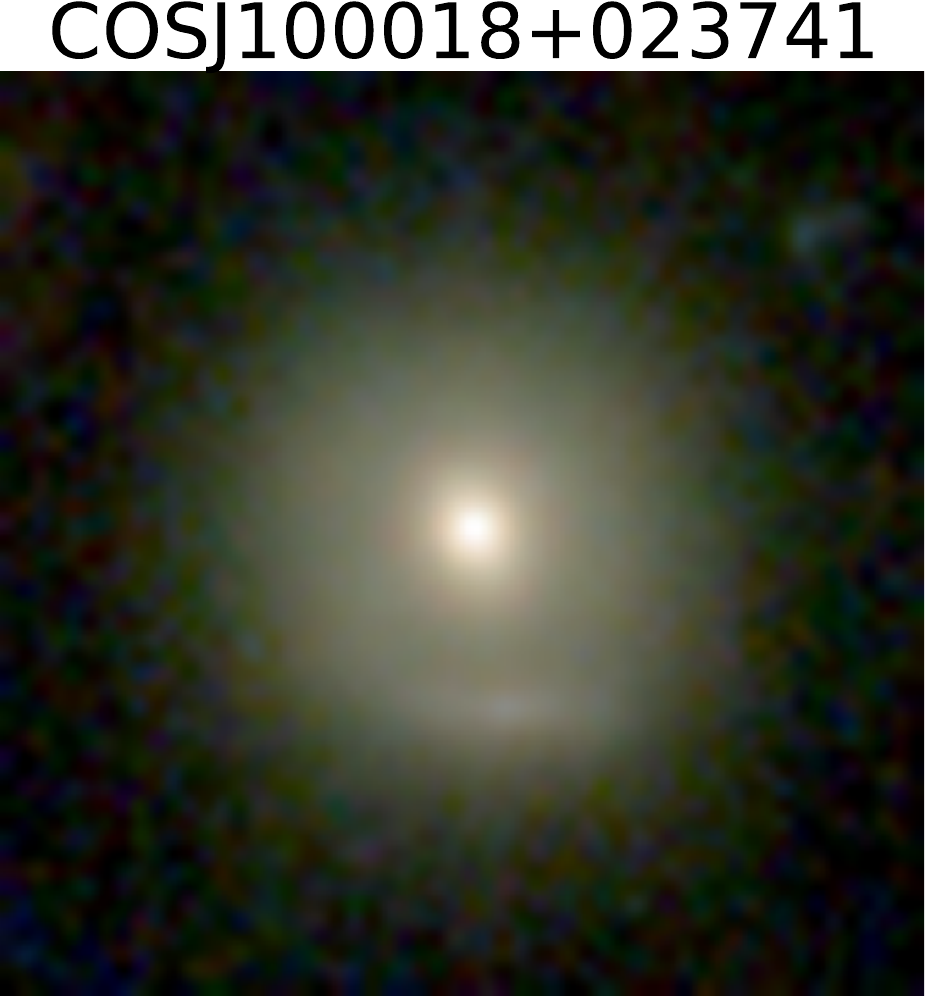}
\includegraphics[width=0.12\textwidth]{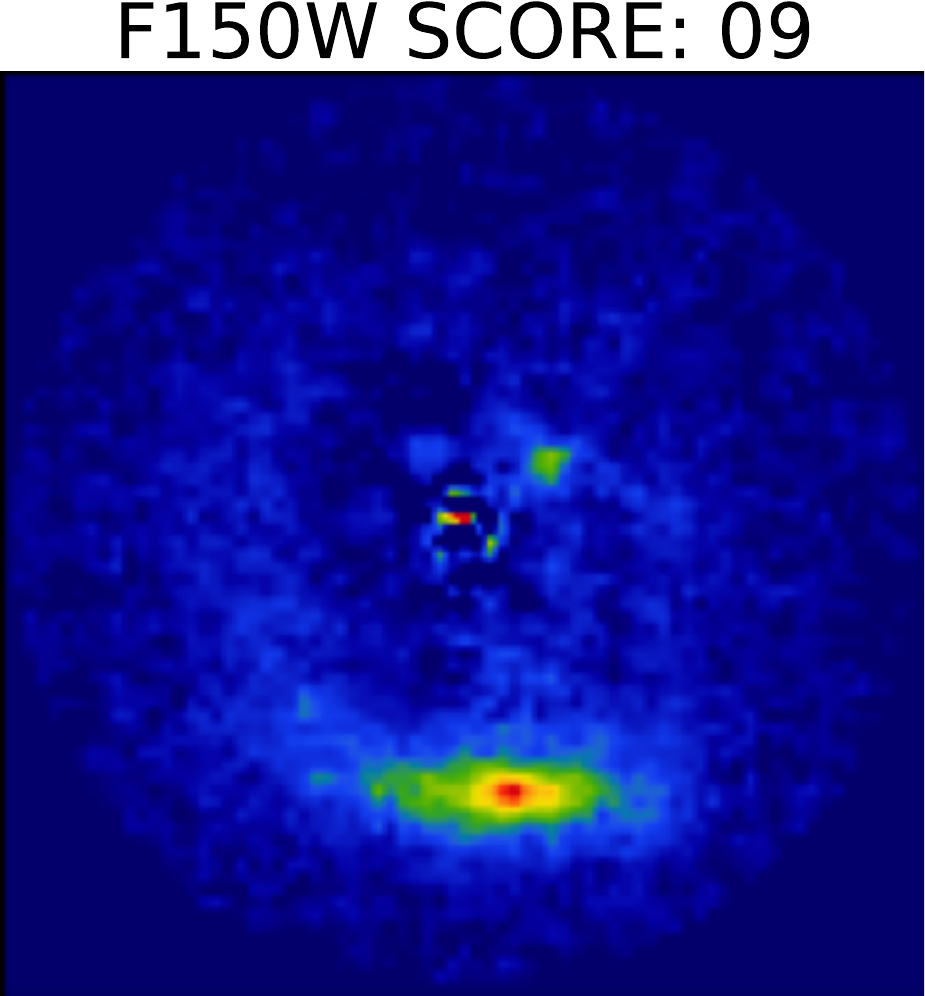}
\includegraphics[width=0.12\textwidth]{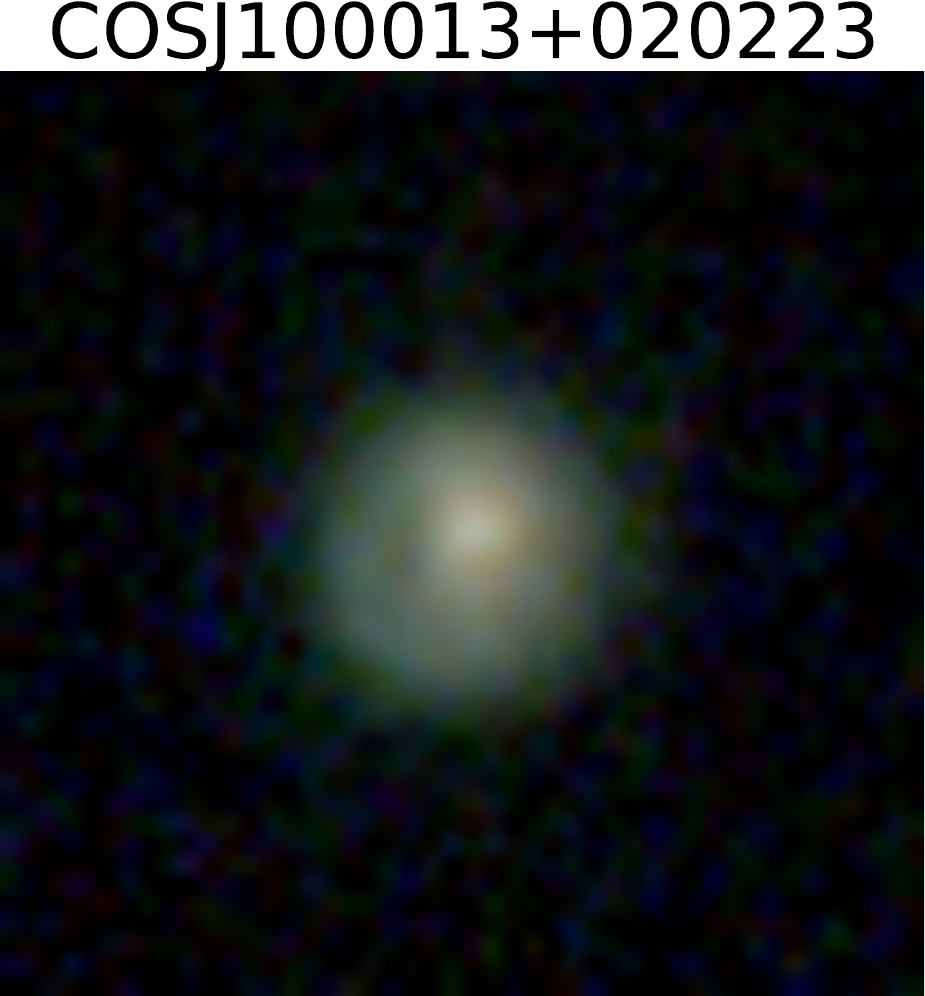}
\includegraphics[width=0.12\textwidth]{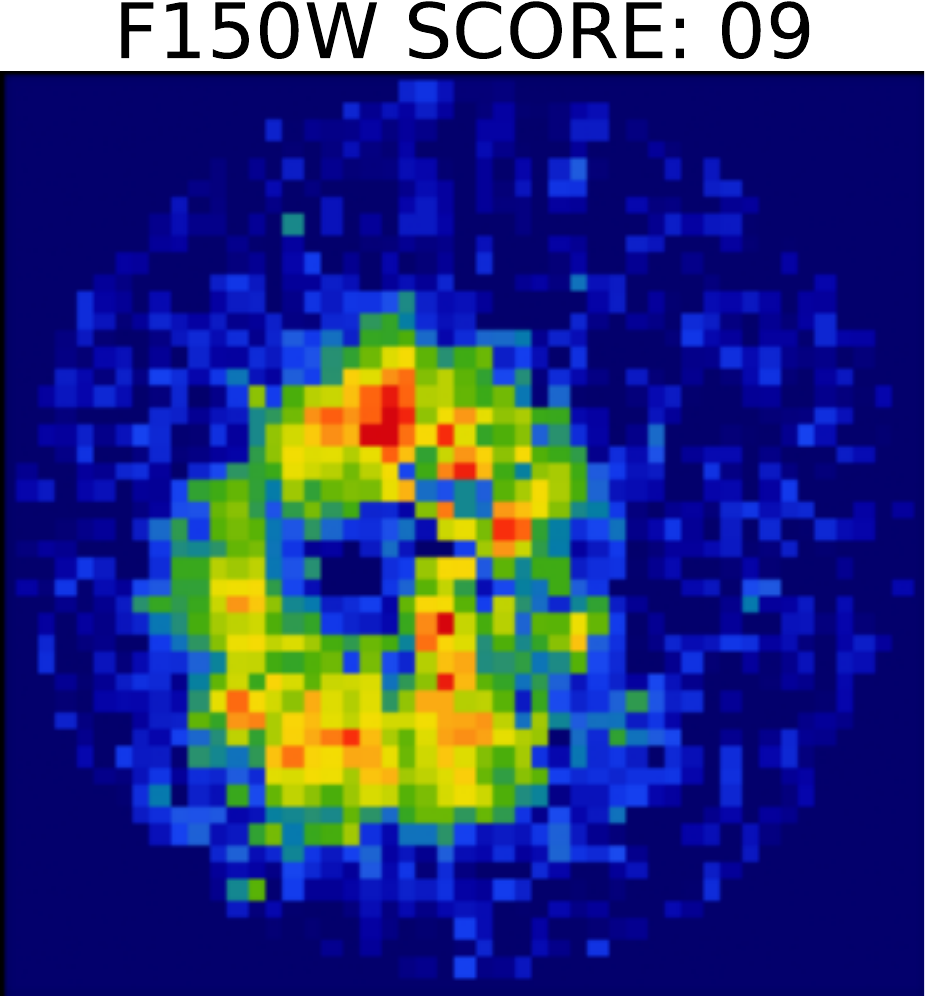}
\includegraphics[width=0.12\textwidth]{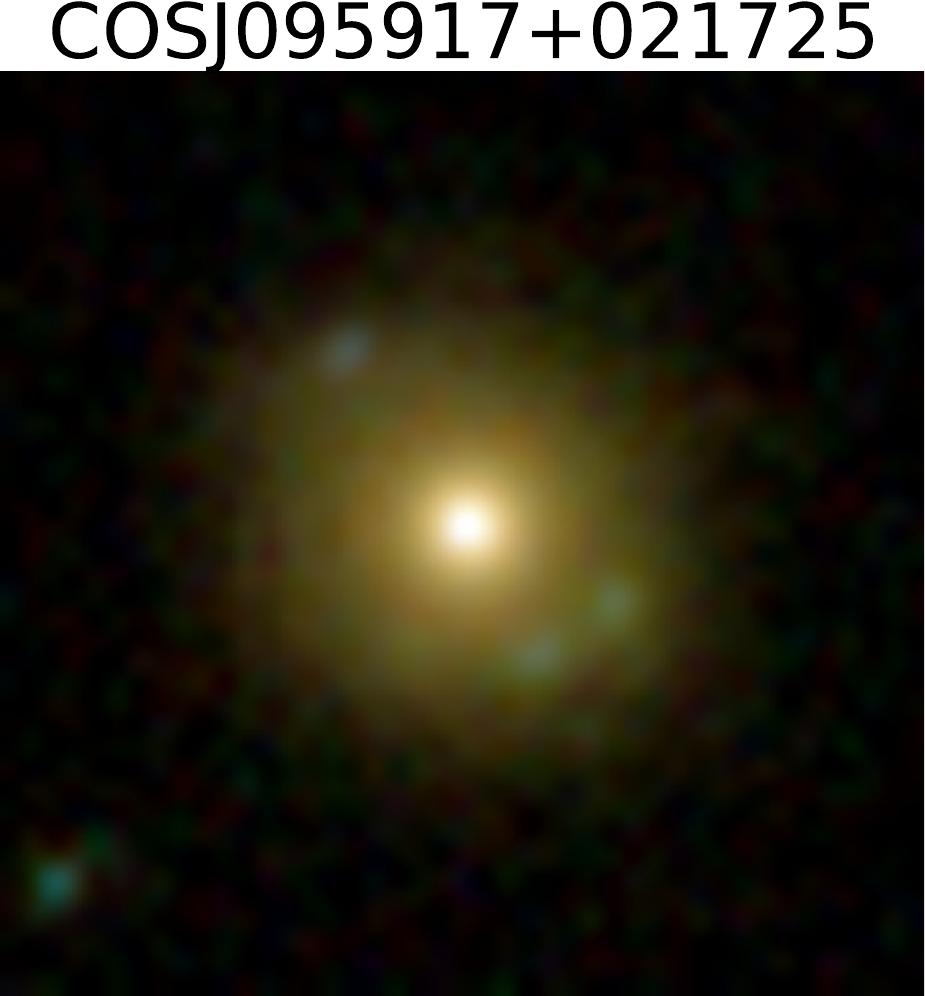}
\includegraphics[width=0.12\textwidth]{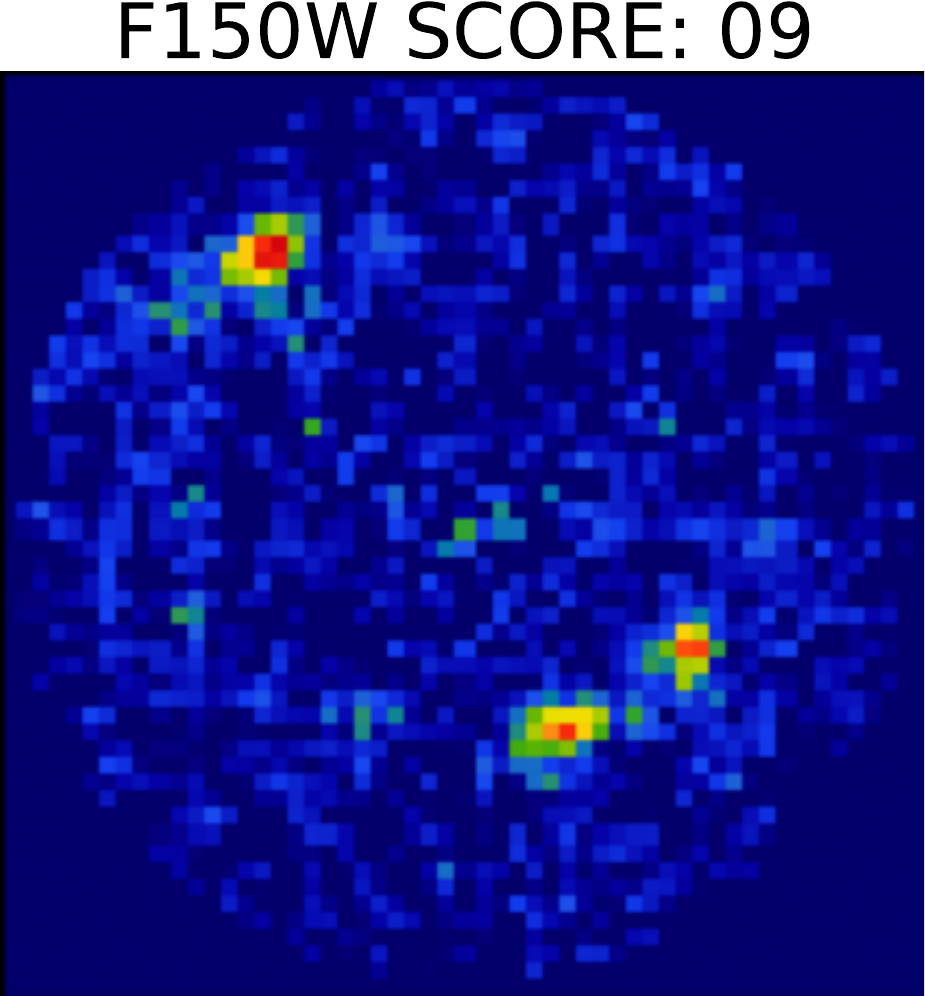}
\includegraphics[width=0.12\textwidth]{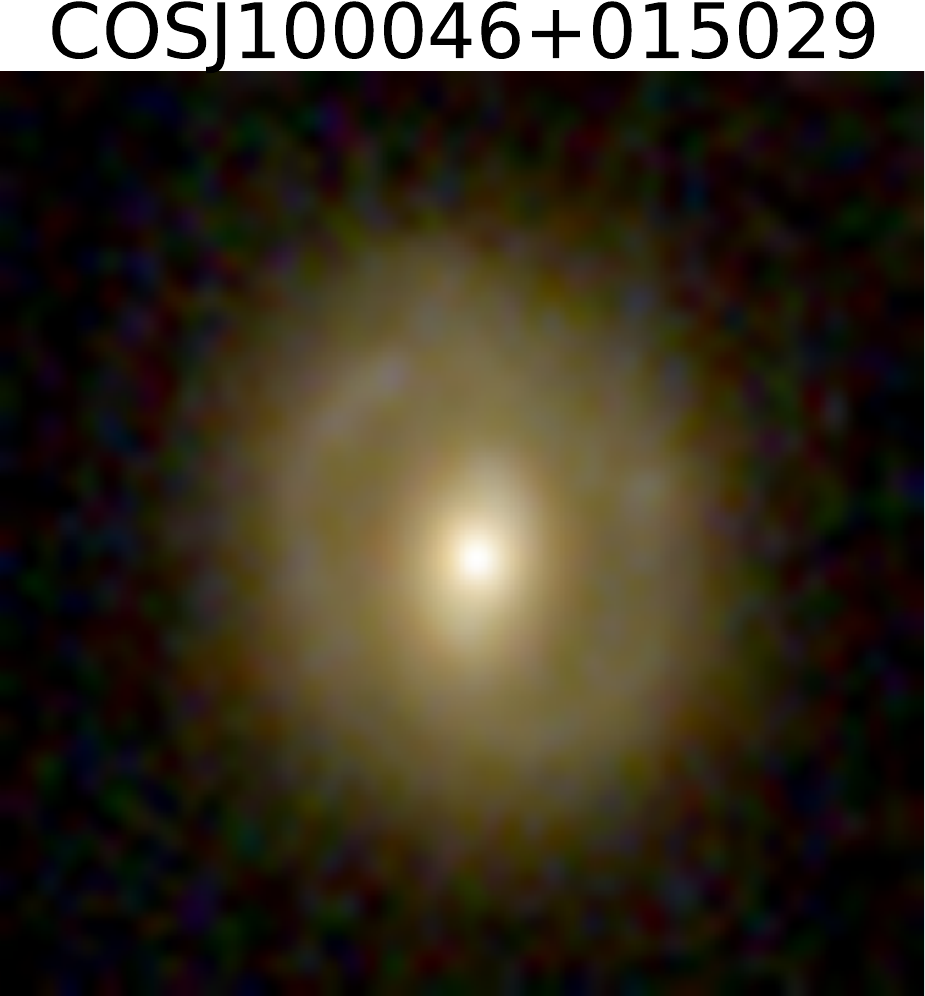}
\includegraphics[width=0.12\textwidth]{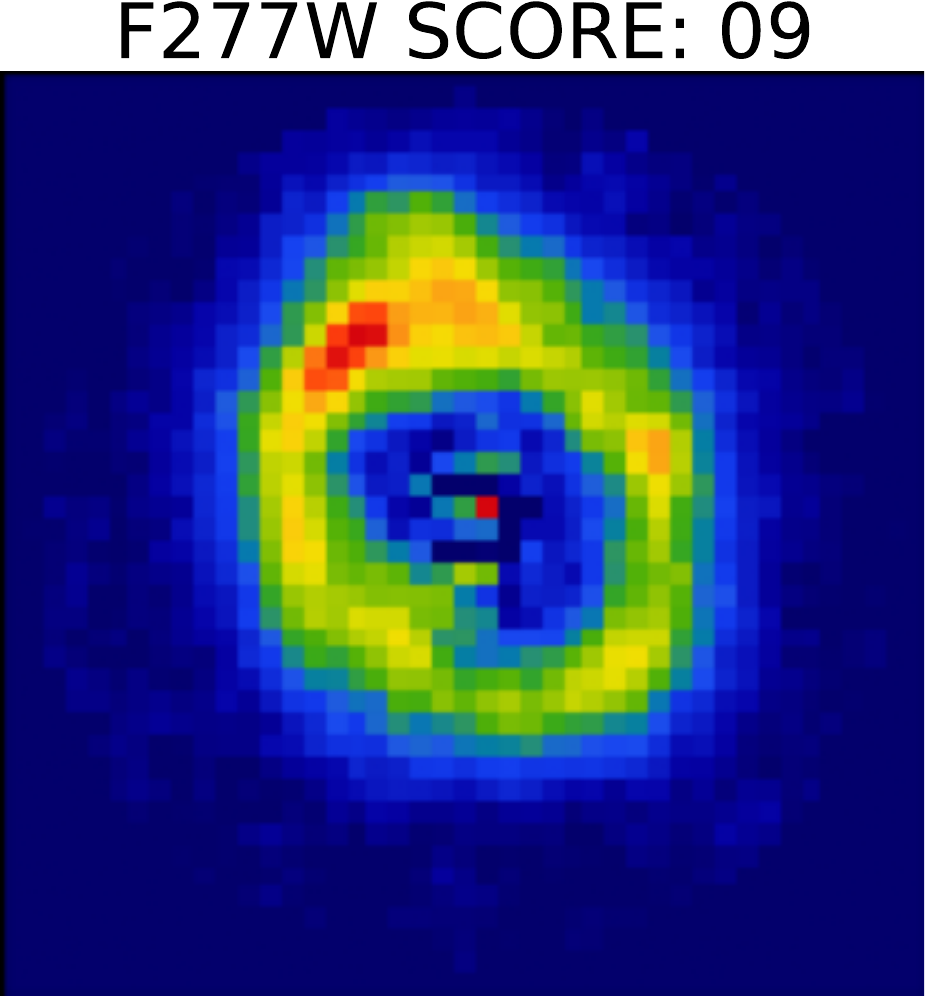}
\includegraphics[width=0.12\textwidth]{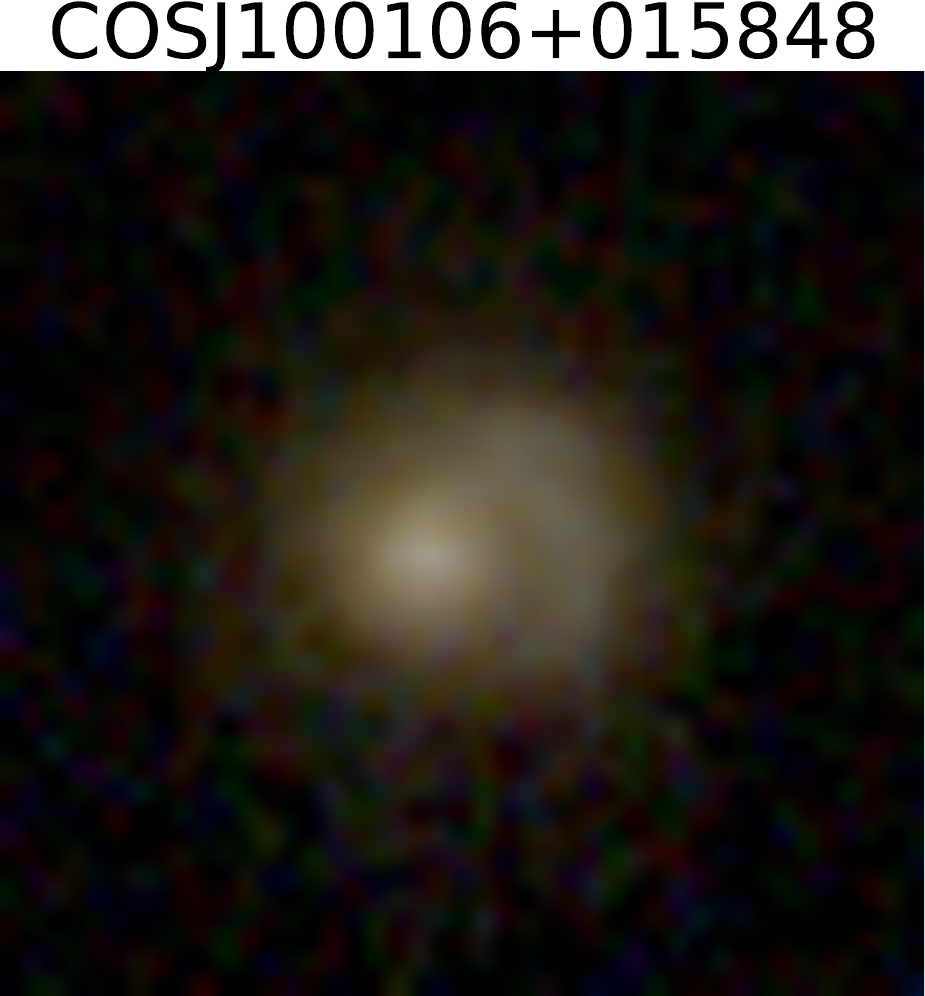}
\includegraphics[width=0.12\textwidth]{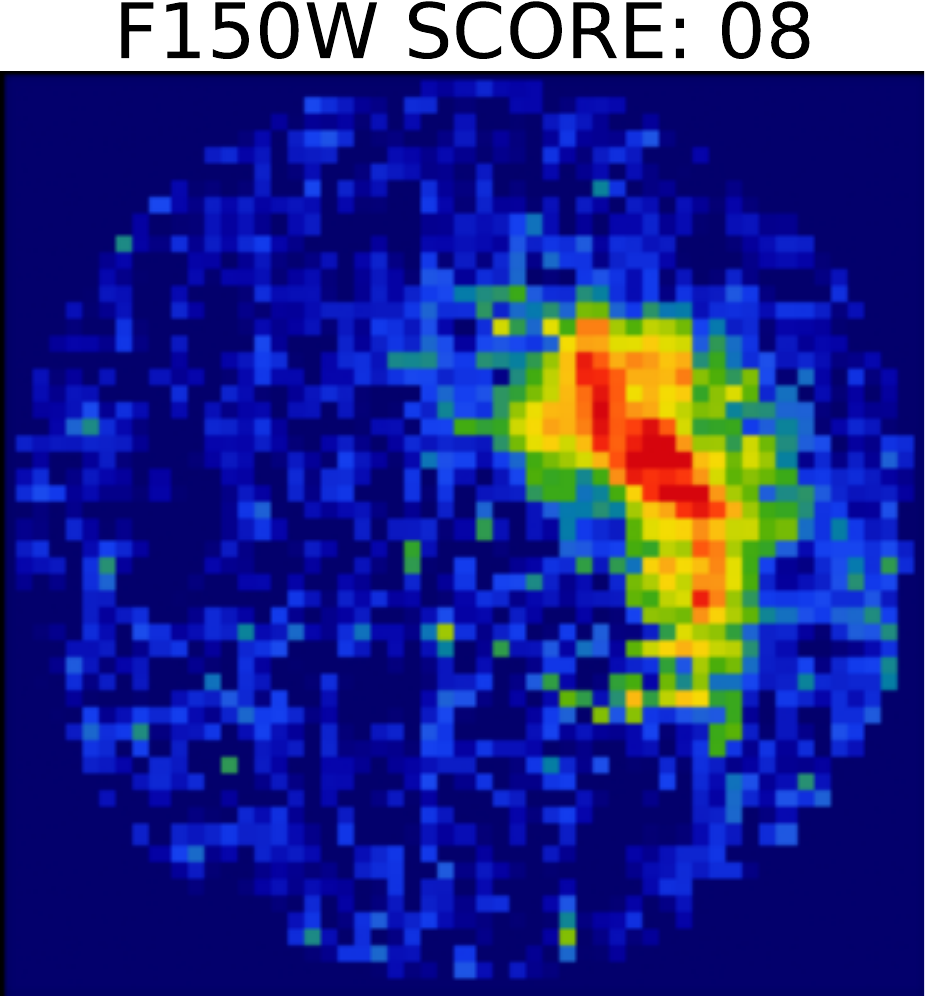}
\includegraphics[width=0.12\textwidth]{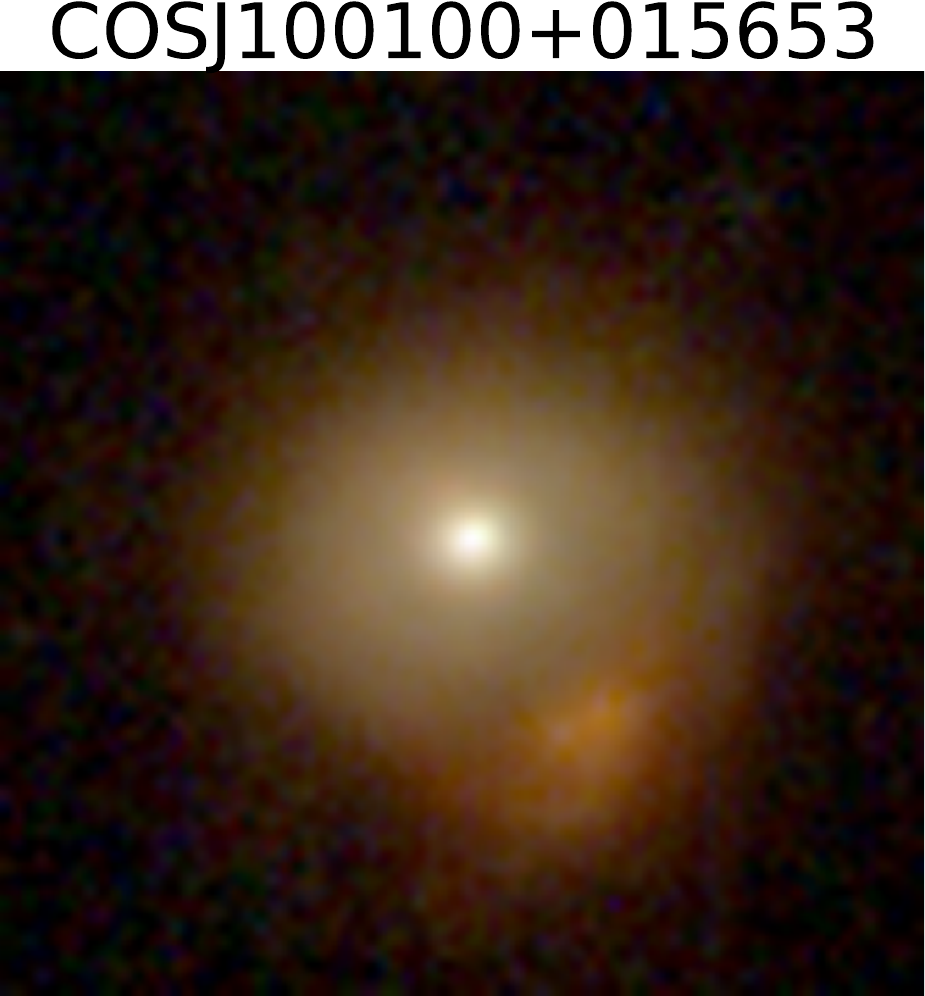}
\includegraphics[width=0.12\textwidth]{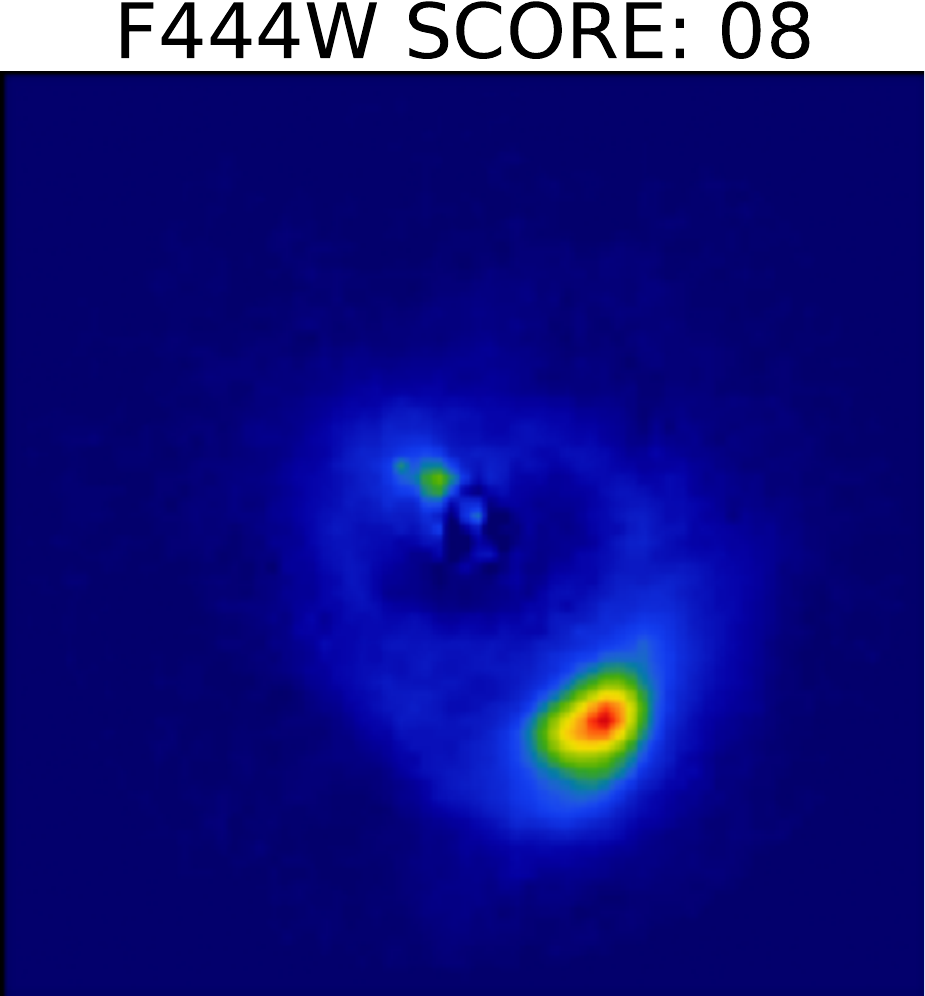}
\includegraphics[width=0.12\textwidth]{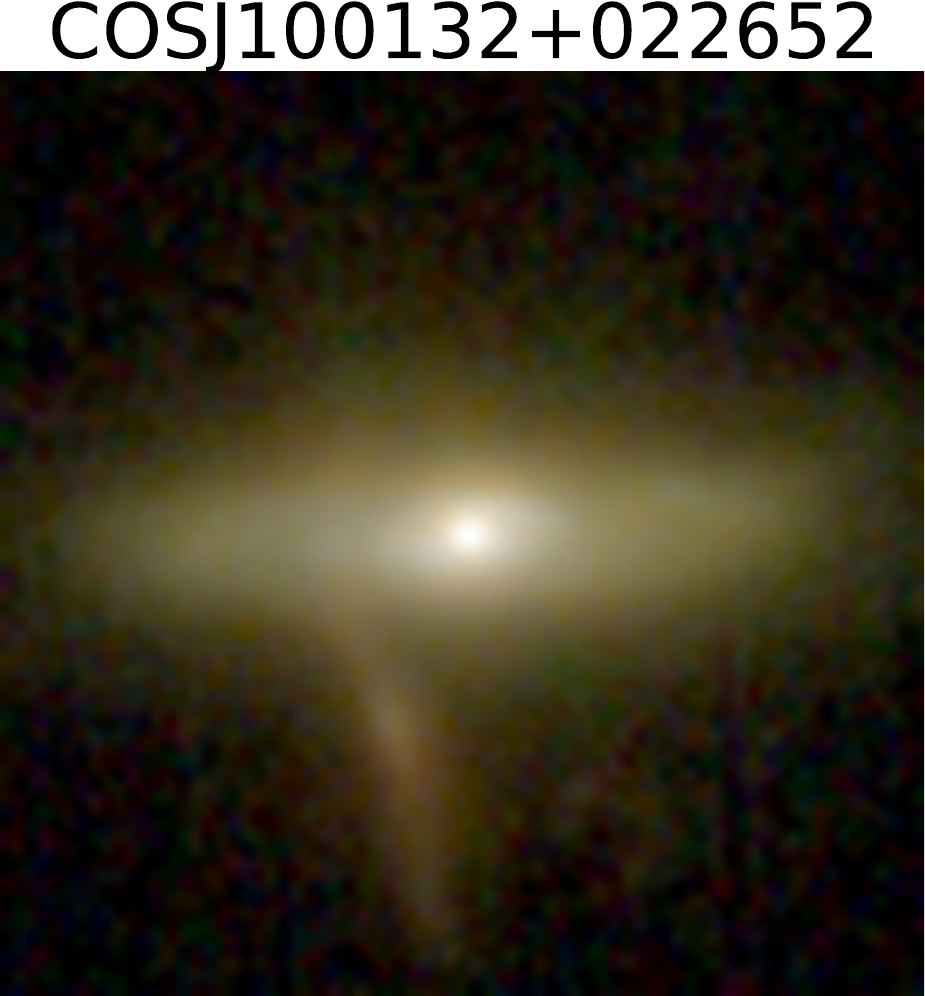}
\includegraphics[width=0.12\textwidth]{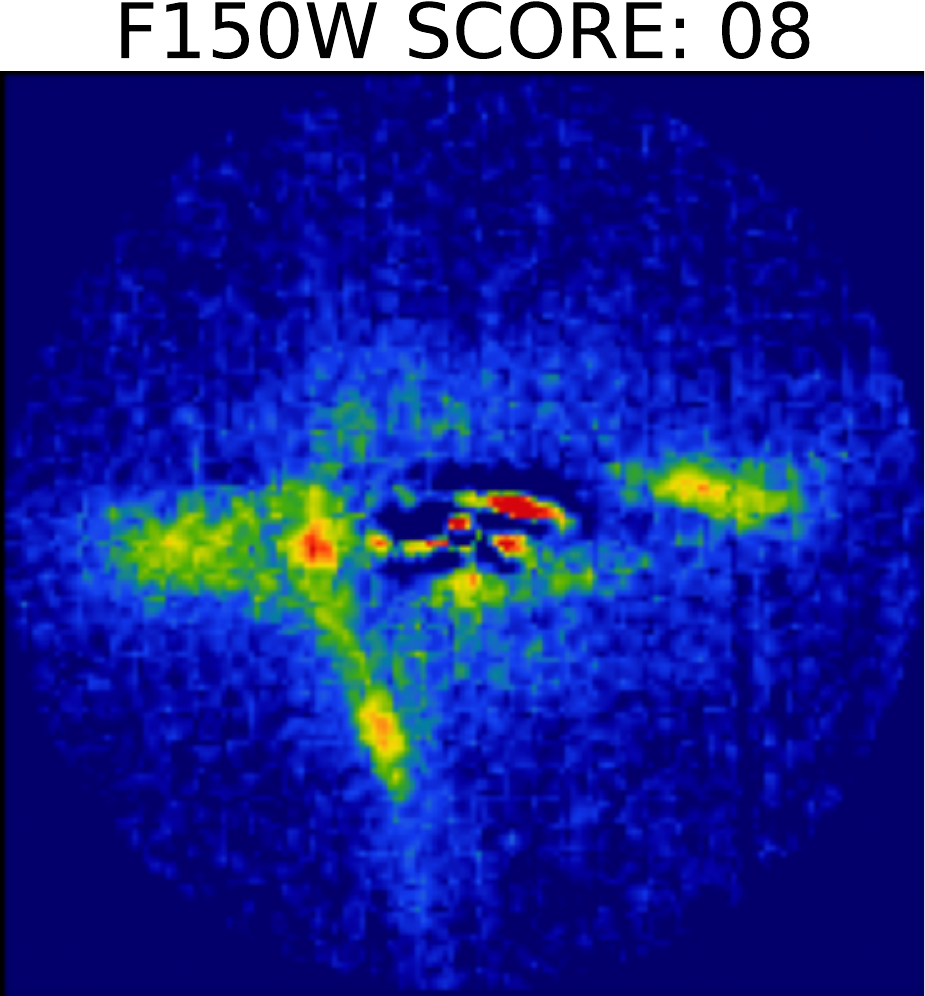}
\includegraphics[width=0.12\textwidth]{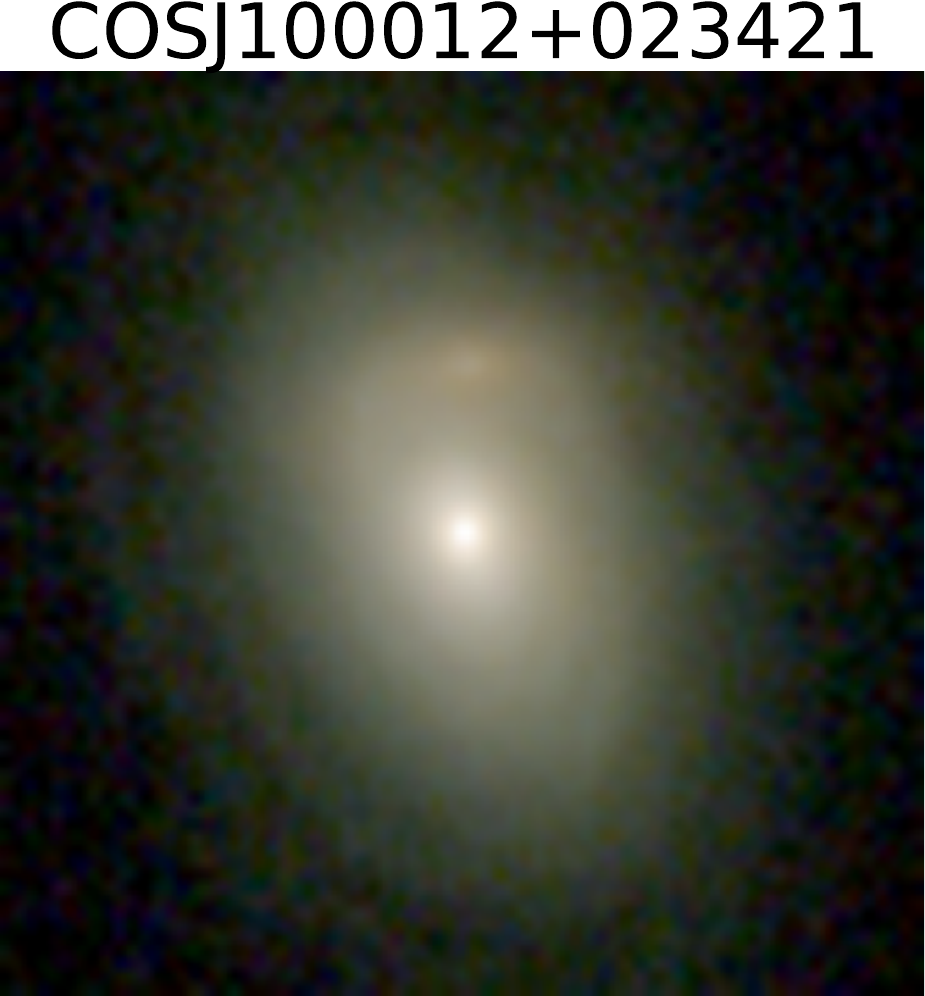}
\includegraphics[width=0.12\textwidth]{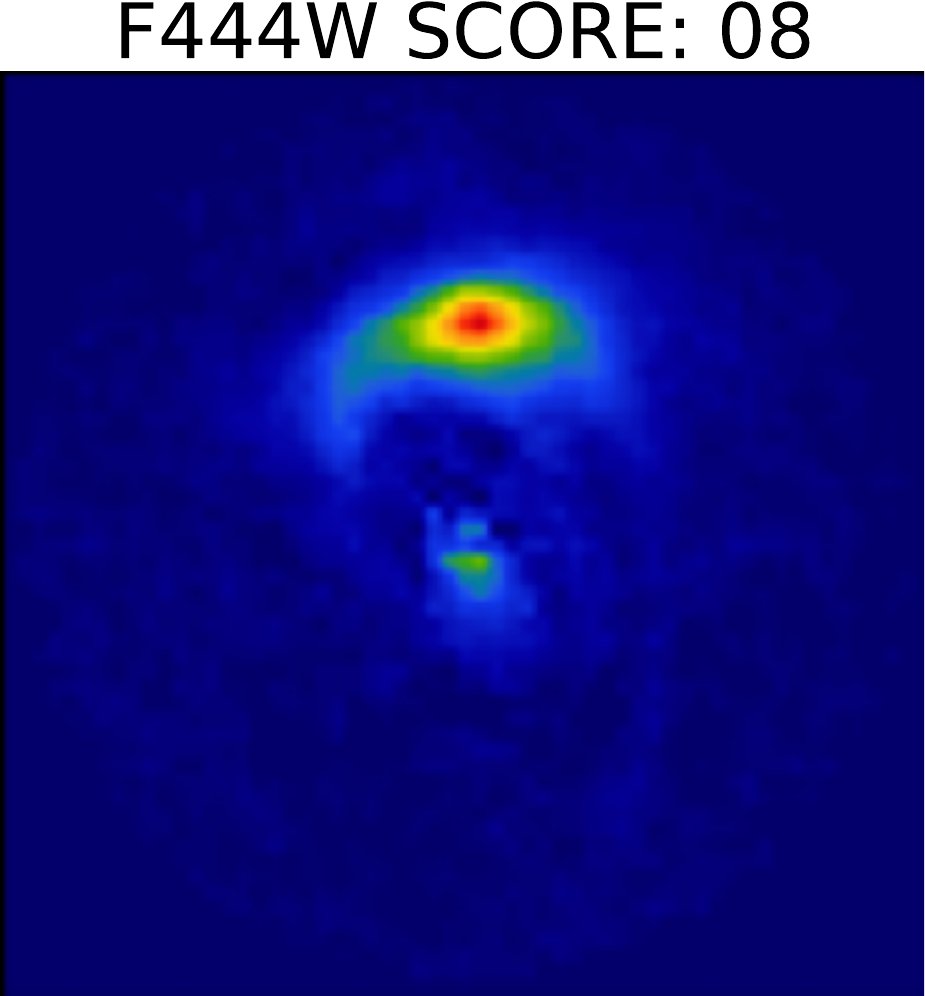}
\includegraphics[width=0.12\textwidth]{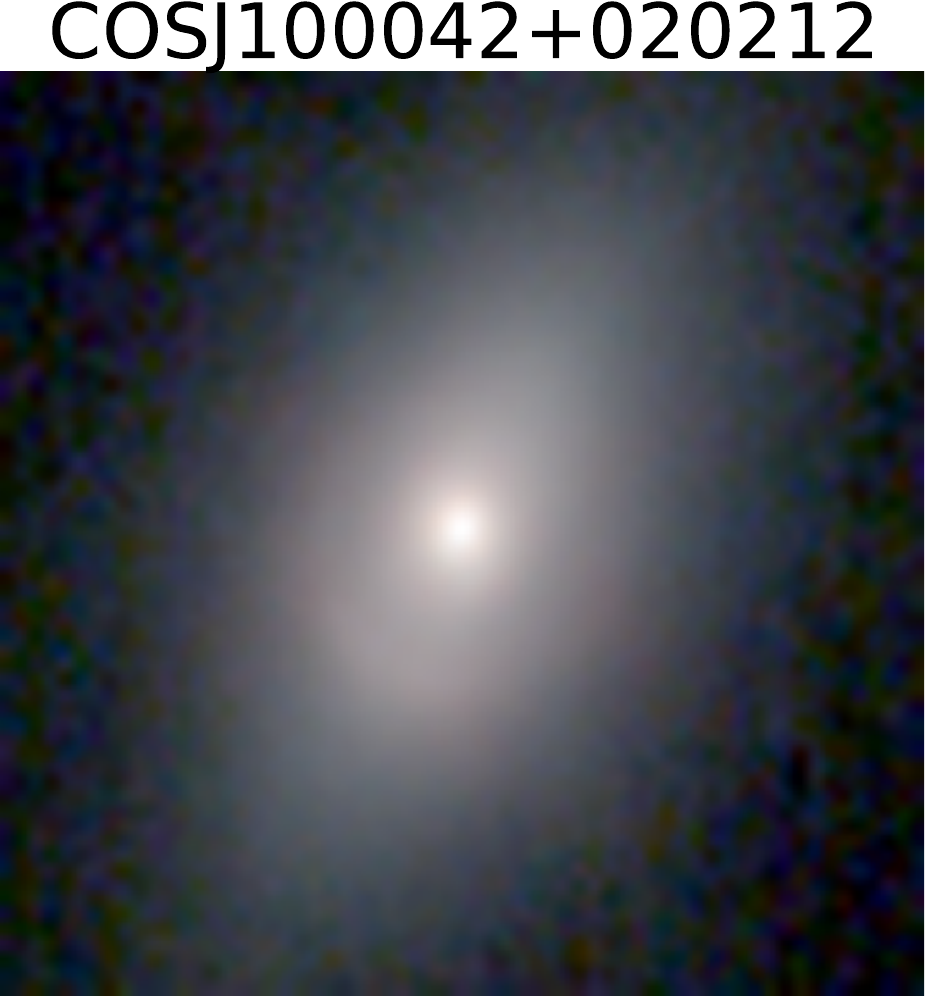}
\includegraphics[width=0.12\textwidth]{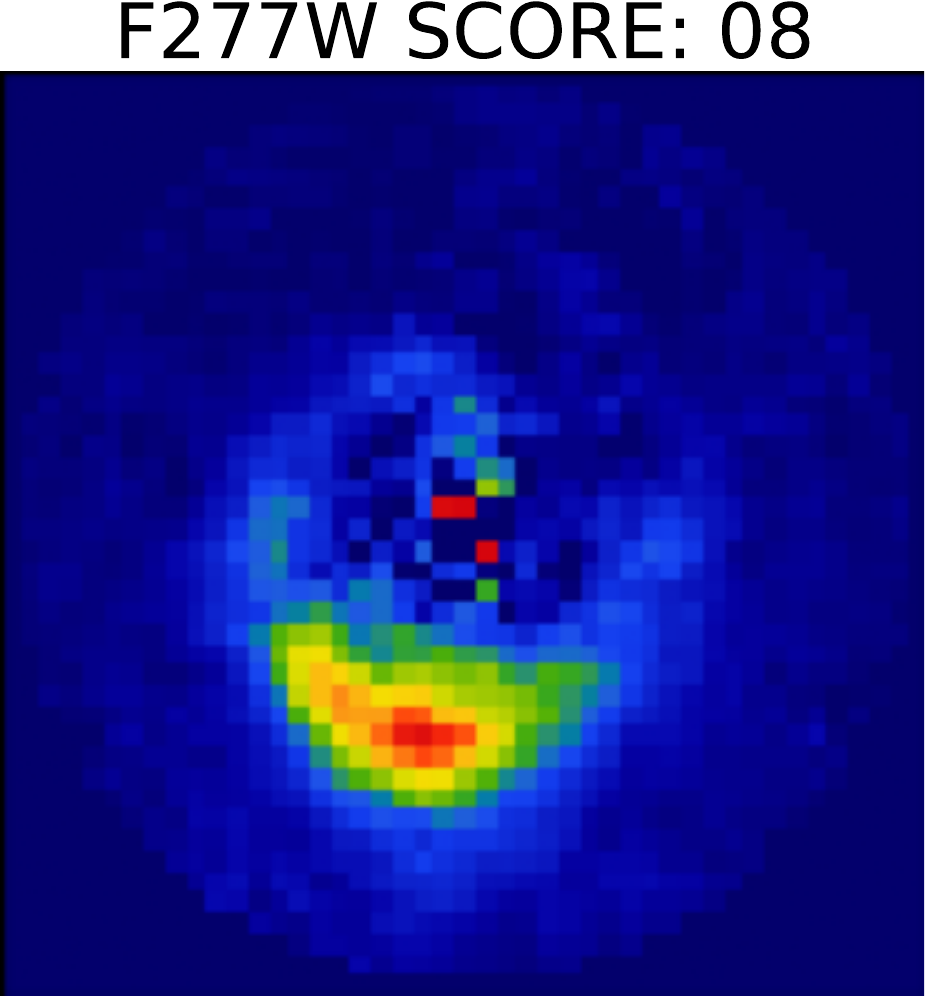}
\includegraphics[width=0.12\textwidth]{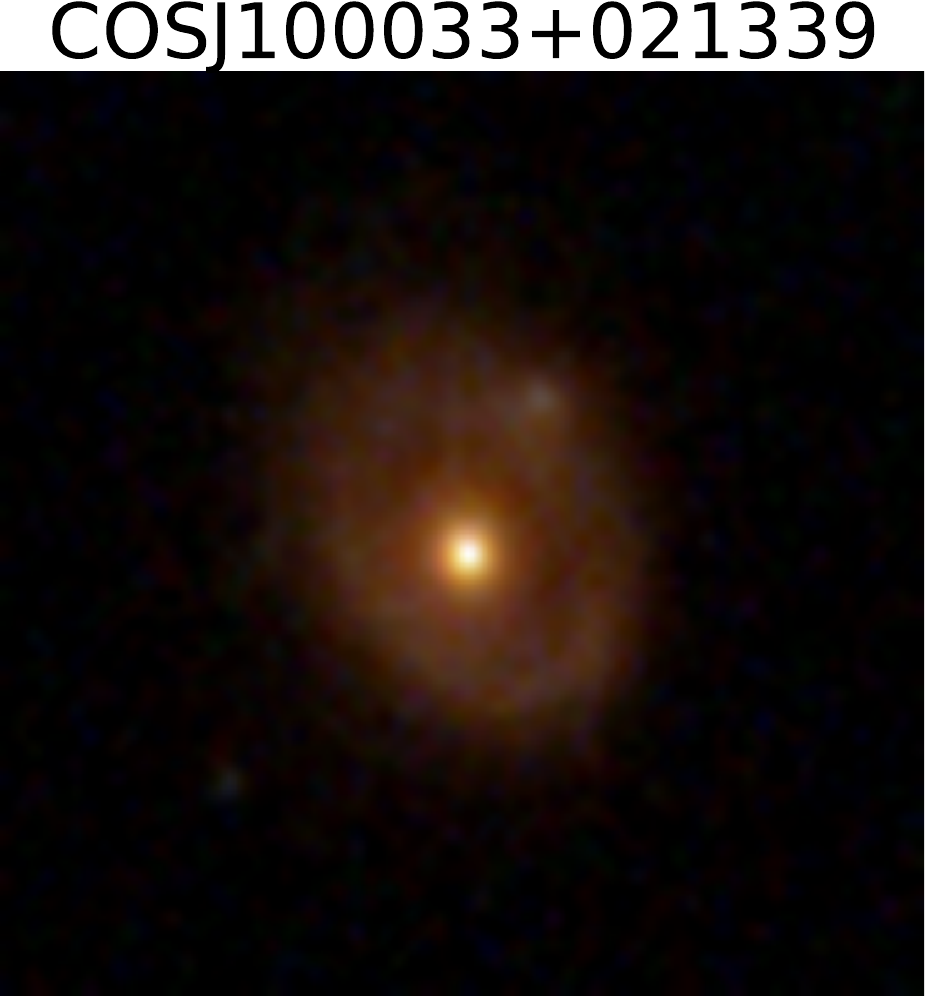}
\includegraphics[width=0.12\textwidth]{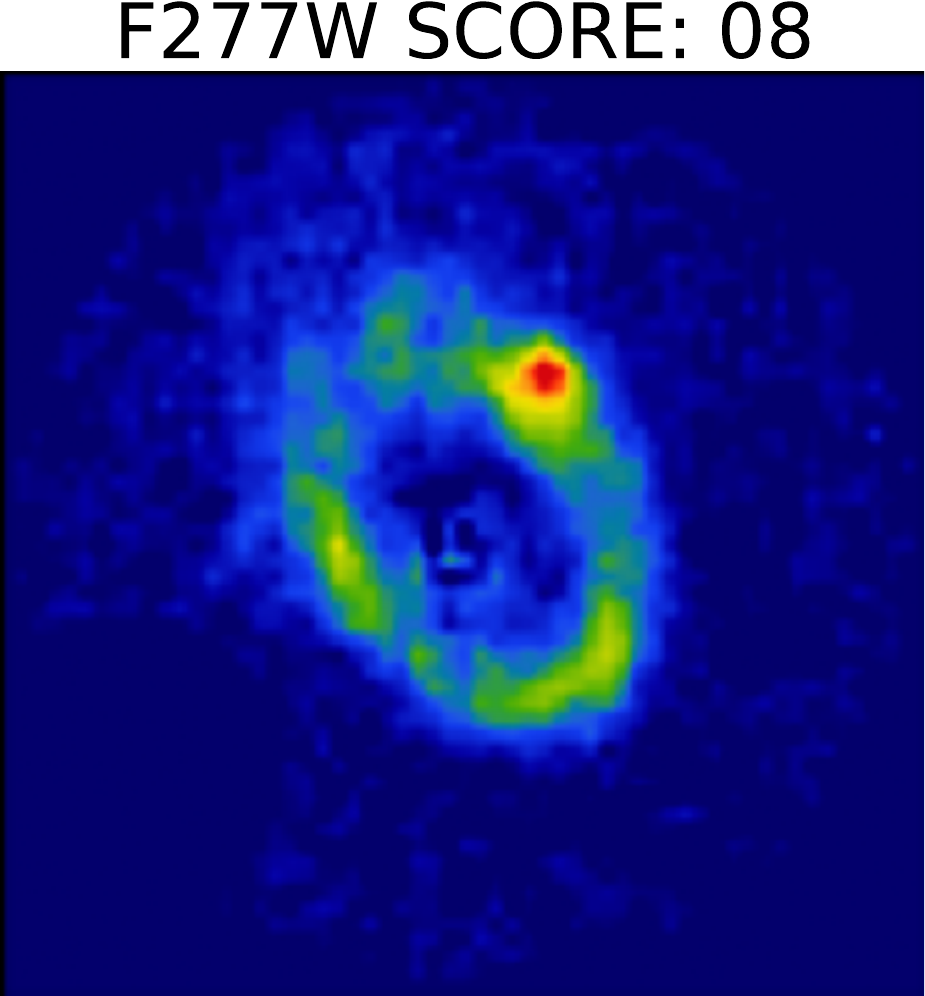}
\includegraphics[width=0.12\textwidth]{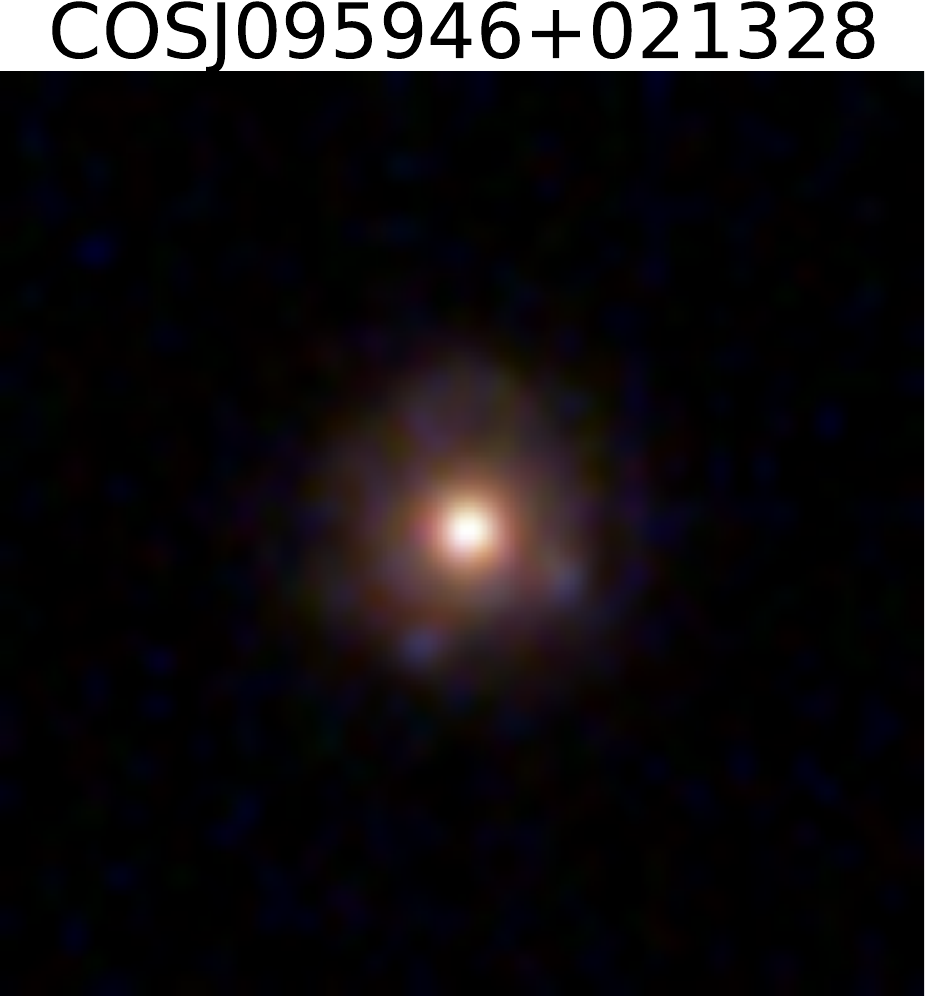}
\includegraphics[width=0.12\textwidth]{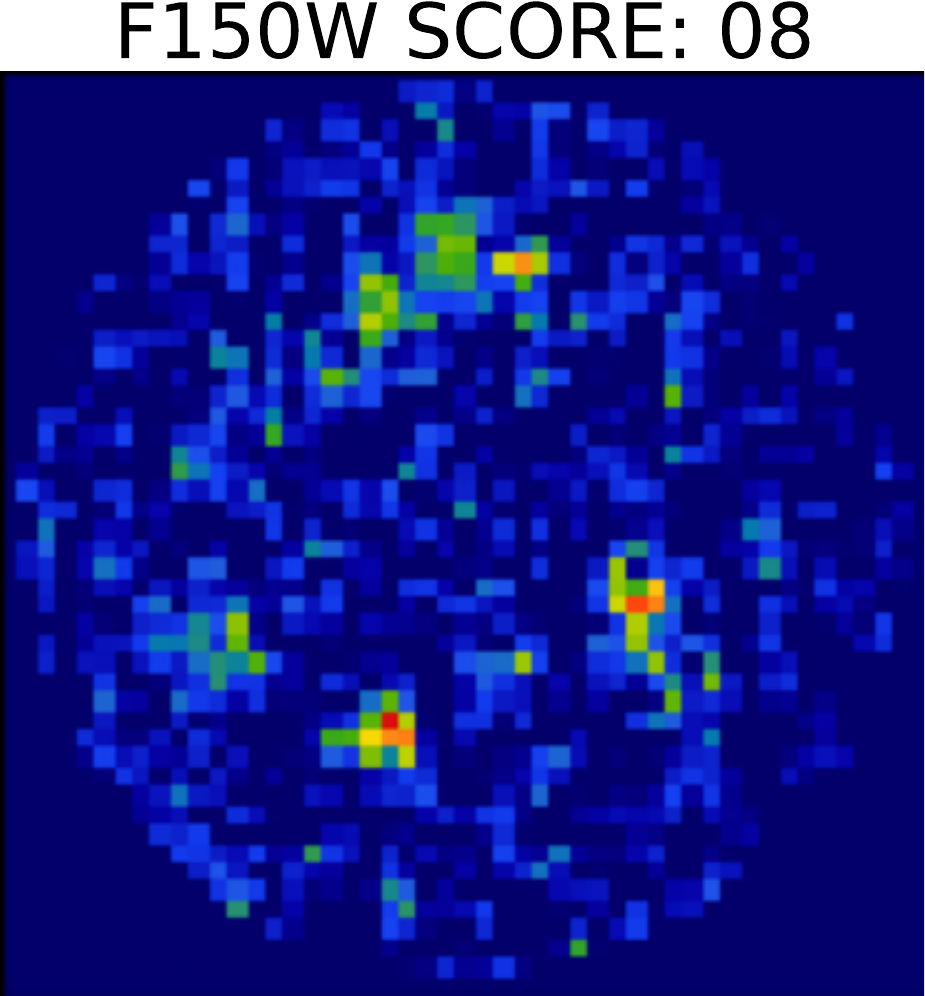}
\includegraphics[width=0.12\textwidth]{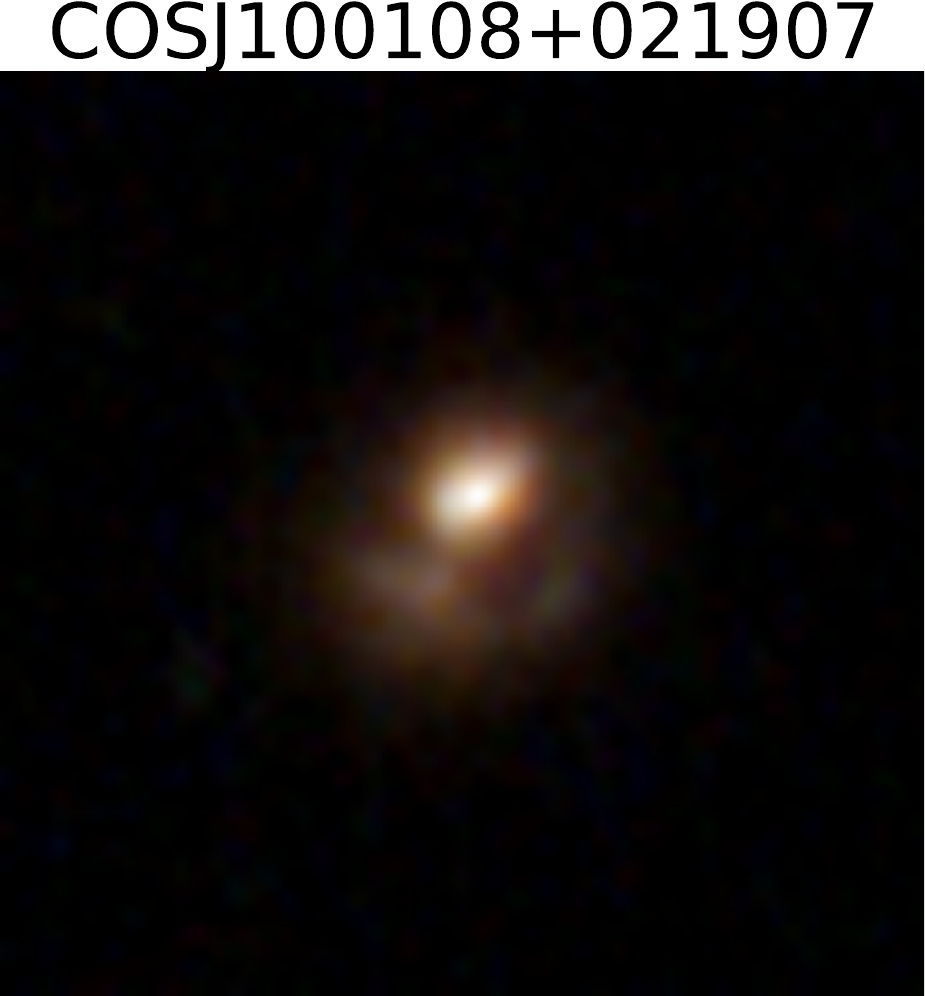}
\includegraphics[width=0.12\textwidth]{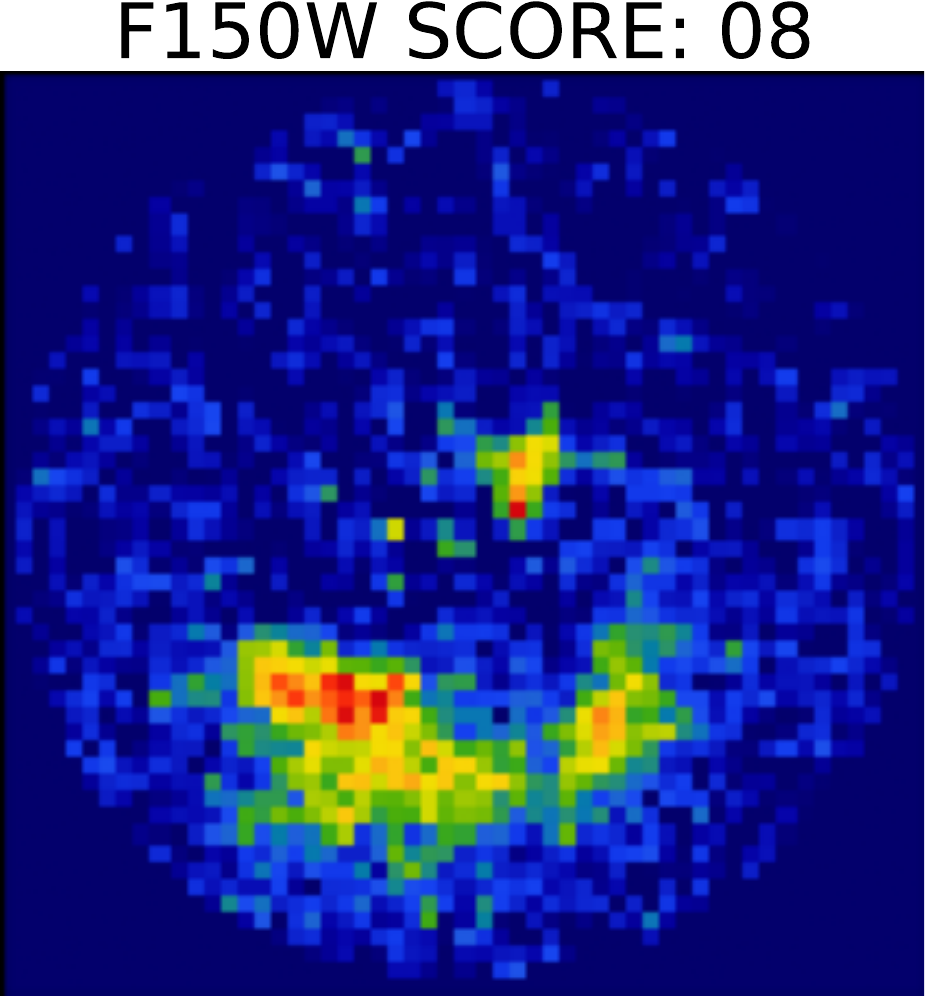}
\includegraphics[width=0.12\textwidth]{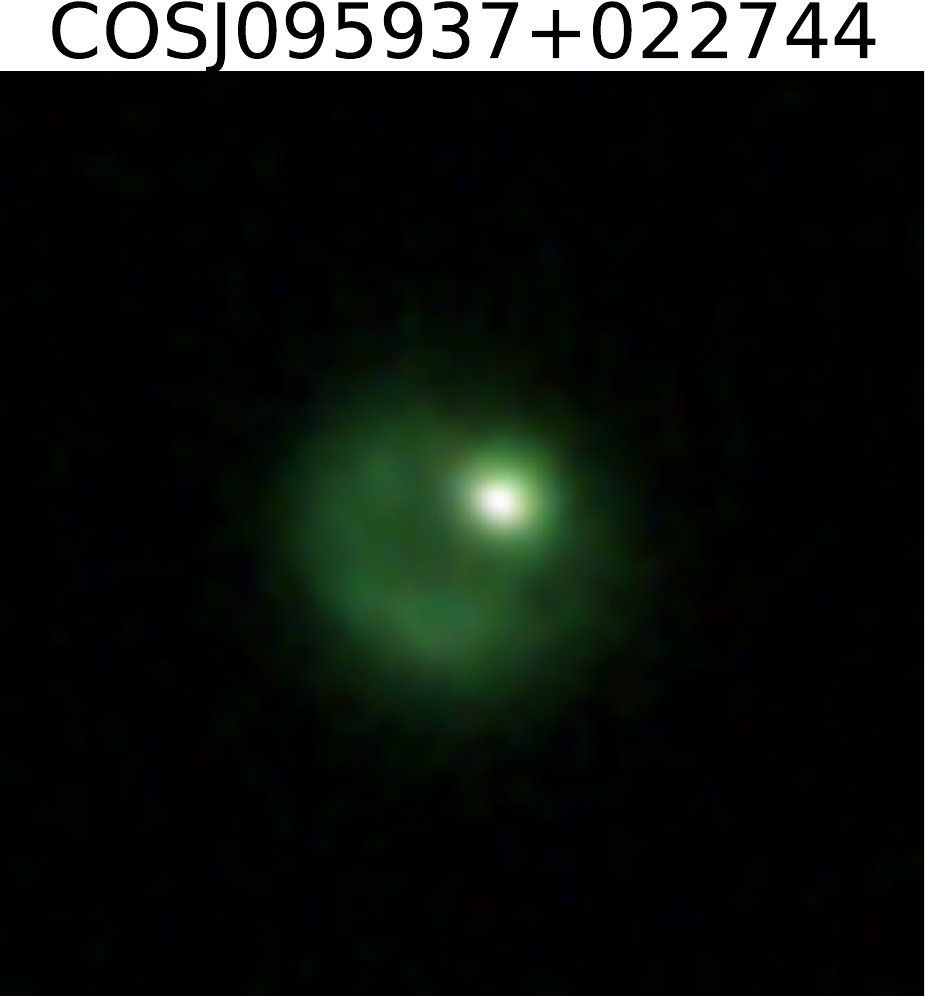}
\includegraphics[width=0.12\textwidth]{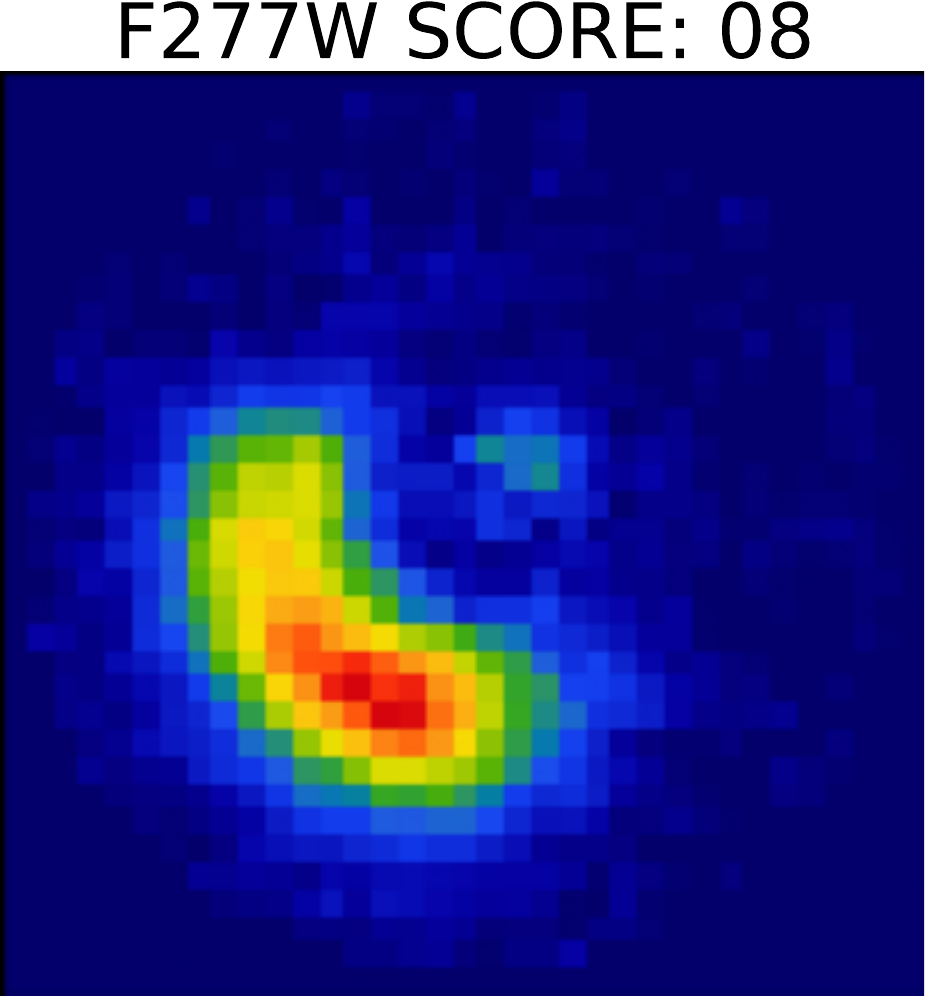}
\includegraphics[width=0.12\textwidth]{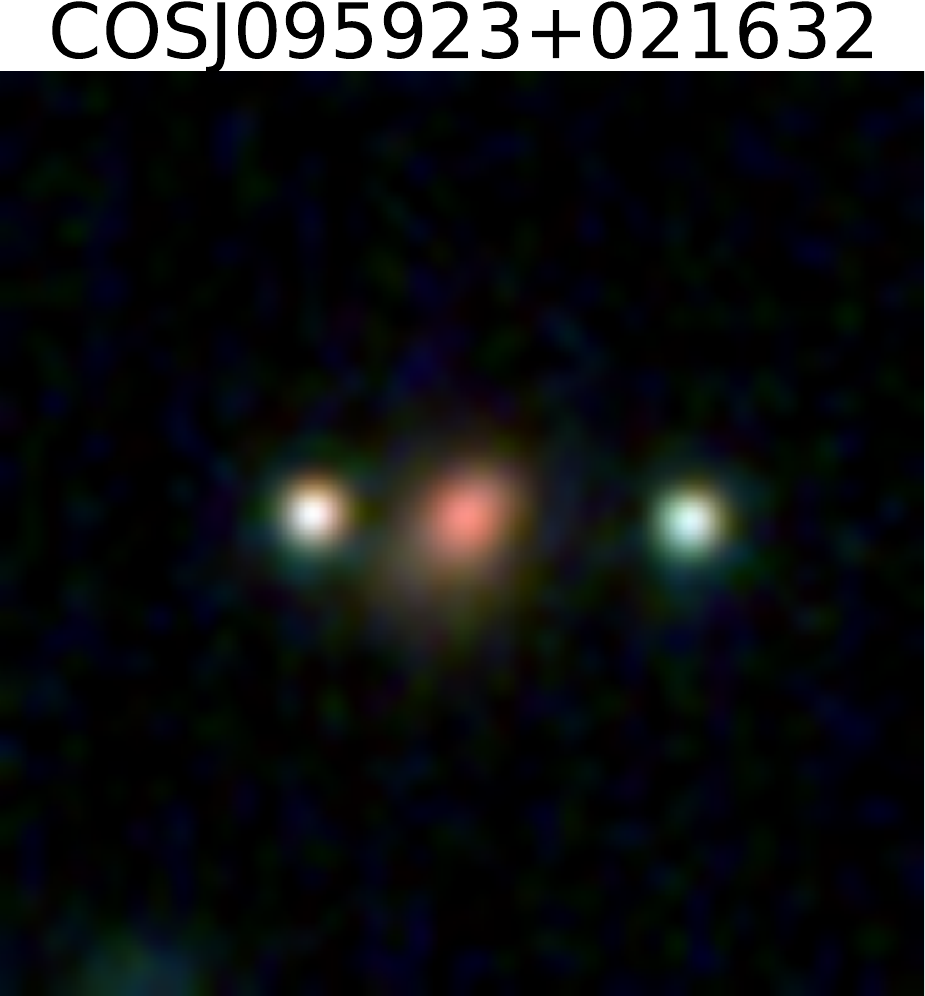}
\includegraphics[width=0.12\textwidth]{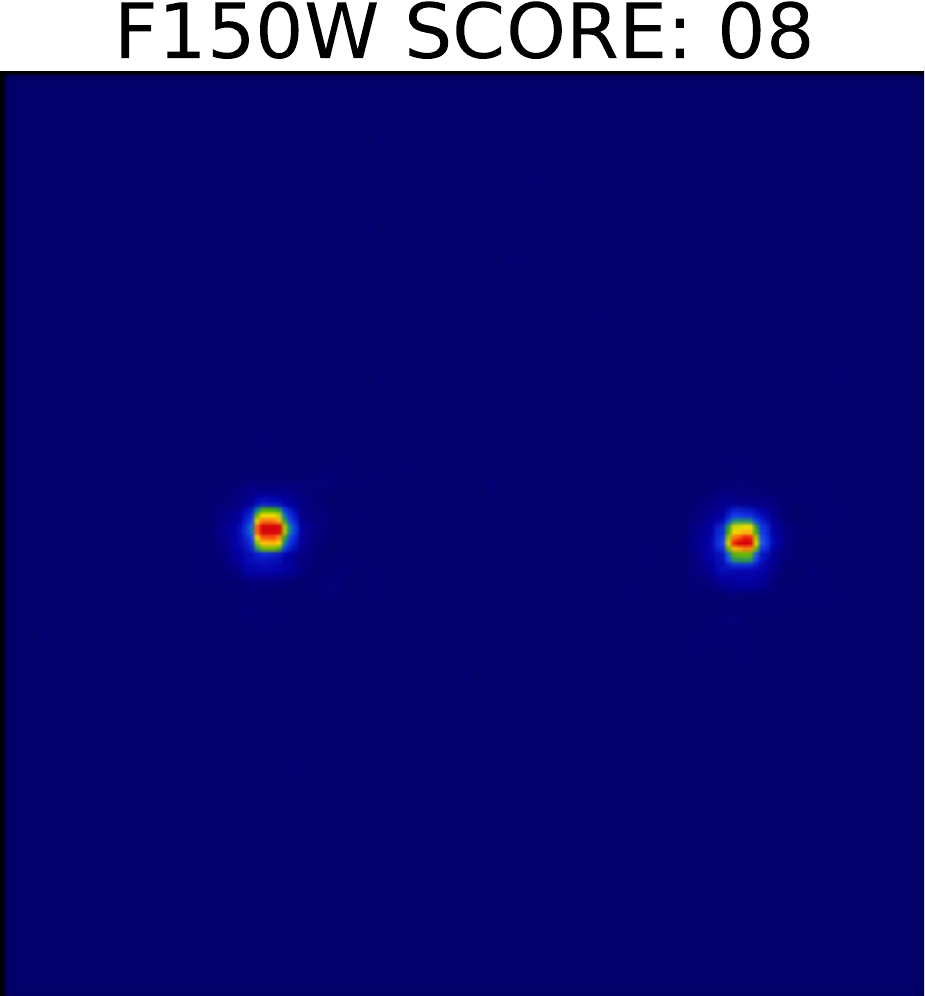}
\includegraphics[width=0.12\textwidth]{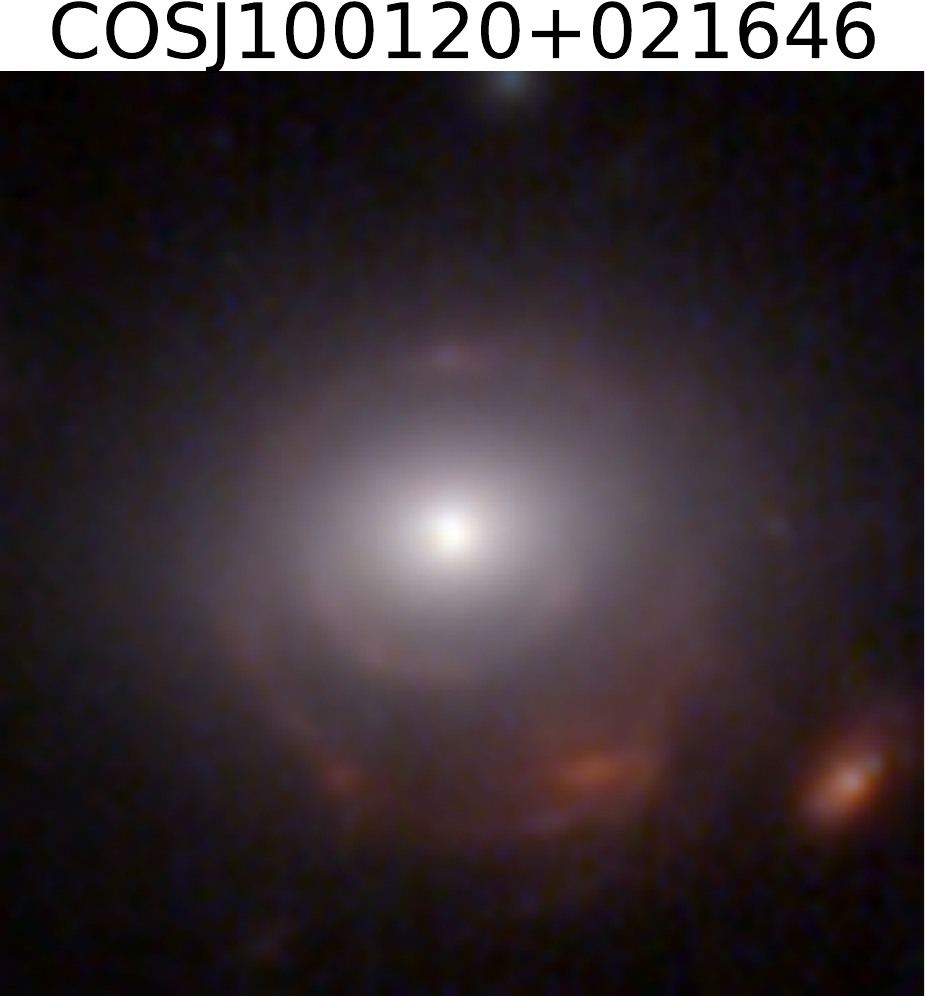}
\includegraphics[width=0.12\textwidth]{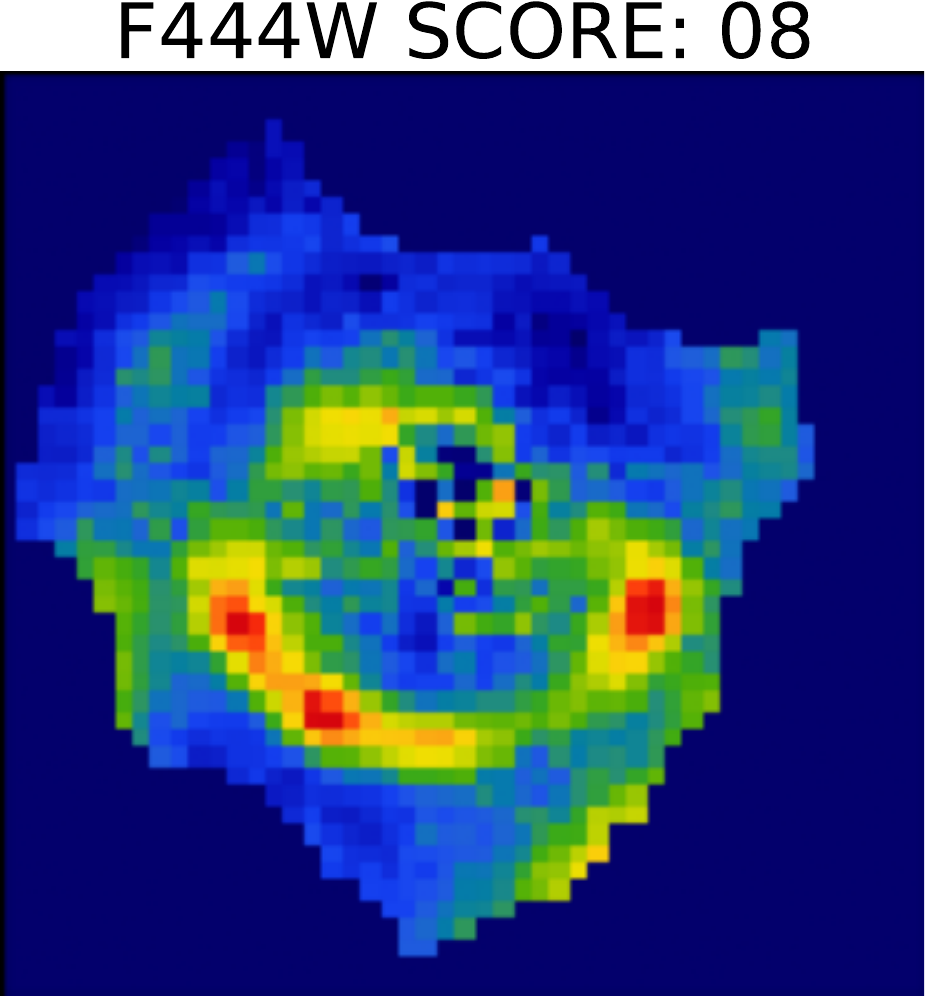}
\includegraphics[width=0.12\textwidth]{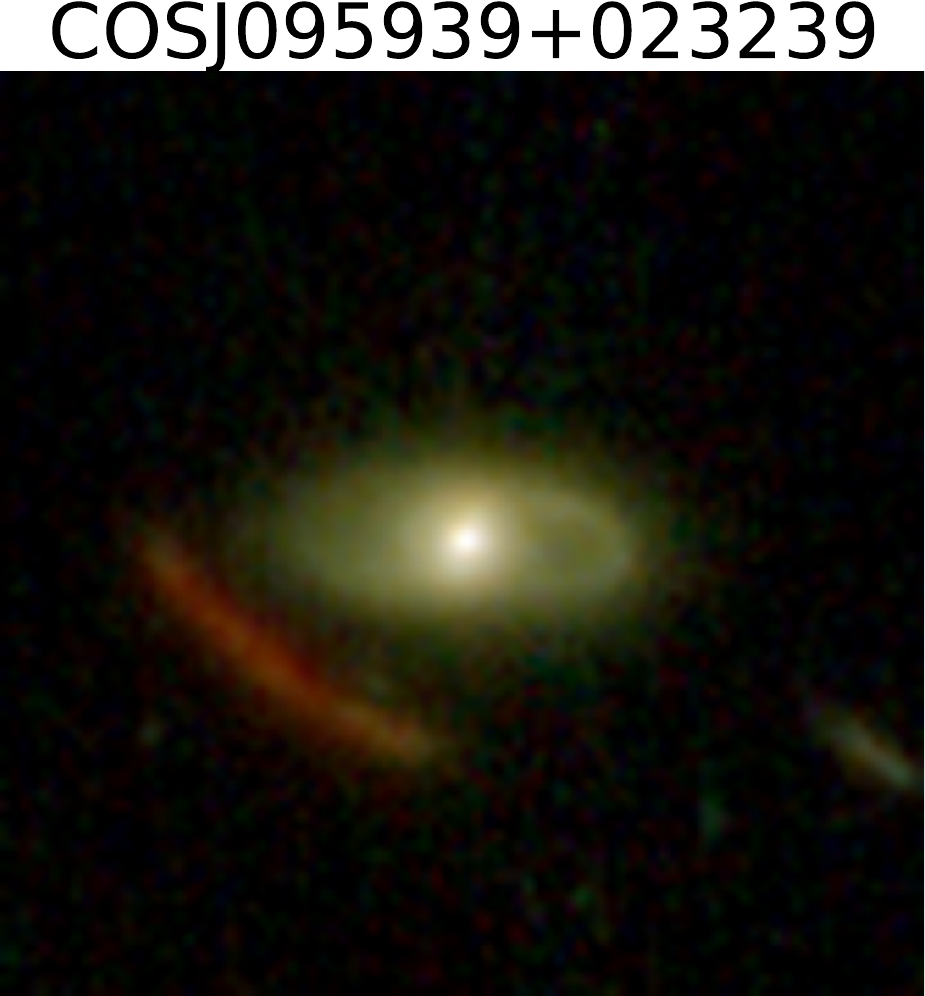}
\includegraphics[width=0.12\textwidth]{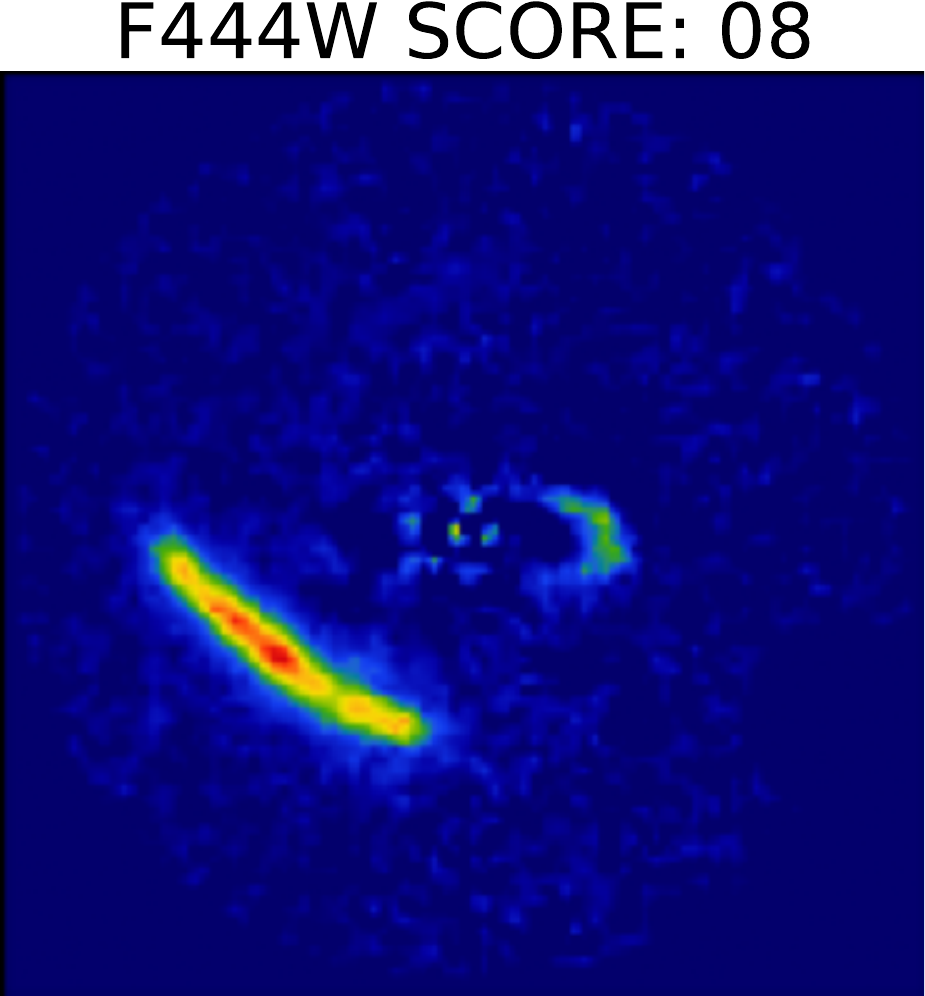}
\includegraphics[width=0.12\textwidth]{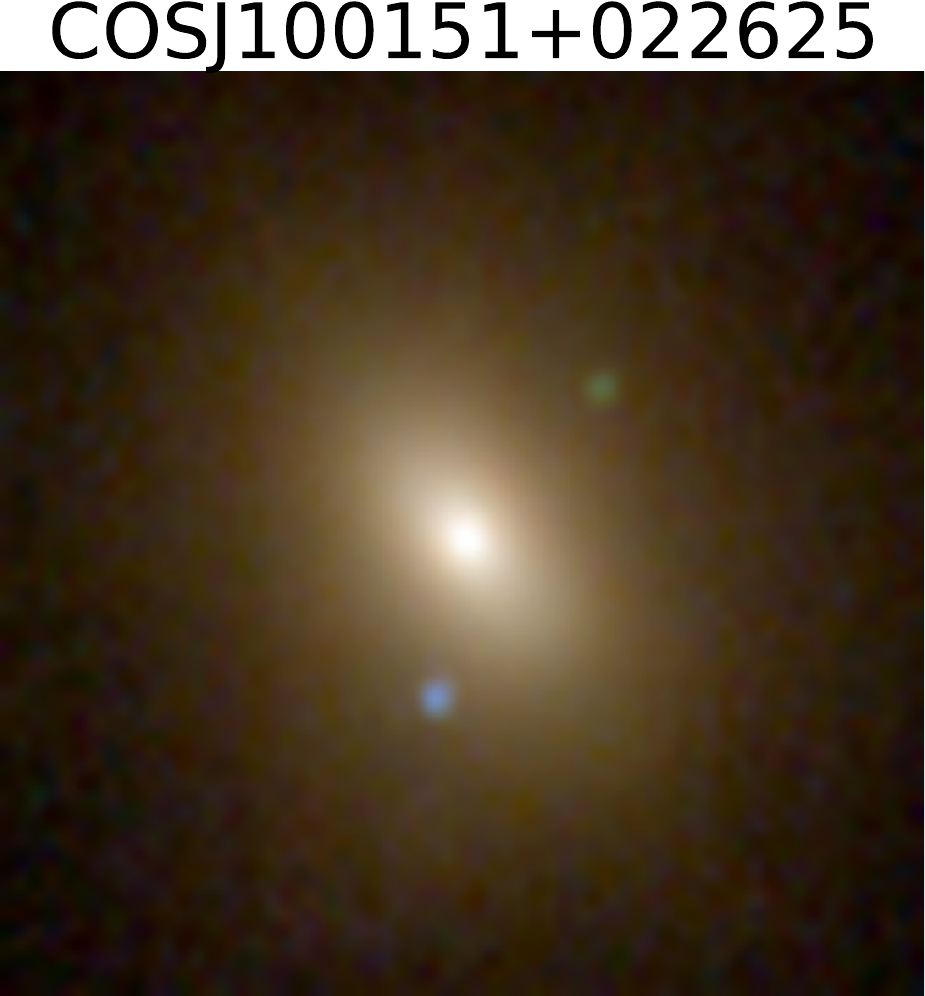}
\includegraphics[width=0.12\textwidth]{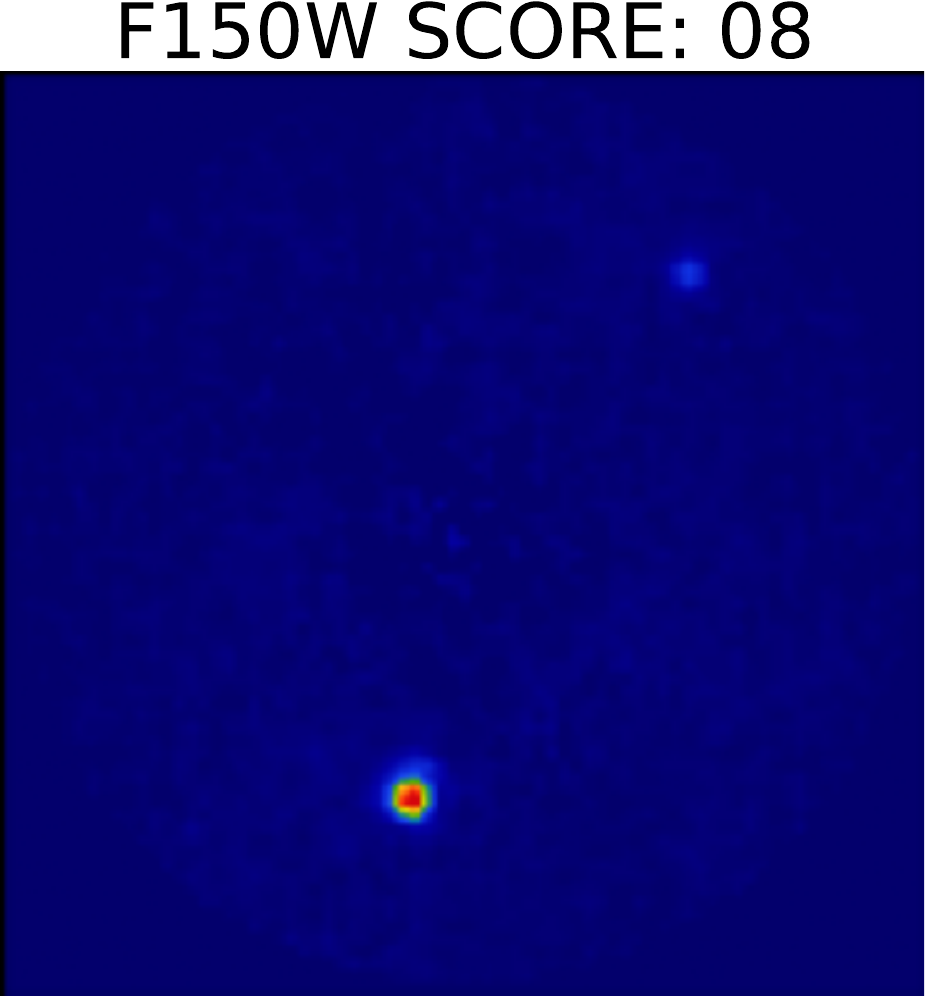}
\includegraphics[width=0.12\textwidth]{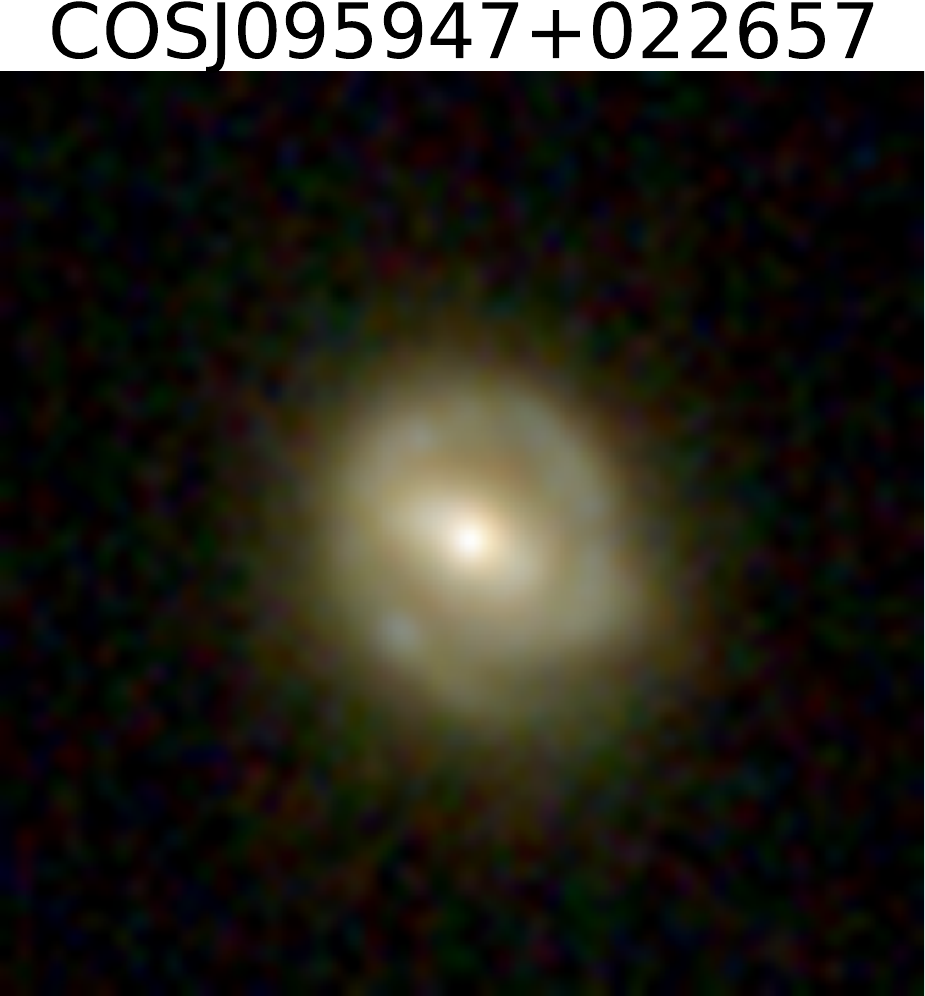}
\includegraphics[width=0.12\textwidth]{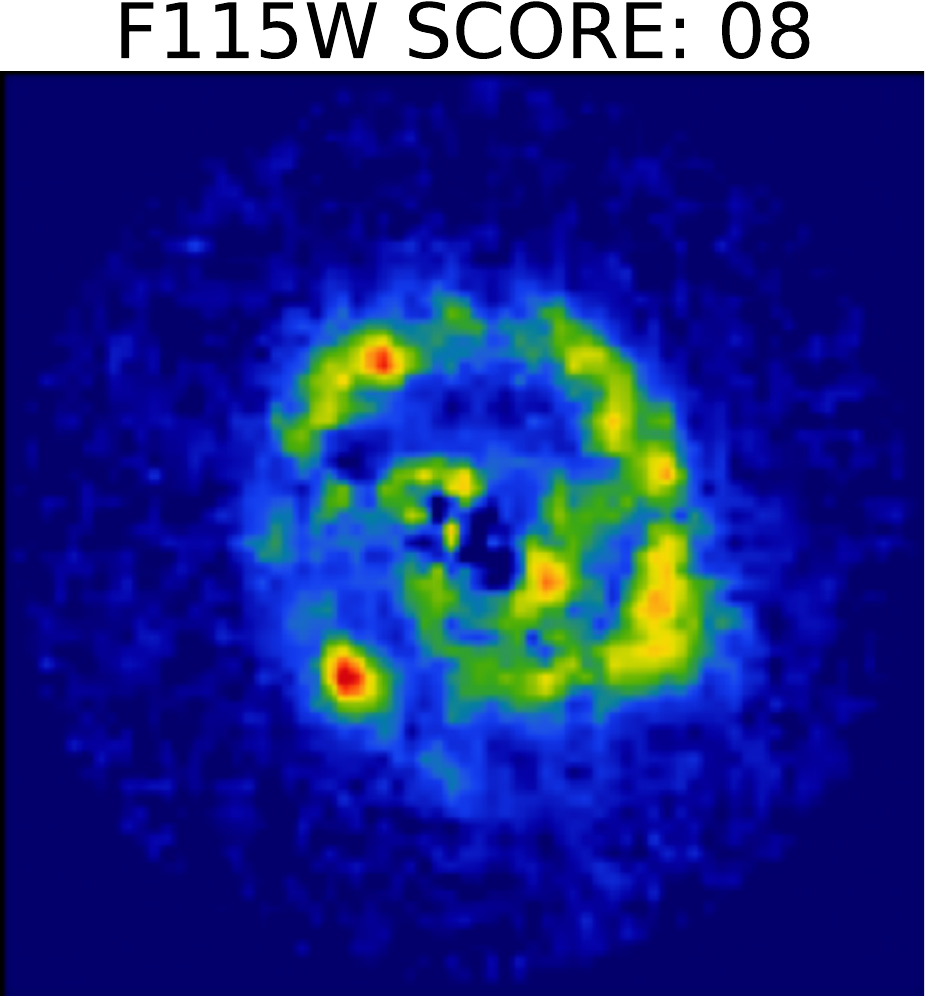}
\includegraphics[width=0.12\textwidth]{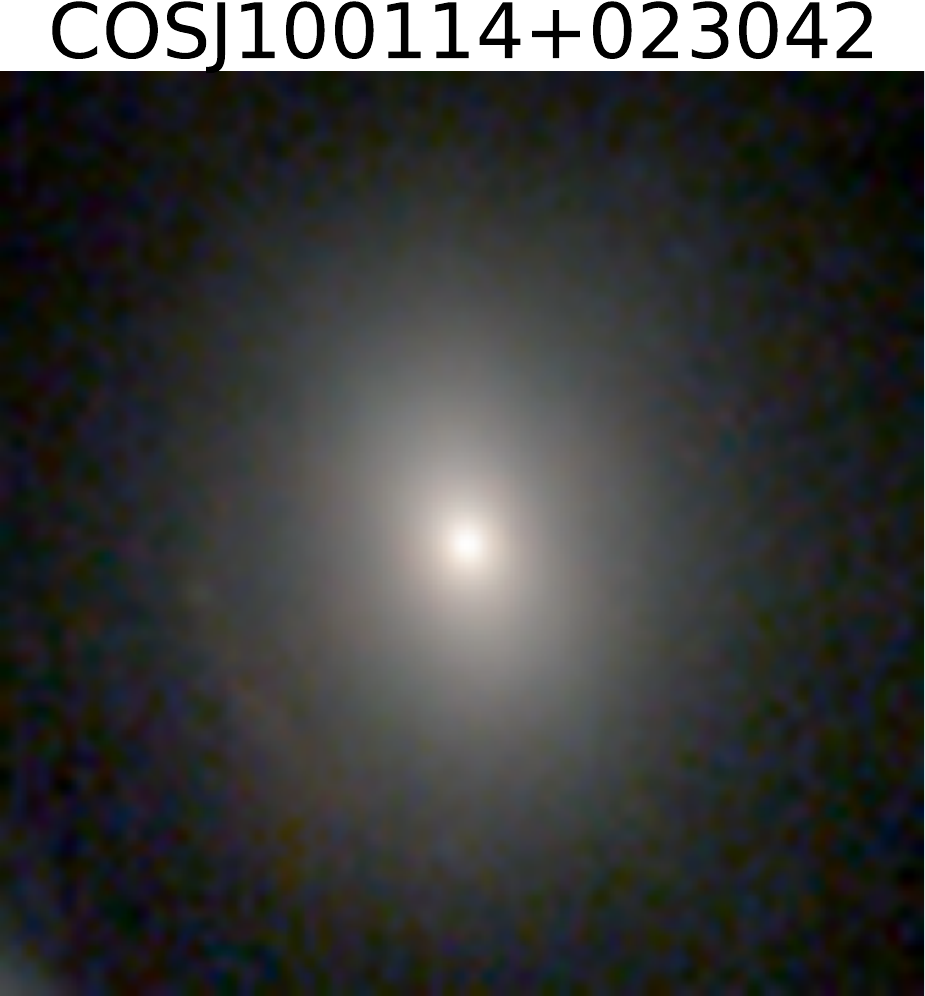}
\includegraphics[width=0.12\textwidth]{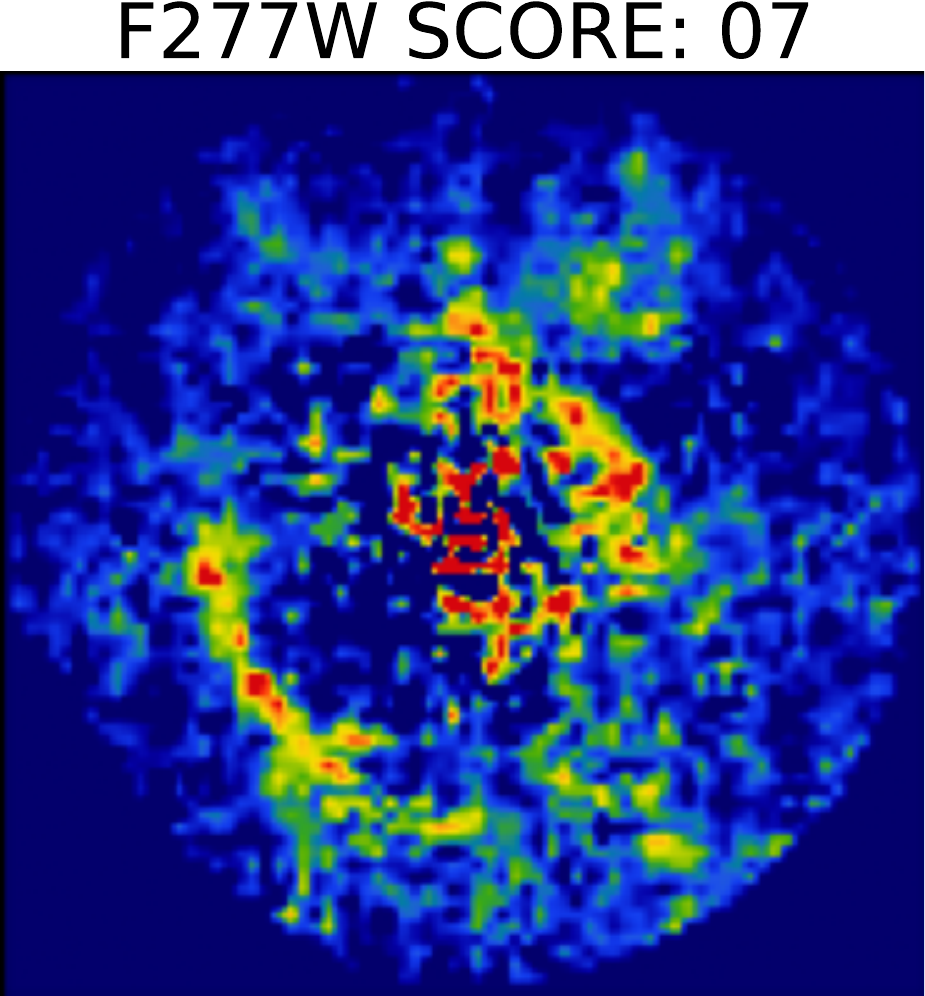}
\includegraphics[width=0.12\textwidth]{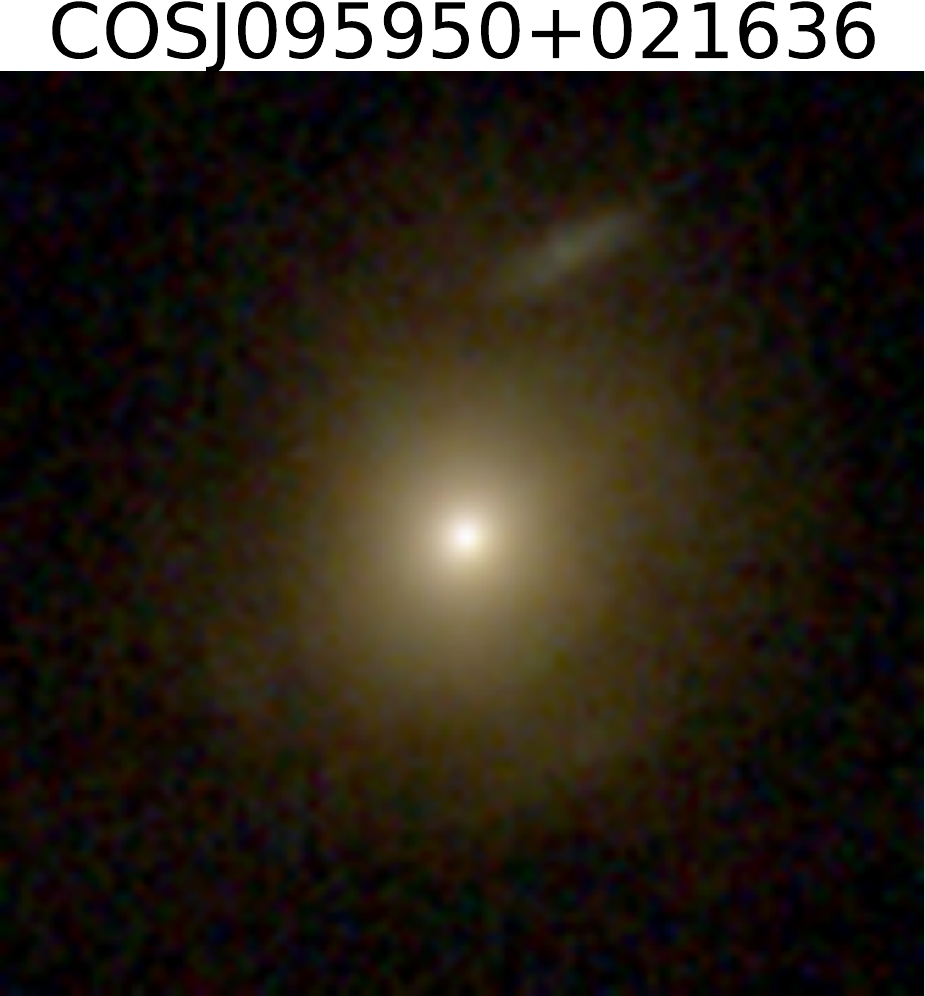}
\includegraphics[width=0.12\textwidth]{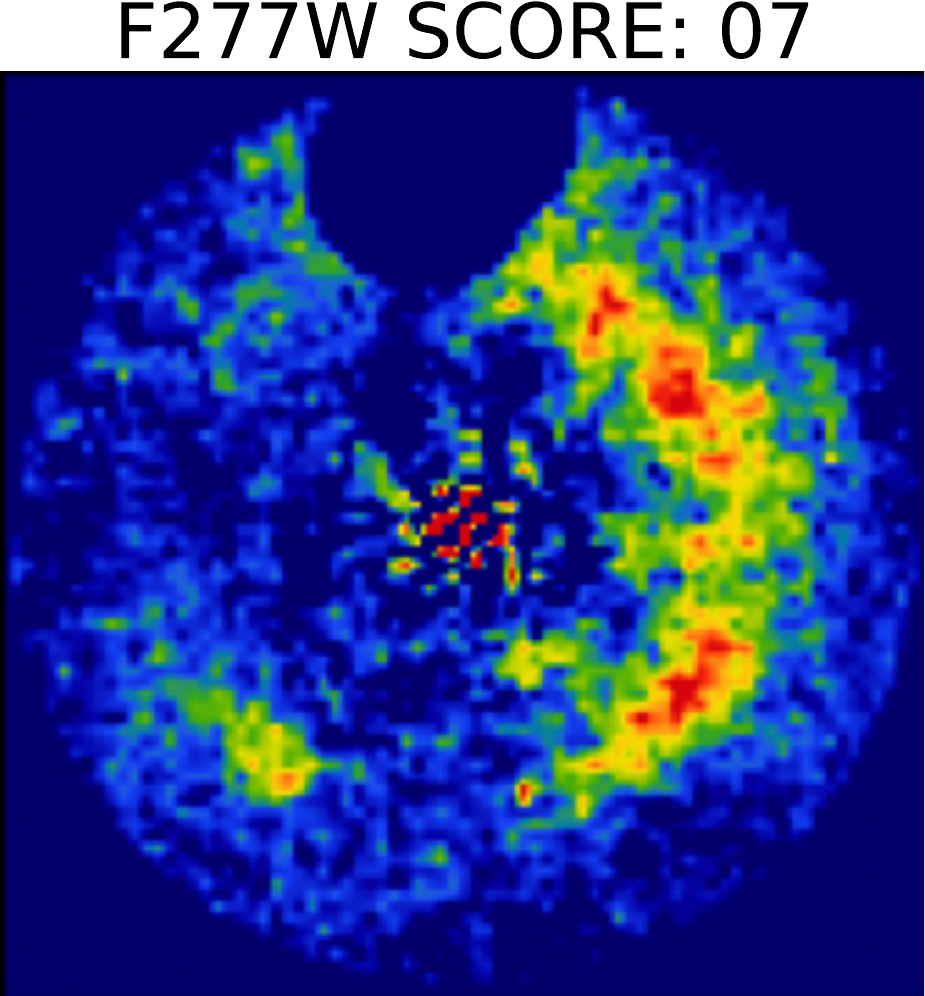}
\includegraphics[width=0.12\textwidth]{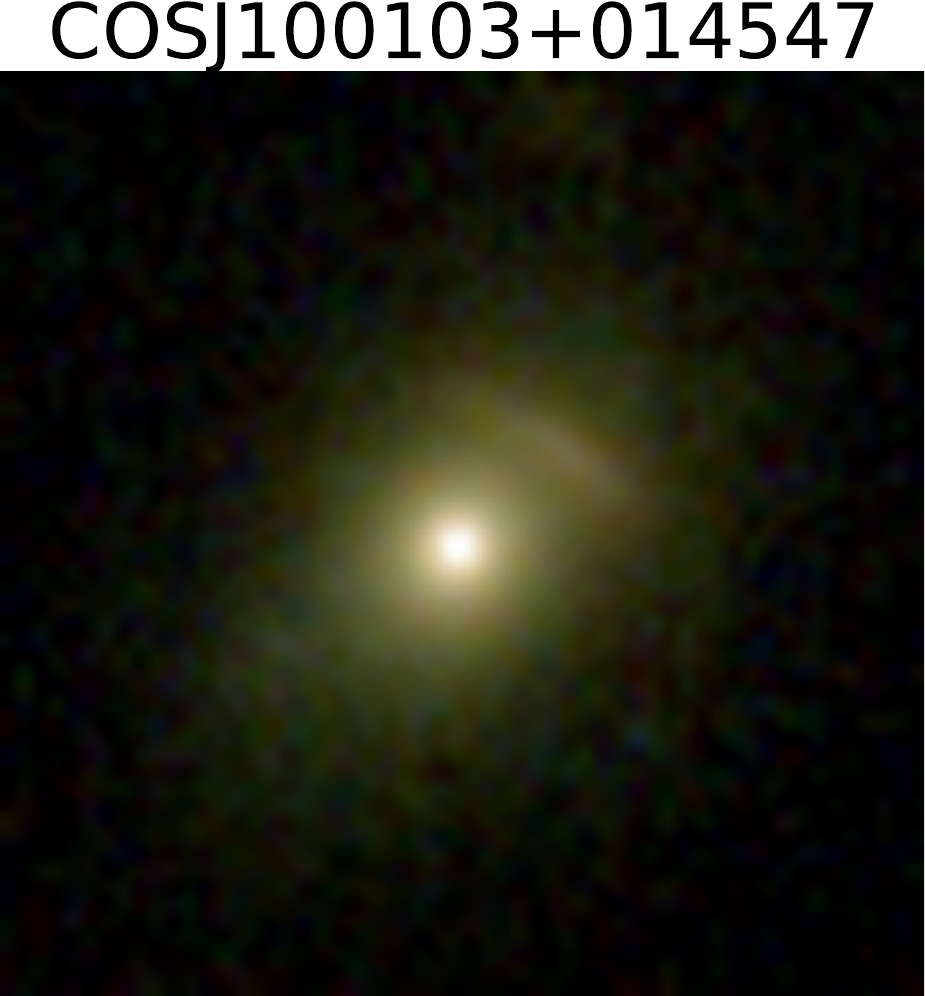}
\includegraphics[width=0.12\textwidth]{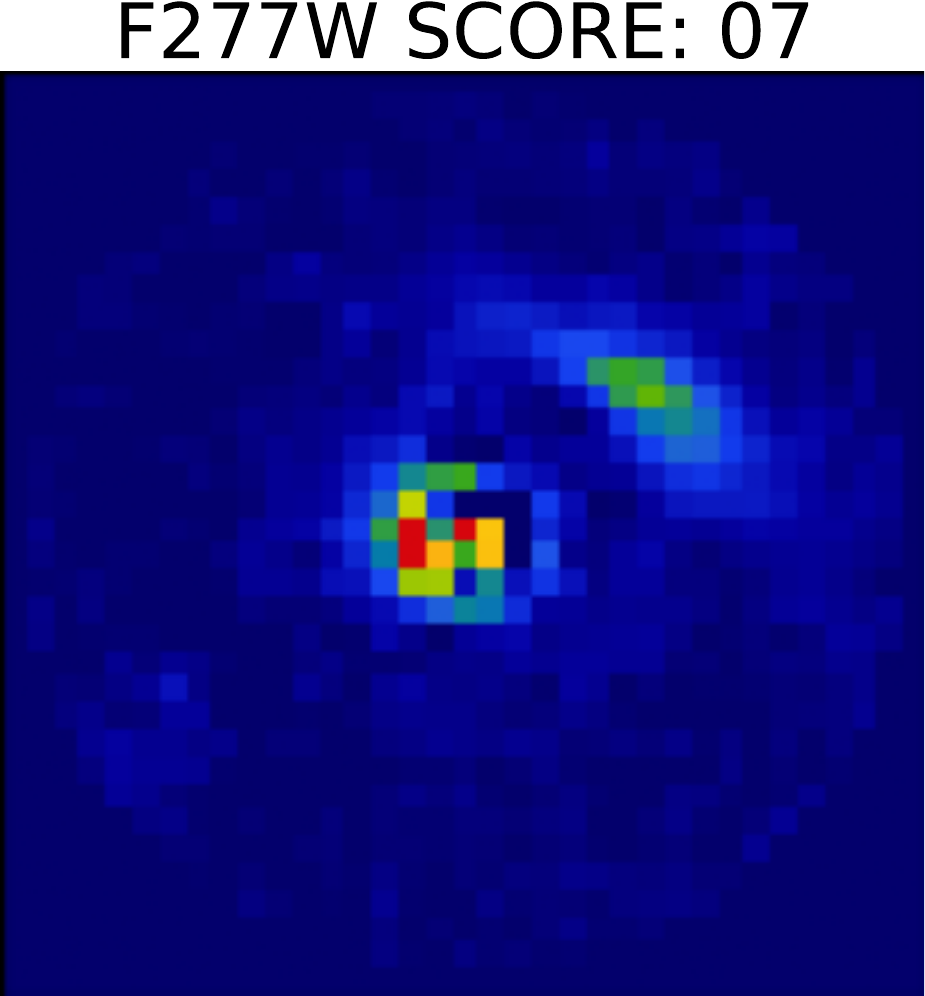}
\includegraphics[width=0.12\textwidth]{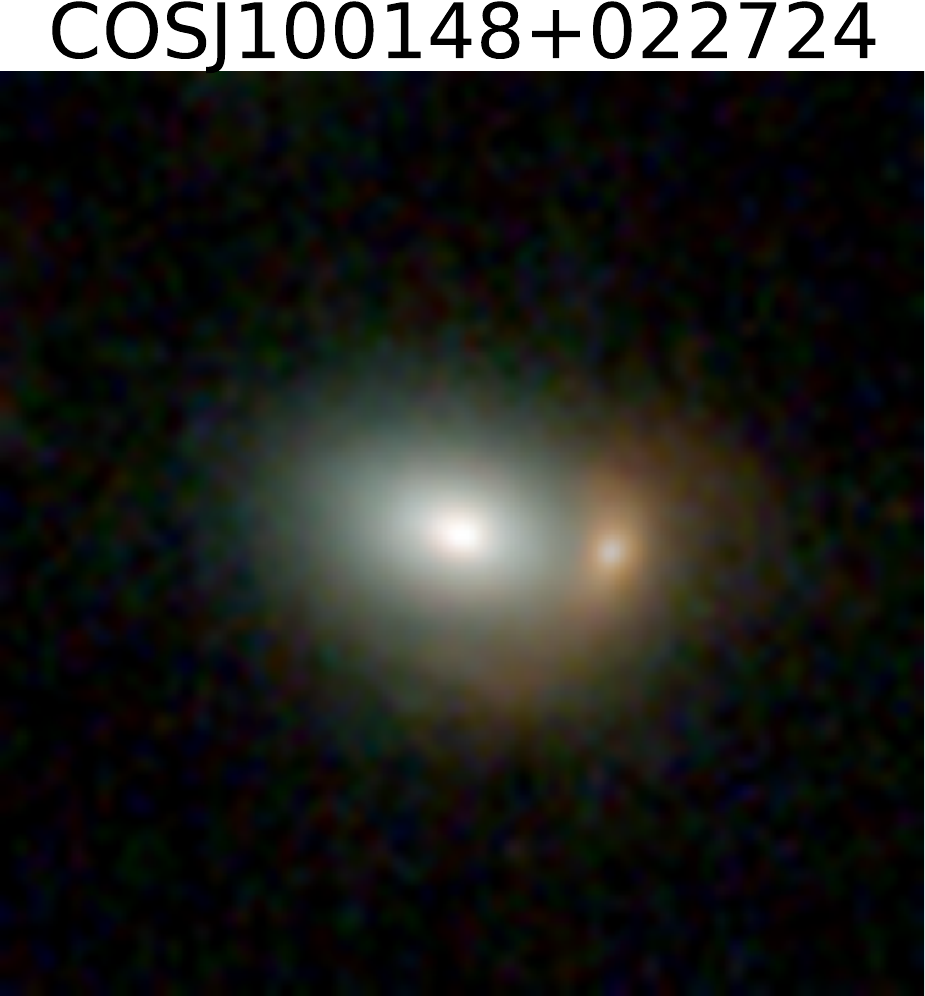}
\includegraphics[width=0.12\textwidth]{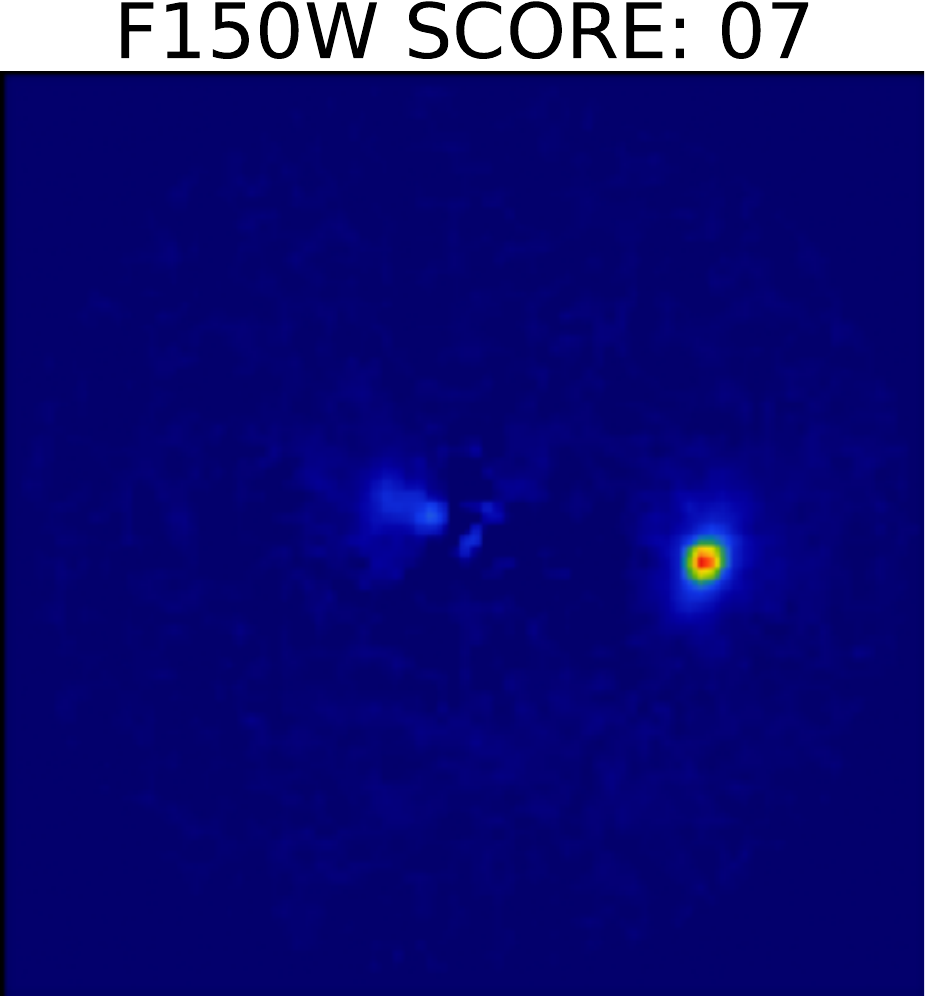}
\includegraphics[width=0.12\textwidth]{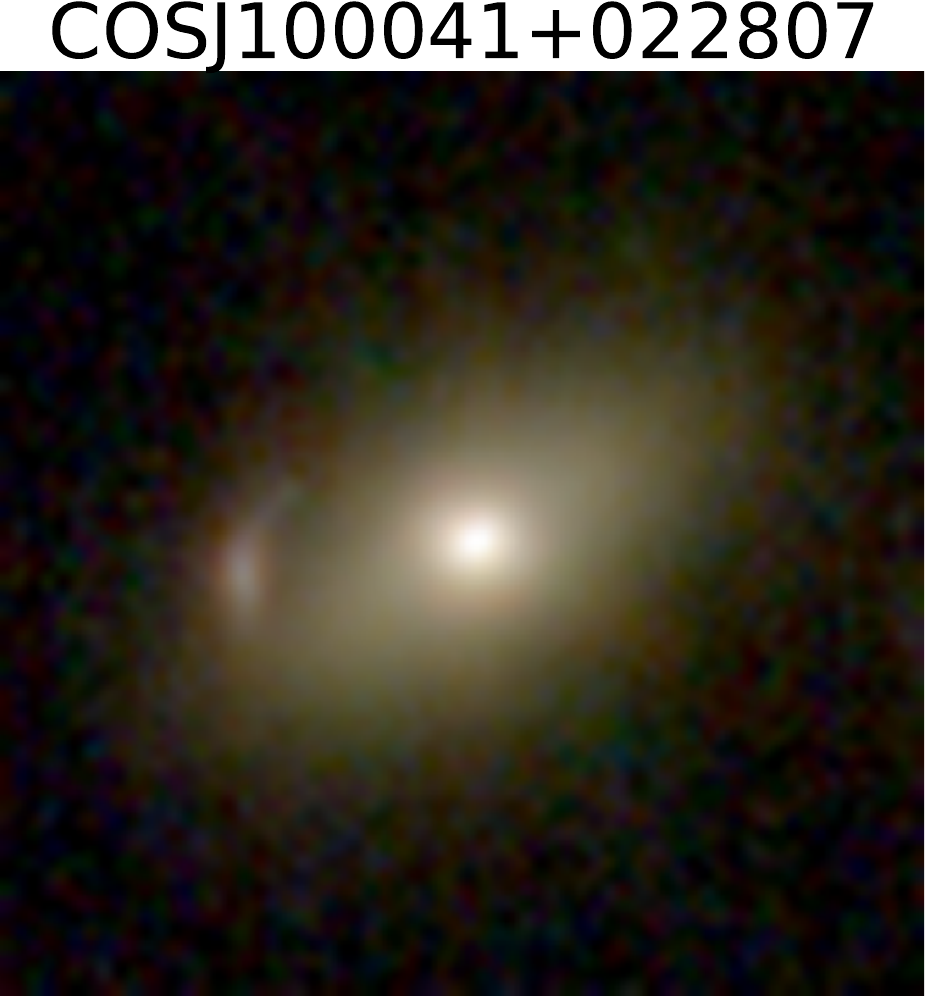}
\includegraphics[width=0.12\textwidth]{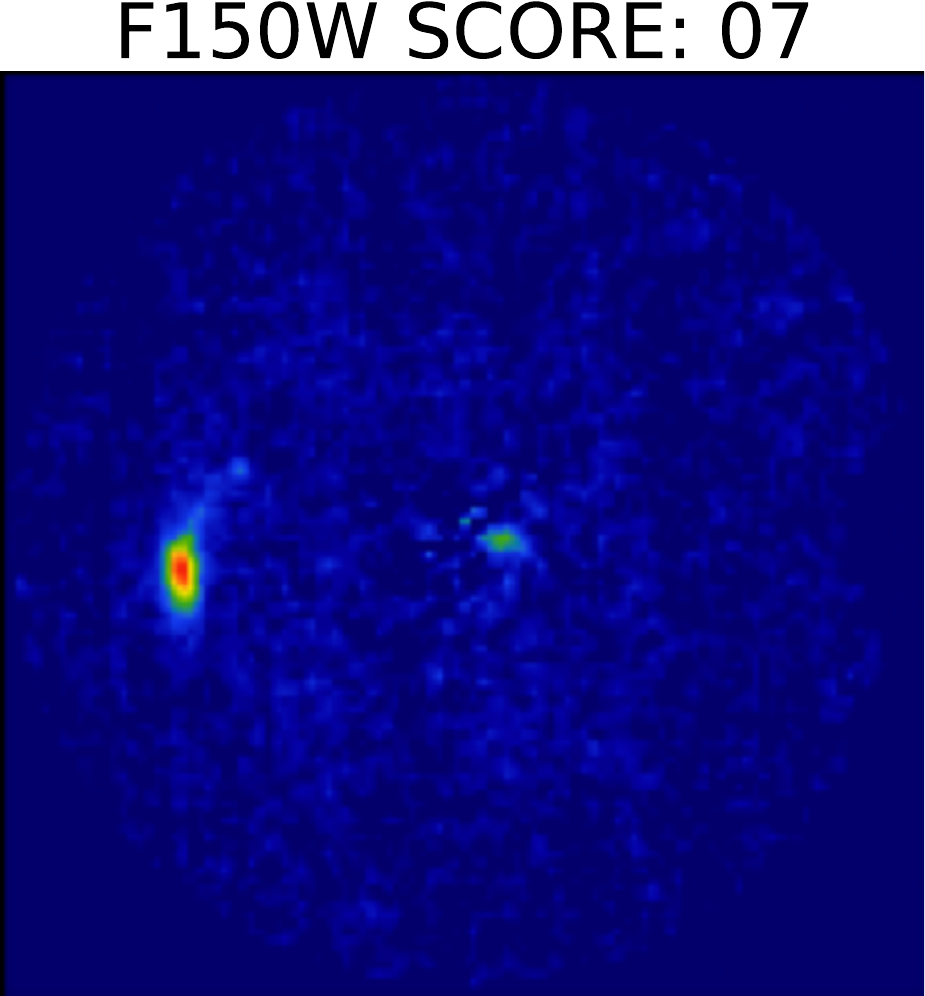}
\includegraphics[width=0.12\textwidth]{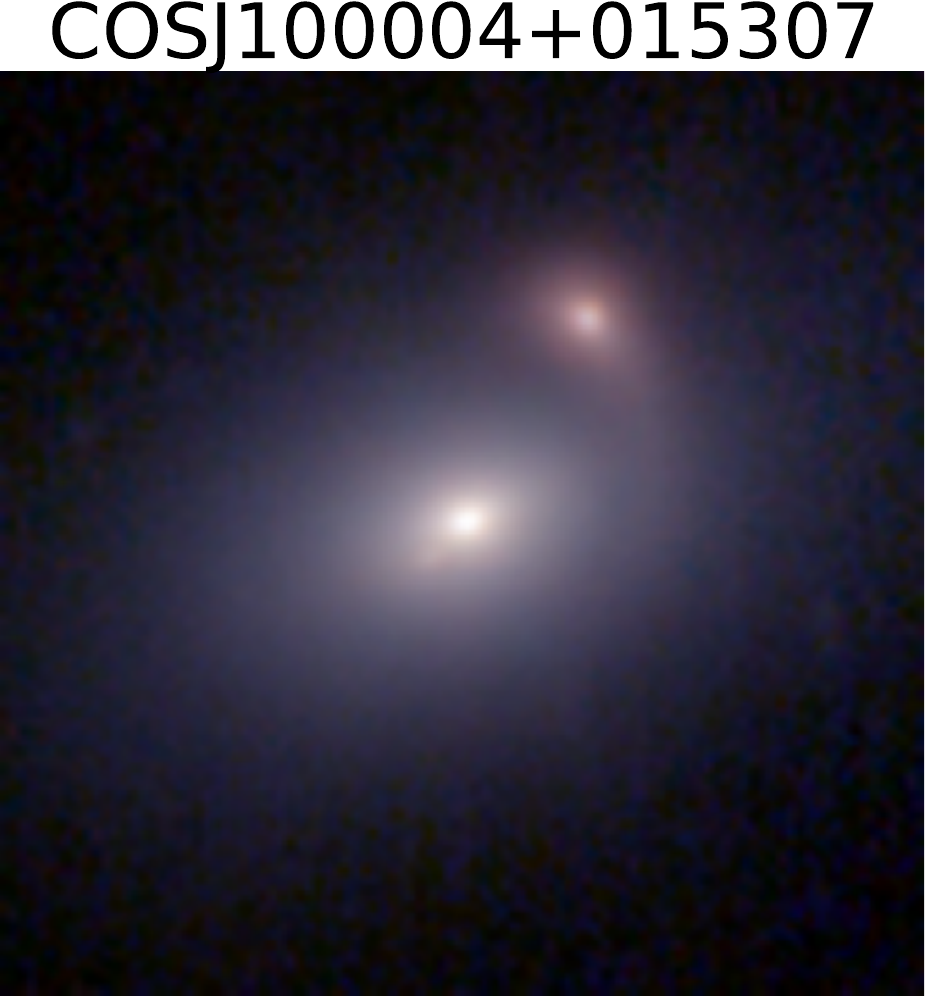}
\includegraphics[width=0.12\textwidth]{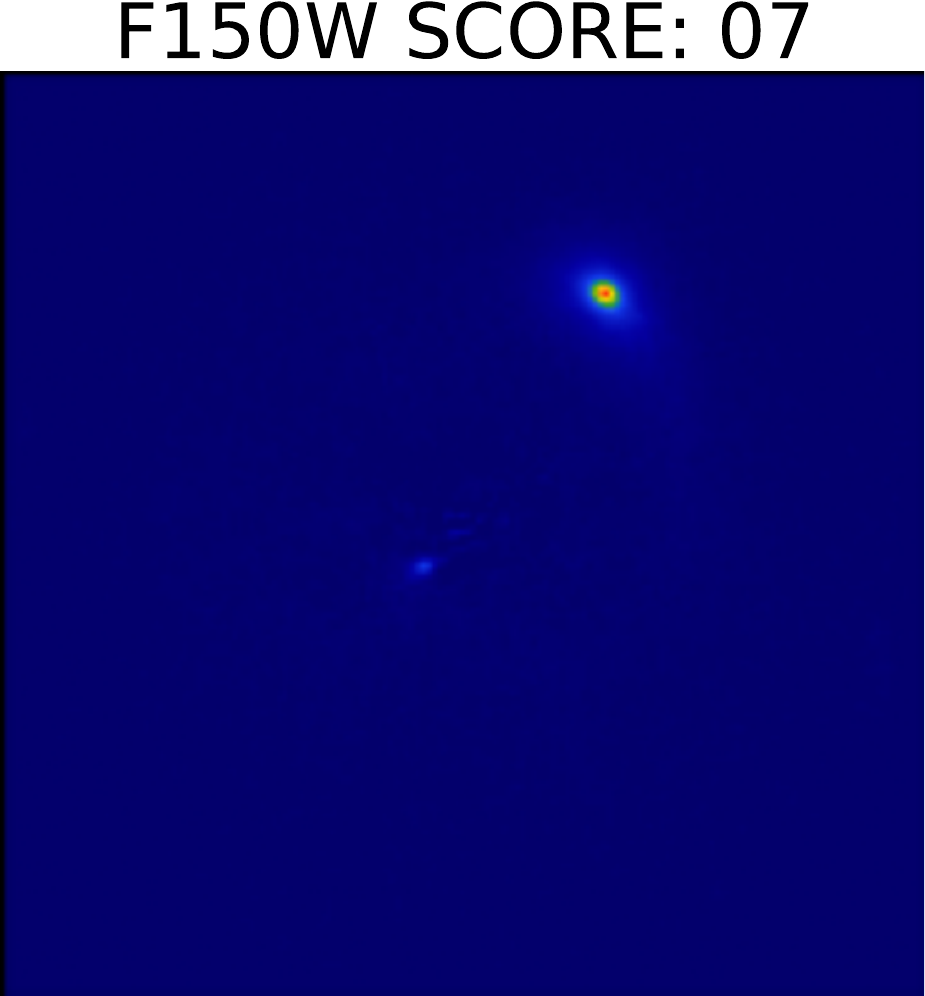}
\includegraphics[width=0.12\textwidth]{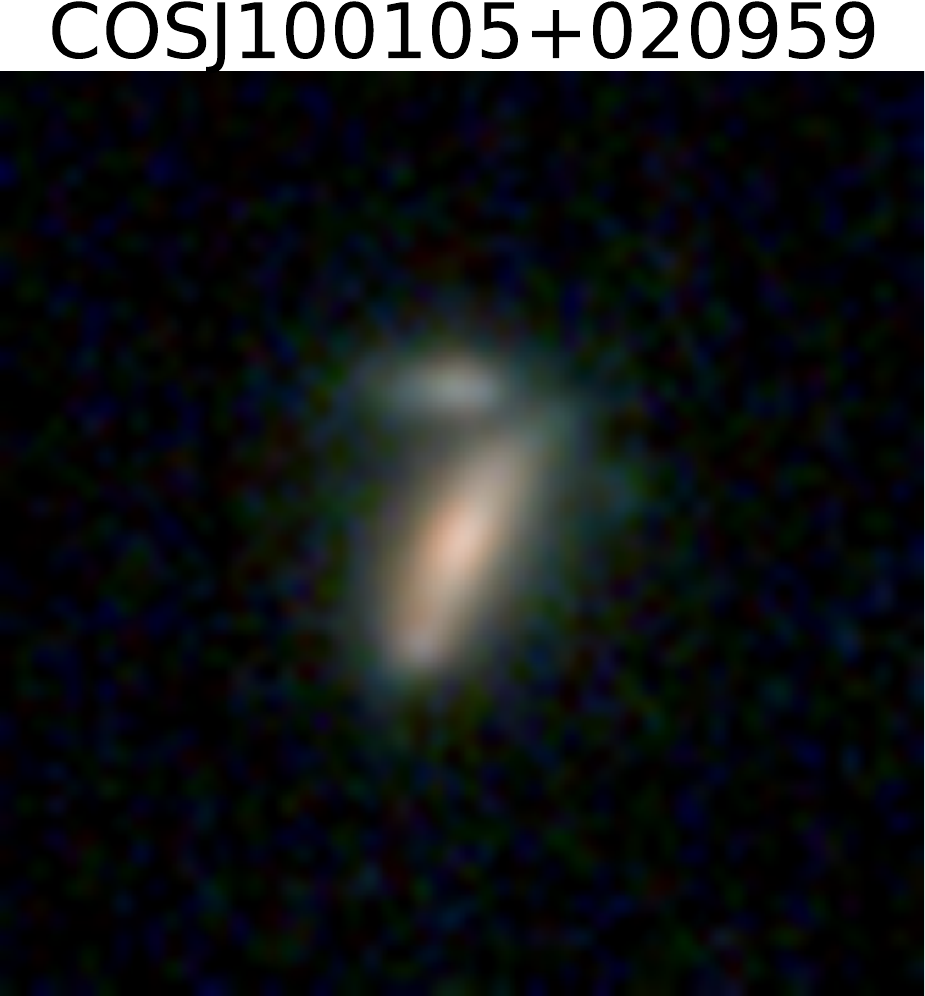}
\includegraphics[width=0.12\textwidth]{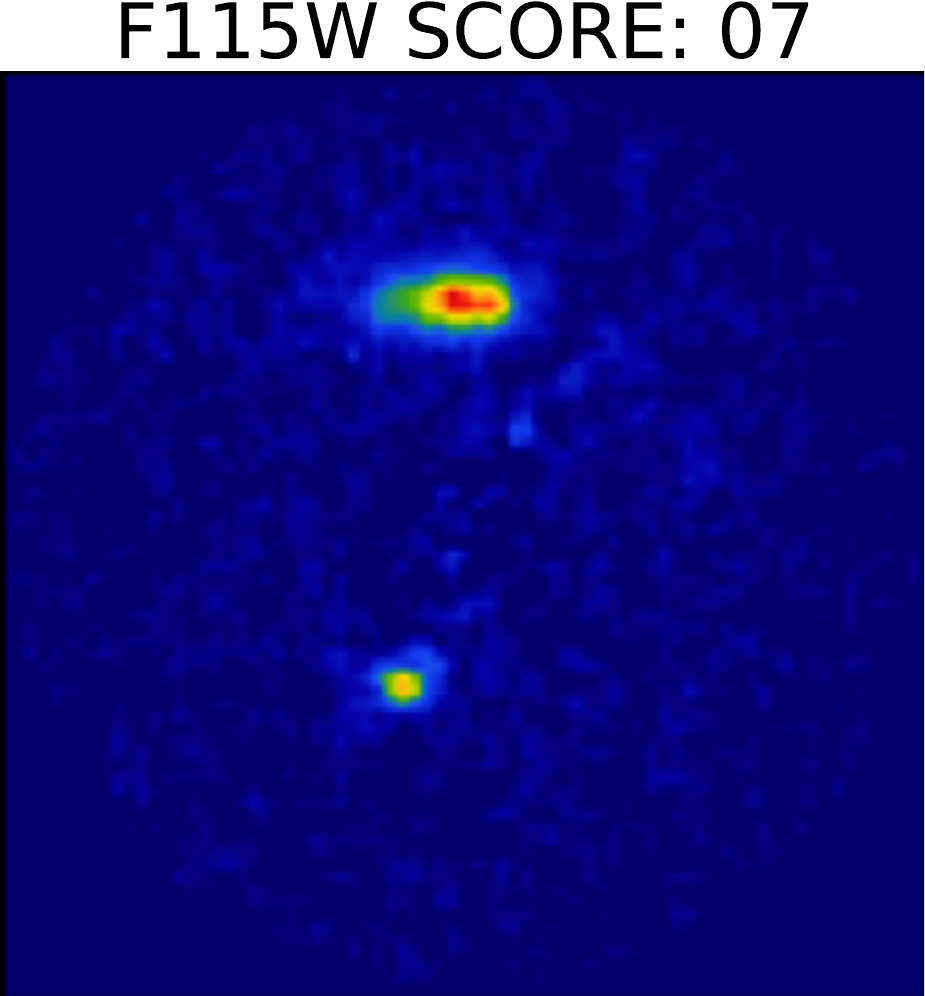}
\includegraphics[width=0.12\textwidth]{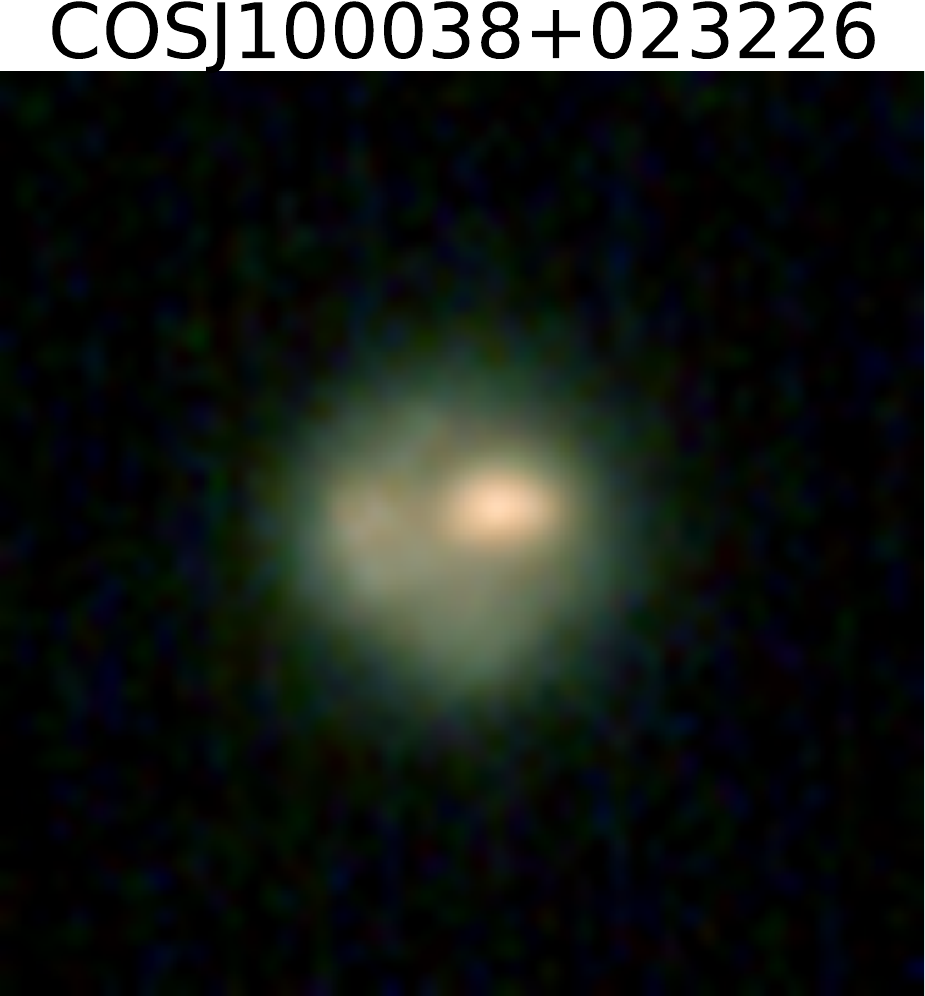}
\includegraphics[width=0.12\textwidth]{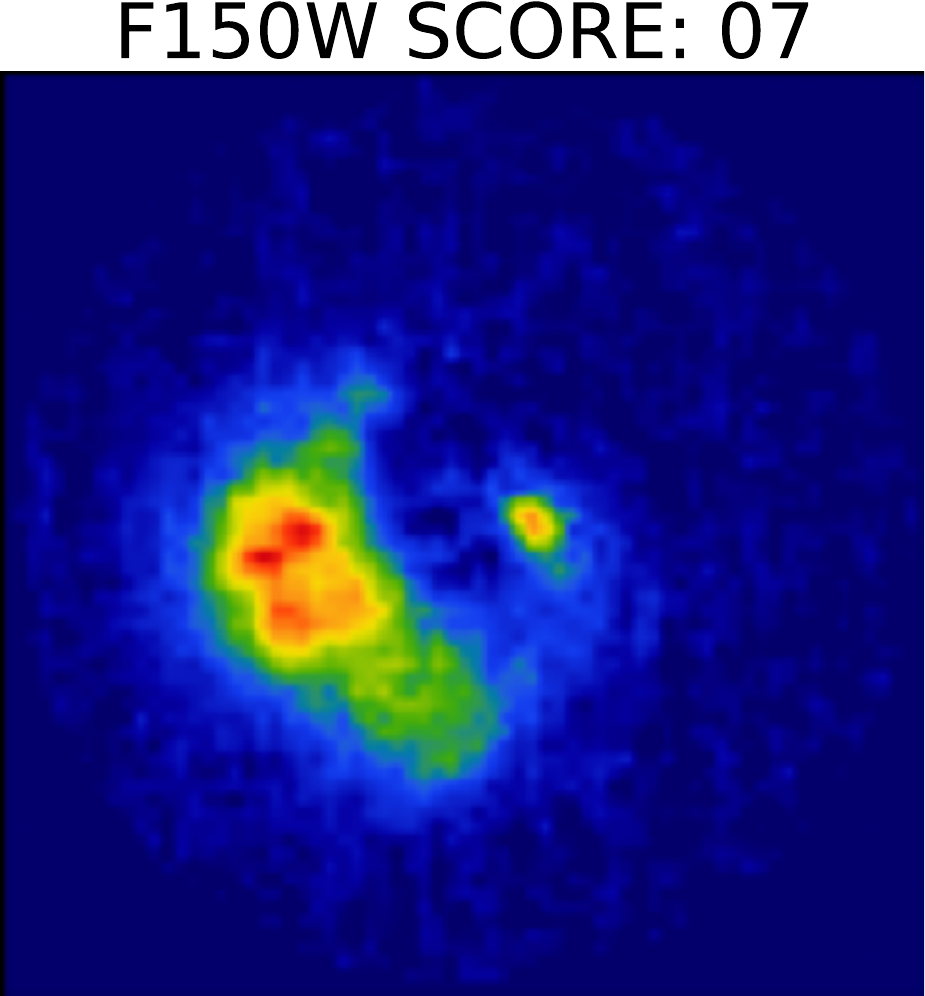}
\includegraphics[width=0.12\textwidth]{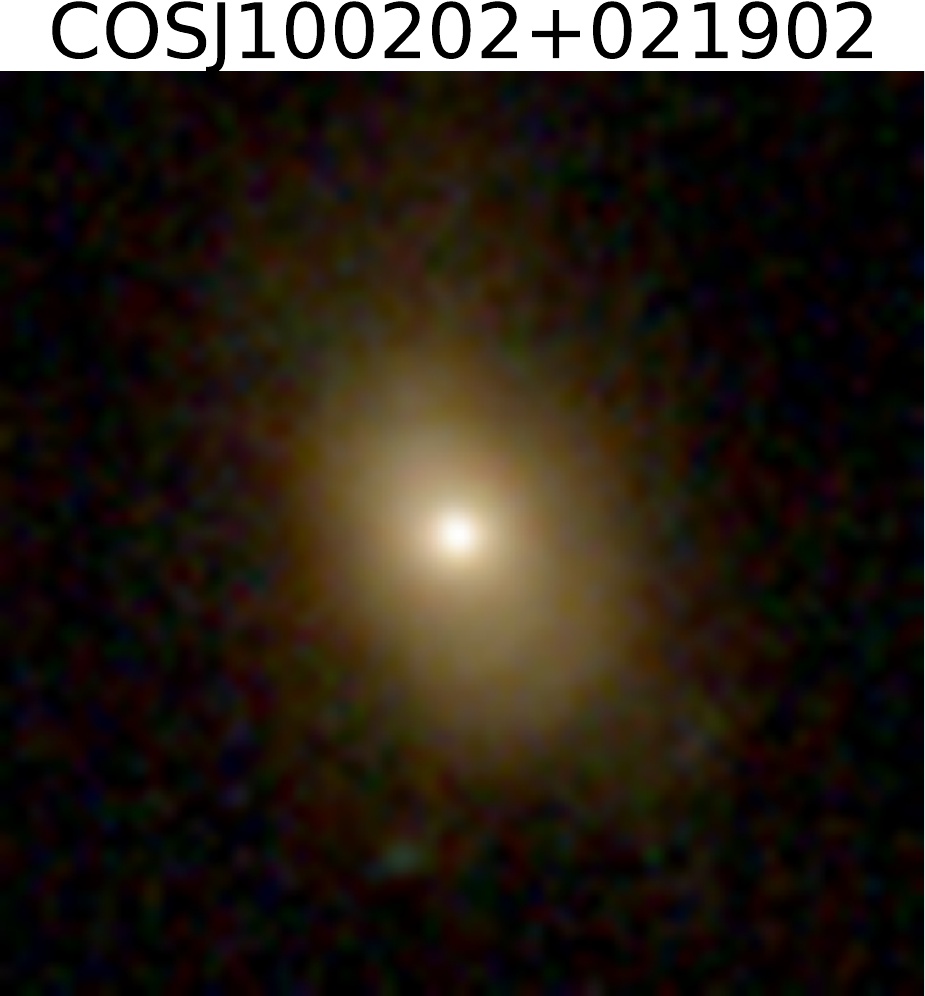}
\includegraphics[width=0.12\textwidth]{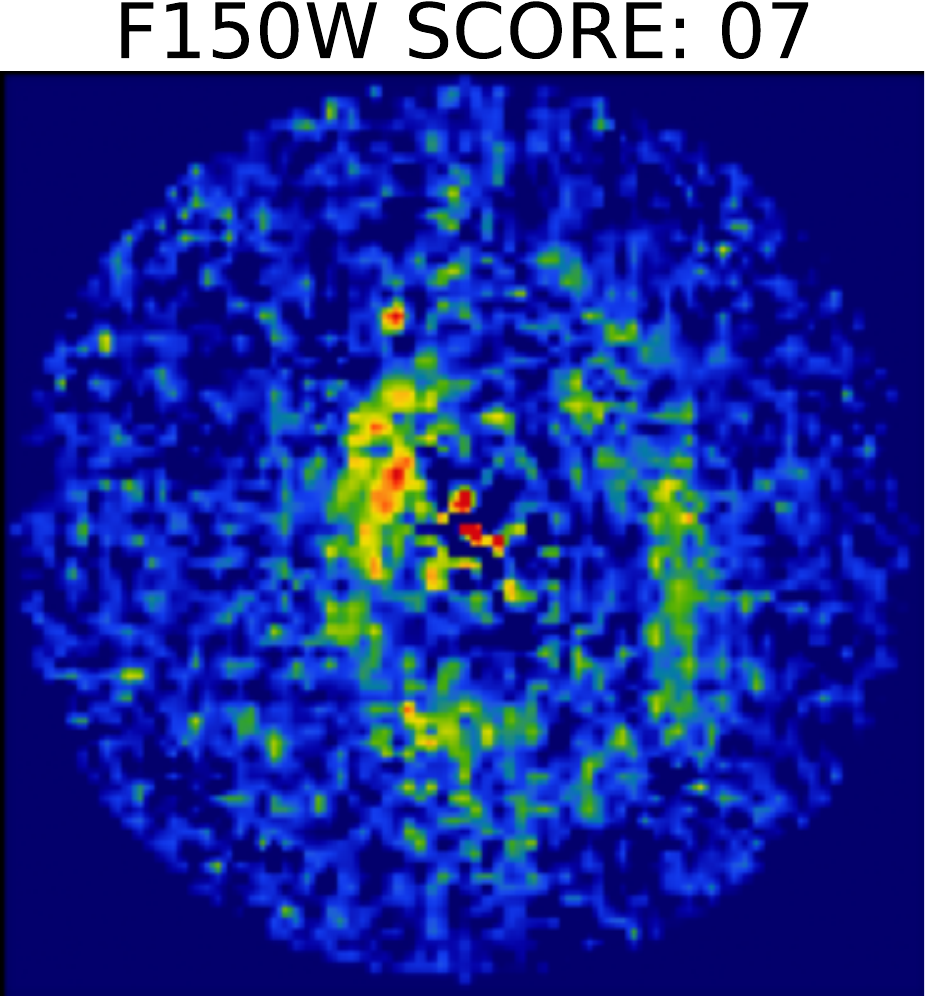}
\includegraphics[width=0.12\textwidth]{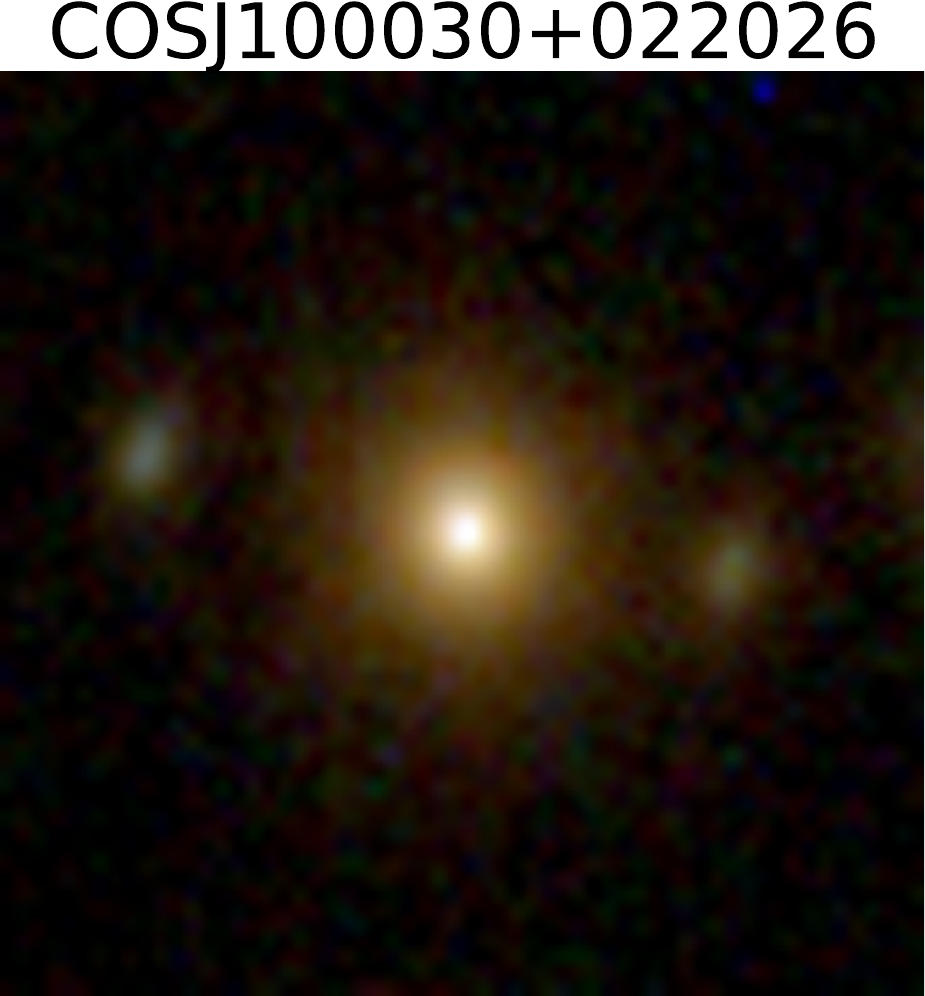}
\includegraphics[width=0.12\textwidth]{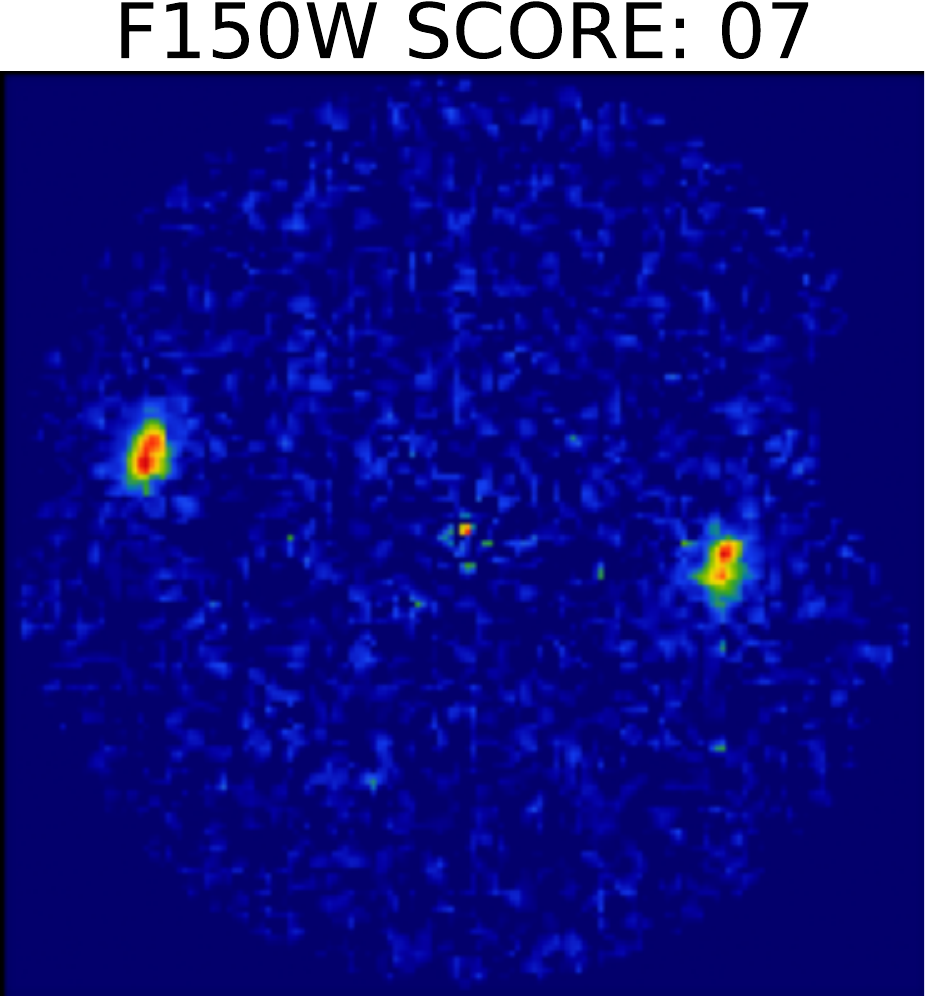}
\includegraphics[width=0.12\textwidth]{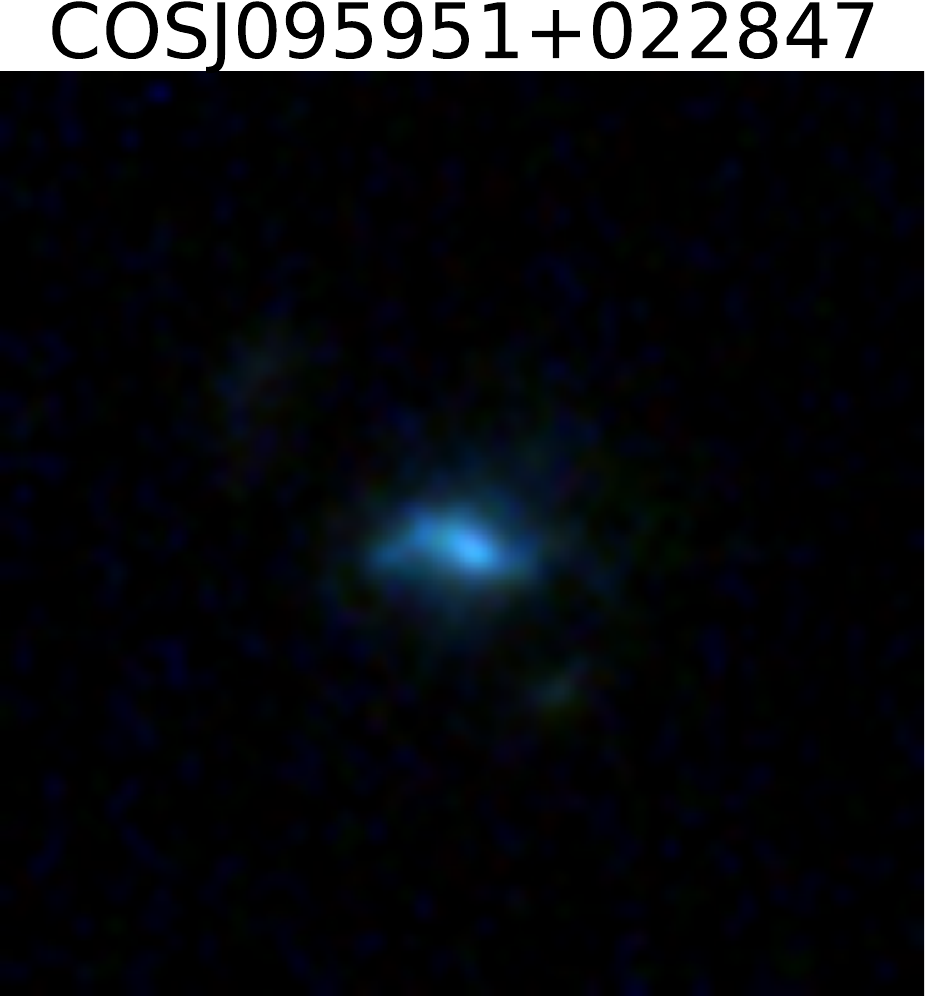}
\includegraphics[width=0.12\textwidth]{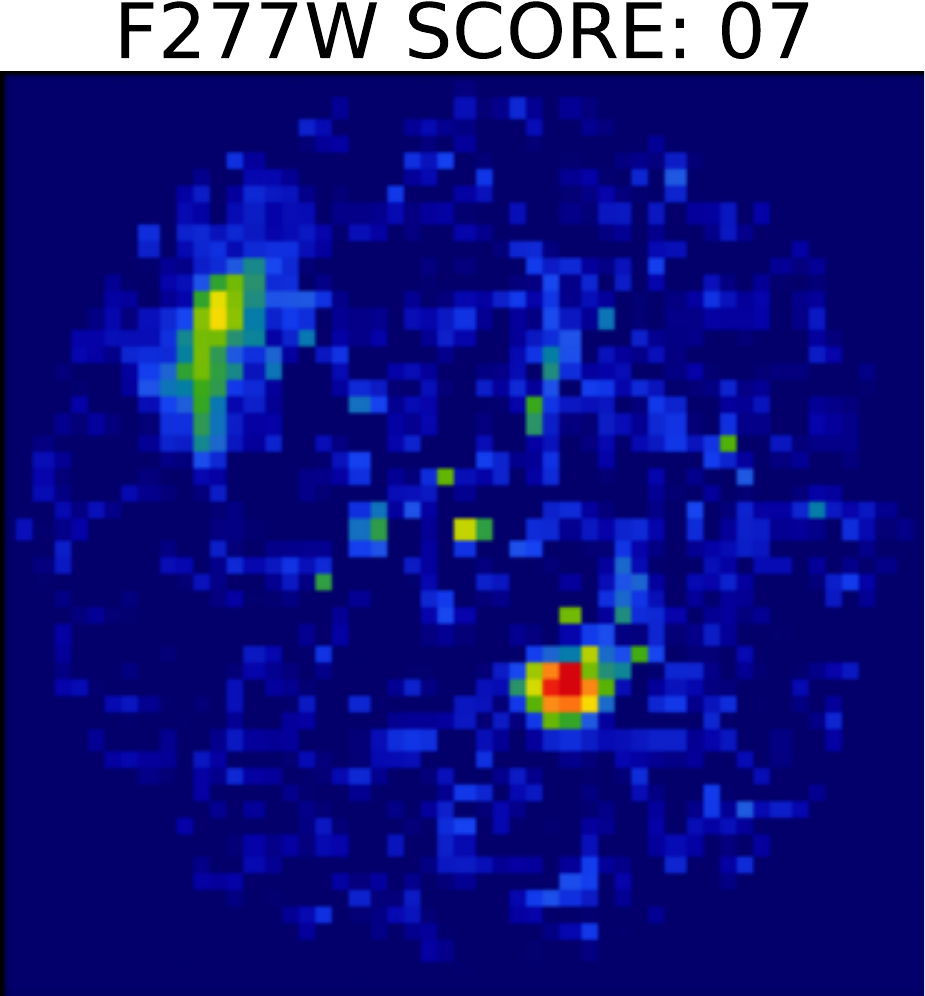}
\includegraphics[width=0.12\textwidth]{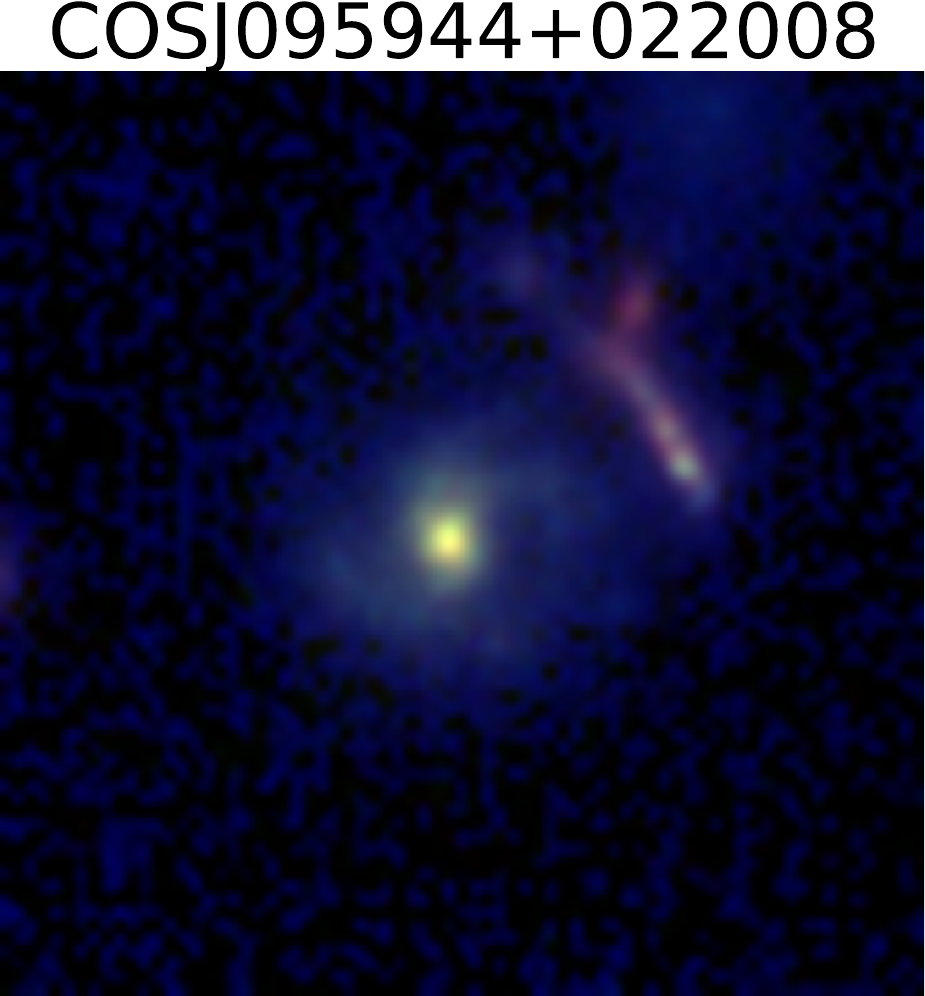}
\includegraphics[width=0.12\textwidth]{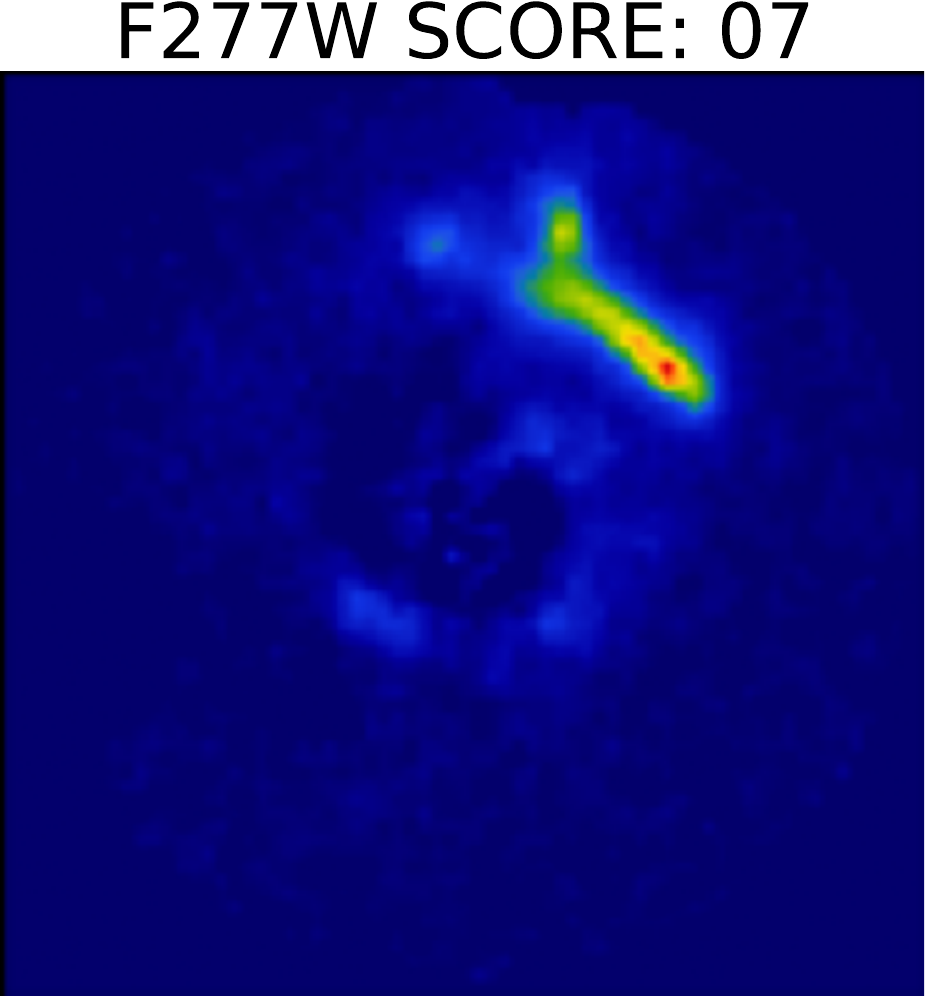}
\includegraphics[width=0.12\textwidth]{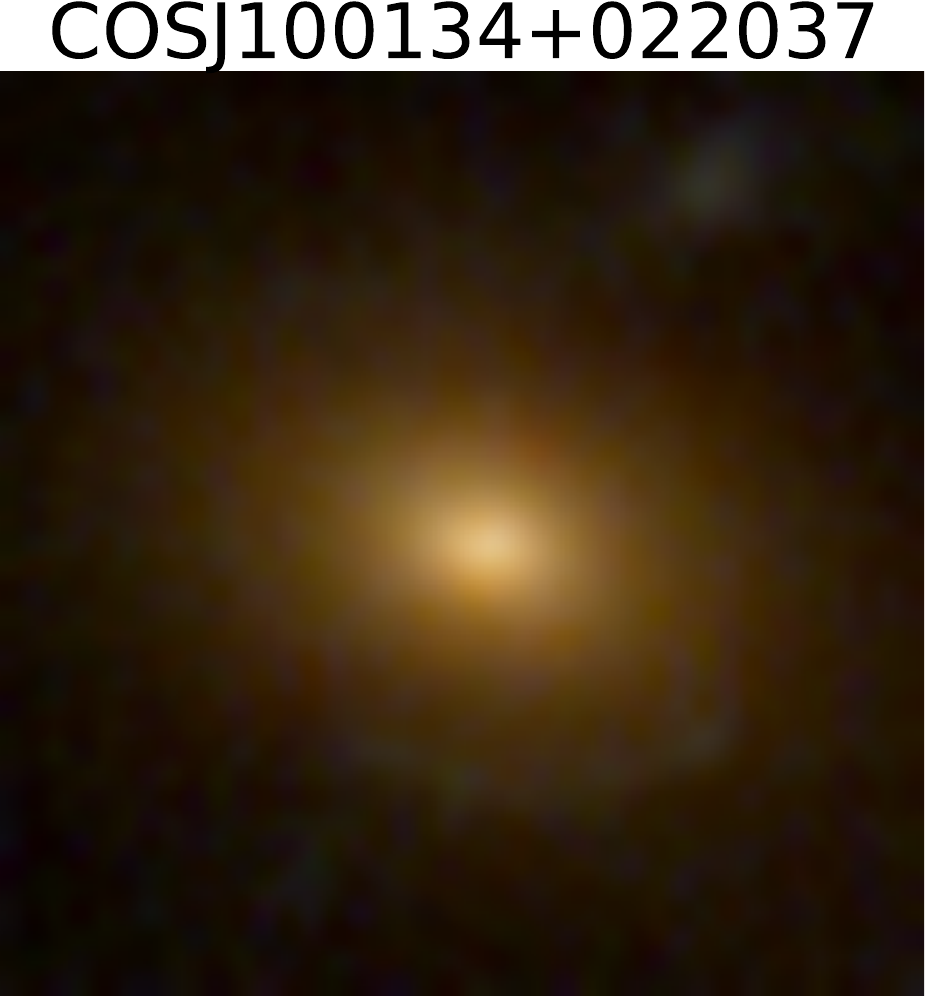}
\includegraphics[width=0.12\textwidth]{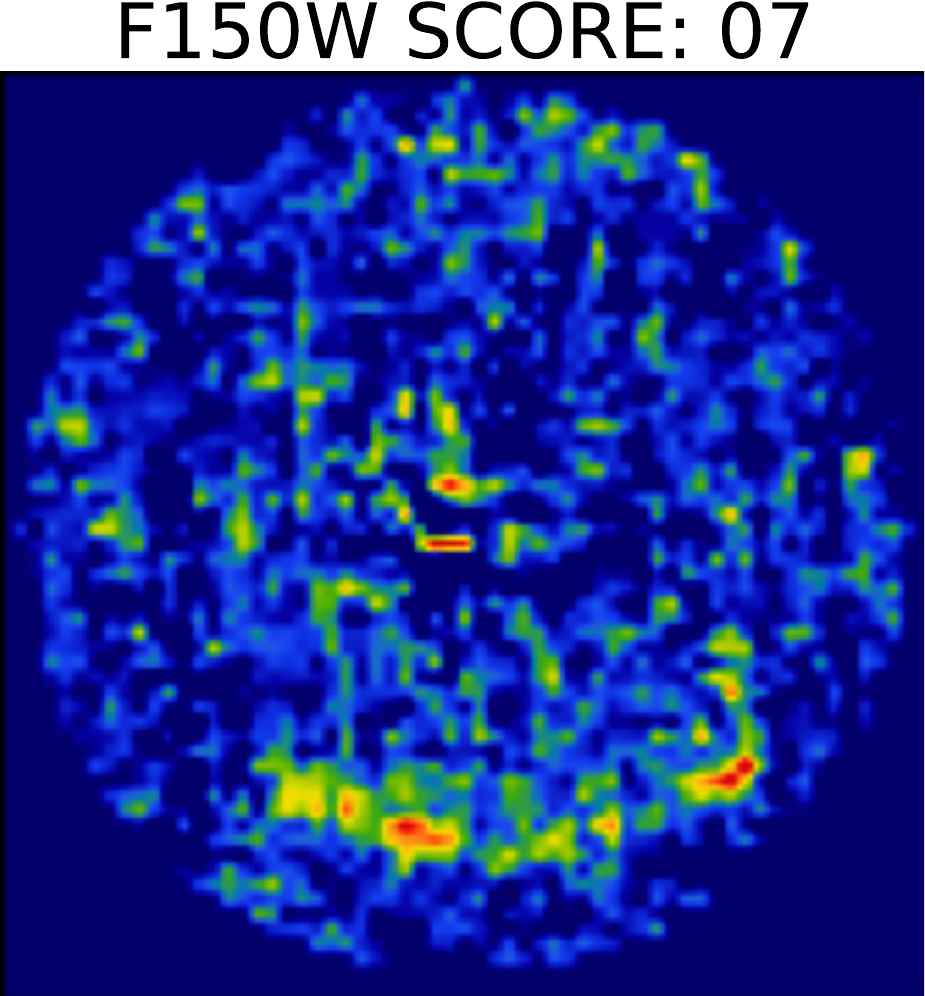}
\includegraphics[width=0.12\textwidth]{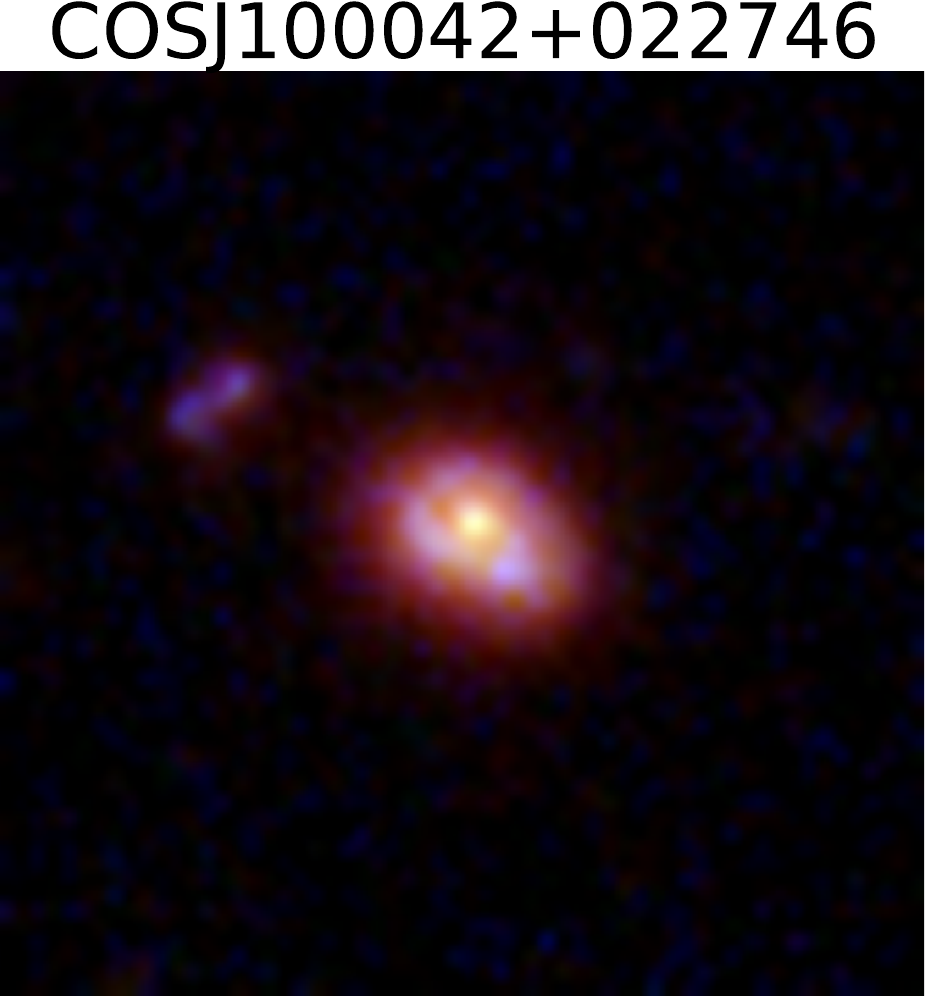}
\includegraphics[width=0.12\textwidth]{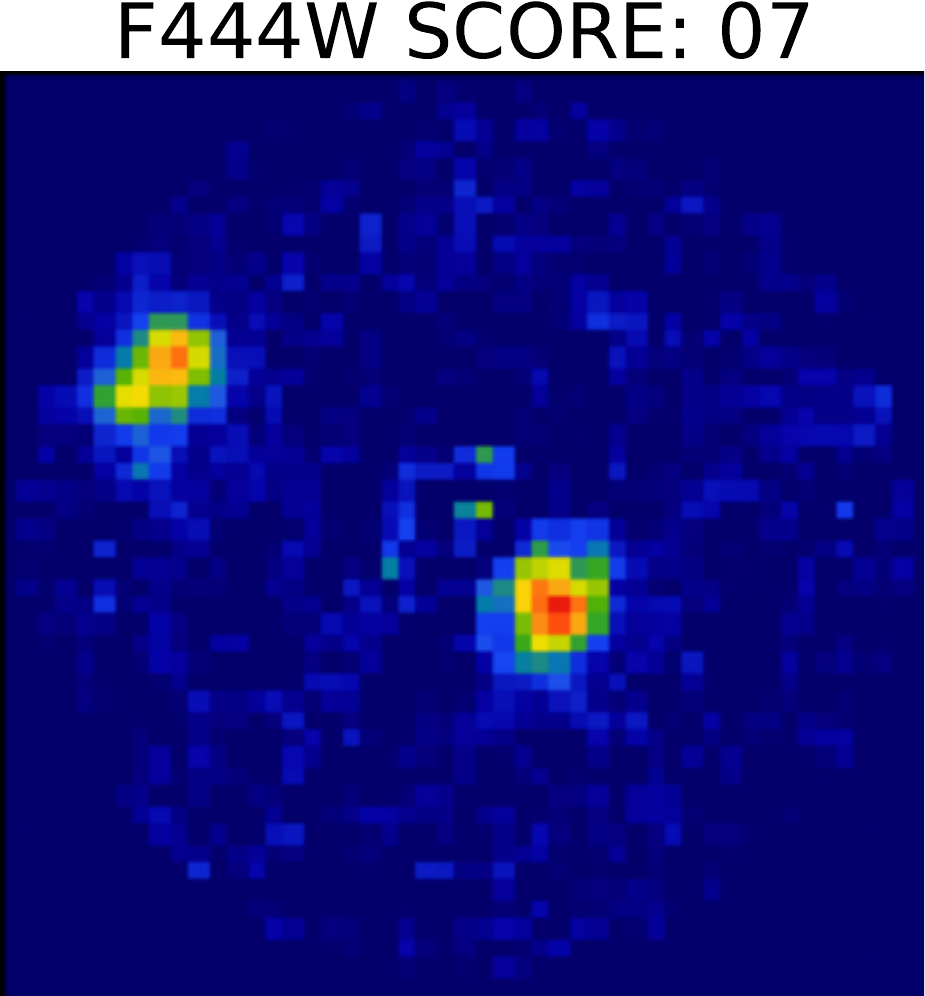}
\includegraphics[width=0.12\textwidth]{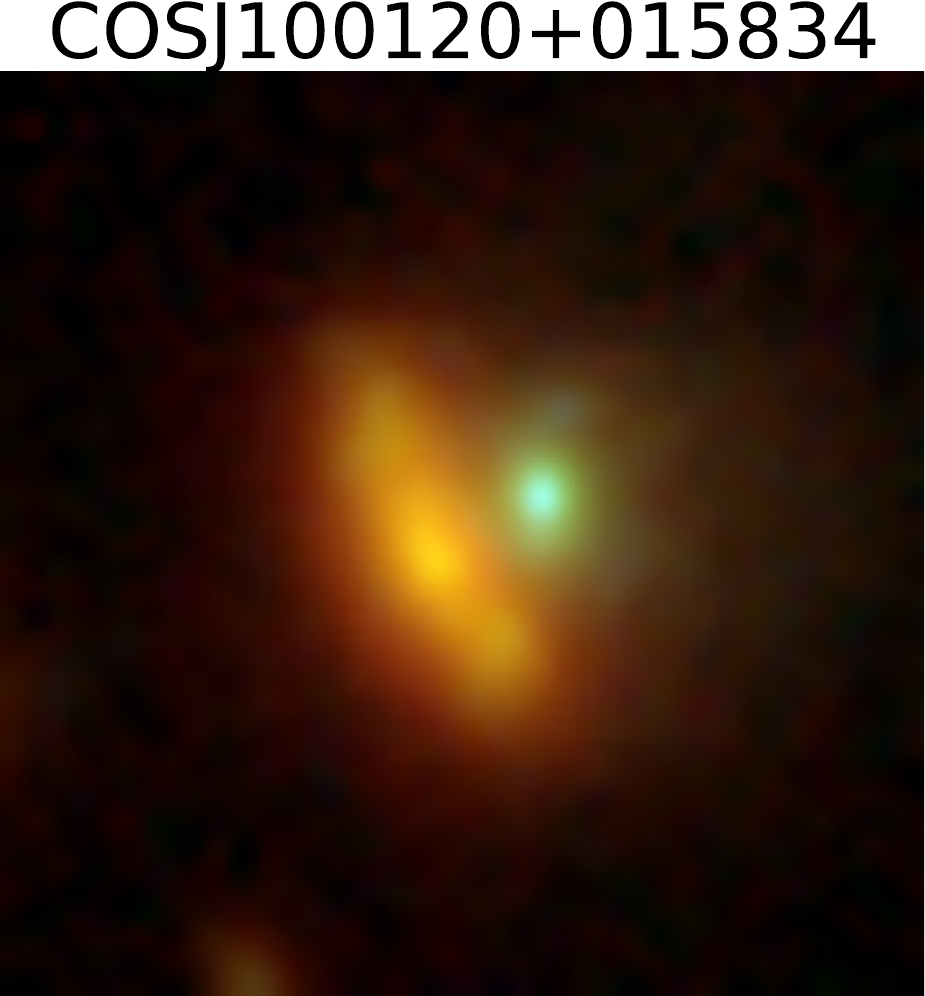}
\includegraphics[width=0.12\textwidth]{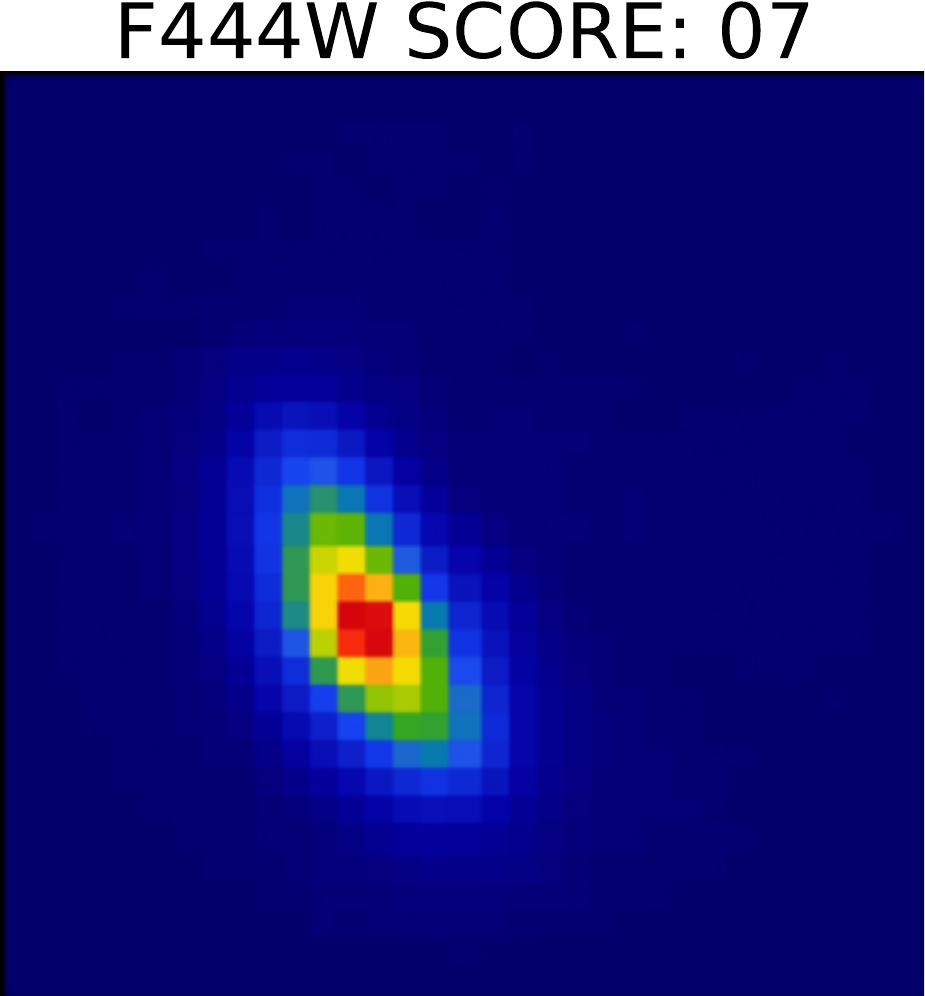}
\includegraphics[width=0.12\textwidth]{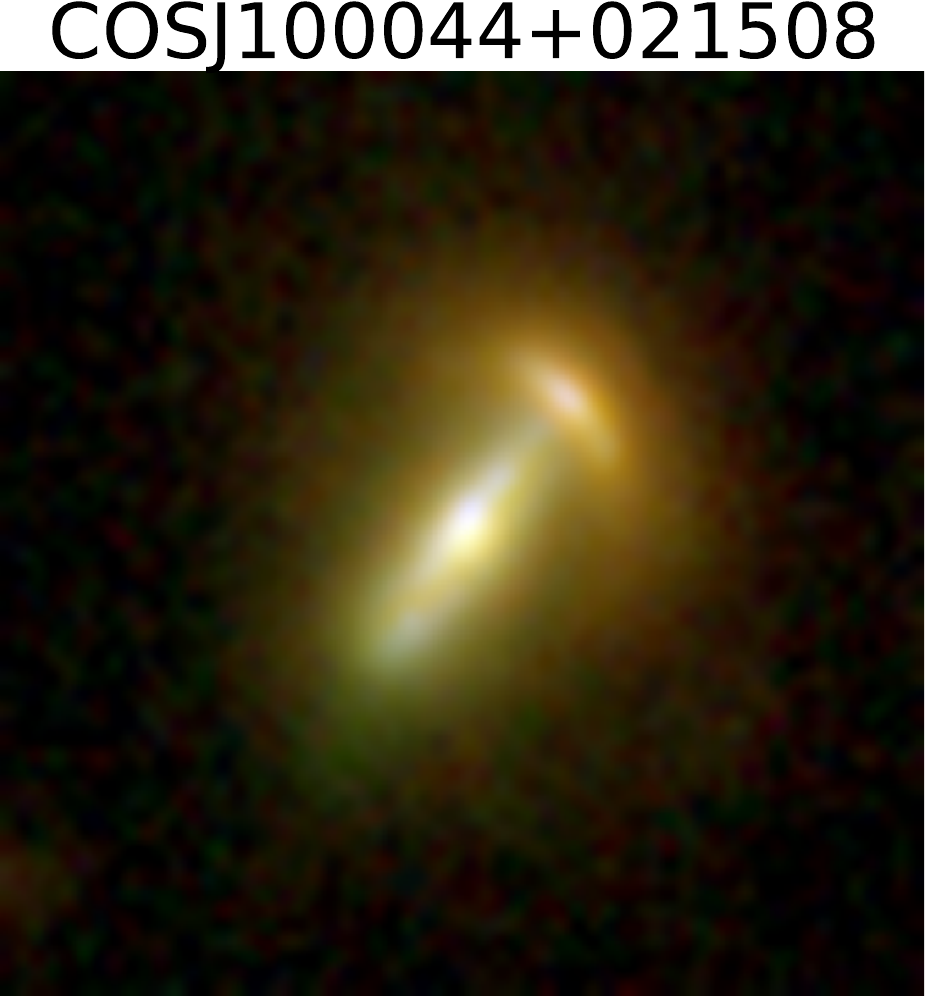}
\includegraphics[width=0.12\textwidth]{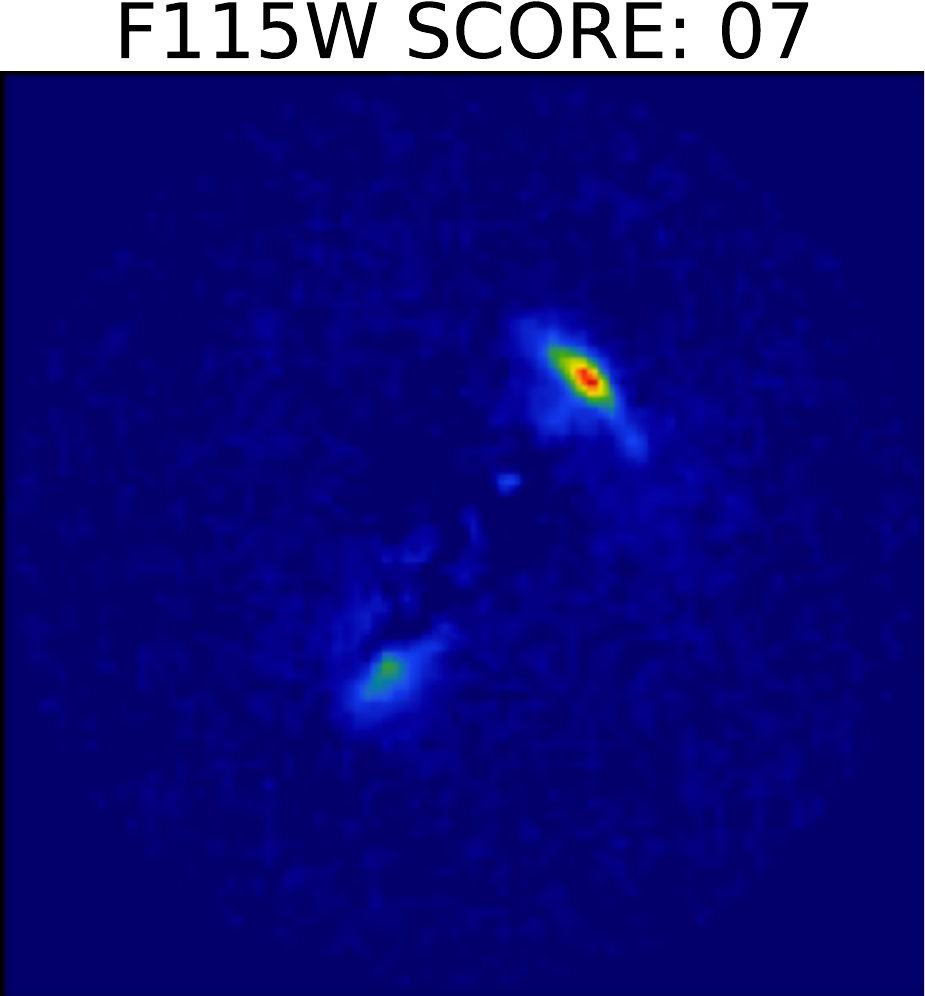}
\includegraphics[width=0.12\textwidth]{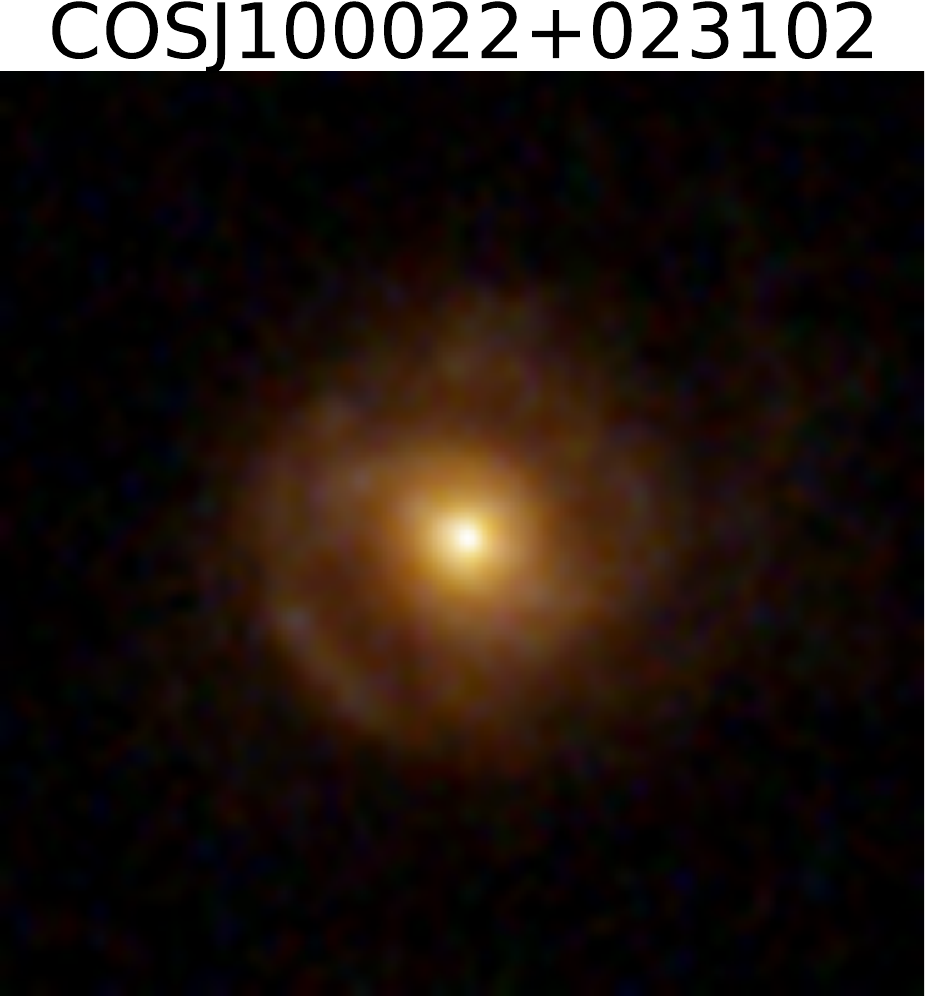}
\includegraphics[width=0.12\textwidth]{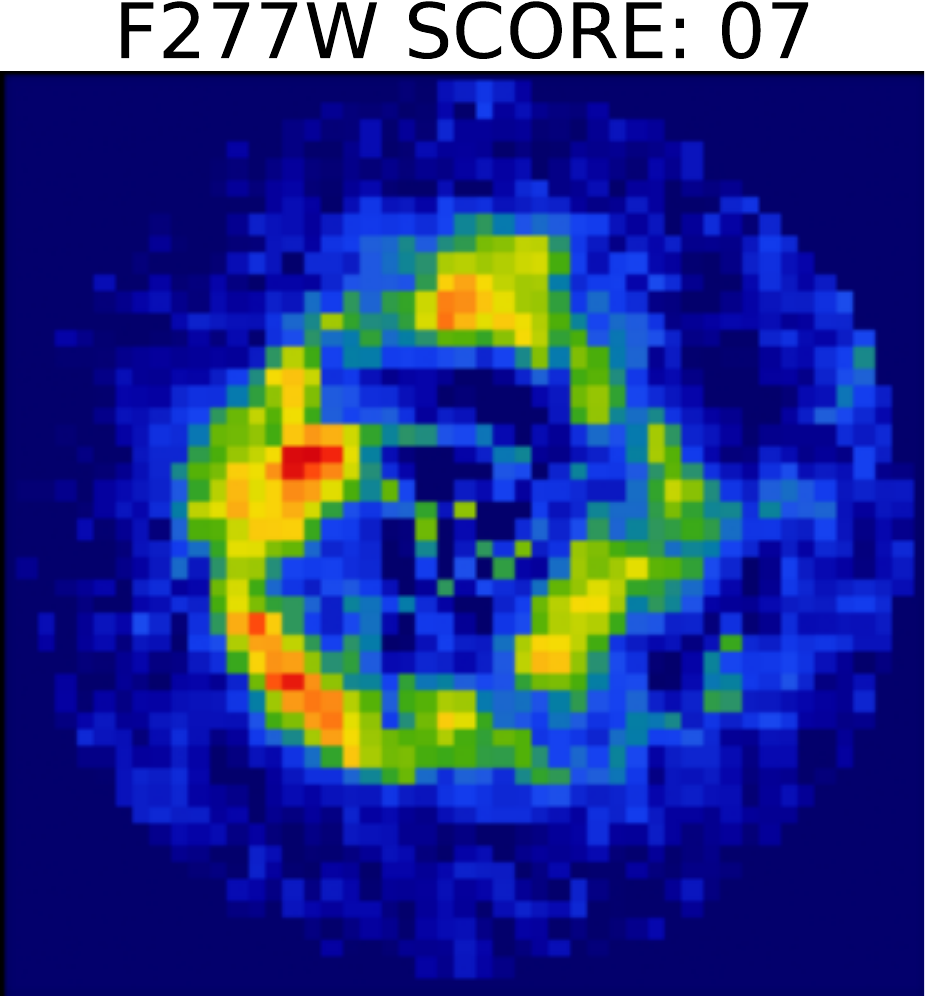}
\includegraphics[width=0.12\textwidth]{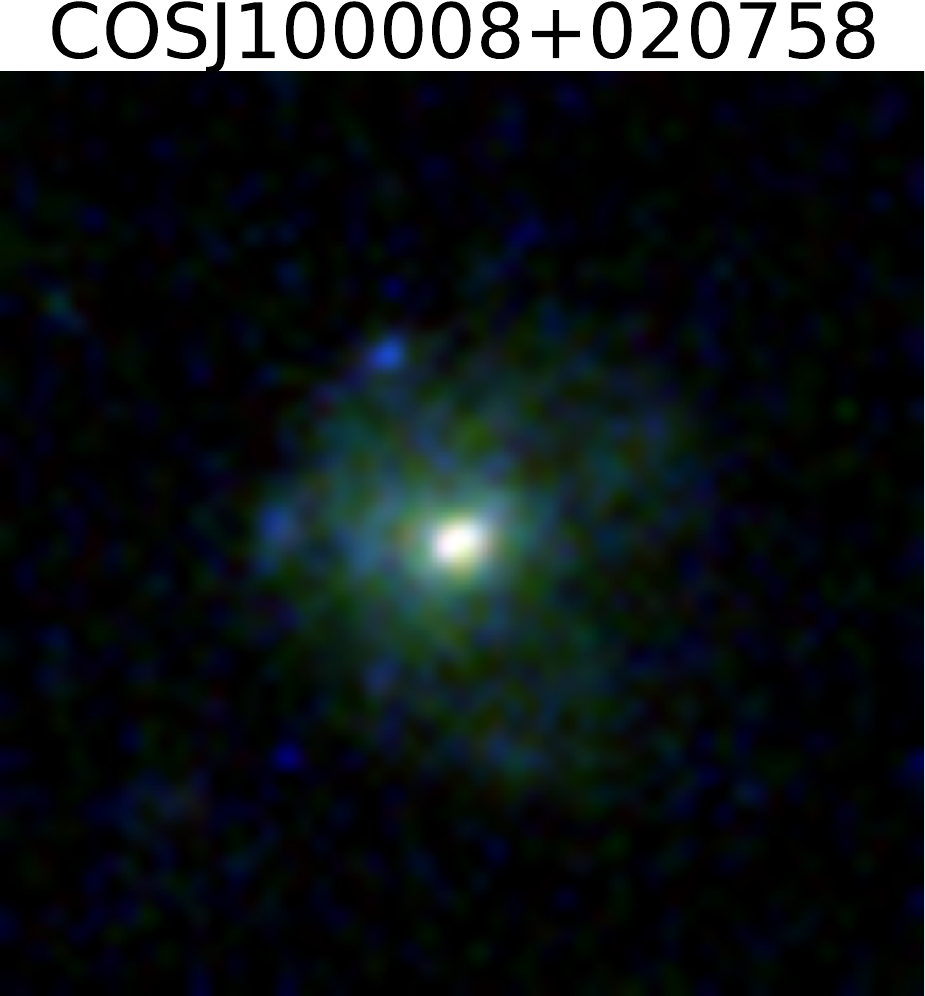}
\includegraphics[width=0.12\textwidth]{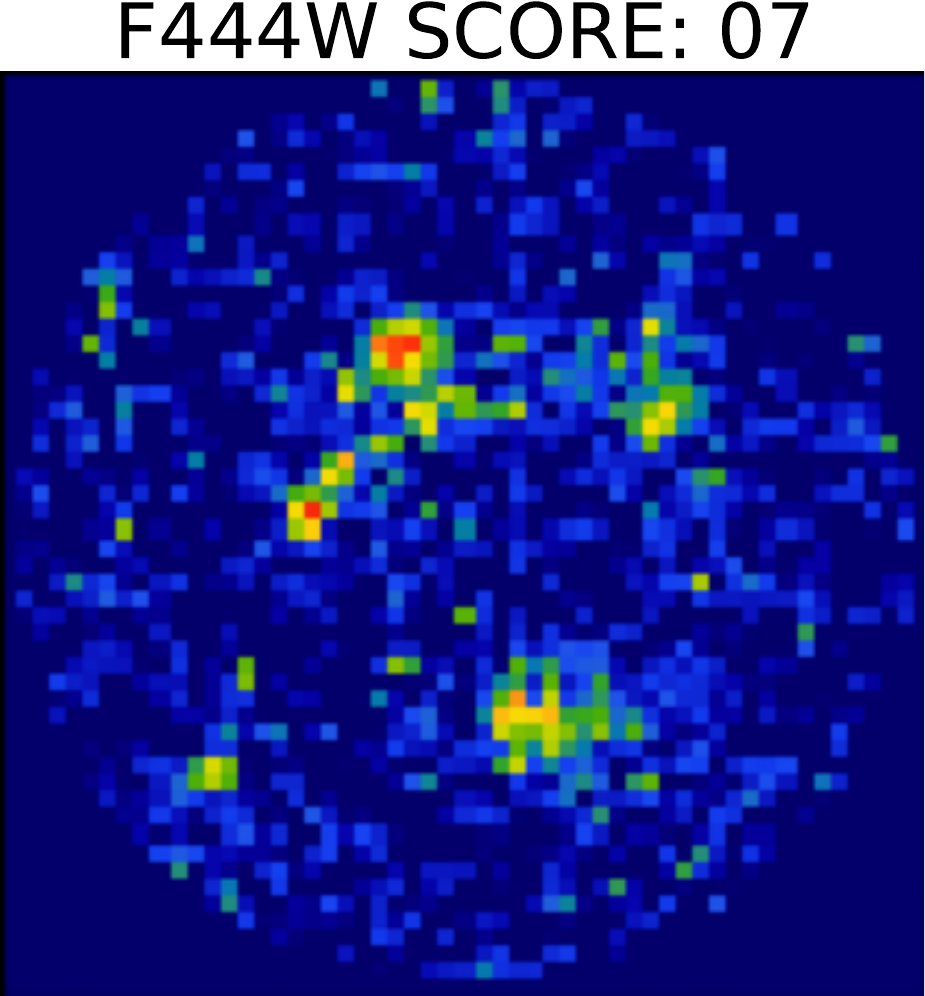}
\includegraphics[width=0.12\textwidth]{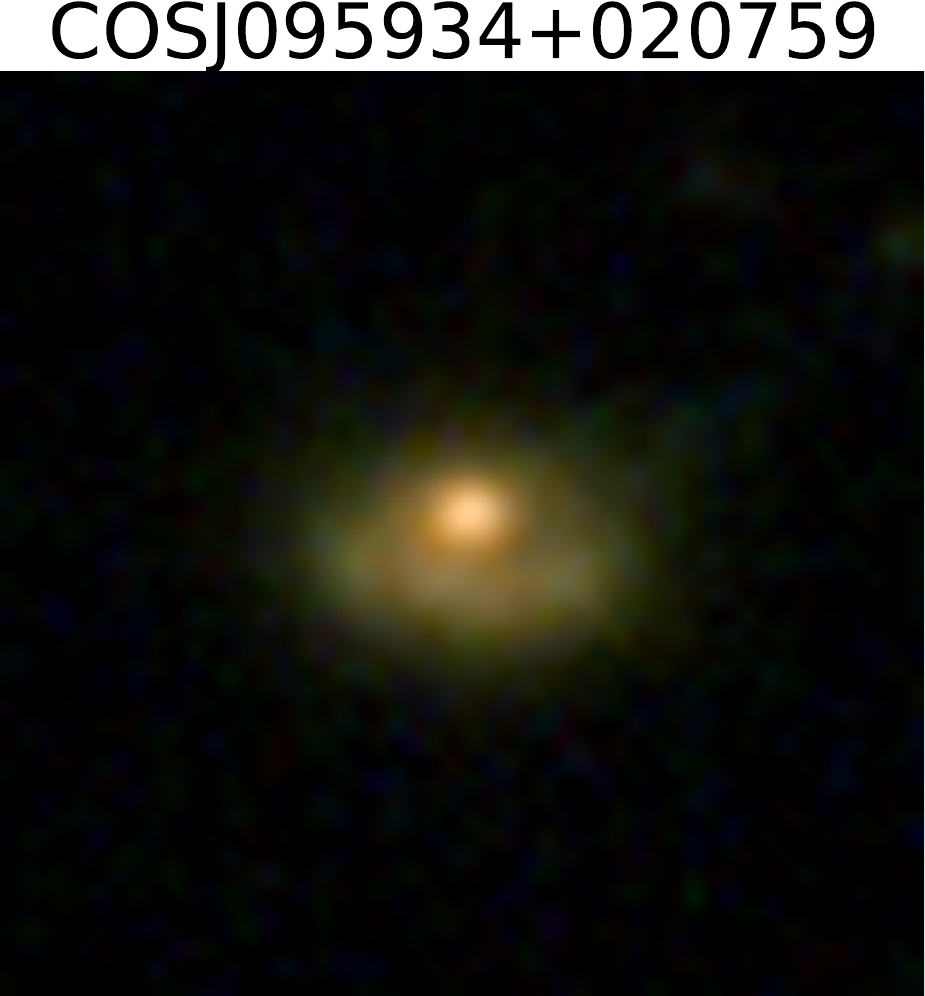}
\includegraphics[width=0.12\textwidth]{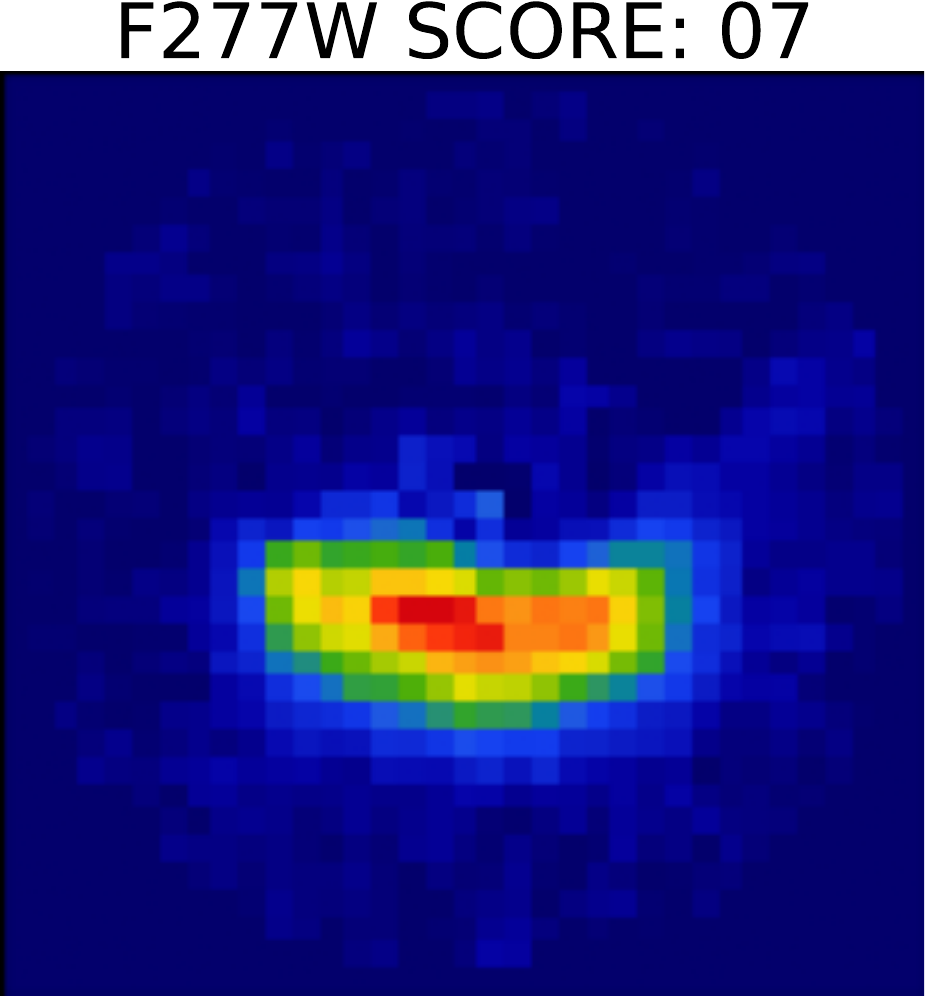}
\caption{
Figure \ref{figure:CutoutA} continued.
}
\label{figure:CutoutA2}
\vspace{-9pt} 
\end{figure*}

\begin{figure*}\ContinuedFloat
\centering
\includegraphics[width=0.12\textwidth]{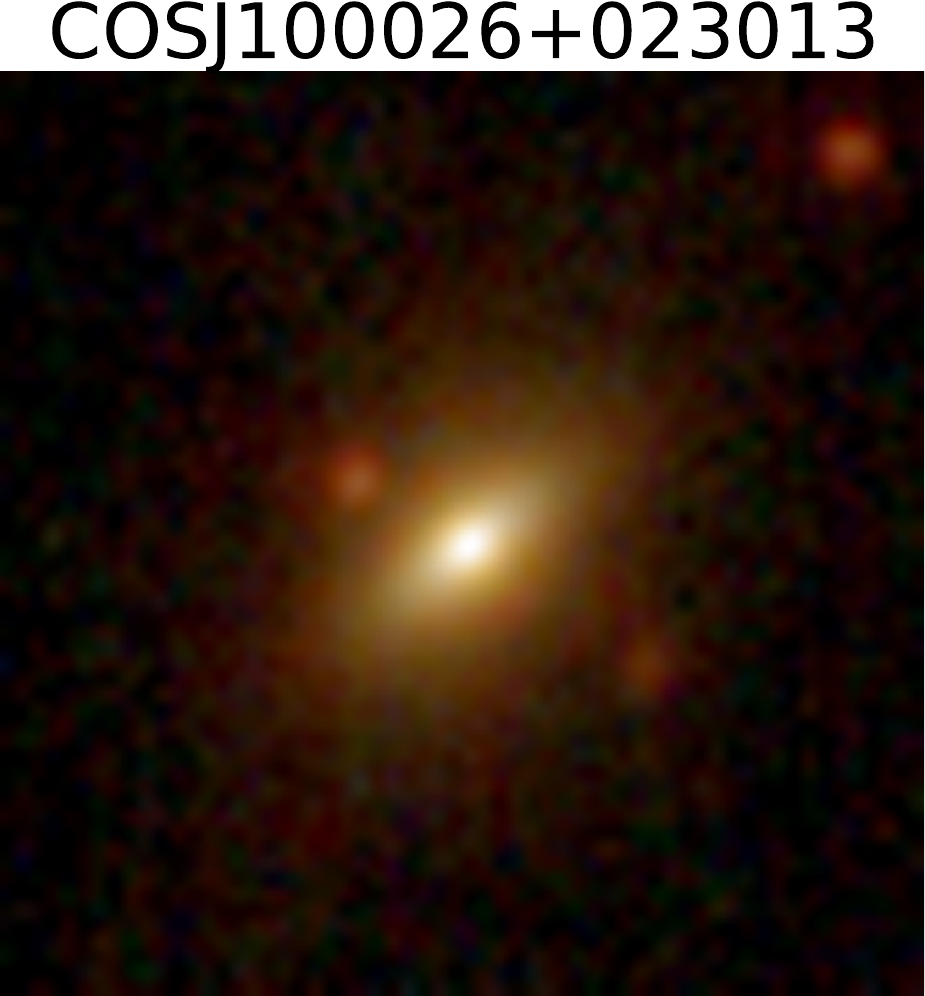}
\includegraphics[width=0.12\textwidth]{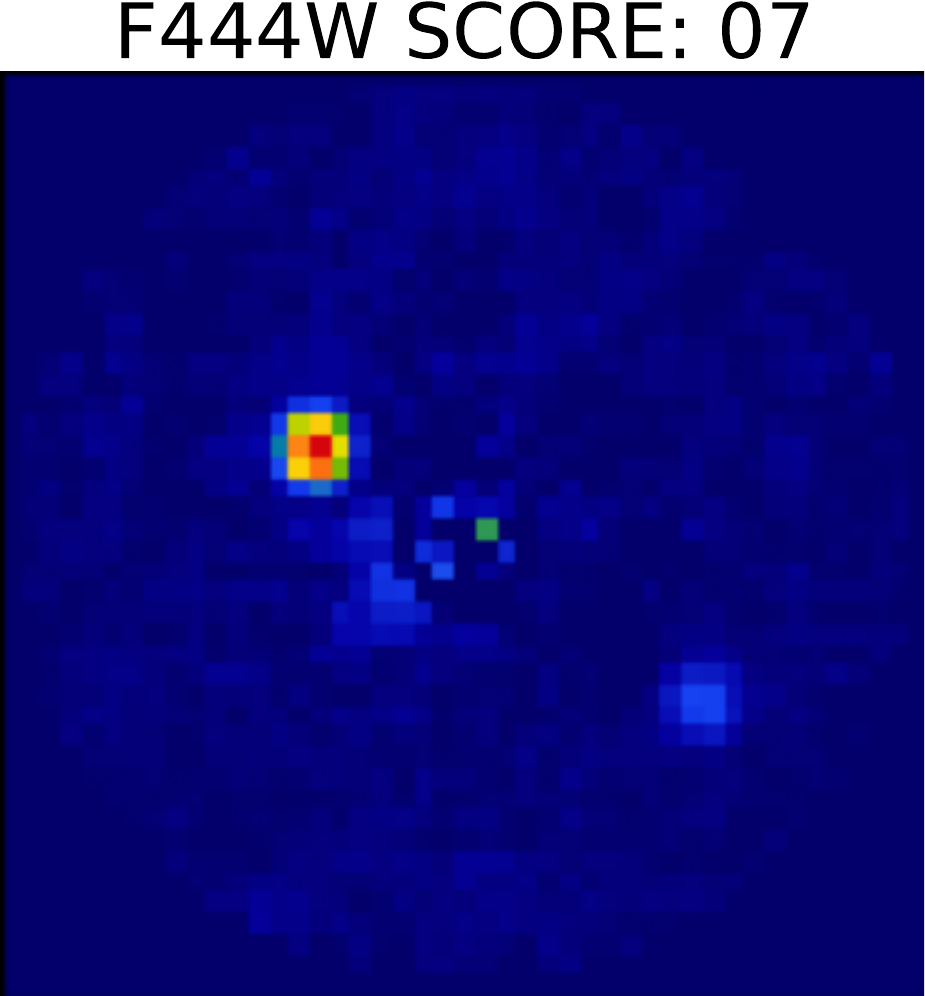}
\includegraphics[width=0.12\textwidth]{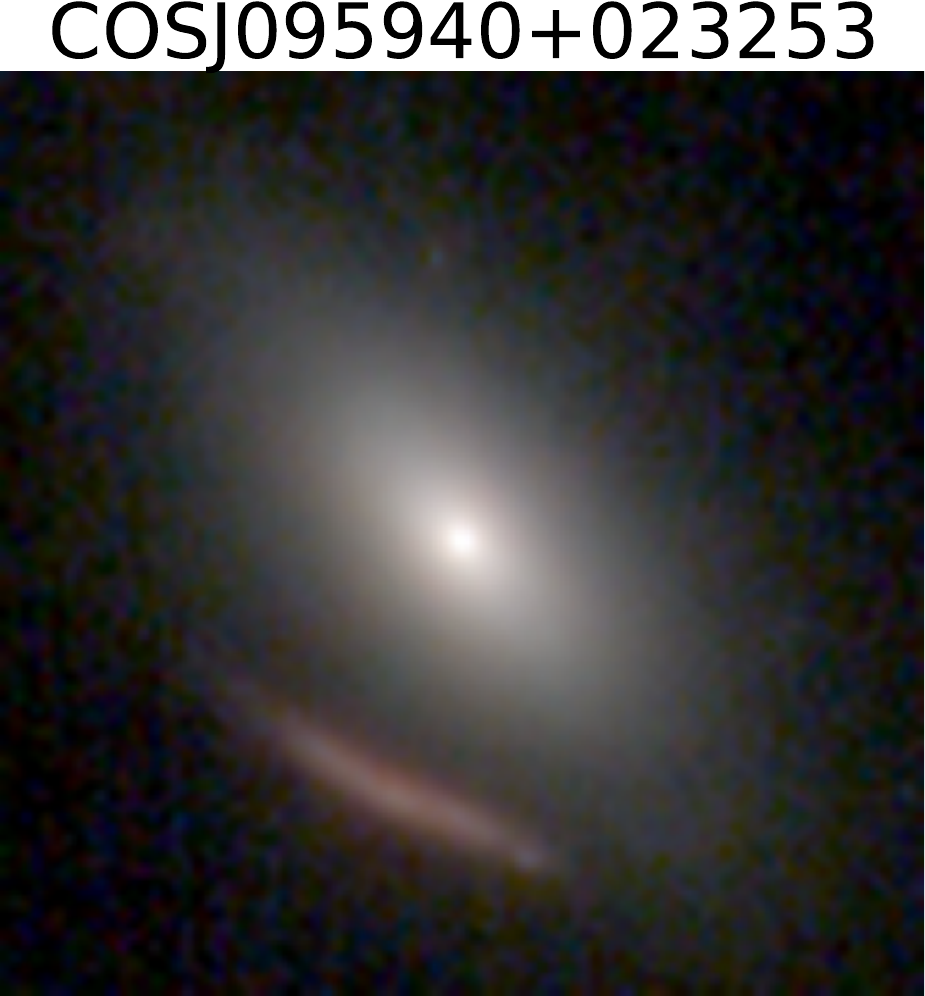}
\includegraphics[width=0.12\textwidth]{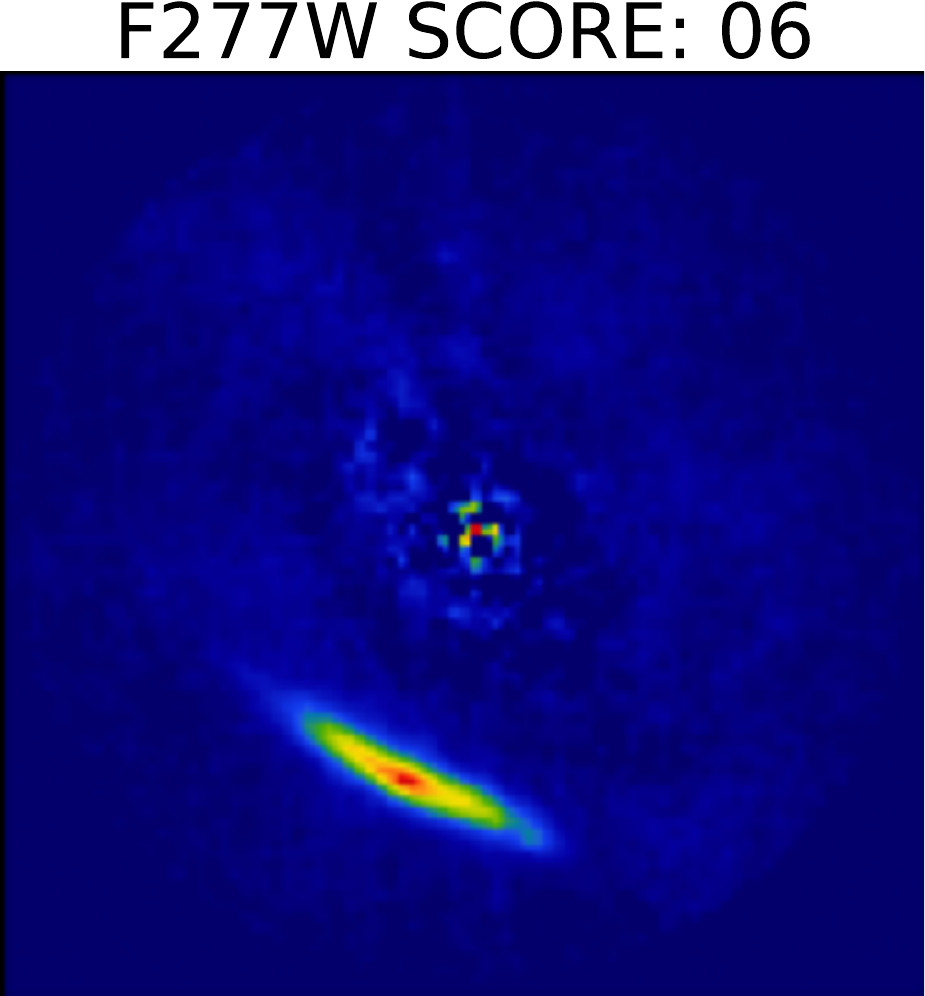}
\includegraphics[width=0.12\textwidth]{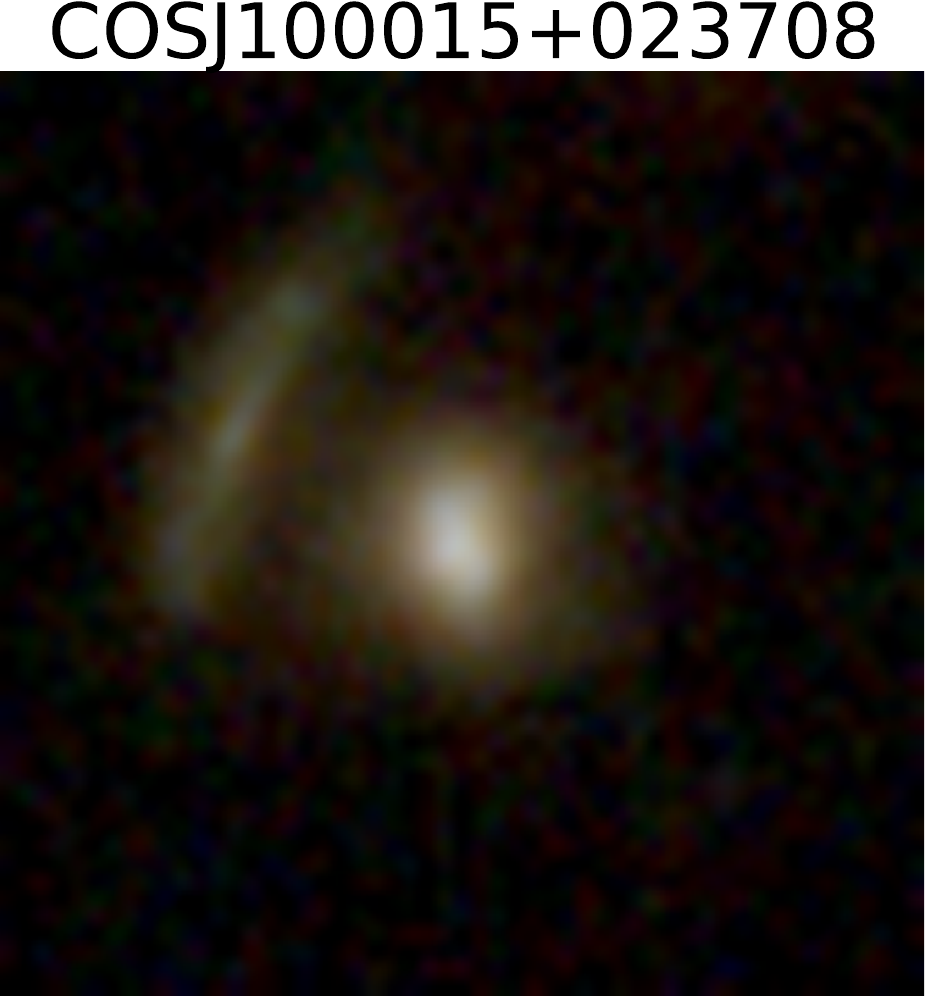}
\includegraphics[width=0.12\textwidth]{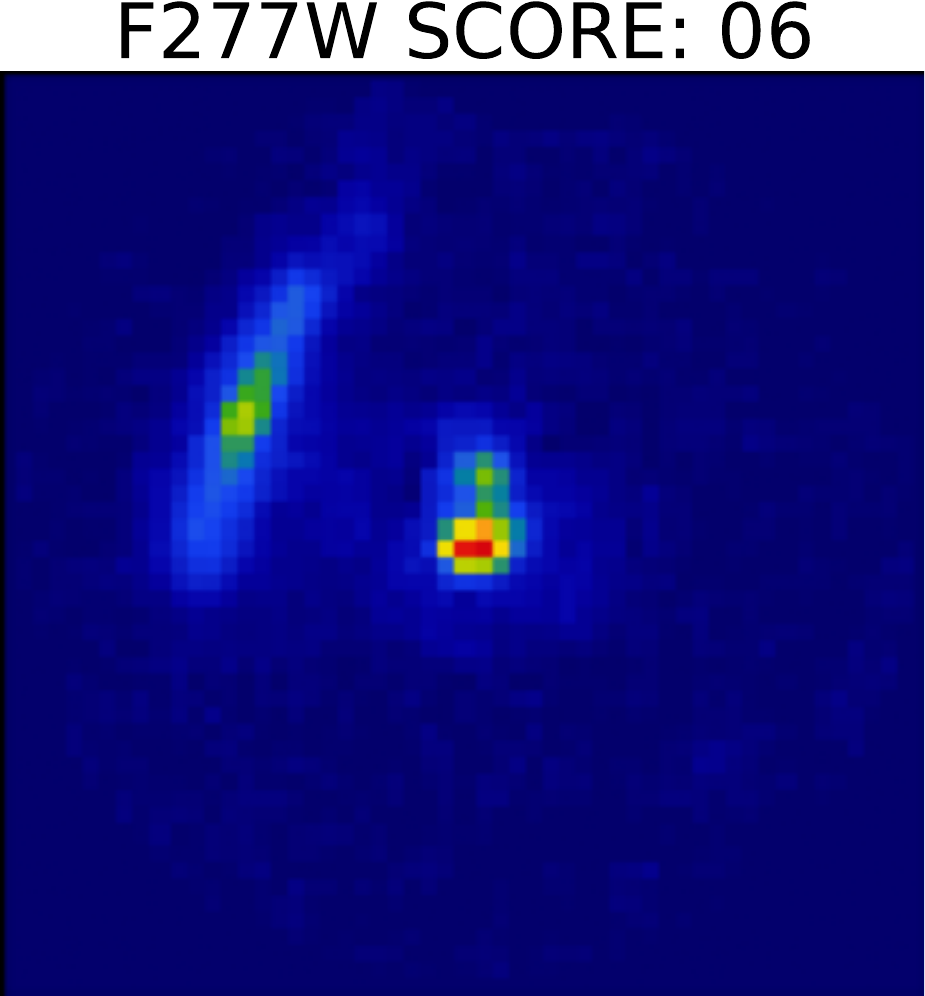}
\includegraphics[width=0.12\textwidth]{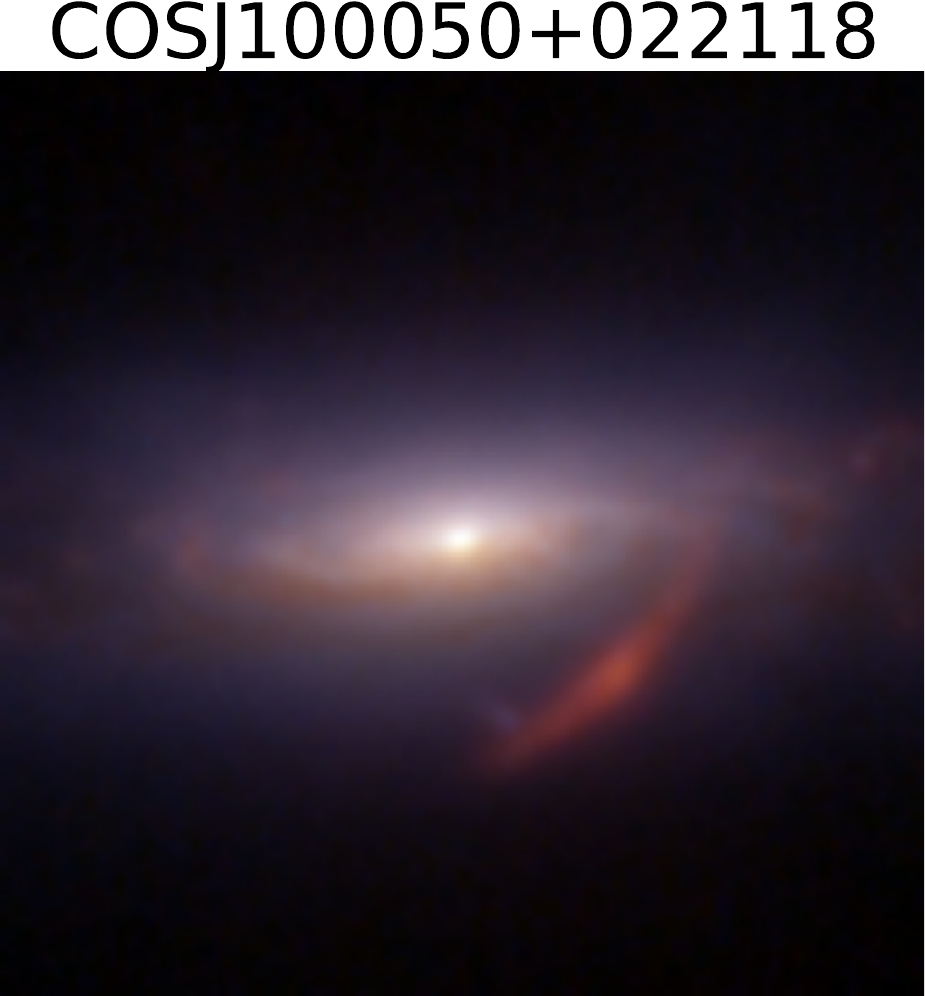}
\includegraphics[width=0.12\textwidth]{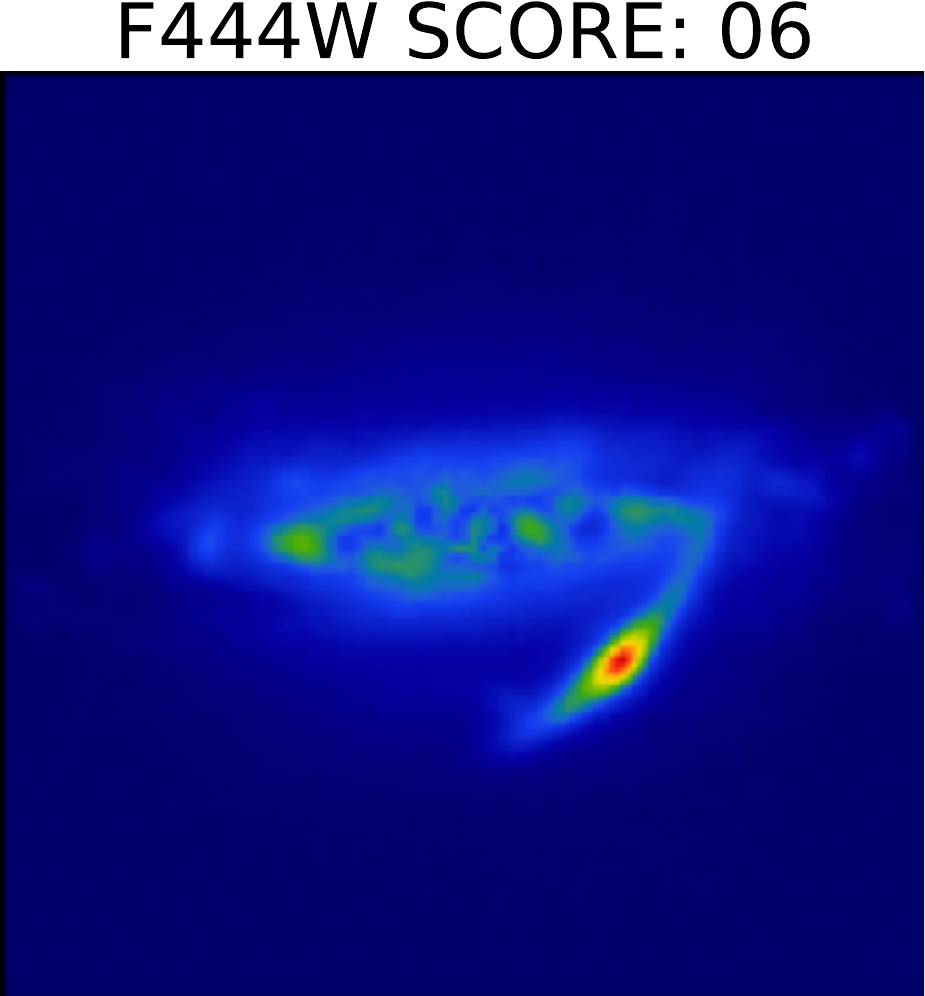}
\includegraphics[width=0.12\textwidth]{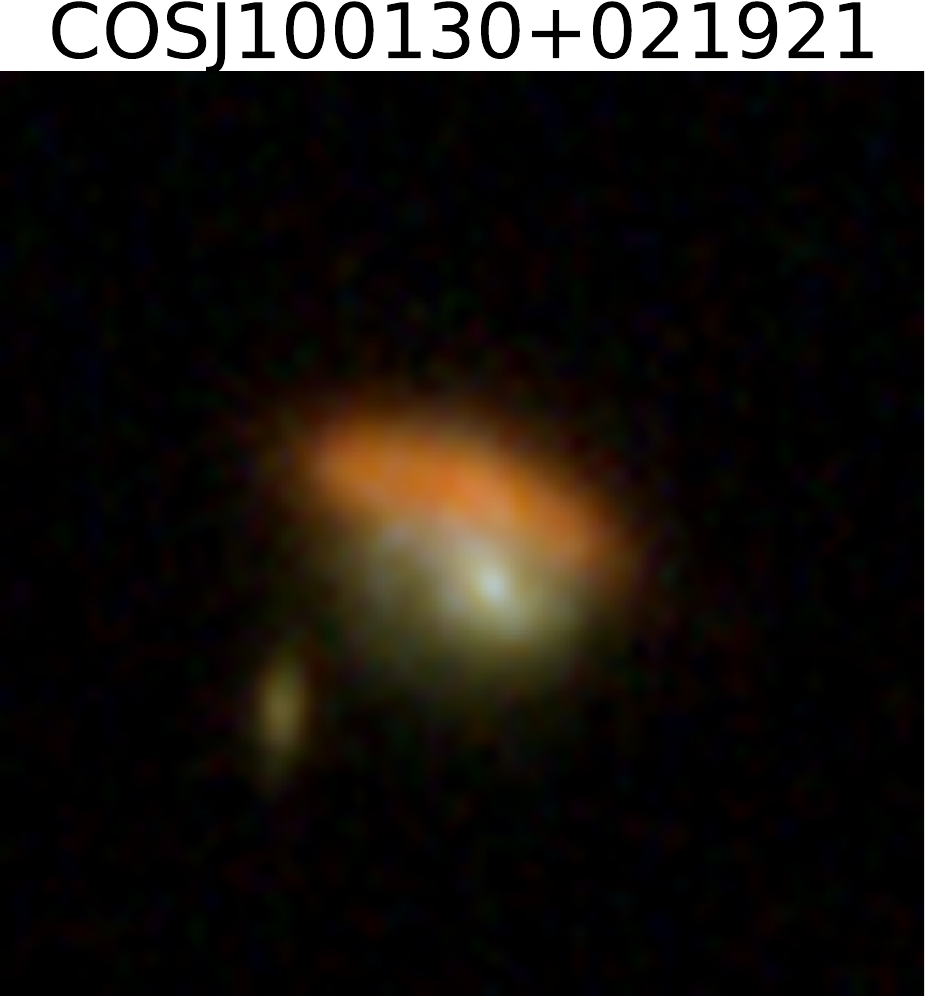}
\includegraphics[width=0.12\textwidth]{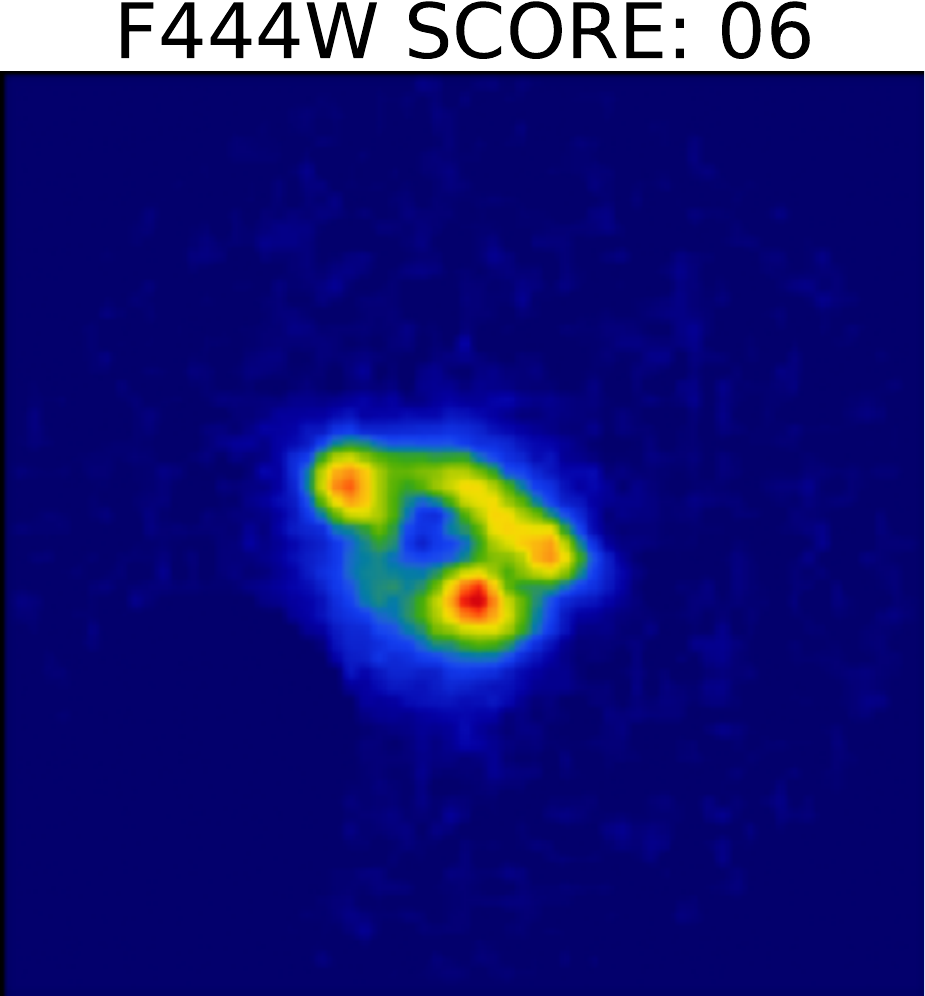}
\includegraphics[width=0.12\textwidth]{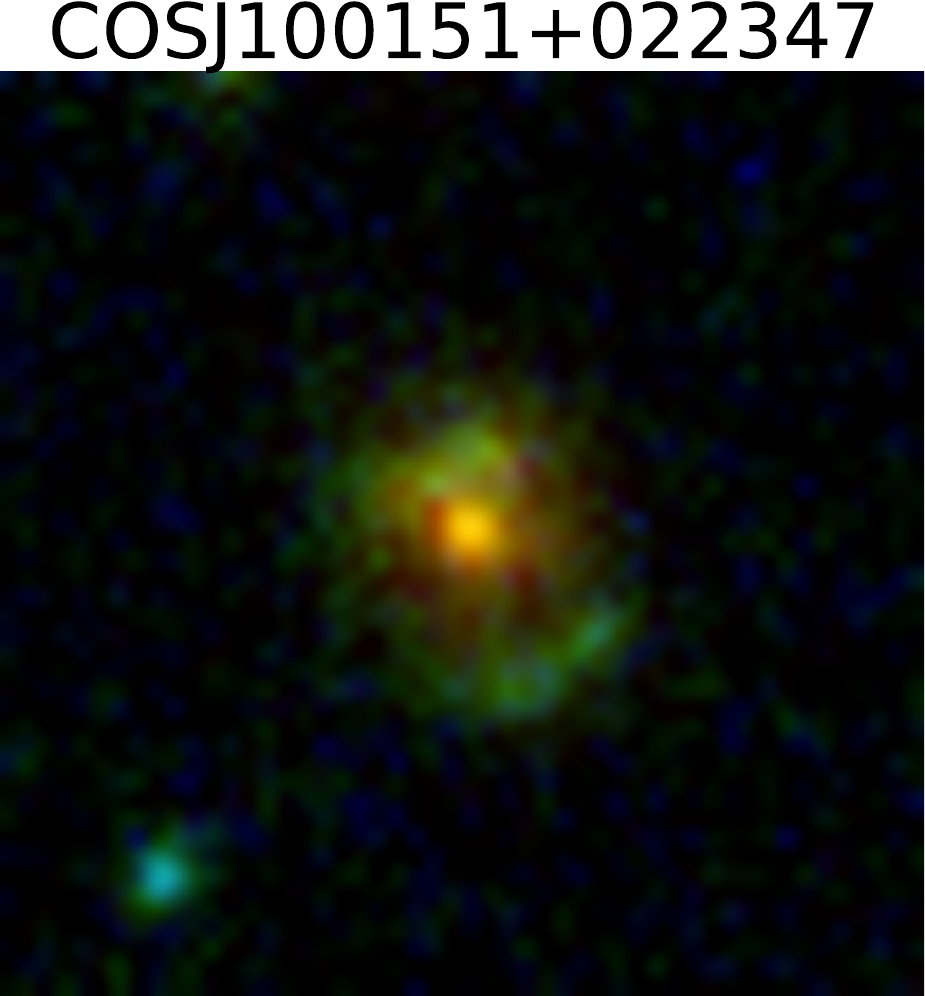}
\includegraphics[width=0.12\textwidth]{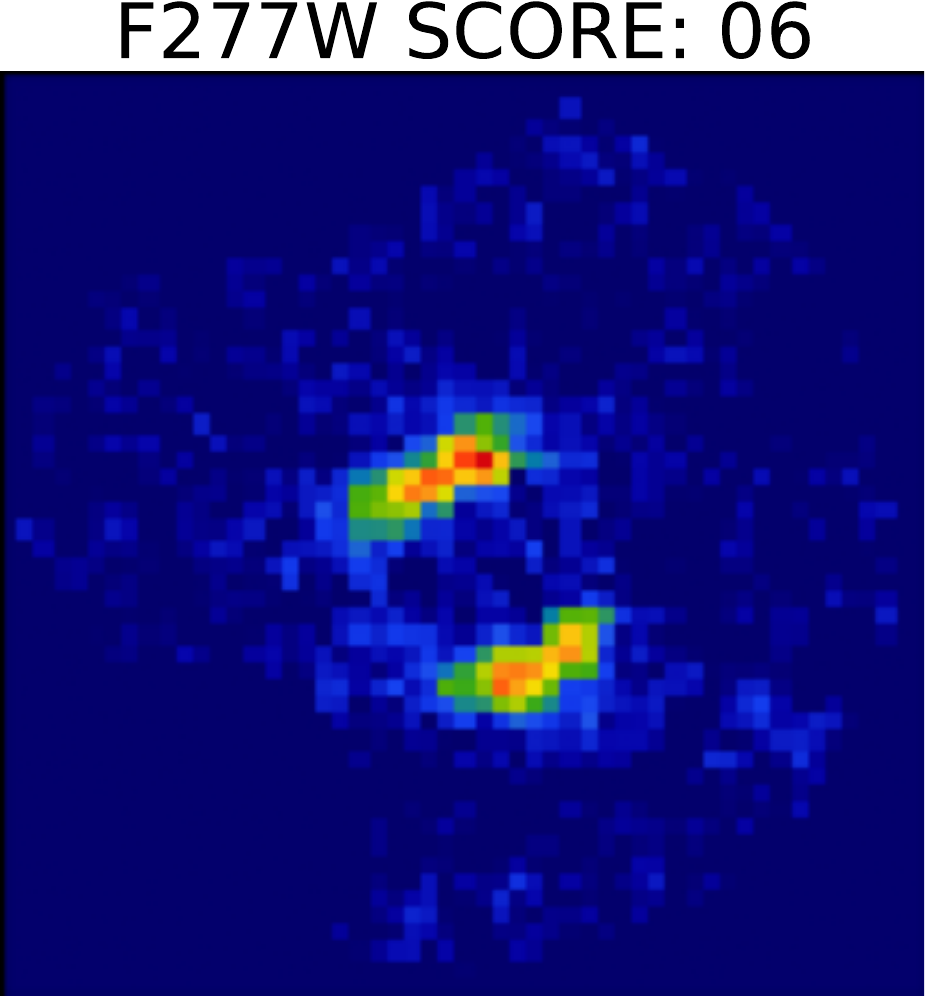}
\includegraphics[width=0.12\textwidth]{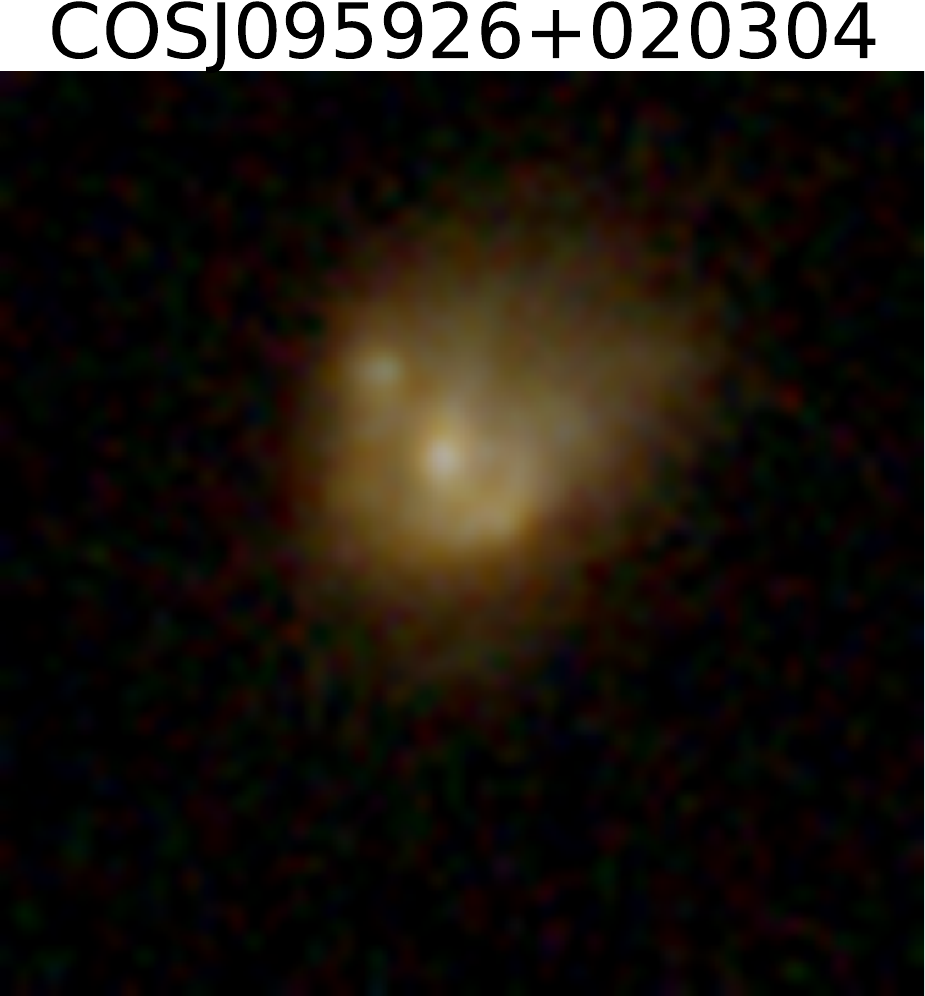}
\includegraphics[width=0.12\textwidth]{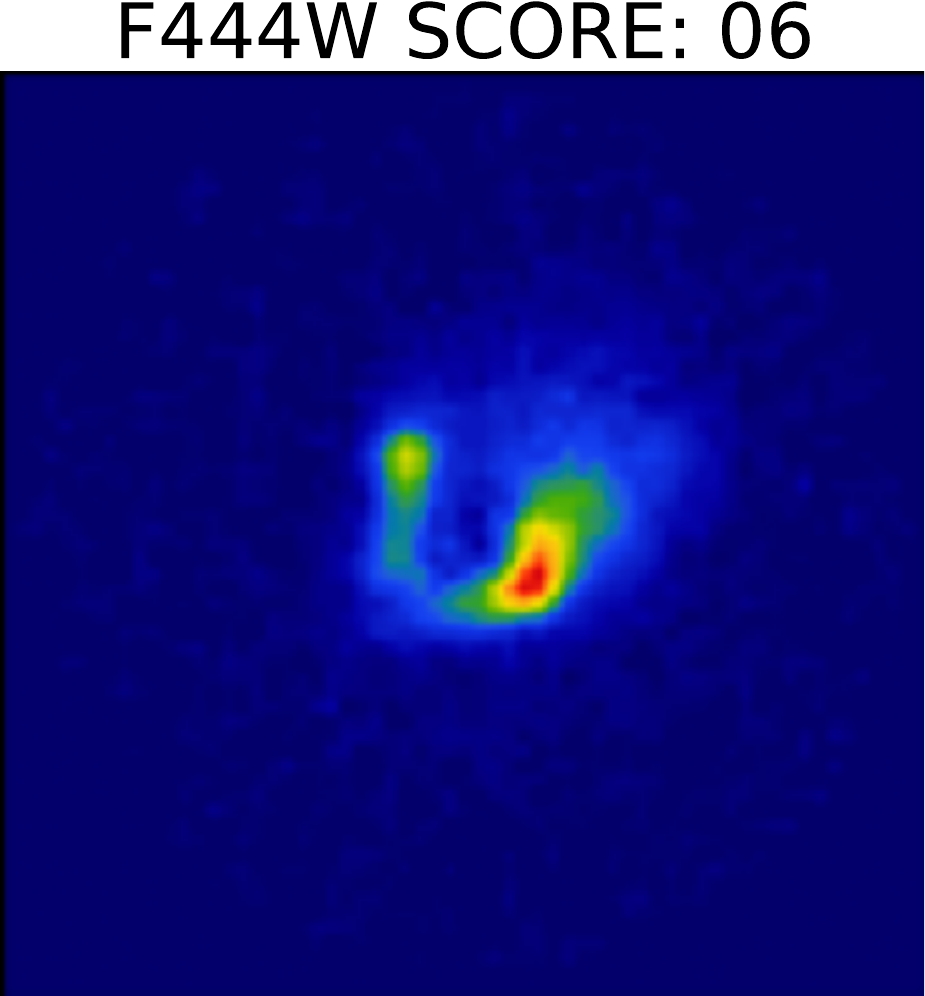}
\includegraphics[width=0.12\textwidth]{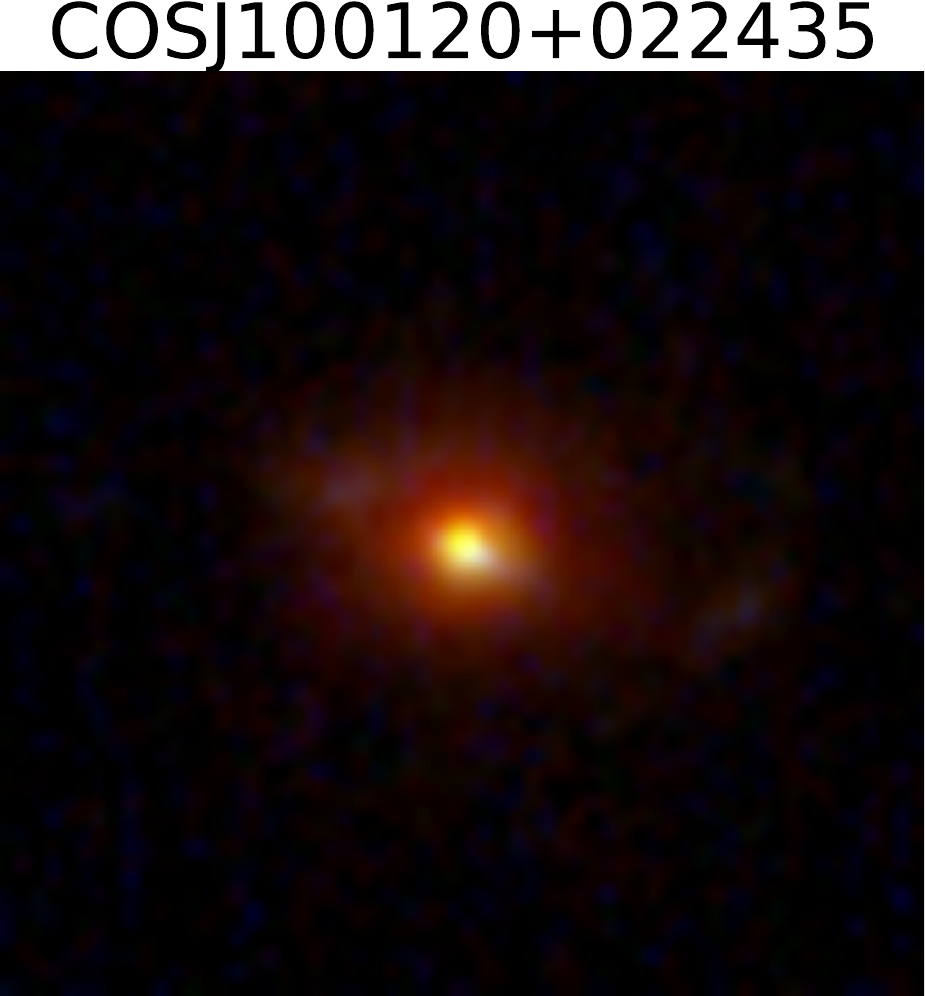}
\includegraphics[width=0.12\textwidth]{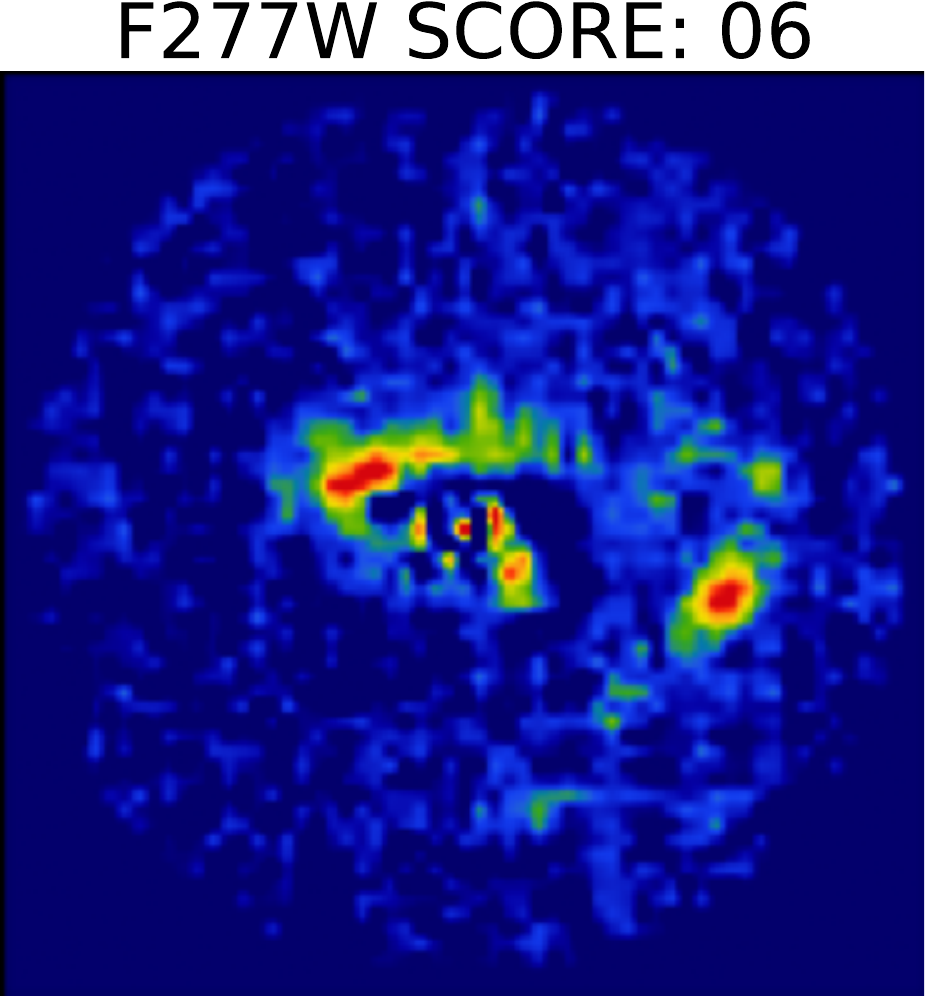}
\includegraphics[width=0.12\textwidth]{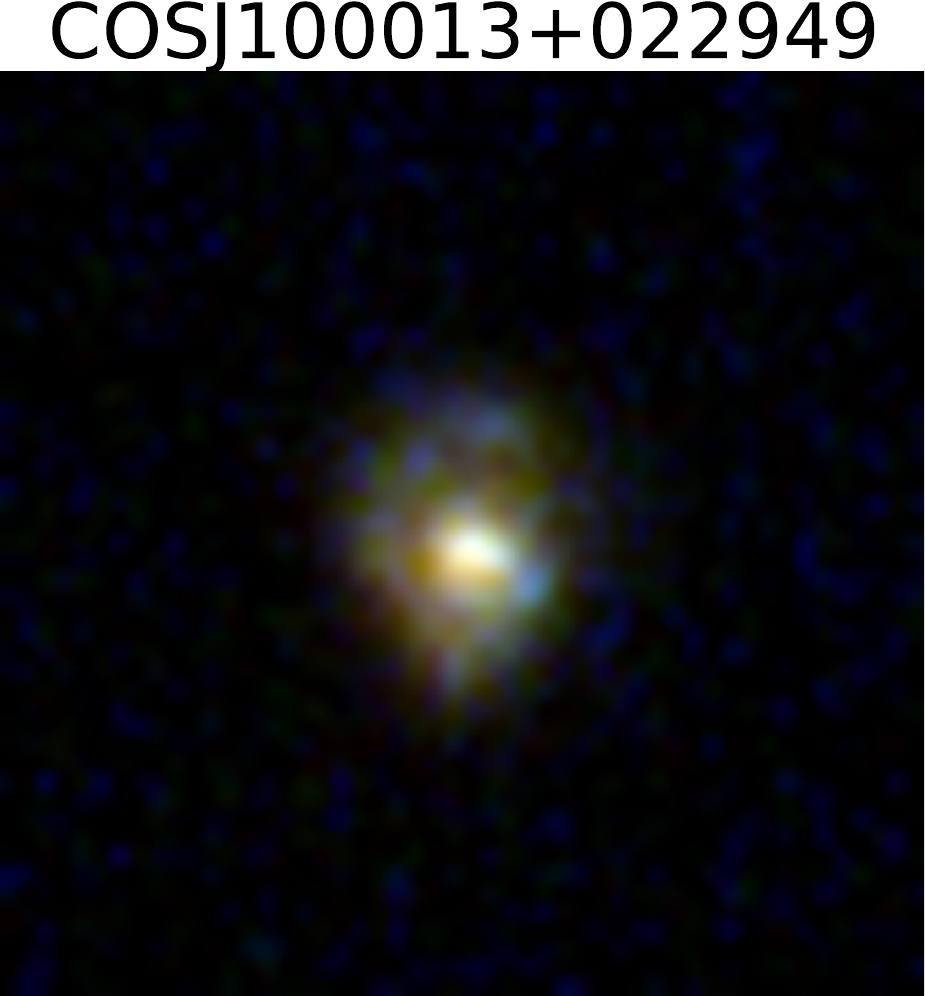}
\includegraphics[width=0.12\textwidth]{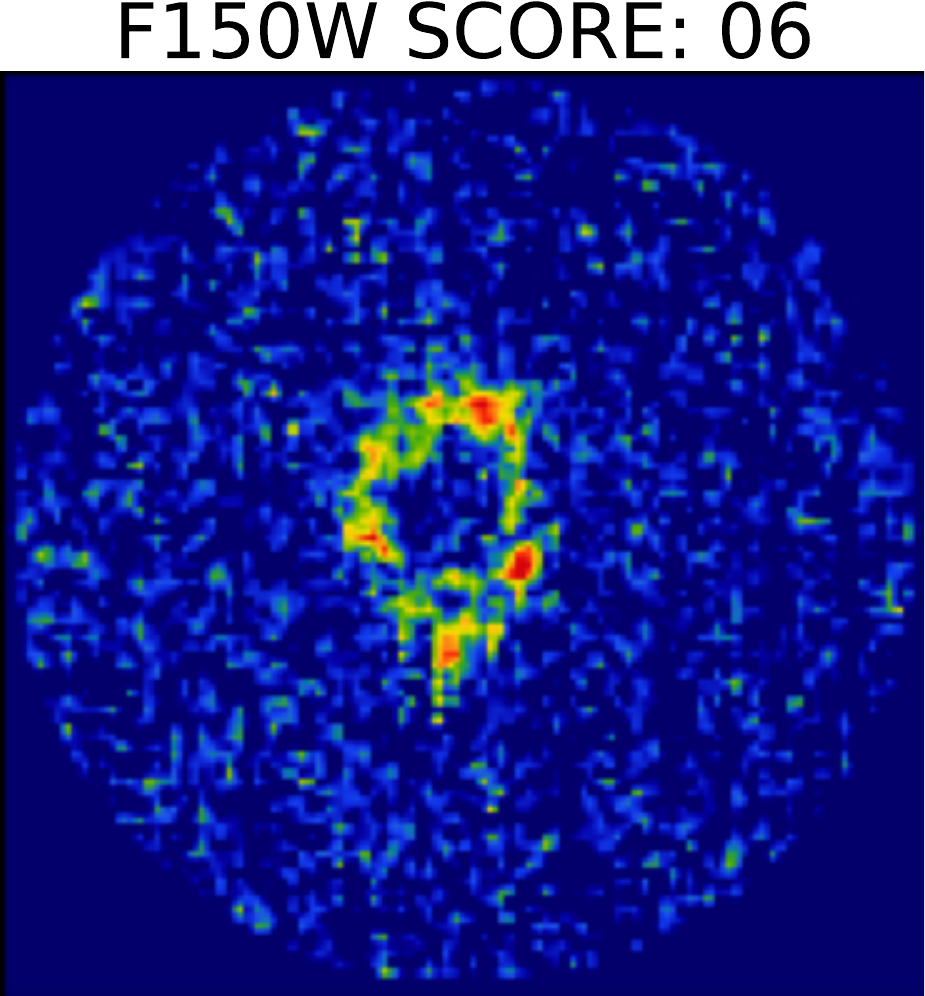}
\includegraphics[width=0.12\textwidth]{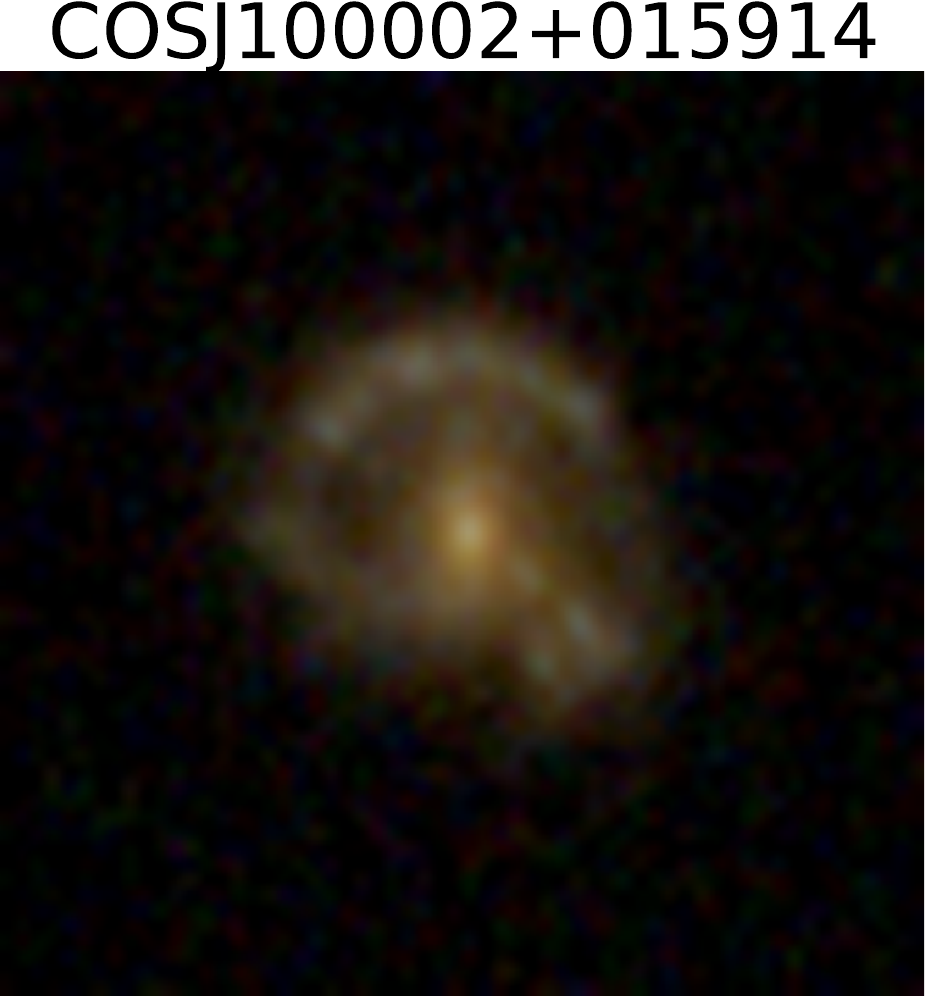}
\includegraphics[width=0.12\textwidth]{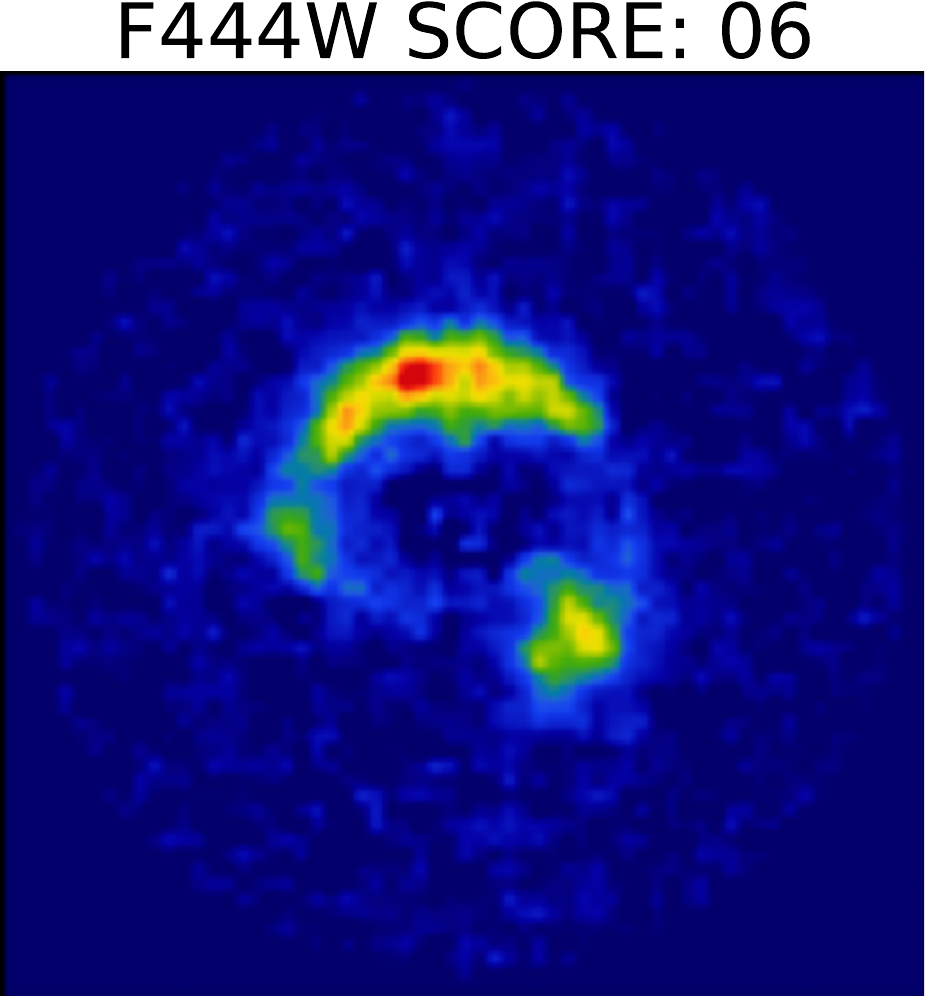}
\includegraphics[width=0.12\textwidth]{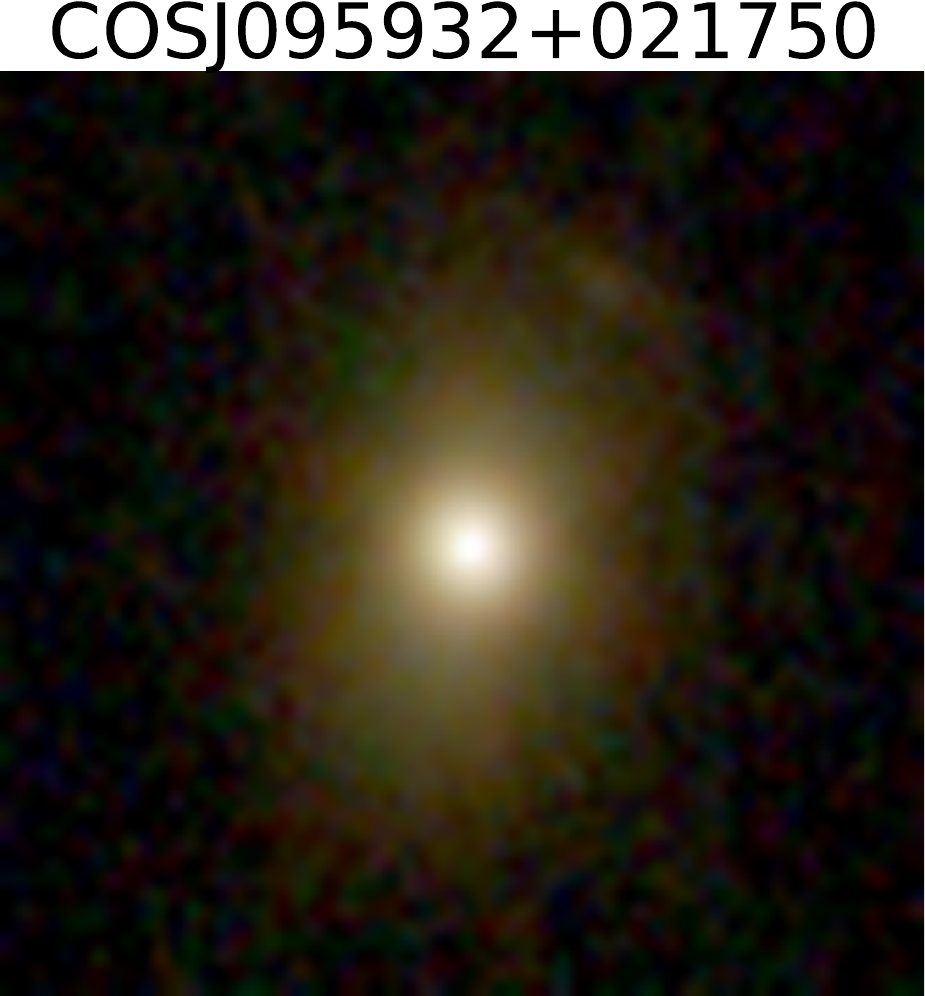}
\includegraphics[width=0.12\textwidth]{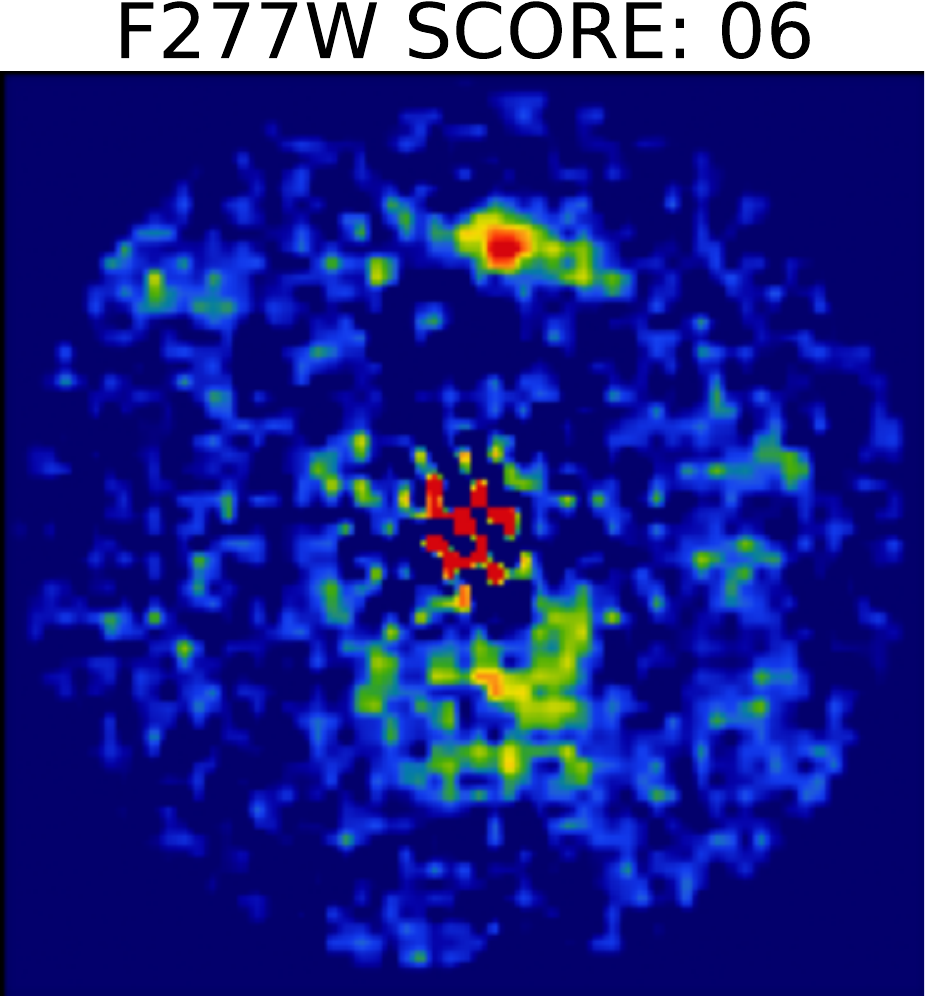}
\includegraphics[width=0.12\textwidth]{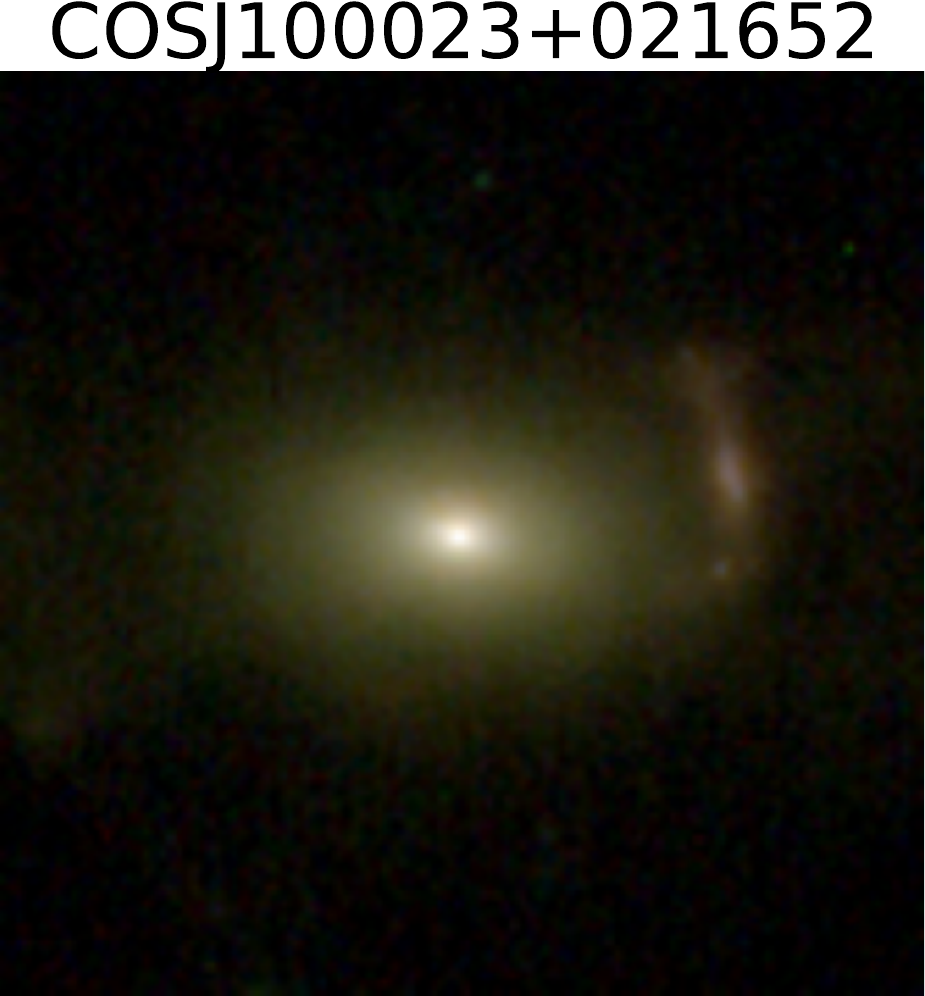}
\includegraphics[width=0.12\textwidth]{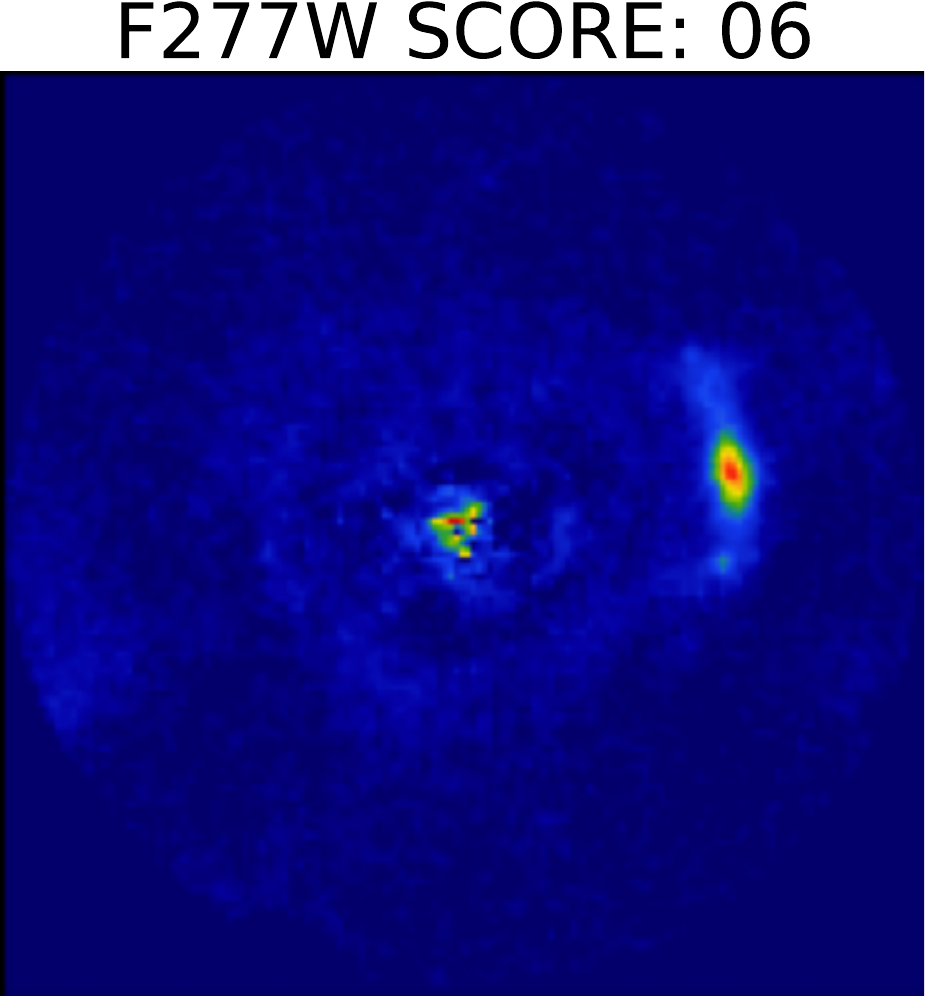}
\includegraphics[width=0.12\textwidth]{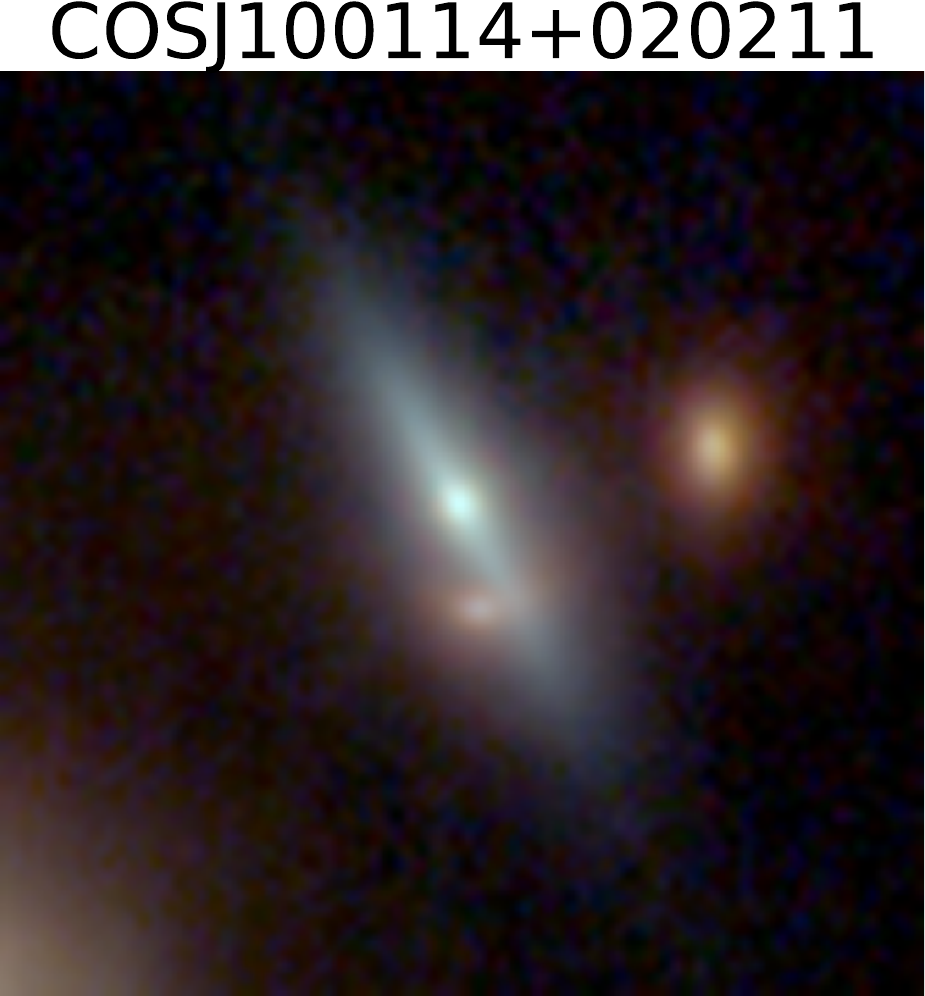}
\includegraphics[width=0.12\textwidth]{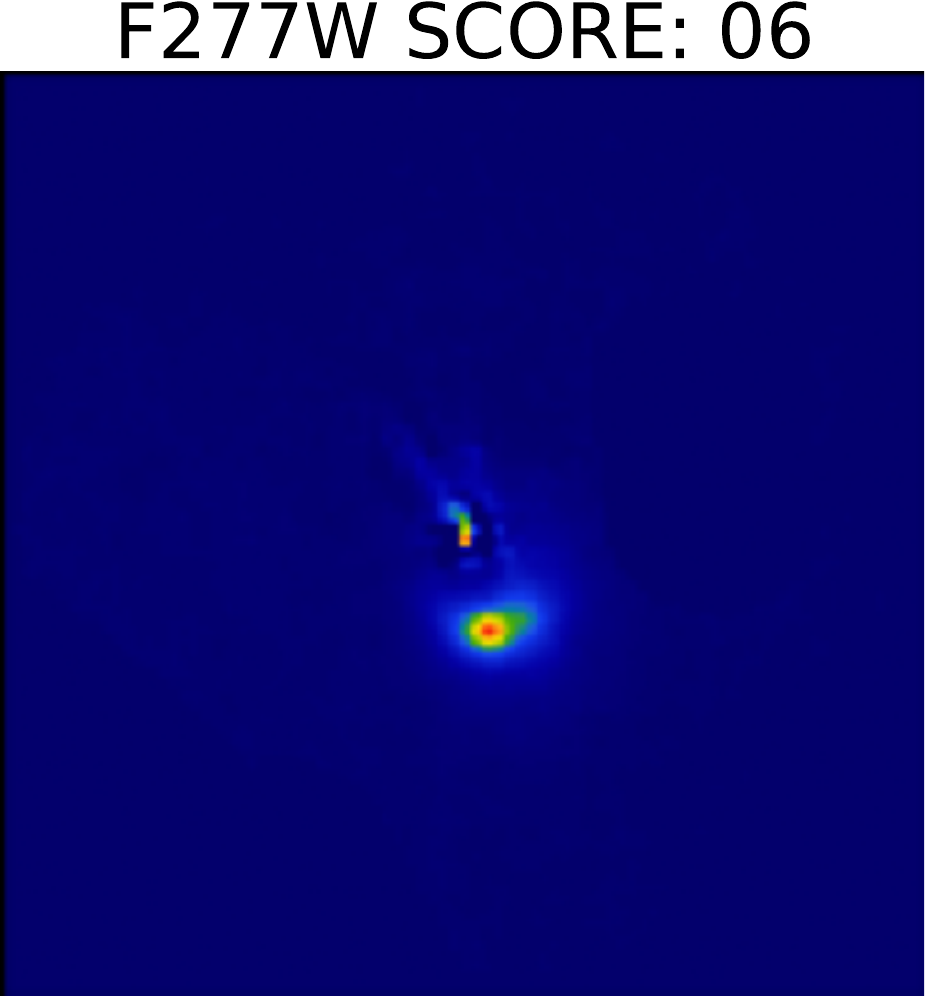}
\includegraphics[width=0.12\textwidth]{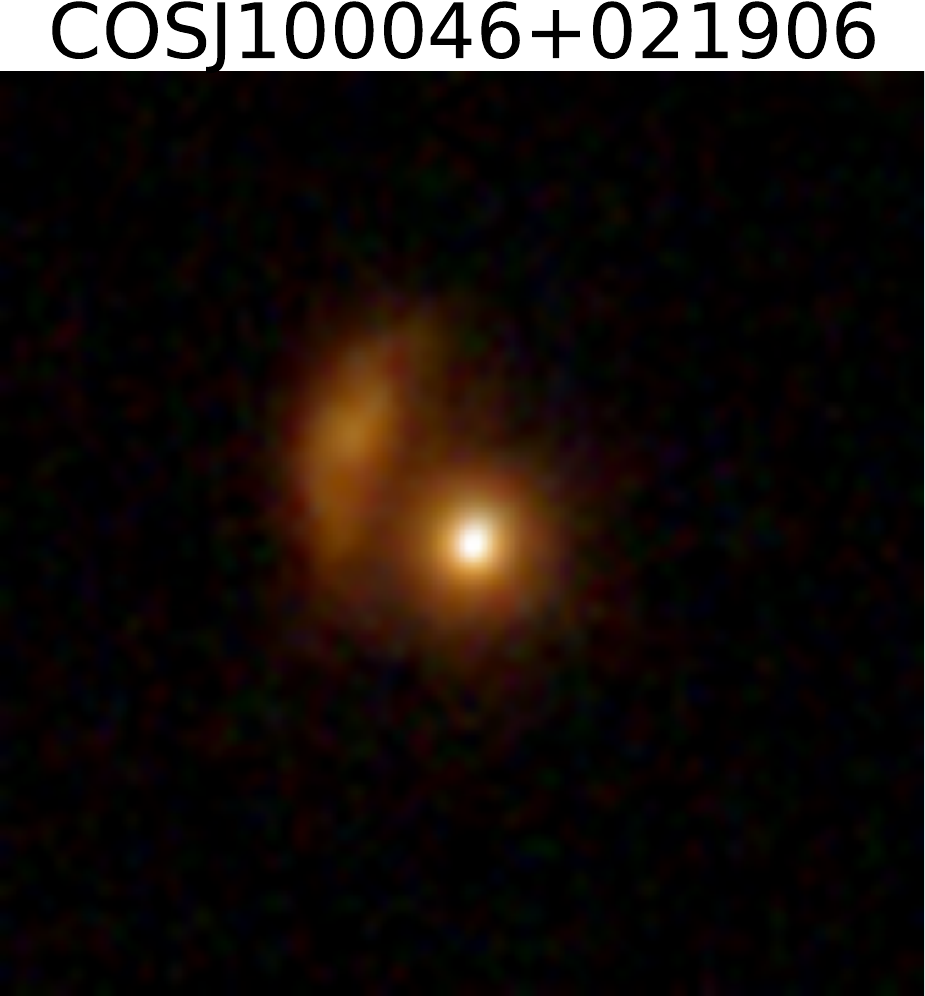}
\includegraphics[width=0.12\textwidth]{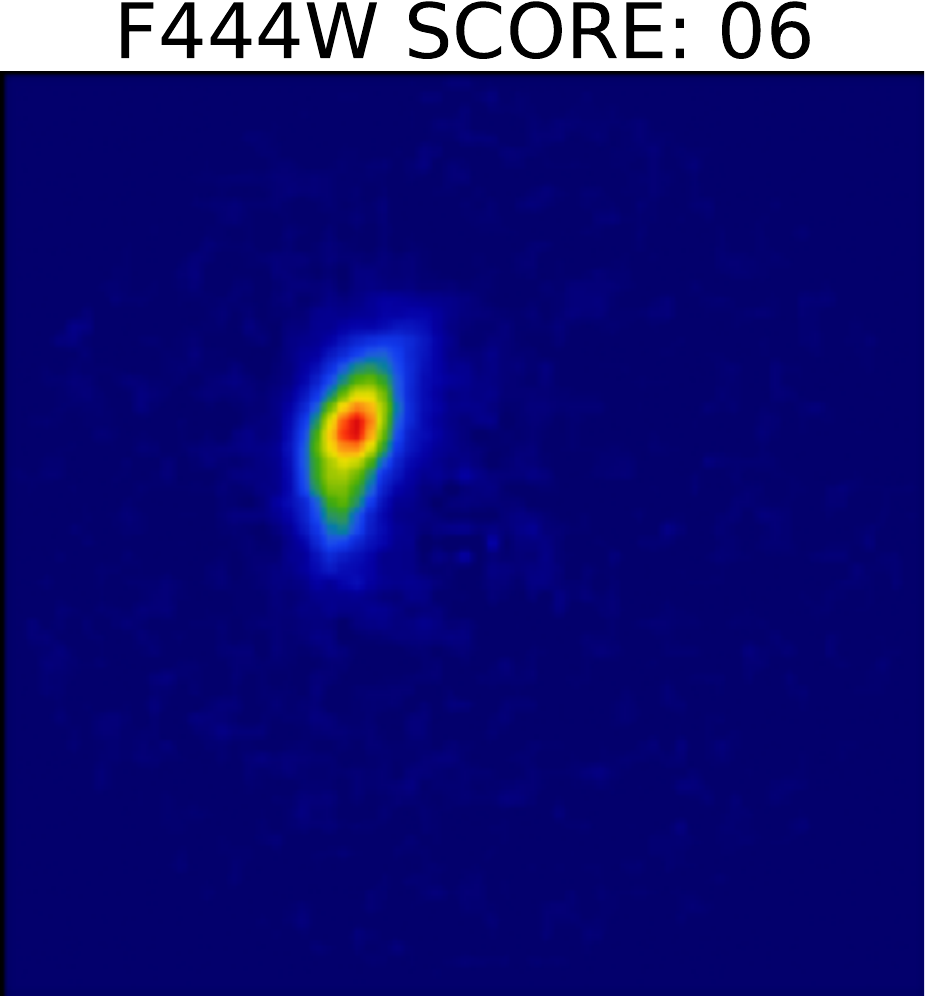}
\includegraphics[width=0.12\textwidth]{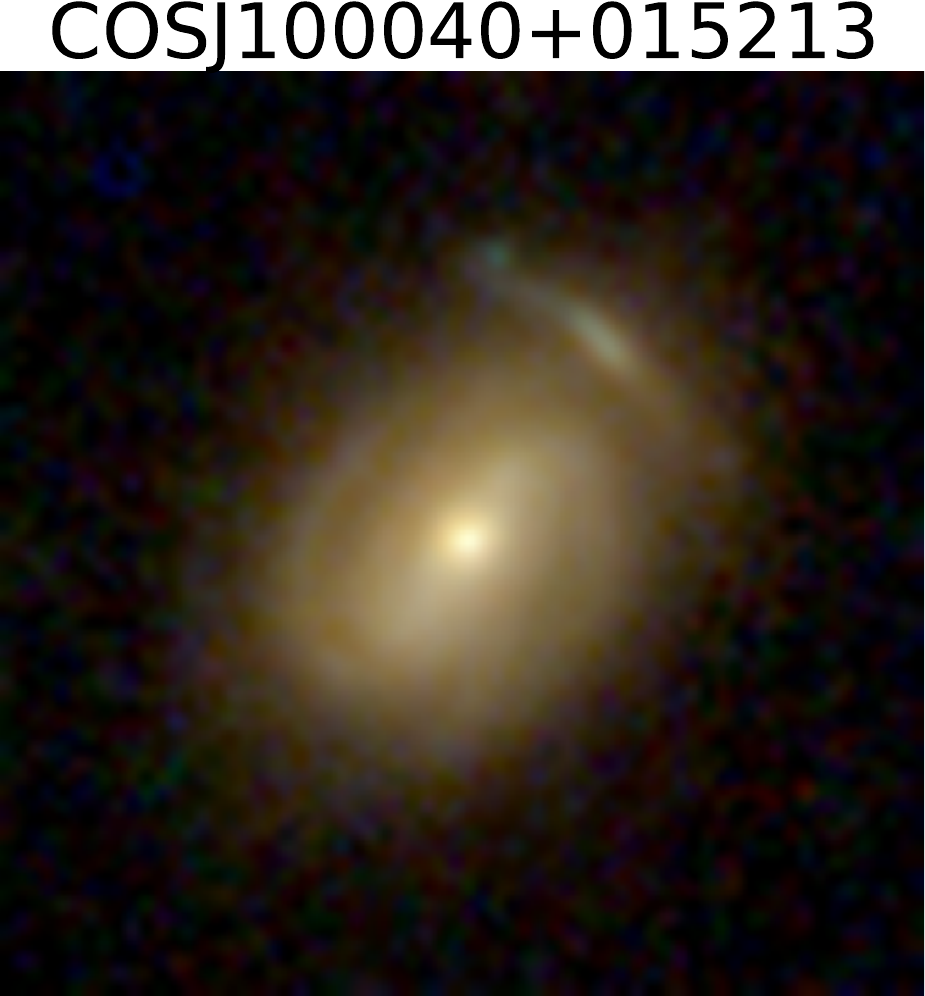}
\includegraphics[width=0.12\textwidth]{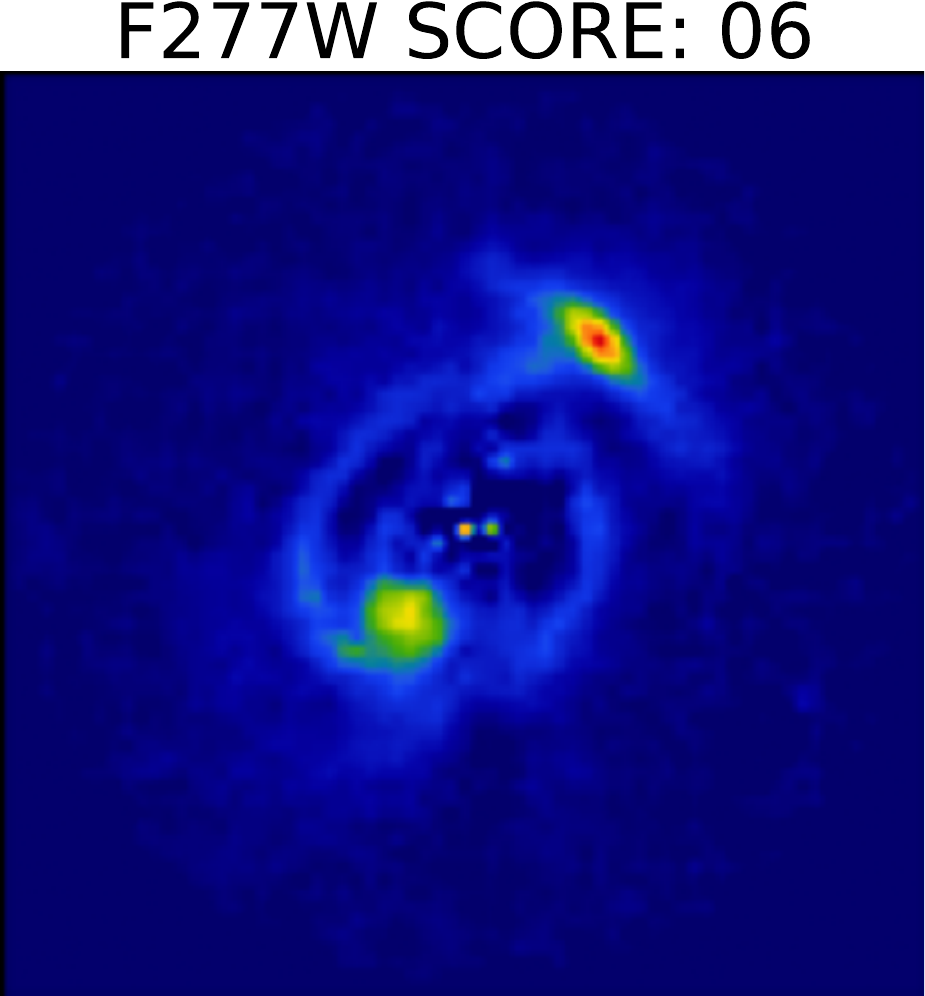}
\includegraphics[width=0.12\textwidth]{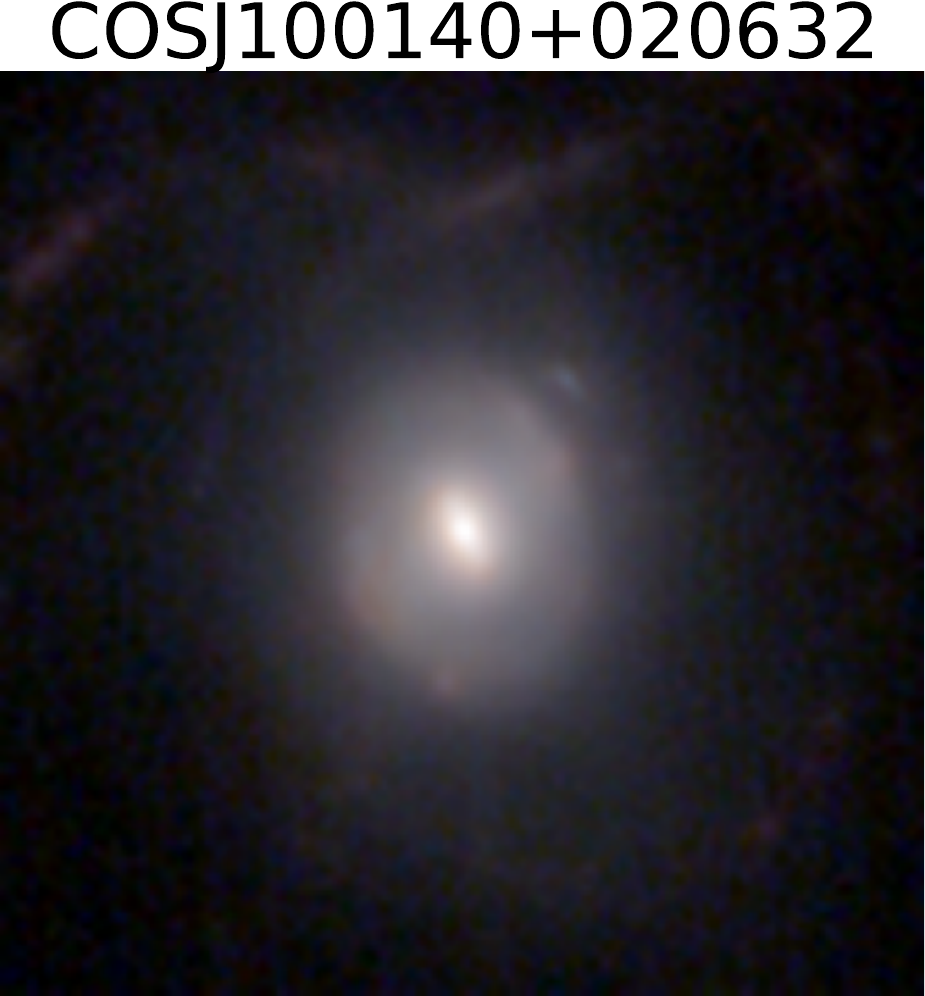}
\includegraphics[width=0.12\textwidth]{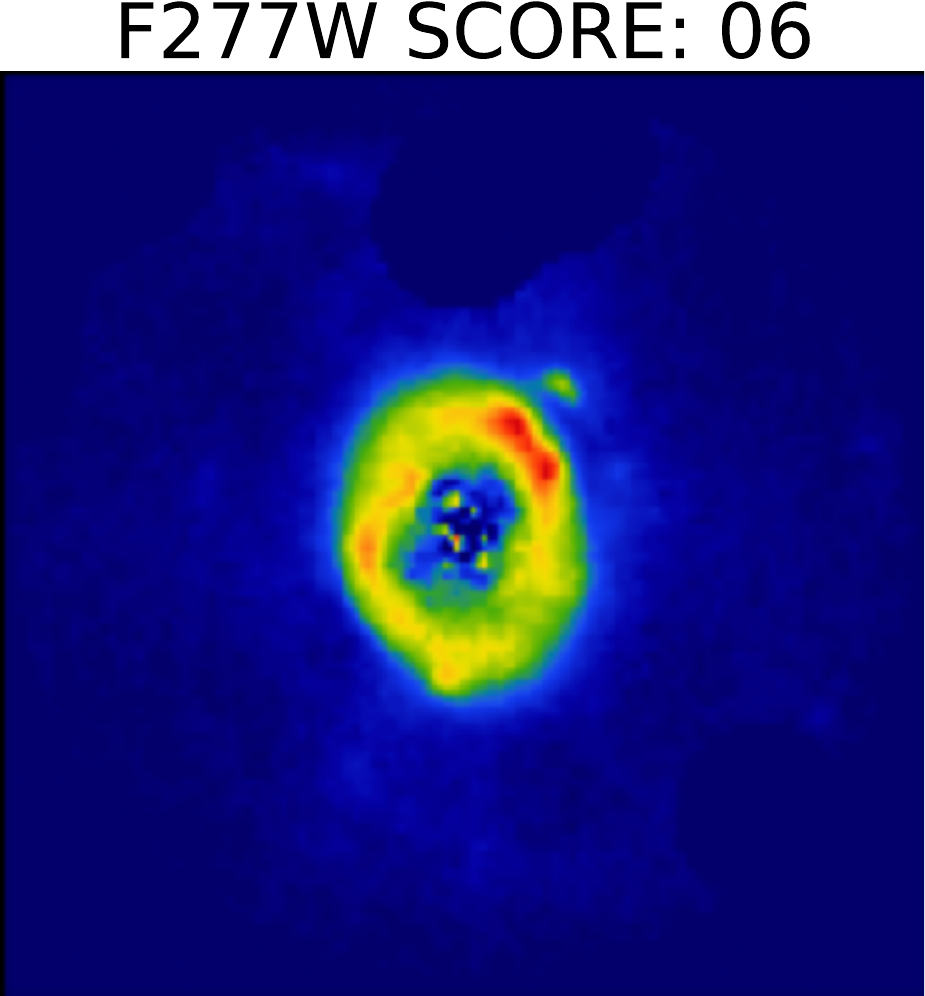}
\includegraphics[width=0.12\textwidth]{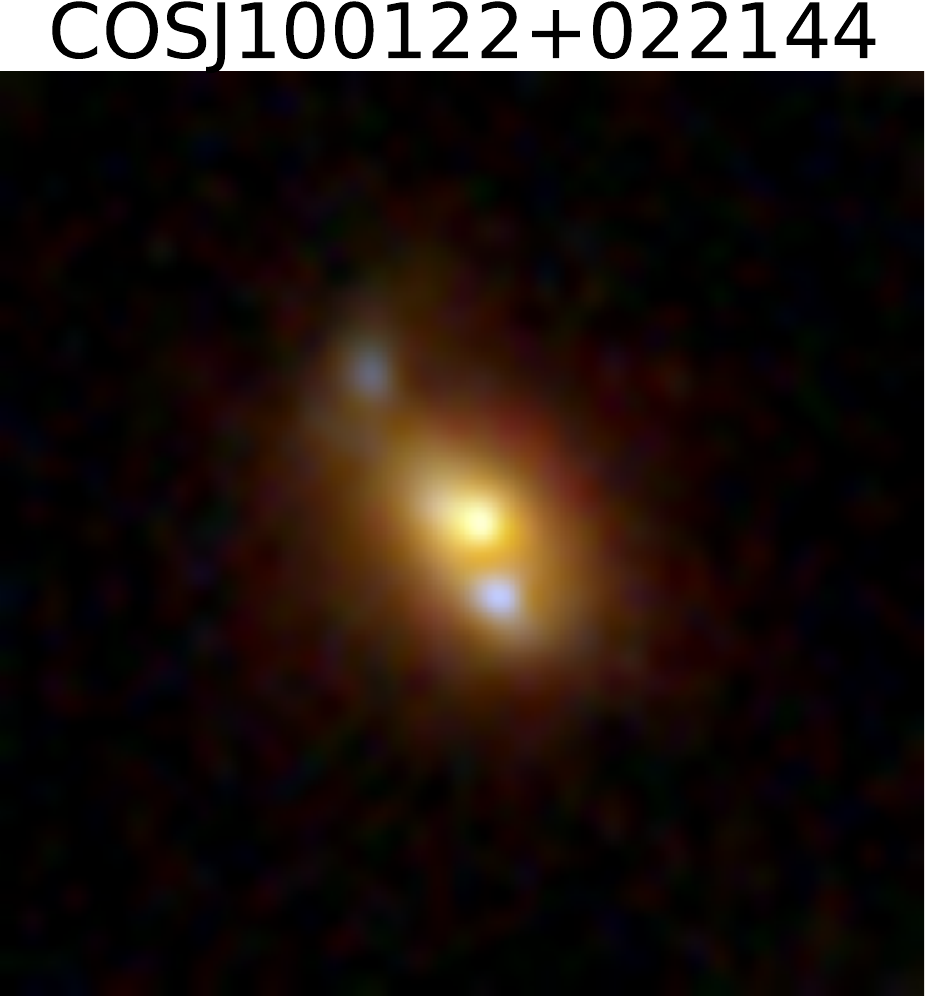}
\includegraphics[width=0.12\textwidth]{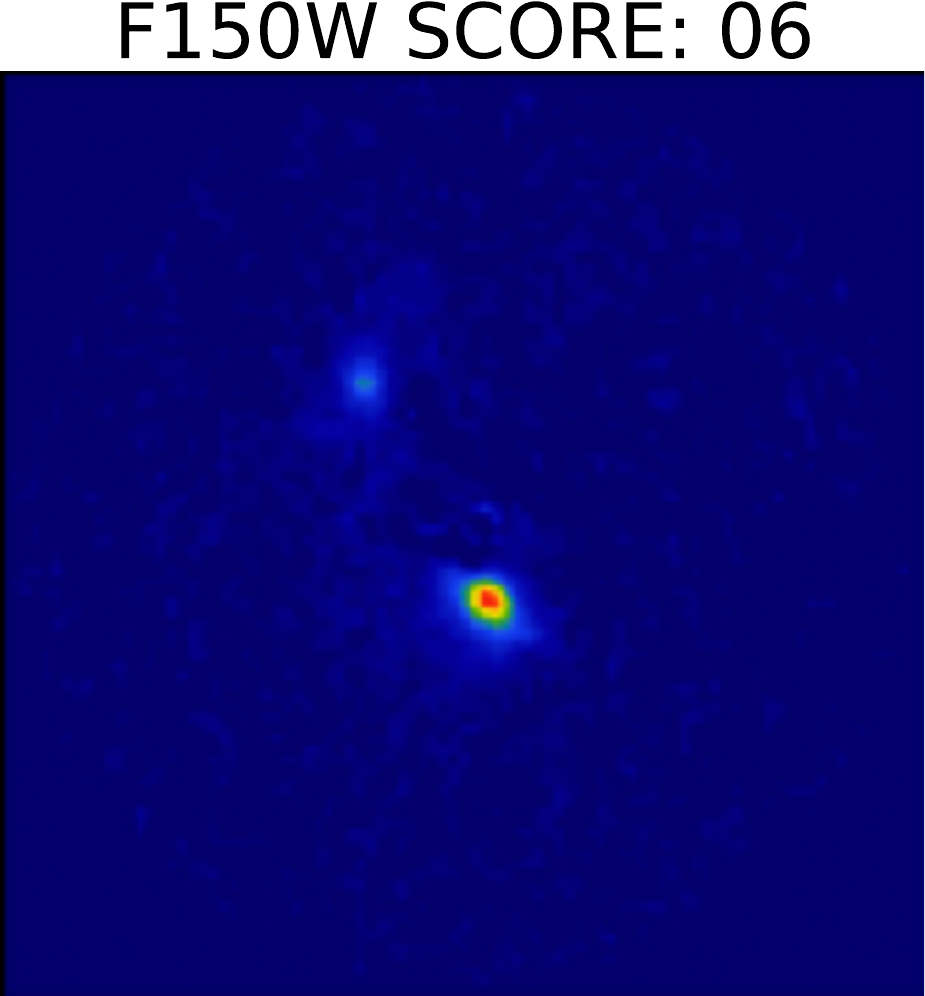}
\includegraphics[width=0.12\textwidth]{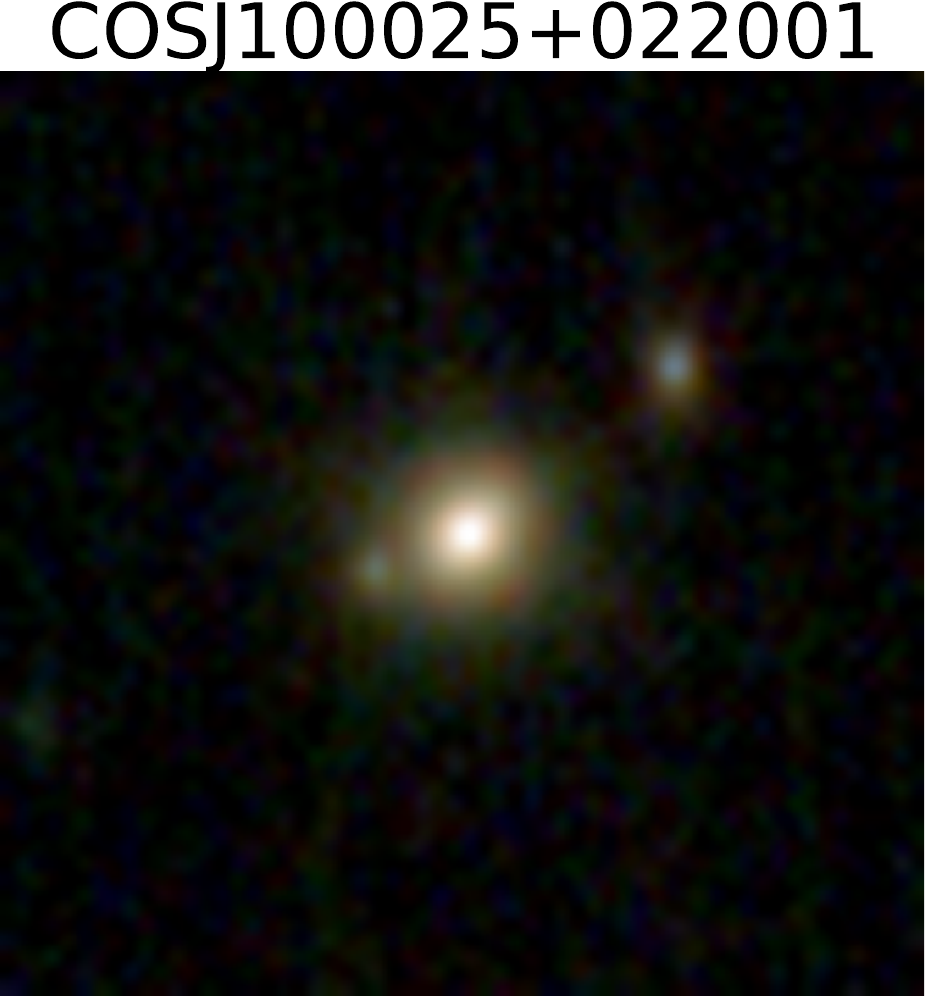}
\includegraphics[width=0.12\textwidth]{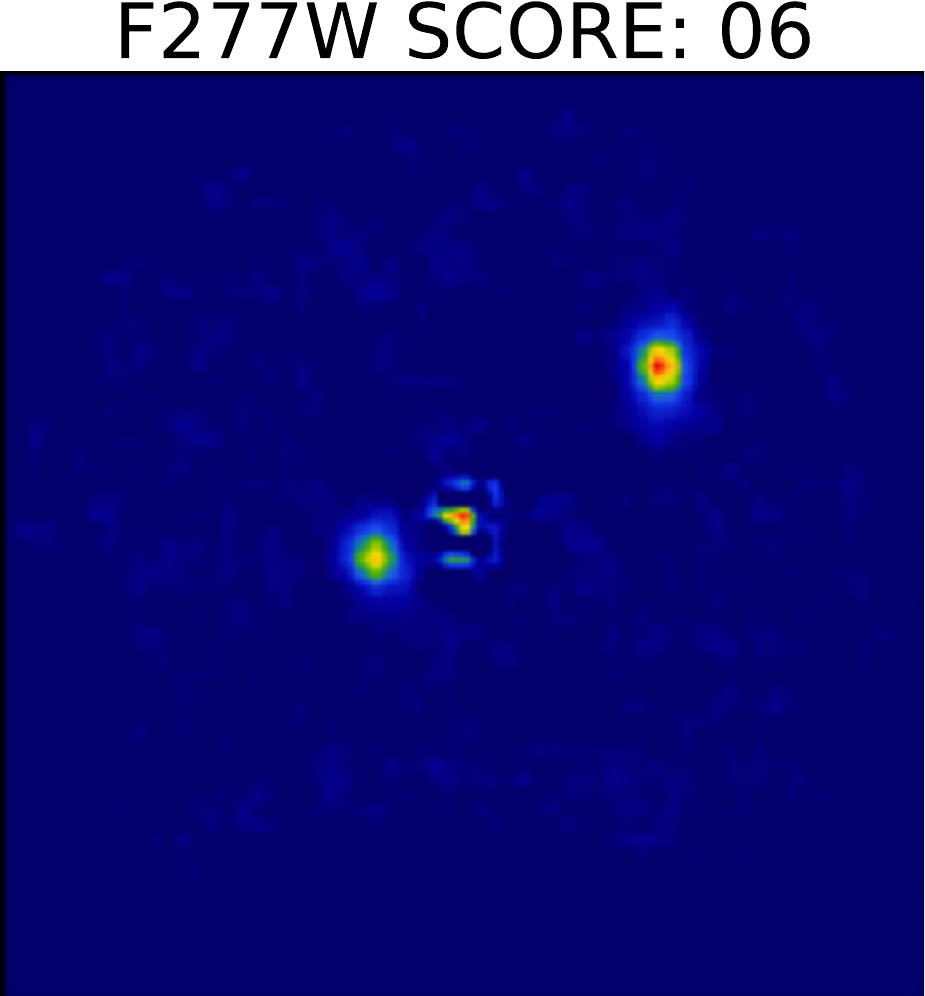}
\includegraphics[width=0.12\textwidth]{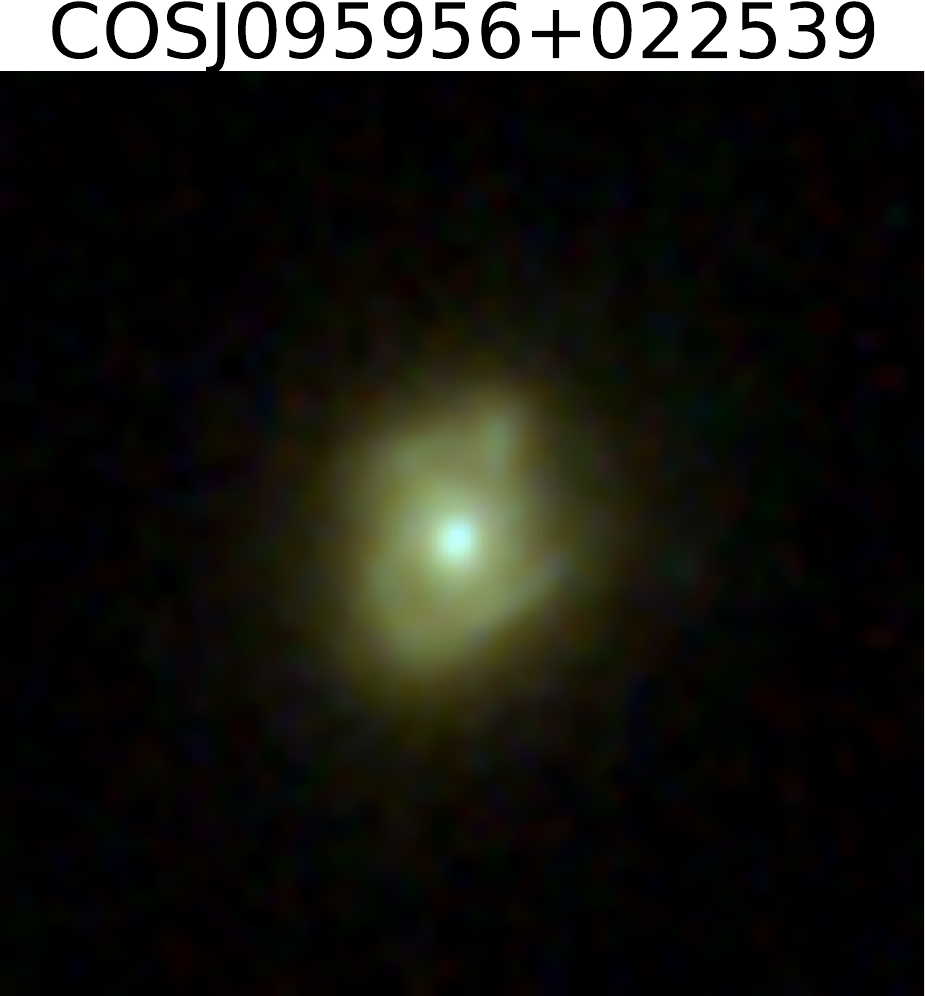}
\includegraphics[width=0.12\textwidth]{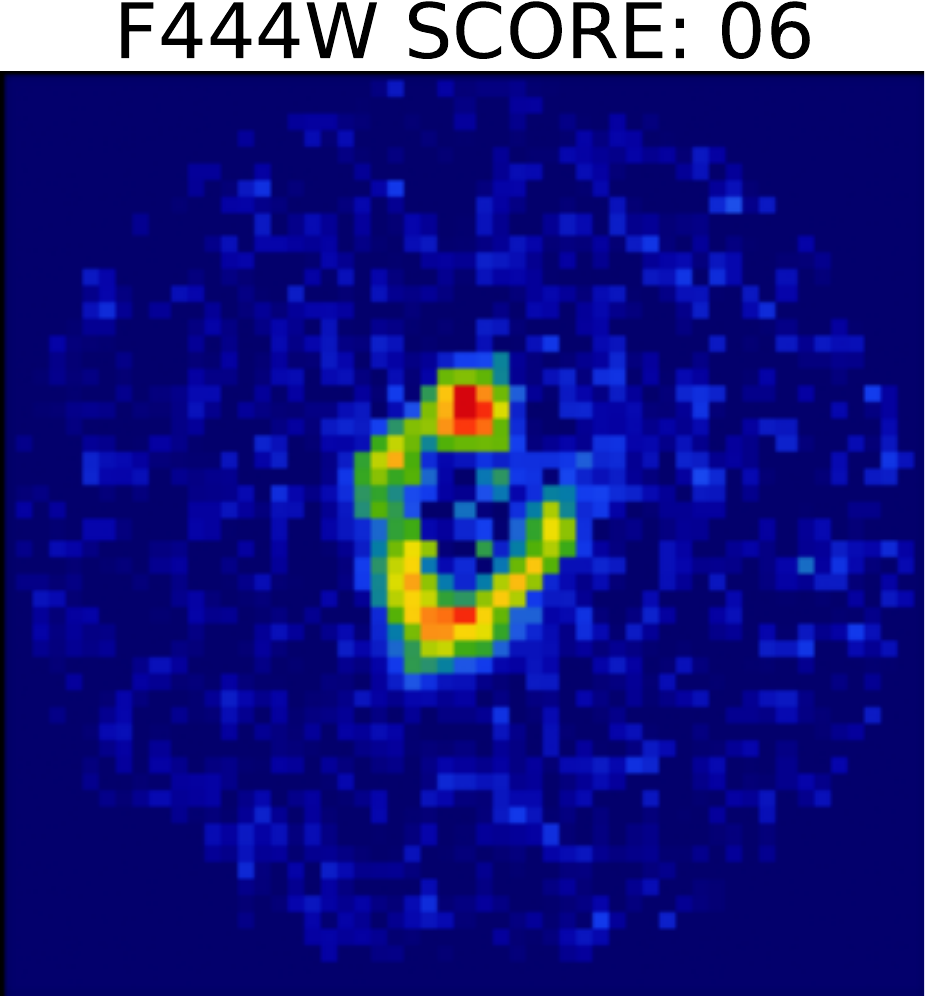}
\includegraphics[width=0.12\textwidth]{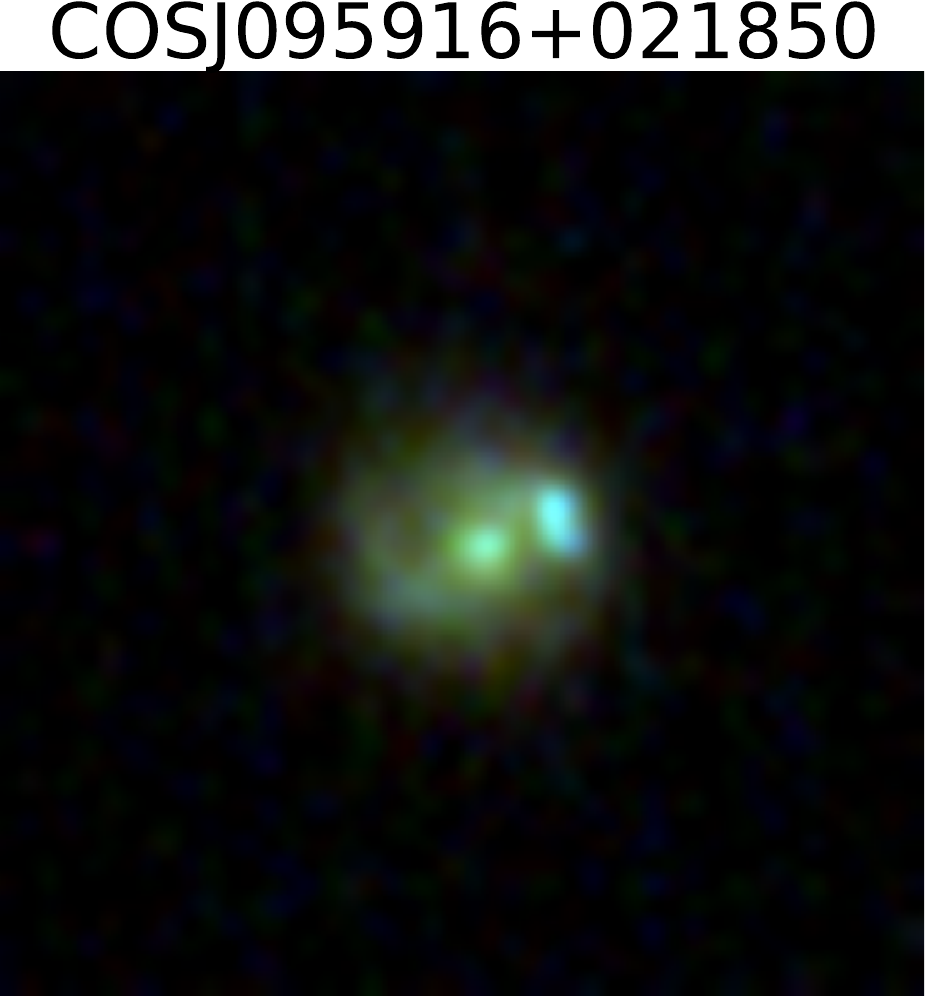}
\includegraphics[width=0.12\textwidth]{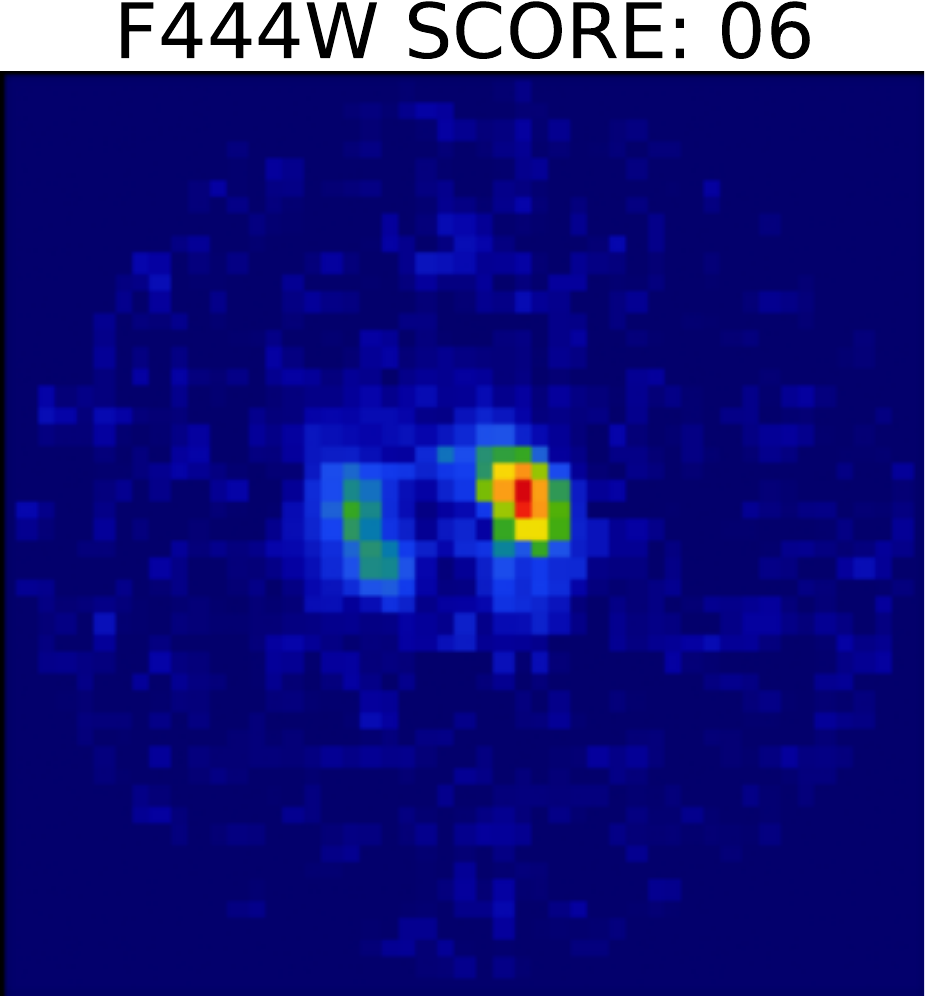}
\includegraphics[width=0.12\textwidth]{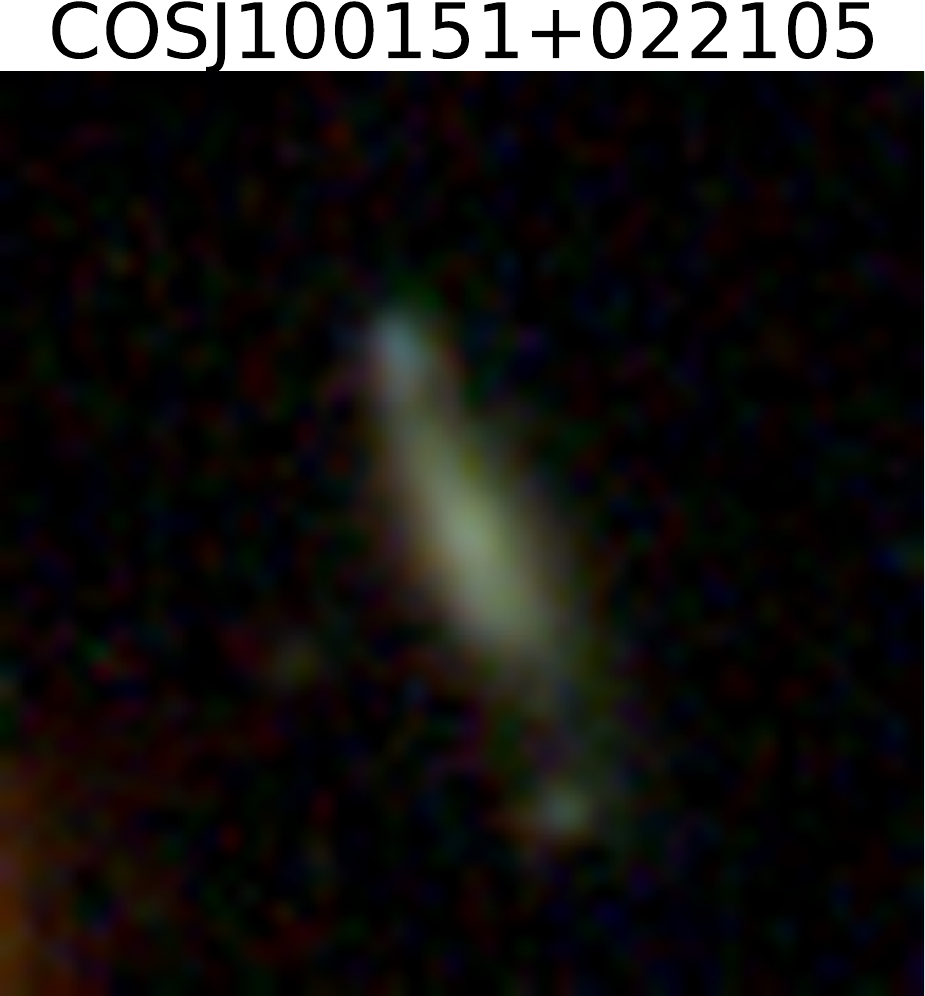}
\includegraphics[width=0.12\textwidth]{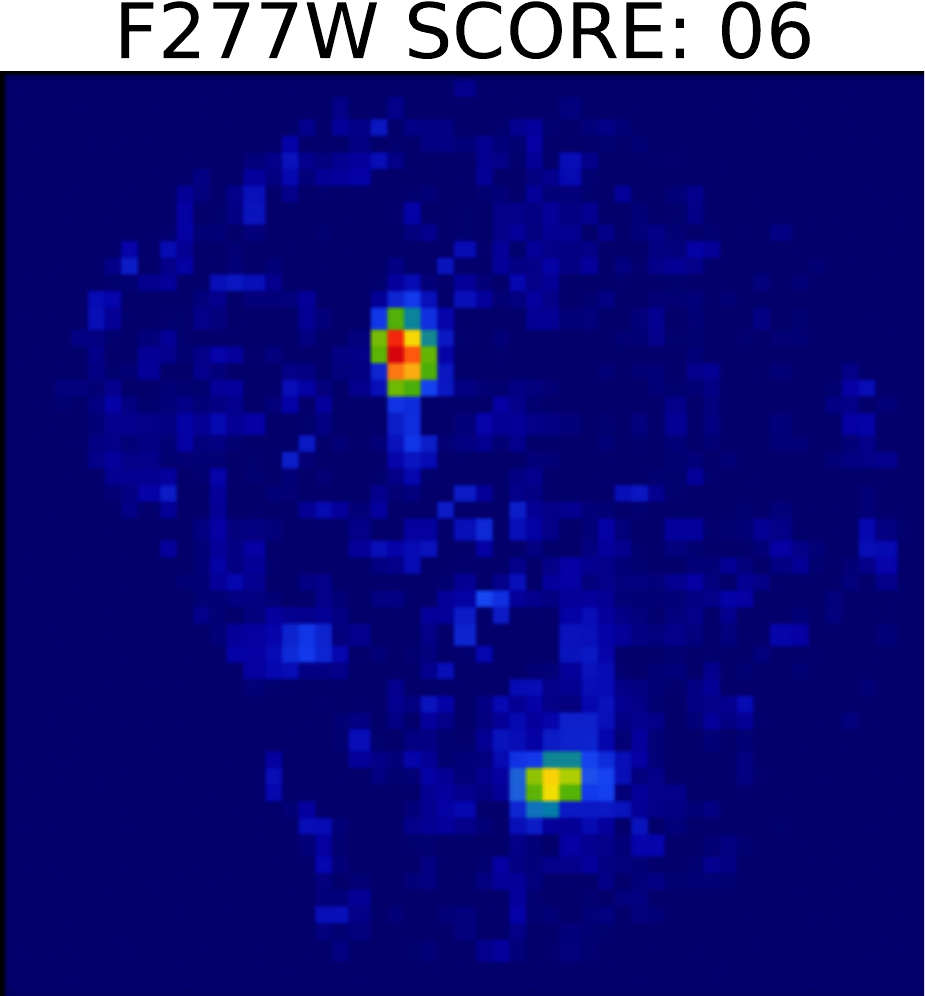}
\includegraphics[width=0.12\textwidth]{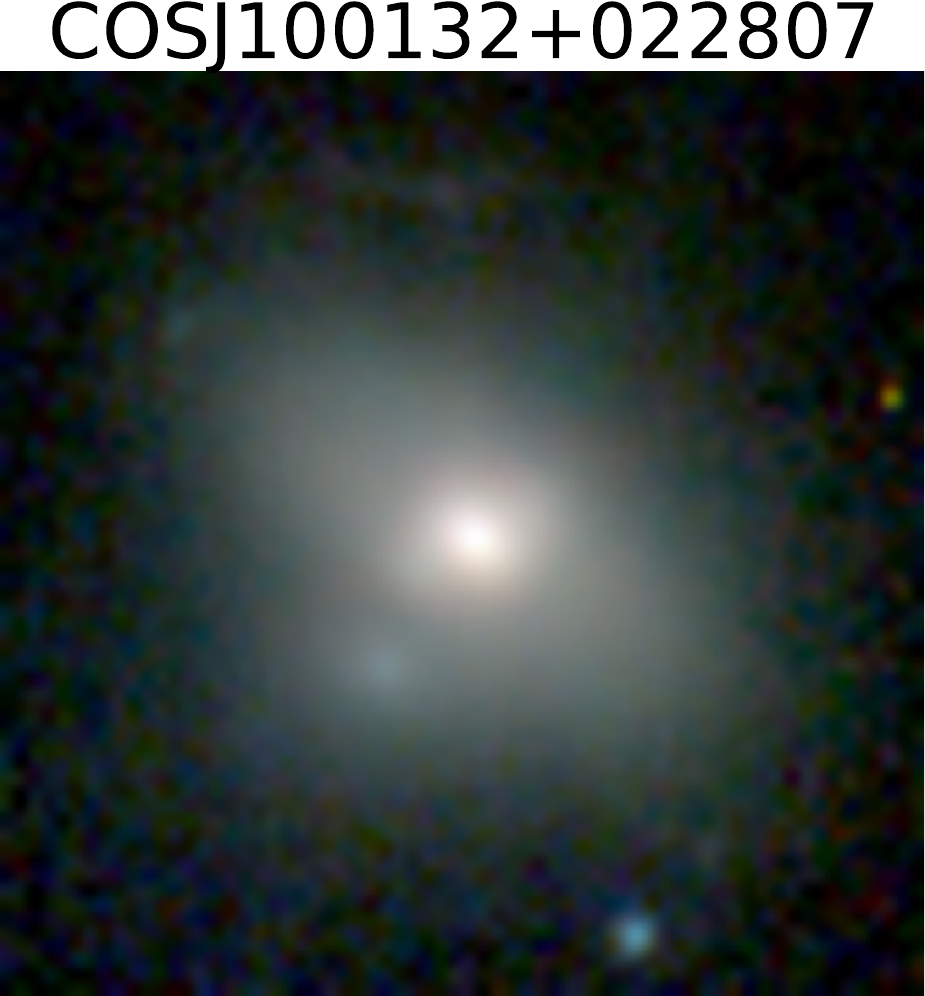}
\includegraphics[width=0.12\textwidth]{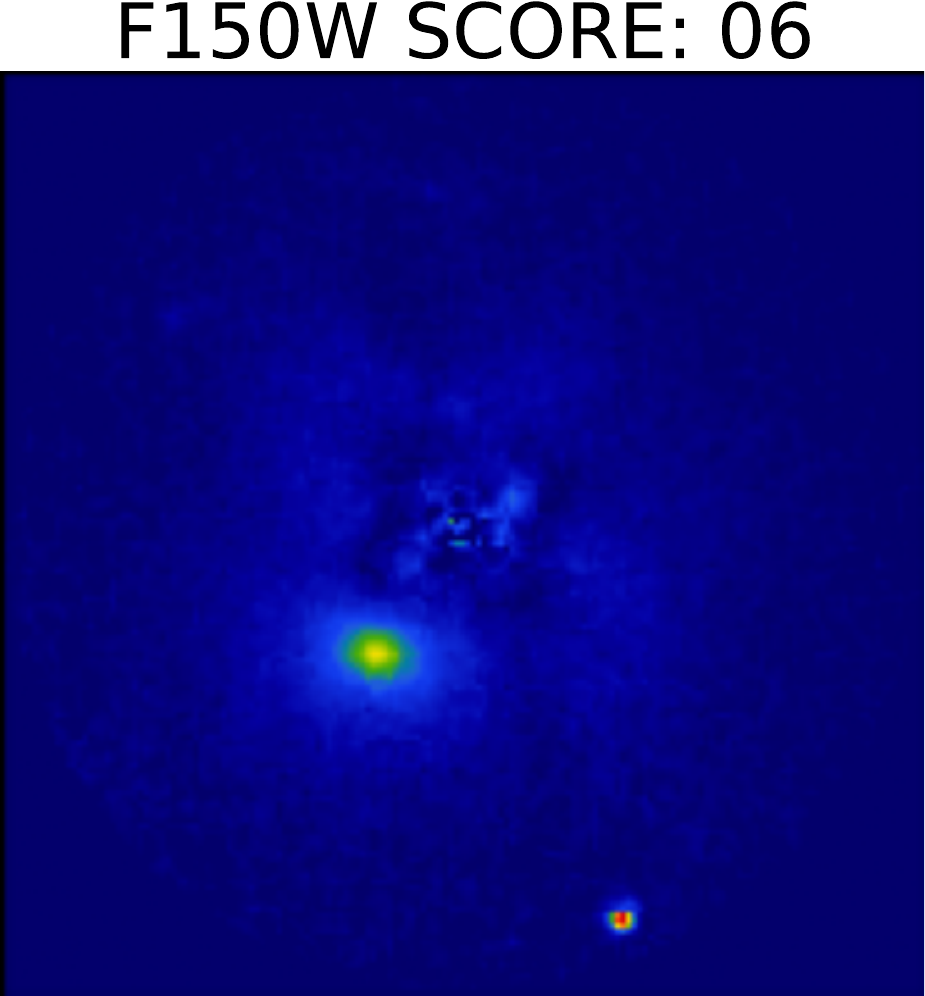}
\includegraphics[width=0.12\textwidth]{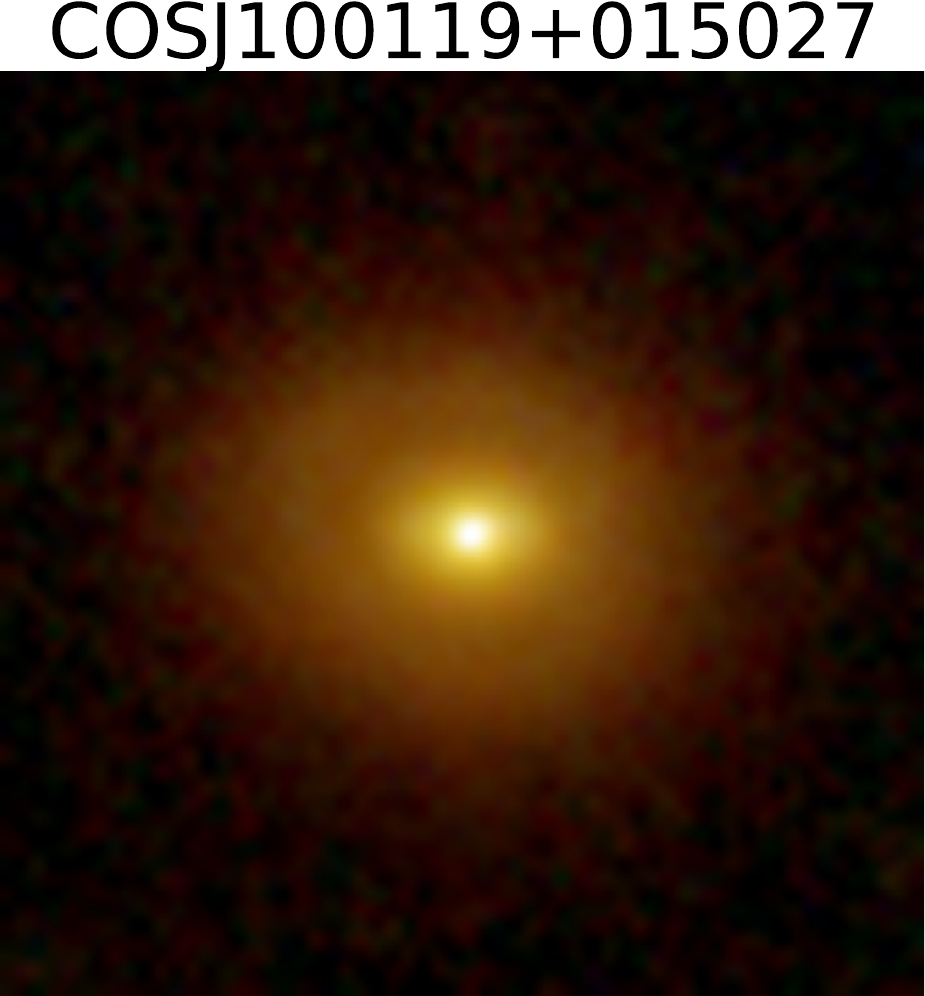}
\includegraphics[width=0.12\textwidth]{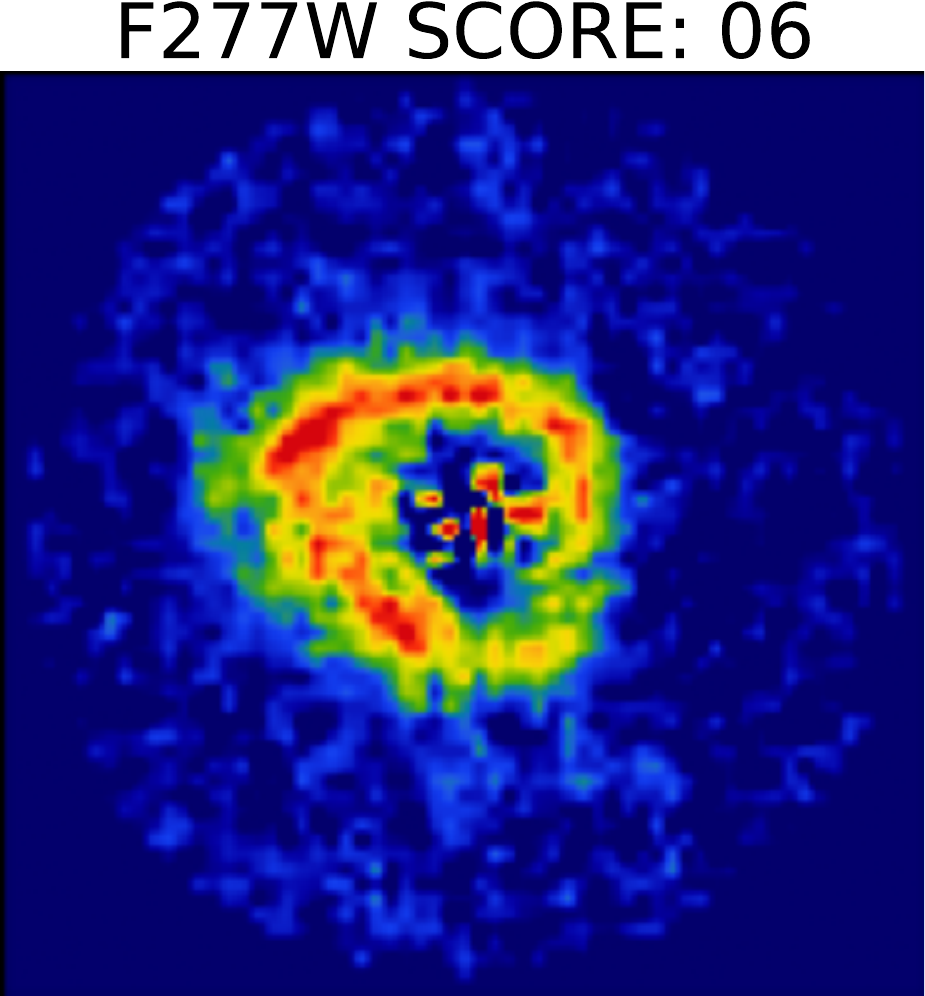}
\includegraphics[width=0.12\textwidth]{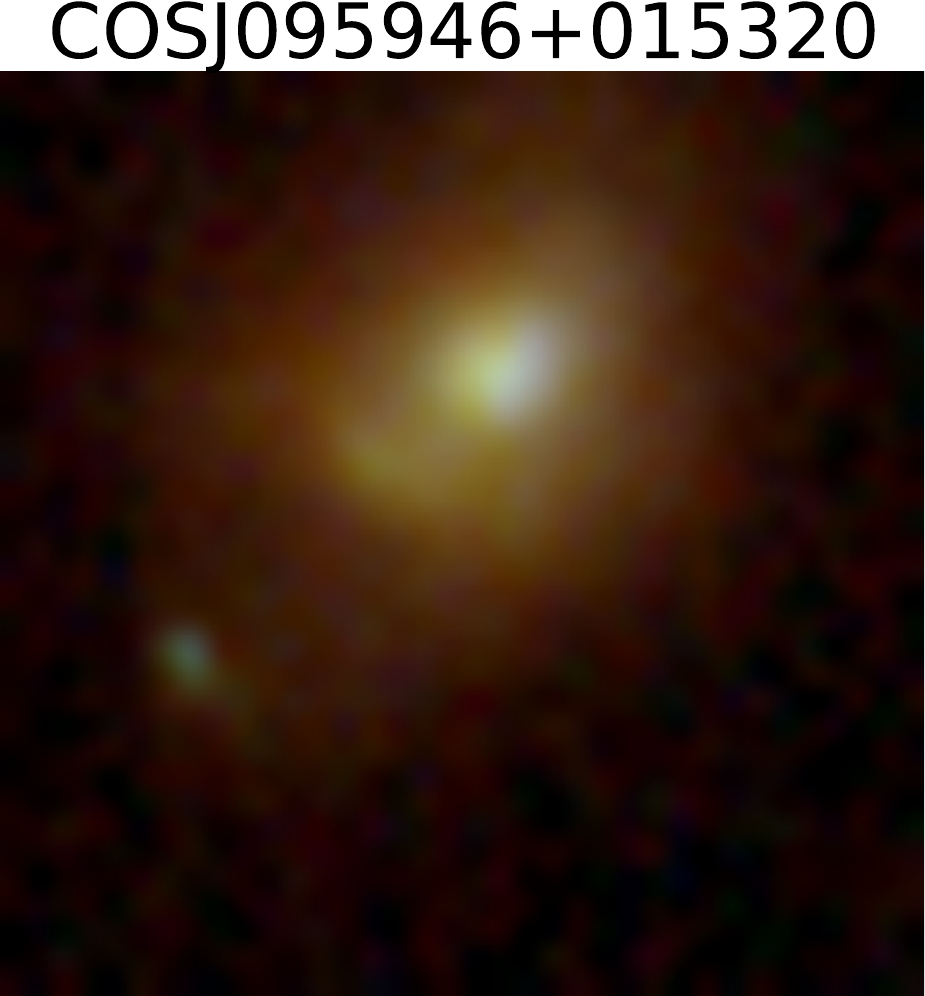}
\includegraphics[width=0.12\textwidth]{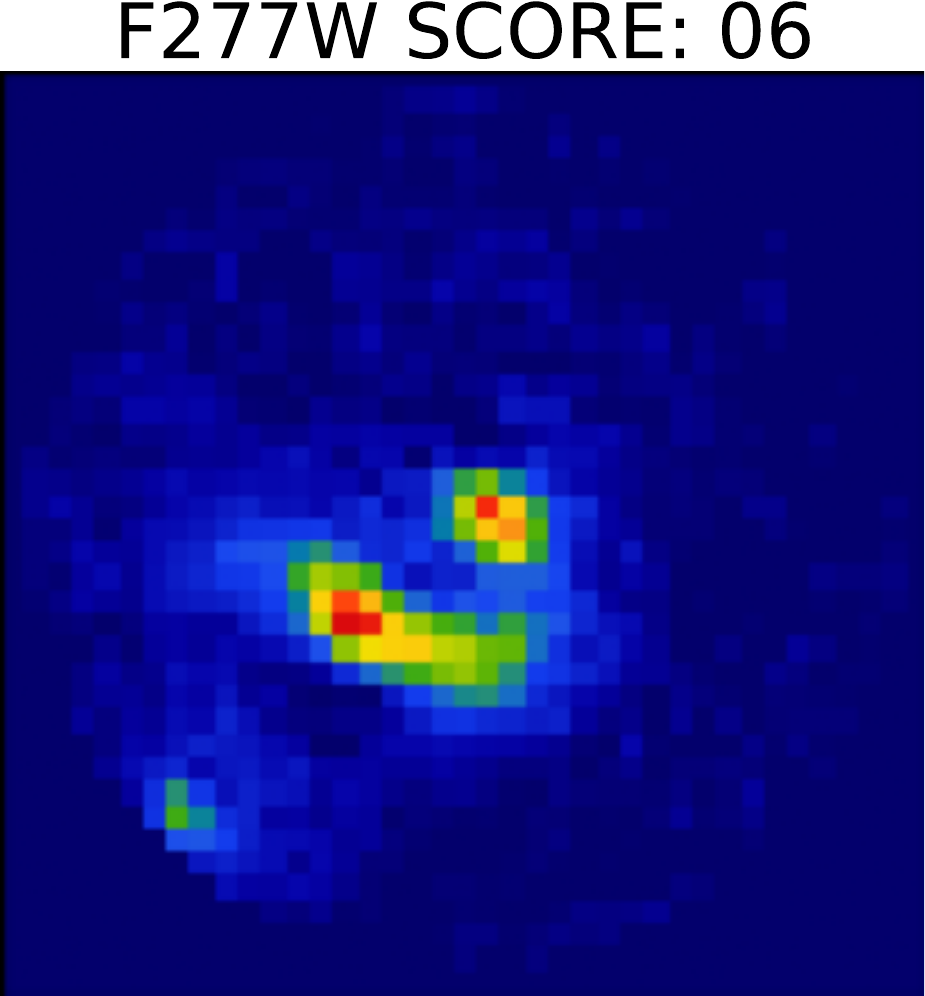}
\includegraphics[width=0.12\textwidth]{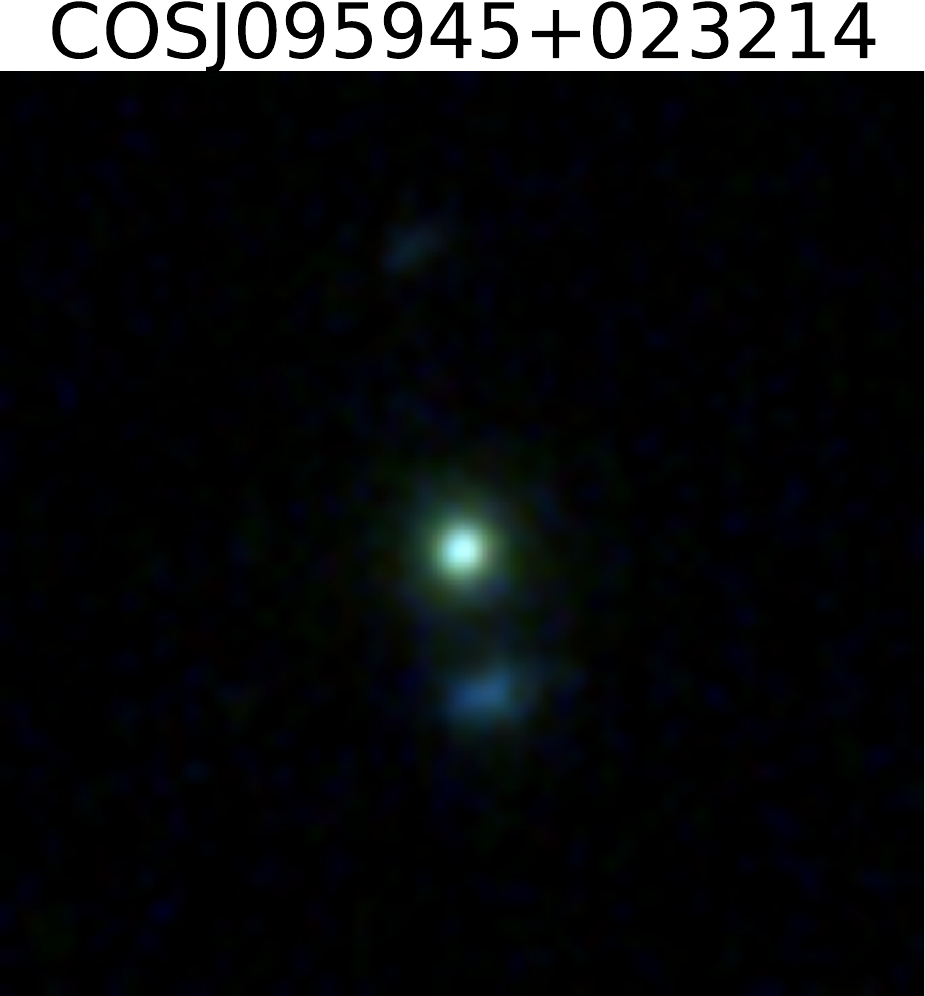}
\includegraphics[width=0.12\textwidth]{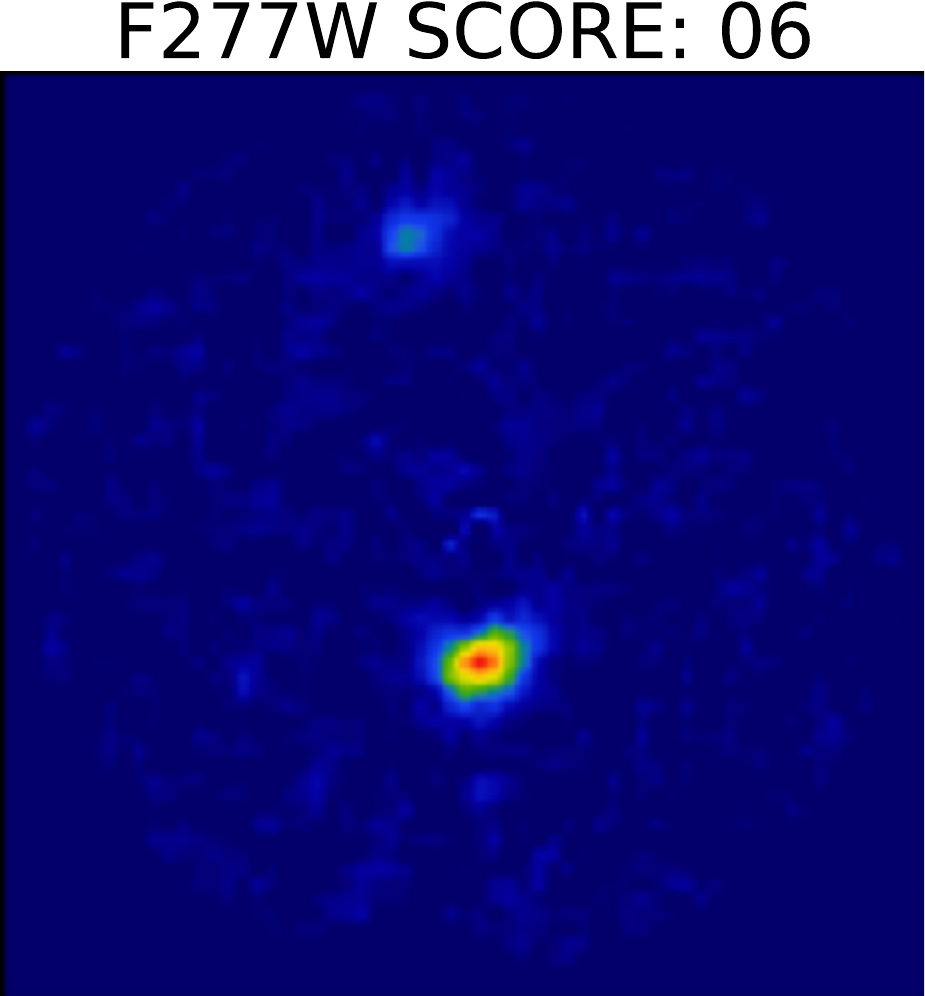}
\includegraphics[width=0.12\textwidth]{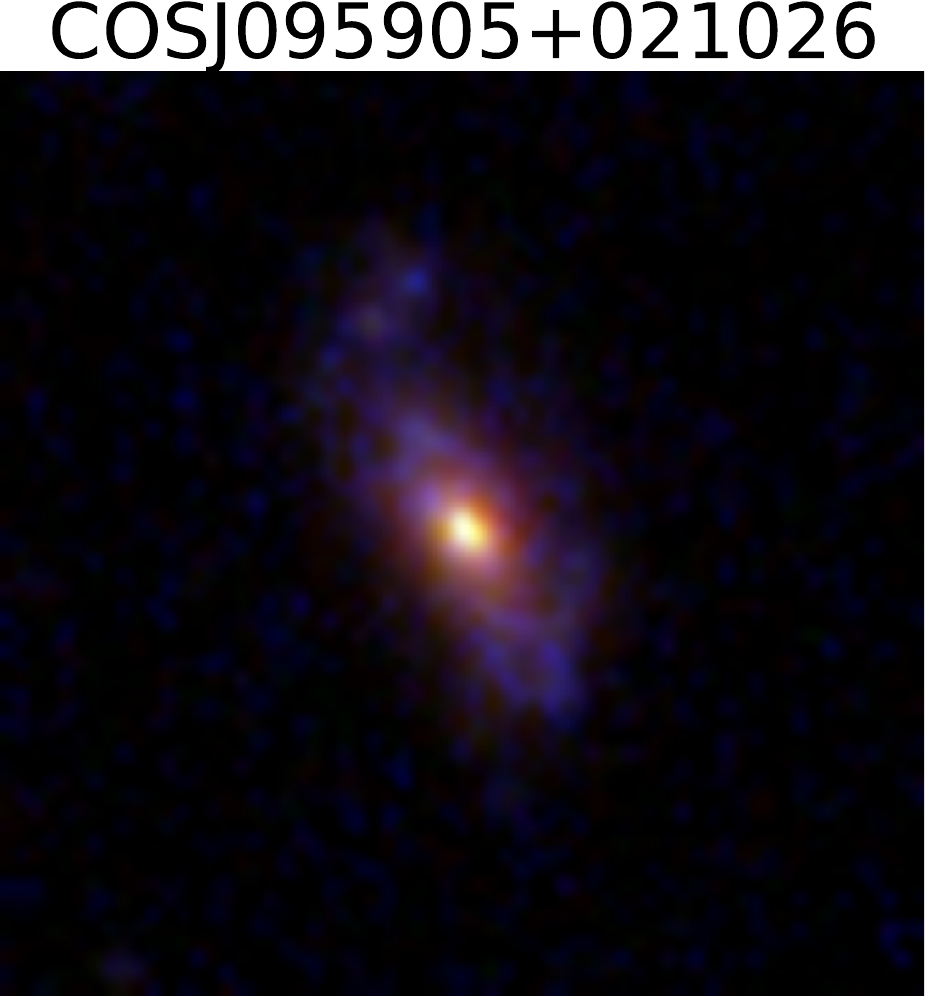}
\includegraphics[width=0.12\textwidth]{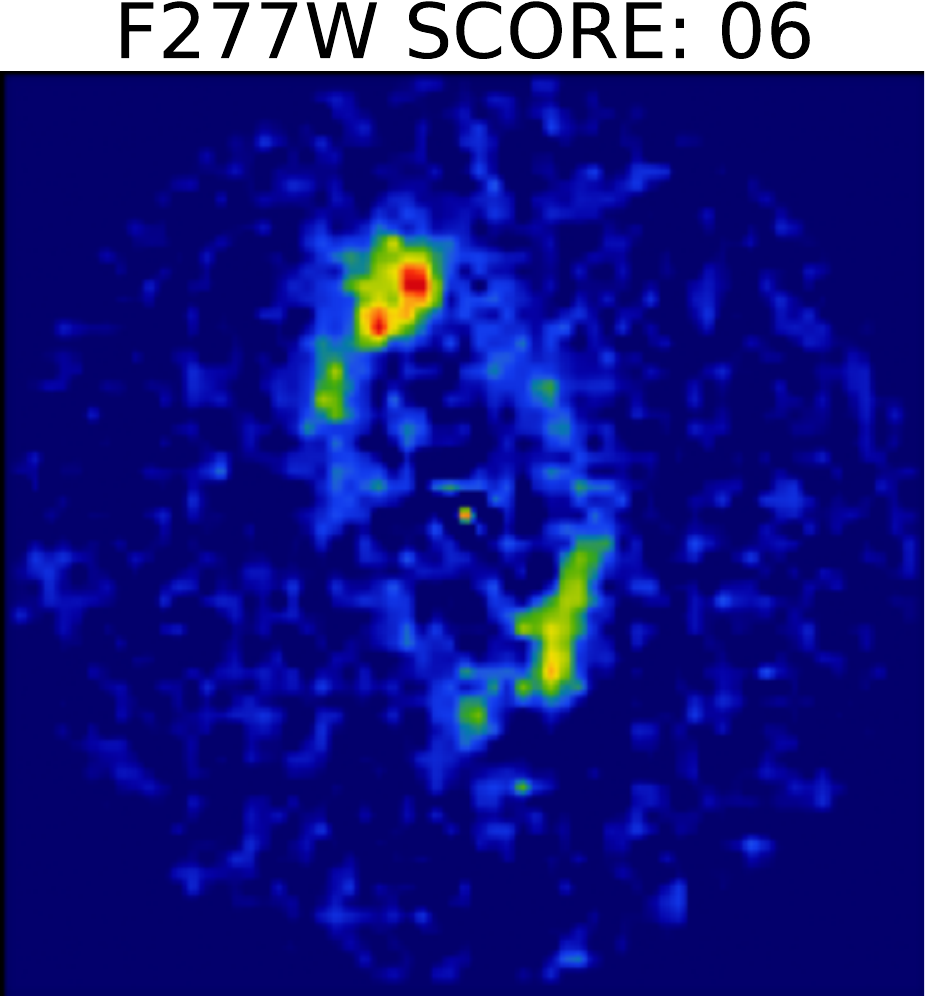}
\includegraphics[width=0.12\textwidth]{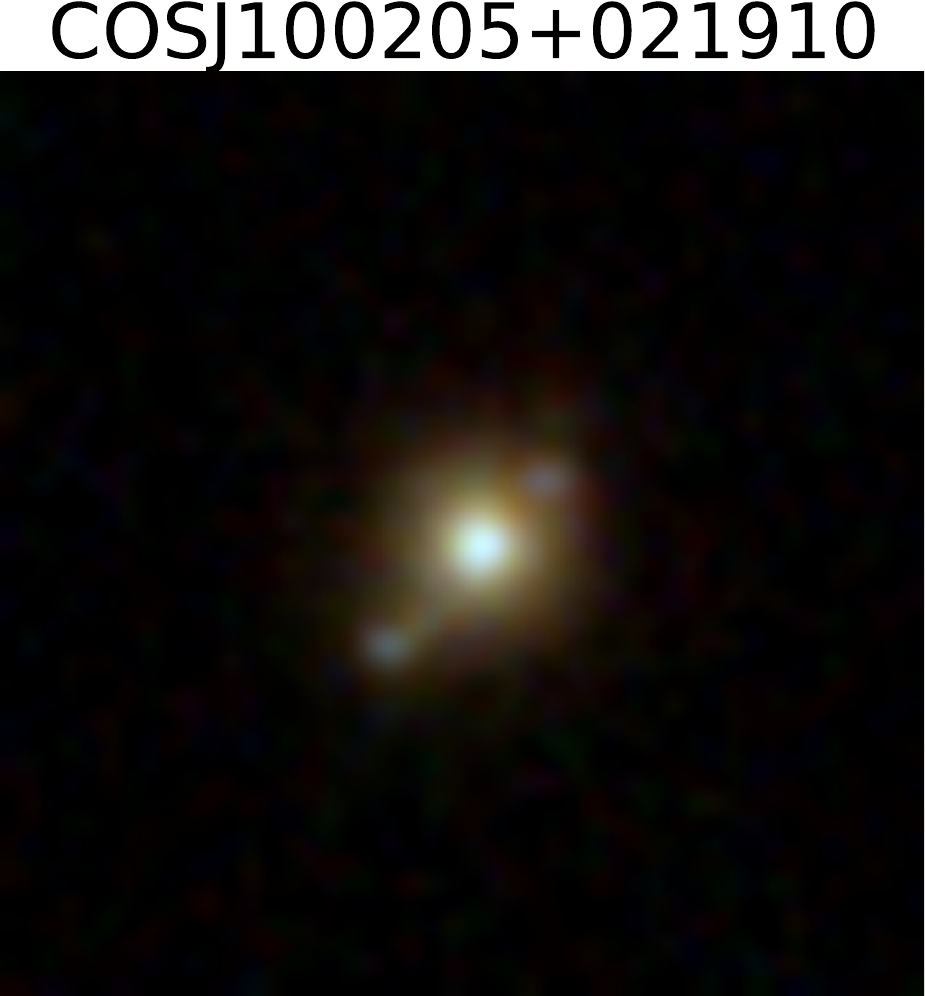}
\includegraphics[width=0.12\textwidth]{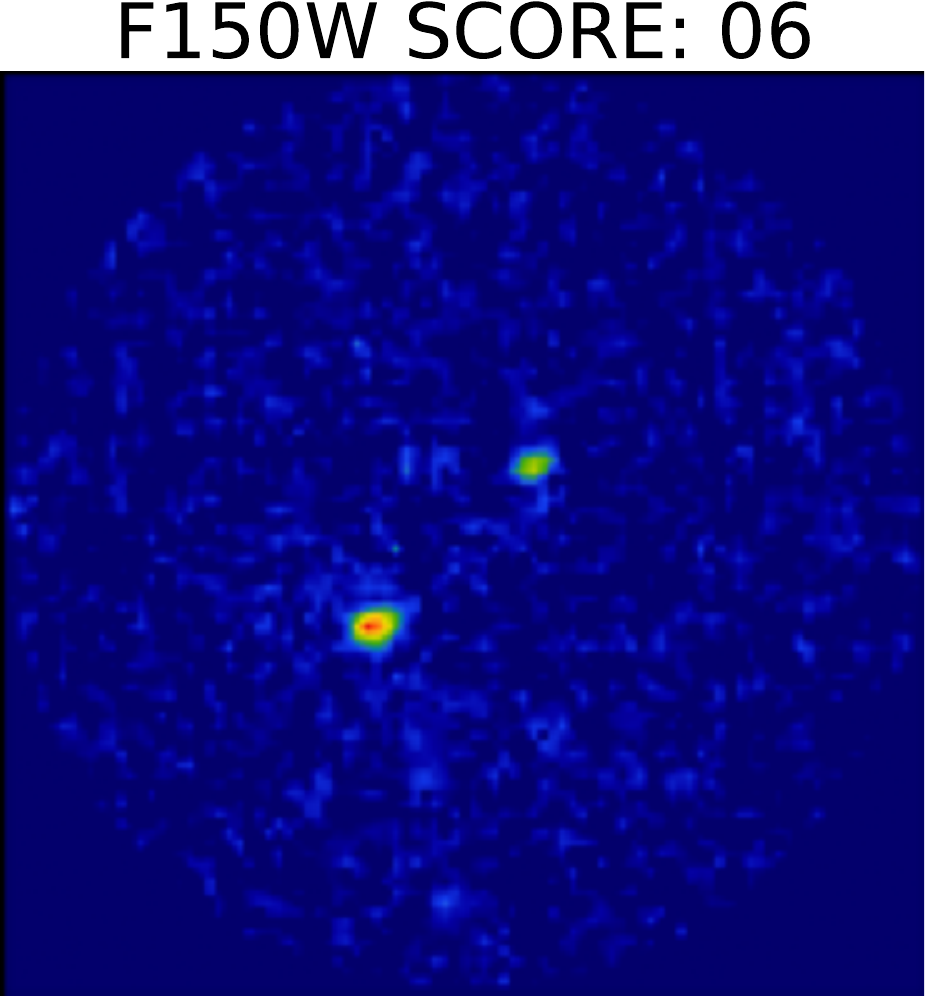}
\includegraphics[width=0.12\textwidth]{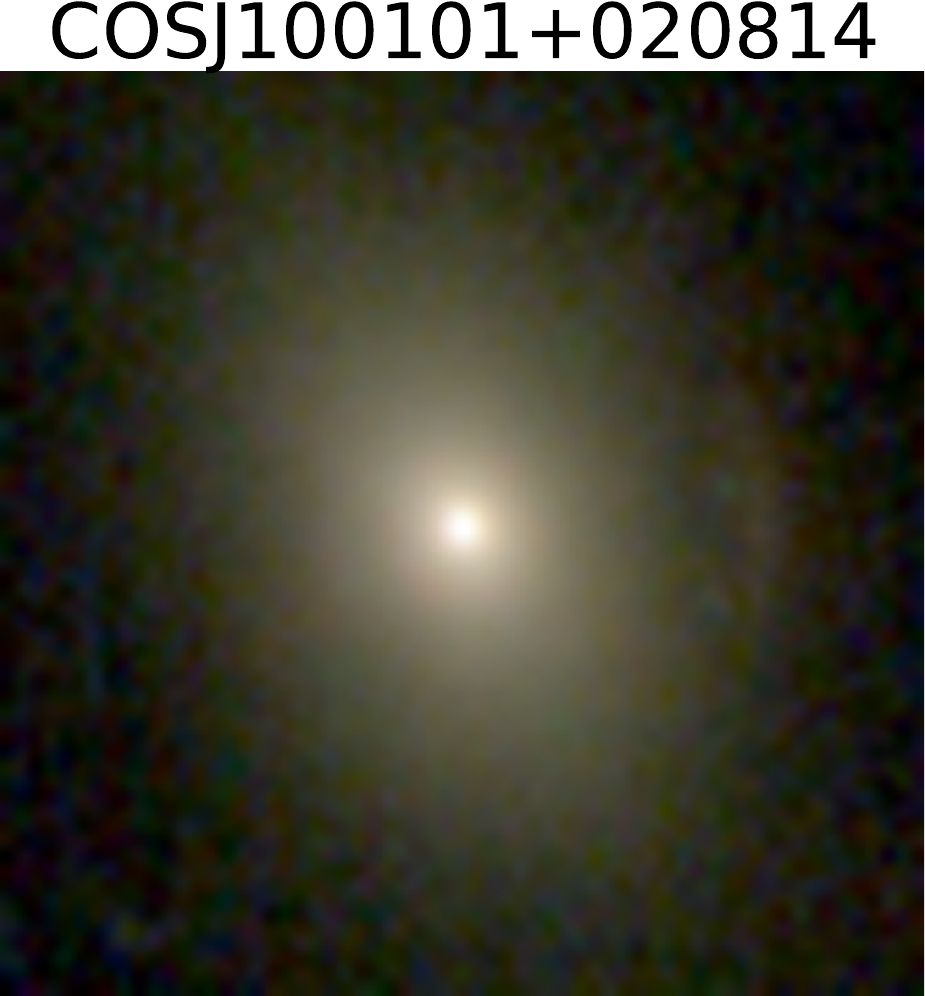}
\includegraphics[width=0.12\textwidth]{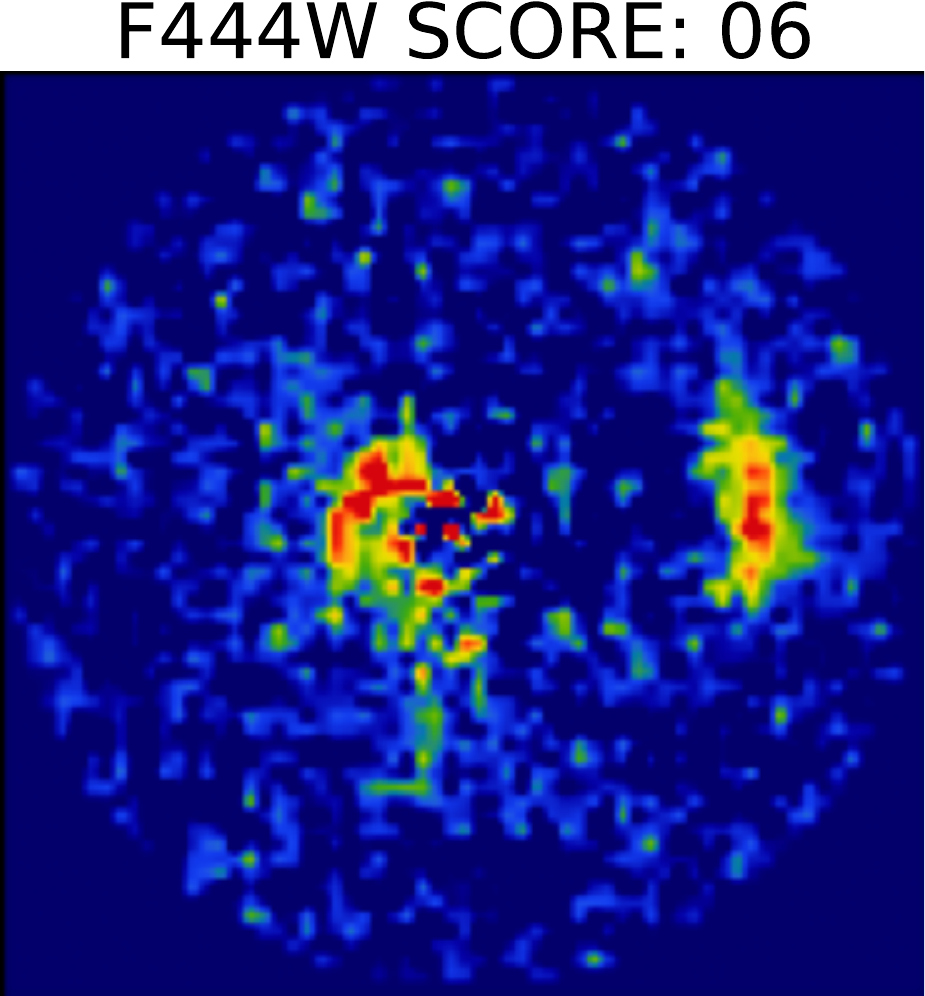}
\includegraphics[width=0.12\textwidth]{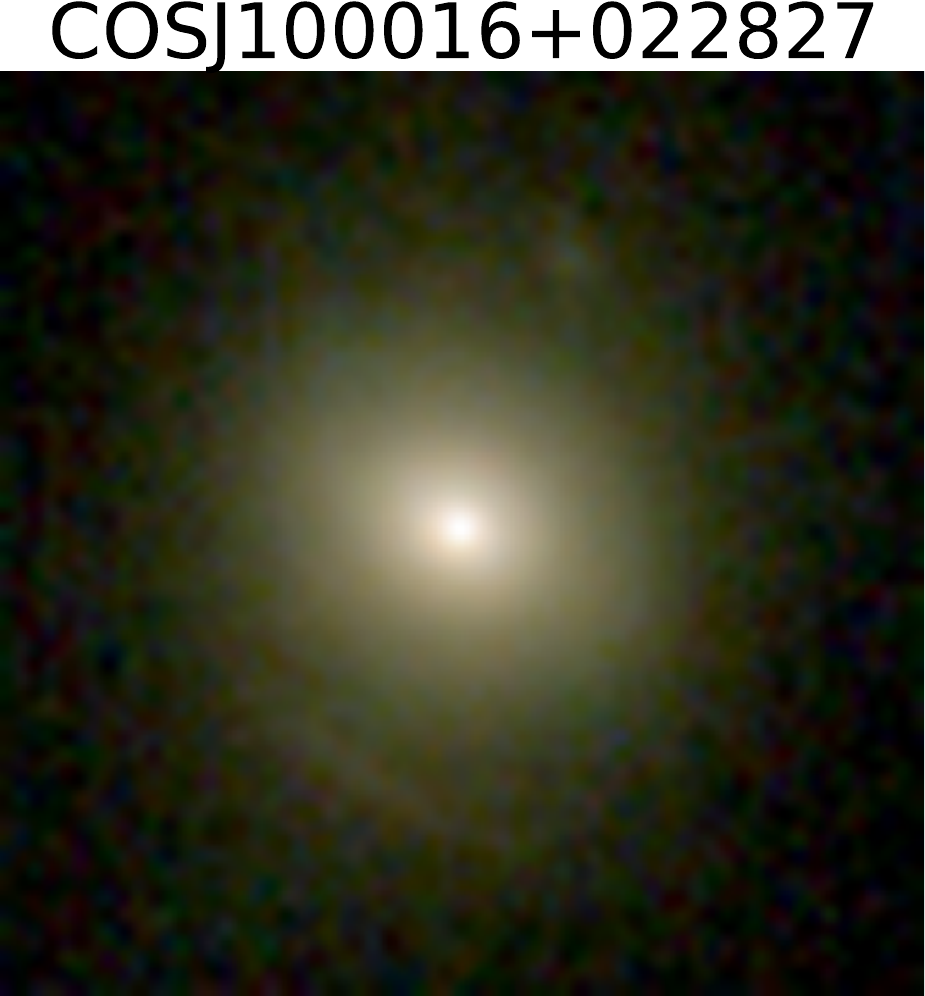}
\includegraphics[width=0.12\textwidth]{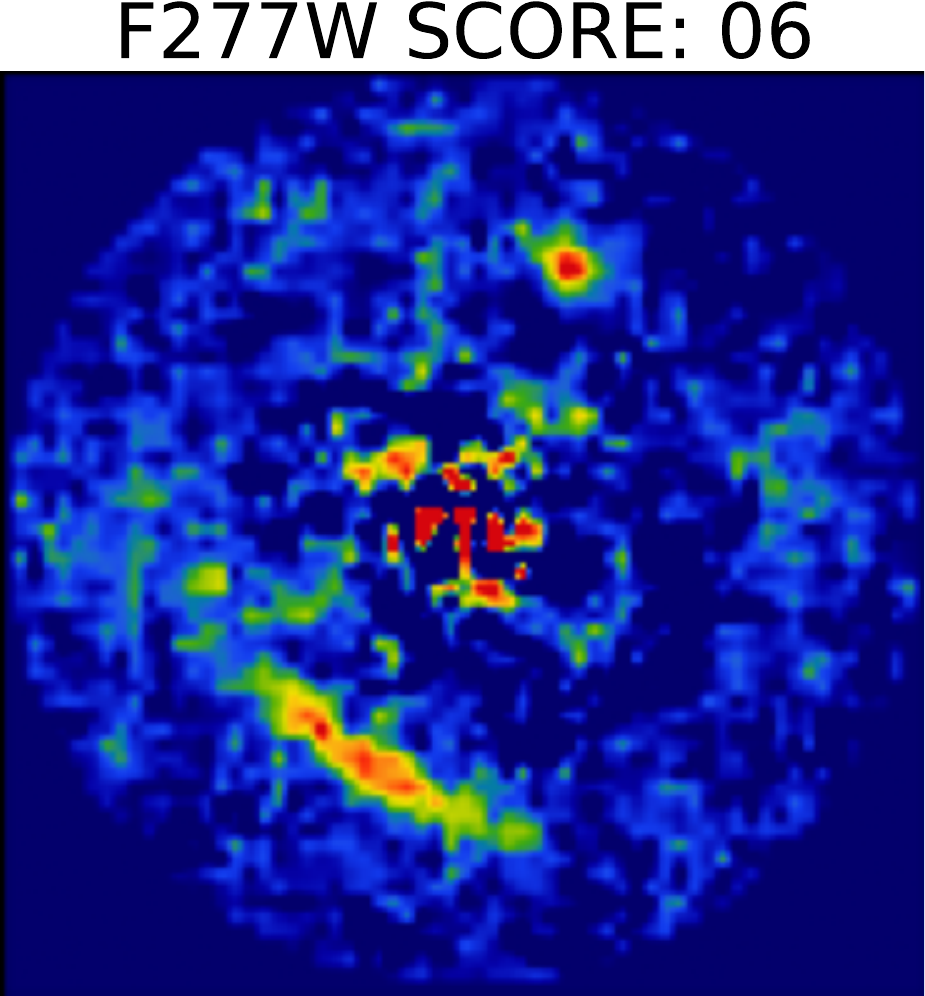}
\includegraphics[width=0.12\textwidth]{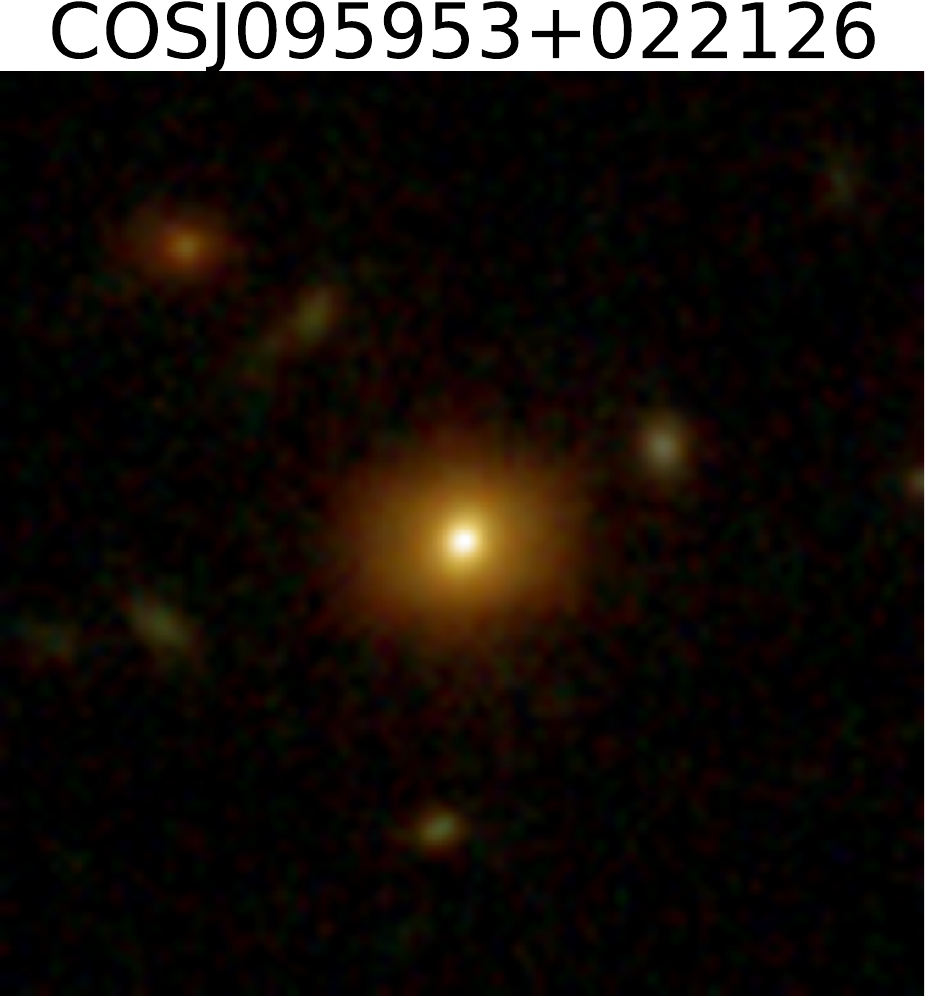}
\includegraphics[width=0.12\textwidth]{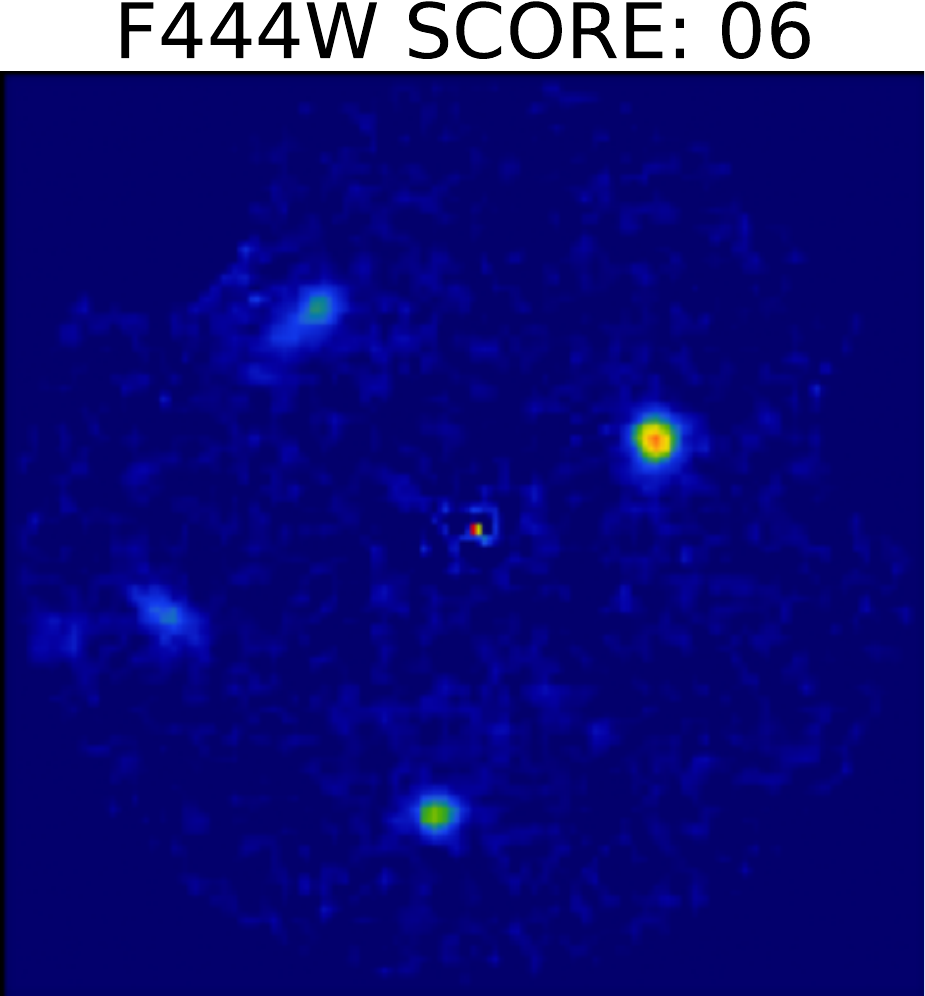}
\includegraphics[width=0.12\textwidth]{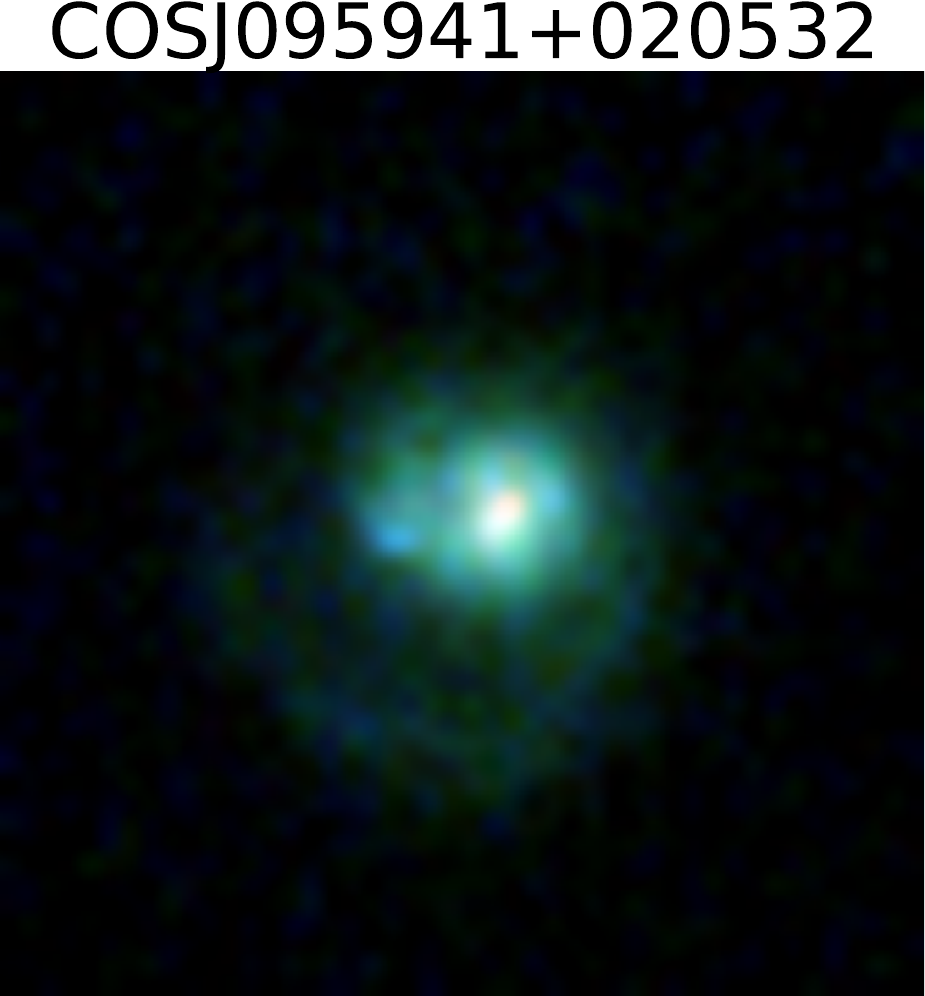}
\includegraphics[width=0.12\textwidth]{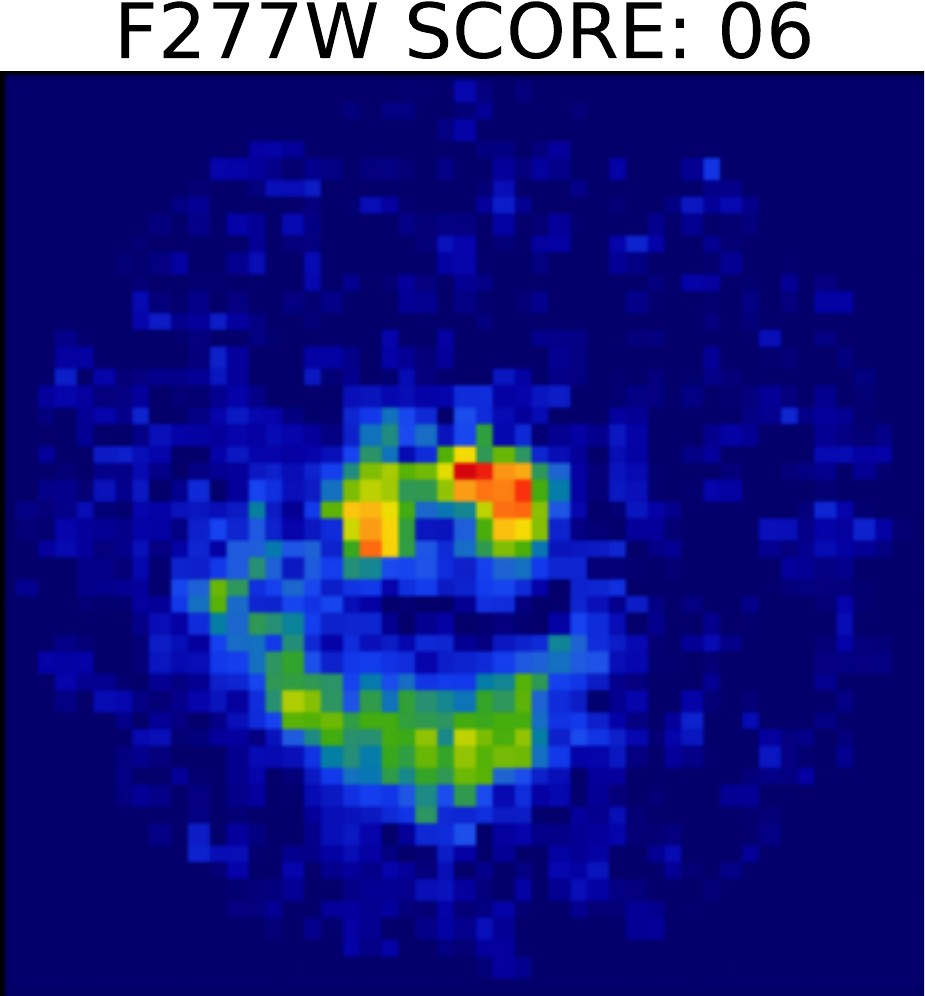}
\includegraphics[width=0.12\textwidth]{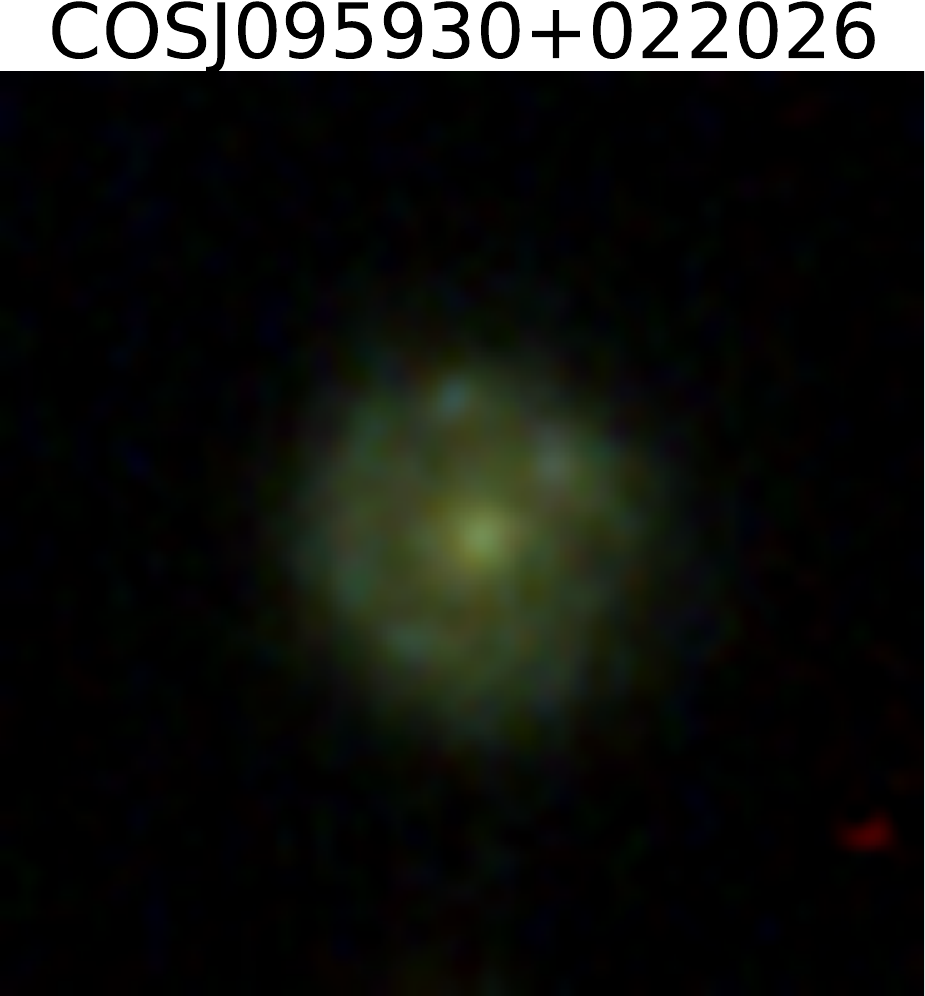}
\includegraphics[width=0.12\textwidth]{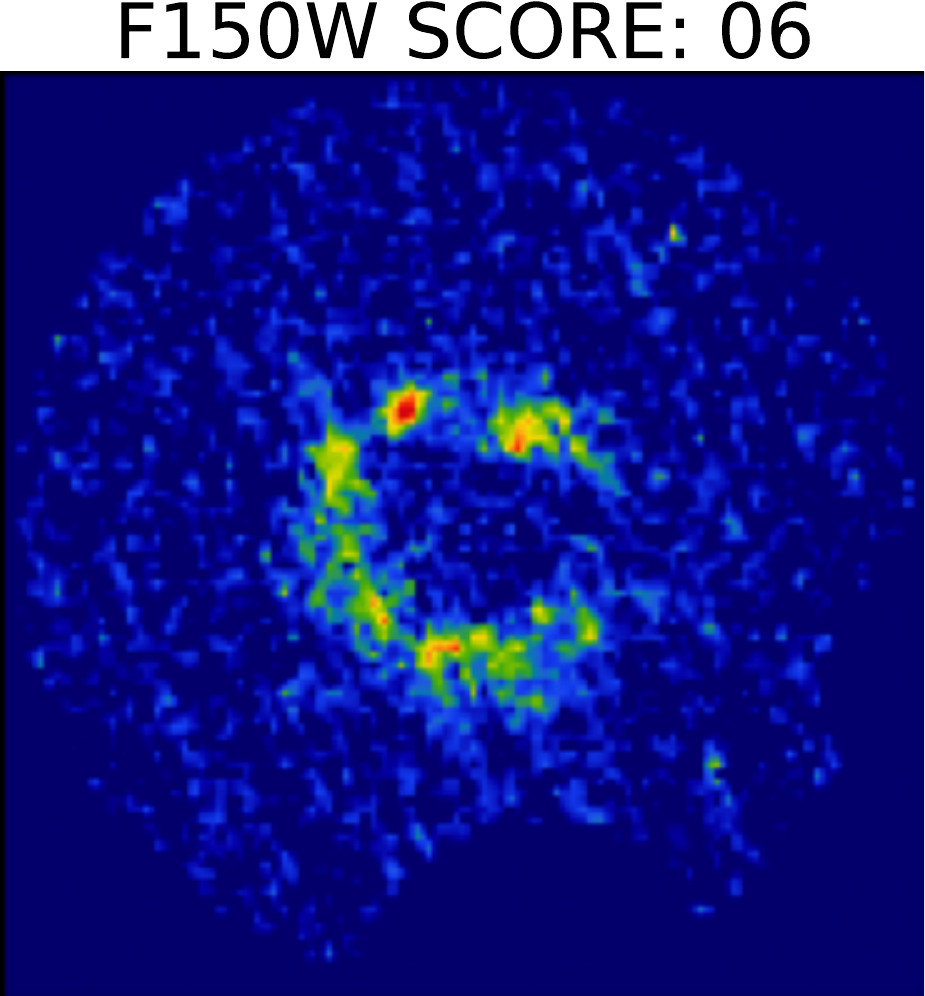}
\includegraphics[width=0.12\textwidth]{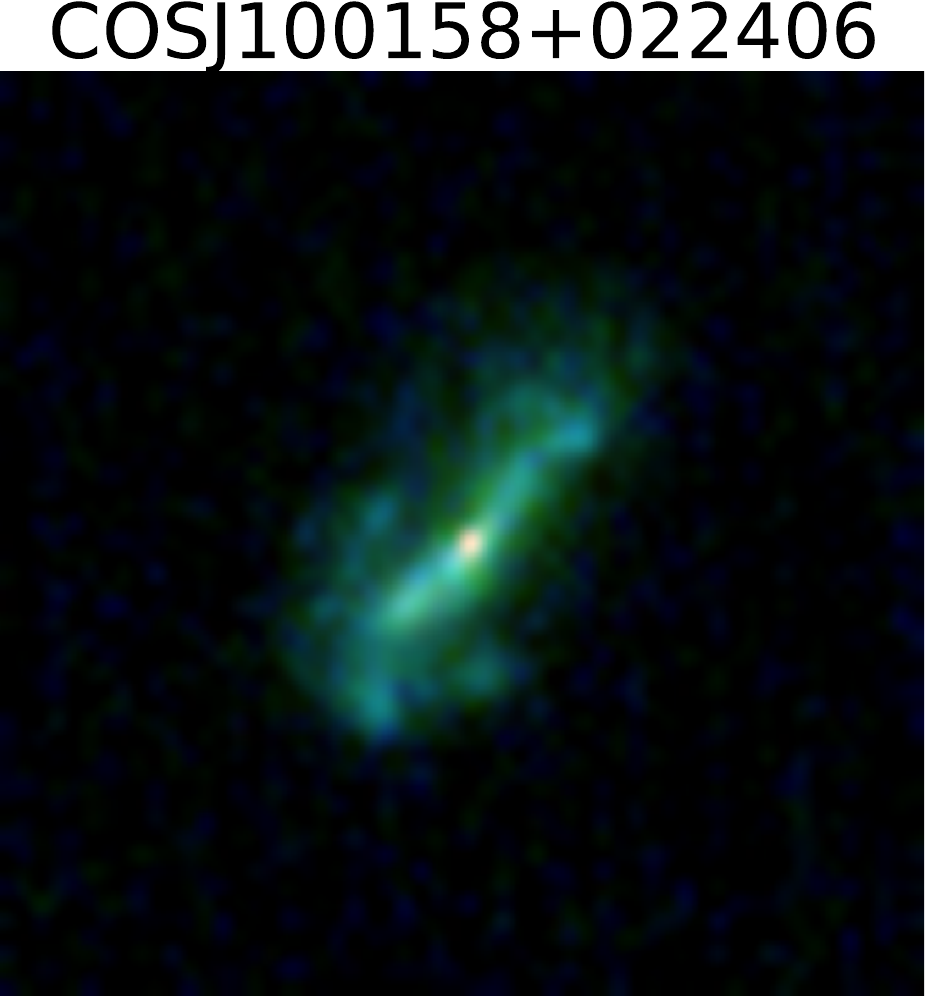}
\includegraphics[width=0.12\textwidth]{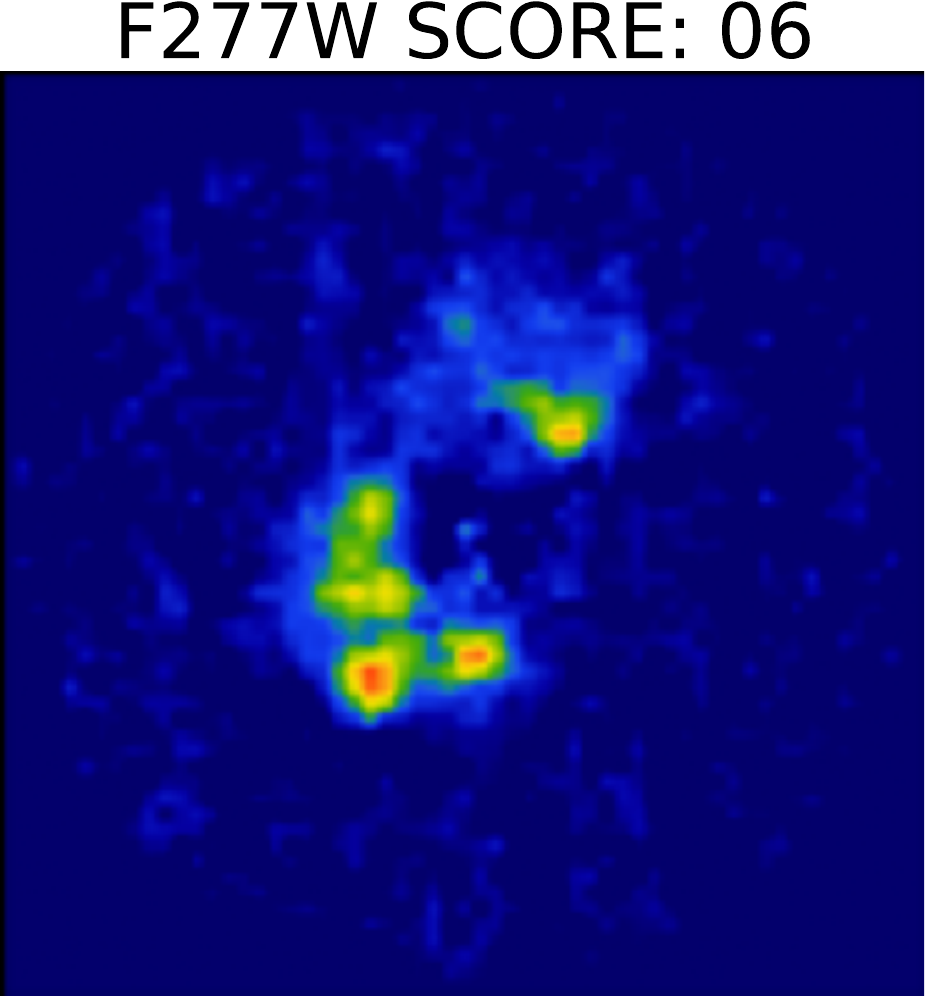}
\includegraphics[width=0.12\textwidth]{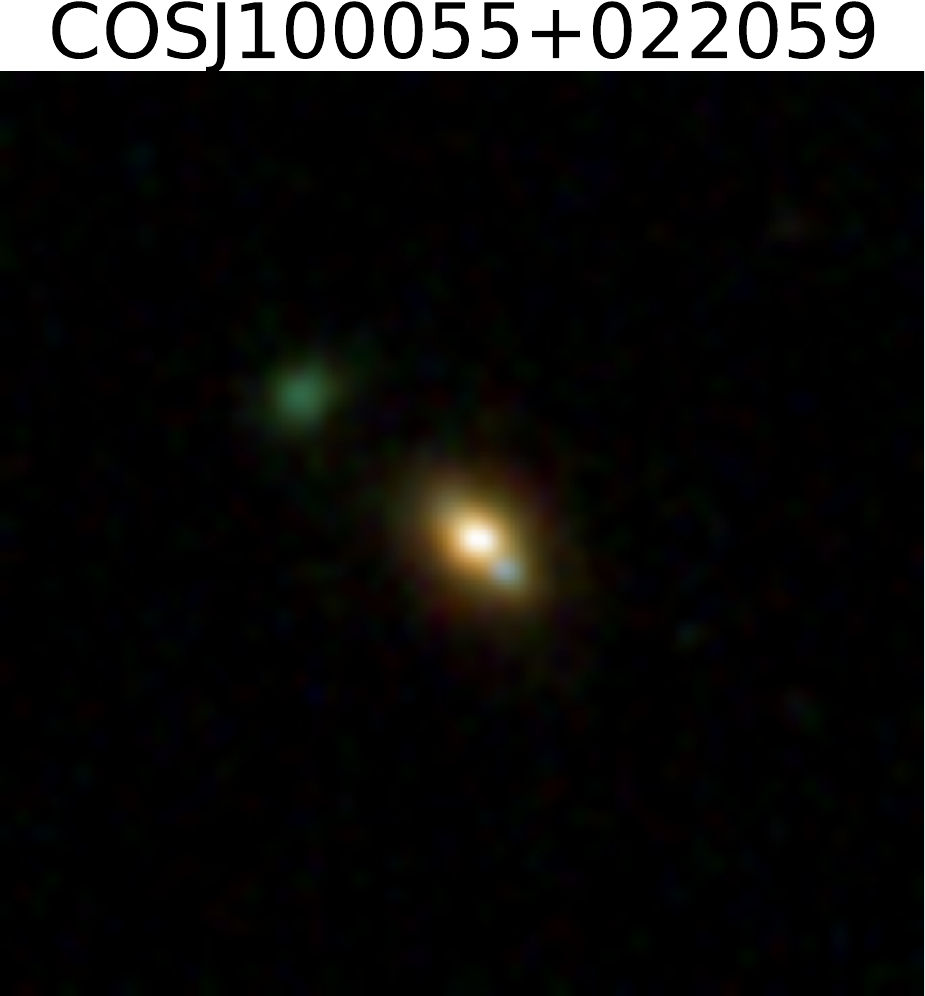}
\includegraphics[width=0.12\textwidth]{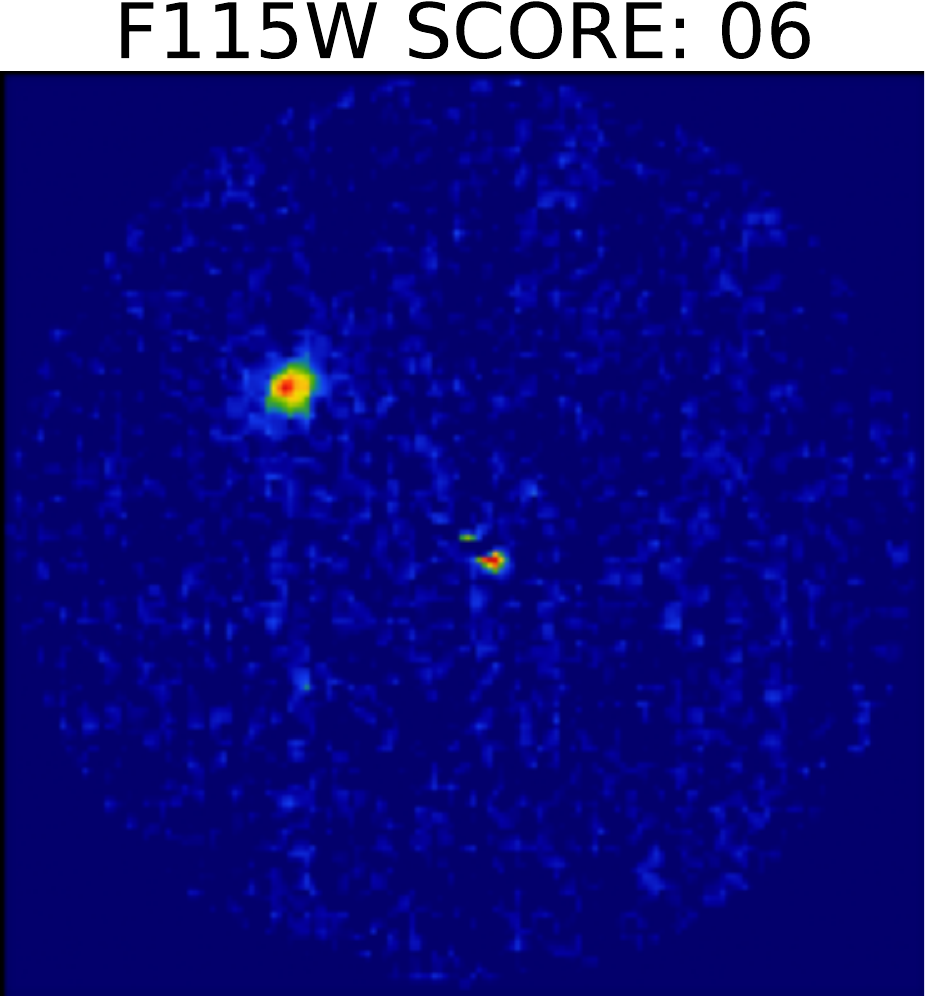}
\includegraphics[width=0.12\textwidth]{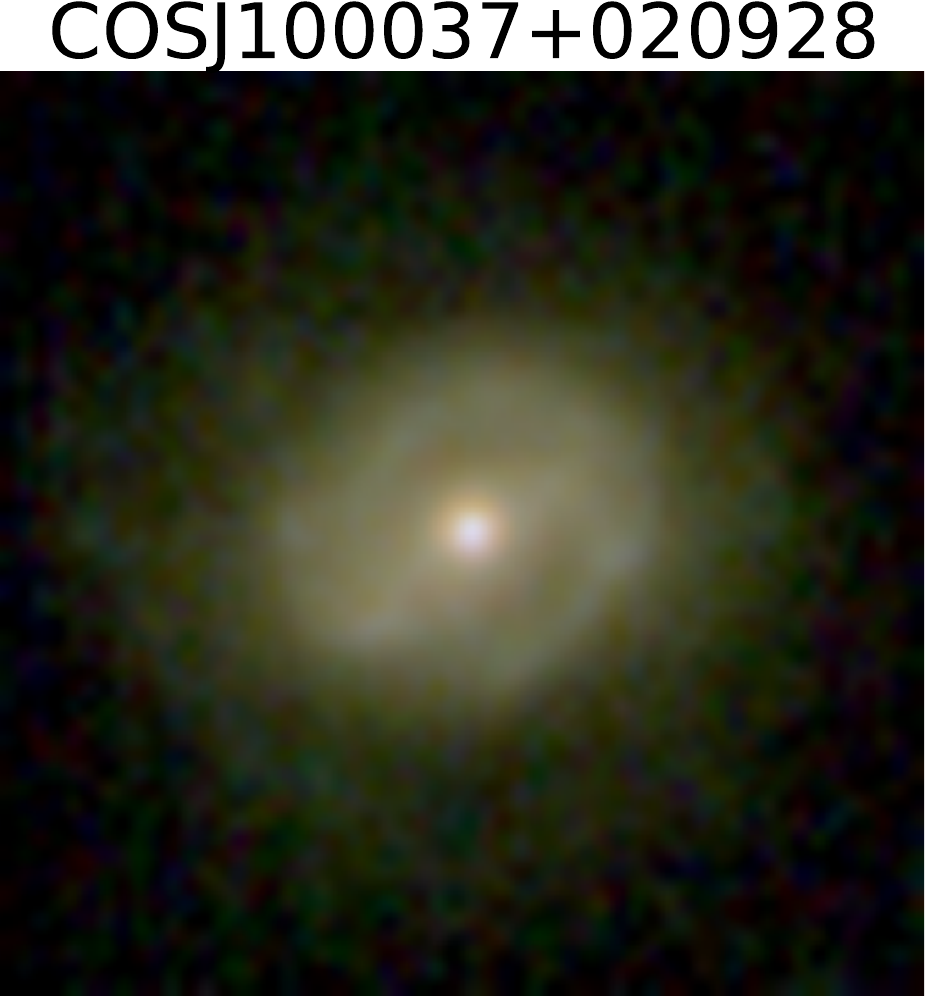}
\includegraphics[width=0.12\textwidth]{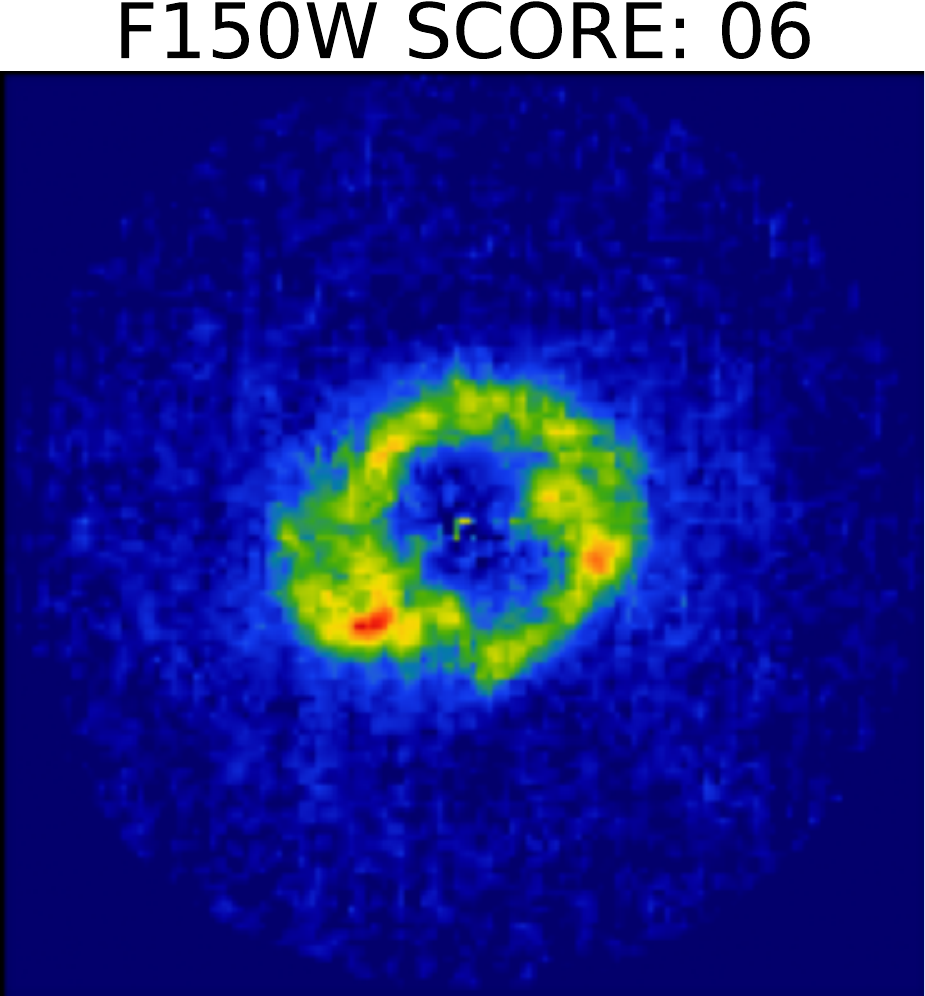}
\includegraphics[width=0.12\textwidth]{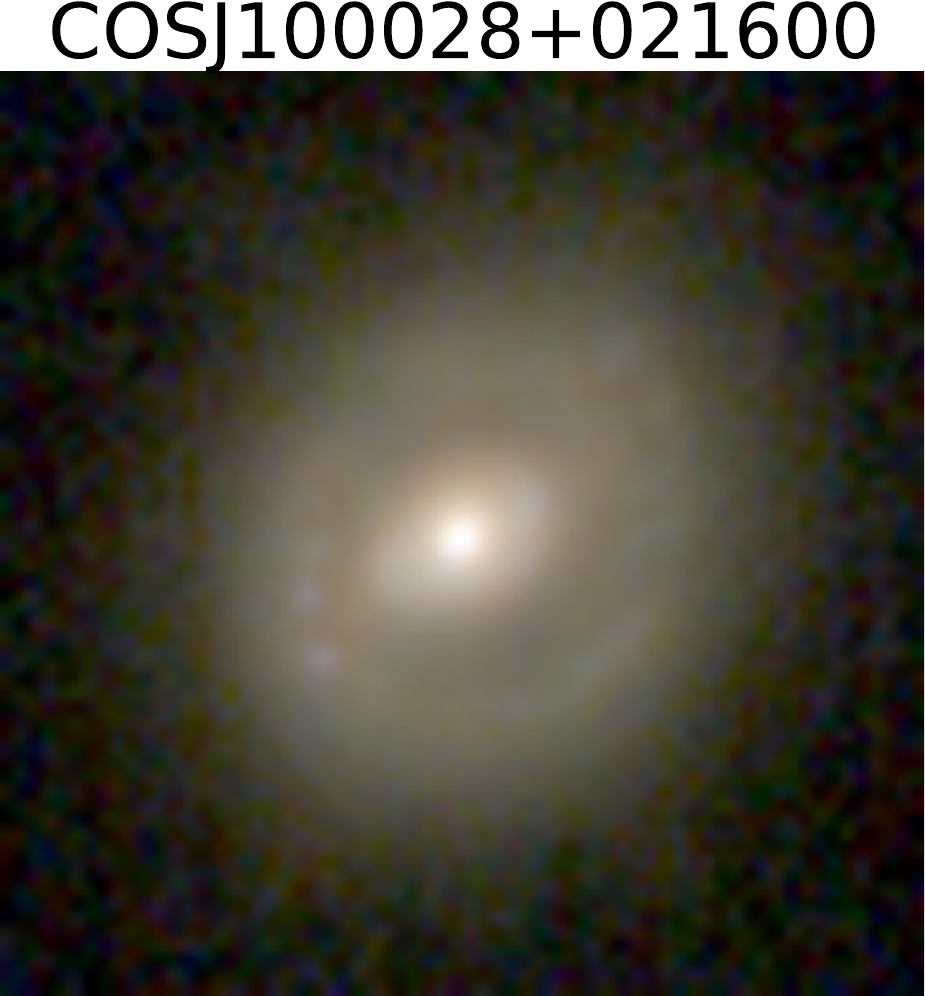}
\includegraphics[width=0.12\textwidth]{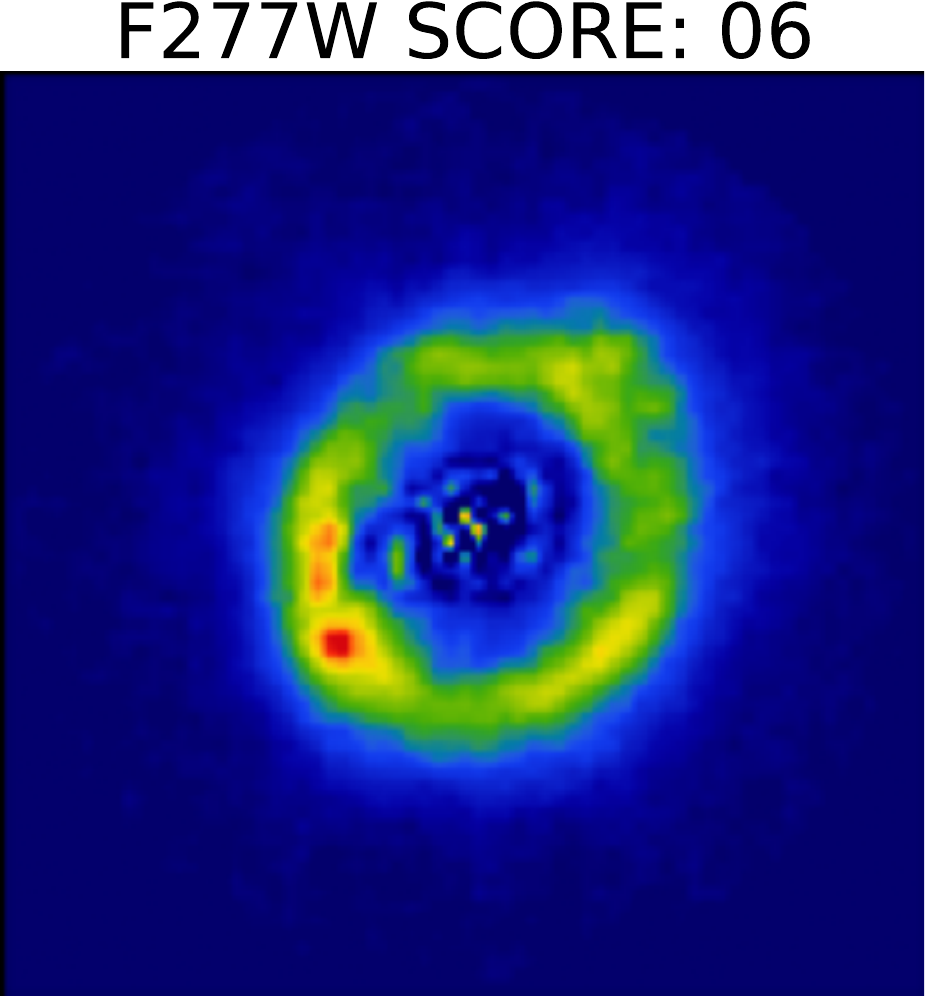}
\includegraphics[width=0.12\textwidth]{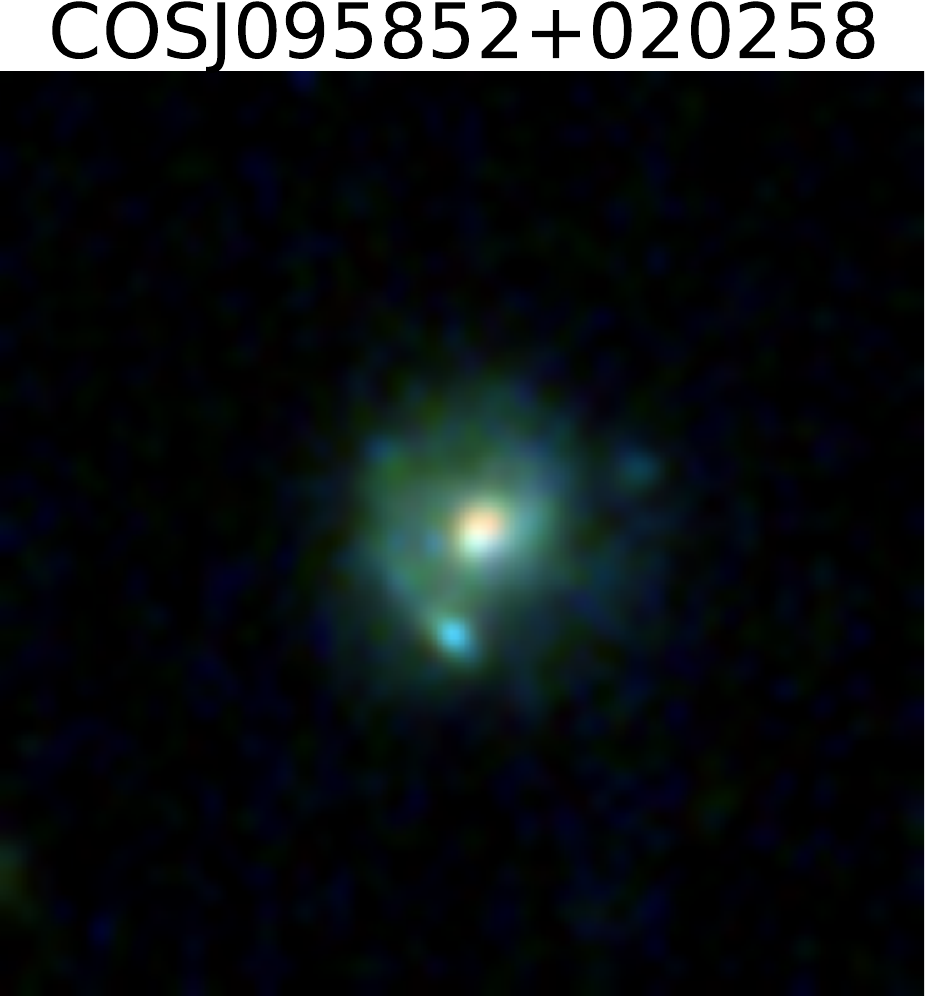}
\includegraphics[width=0.12\textwidth]{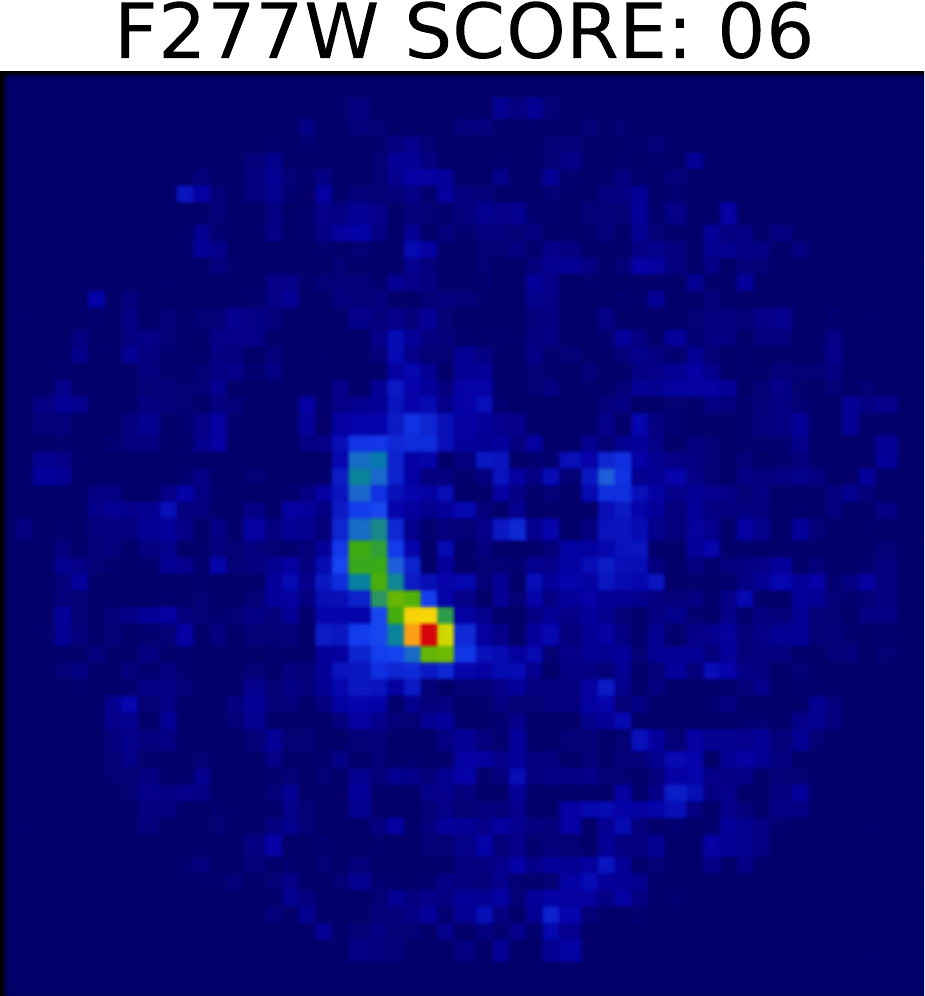}
\includegraphics[width=0.12\textwidth]{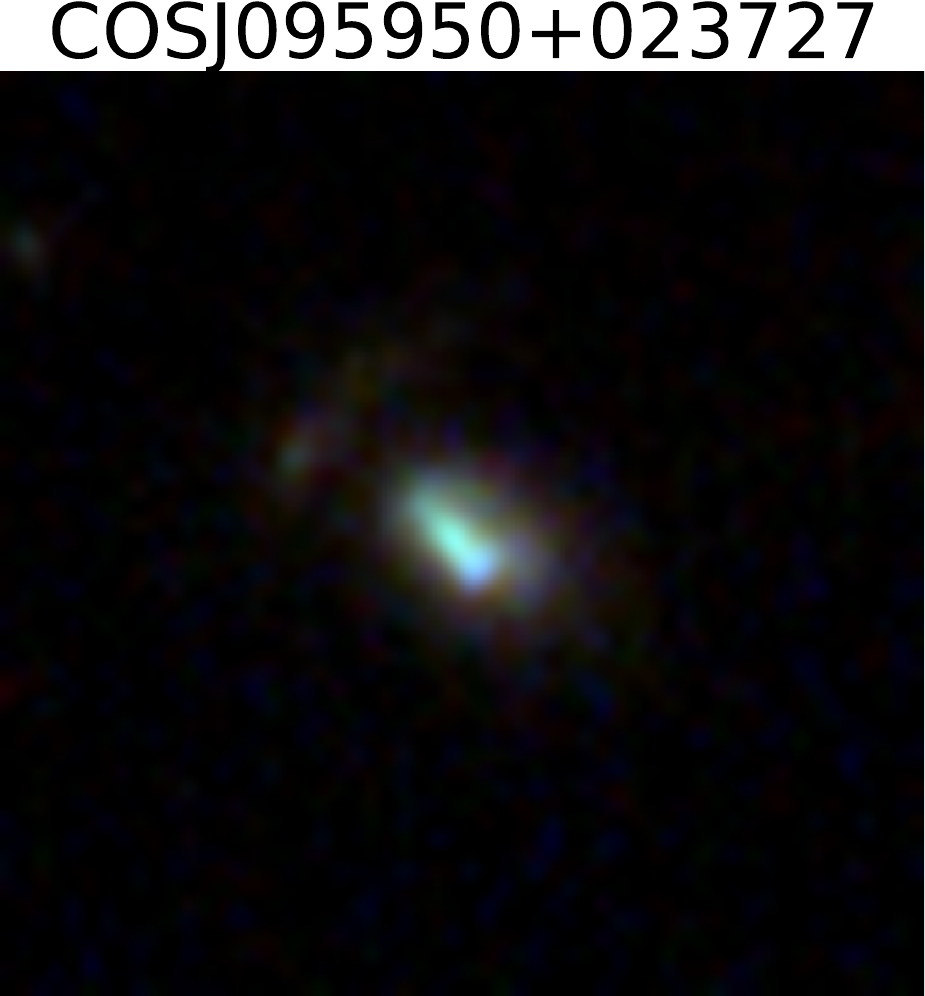}
\includegraphics[width=0.12\textwidth]{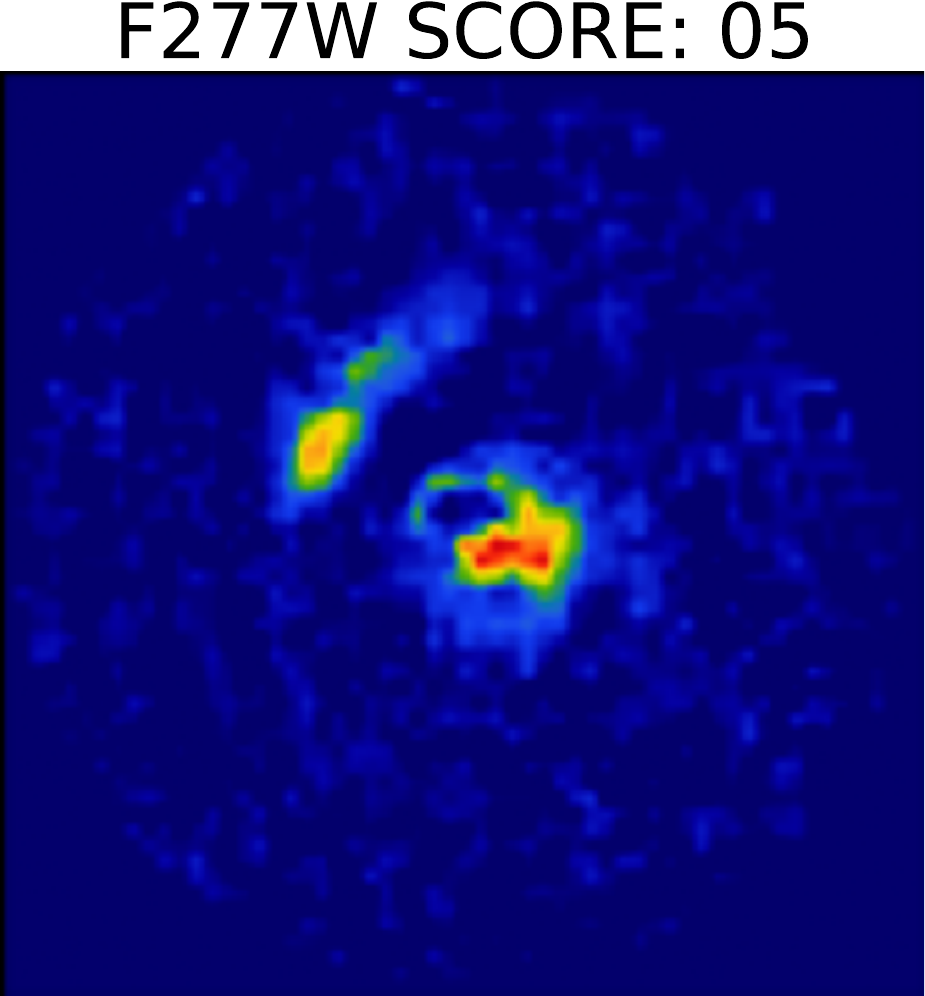}
\includegraphics[width=0.12\textwidth]{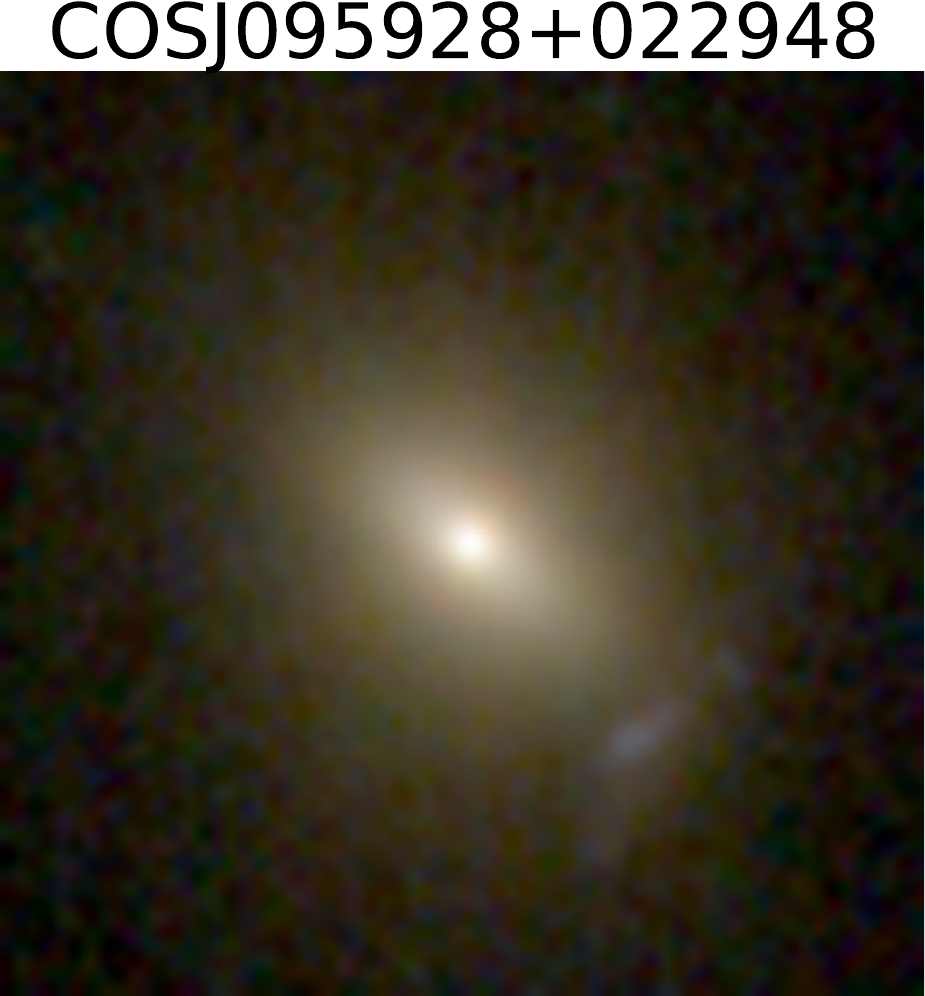}
\includegraphics[width=0.12\textwidth]{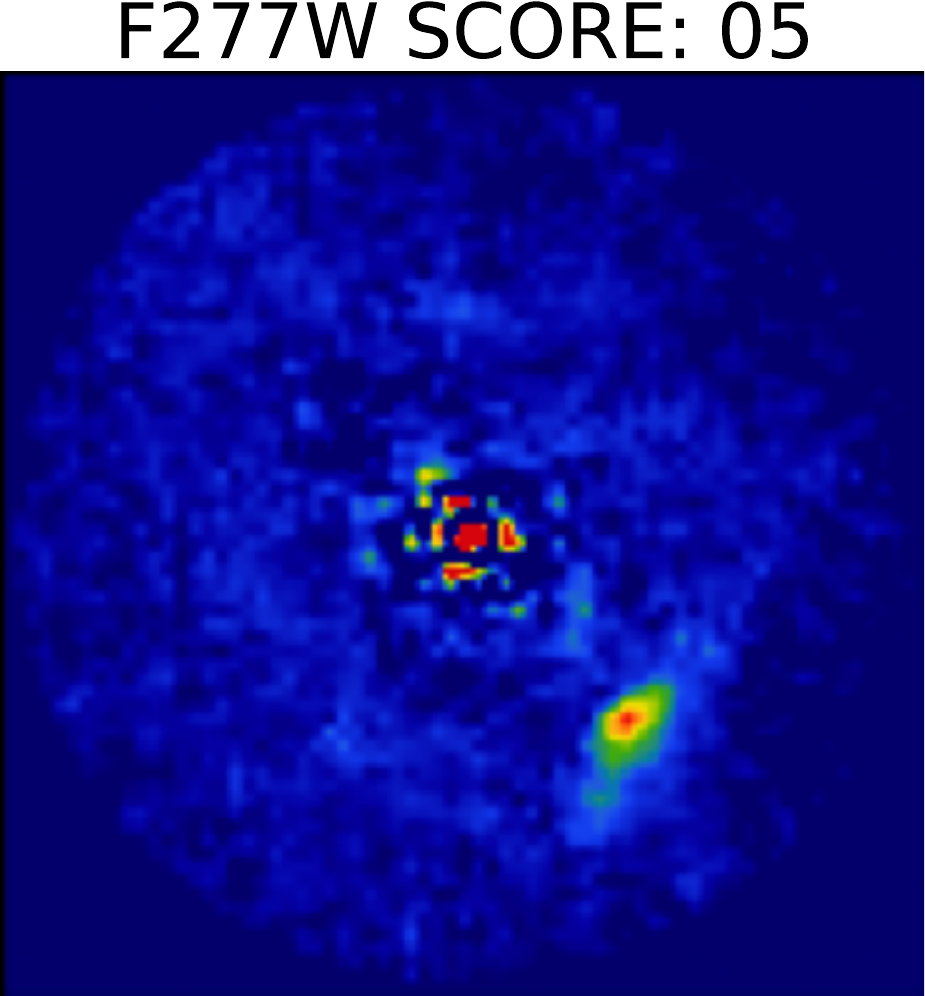}
\includegraphics[width=0.12\textwidth]{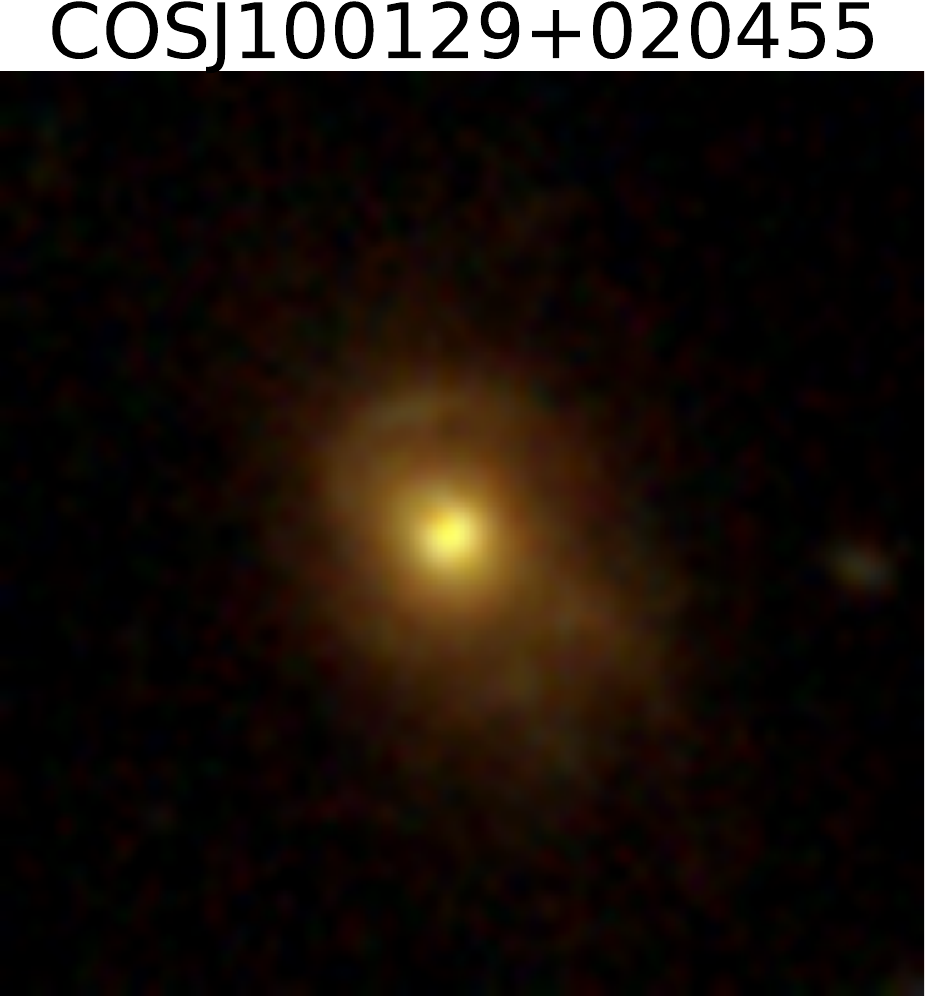}
\includegraphics[width=0.12\textwidth]{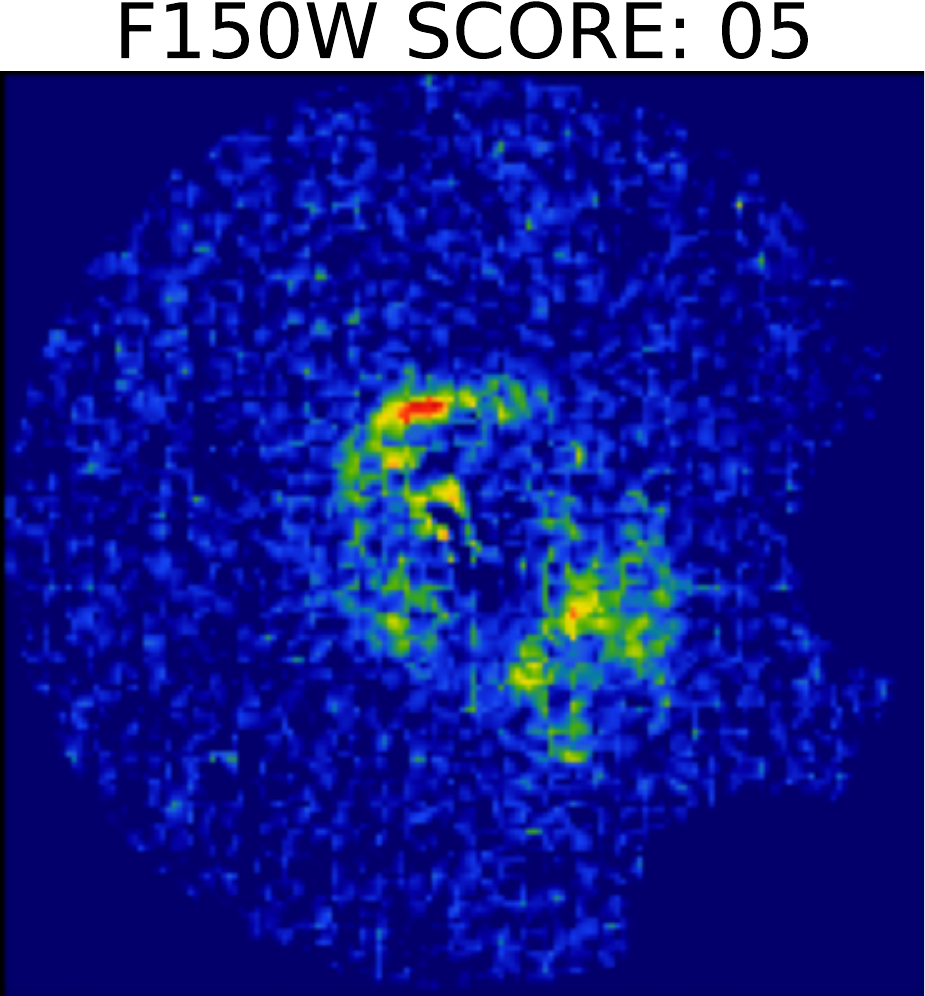}
\caption{
Figure \ref{figure:CutoutA} continued.
}
\label{figure:CutoutA3}
\vspace{-9pt} 
\end{figure*}

\begin{figure*}\ContinuedFloat
\centering
\includegraphics[width=0.12\textwidth]{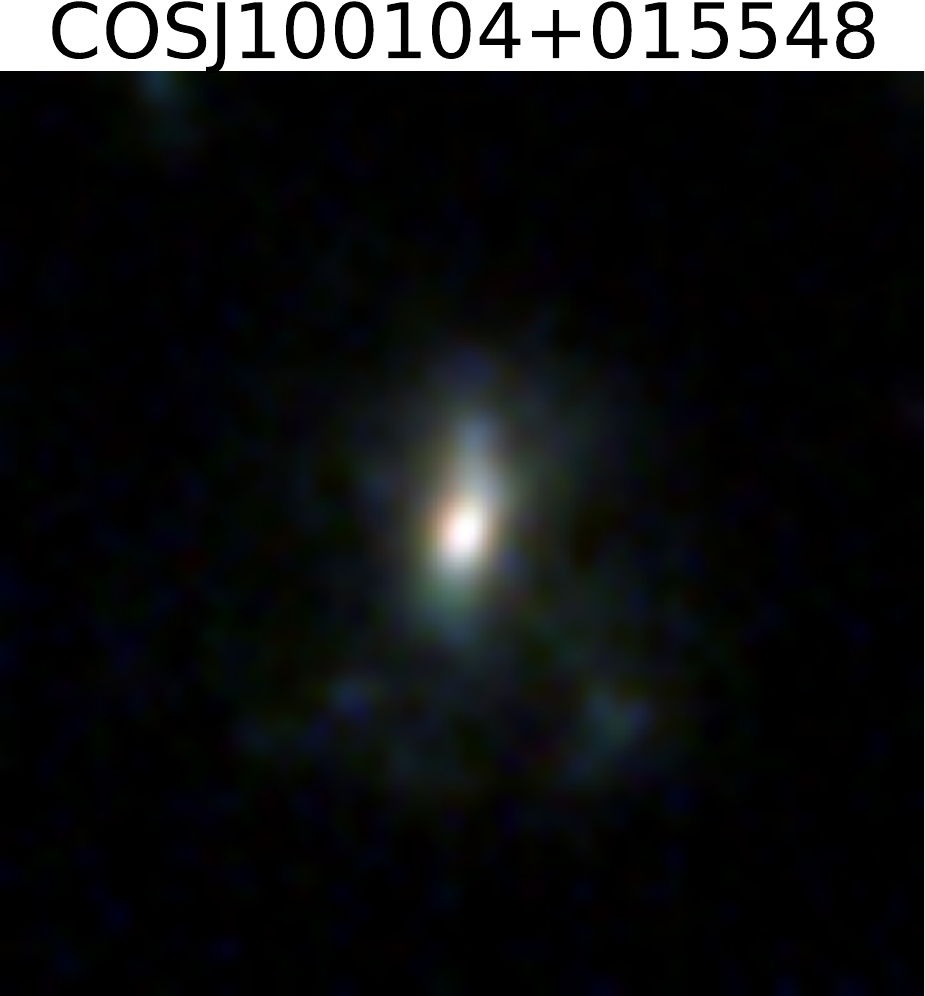}
\includegraphics[width=0.12\textwidth]{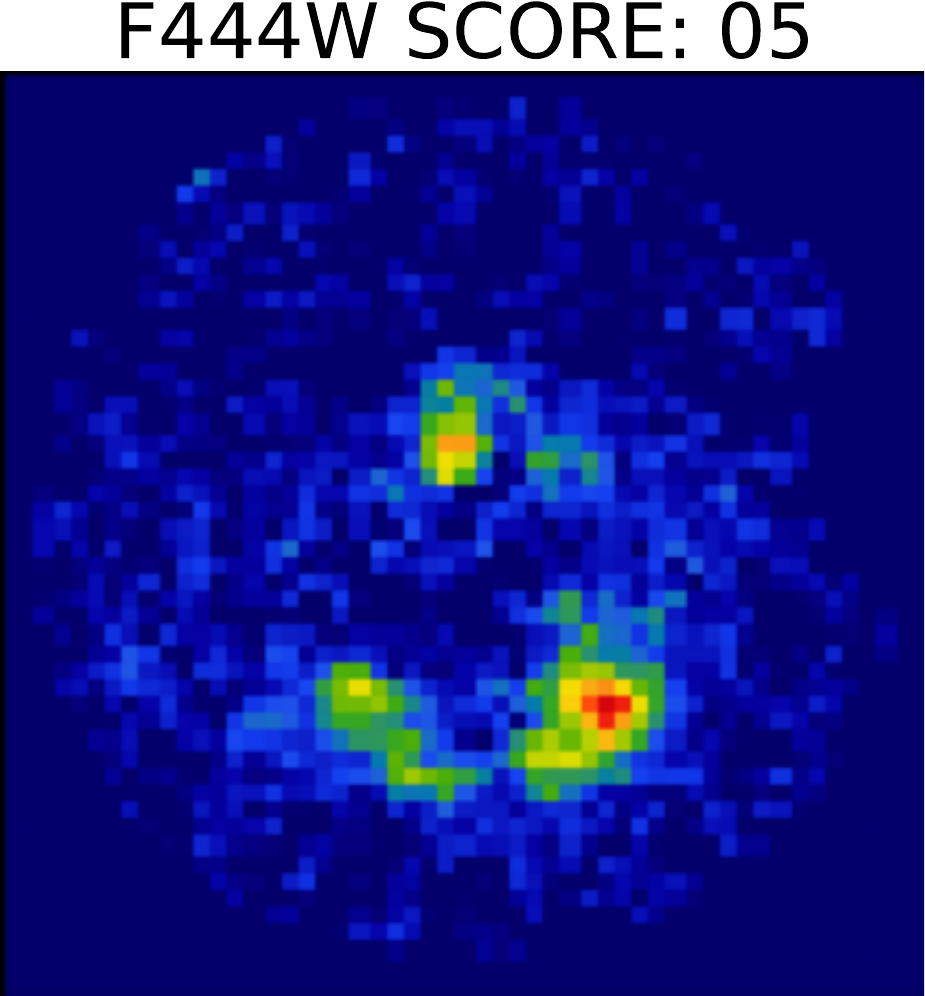}
\includegraphics[width=0.12\textwidth]{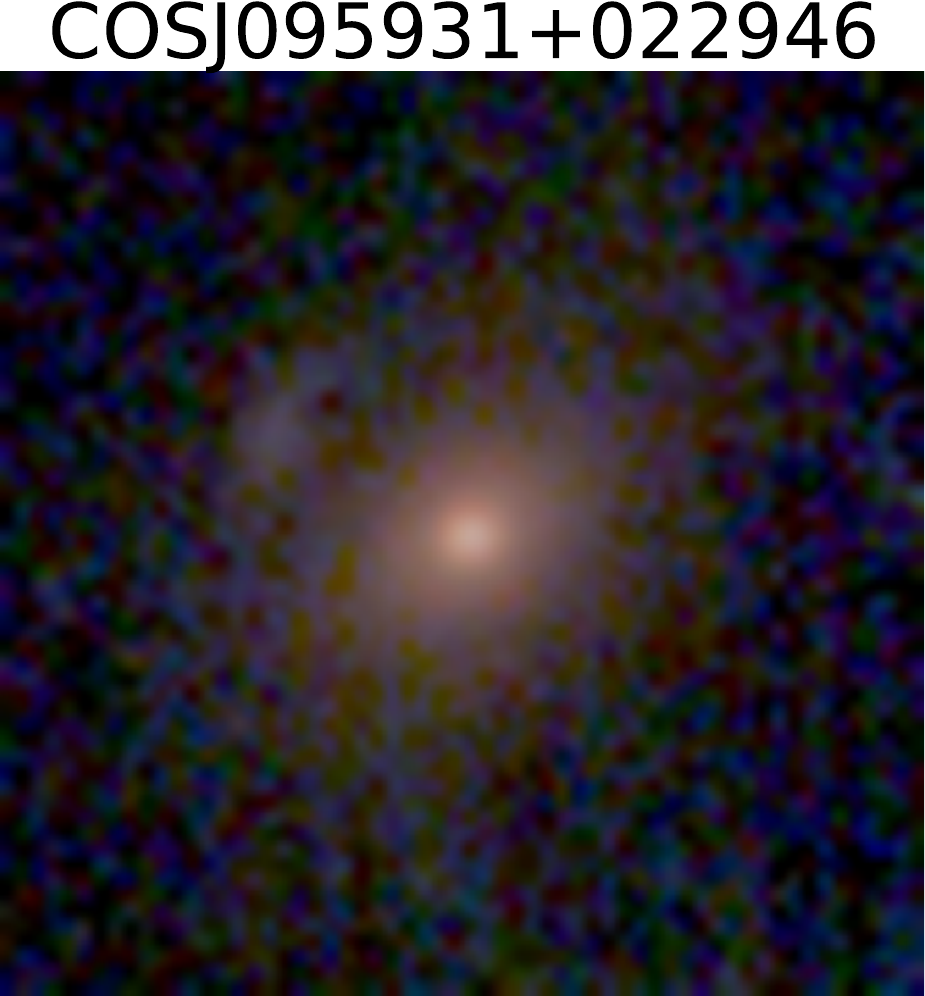}
\includegraphics[width=0.12\textwidth]{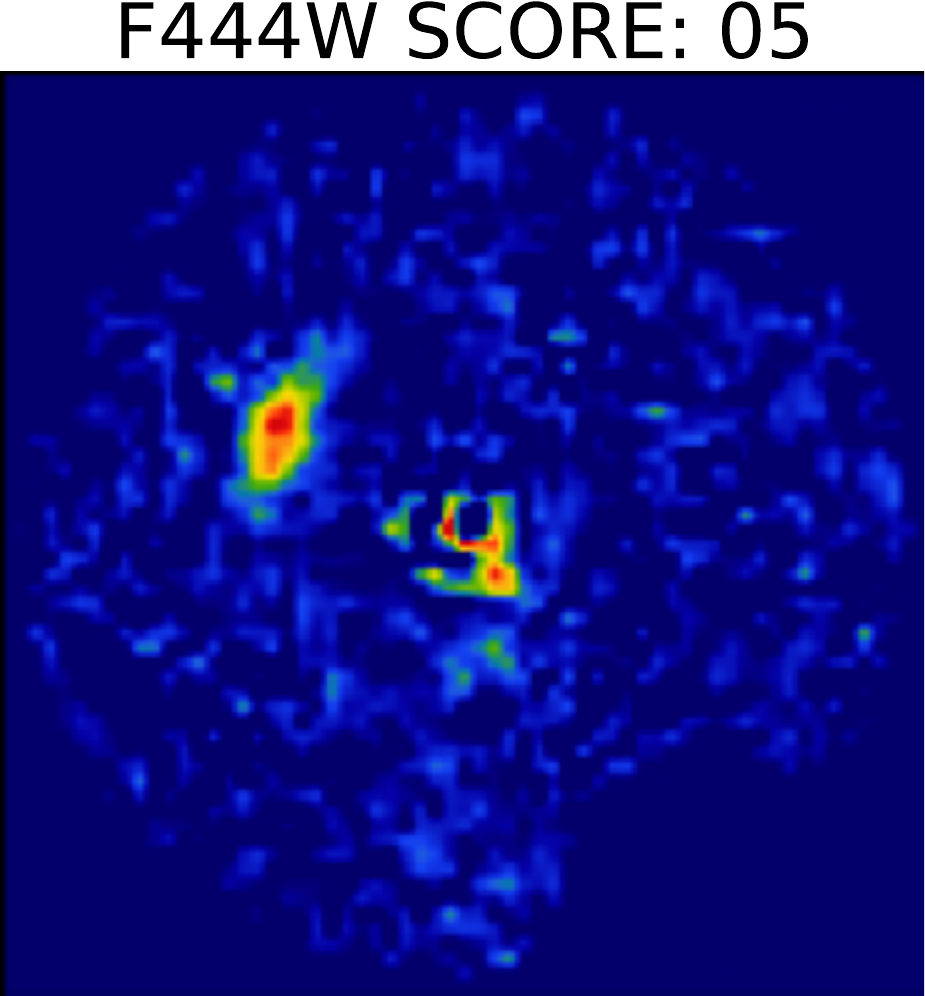}
\includegraphics[width=0.12\textwidth]{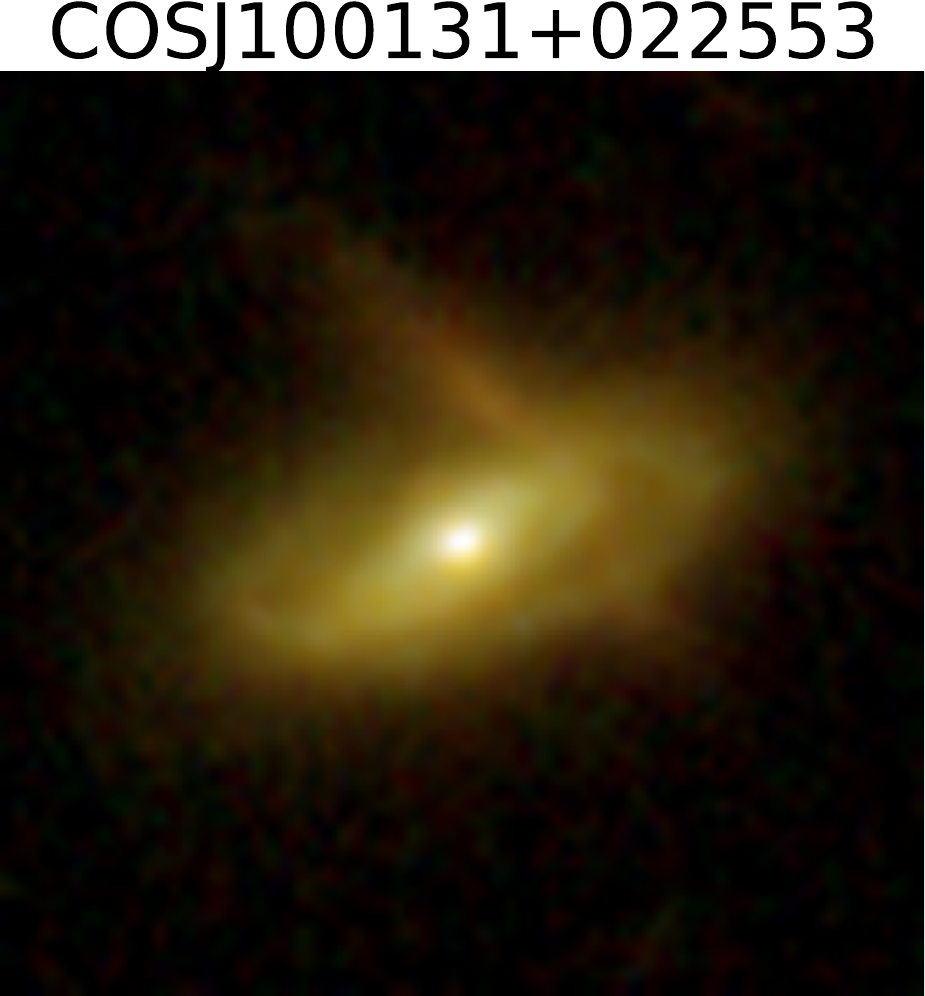}
\includegraphics[width=0.12\textwidth]{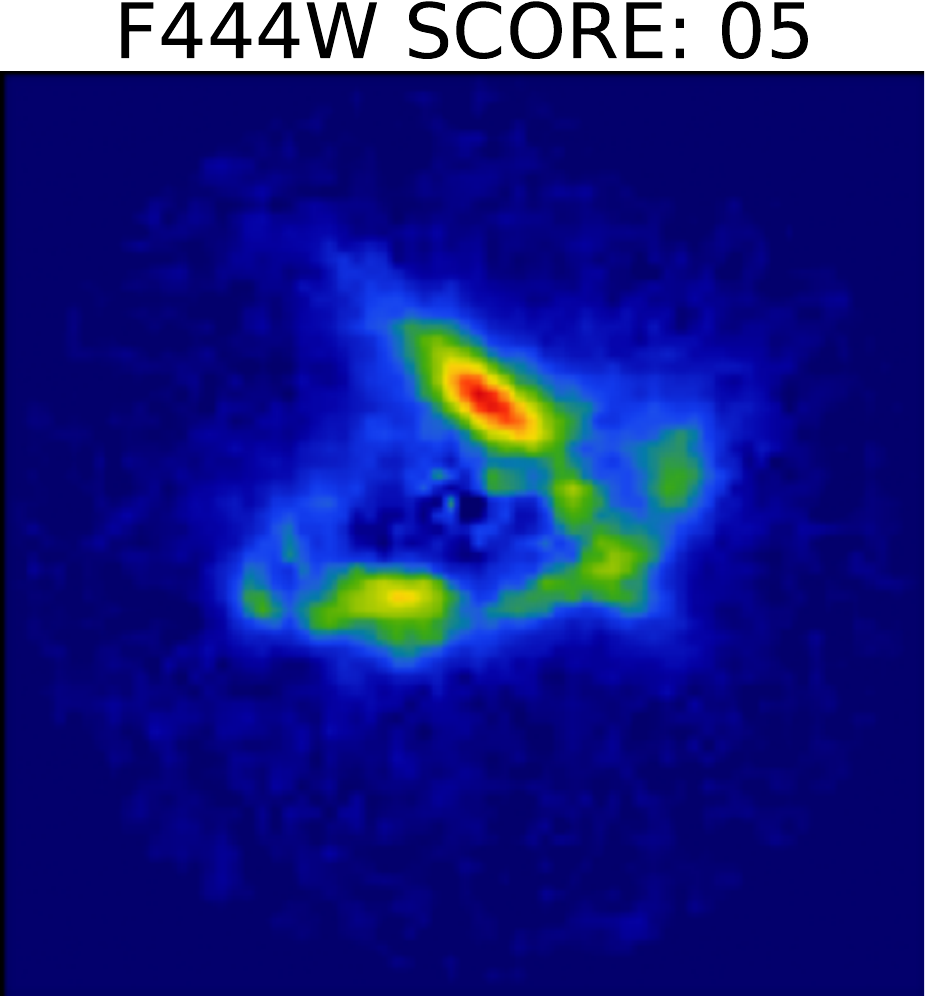}
\includegraphics[width=0.12\textwidth]{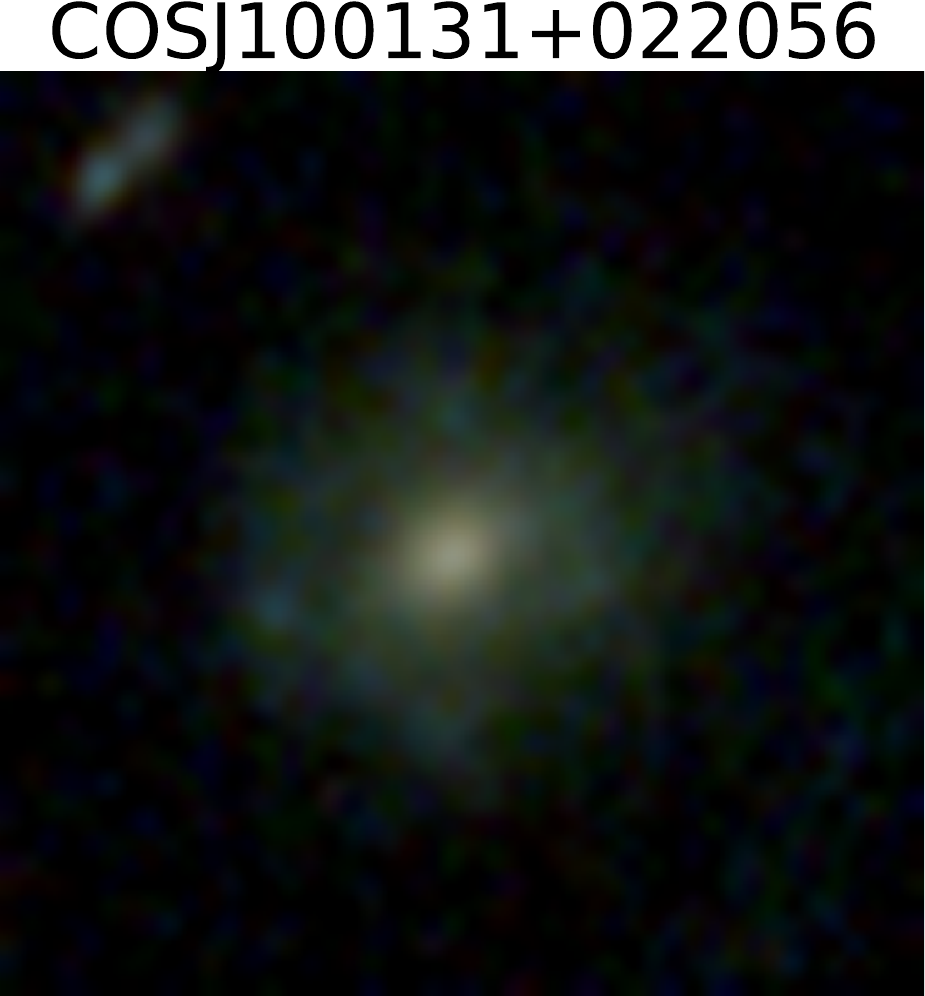}
\includegraphics[width=0.12\textwidth]{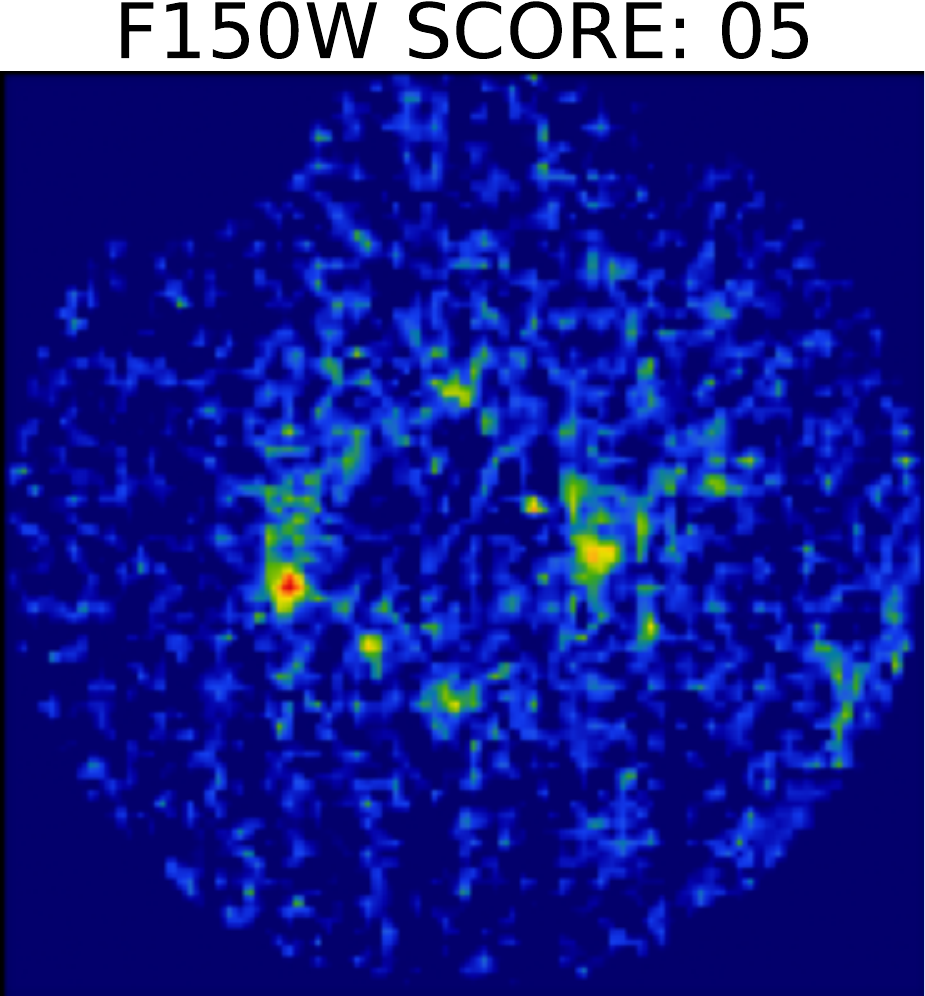}
\includegraphics[width=0.12\textwidth]{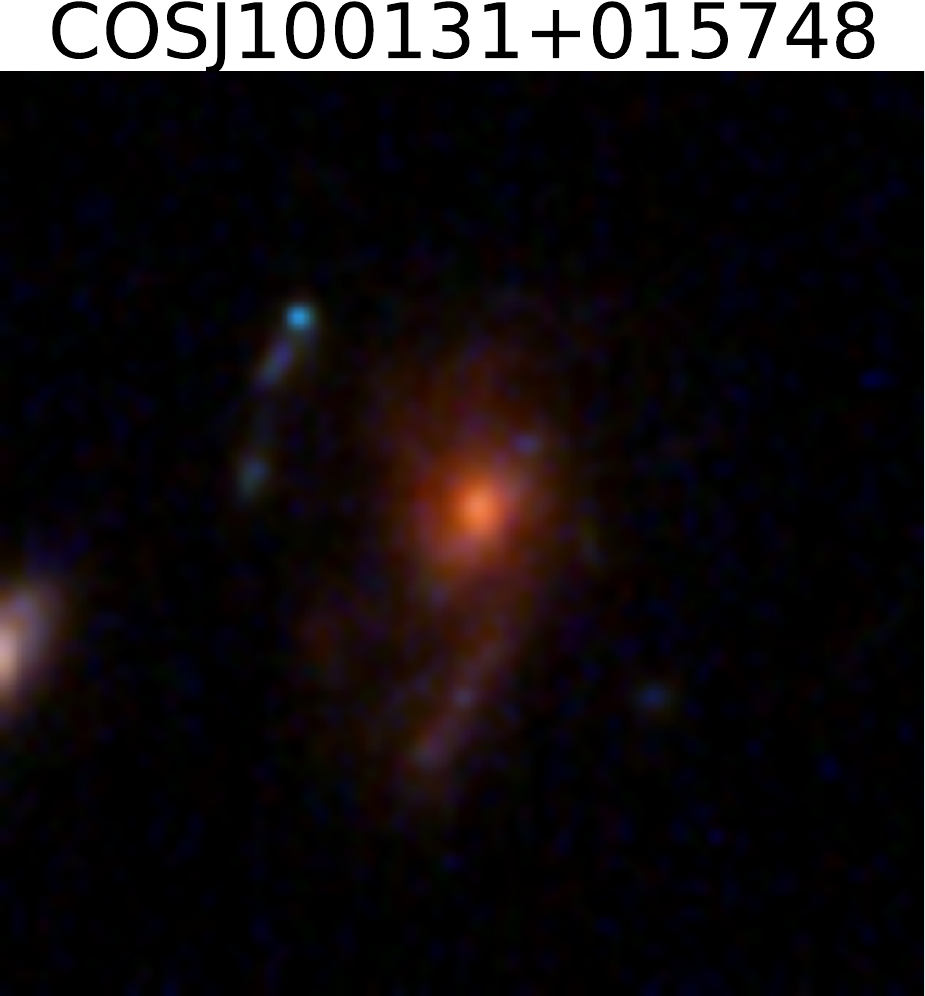}
\includegraphics[width=0.12\textwidth]{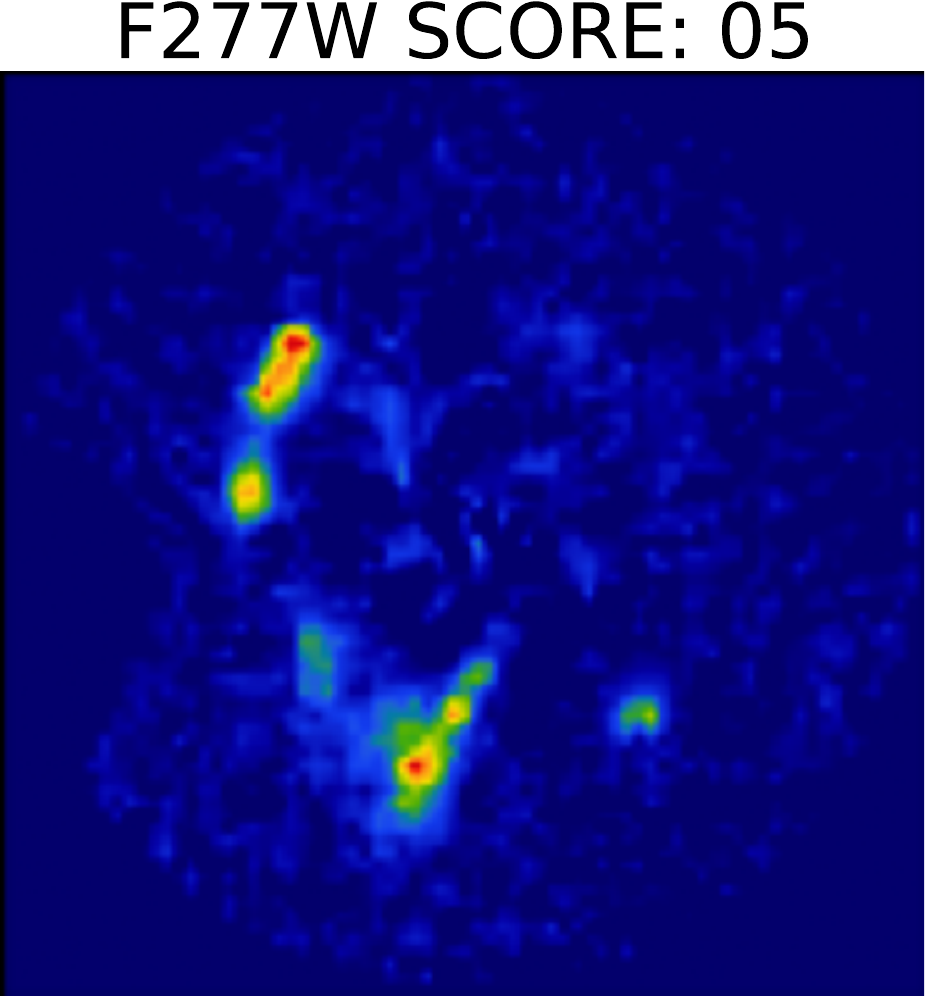}
\includegraphics[width=0.12\textwidth]{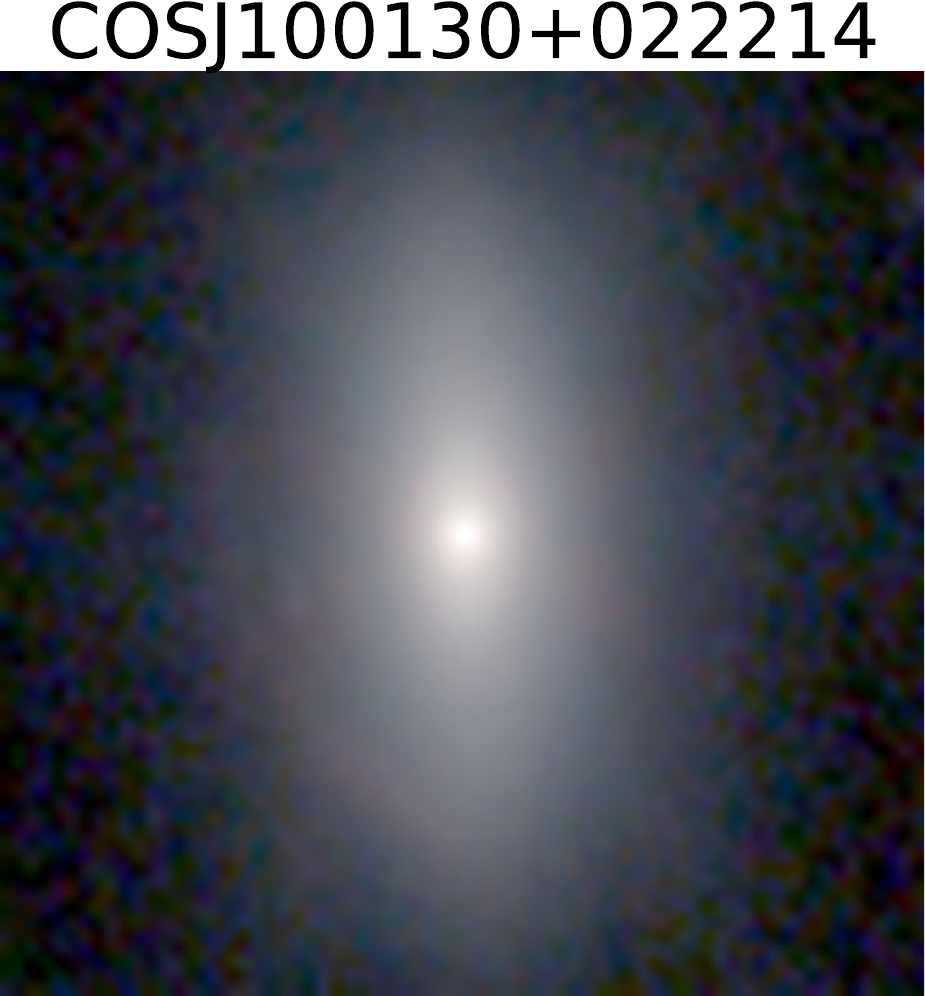}
\includegraphics[width=0.12\textwidth]{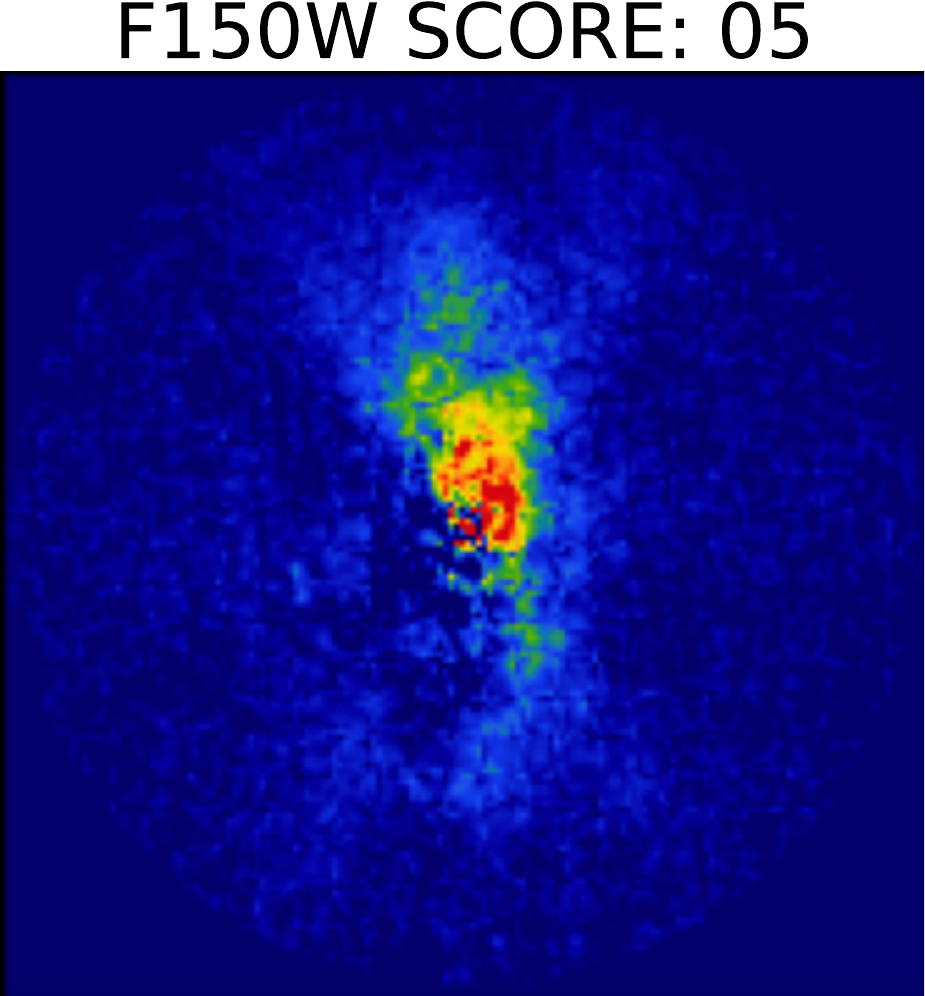}
\includegraphics[width=0.12\textwidth]{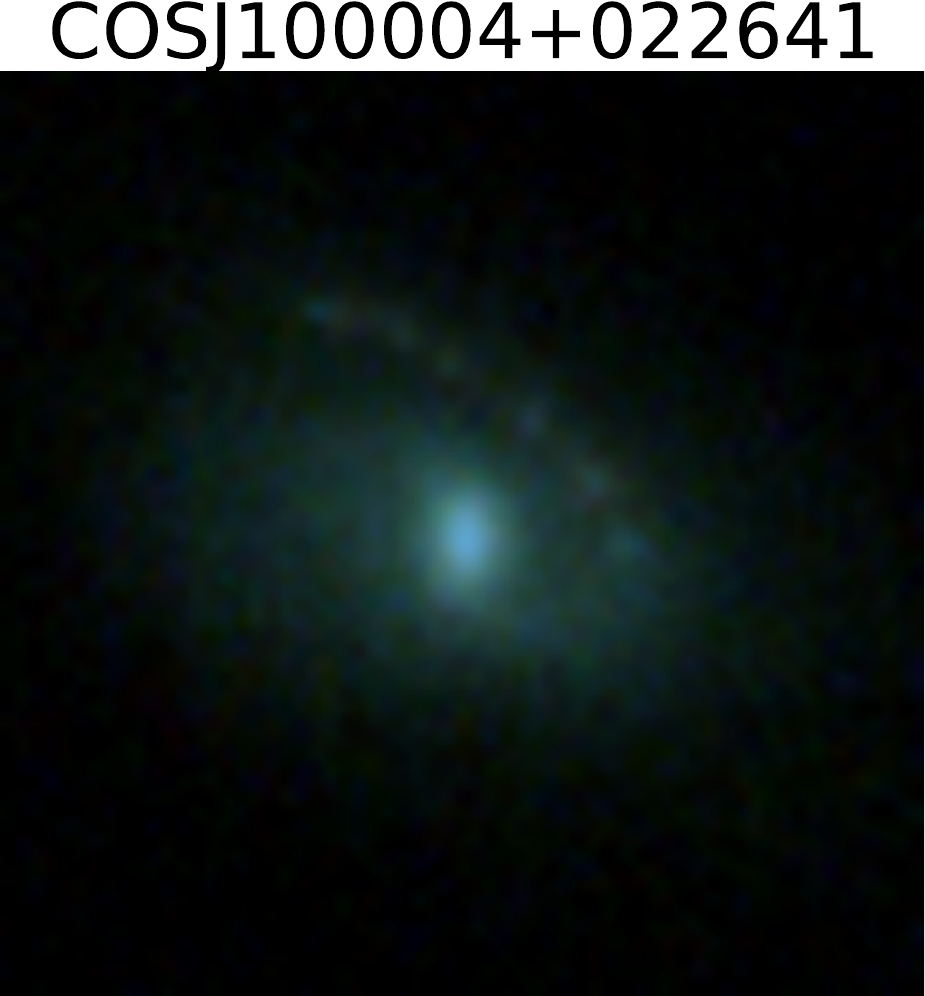}
\includegraphics[width=0.12\textwidth]{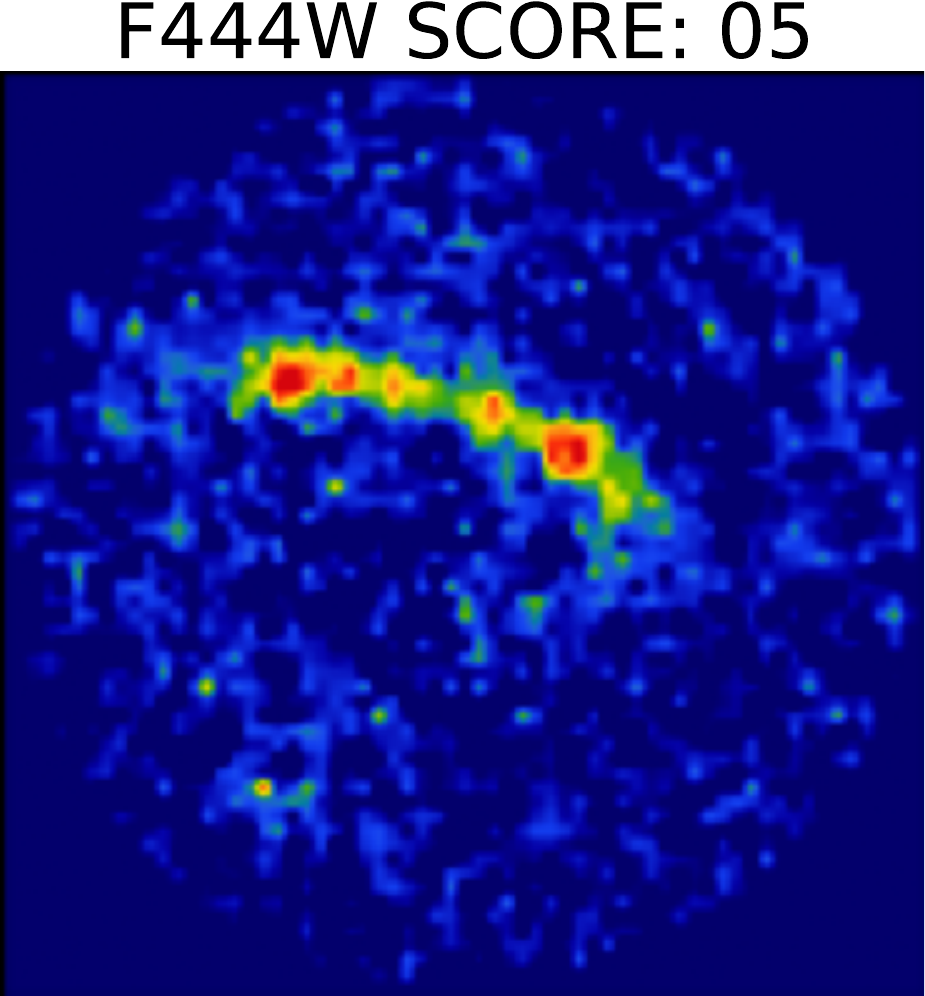}
\includegraphics[width=0.12\textwidth]{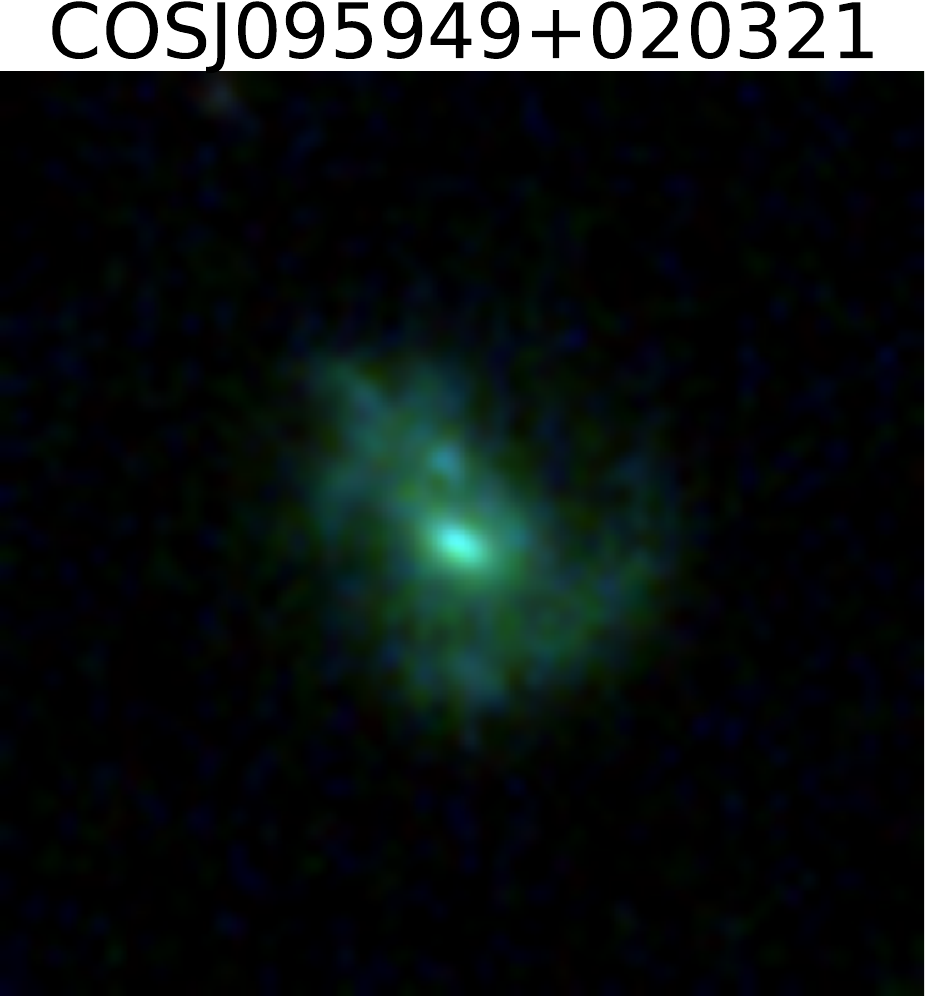}
\includegraphics[width=0.12\textwidth]{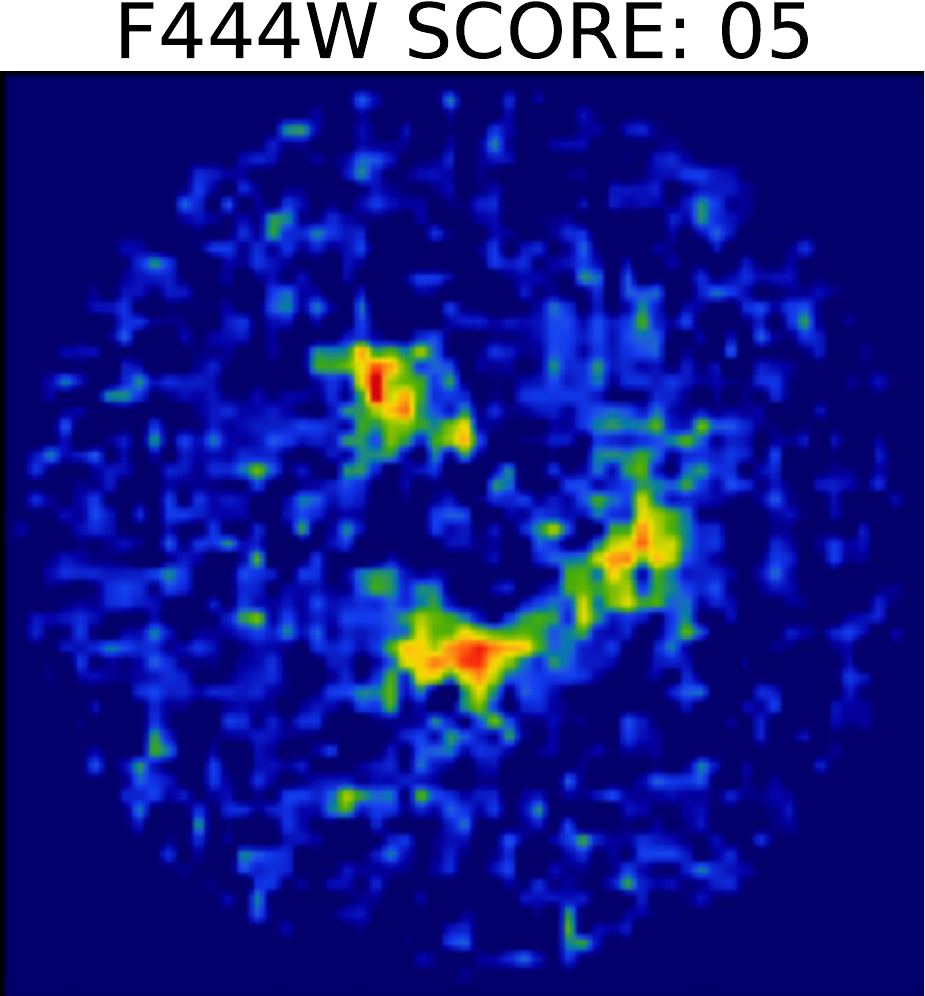}
\includegraphics[width=0.12\textwidth]{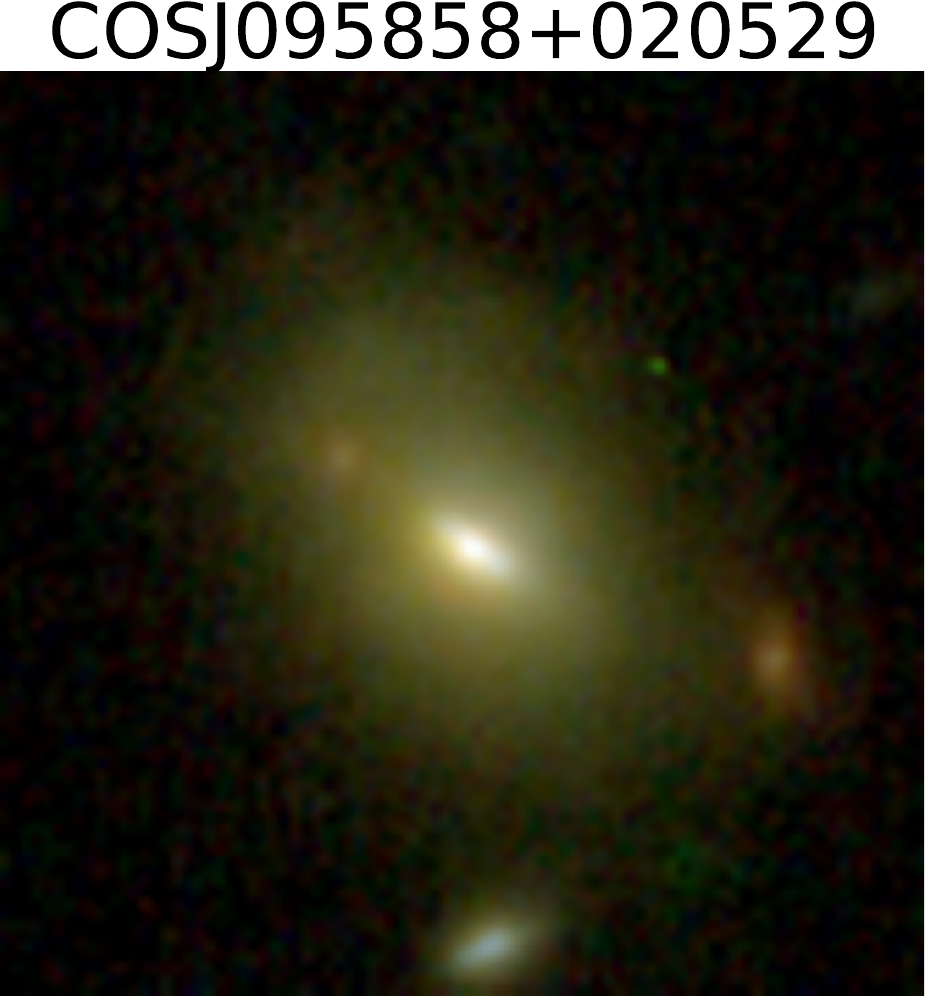}
\includegraphics[width=0.12\textwidth]{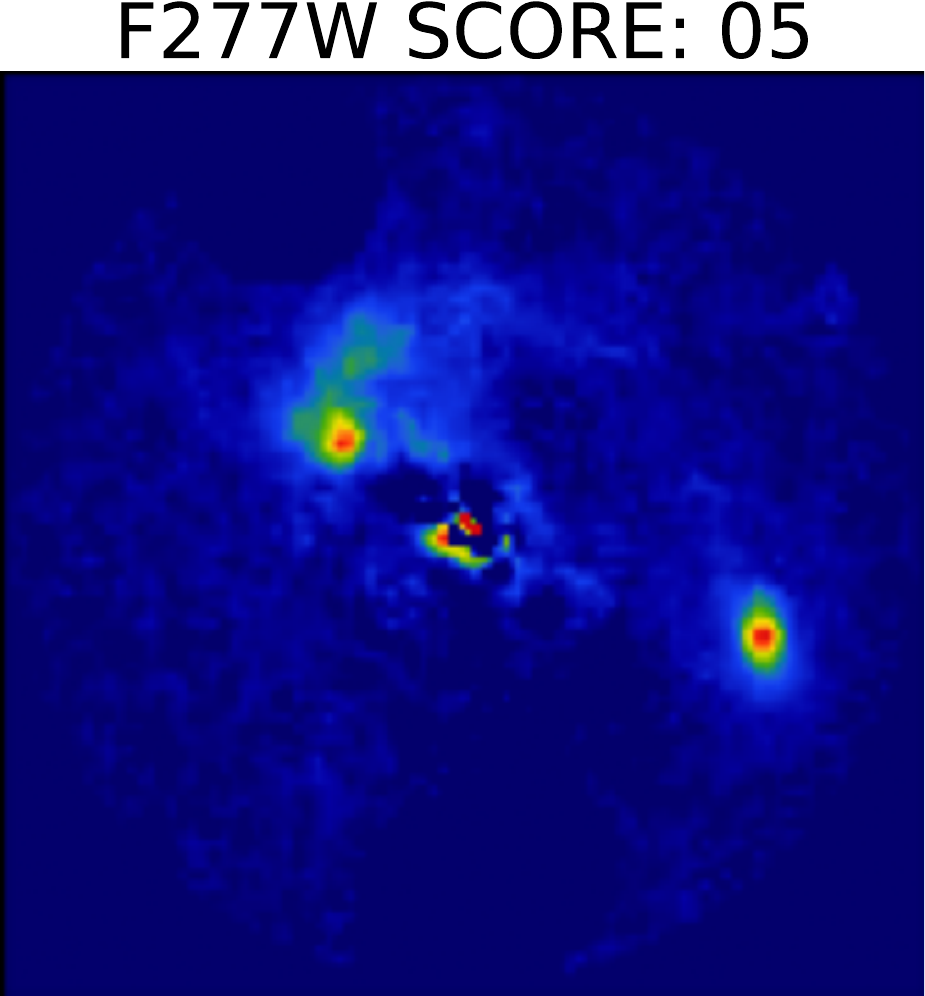}
\includegraphics[width=0.12\textwidth]{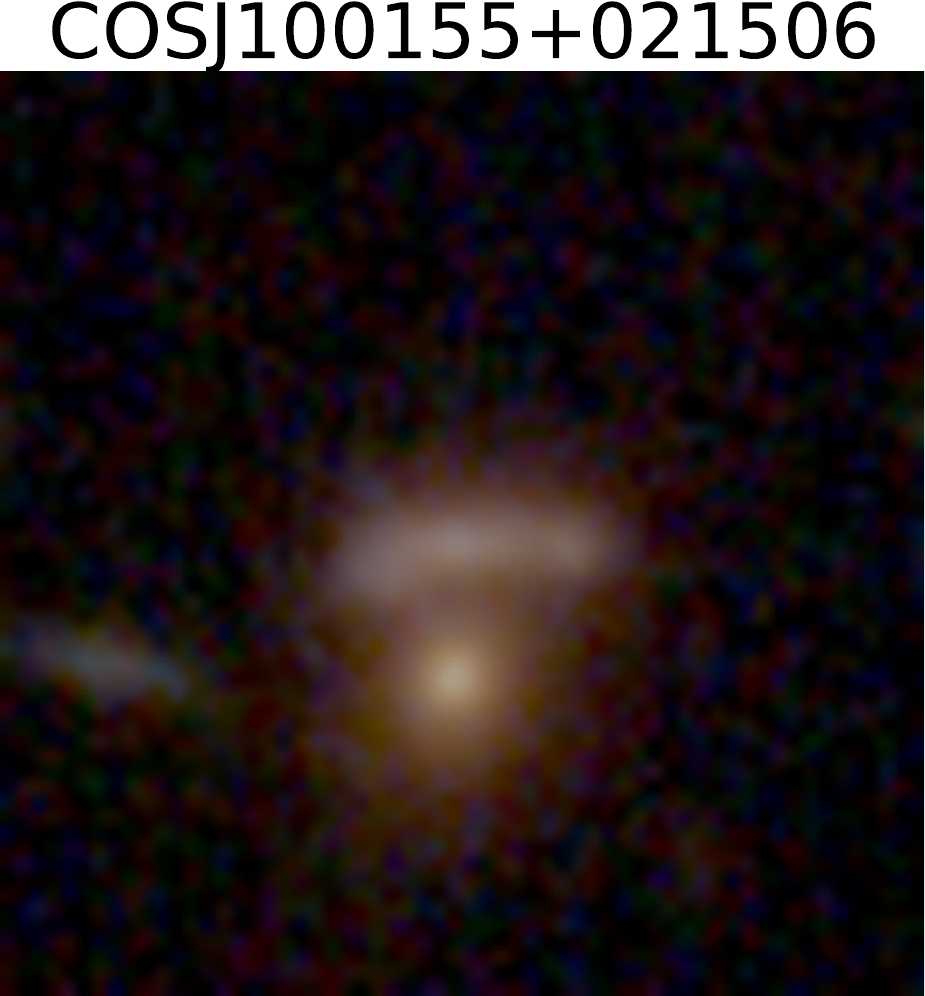}
\includegraphics[width=0.12\textwidth]{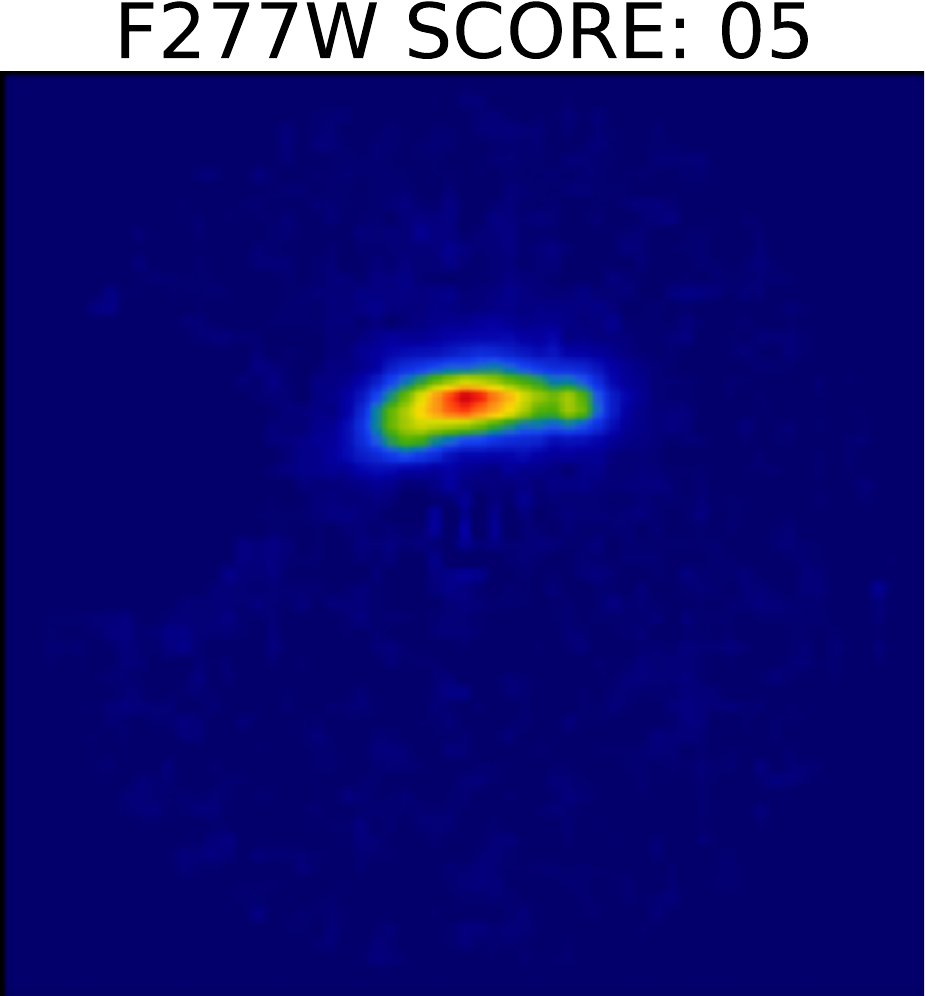}
\includegraphics[width=0.12\textwidth]{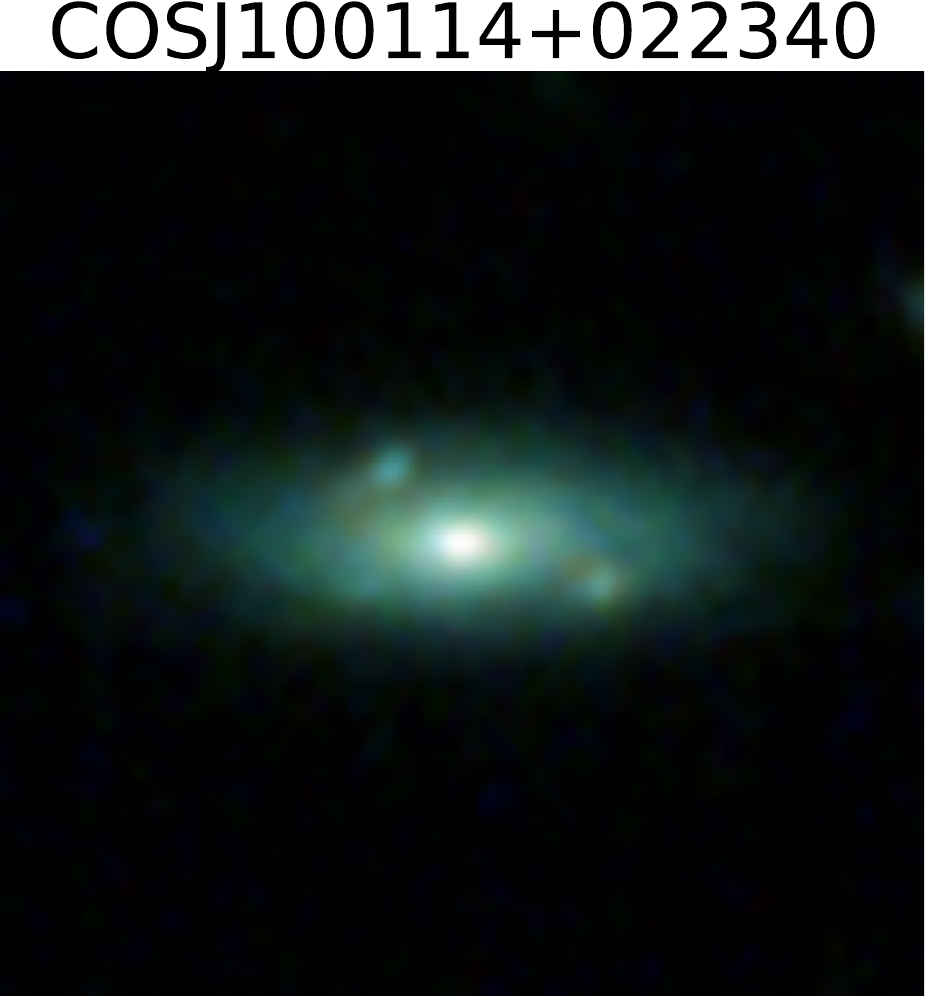}
\includegraphics[width=0.12\textwidth]{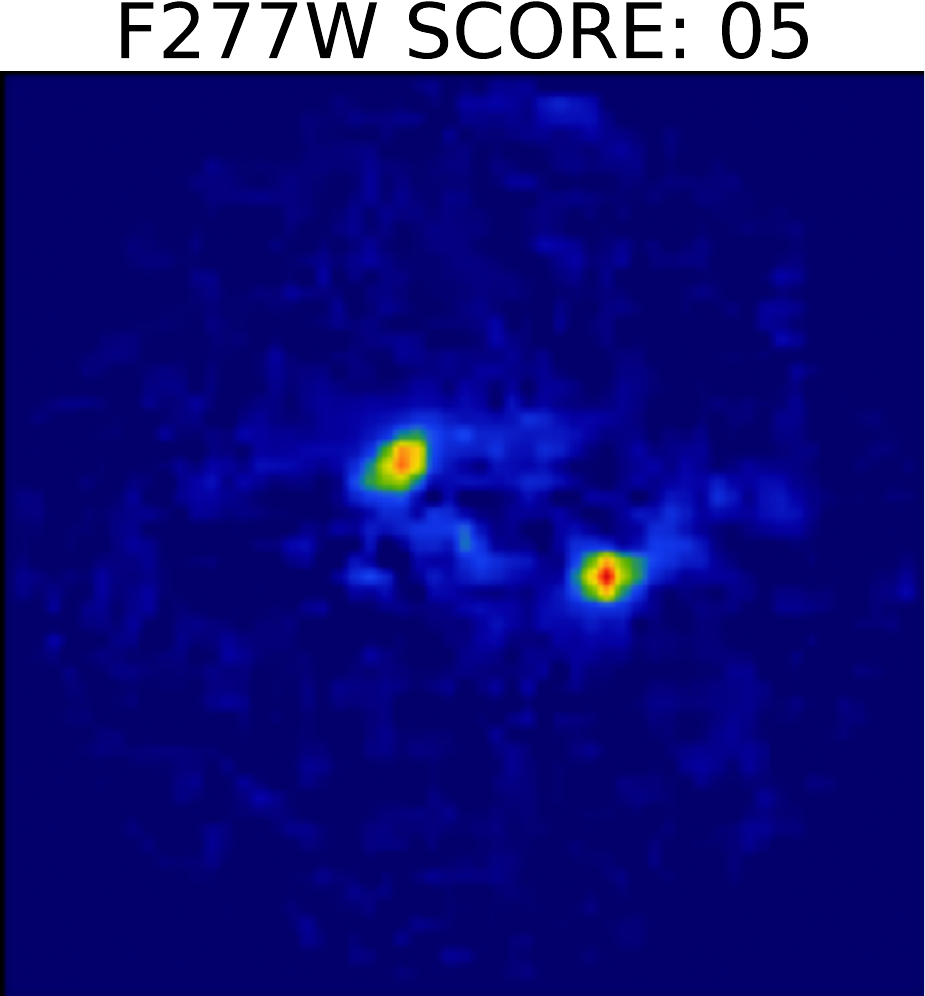}
\includegraphics[width=0.12\textwidth]{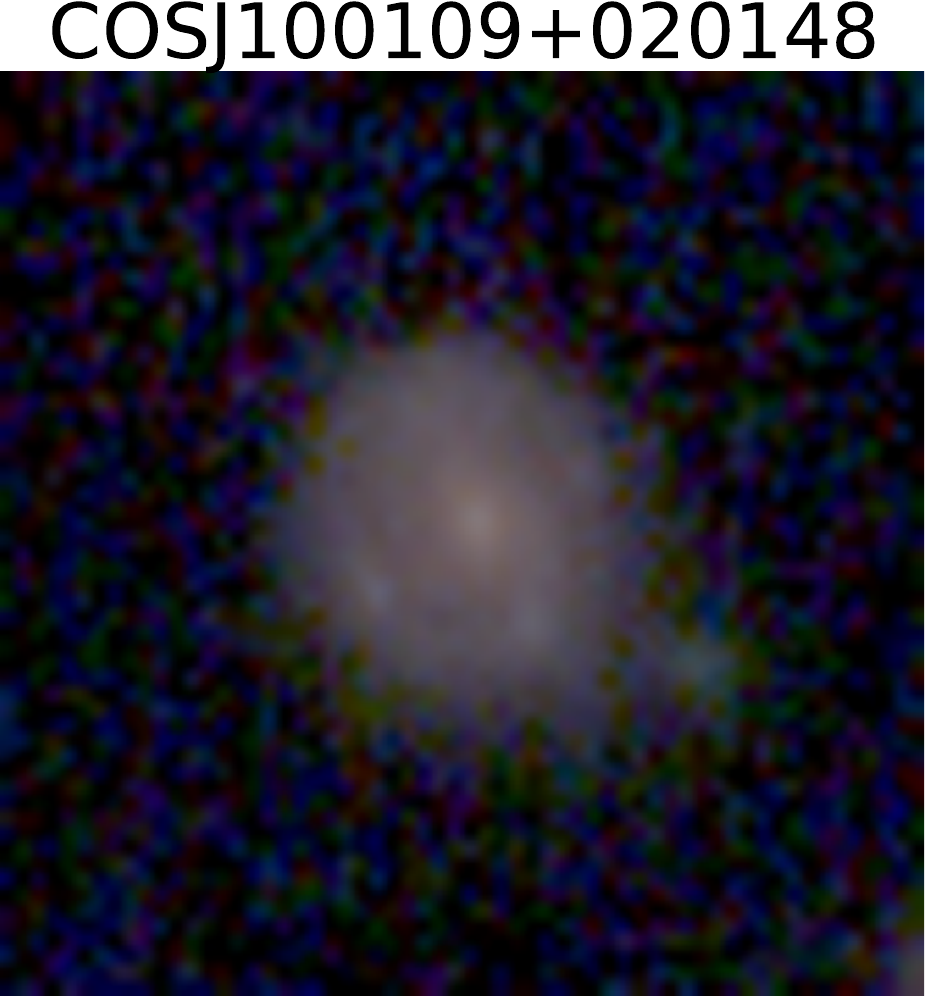}
\includegraphics[width=0.12\textwidth]{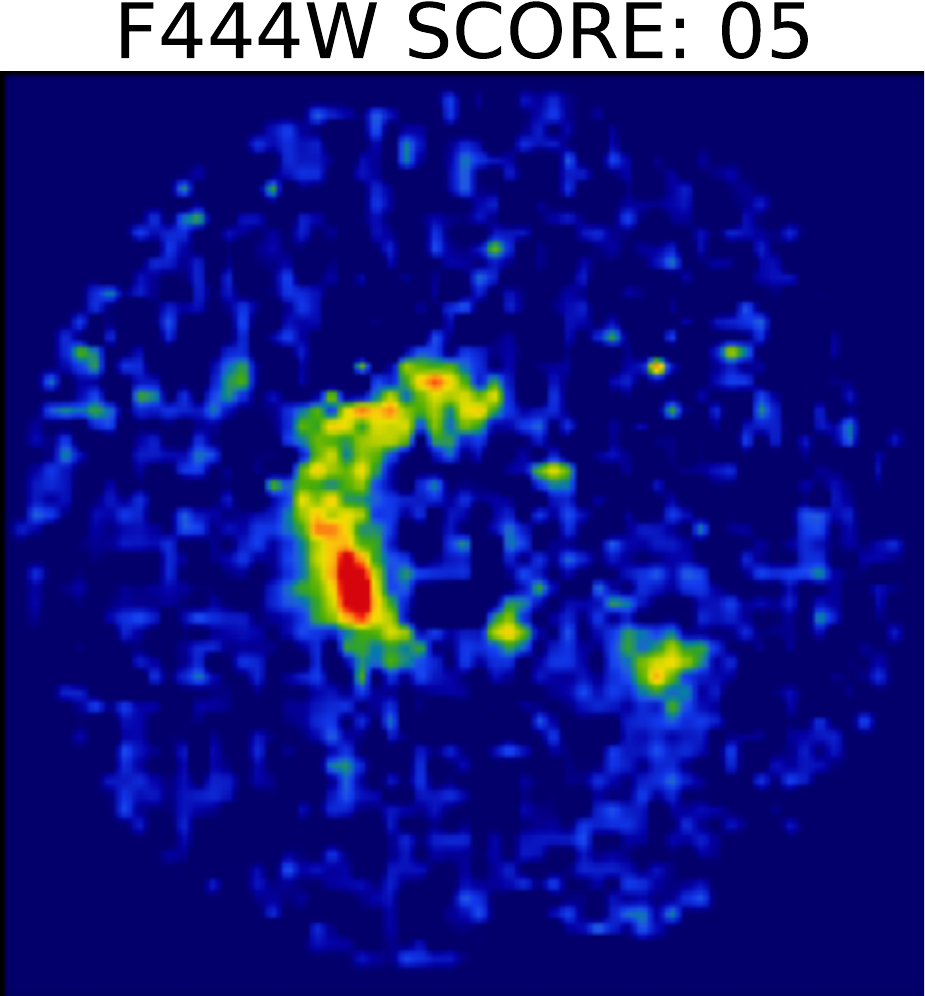}
\includegraphics[width=0.12\textwidth]{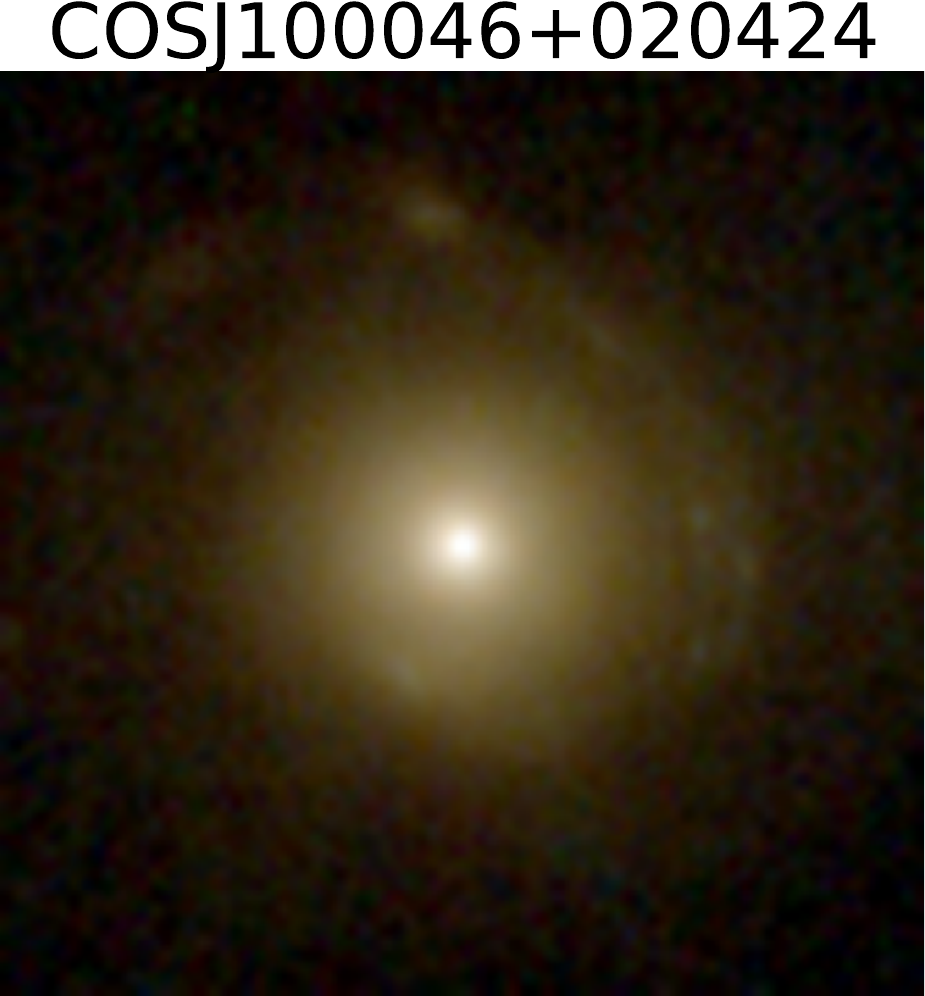}
\includegraphics[width=0.12\textwidth]{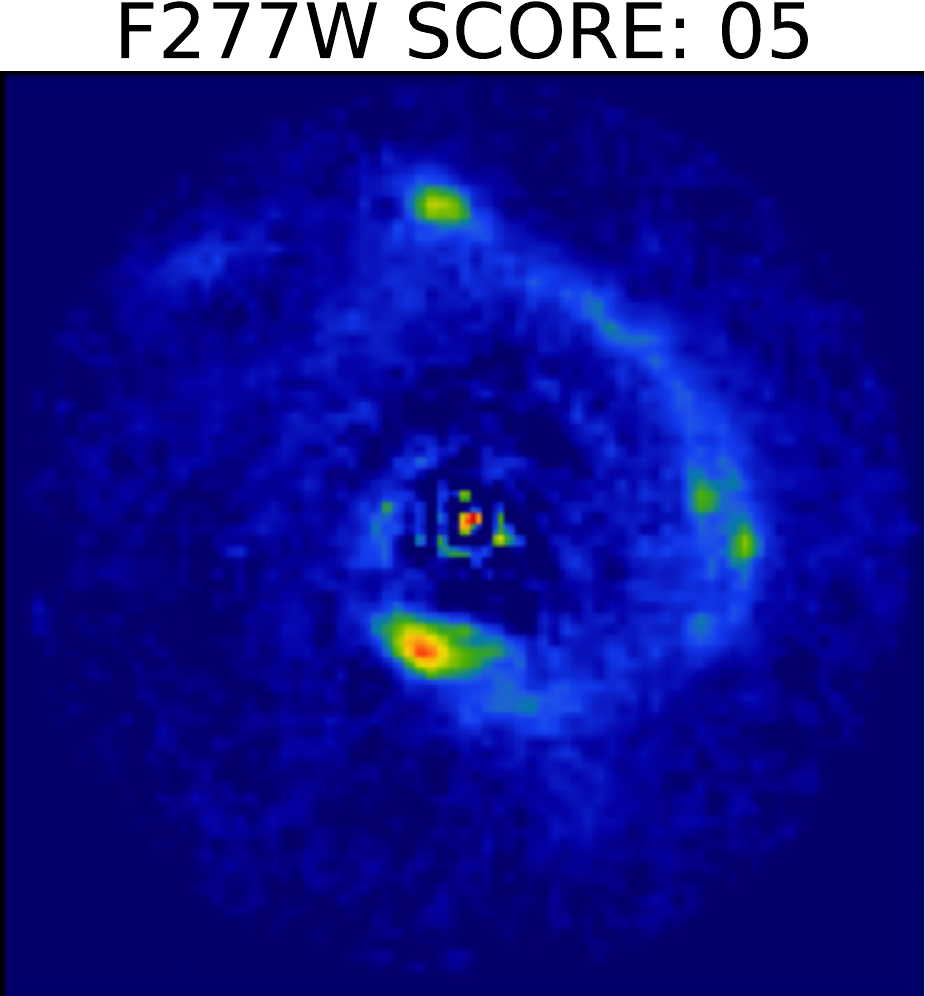}
\includegraphics[width=0.12\textwidth]{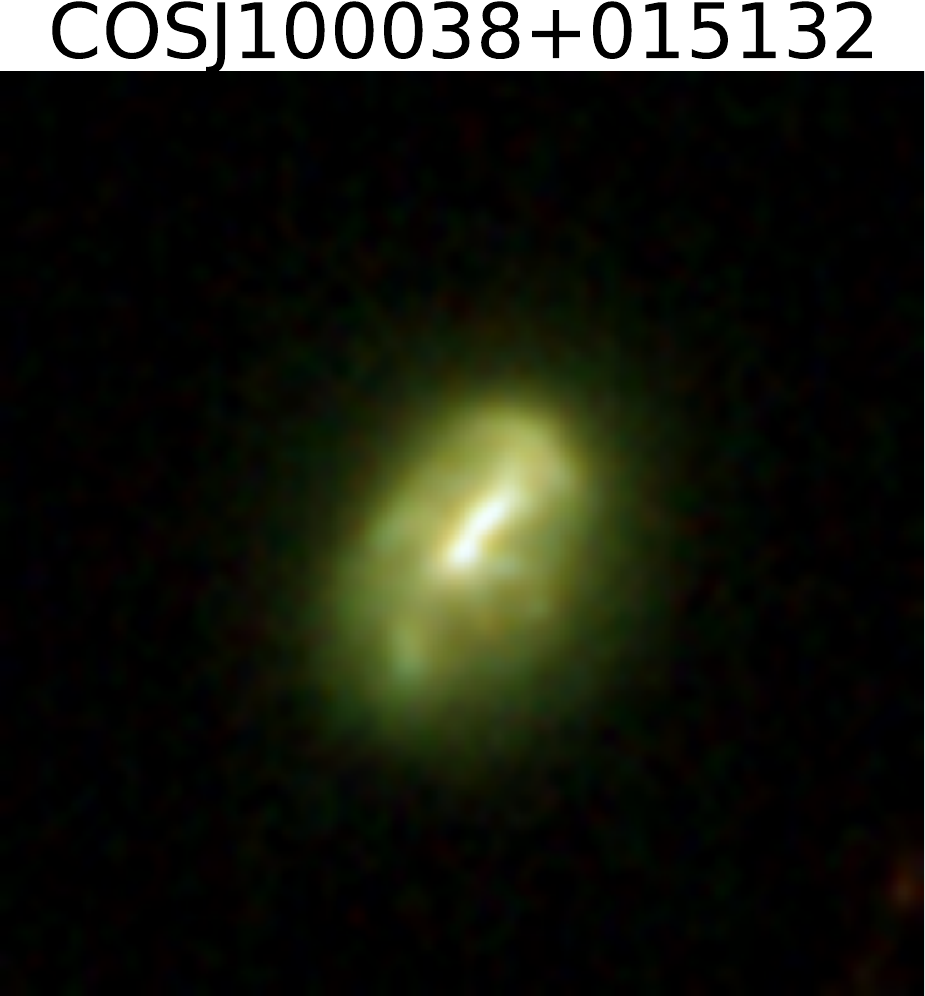}
\includegraphics[width=0.12\textwidth]{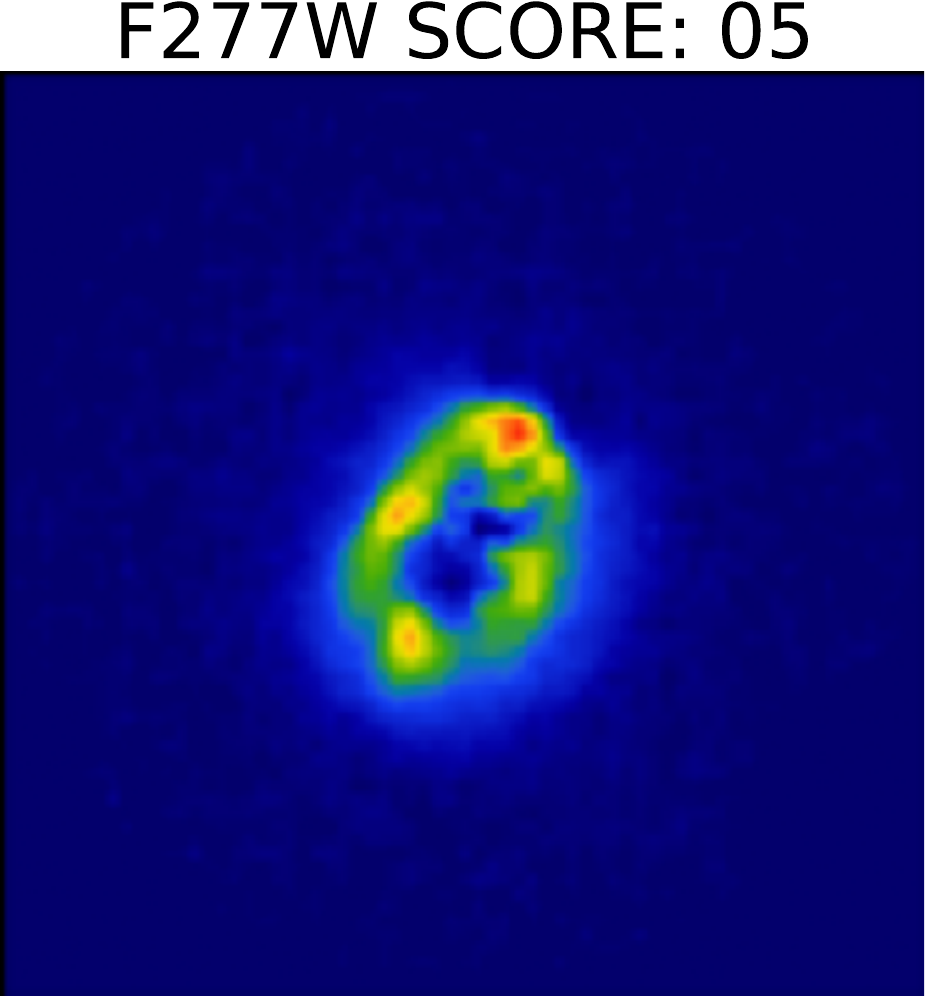}
\includegraphics[width=0.12\textwidth]{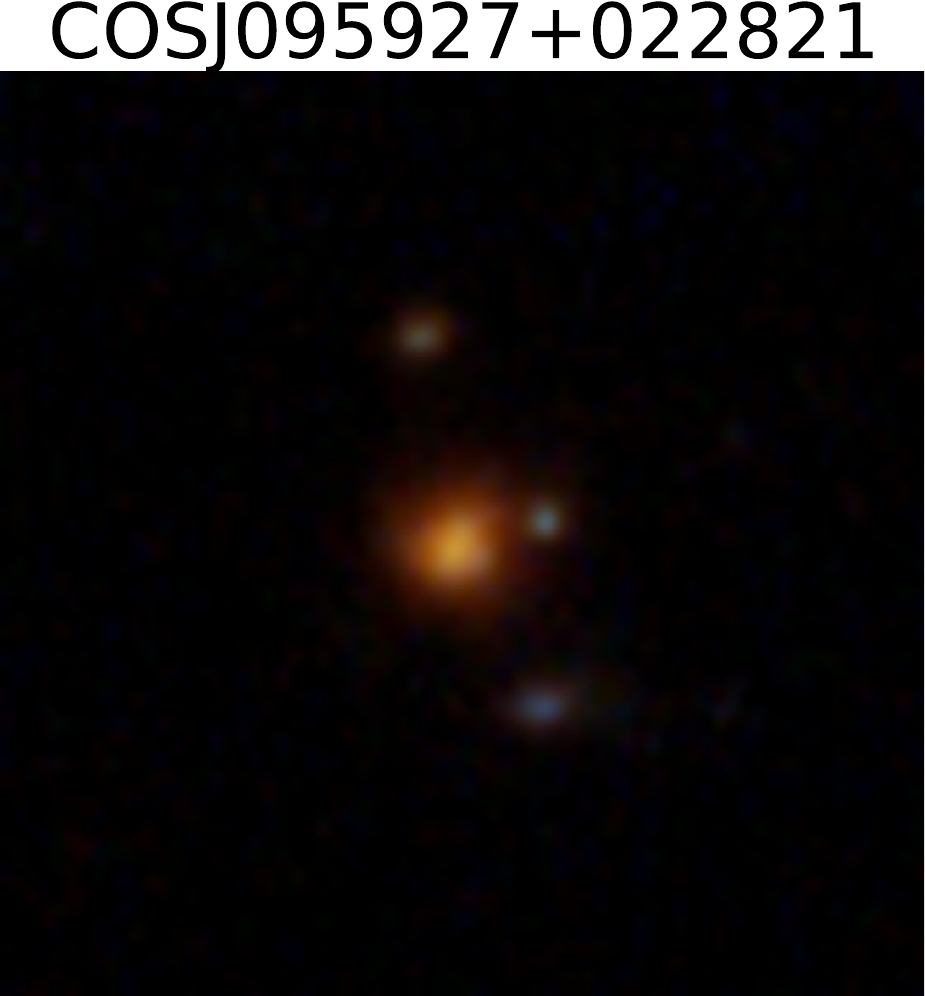}
\includegraphics[width=0.12\textwidth]{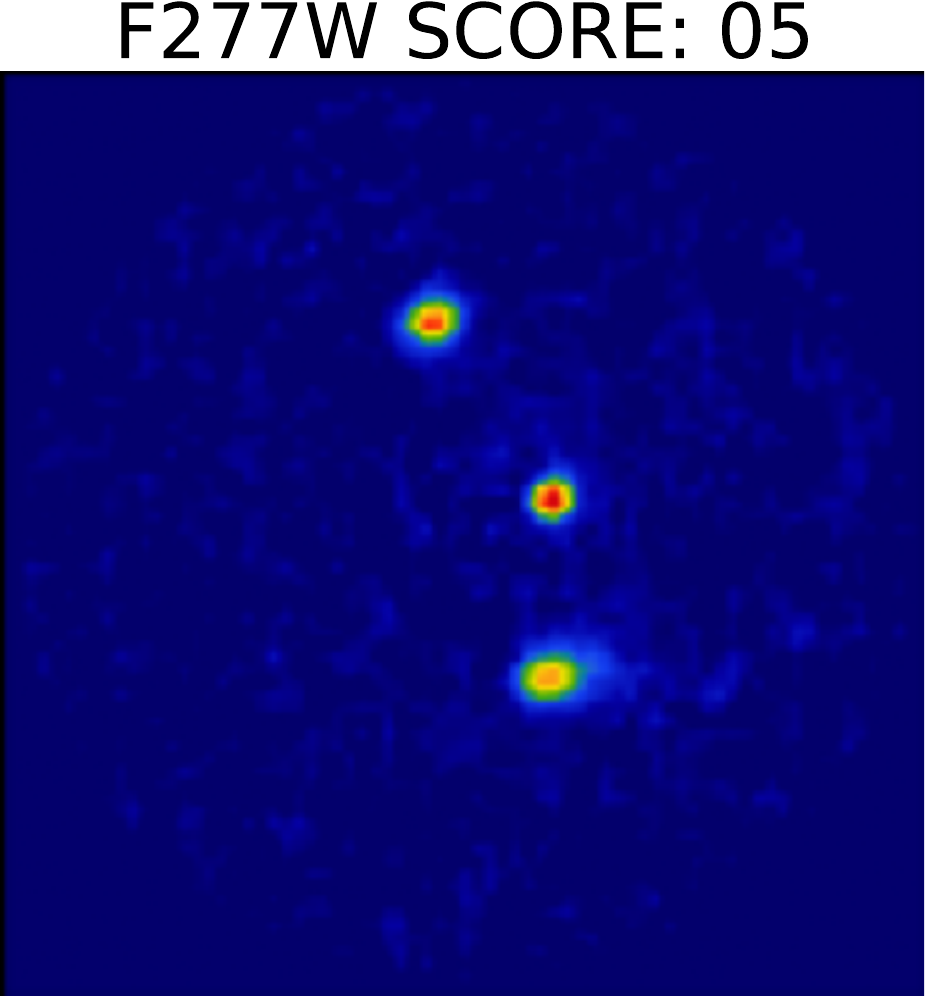}
\includegraphics[width=0.12\textwidth]{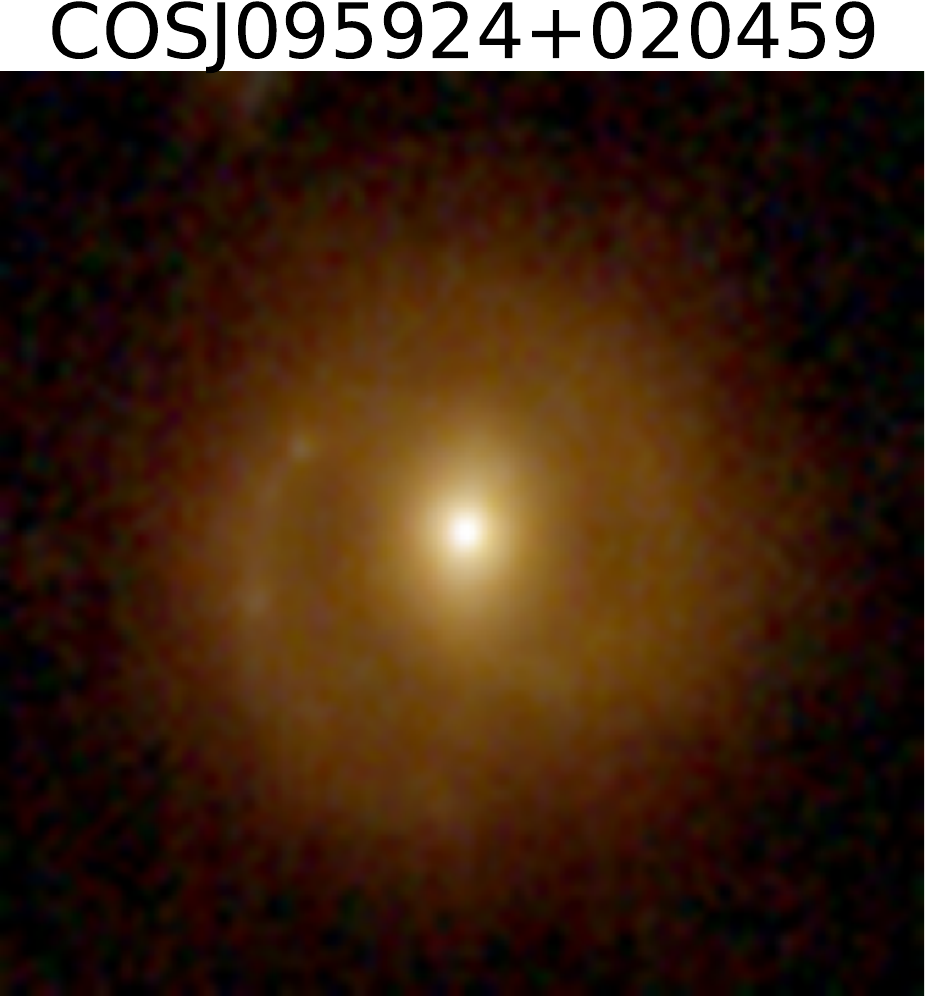}
\includegraphics[width=0.12\textwidth]{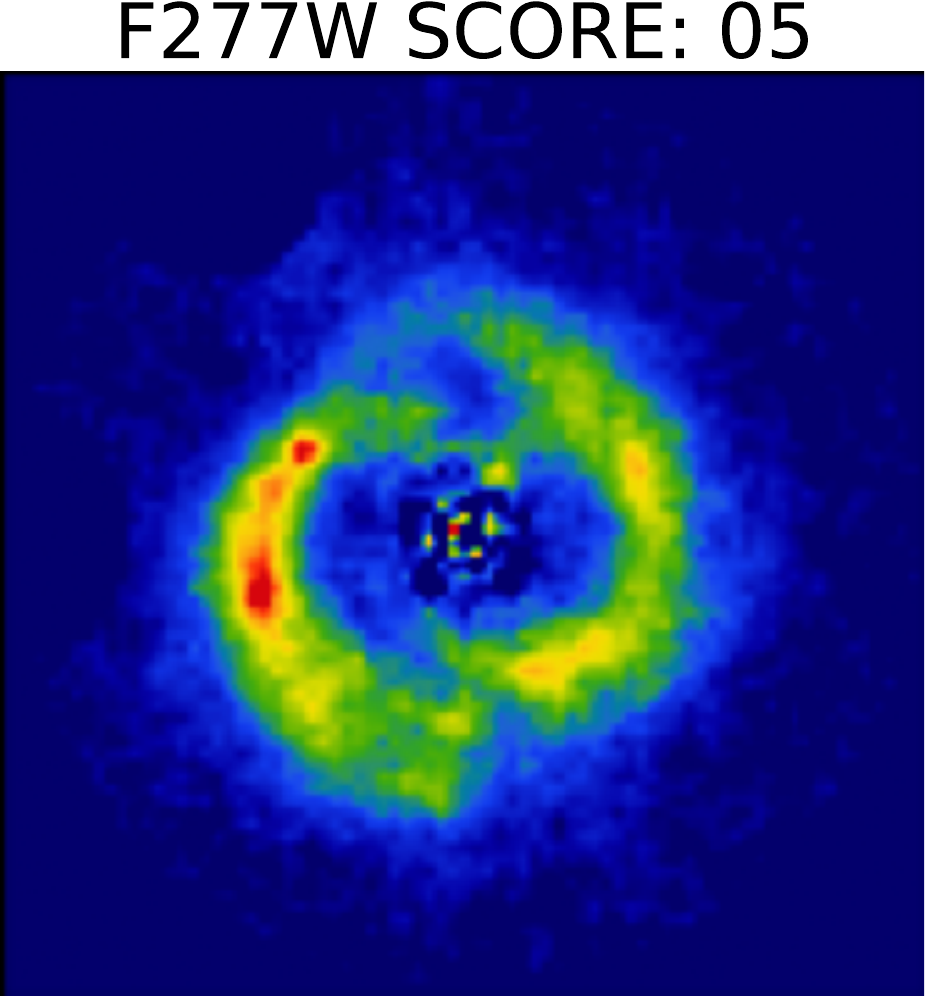}
\includegraphics[width=0.12\textwidth]{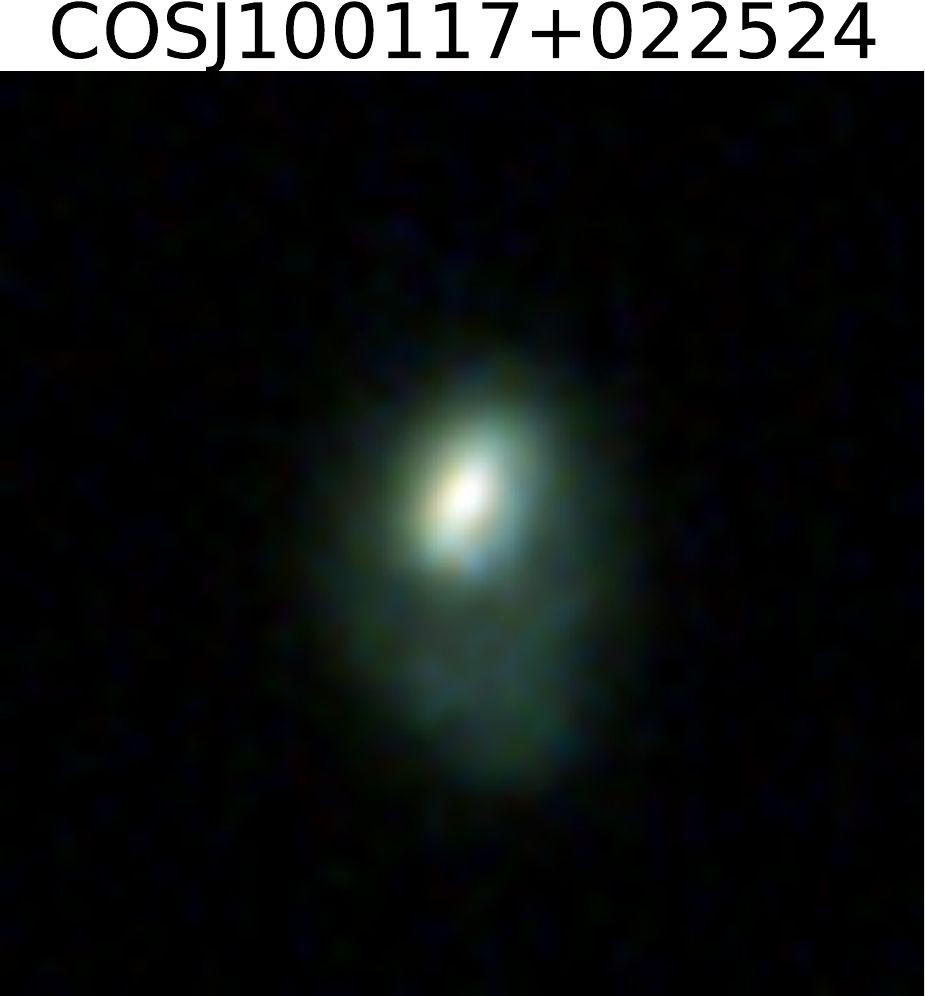}
\includegraphics[width=0.12\textwidth]{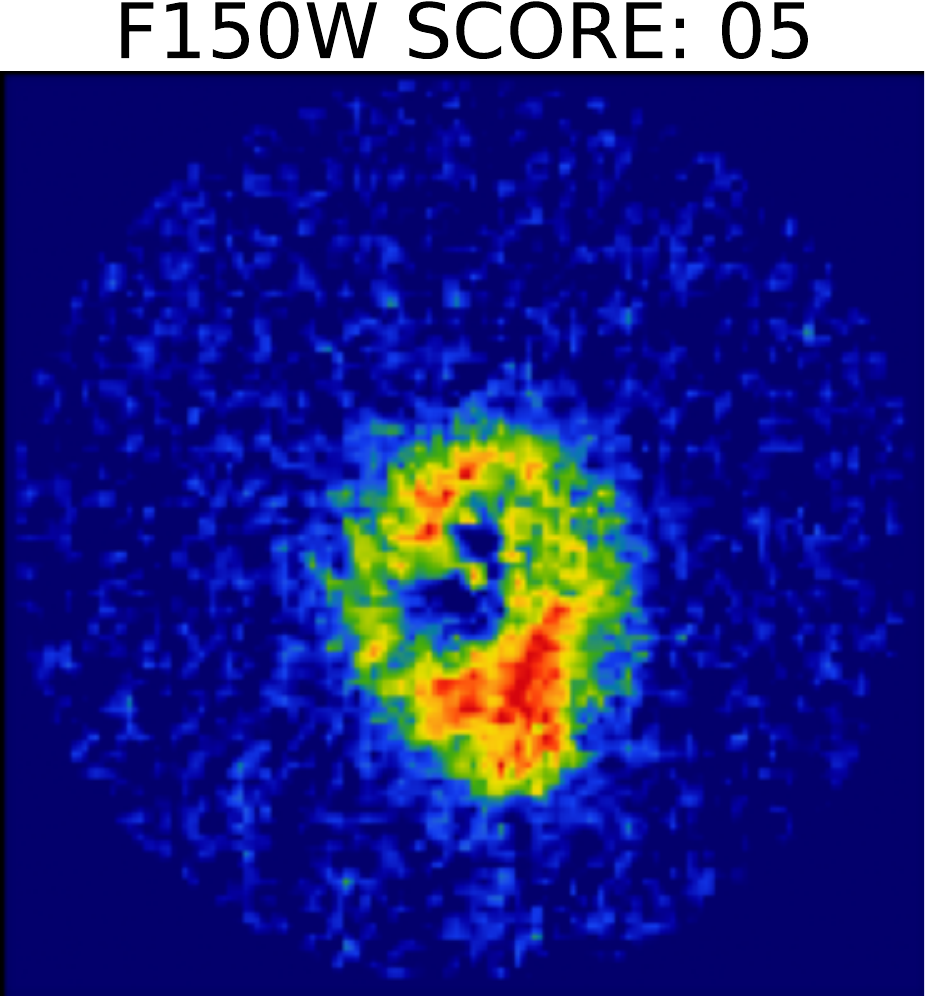}
\includegraphics[width=0.12\textwidth]{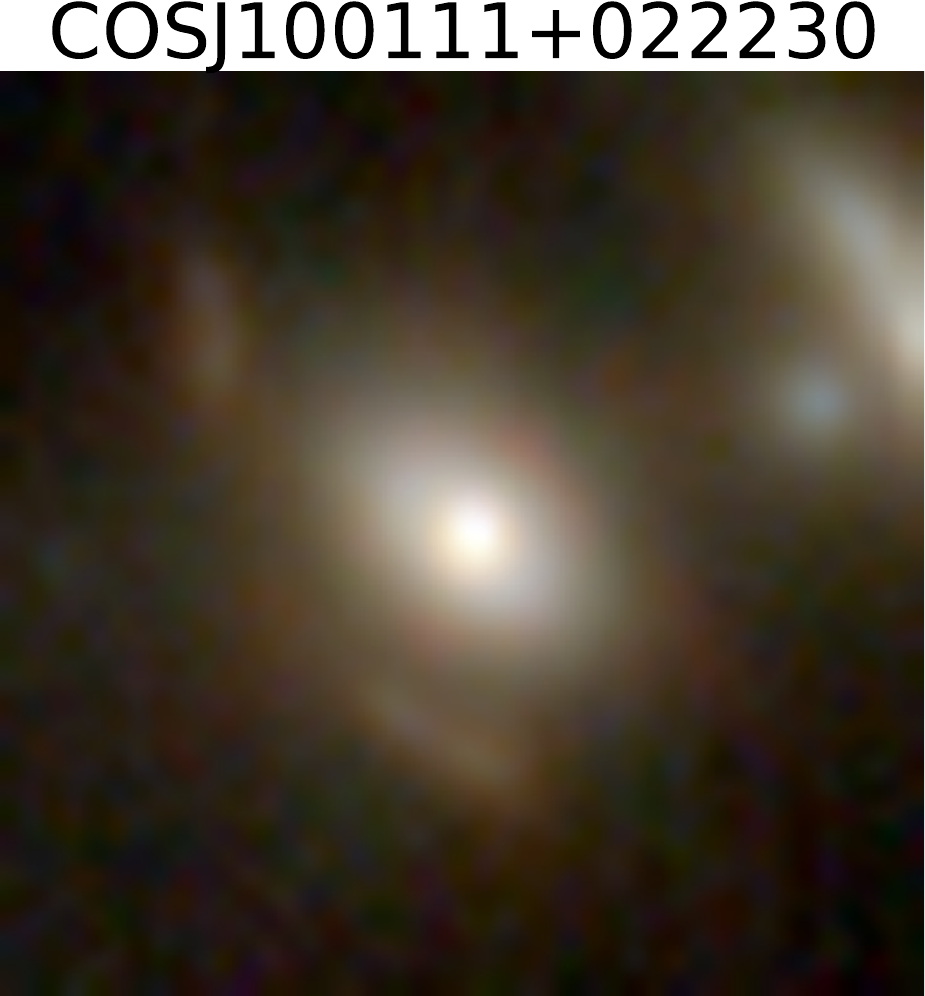}
\includegraphics[width=0.12\textwidth]{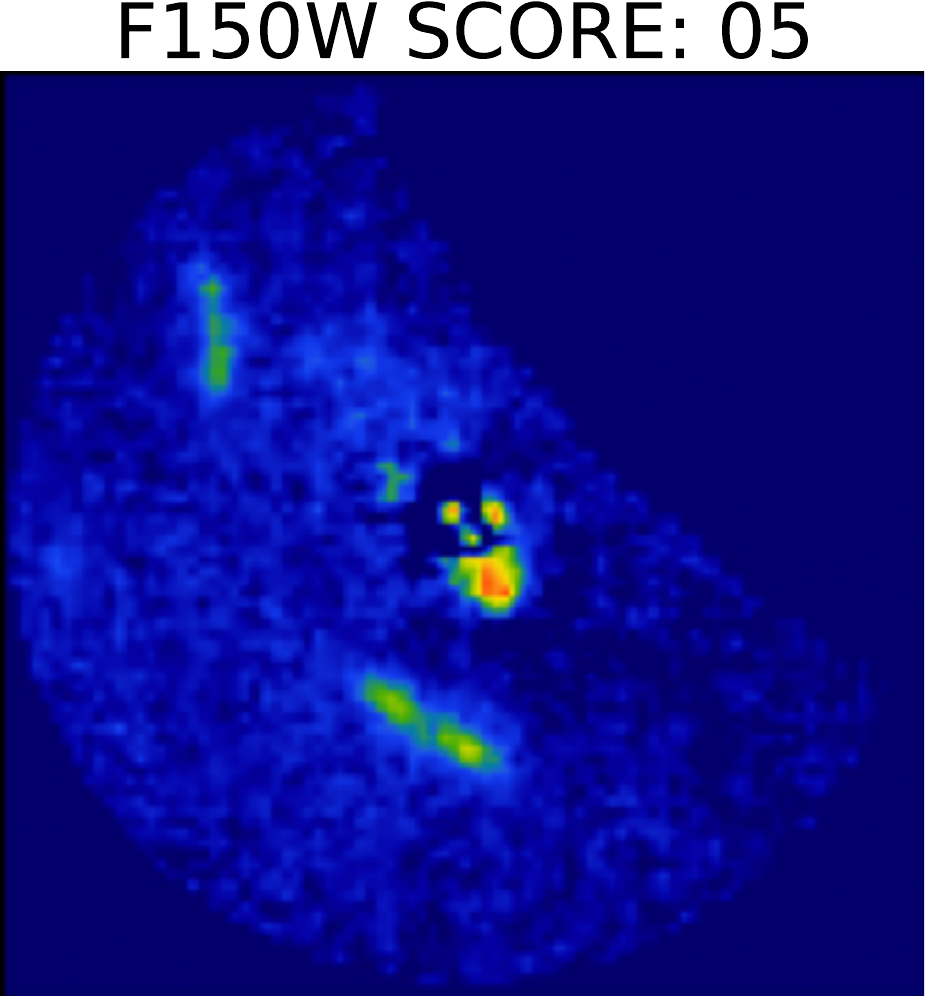}
\includegraphics[width=0.12\textwidth]{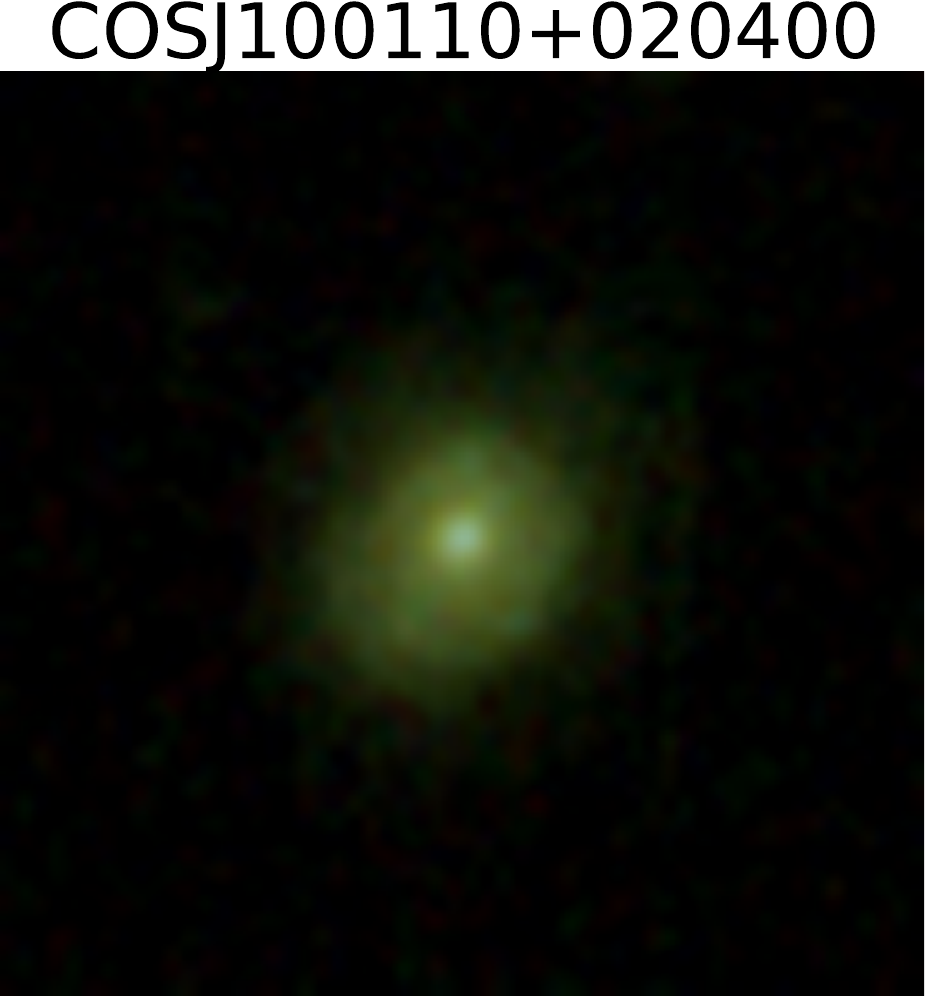}
\includegraphics[width=0.12\textwidth]{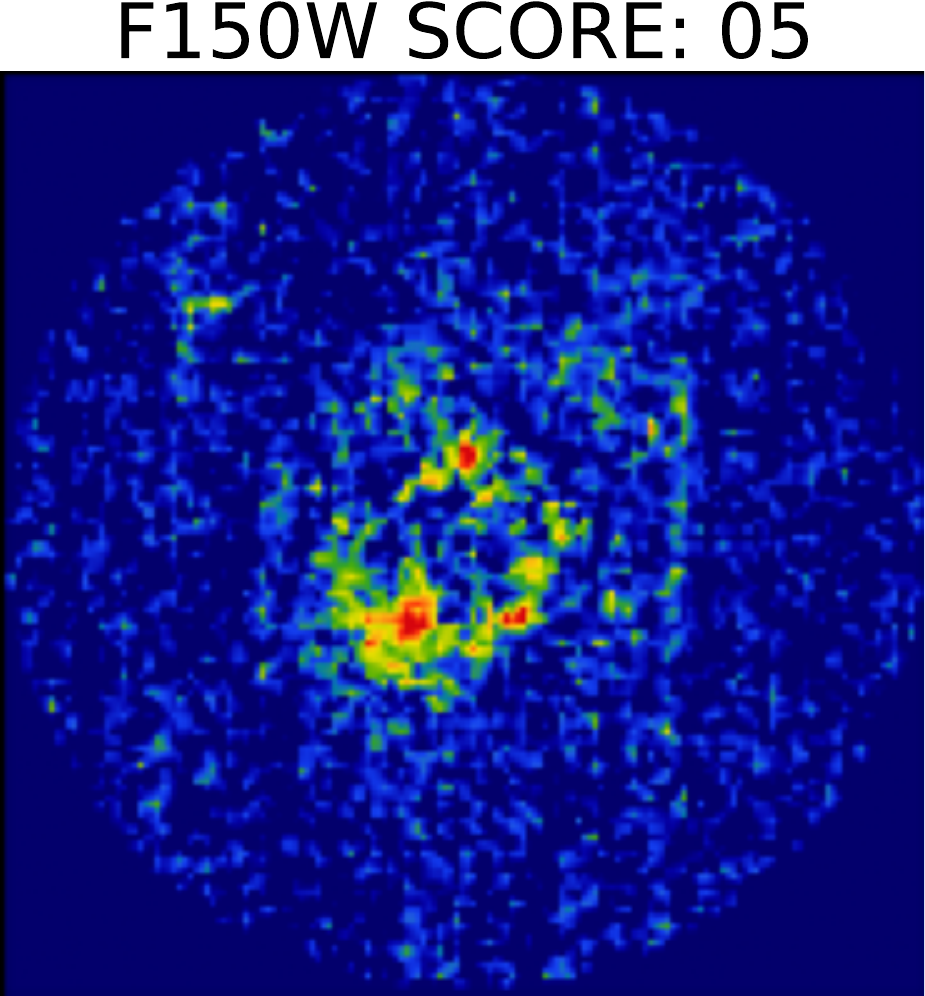}
\includegraphics[width=0.12\textwidth]{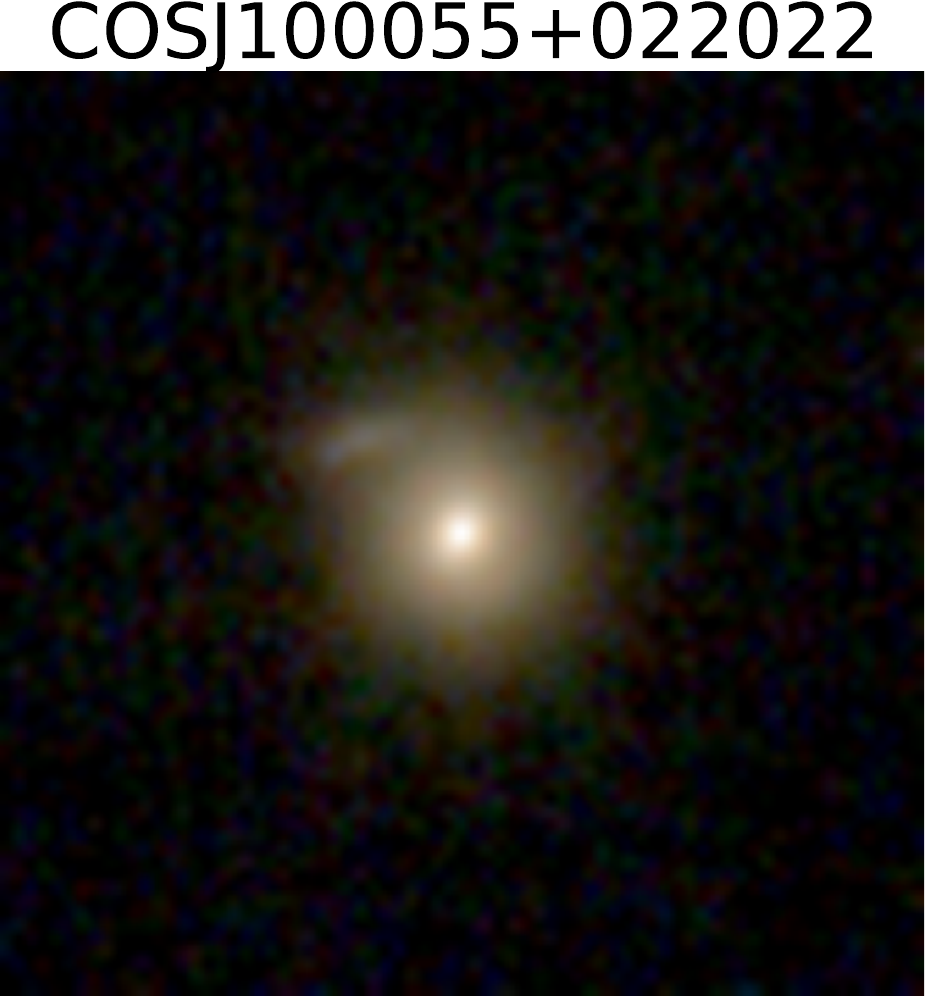}
\includegraphics[width=0.12\textwidth]{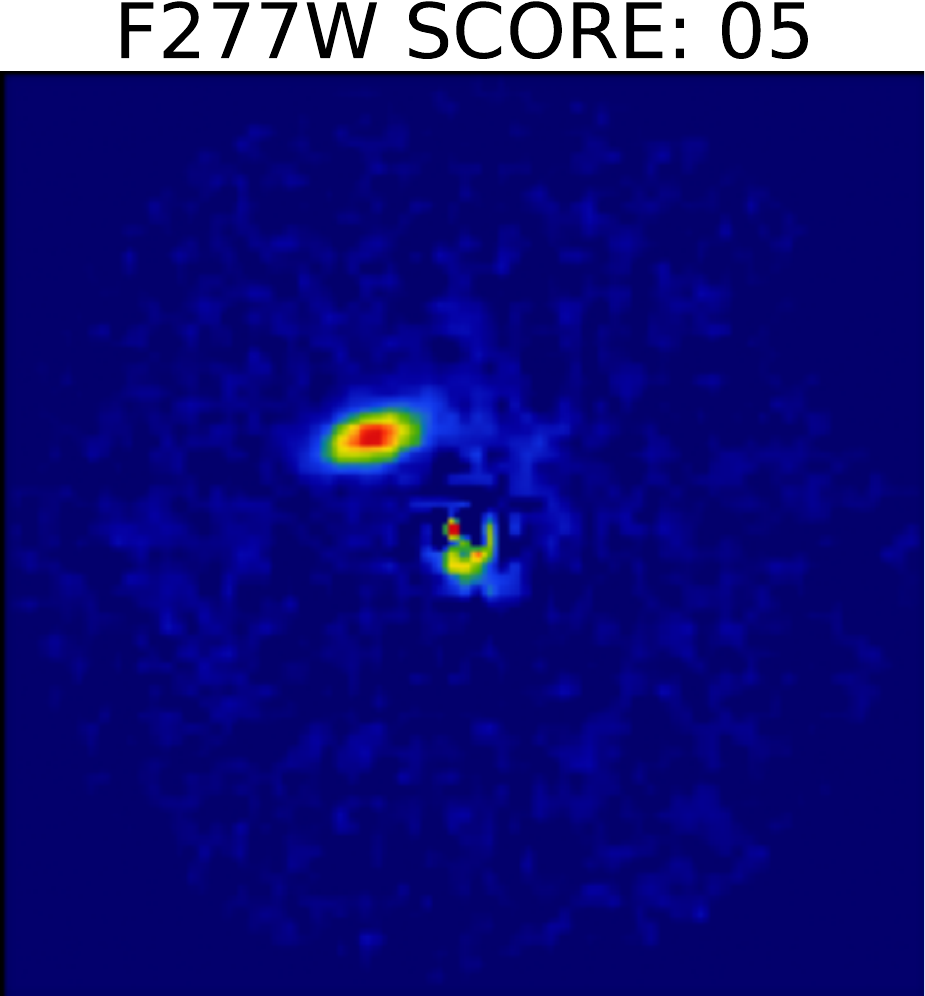}
\includegraphics[width=0.12\textwidth]{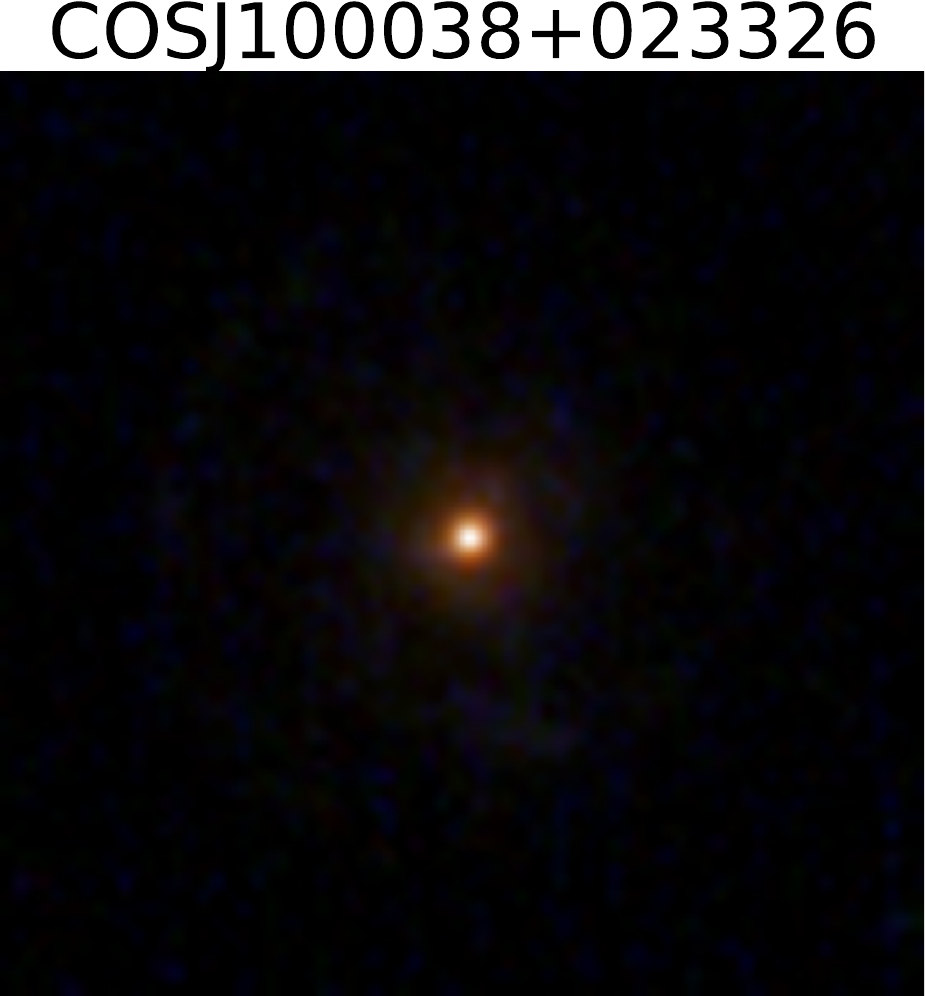}
\includegraphics[width=0.12\textwidth]{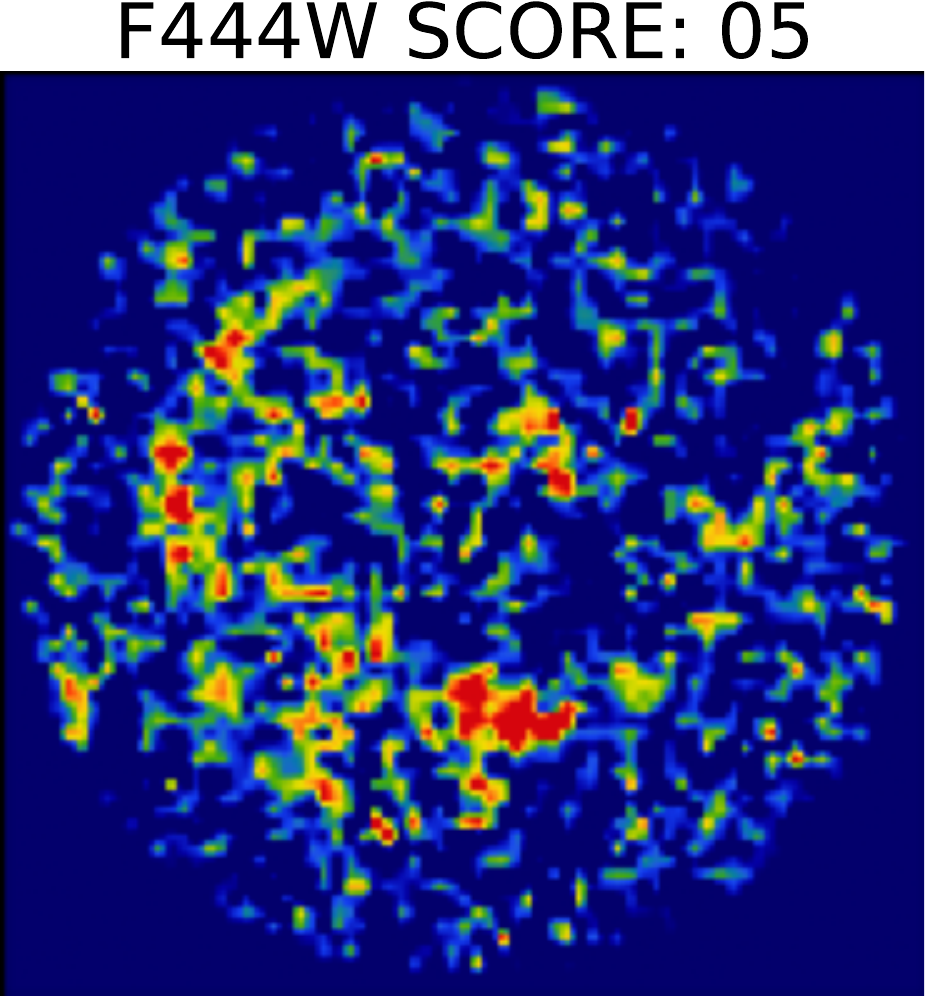}
\includegraphics[width=0.12\textwidth]{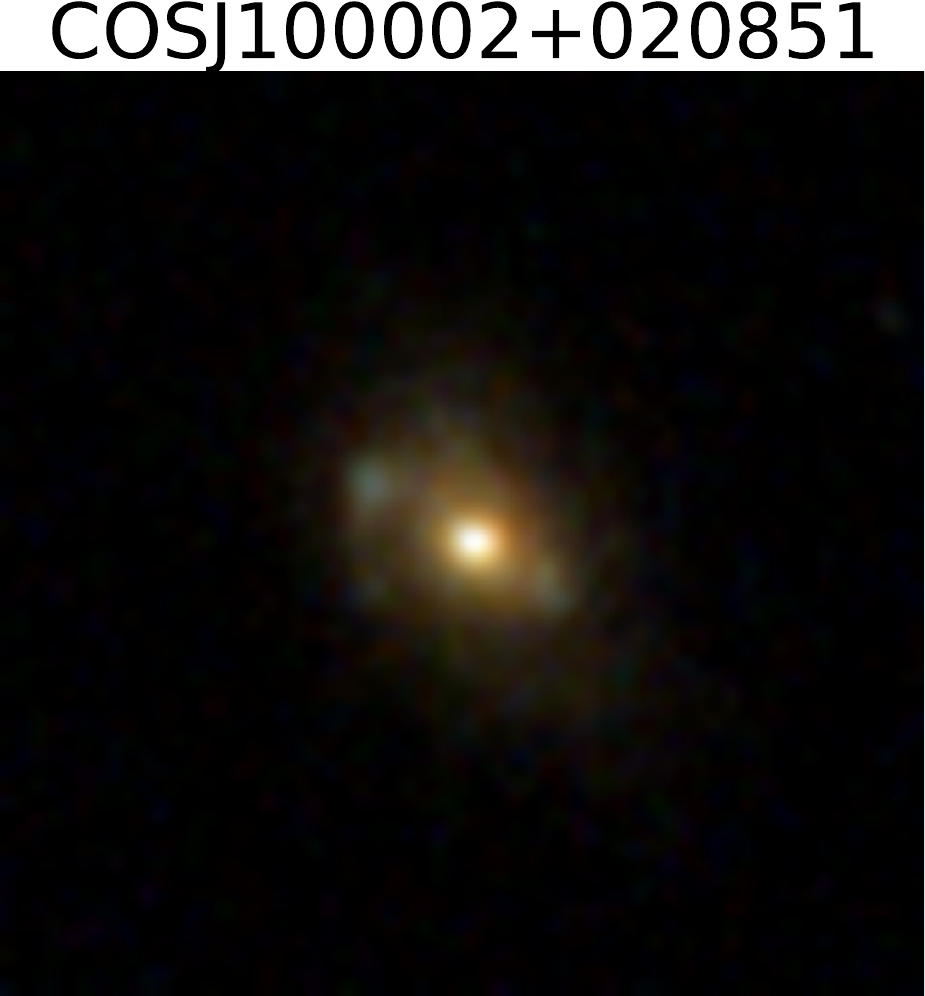}
\includegraphics[width=0.12\textwidth]{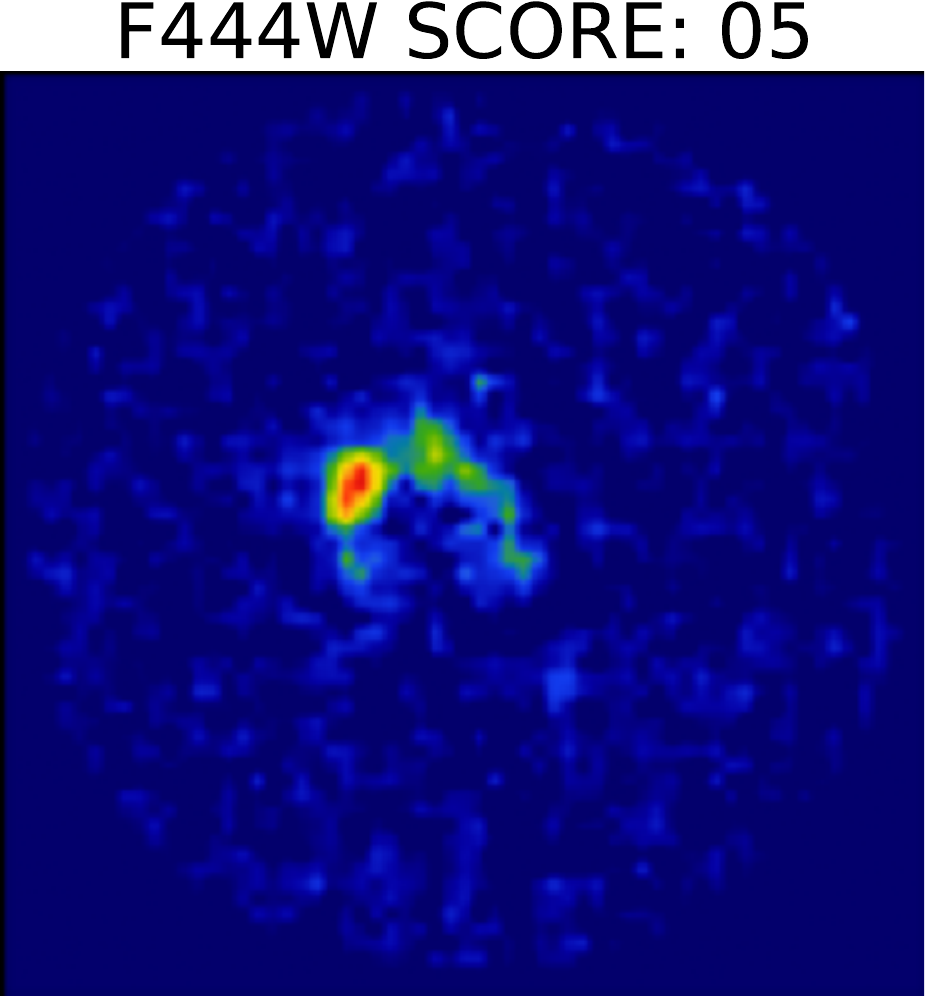}
\includegraphics[width=0.12\textwidth]{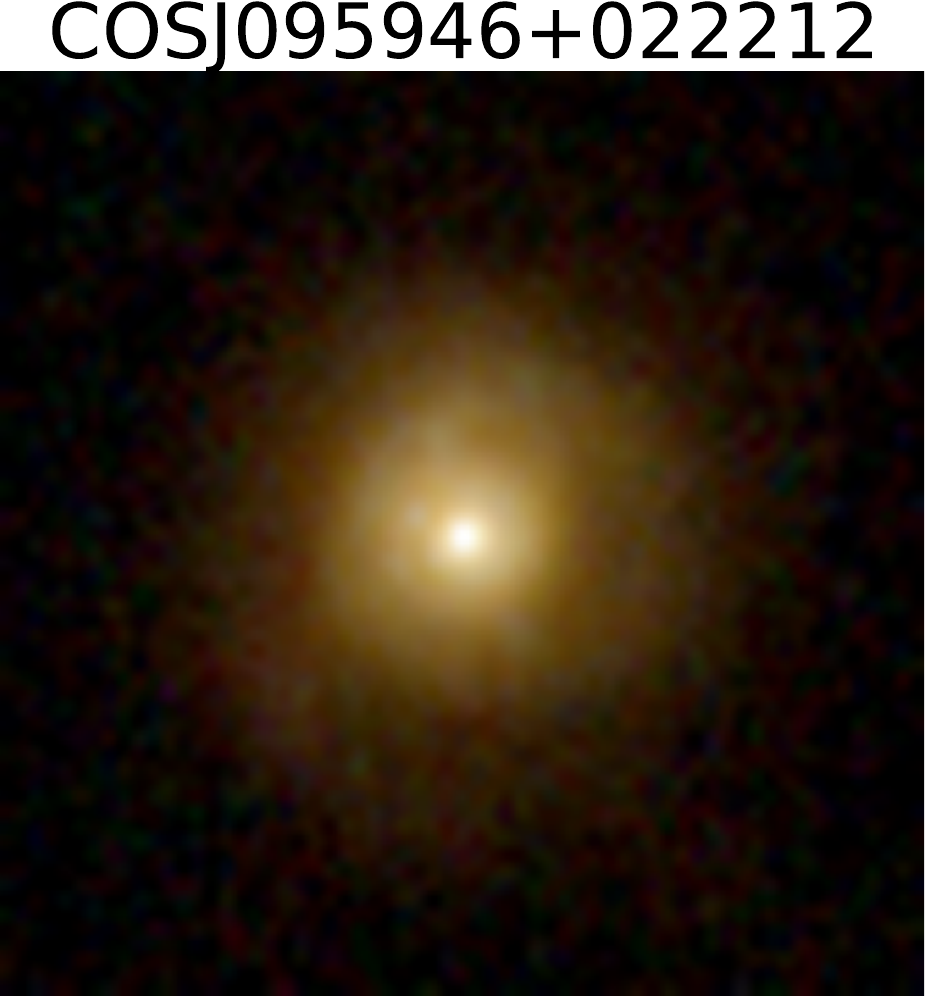}
\includegraphics[width=0.12\textwidth]{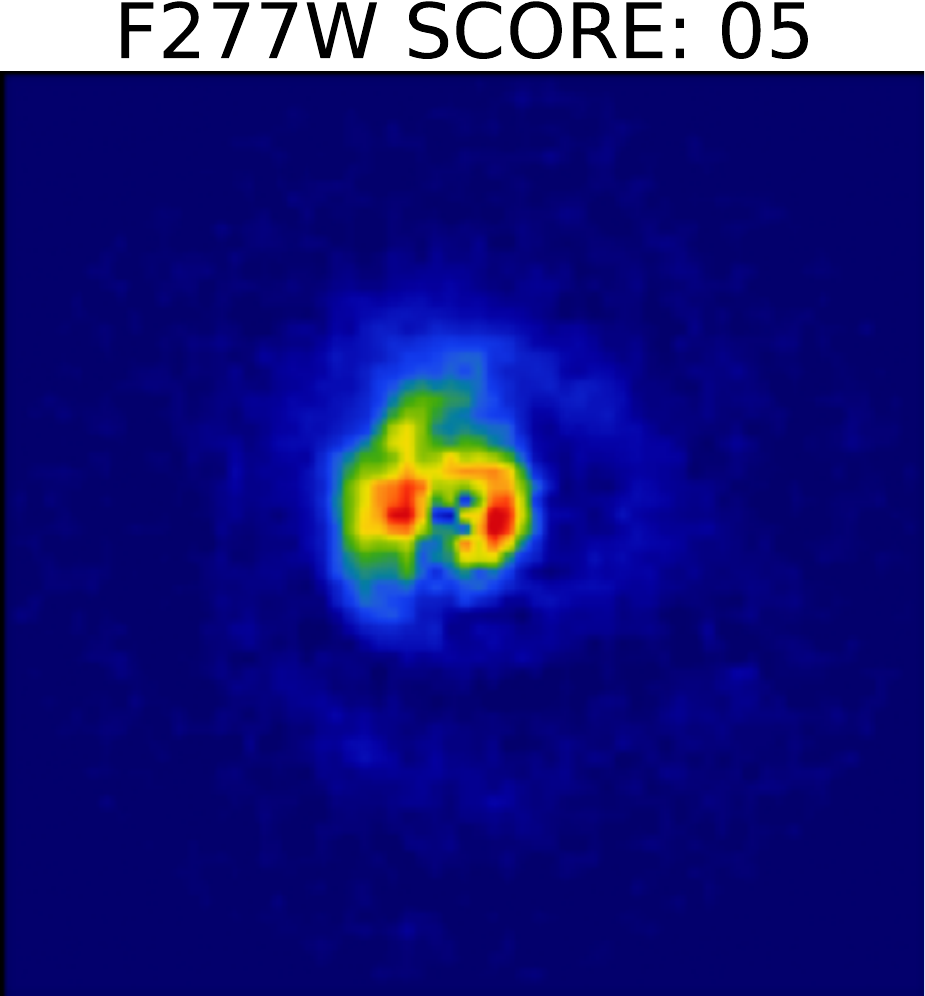}
\includegraphics[width=0.12\textwidth]{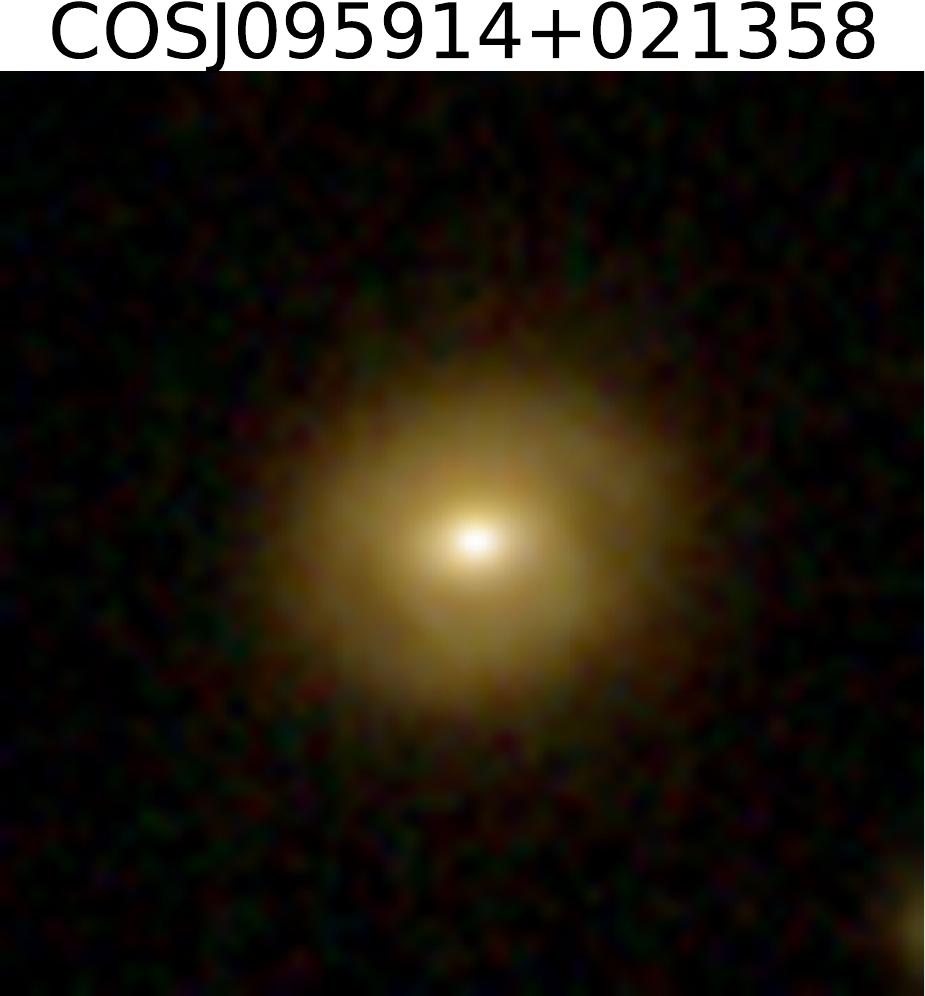}
\includegraphics[width=0.12\textwidth]{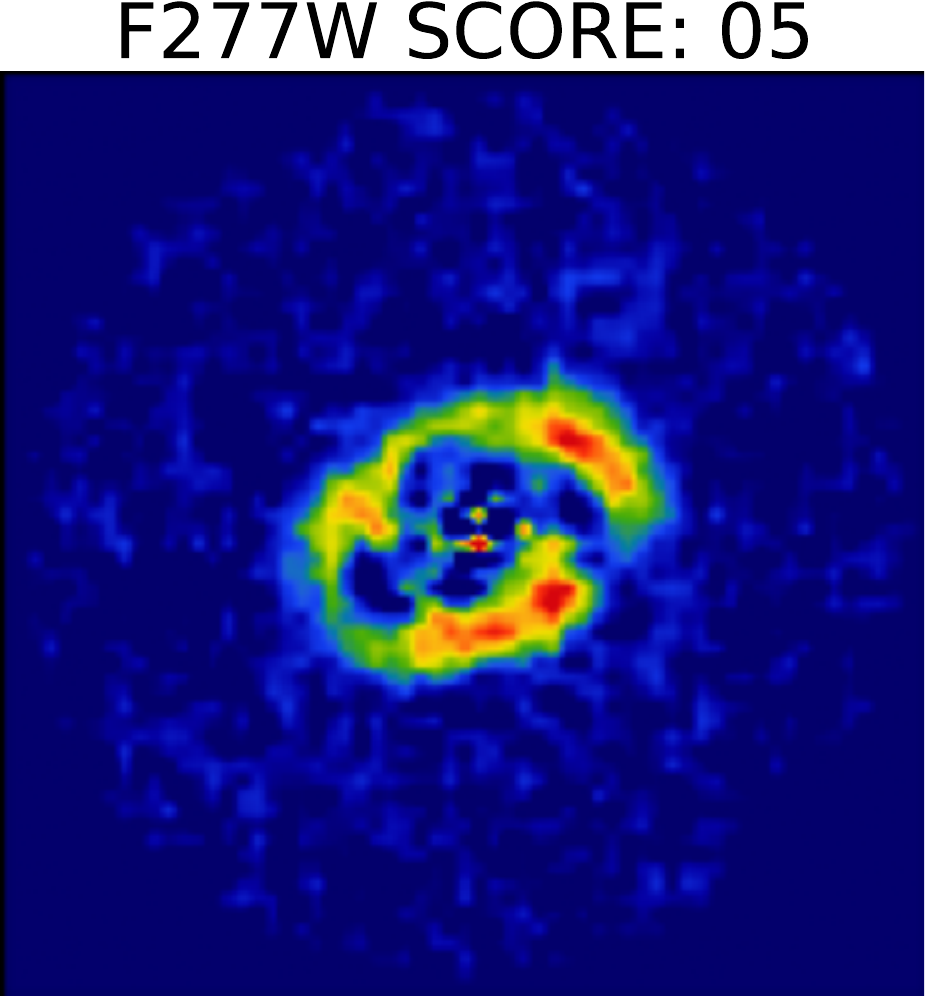}
\includegraphics[width=0.12\textwidth]{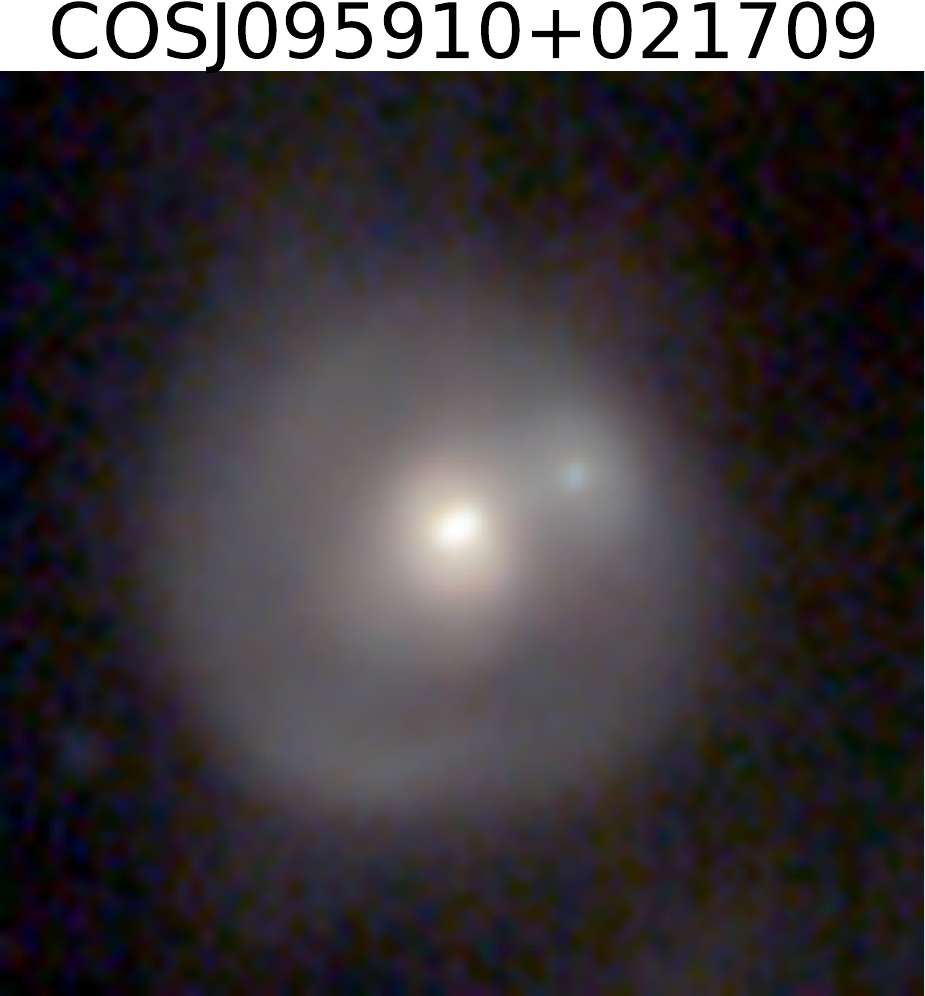}
\includegraphics[width=0.12\textwidth]{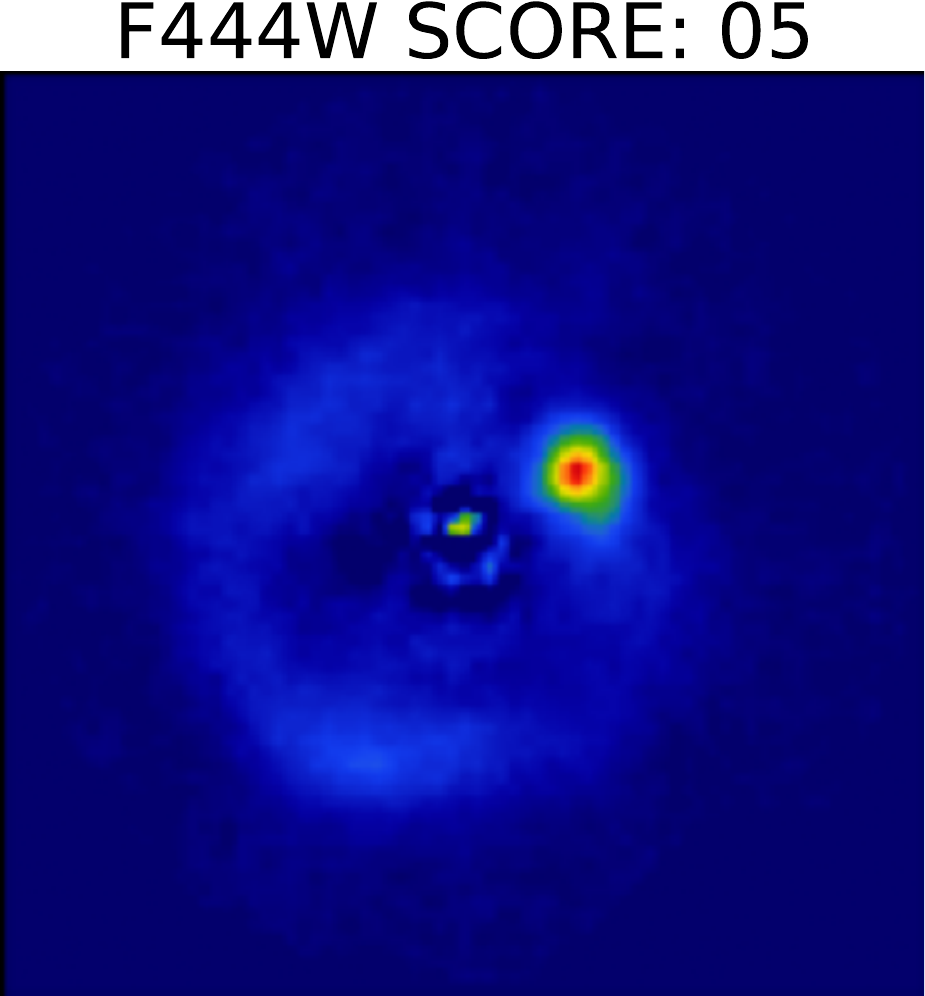}
\includegraphics[width=0.12\textwidth]{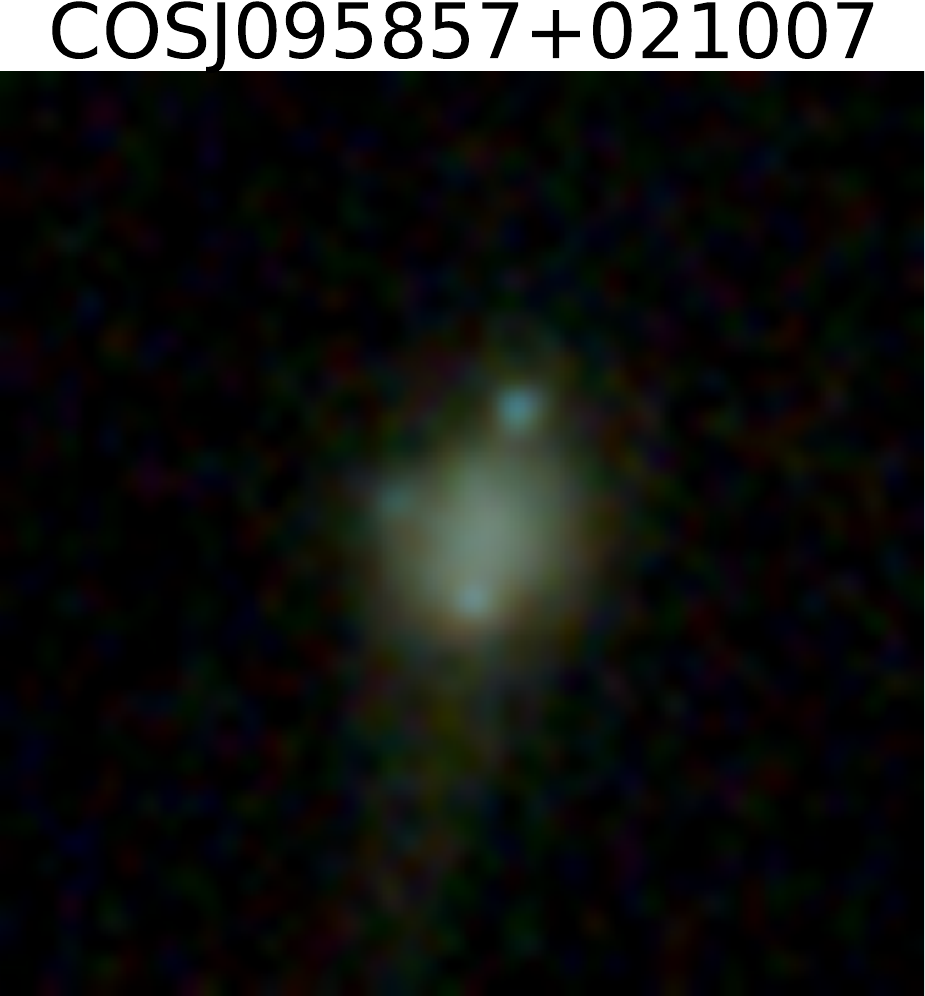}
\includegraphics[width=0.12\textwidth]{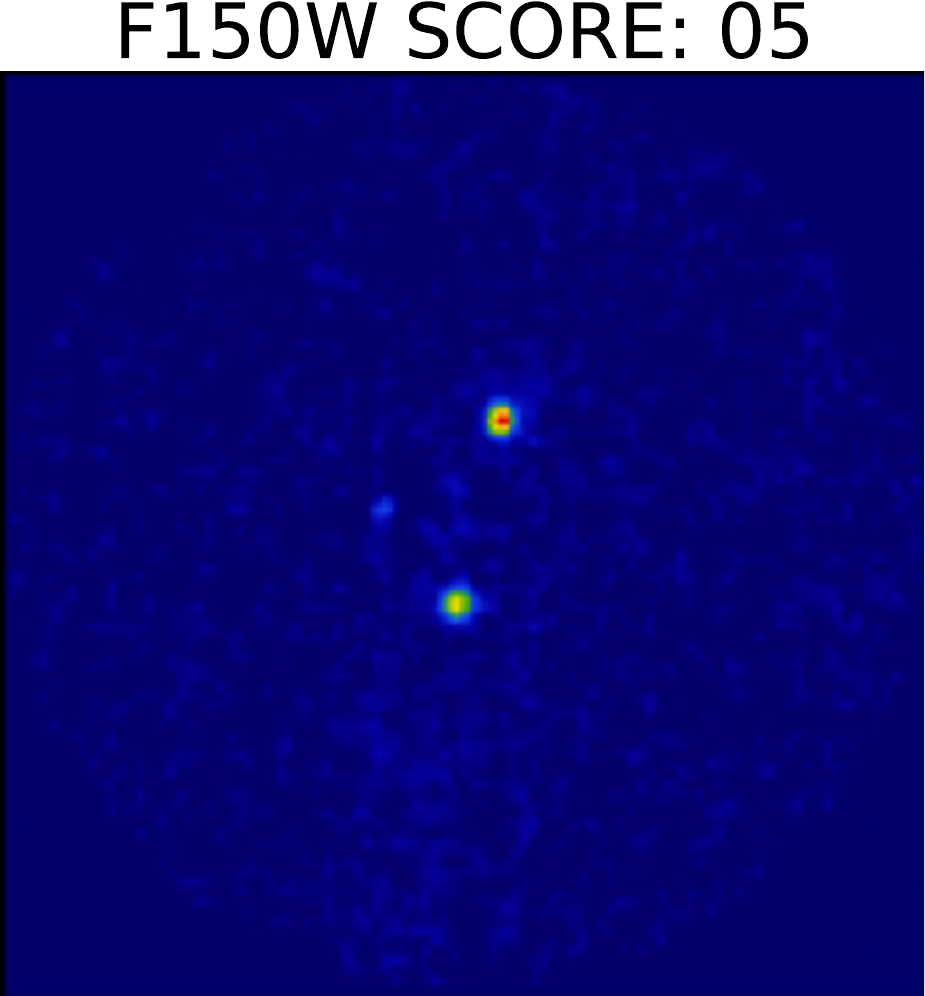}
\includegraphics[width=0.12\textwidth]{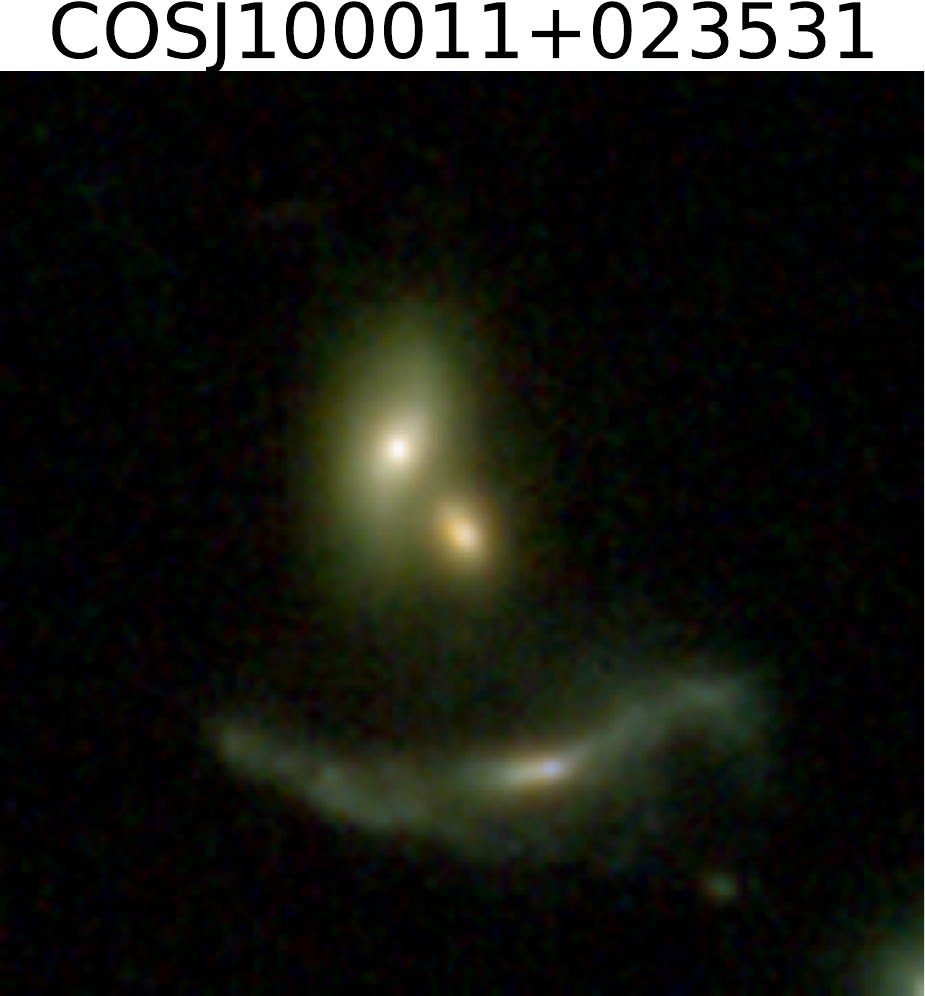}
\includegraphics[width=0.12\textwidth]{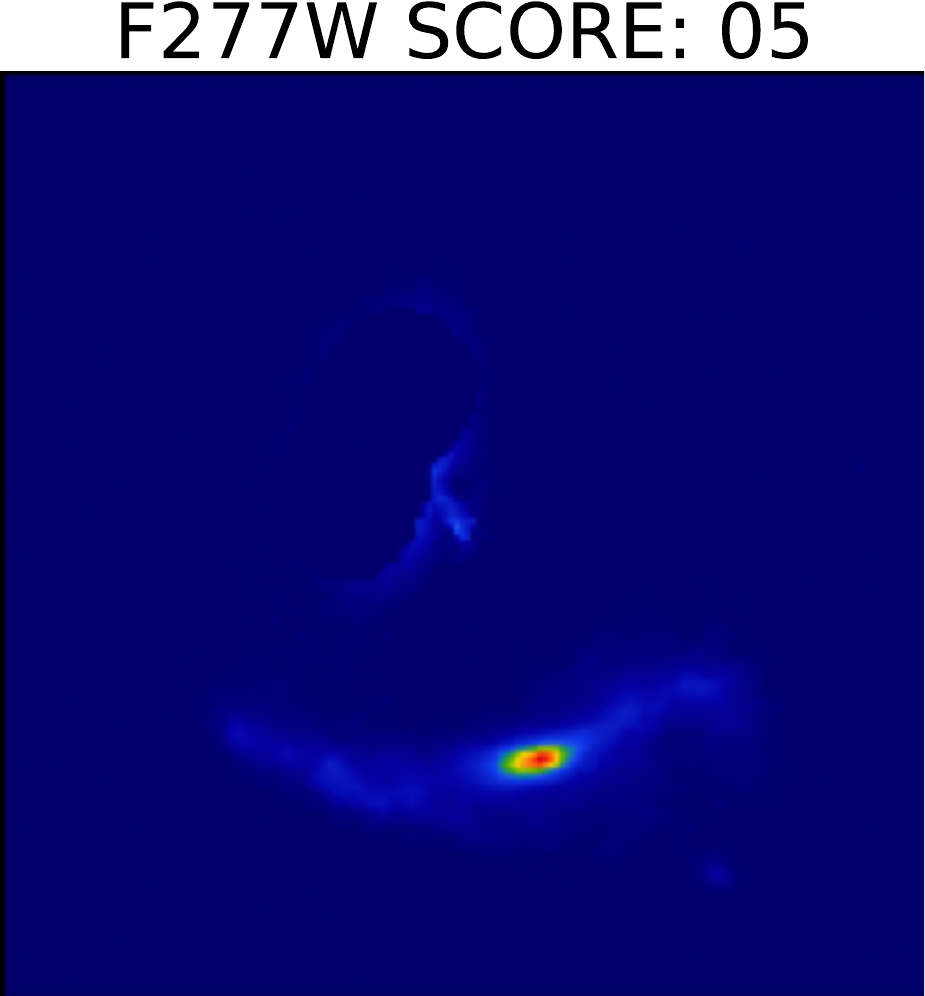}
\includegraphics[width=0.12\textwidth]{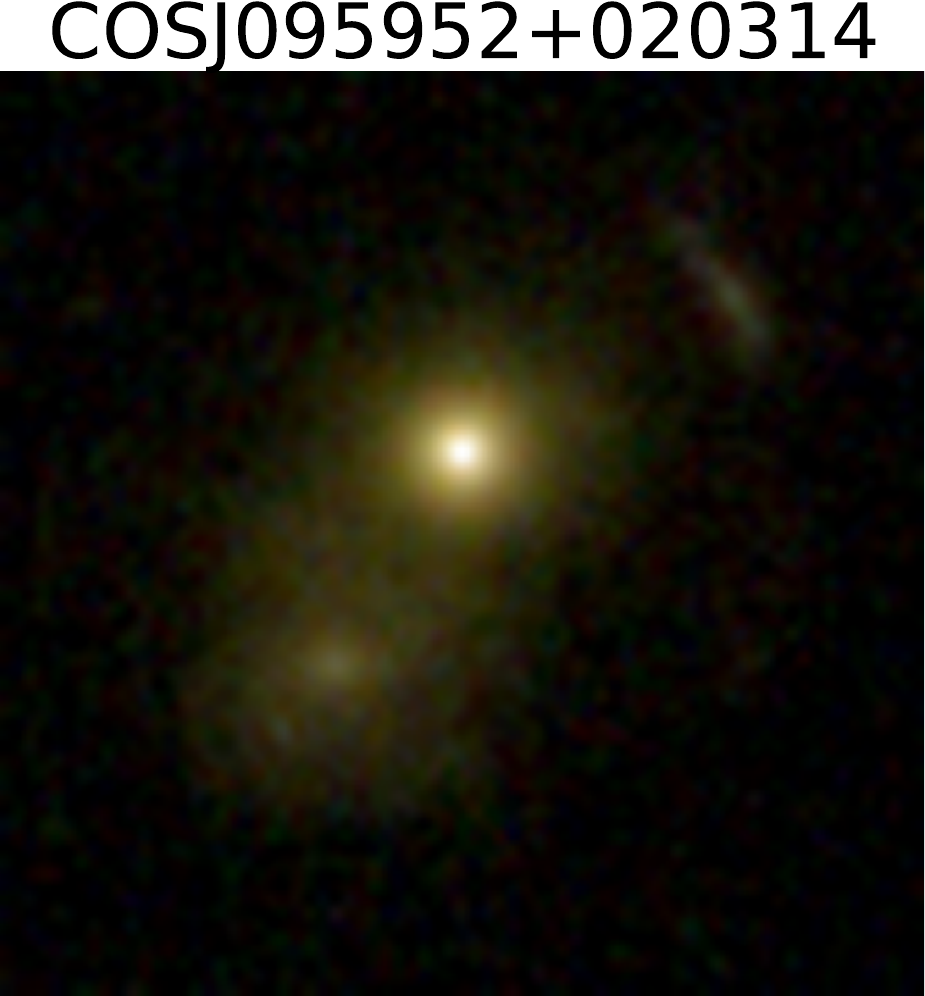}
\includegraphics[width=0.12\textwidth]{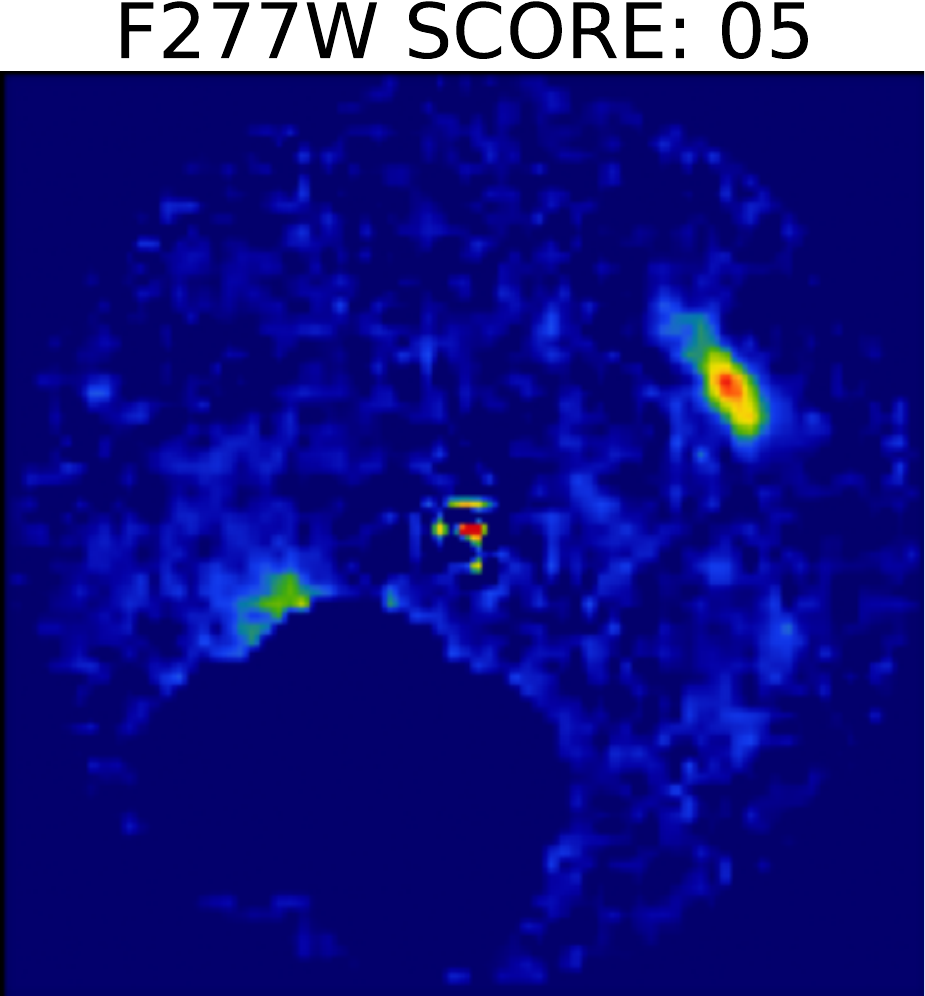}
\includegraphics[width=0.12\textwidth]{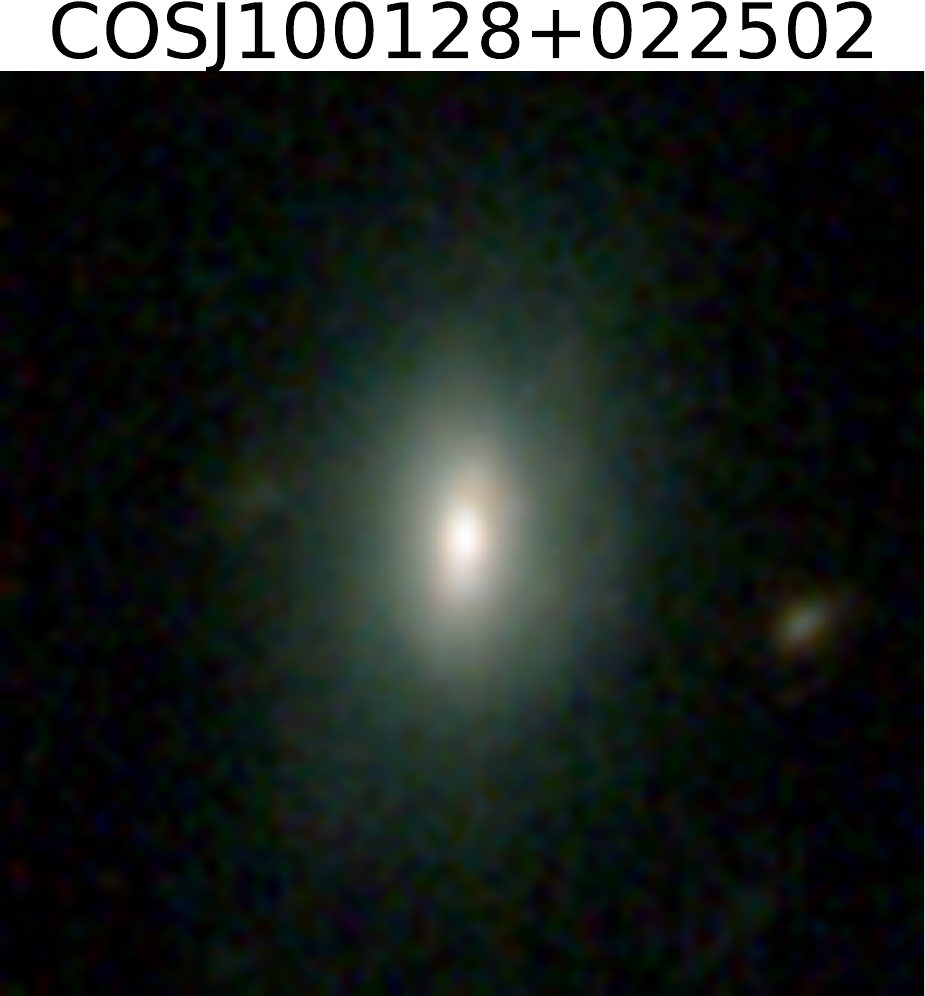}
\includegraphics[width=0.12\textwidth]{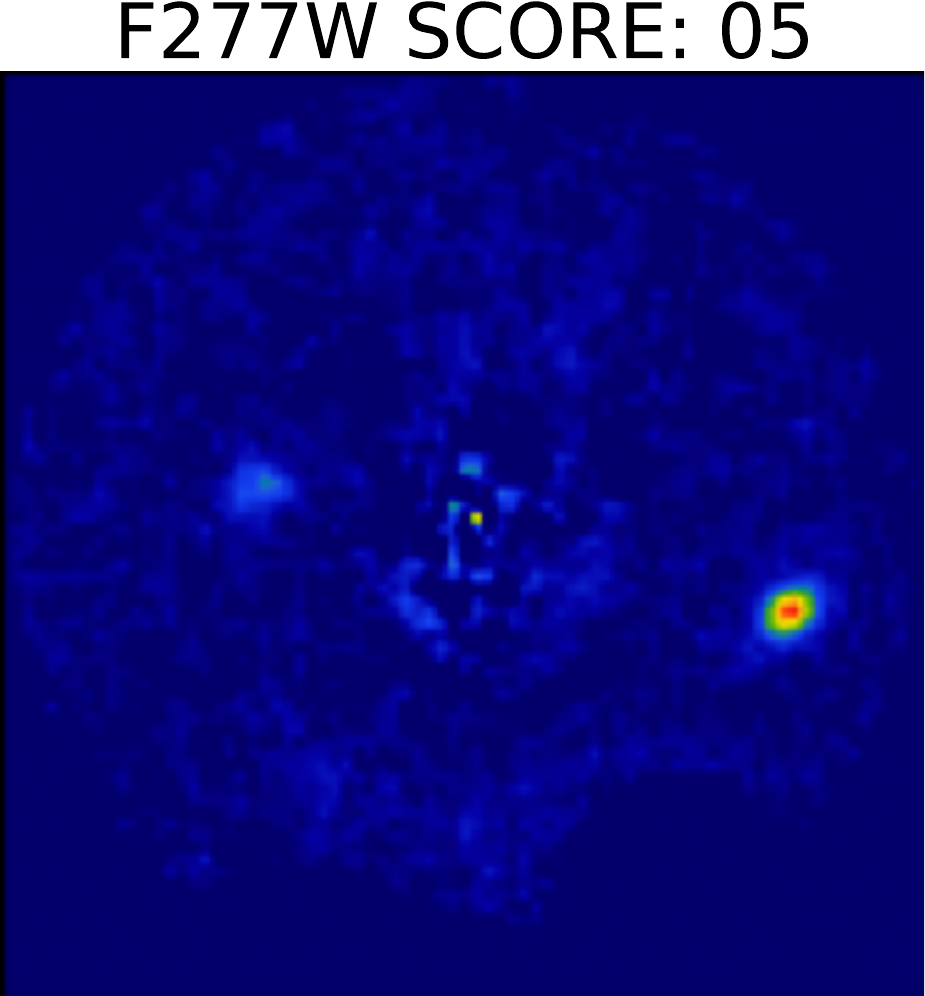}
\includegraphics[width=0.12\textwidth]{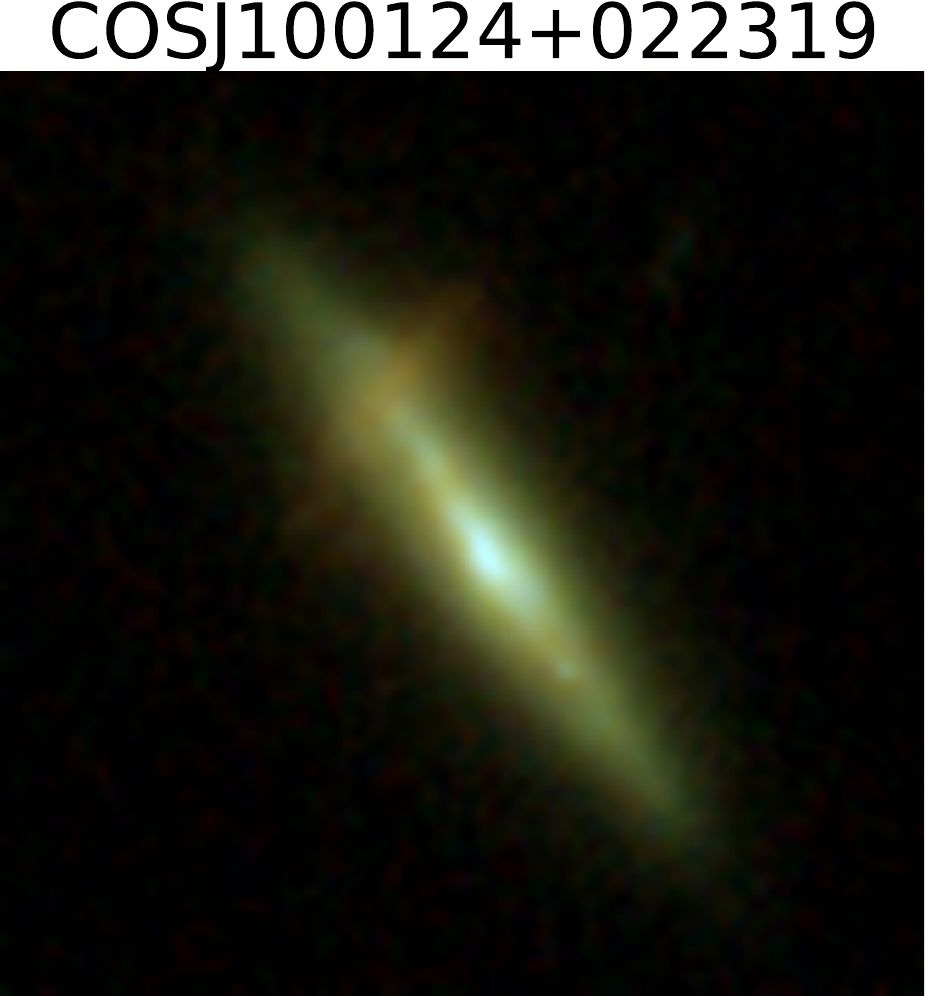}
\includegraphics[width=0.12\textwidth]{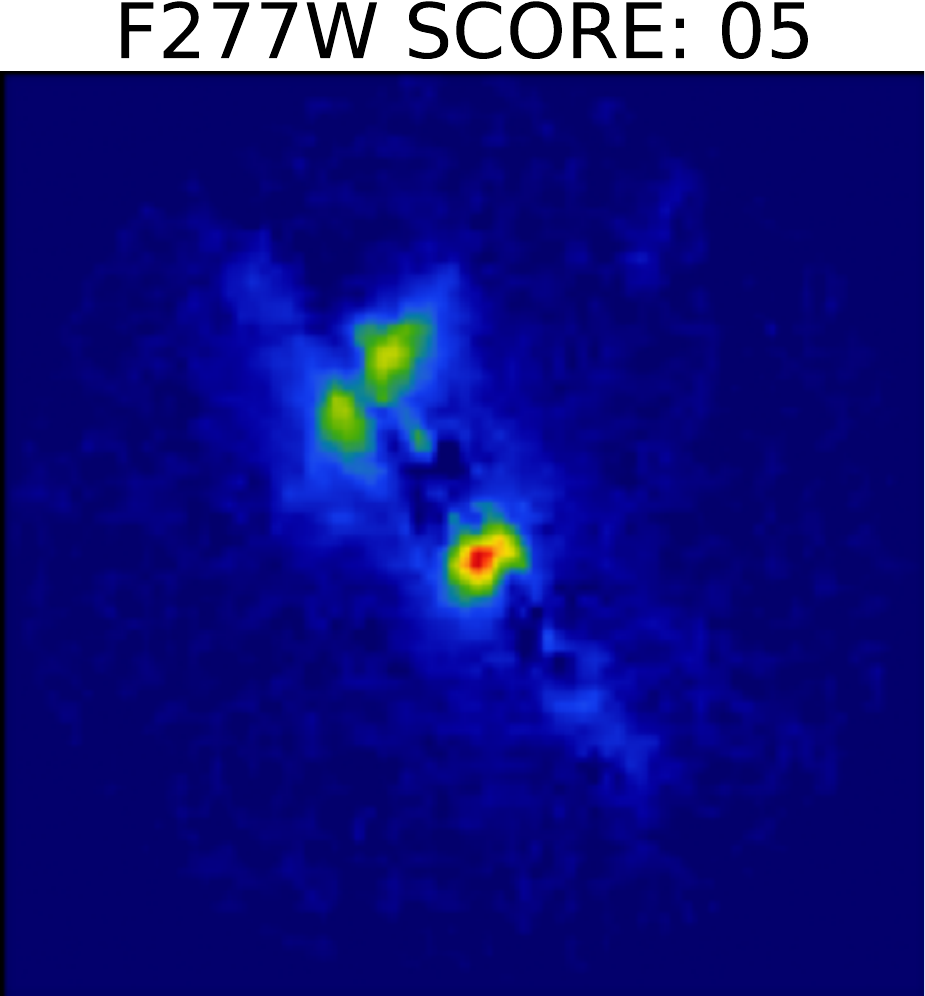}
\includegraphics[width=0.12\textwidth]{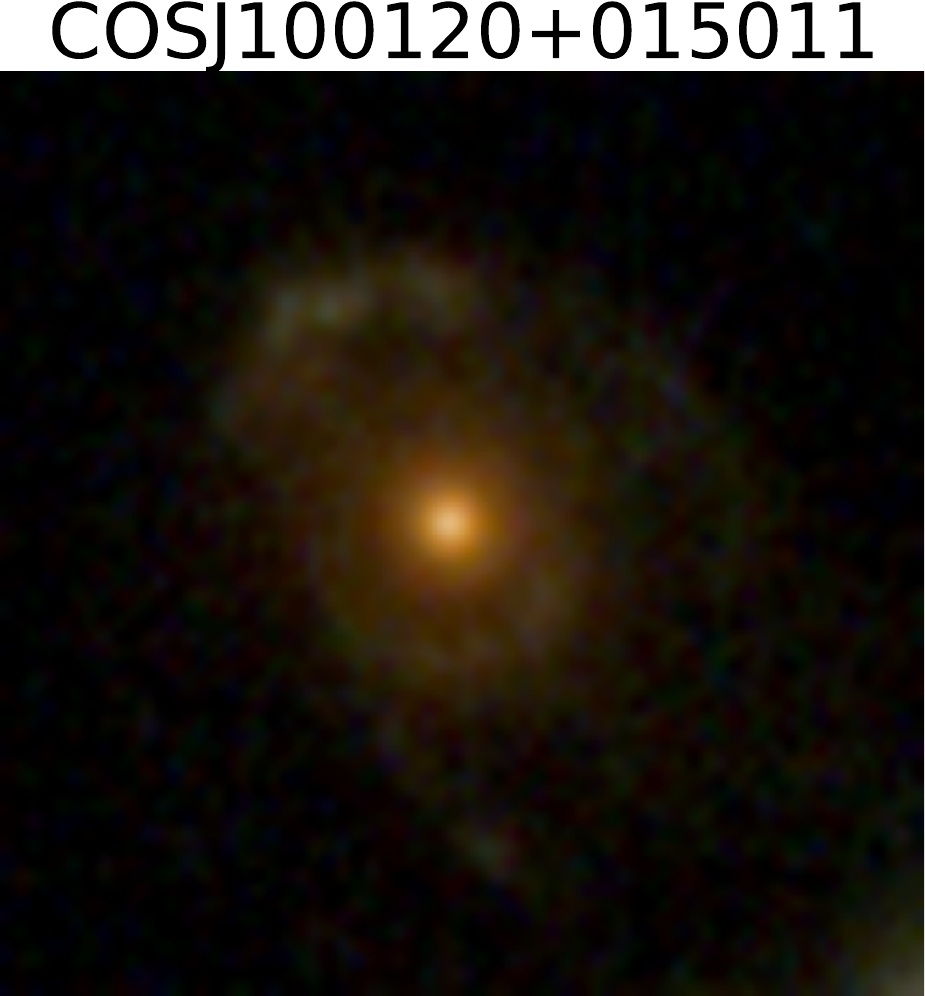}
\includegraphics[width=0.12\textwidth]{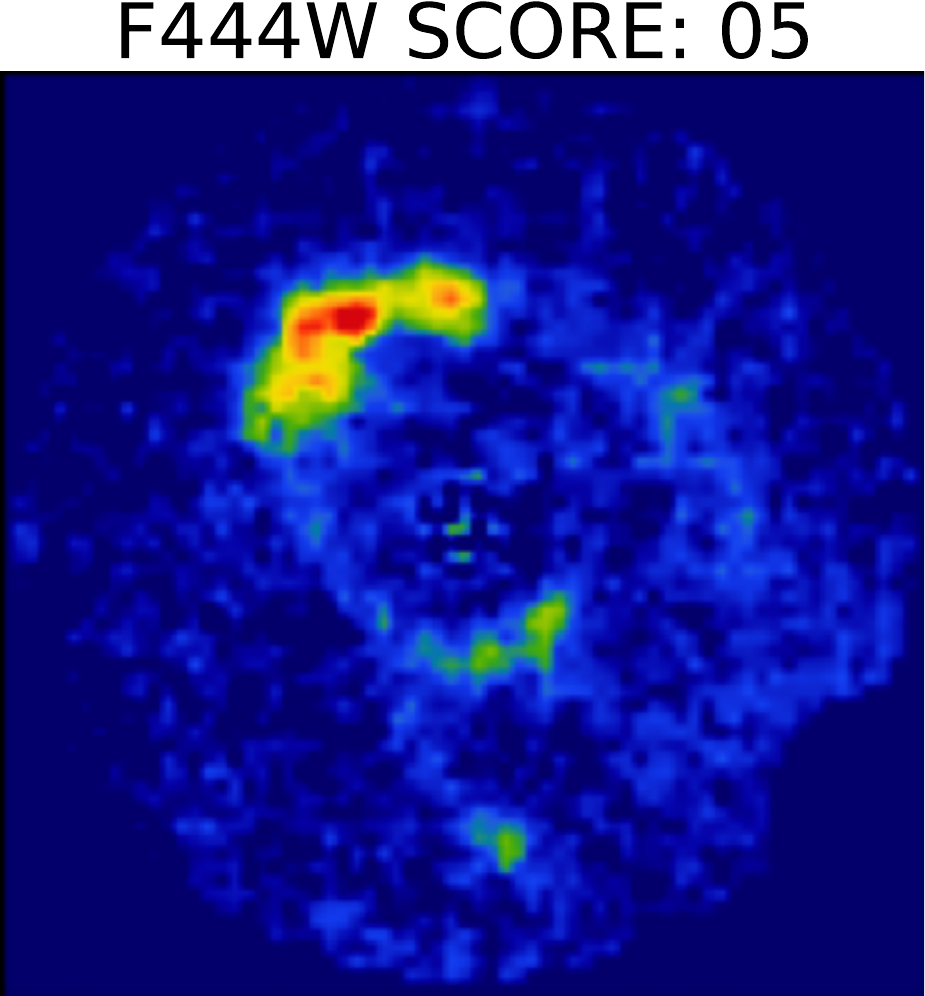}
\includegraphics[width=0.12\textwidth]{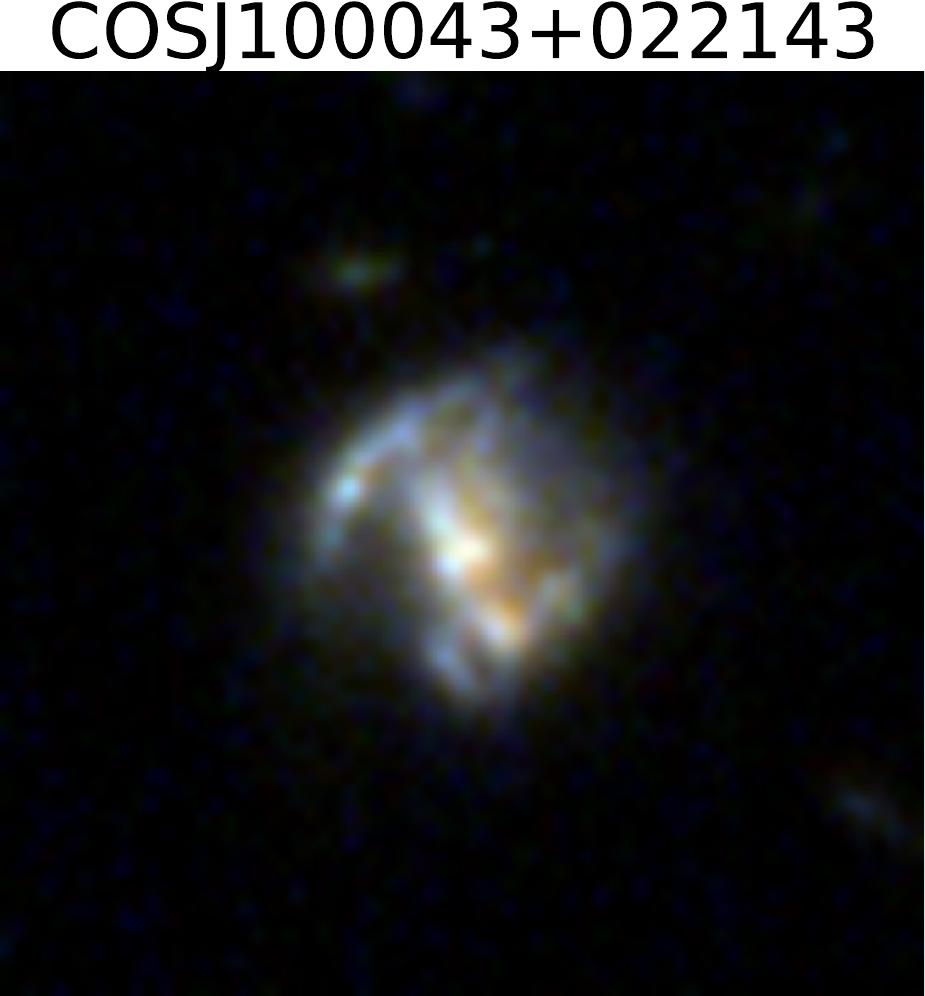}
\includegraphics[width=0.12\textwidth]{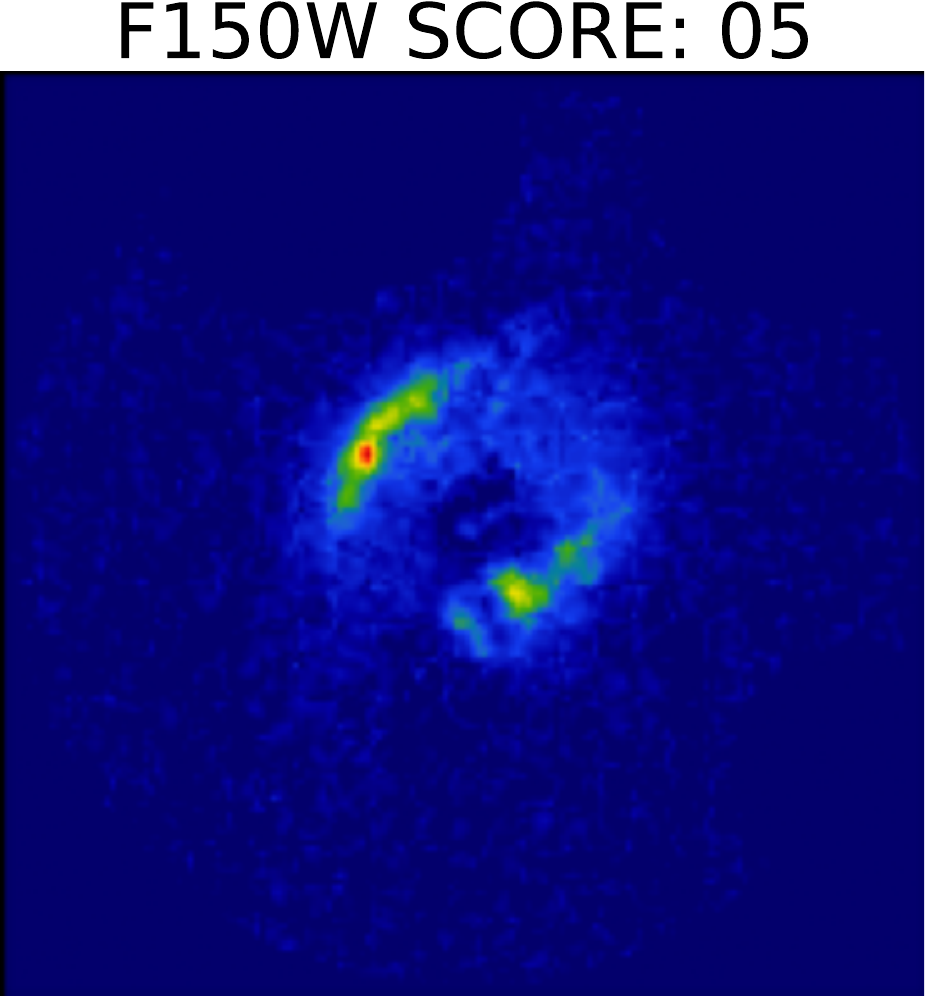}
\includegraphics[width=0.12\textwidth]{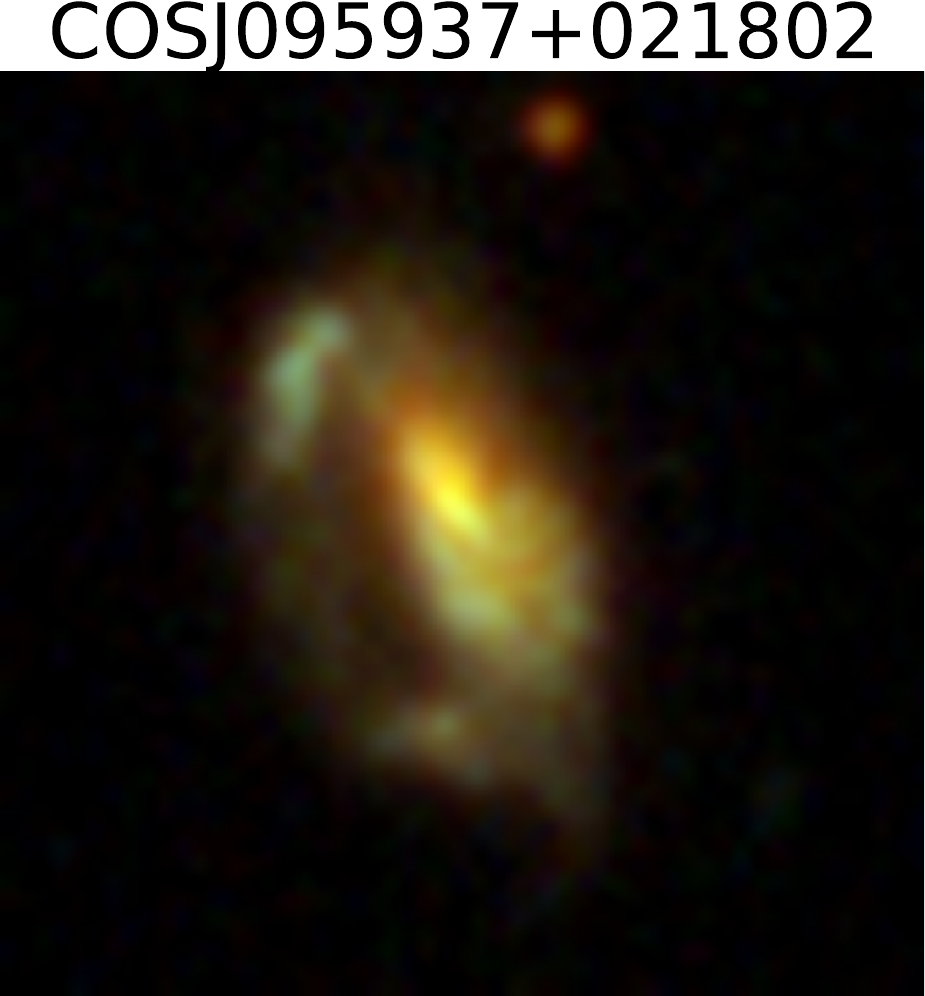}
\includegraphics[width=0.12\textwidth]{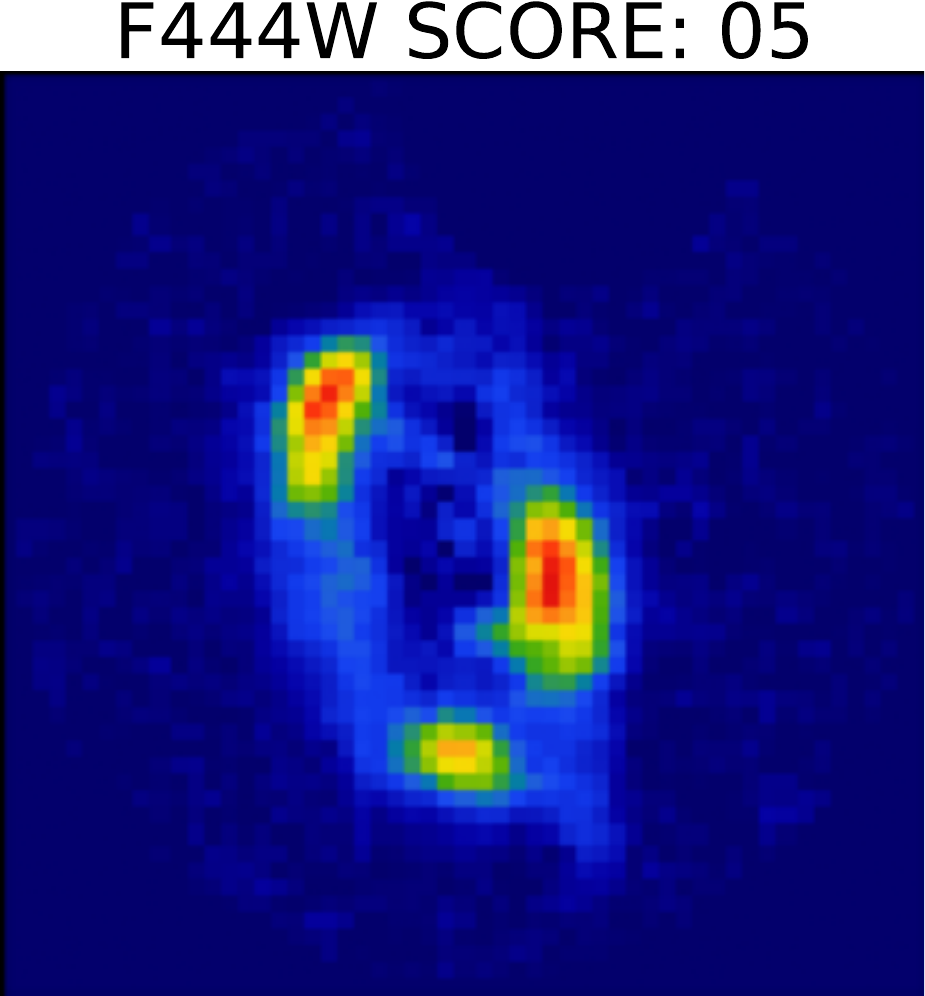}
\includegraphics[width=0.12\textwidth]{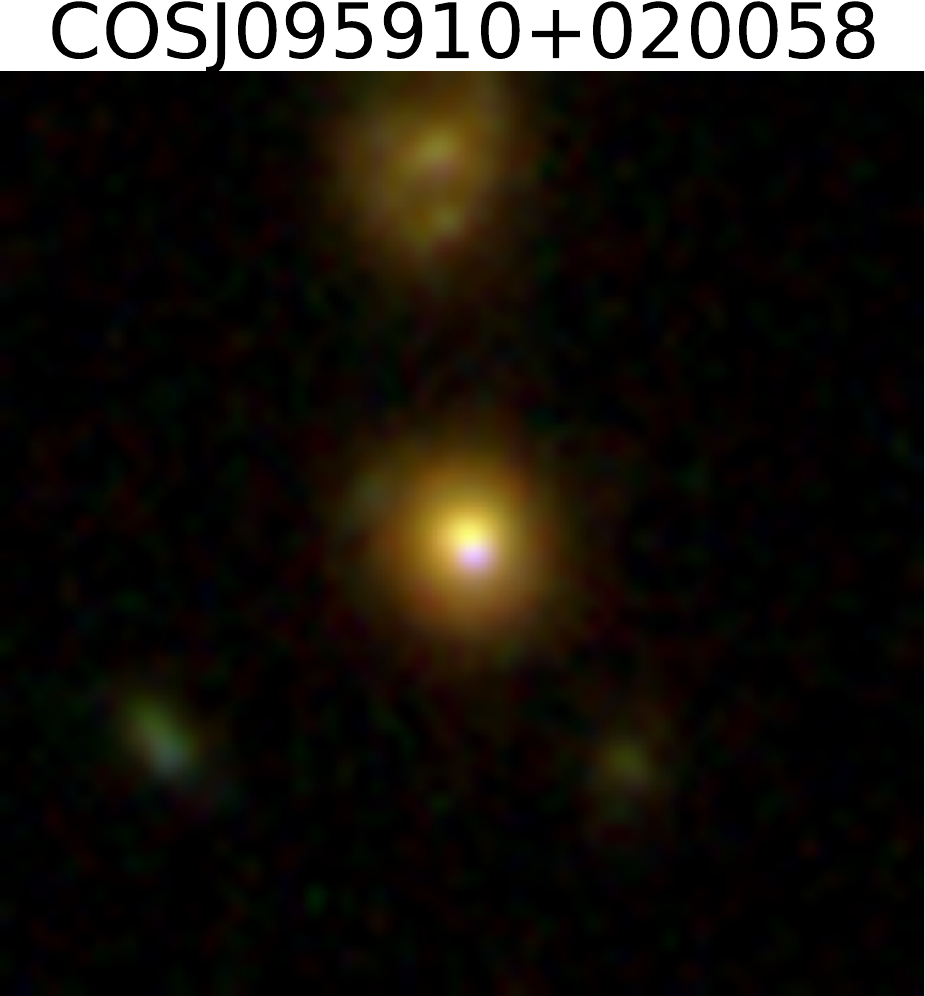}
\includegraphics[width=0.12\textwidth]{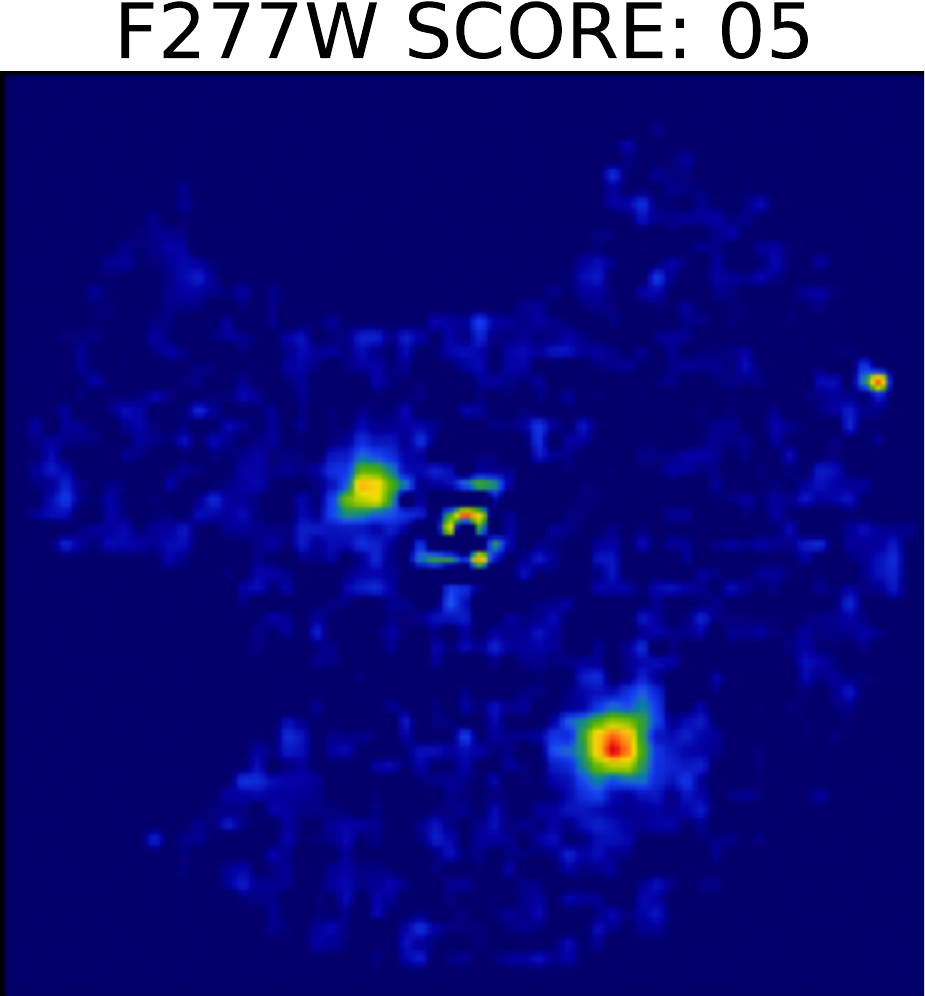}
\includegraphics[width=0.12\textwidth]{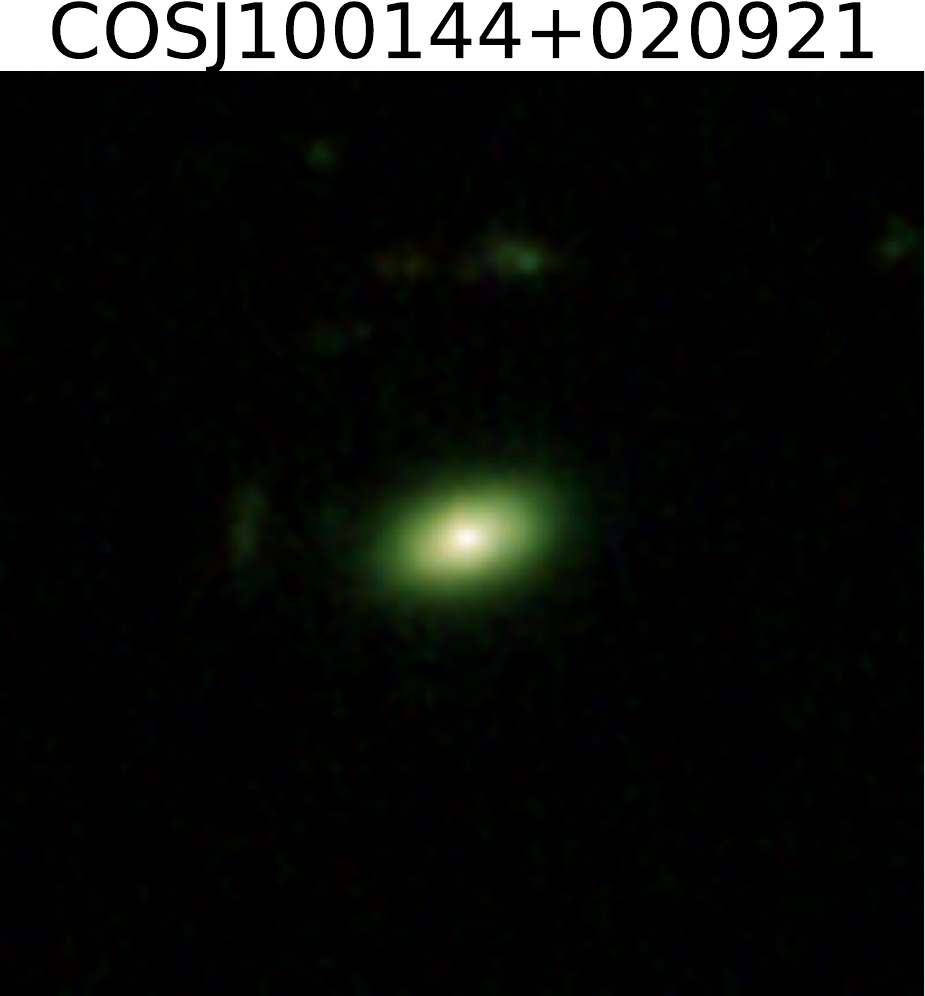}
\includegraphics[width=0.12\textwidth]{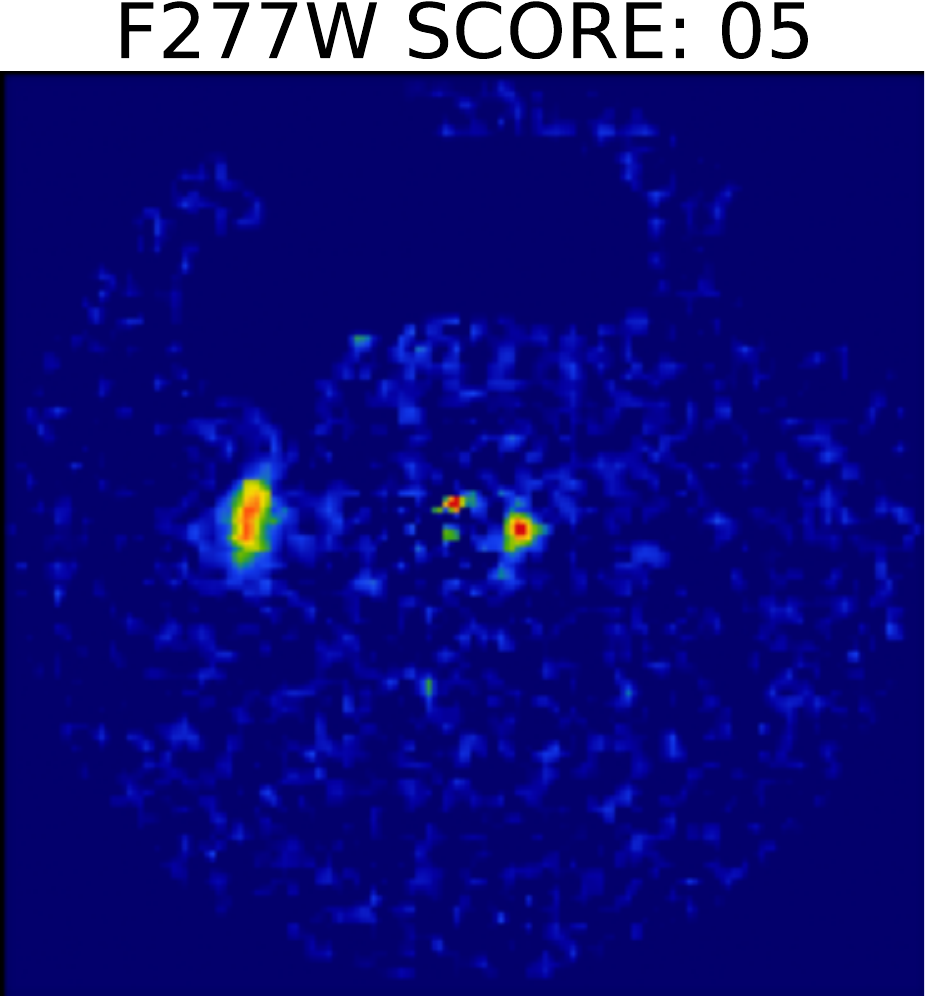}
\includegraphics[width=0.12\textwidth]{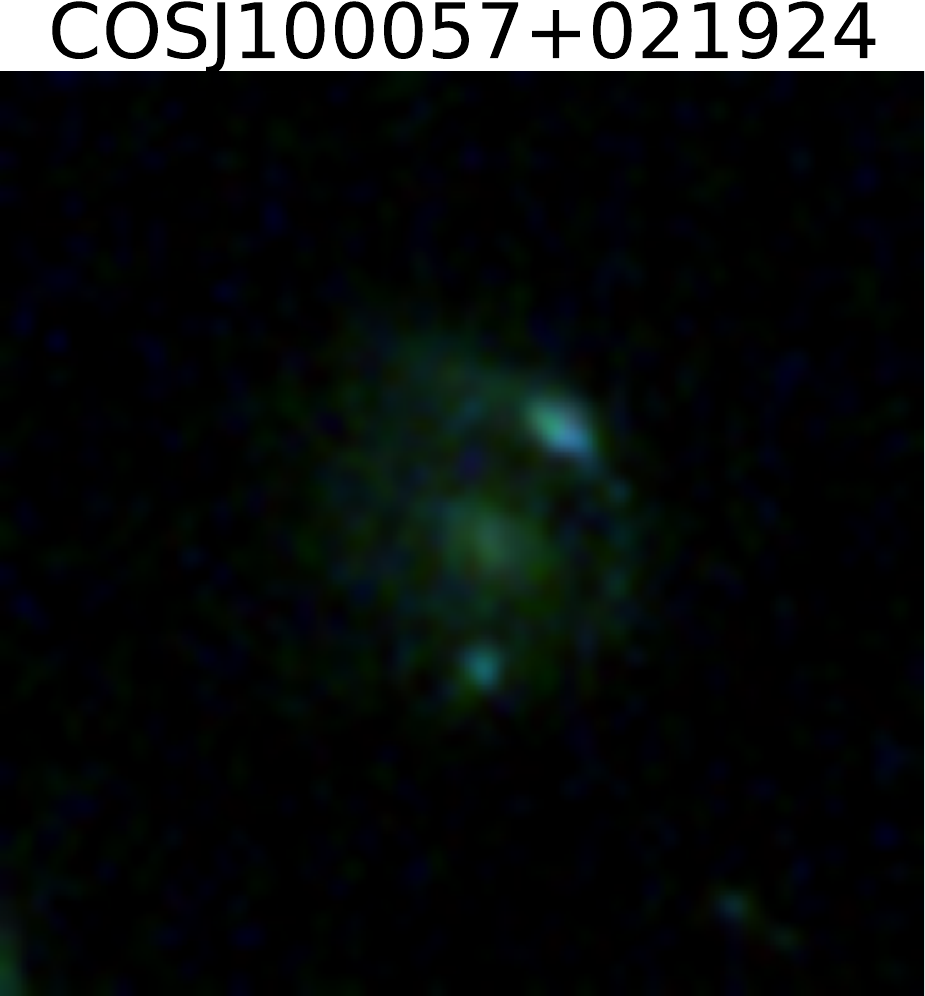}
\includegraphics[width=0.12\textwidth]{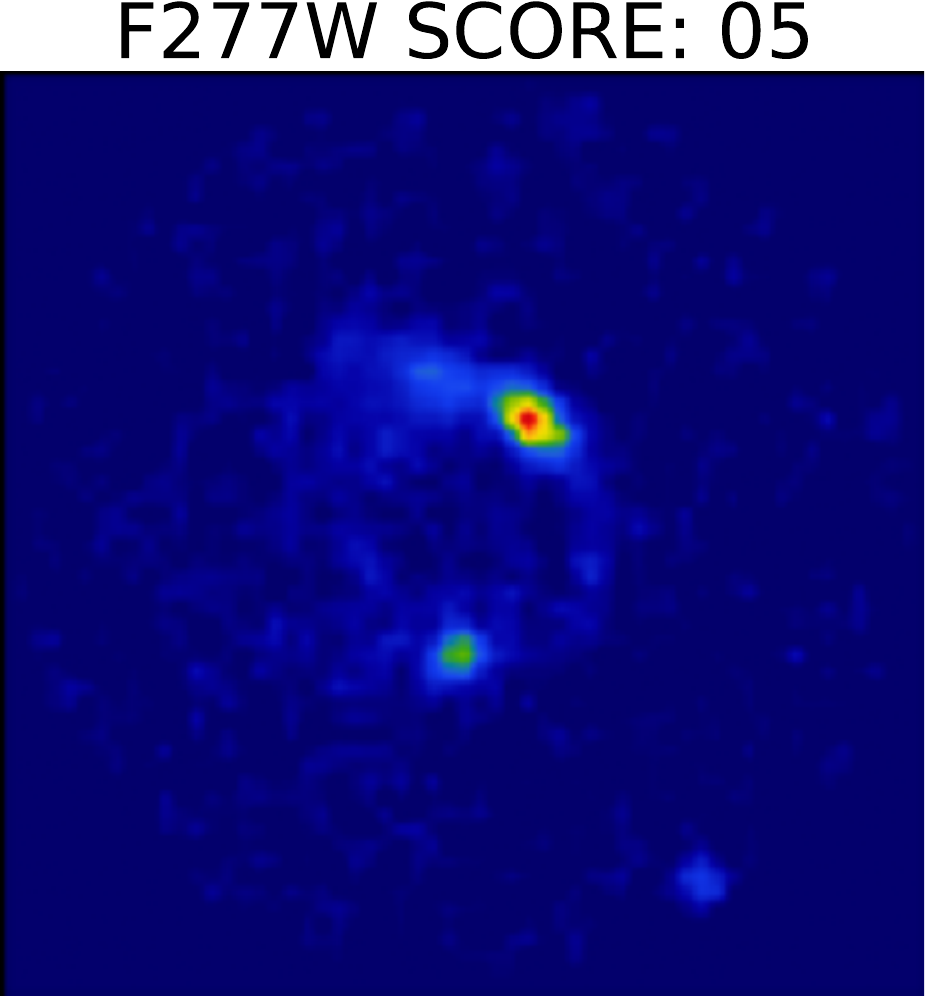}
\includegraphics[width=0.12\textwidth]{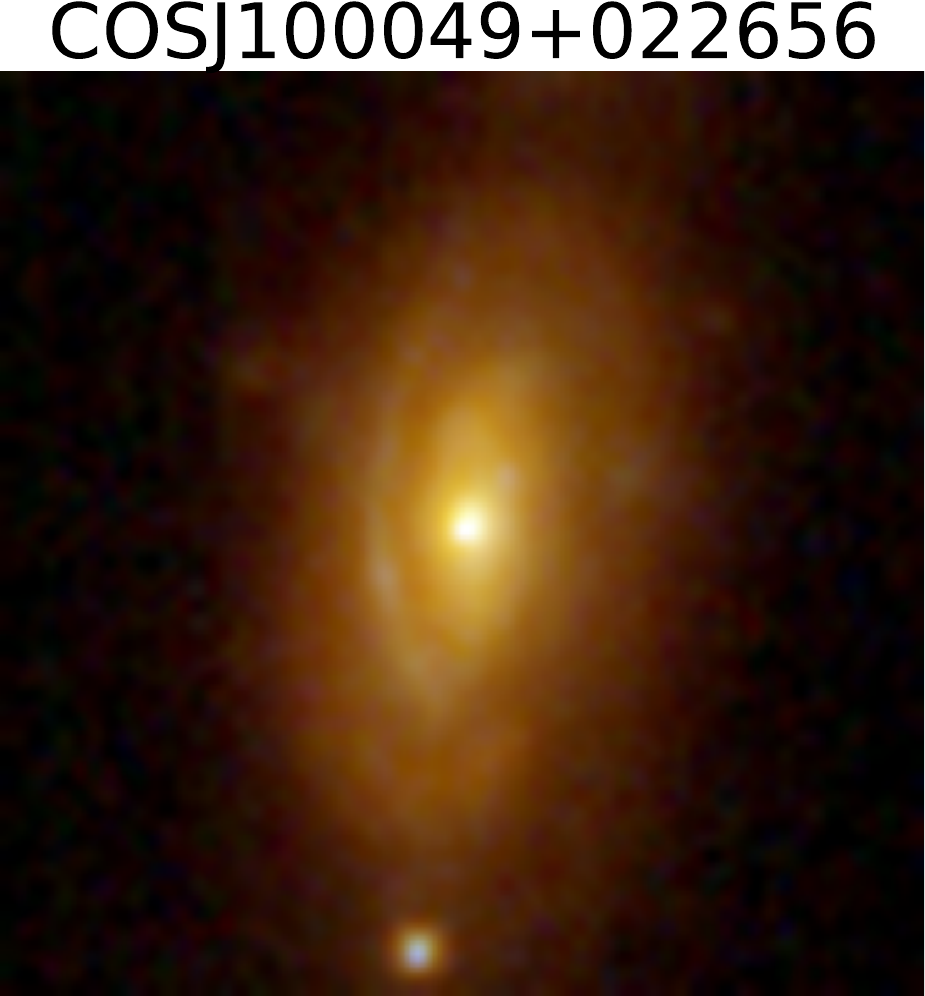}
\includegraphics[width=0.12\textwidth]{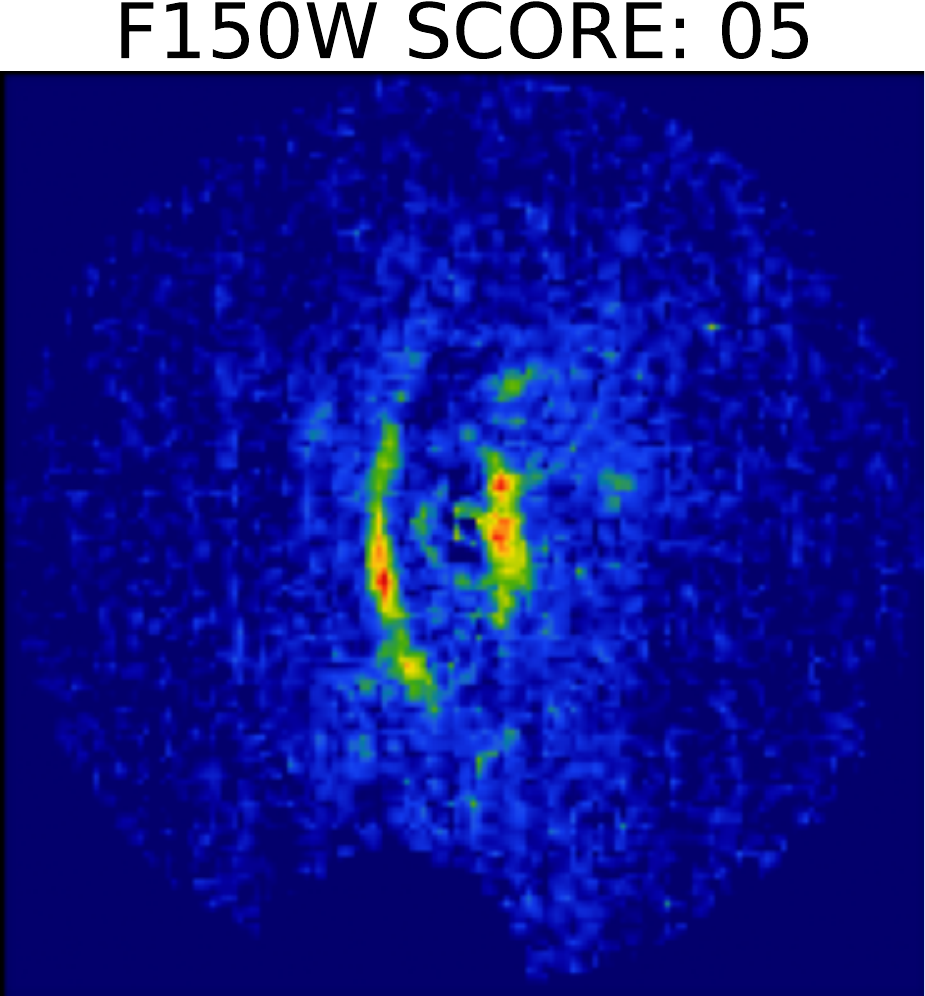}
\includegraphics[width=0.12\textwidth]{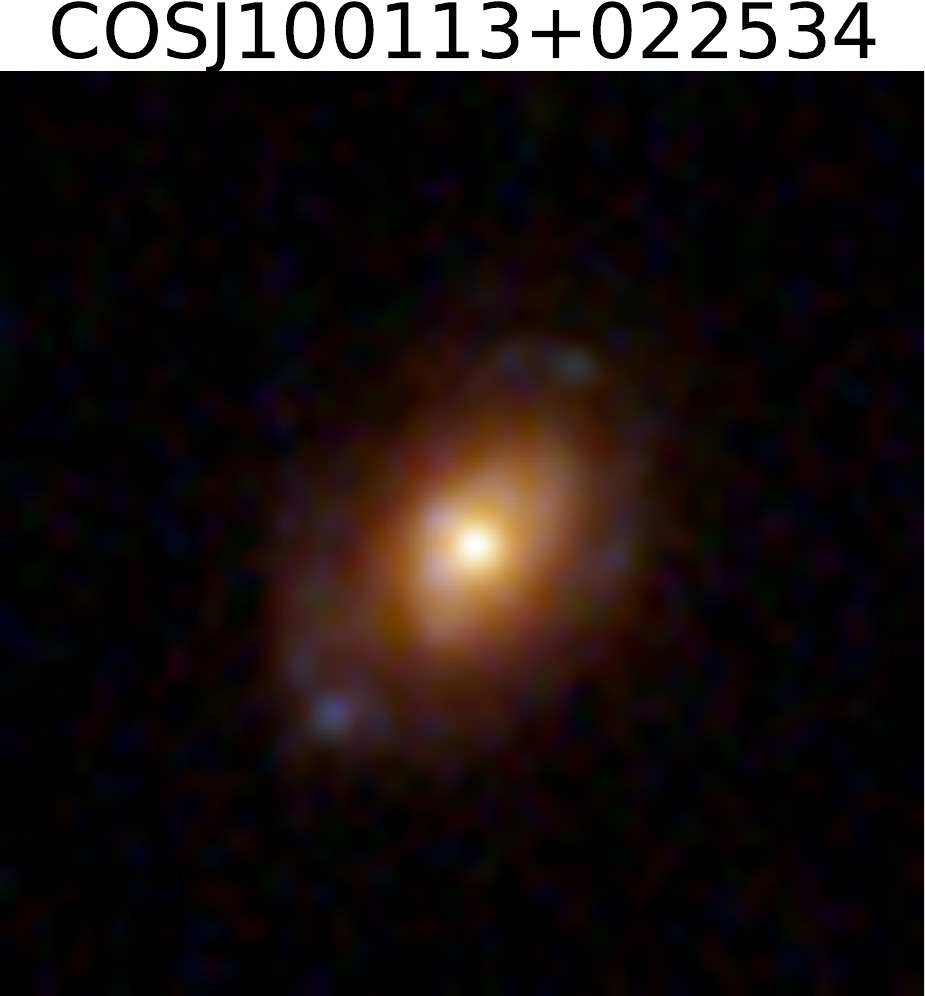}
\includegraphics[width=0.12\textwidth]{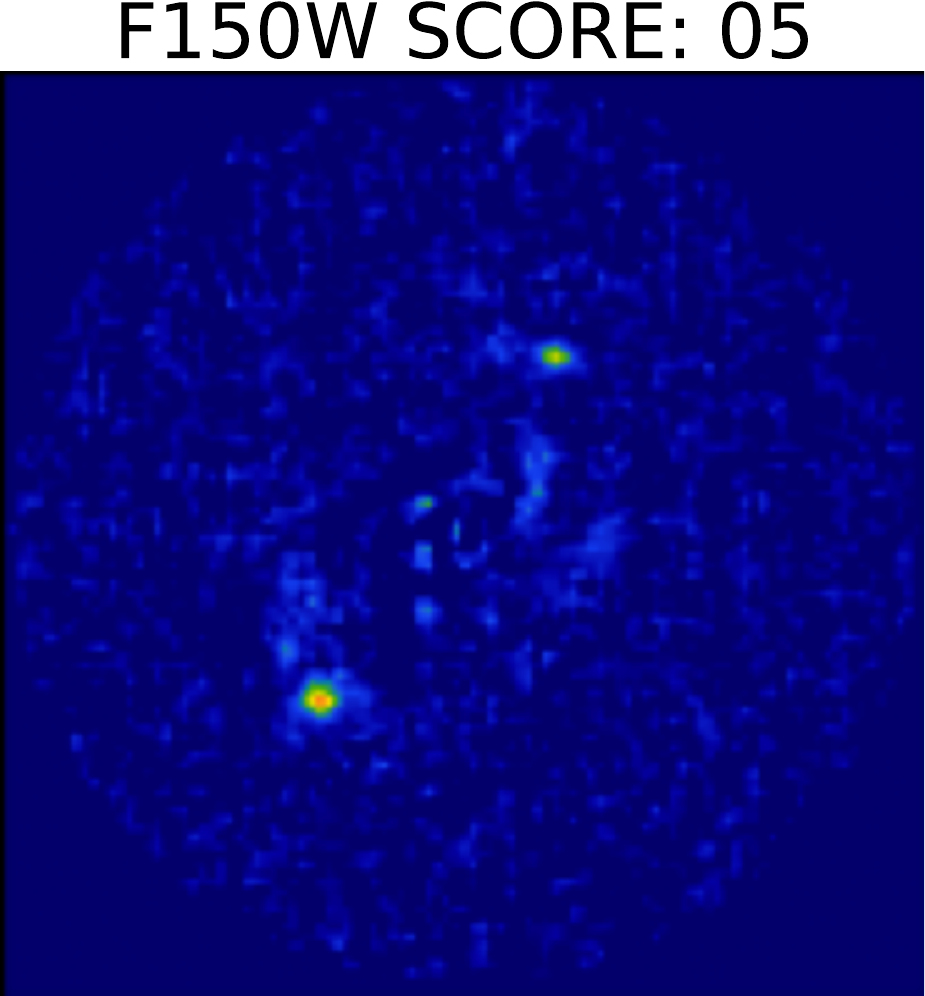}
\includegraphics[width=0.12\textwidth]{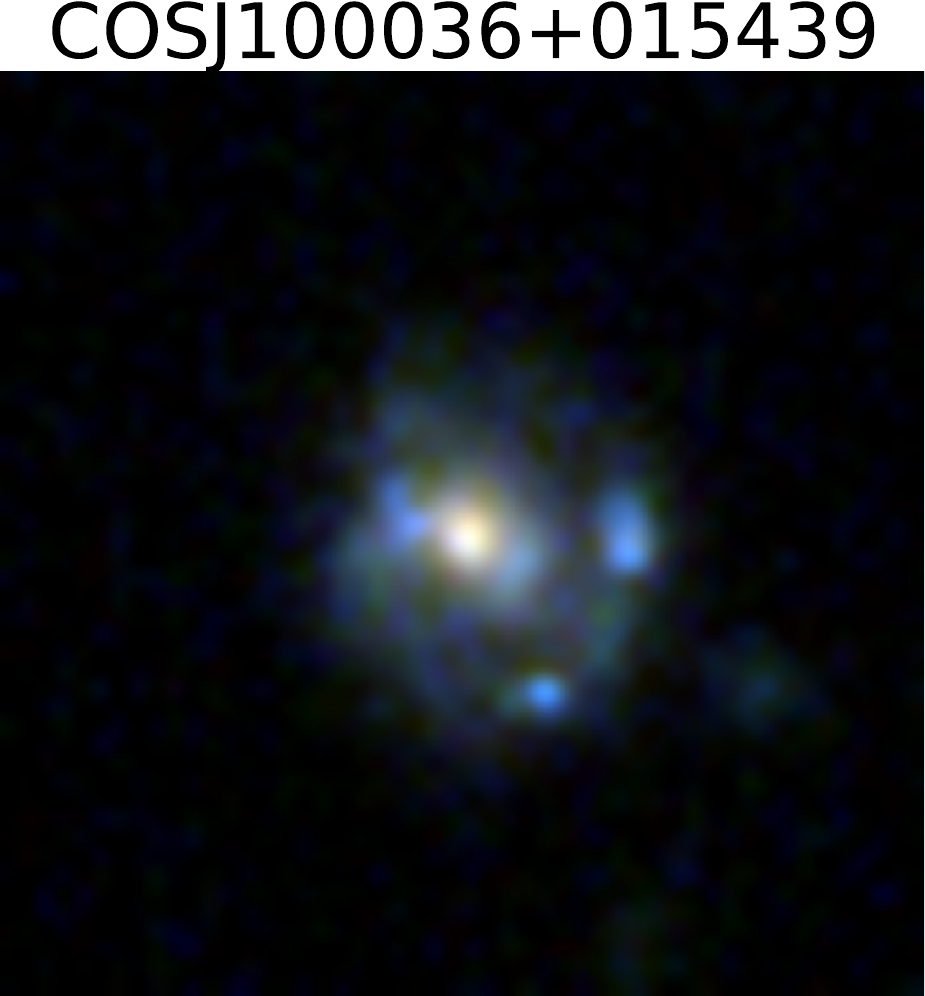}
\includegraphics[width=0.12\textwidth]{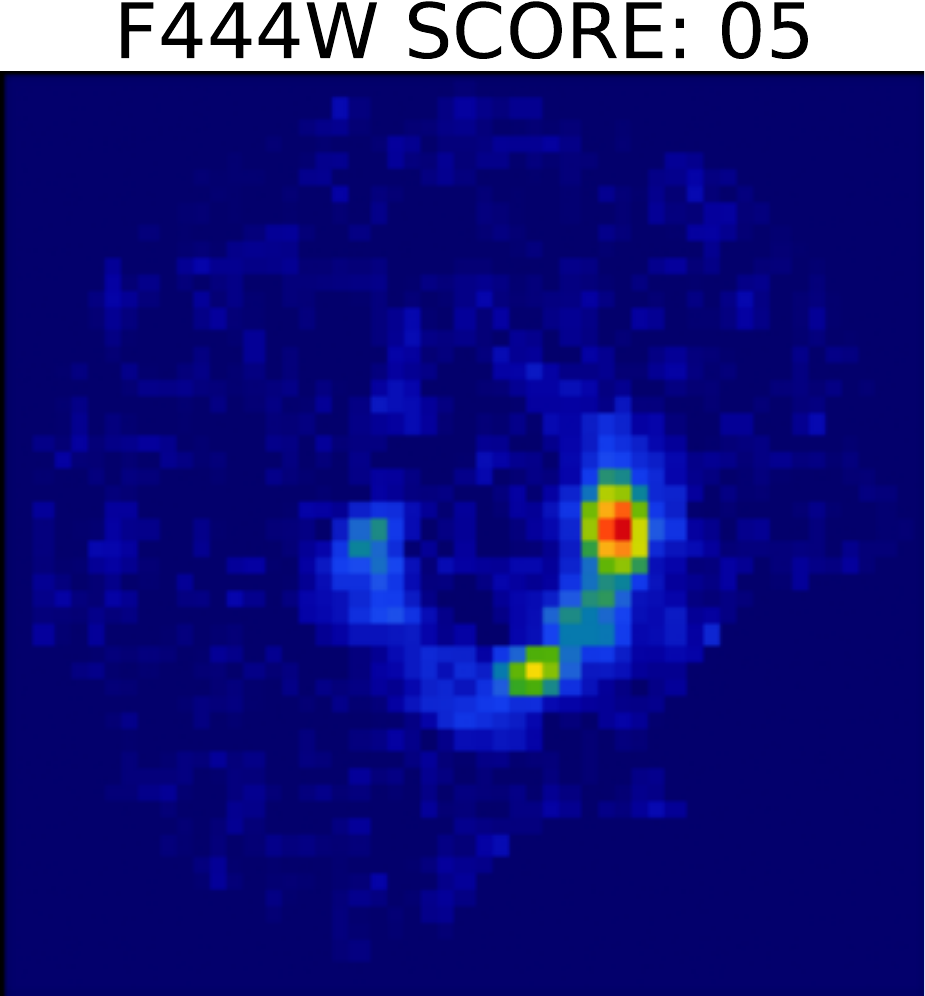}
\includegraphics[width=0.12\textwidth]{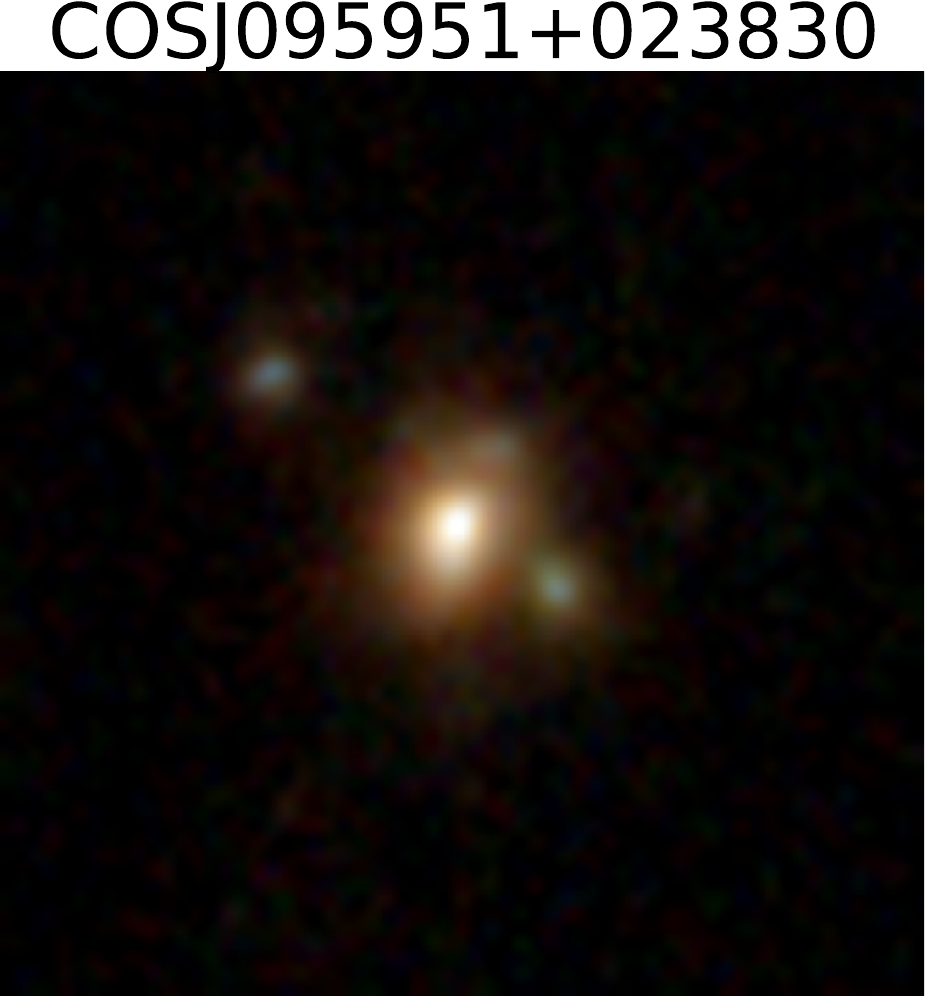}
\includegraphics[width=0.12\textwidth]{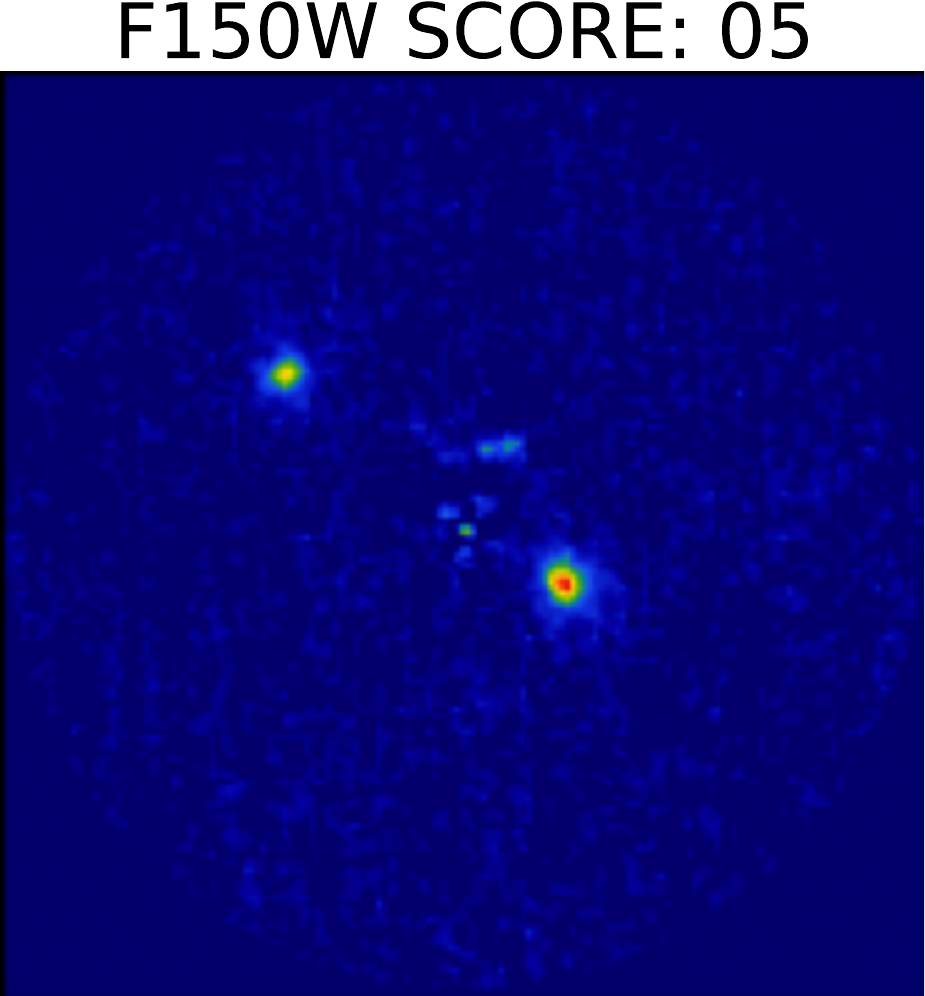}
\caption{
Figure \ref{figure:CutoutA} continued.
}
\label{figure:CutoutA4}
\vspace{-9pt} 
\end{figure*}

Figure \ref{figure:CutoutA} shows RGB colour composite images and foreground galaxy subtracted images inferred via lens modelling for the M25 spectacular lenses and 143 candidates which scored 5 or above. Single wavelength images which most clearly show the candidate lensed source emission are shown. For conciseness, only two postage stamp cut-out images are shown for each candidate, however inspectors had access to more information when grading these lenses, including data from all four wavelengths and lens models. As a result, some objects with high scores might not appear to be convincing strong lenses based solely on these single cut-outs but are more persuasive when viewed with all the available data, and vice versa. All images used for the second round of inspection can be accessed at the following URL:~\github{https://github.com/Jammy2211/COWLS_COSMOS_Web_Lens_Survey}.

Figure \ref{figure:CutoutA} shows the M25 lenses followed by the 143 candidates in descending order of score, from the top left, with score decreasing right and then down. Readers can therefore scan across and down this figure, and assess how the quality of candidates changes with decreasing score. The majority of candidates show features distinctive of strong lensing, for example emission from a central lens galaxy, multiple images in locations consistent with a strong lens model and arcs tangentially stretched relative to the lens galaxy centre. However, especially for the lower scoring candidates, many of these features are hard to discern from complex and irregular morphological structures often seen in deep imaging of high redshift galaxies (e.g. spiral arms, rings).

\begin{figure*}
\centering
\includegraphics[width=0.12\textwidth]{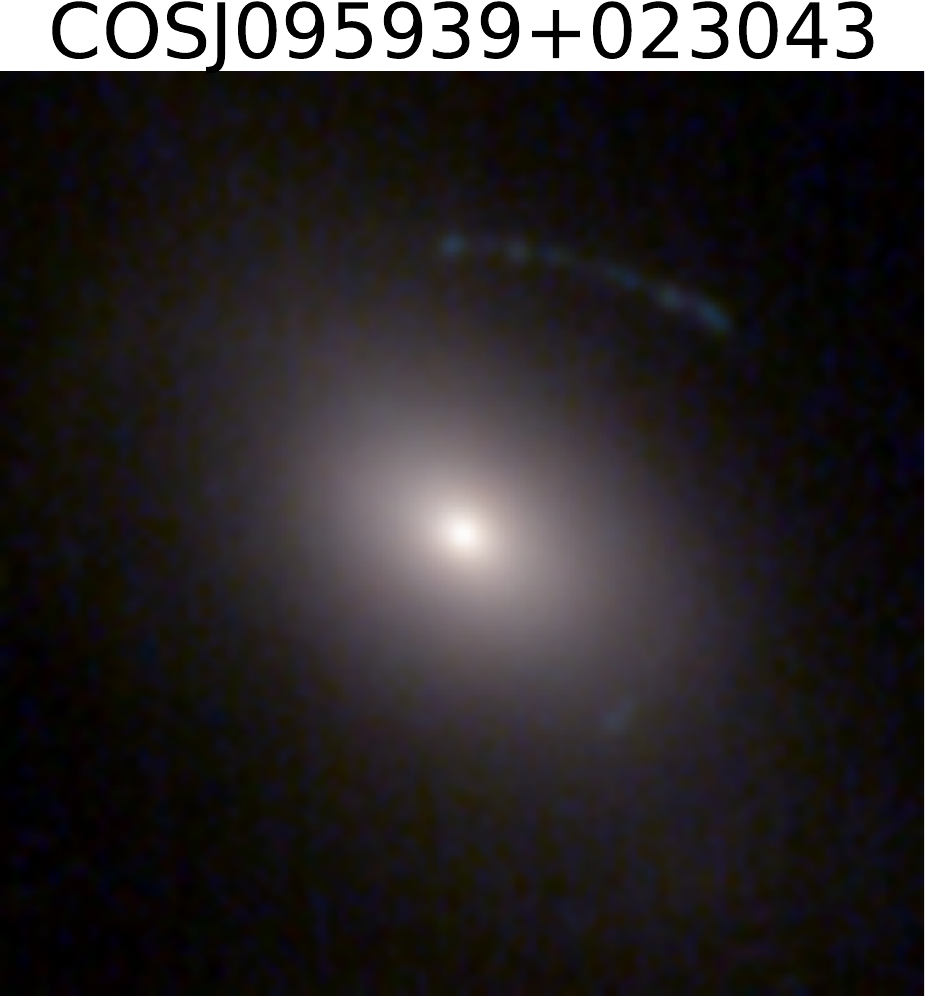}
\includegraphics[width=0.12\textwidth]{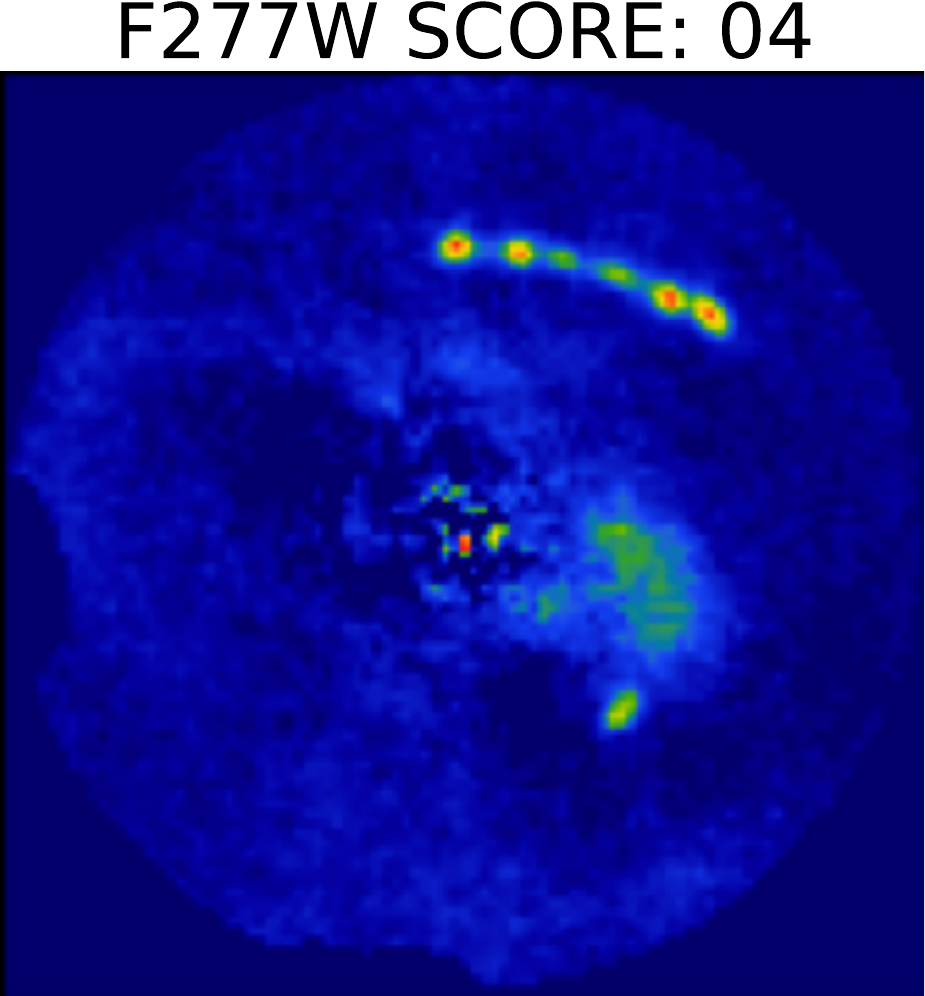}
\includegraphics[width=0.12\textwidth]{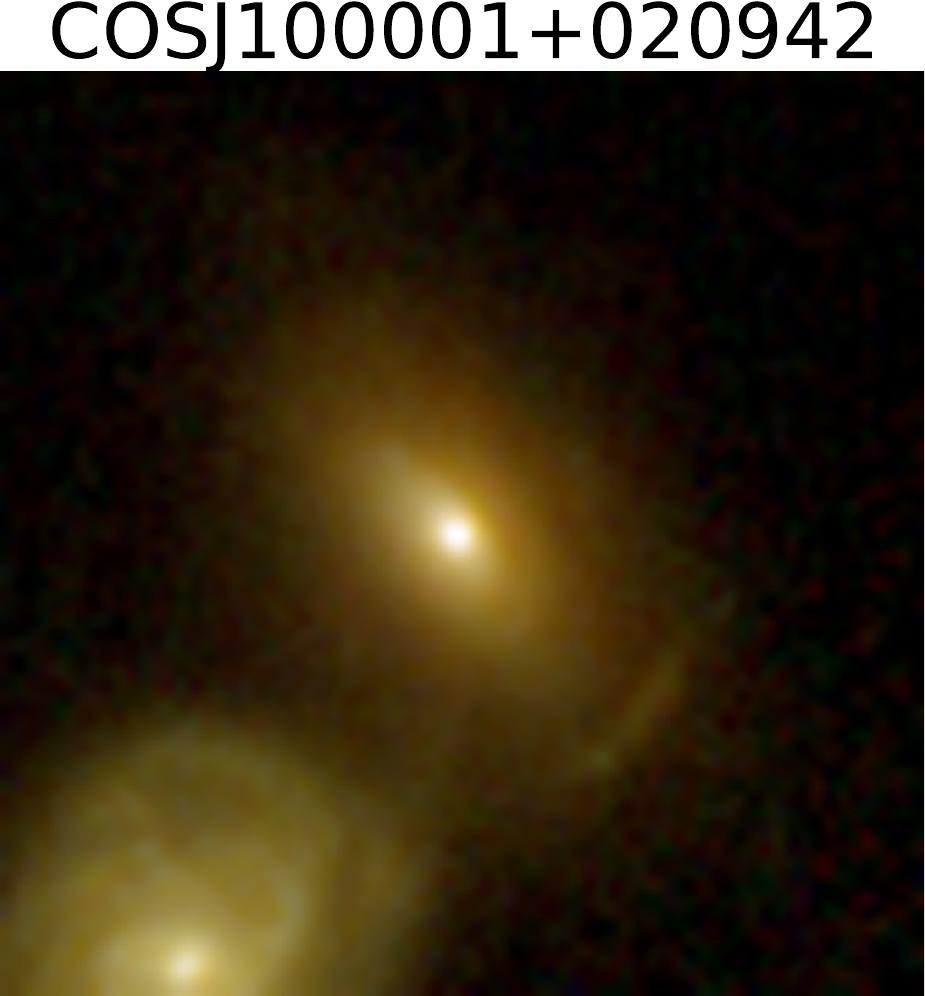}
\includegraphics[width=0.12\textwidth]{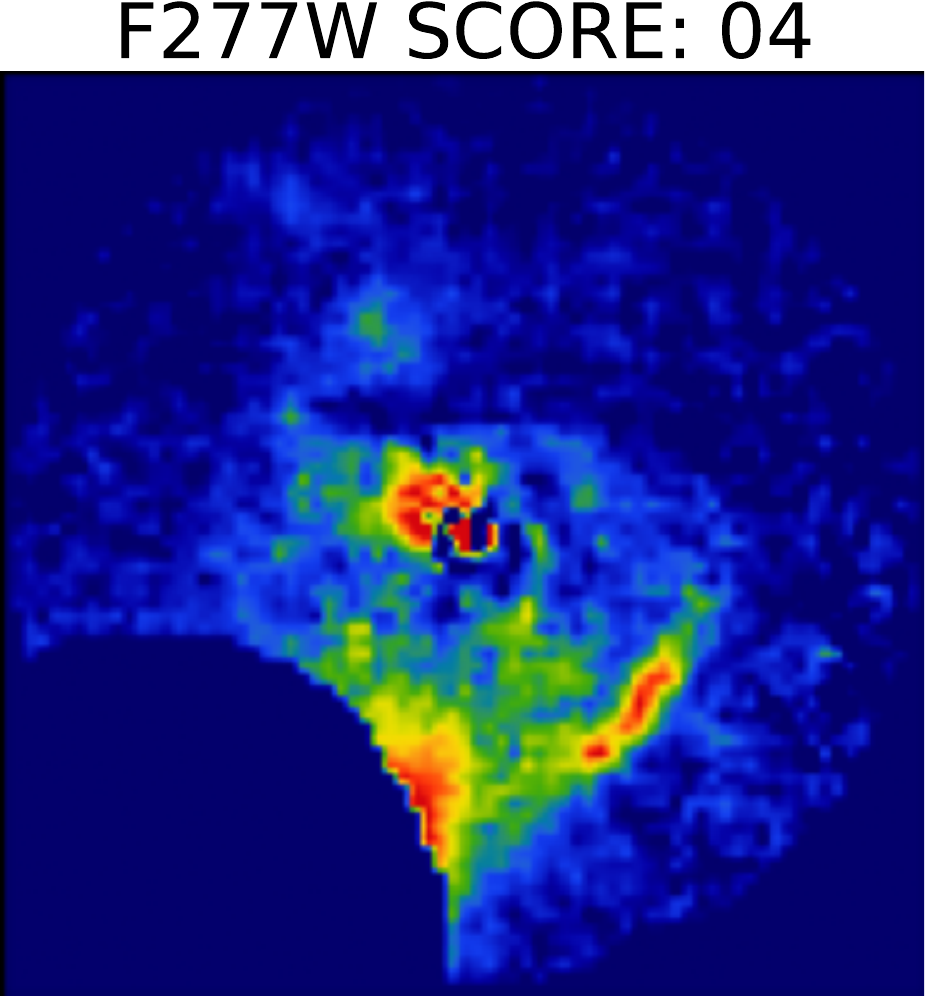}
\includegraphics[width=0.12\textwidth]{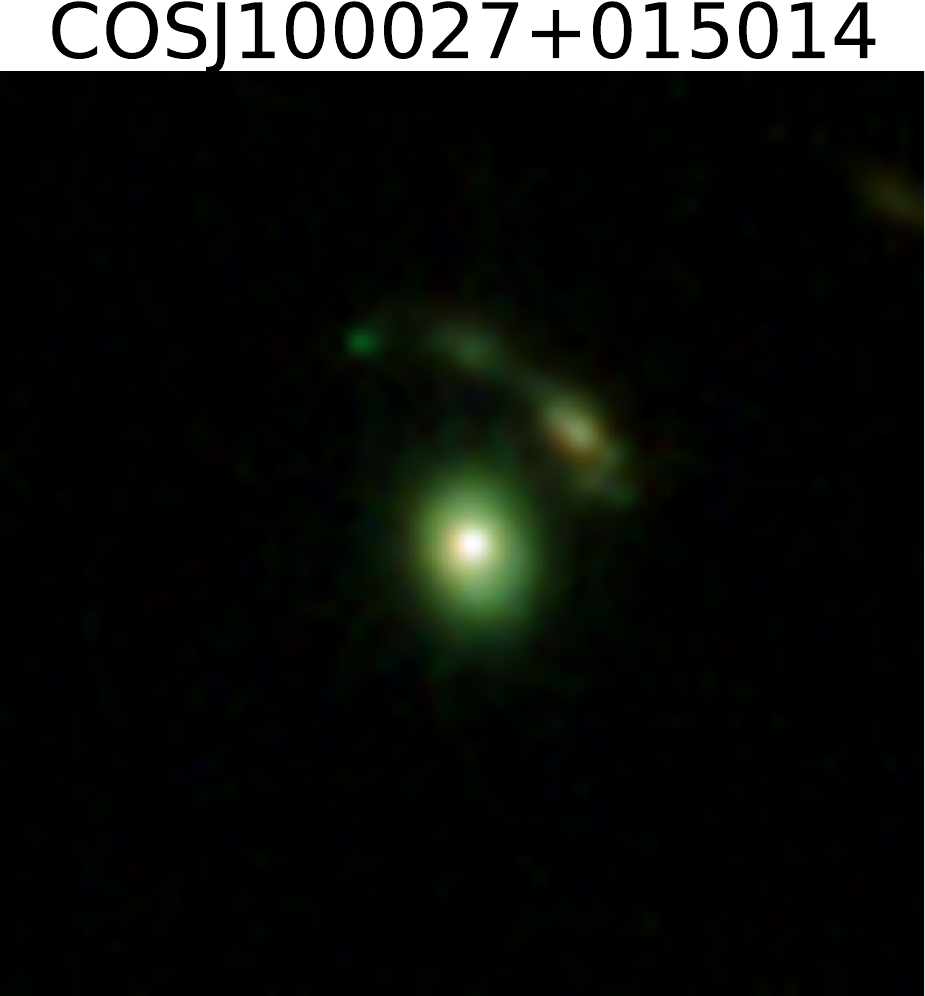}
\includegraphics[width=0.12\textwidth]{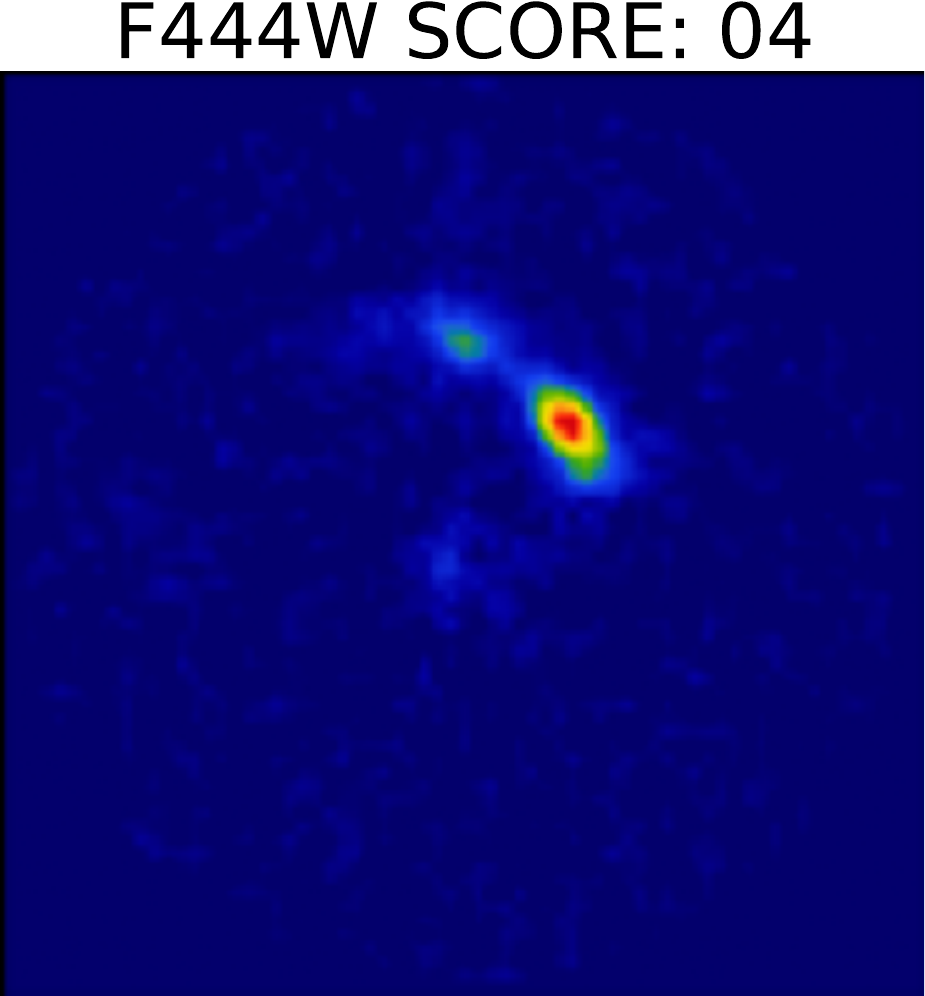}
\includegraphics[width=0.12\textwidth]{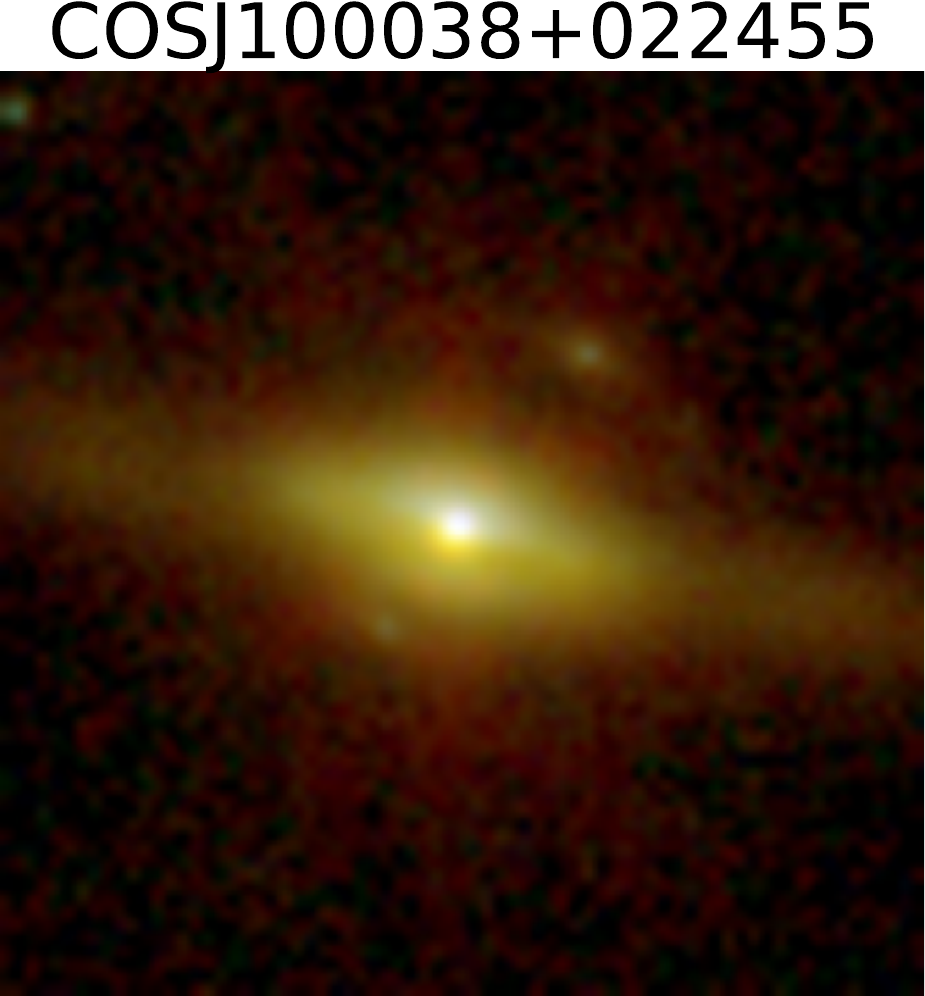}
\includegraphics[width=0.12\textwidth]{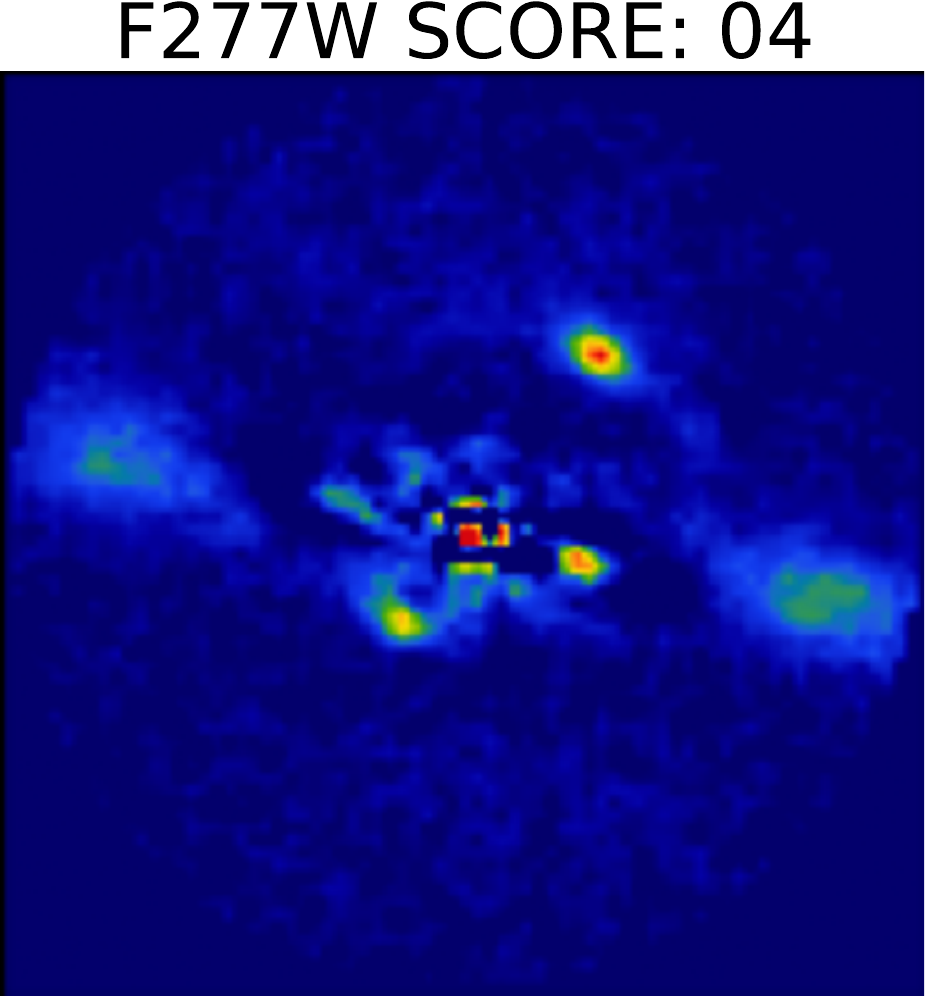}
\includegraphics[width=0.12\textwidth]{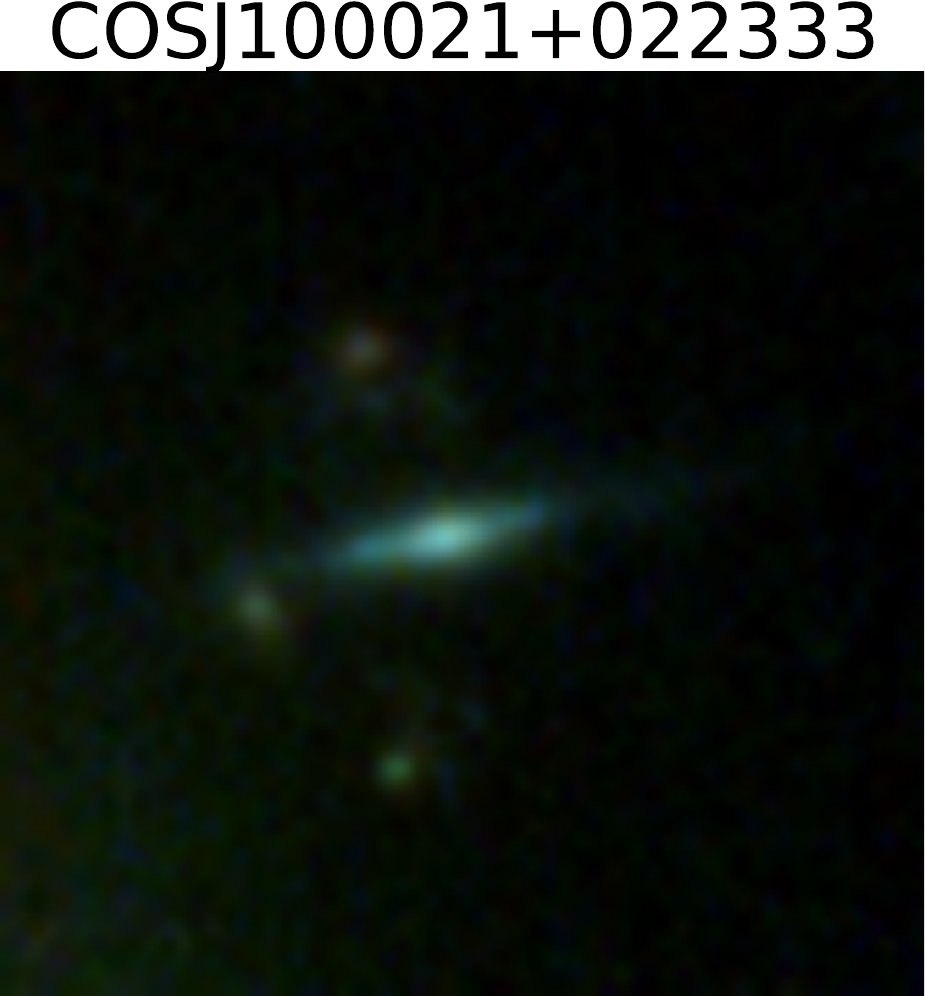}
\includegraphics[width=0.12\textwidth]{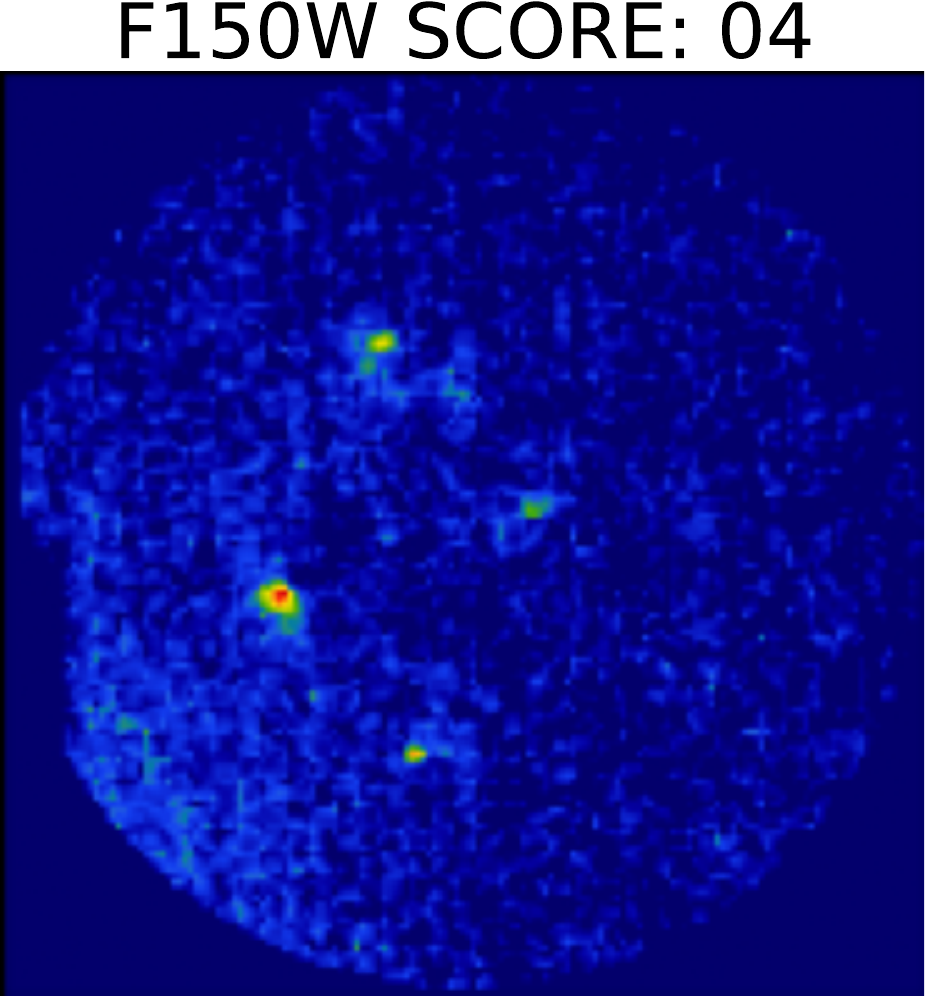}
\includegraphics[width=0.12\textwidth]{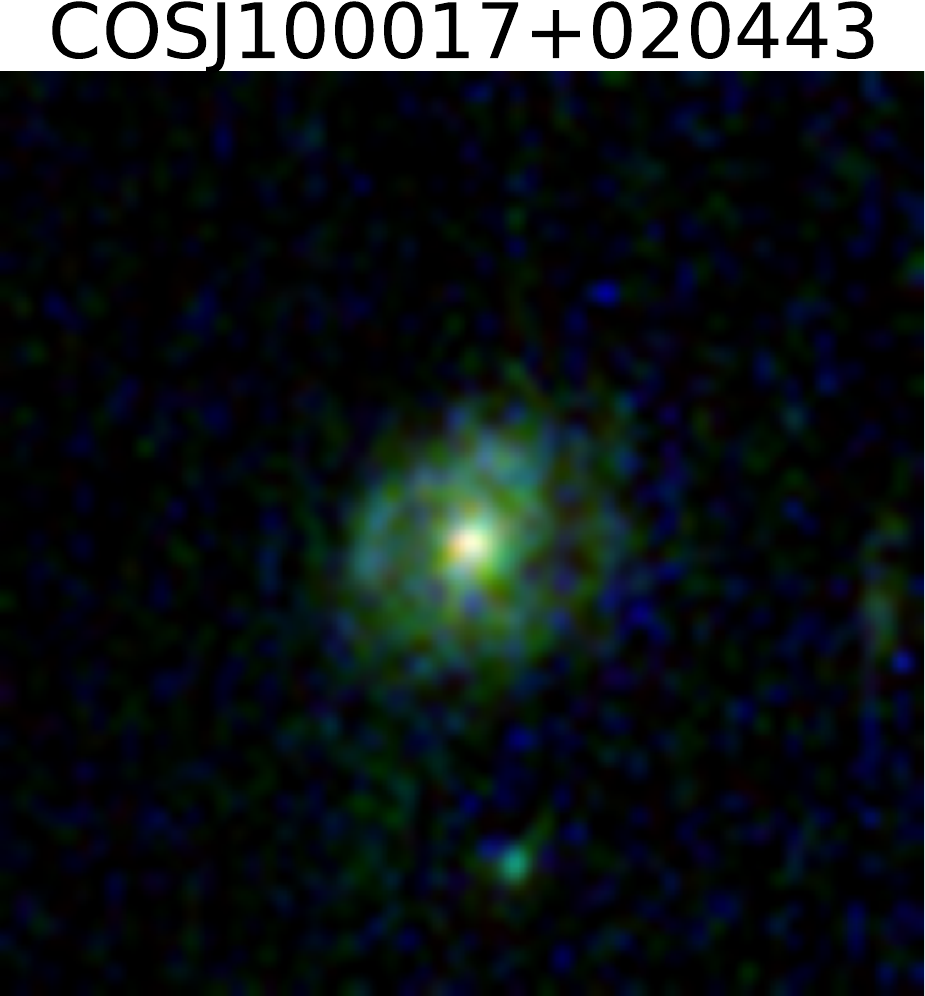}
\includegraphics[width=0.12\textwidth]{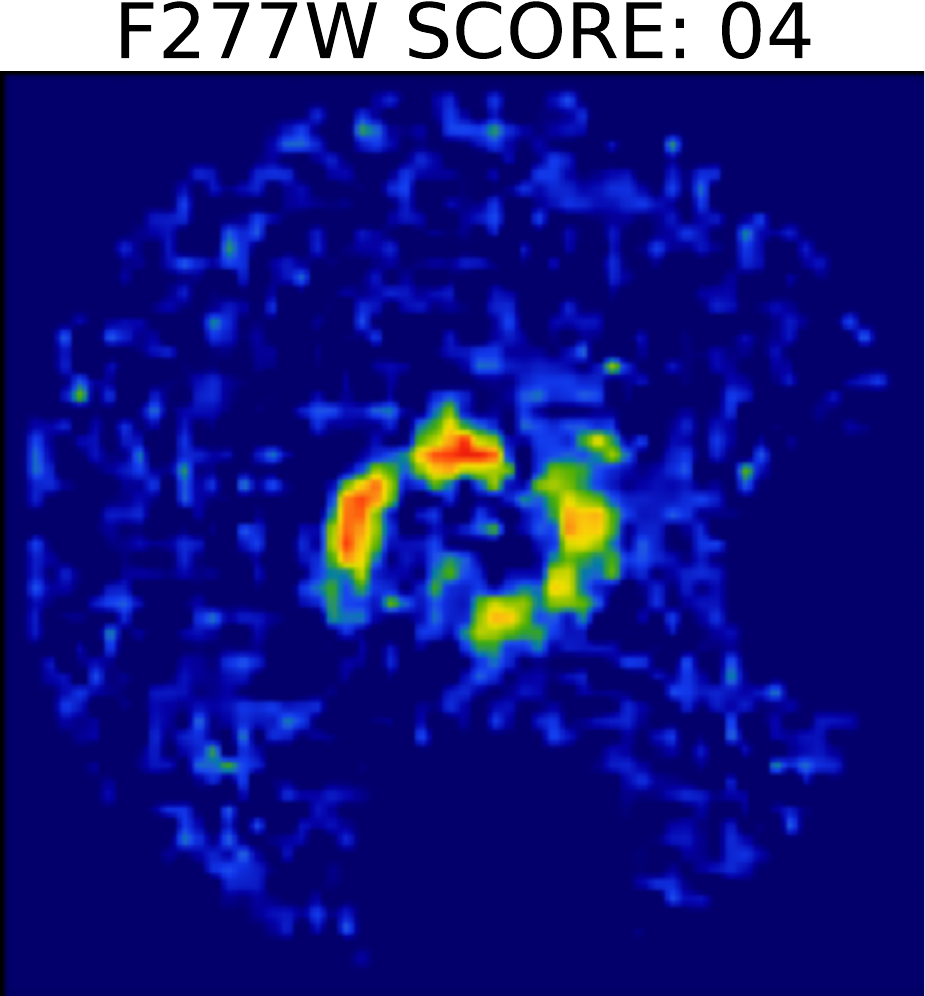}
\includegraphics[width=0.12\textwidth]{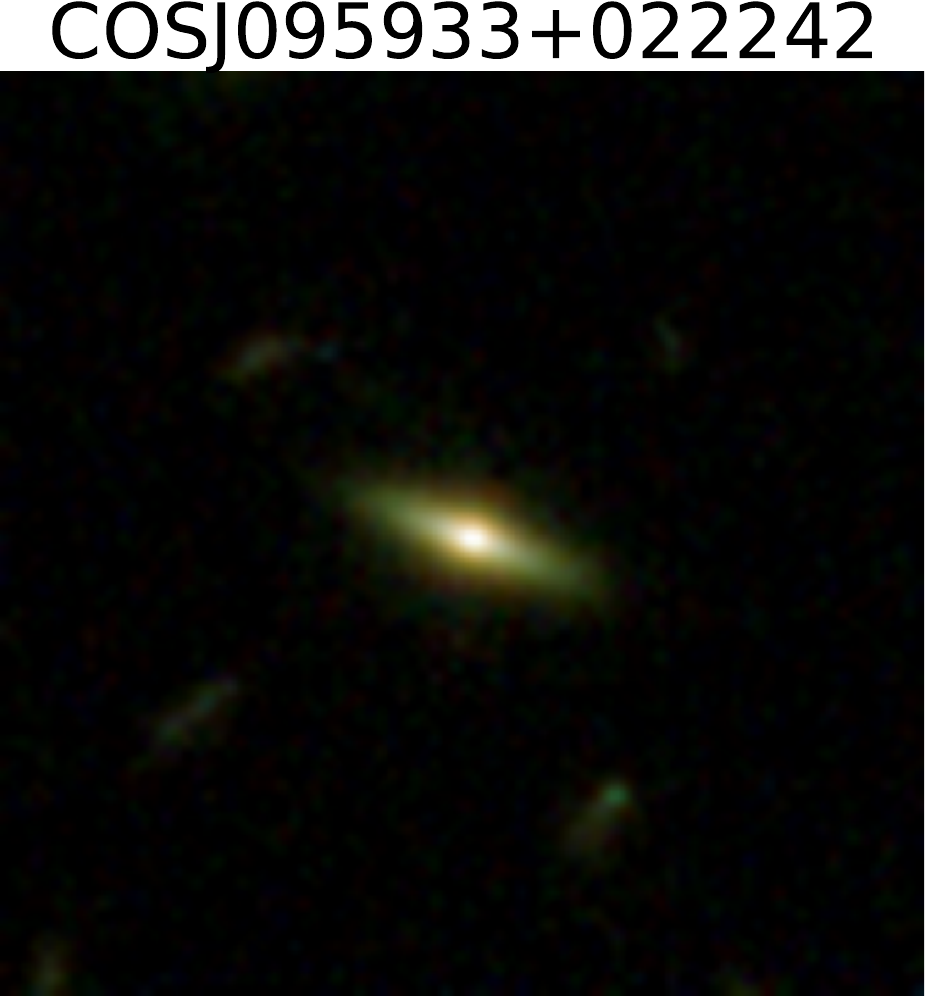}
\includegraphics[width=0.12\textwidth]{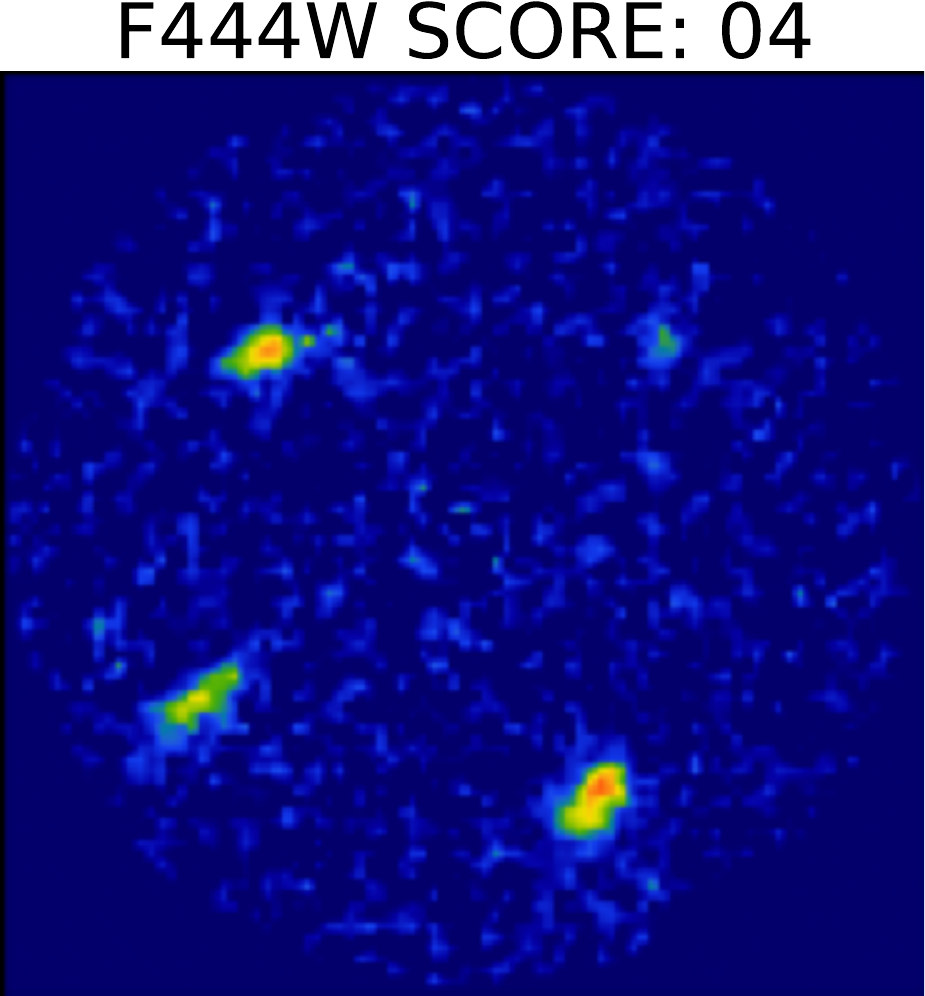}
\includegraphics[width=0.12\textwidth]{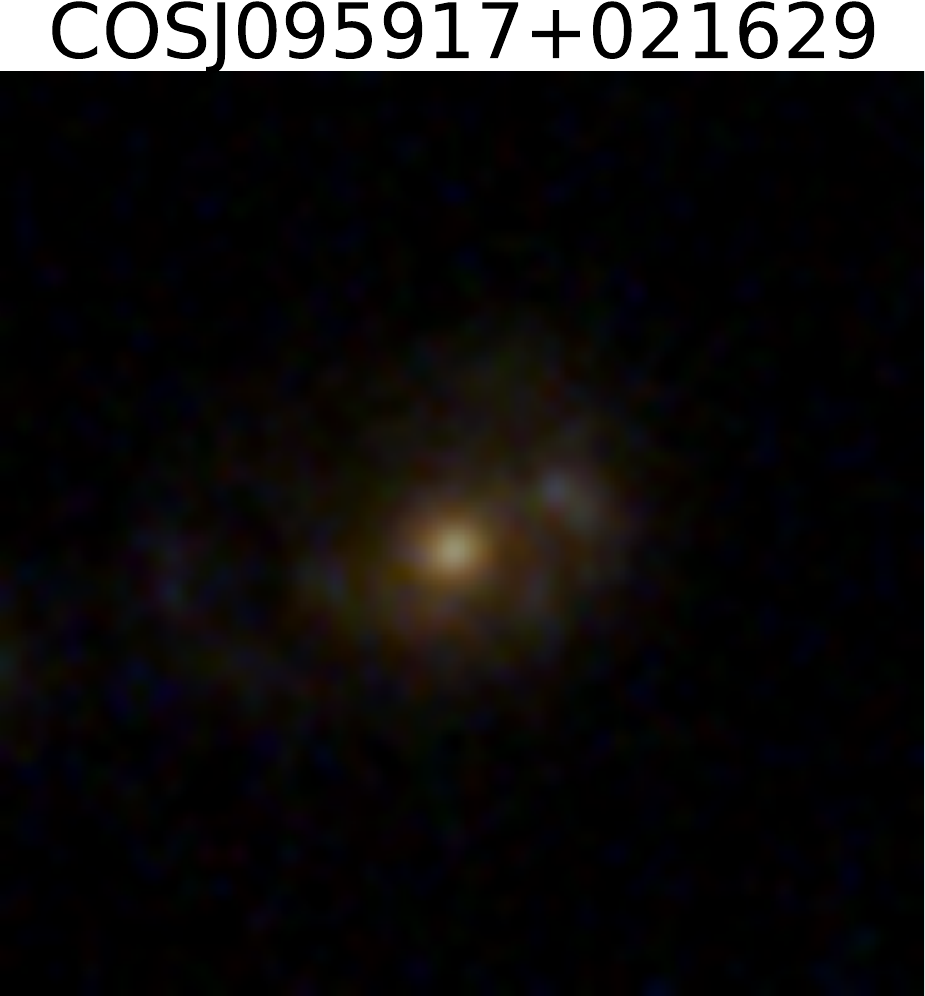}
\includegraphics[width=0.12\textwidth]{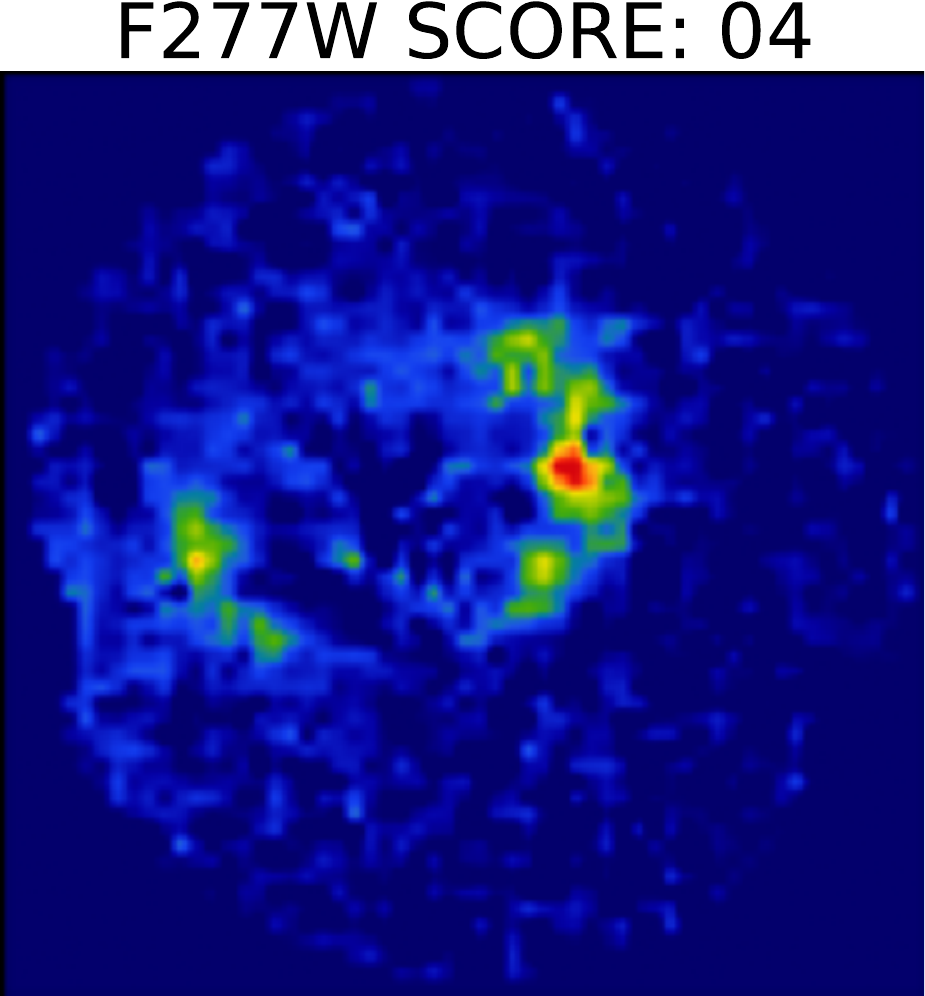}
\includegraphics[width=0.12\textwidth]{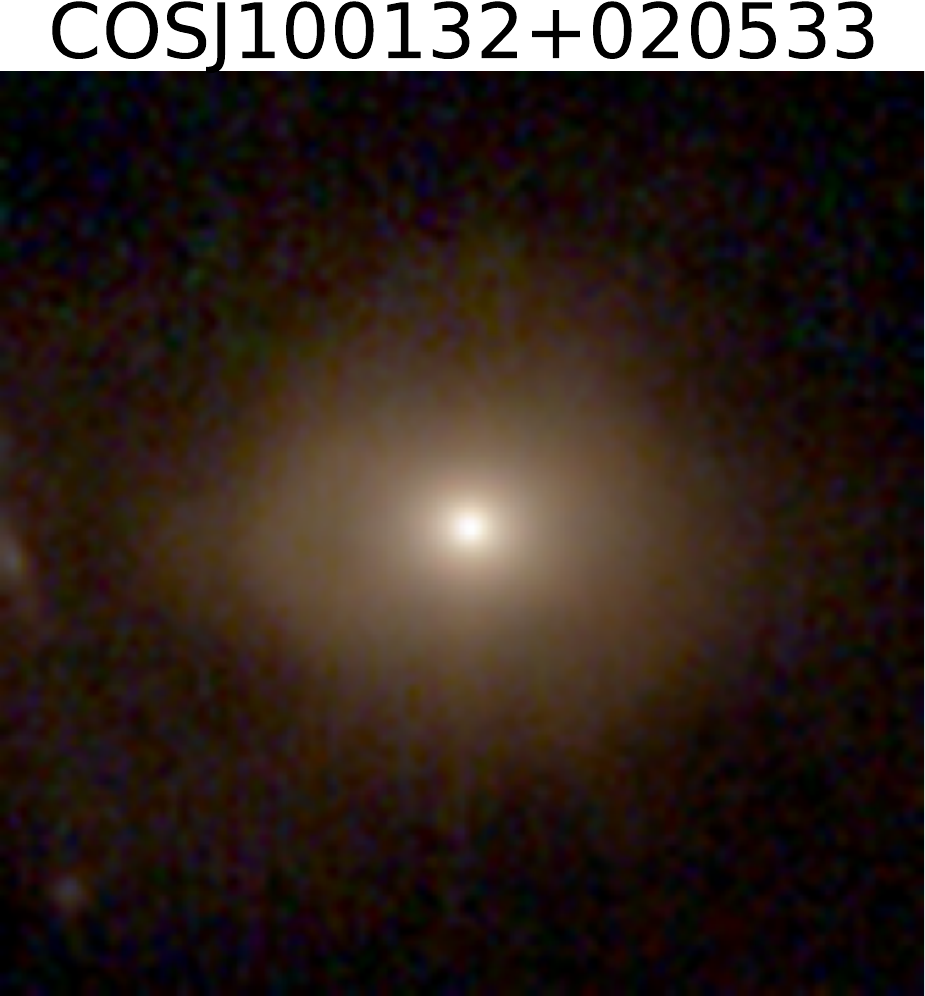}
\includegraphics[width=0.12\textwidth]{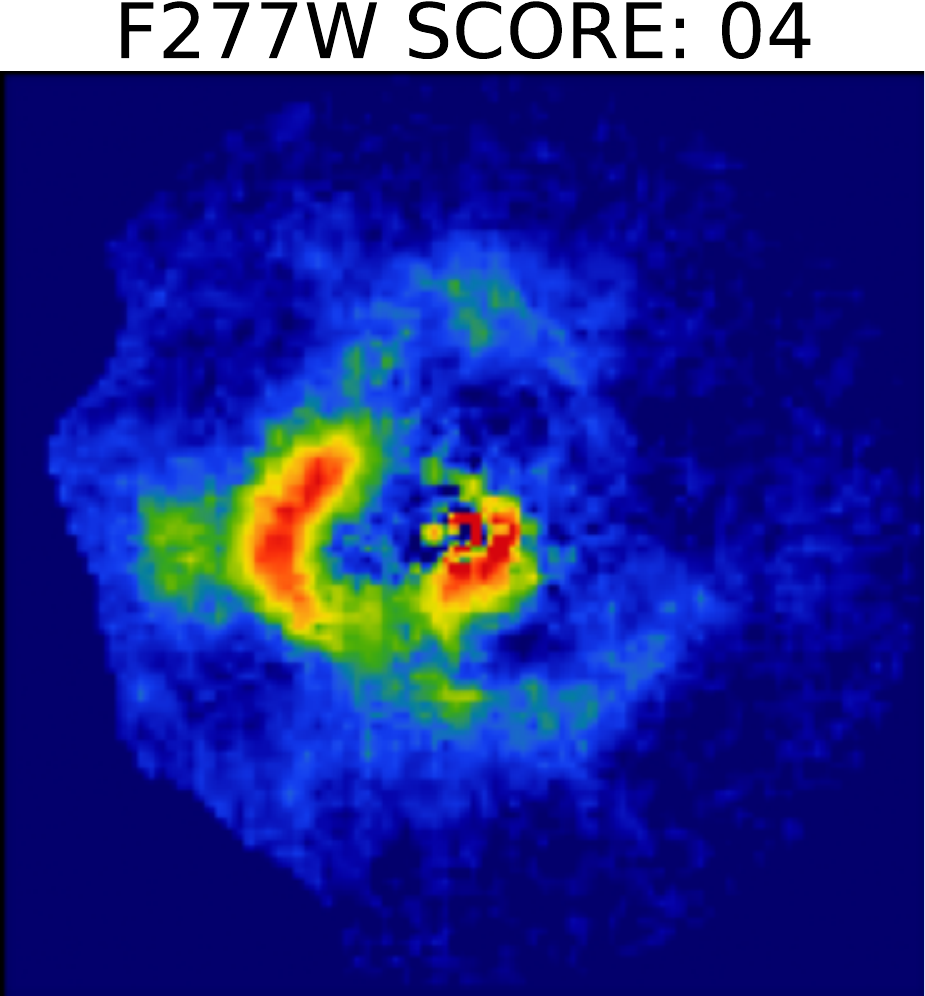}
\includegraphics[width=0.12\textwidth]{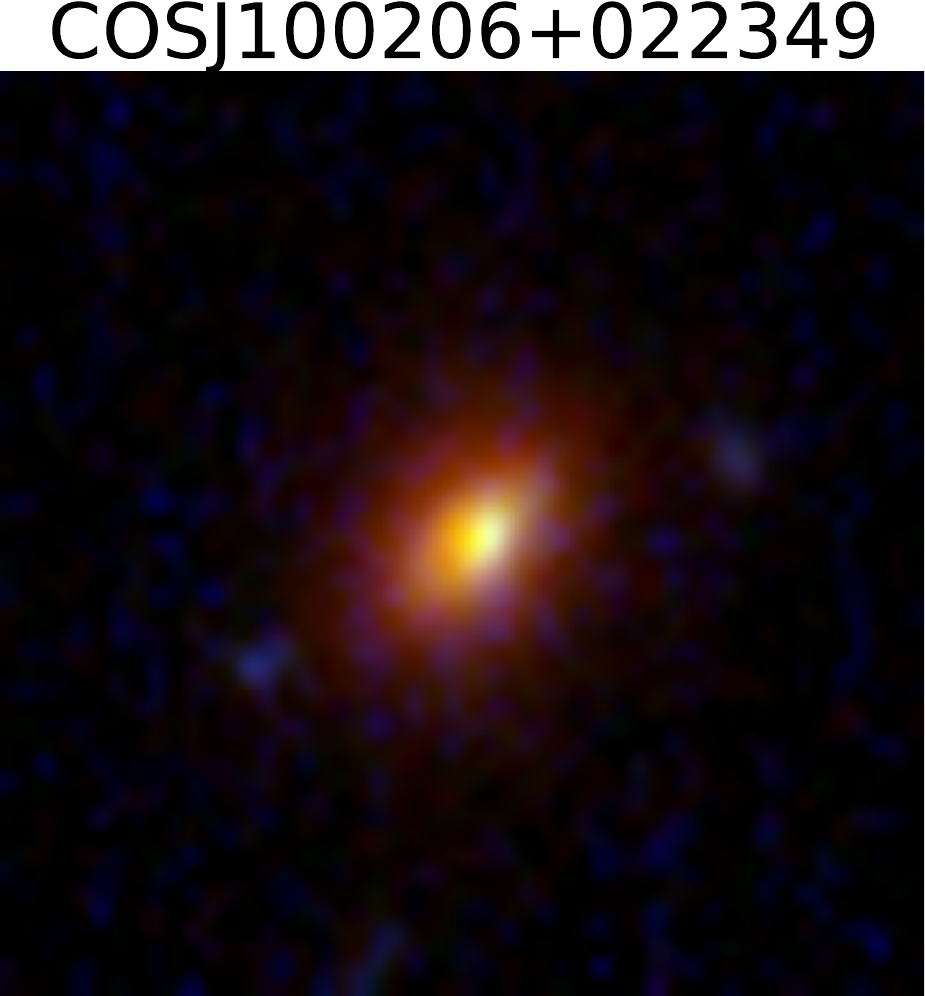}
\includegraphics[width=0.12\textwidth]{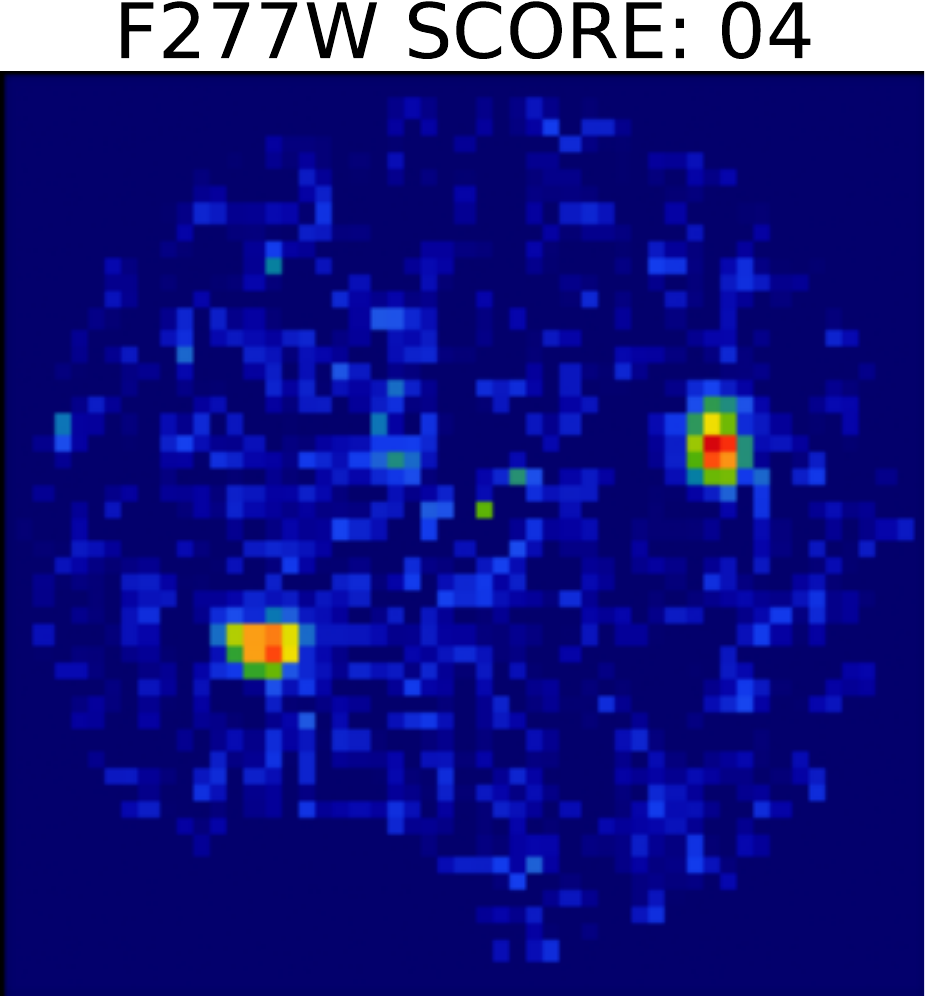}
\includegraphics[width=0.12\textwidth]{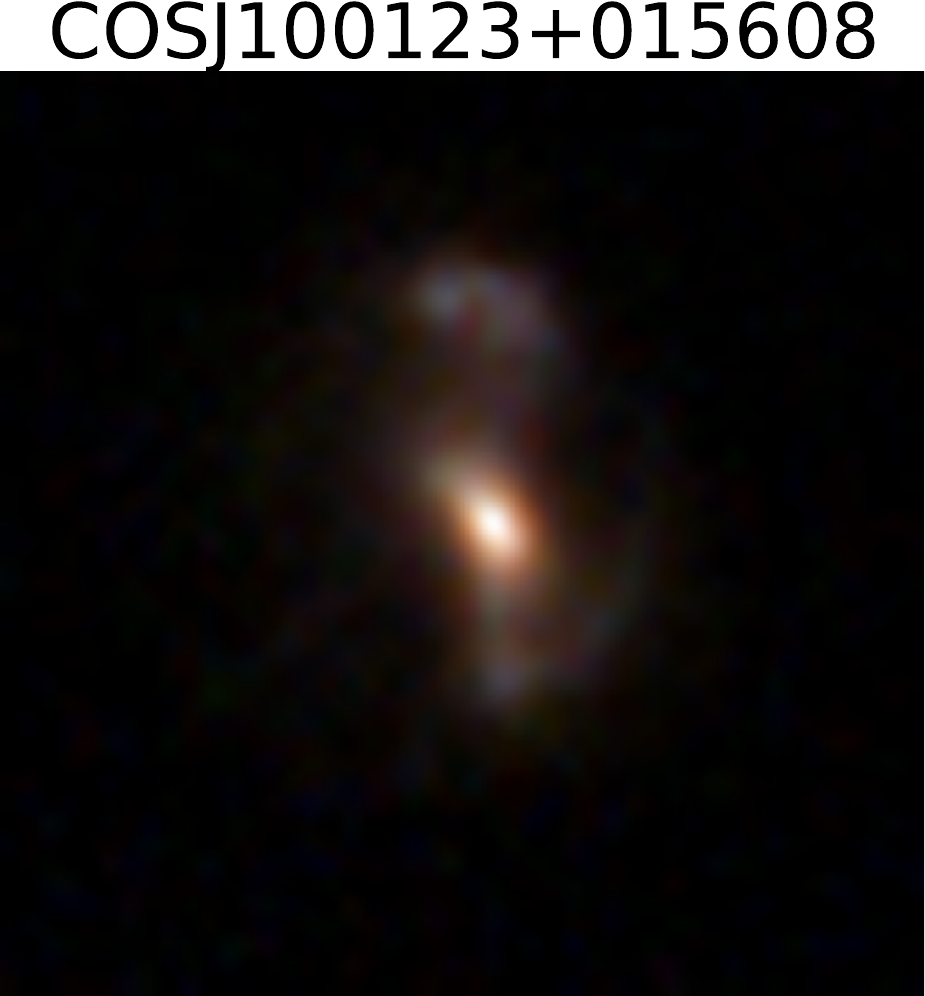}
\includegraphics[width=0.12\textwidth]{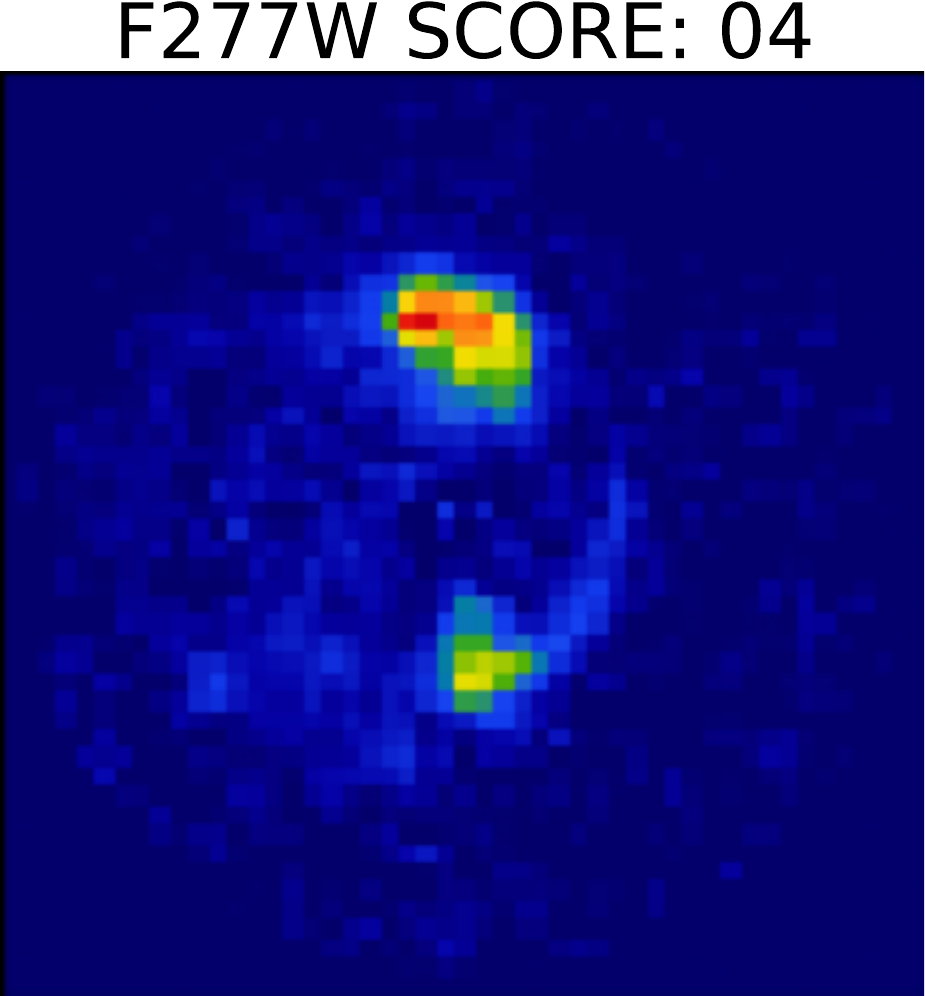}
\includegraphics[width=0.12\textwidth]{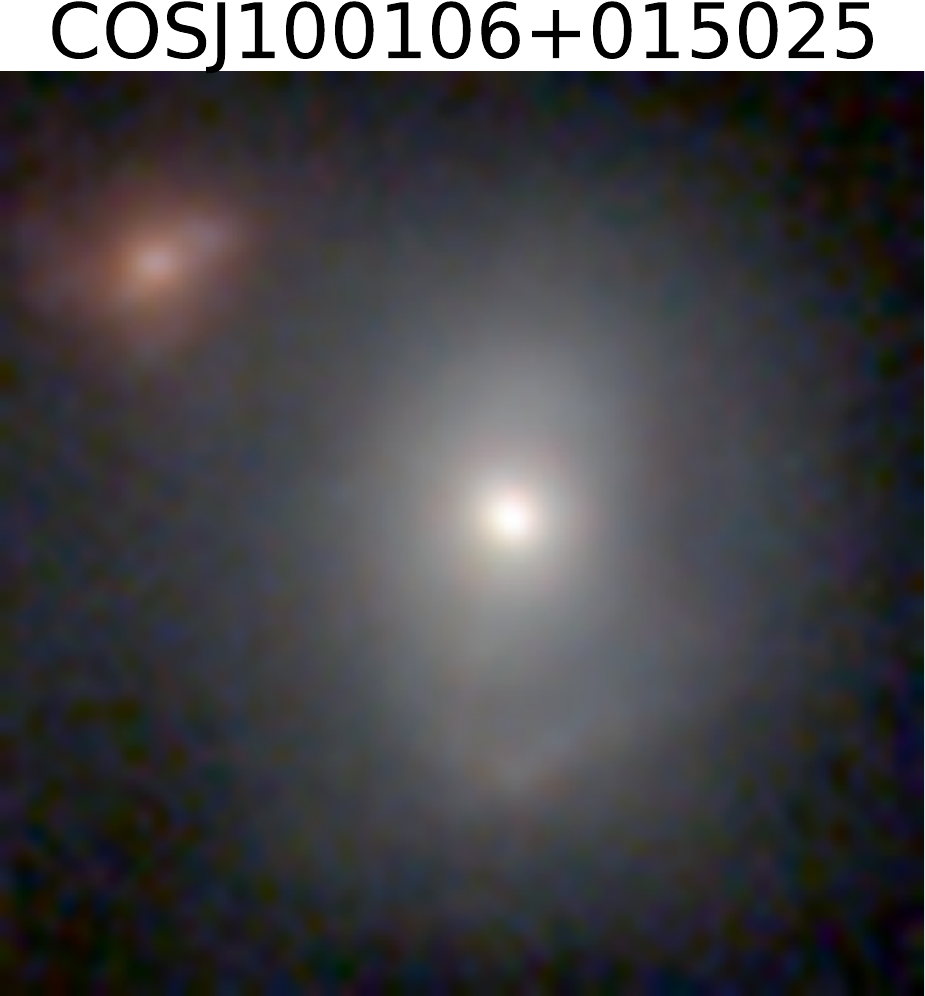}
\includegraphics[width=0.12\textwidth]{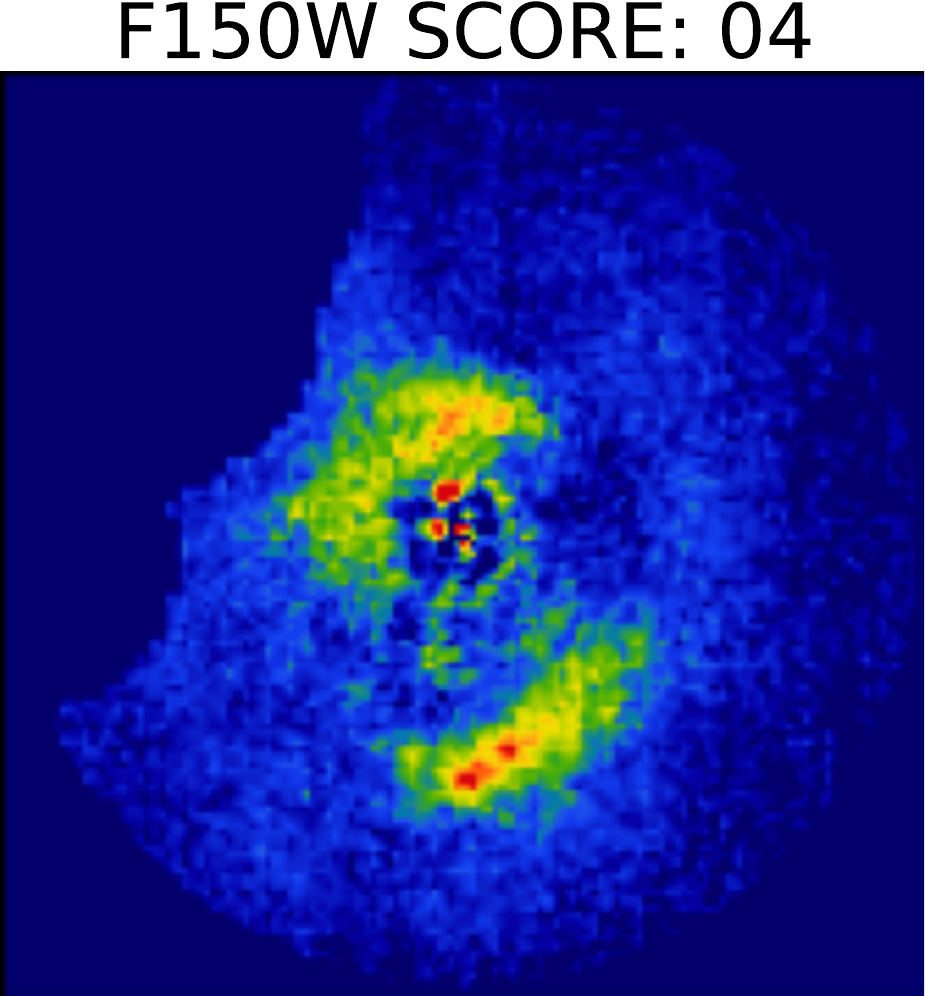}
\includegraphics[width=0.12\textwidth]{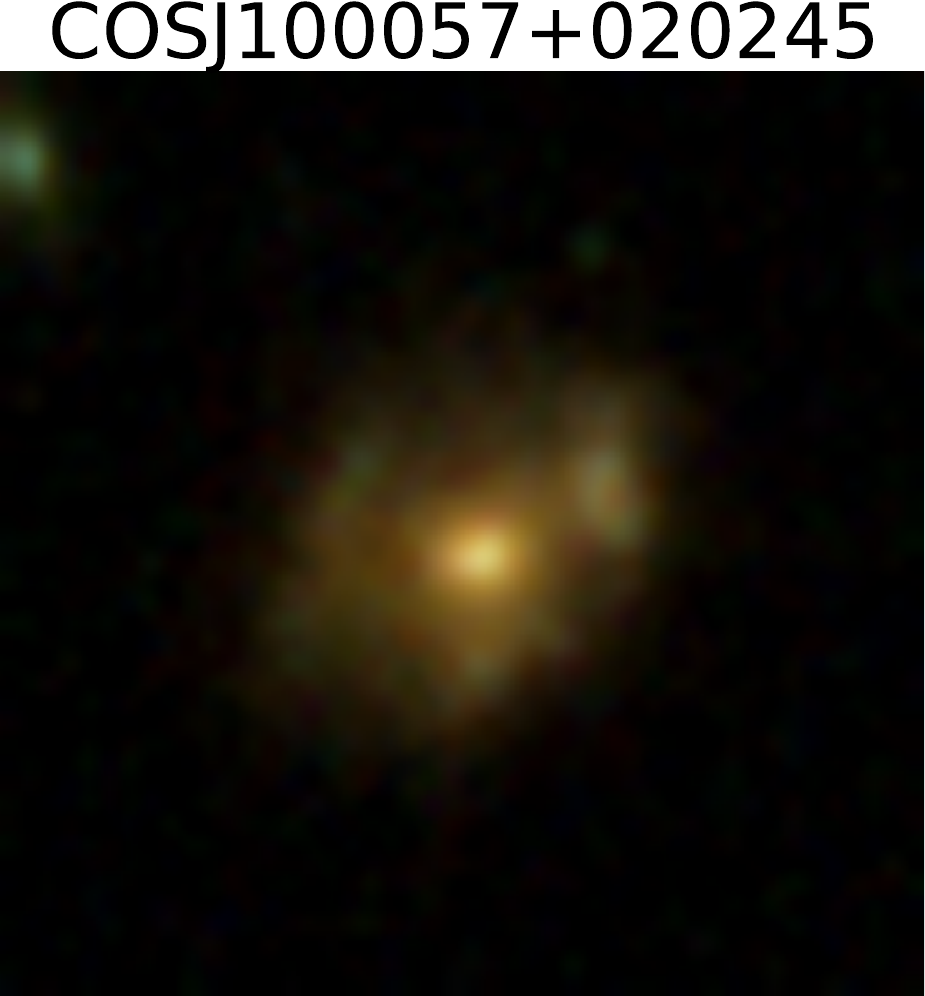}
\includegraphics[width=0.12\textwidth]{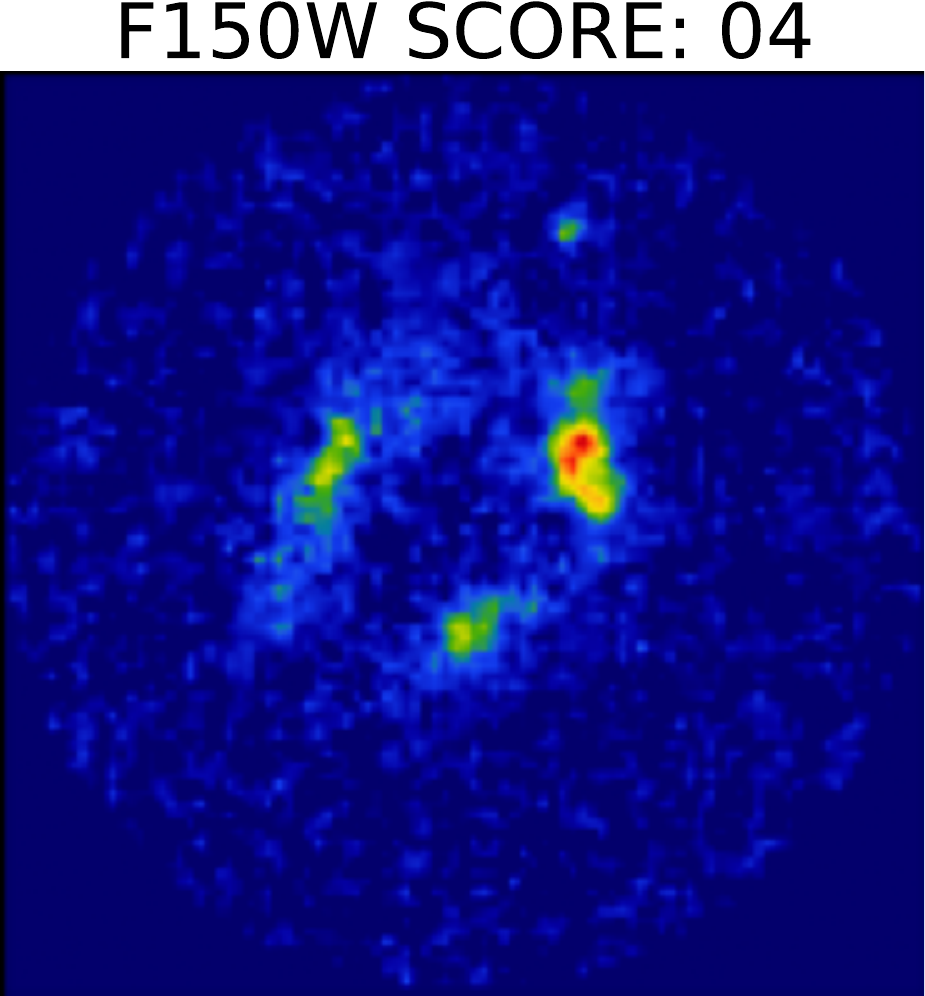}
\includegraphics[width=0.12\textwidth]{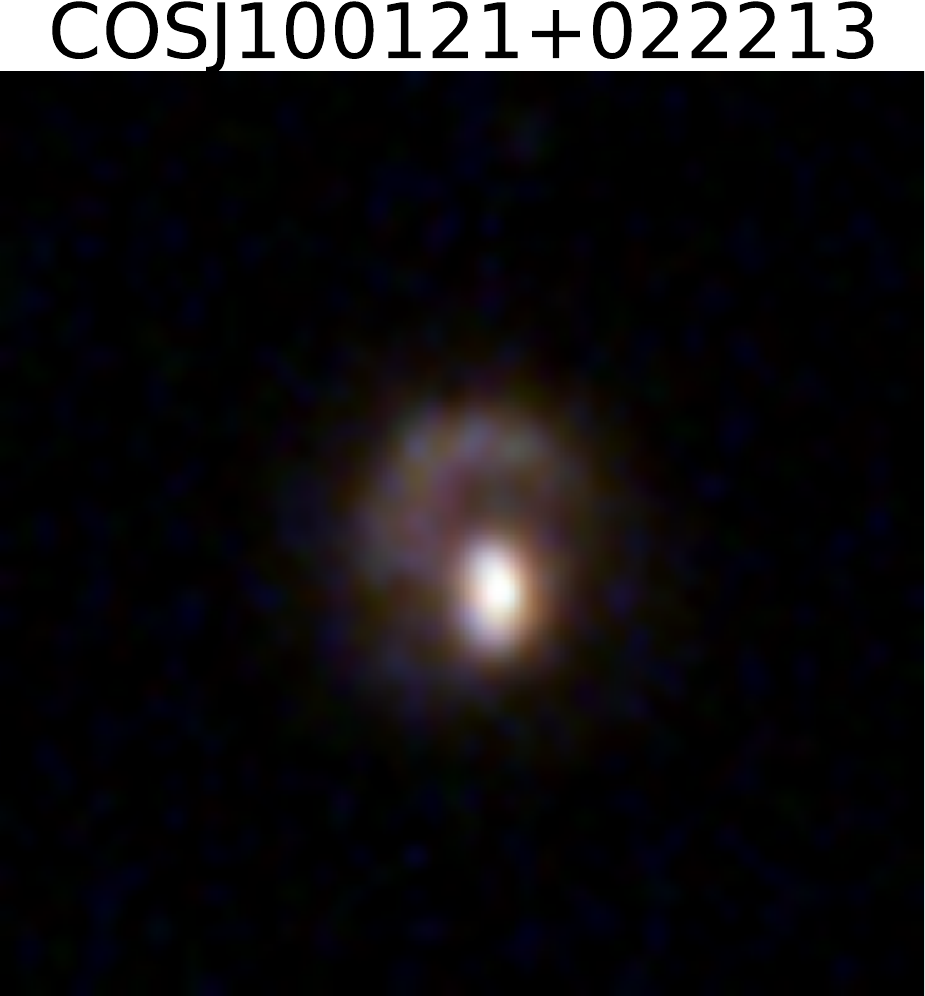}
\includegraphics[width=0.12\textwidth]{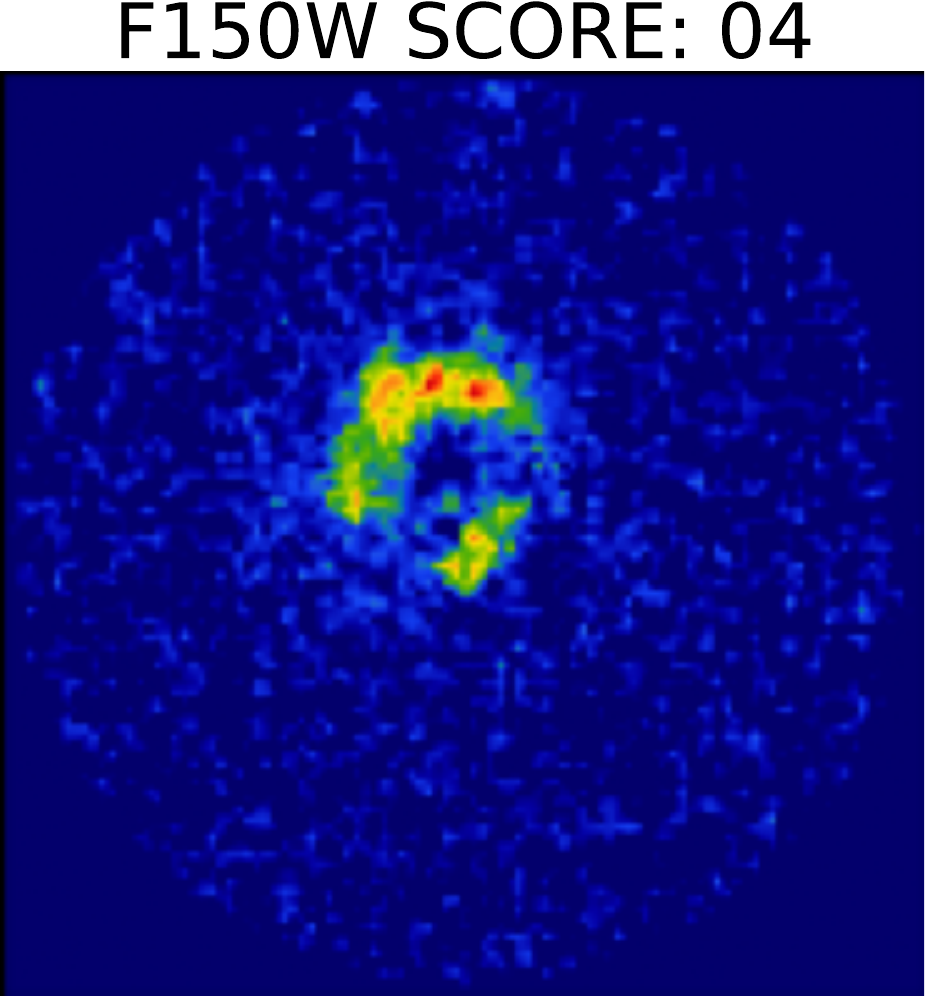}
\includegraphics[width=0.12\textwidth]{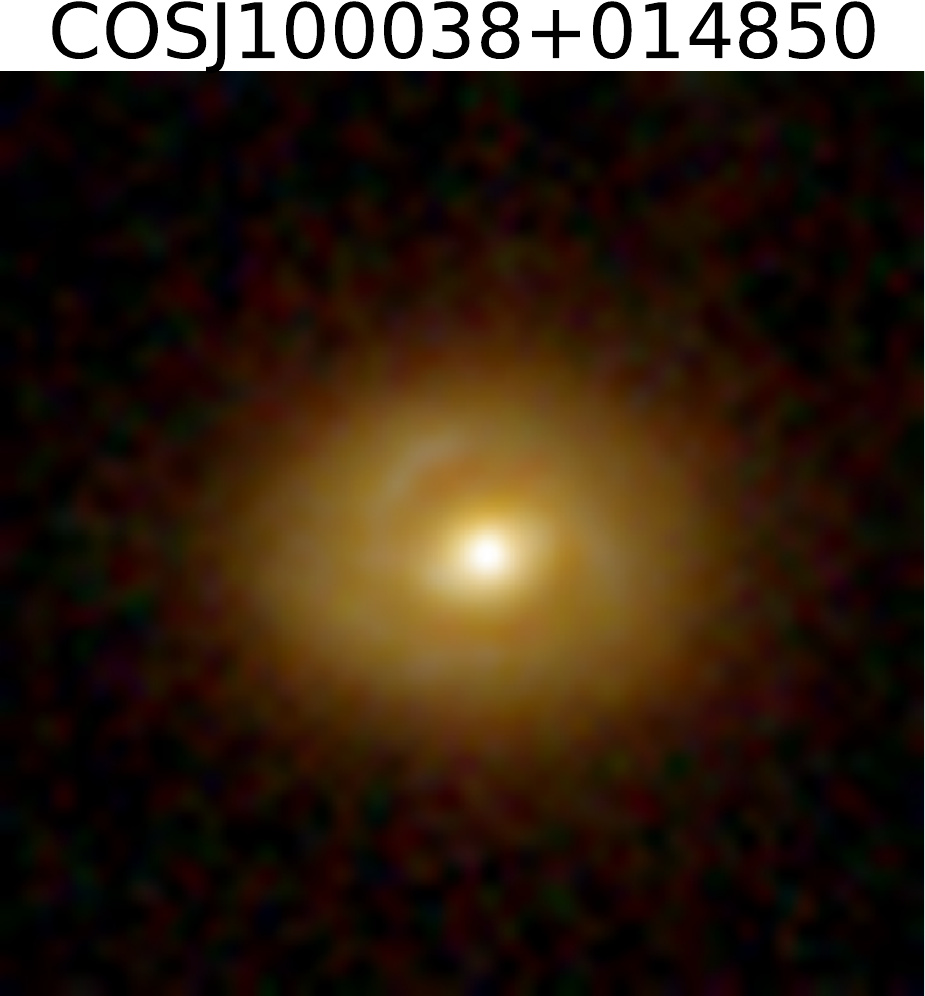}
\includegraphics[width=0.12\textwidth]{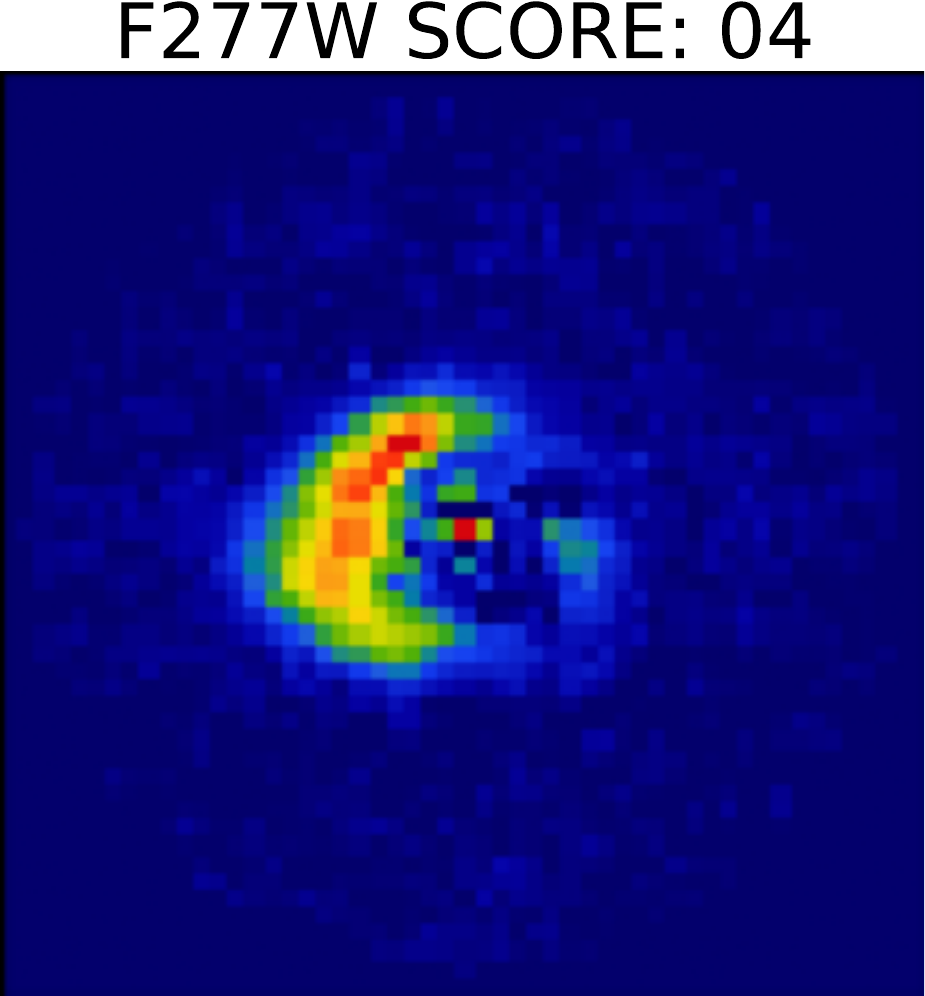}
\includegraphics[width=0.12\textwidth]{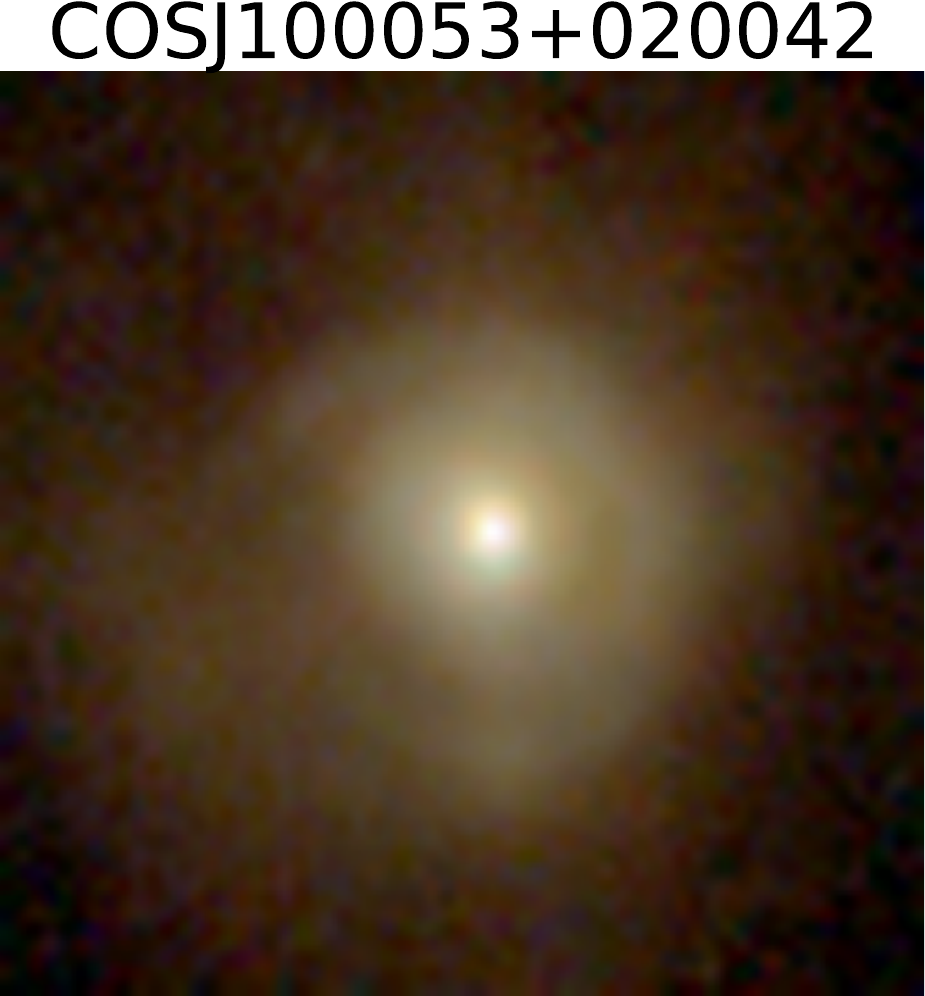}
\includegraphics[width=0.12\textwidth]{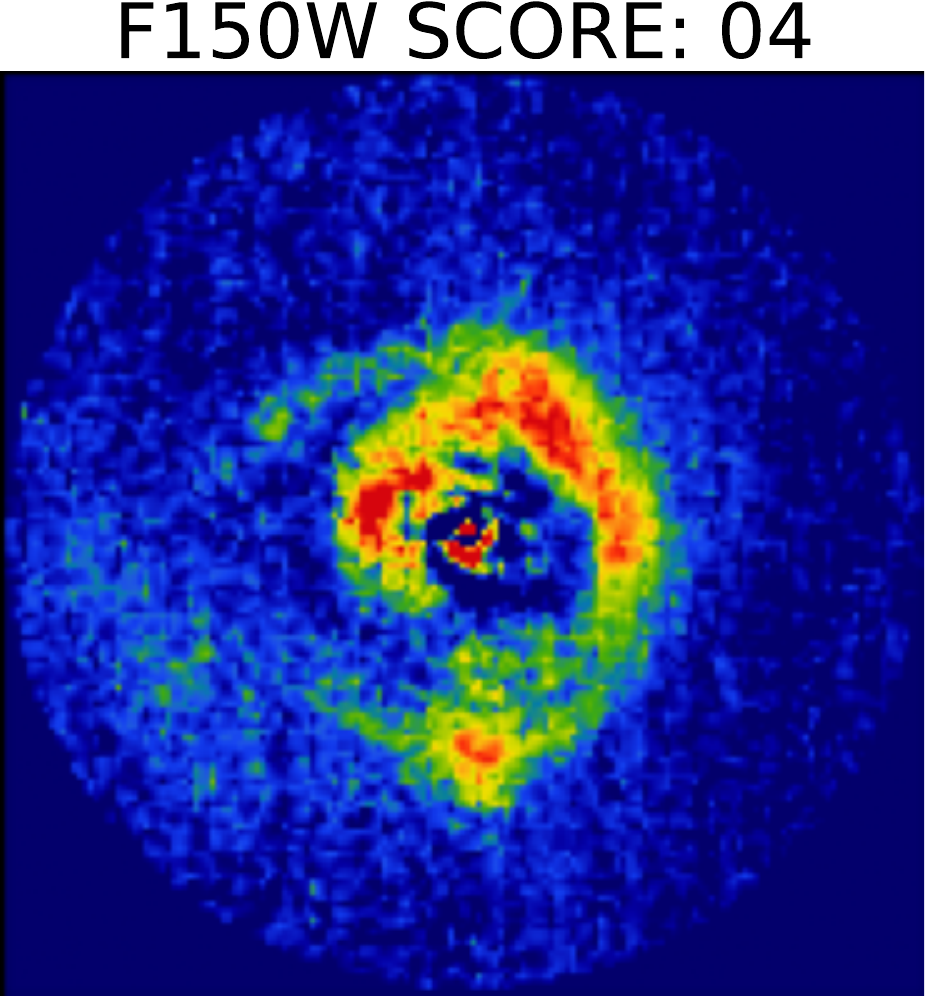}
\includegraphics[width=0.12\textwidth]{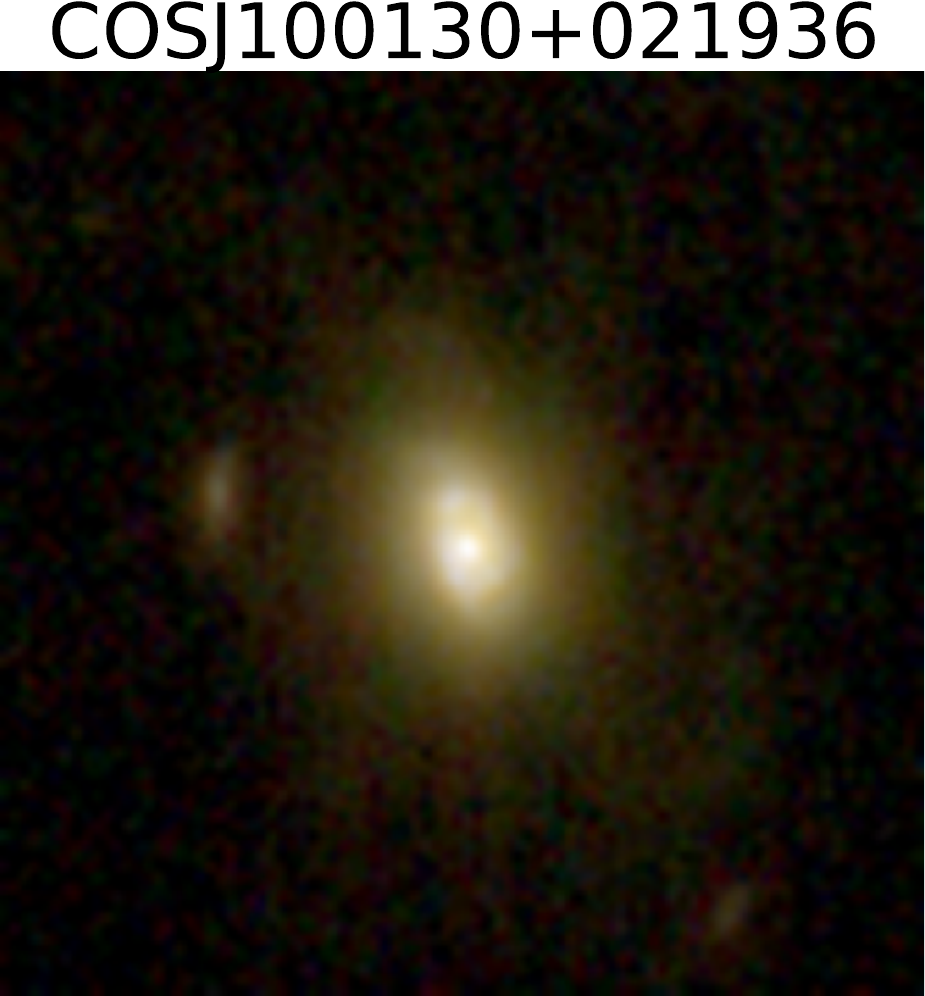}
\includegraphics[width=0.12\textwidth]{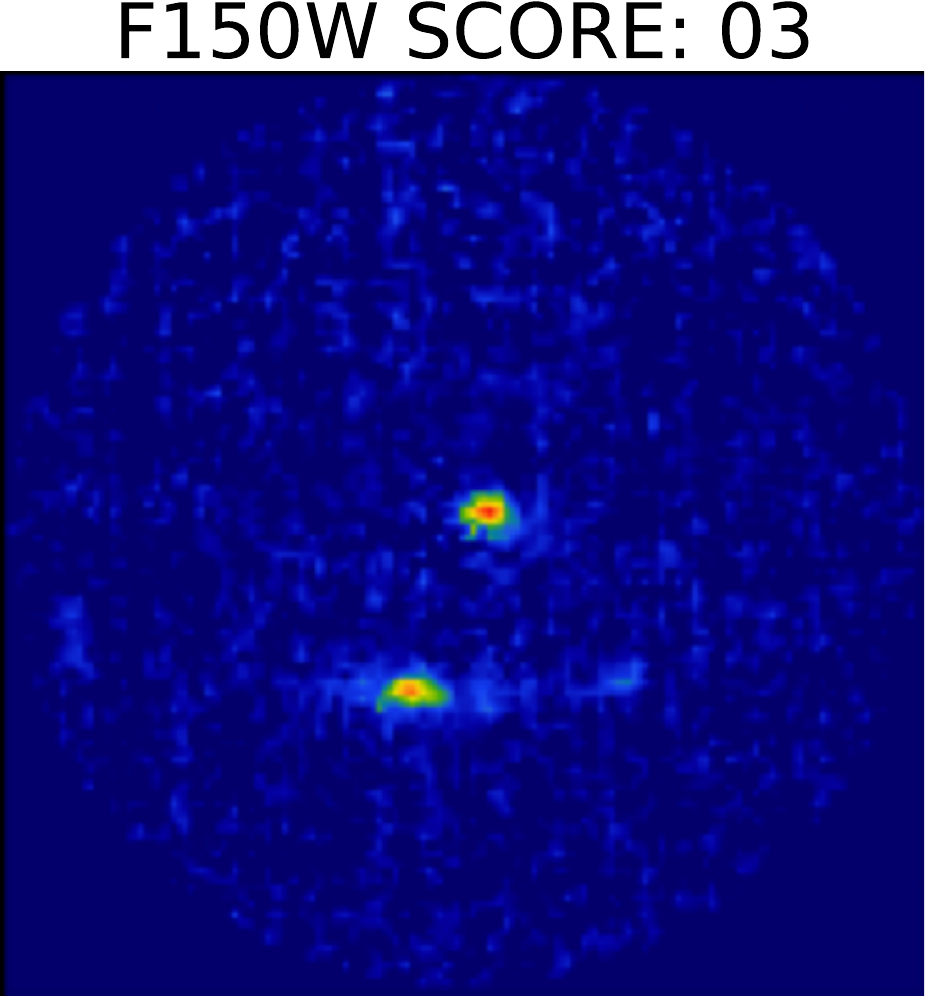}
\includegraphics[width=0.12\textwidth]{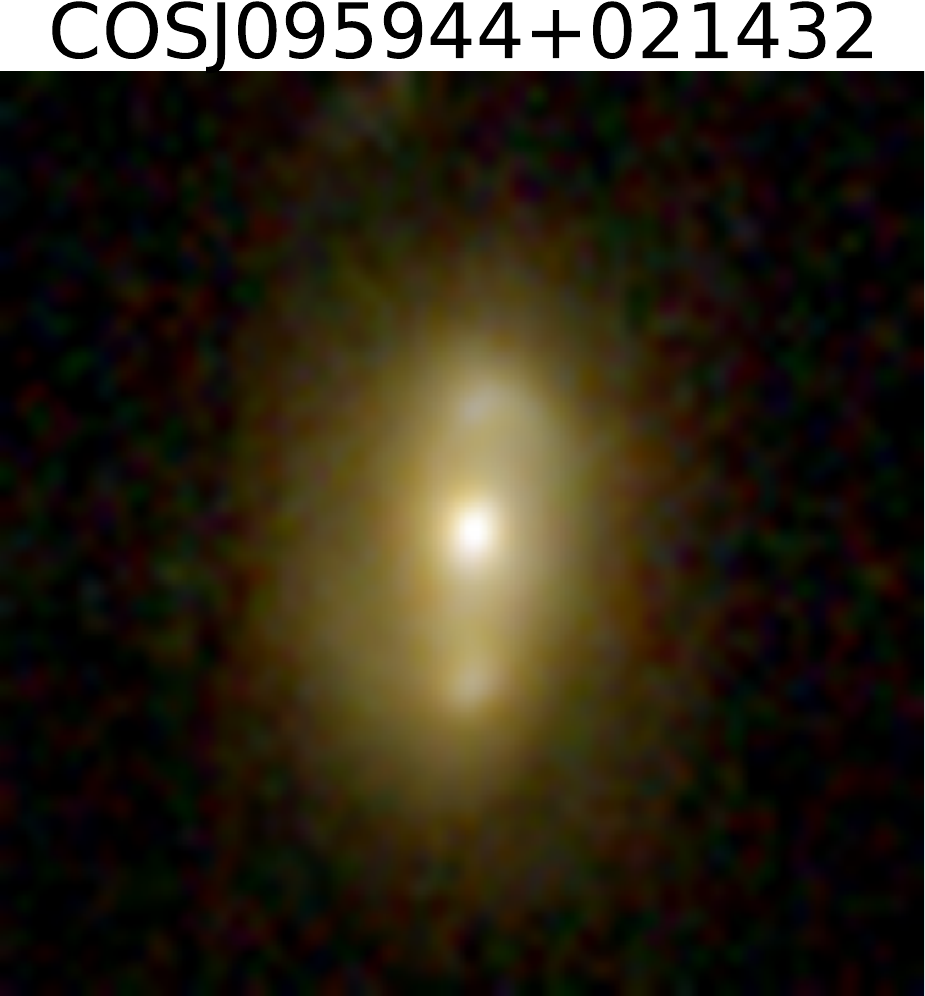}
\includegraphics[width=0.12\textwidth]{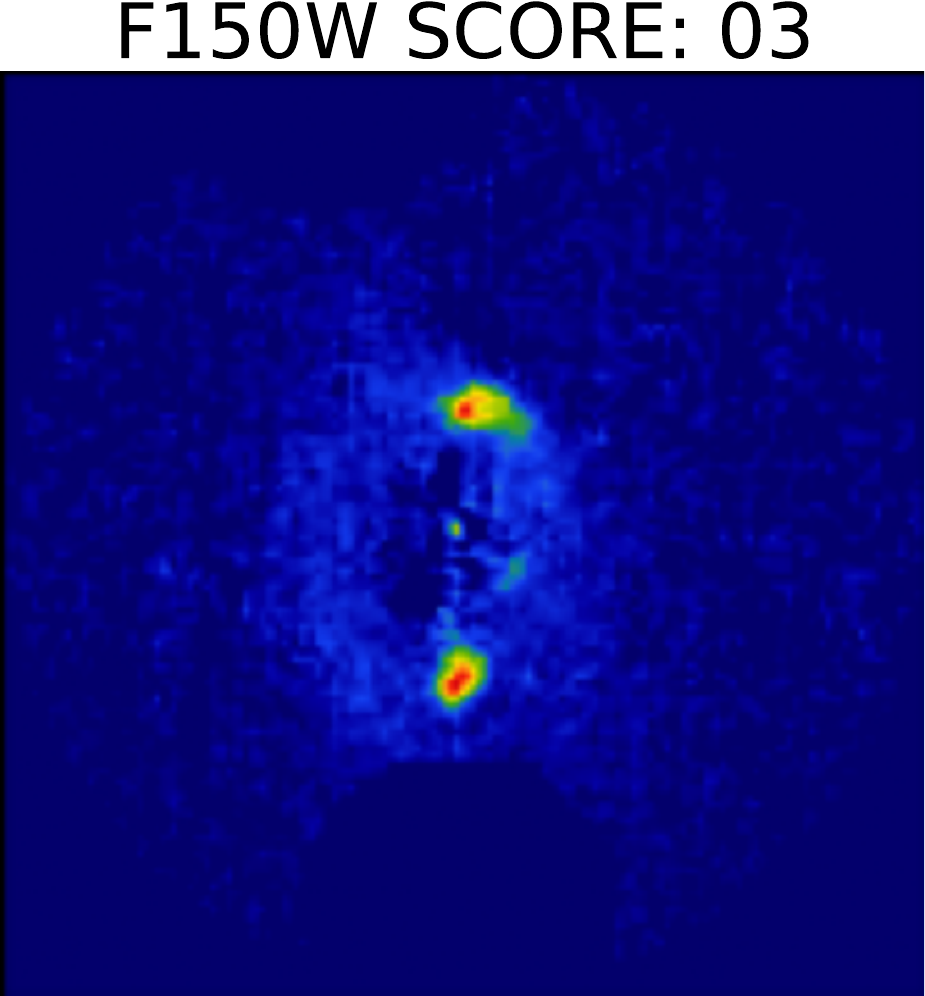}
\includegraphics[width=0.12\textwidth]{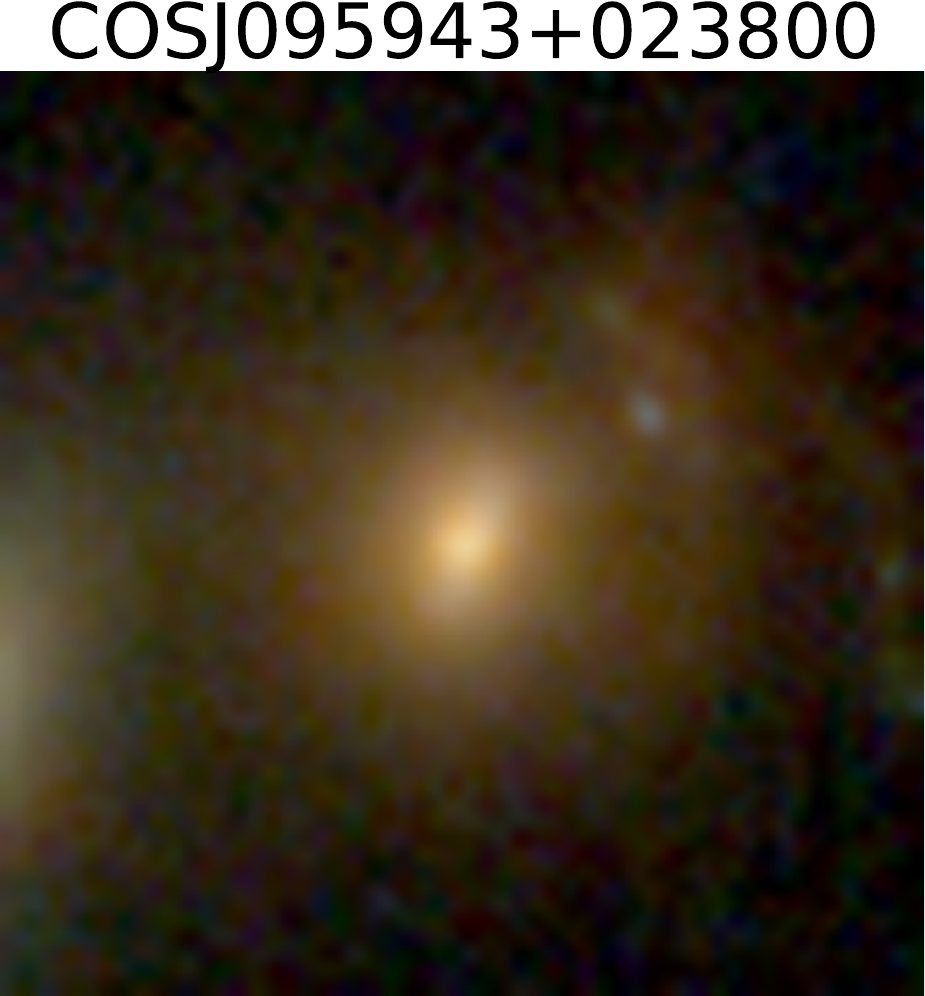}
\includegraphics[width=0.12\textwidth]{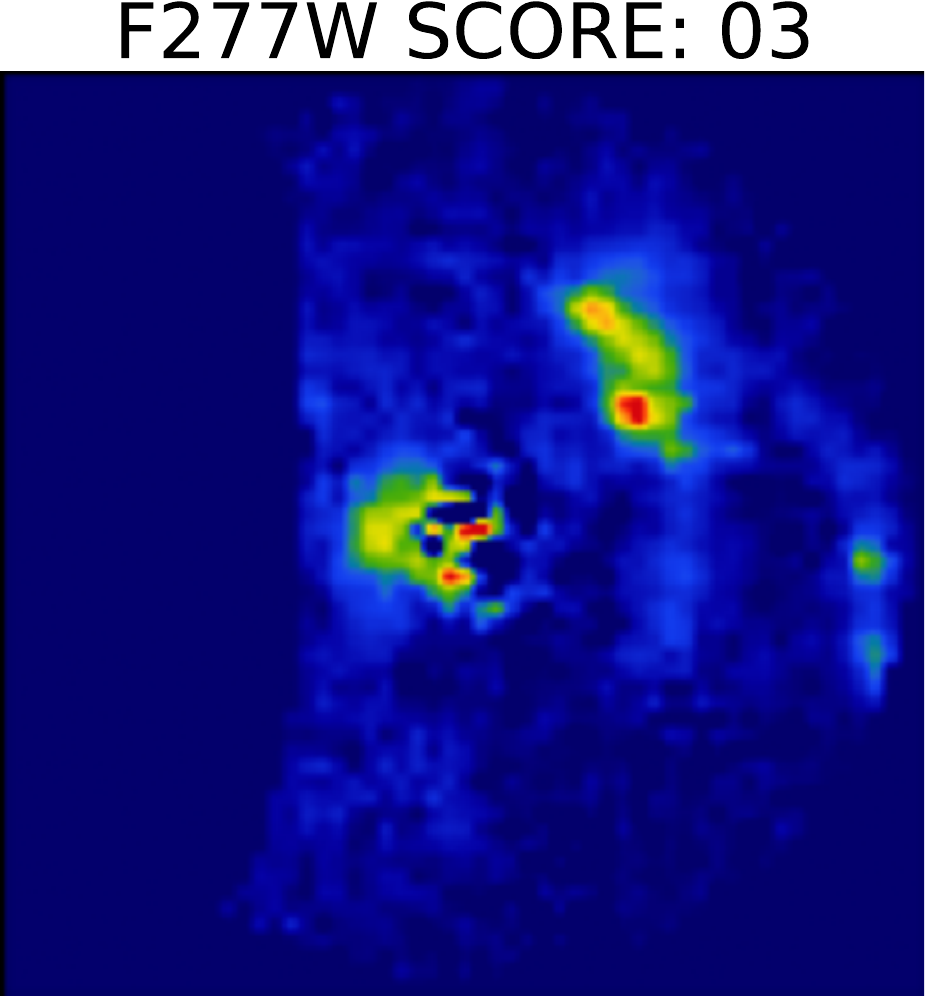}
\includegraphics[width=0.12\textwidth]{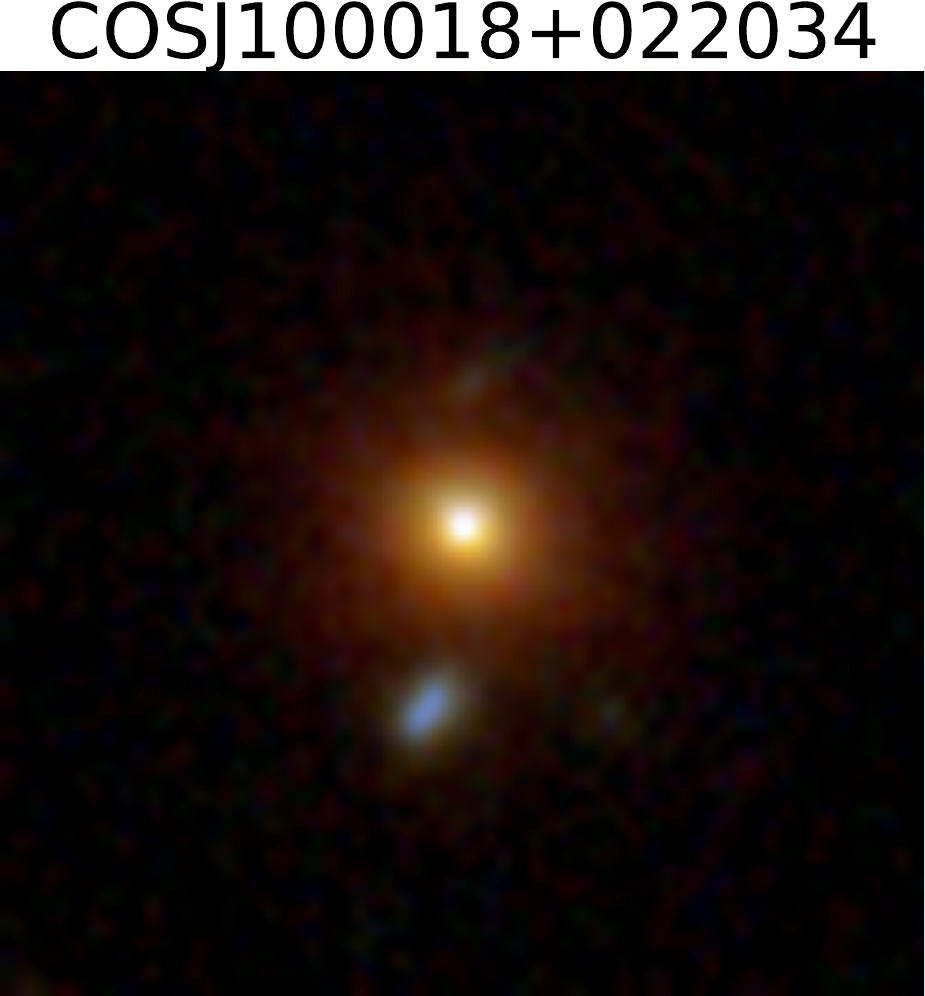}
\includegraphics[width=0.12\textwidth]{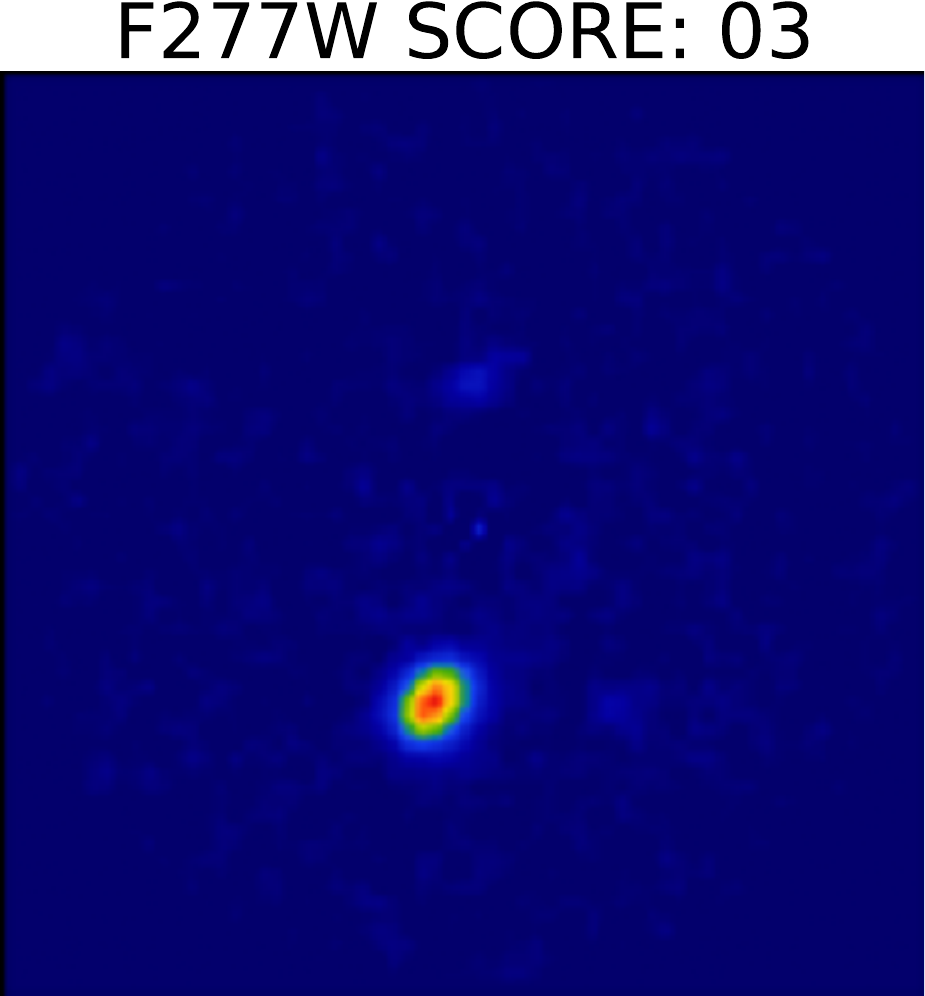}
\includegraphics[width=0.12\textwidth]{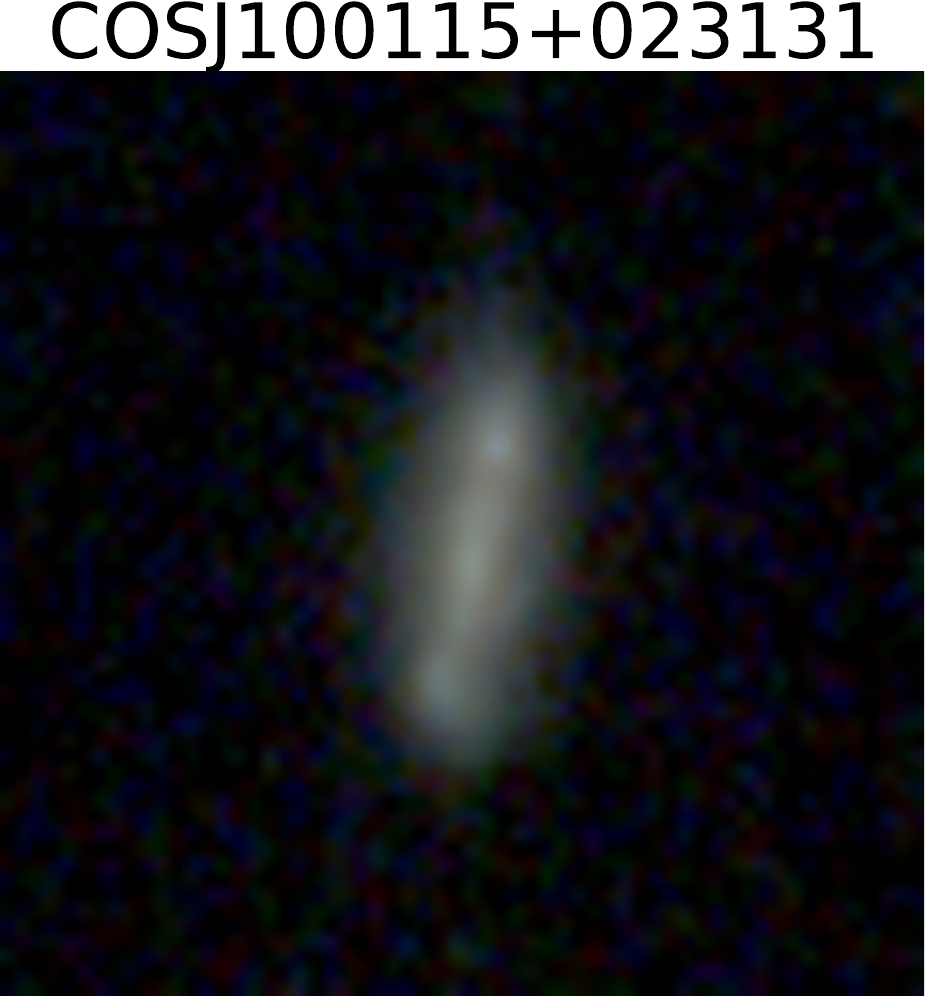}
\includegraphics[width=0.12\textwidth]{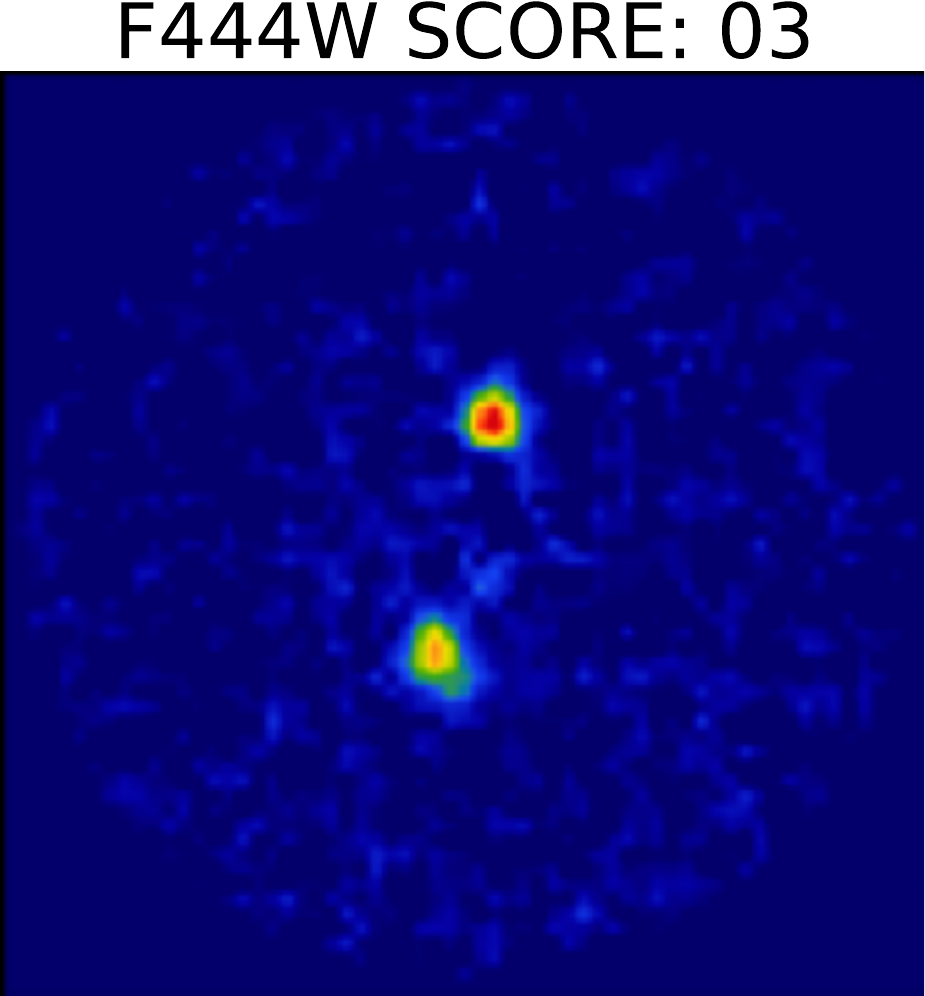}
\includegraphics[width=0.12\textwidth]{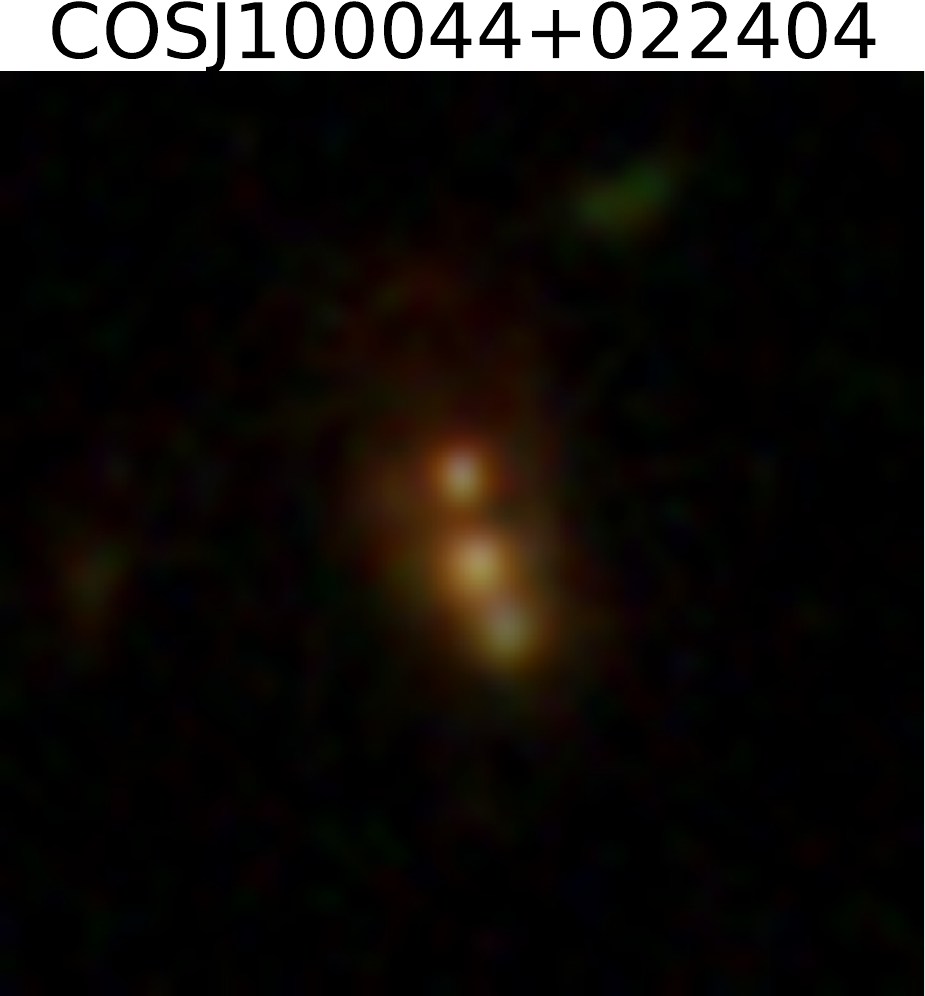}
\includegraphics[width=0.12\textwidth]{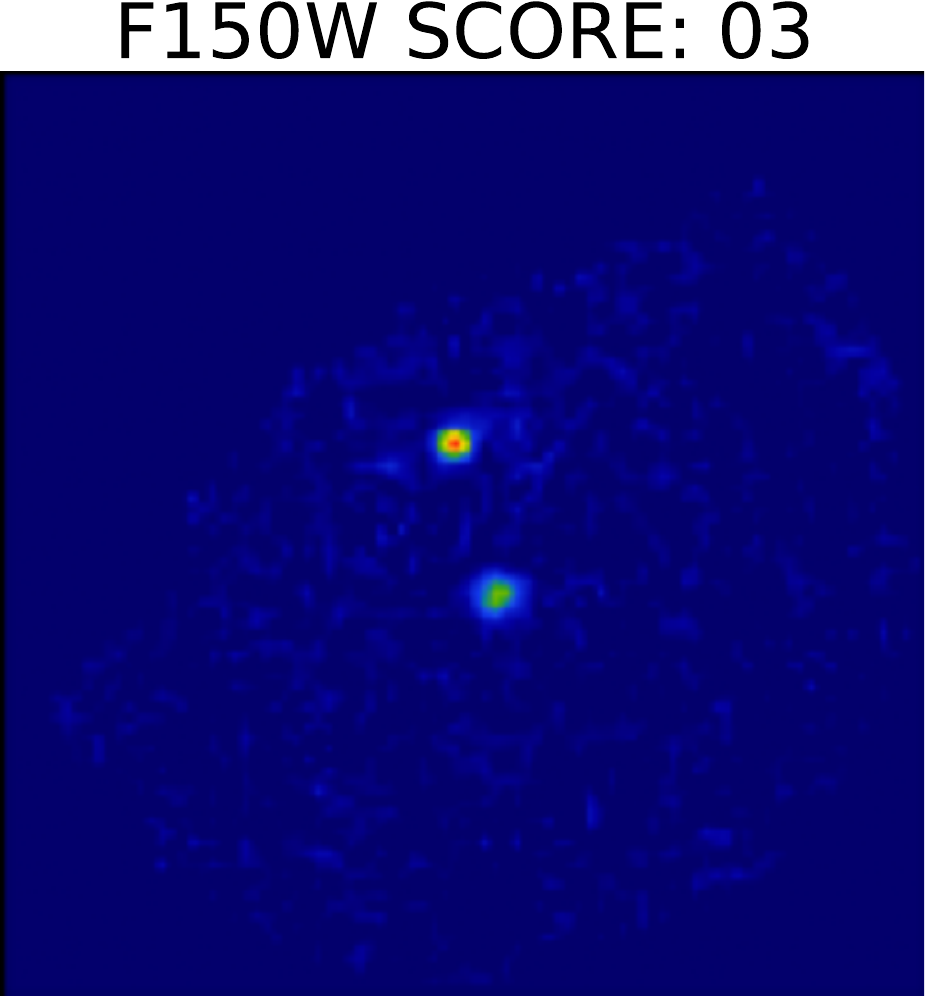}
\includegraphics[width=0.12\textwidth]{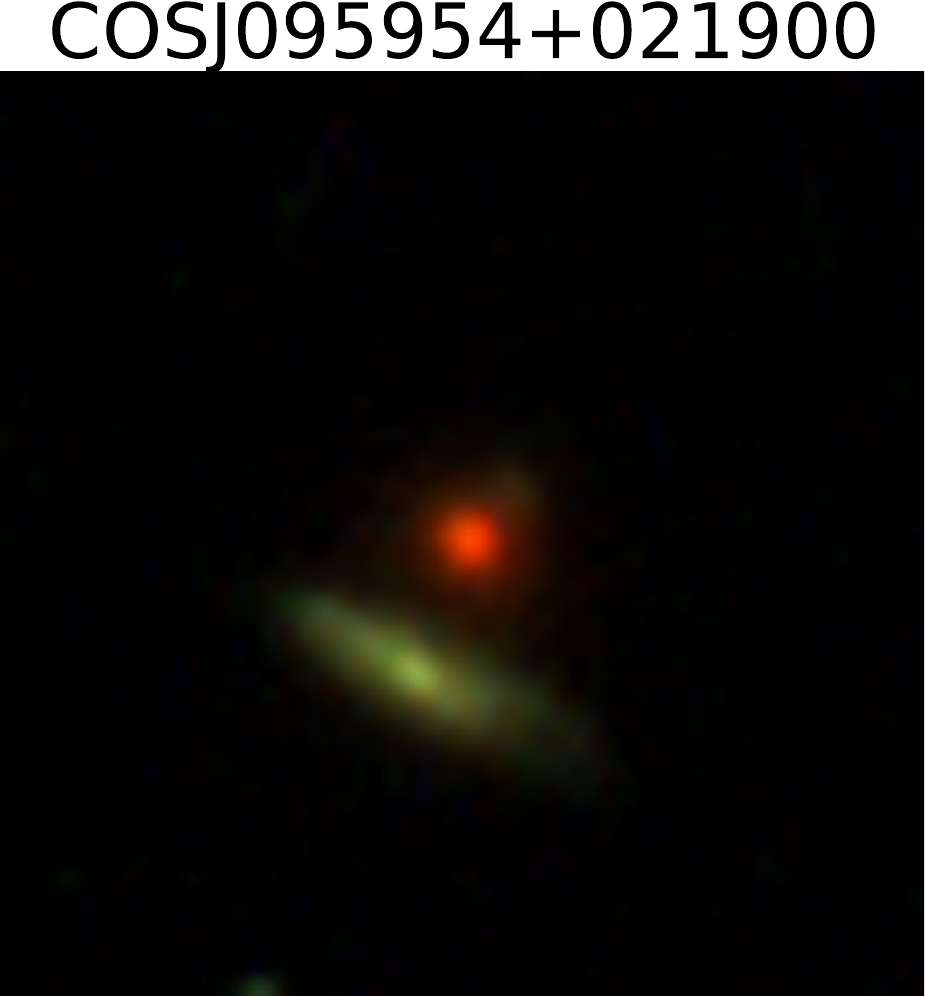}
\includegraphics[width=0.12\textwidth]{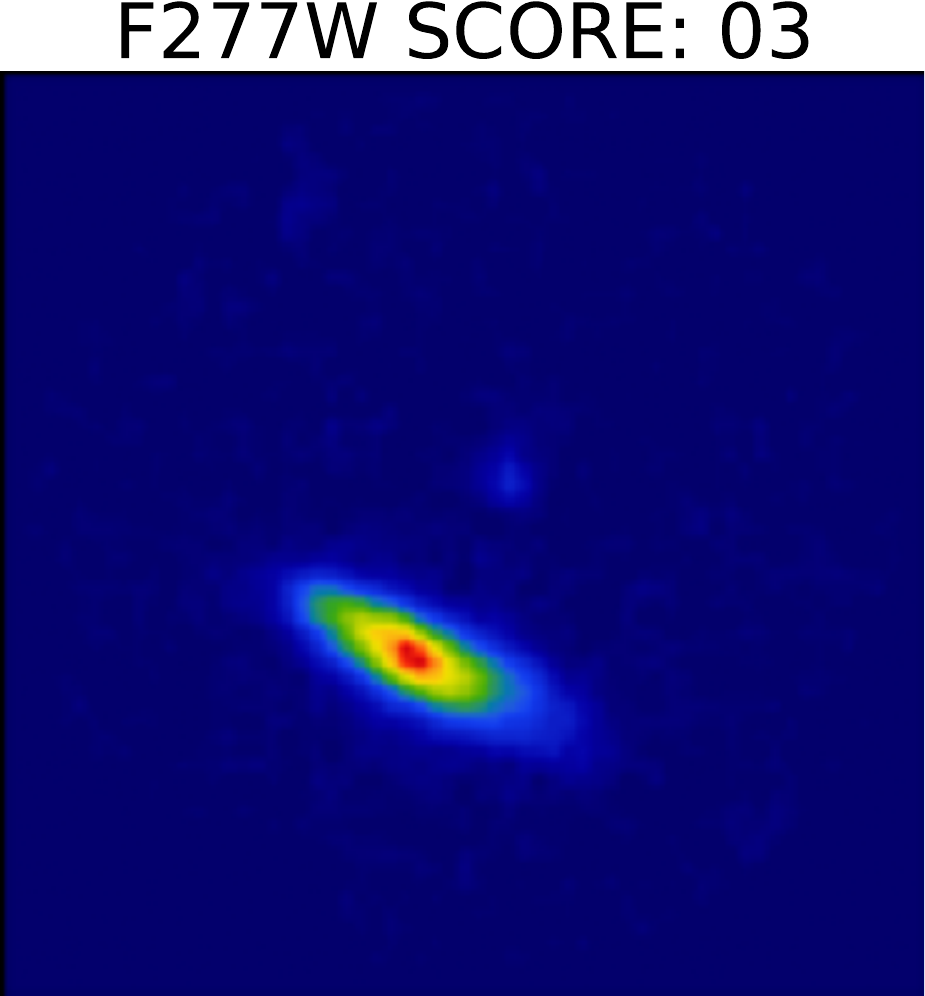}
\includegraphics[width=0.12\textwidth]{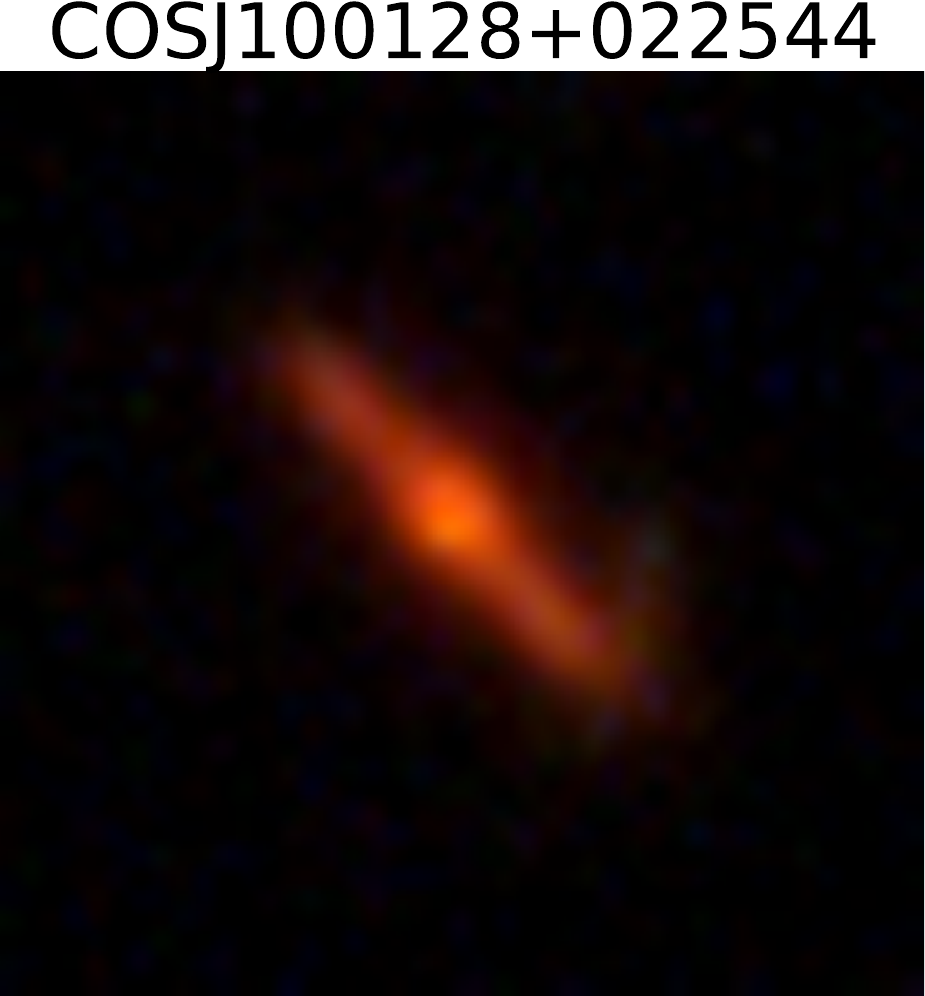}
\includegraphics[width=0.12\textwidth]{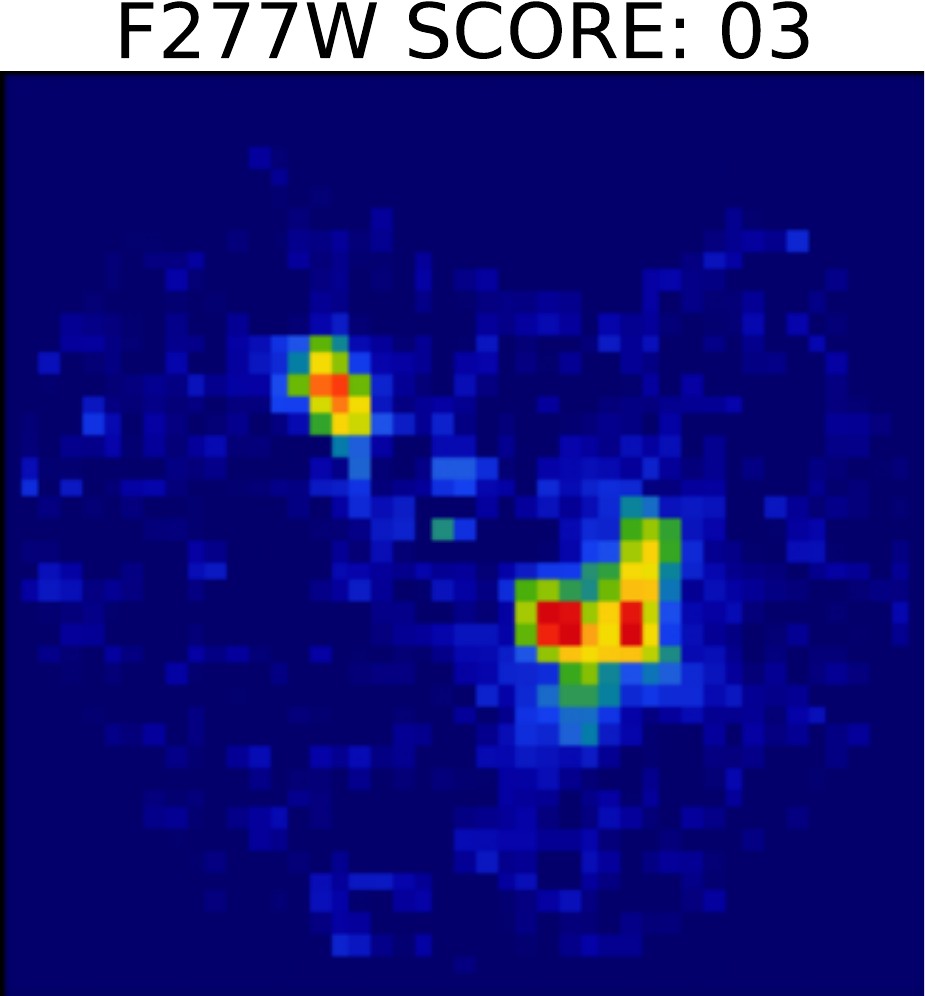}
\includegraphics[width=0.12\textwidth]{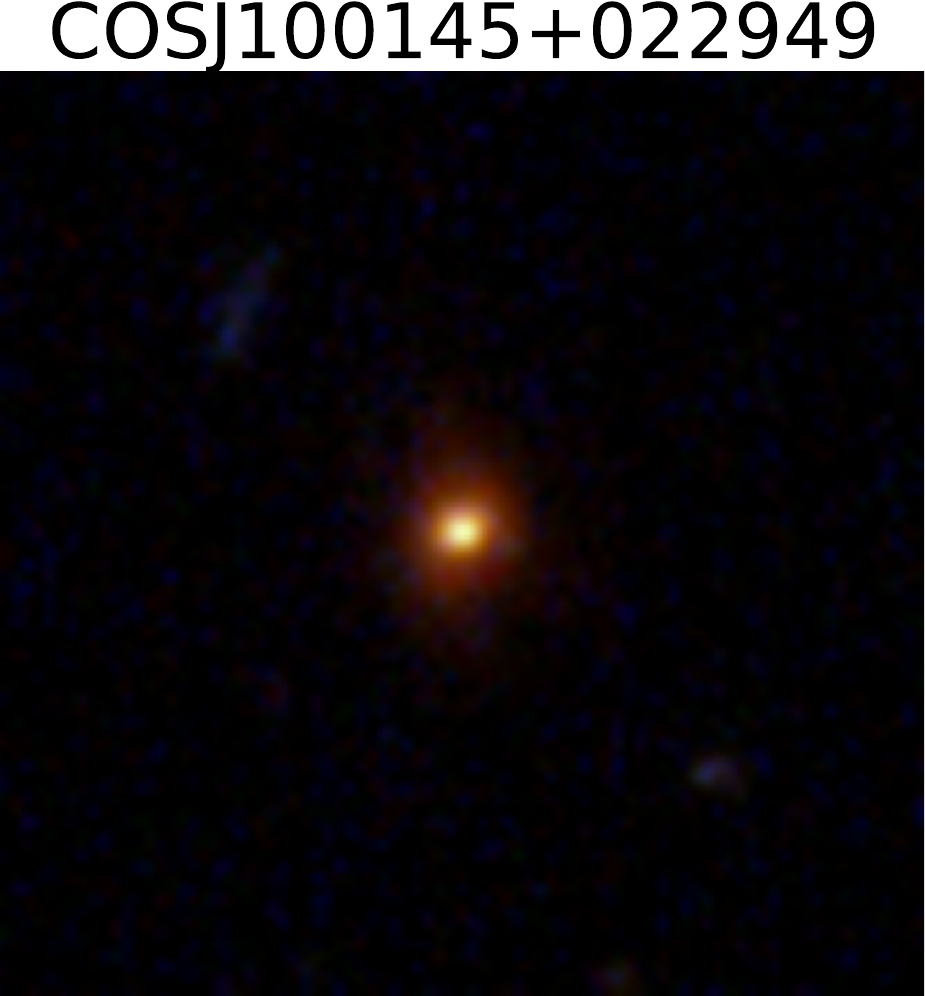}
\includegraphics[width=0.12\textwidth]{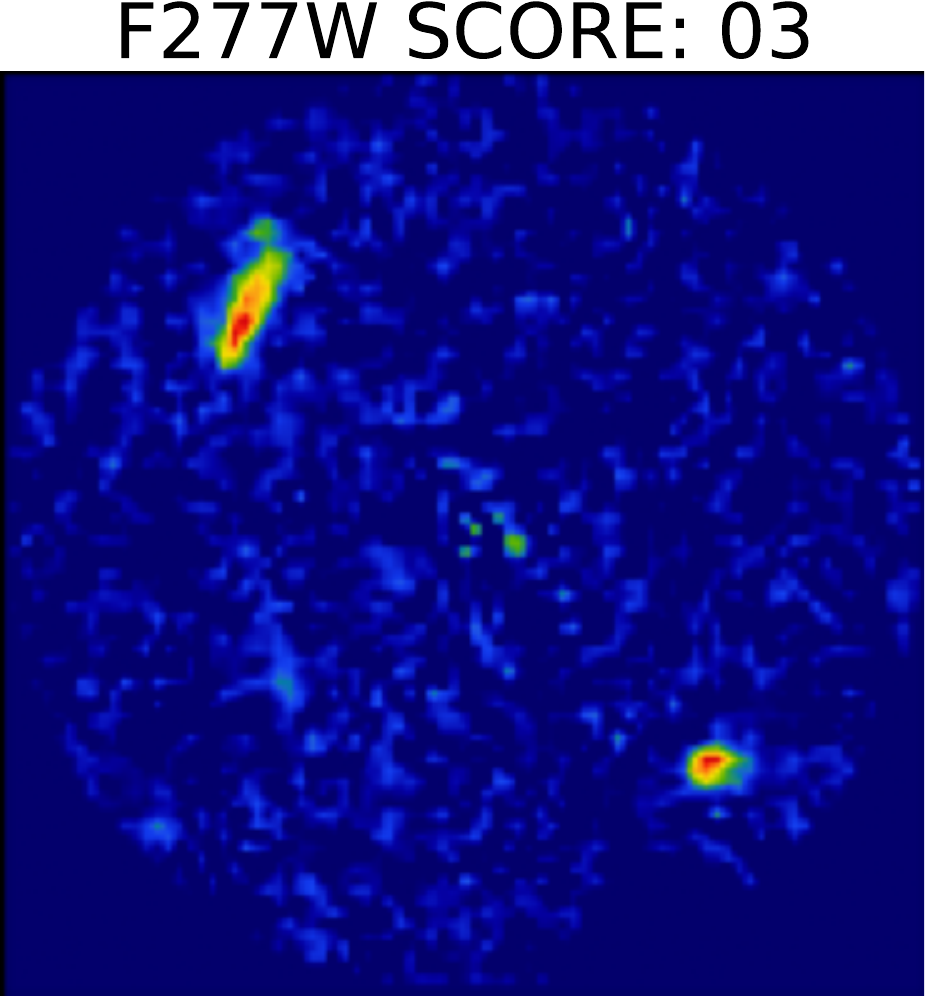}
\includegraphics[width=0.12\textwidth]{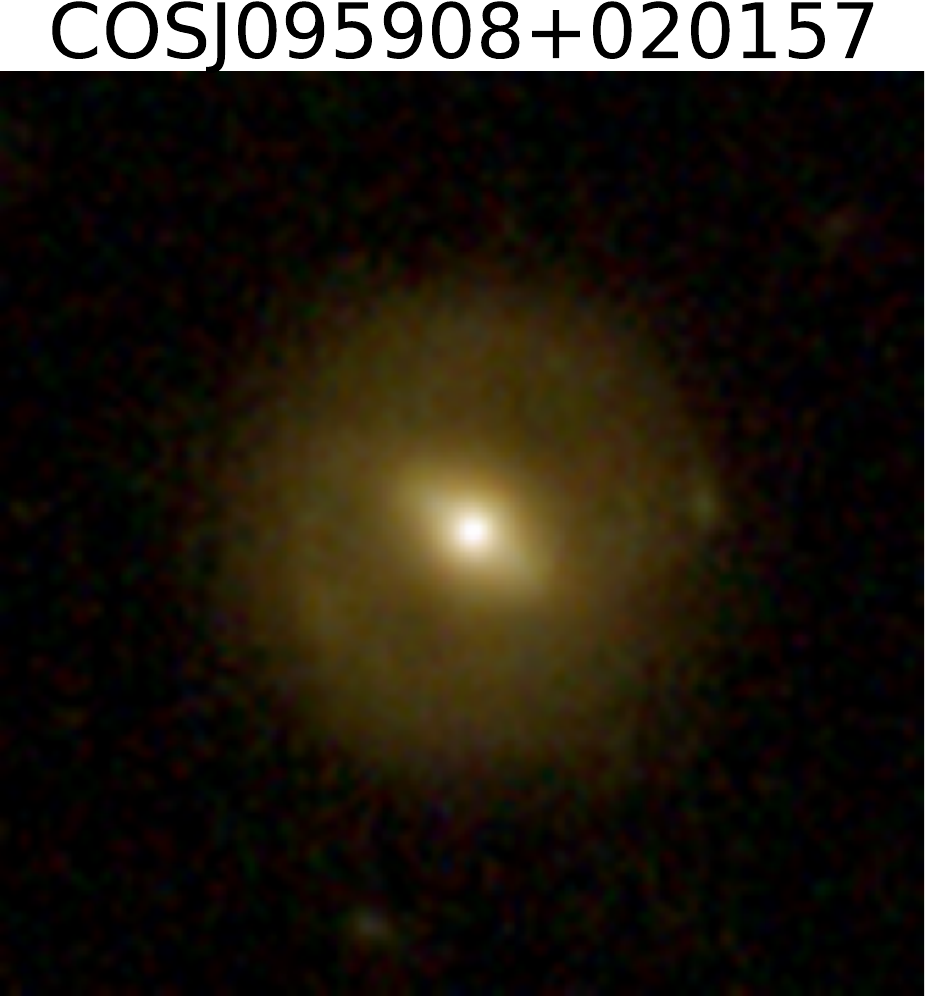}
\includegraphics[width=0.12\textwidth]{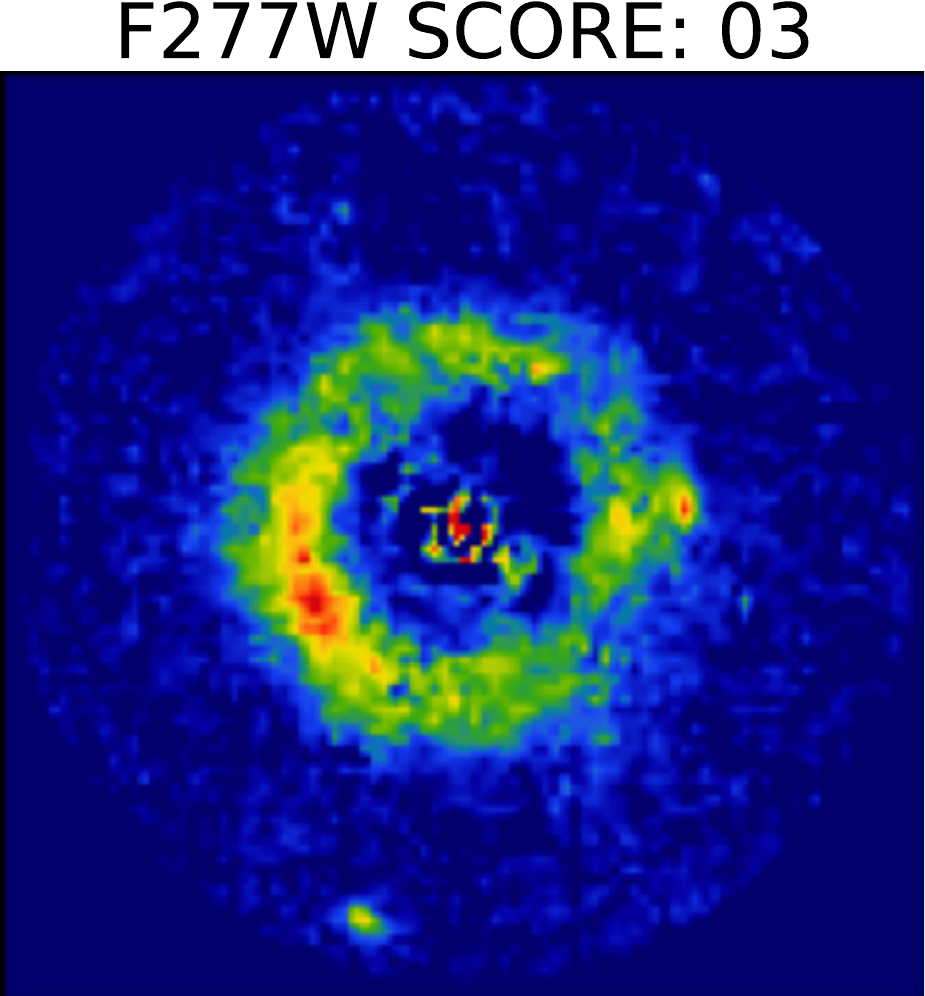}
\includegraphics[width=0.12\textwidth]{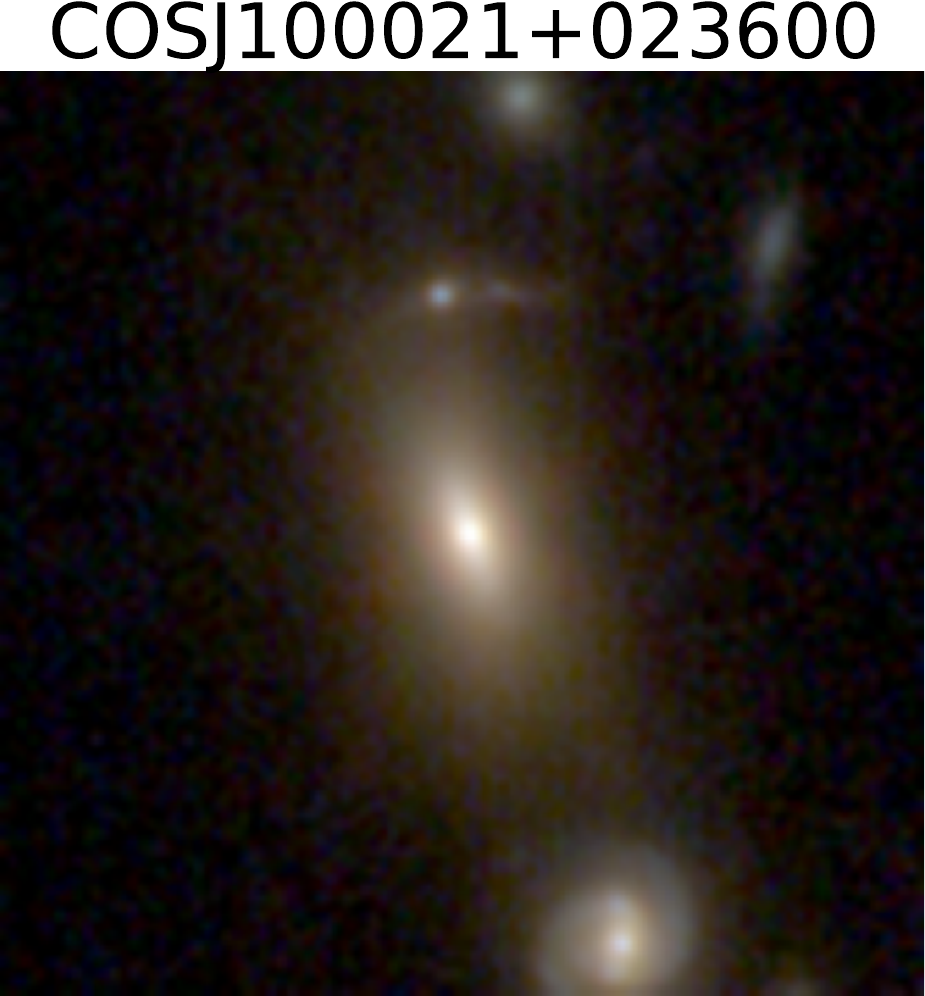}
\includegraphics[width=0.12\textwidth]{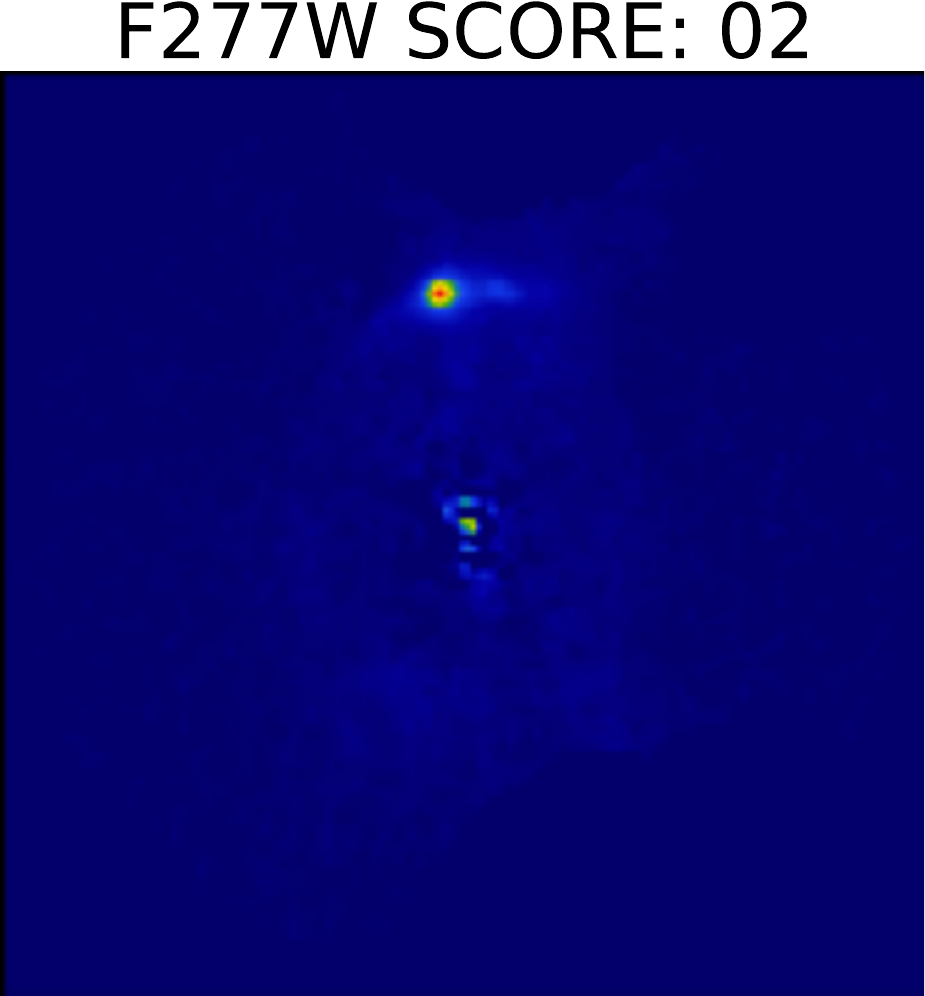}
\includegraphics[width=0.12\textwidth]{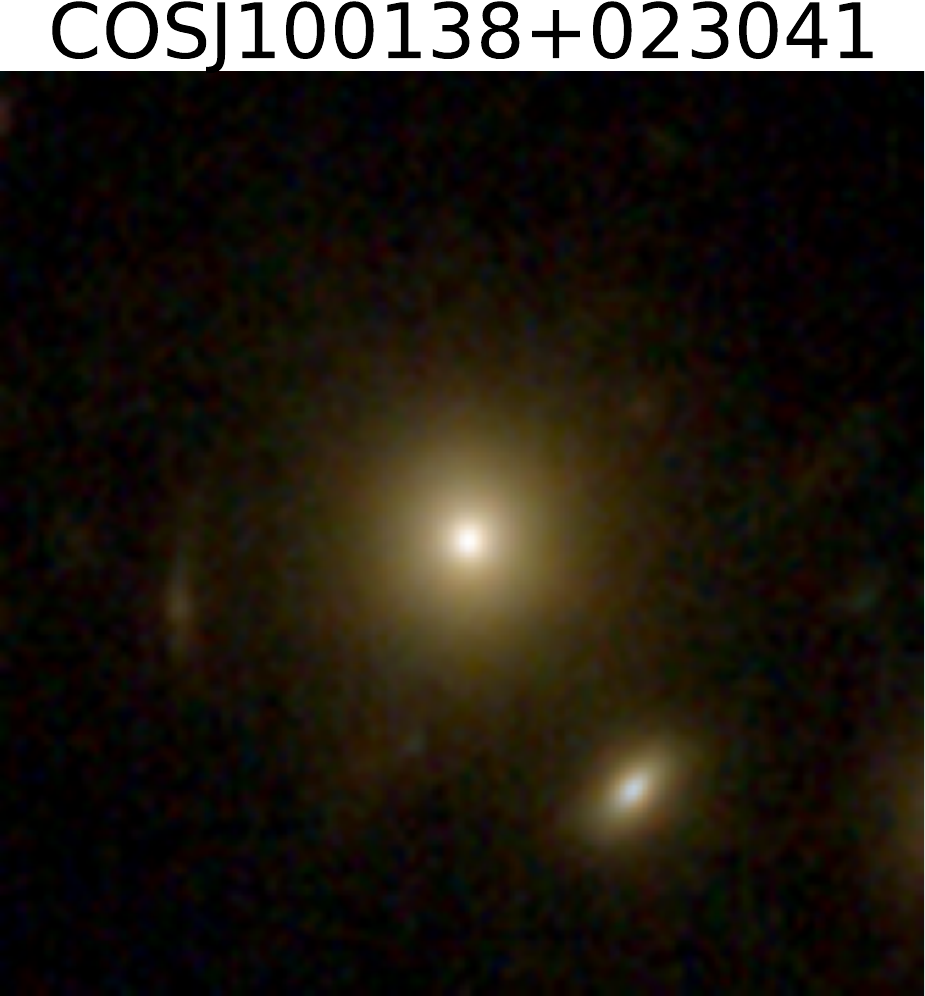}
\includegraphics[width=0.12\textwidth]{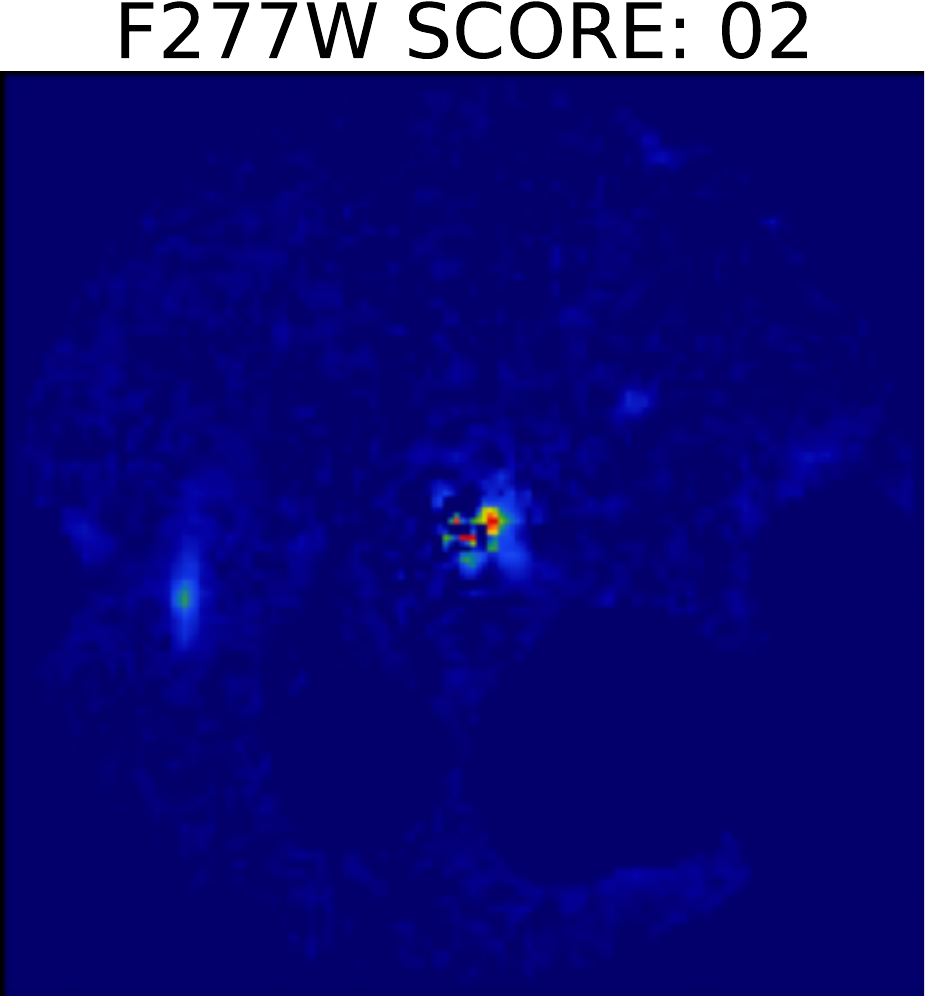}
\includegraphics[width=0.12\textwidth]{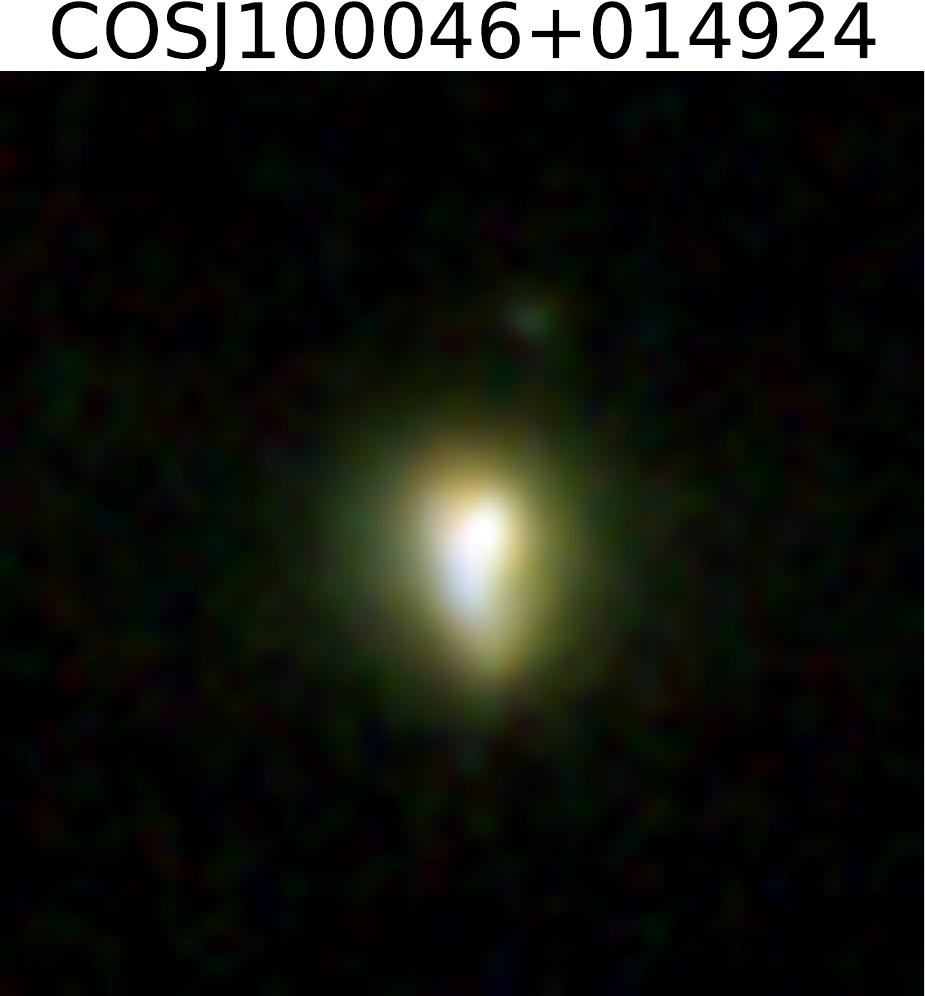}
\includegraphics[width=0.12\textwidth]{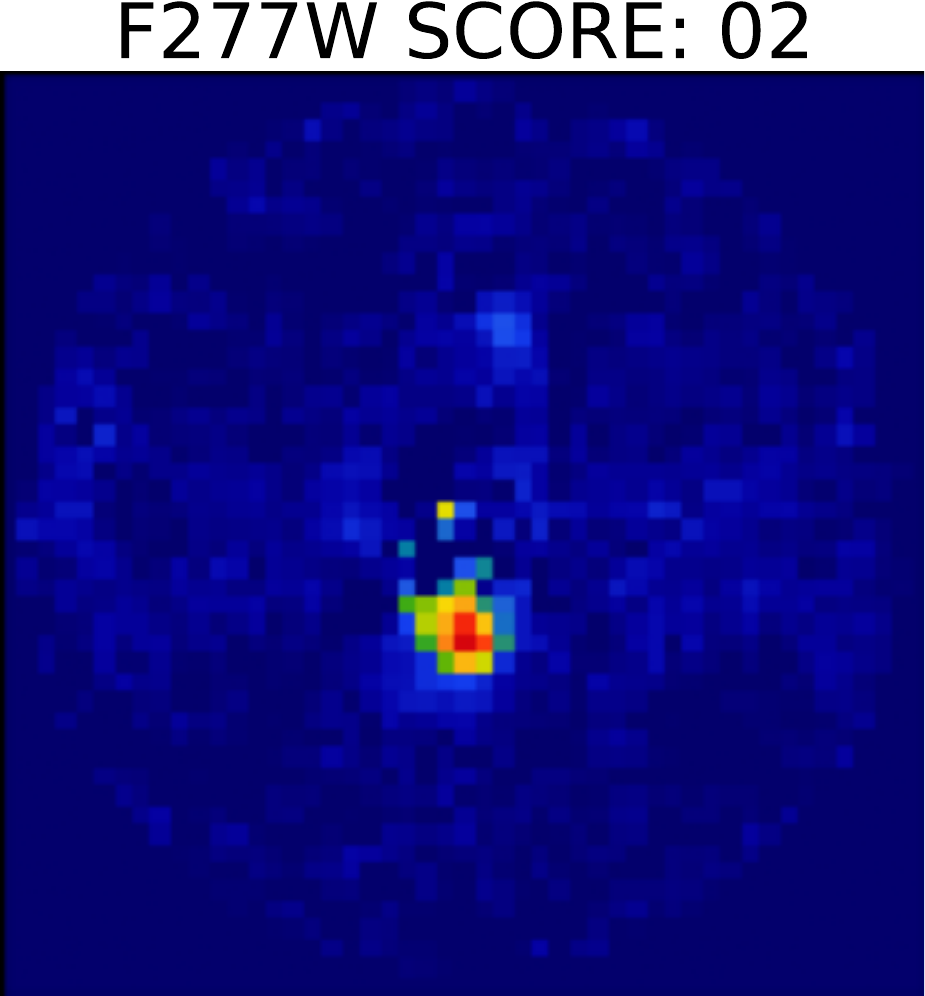}
\includegraphics[width=0.12\textwidth]{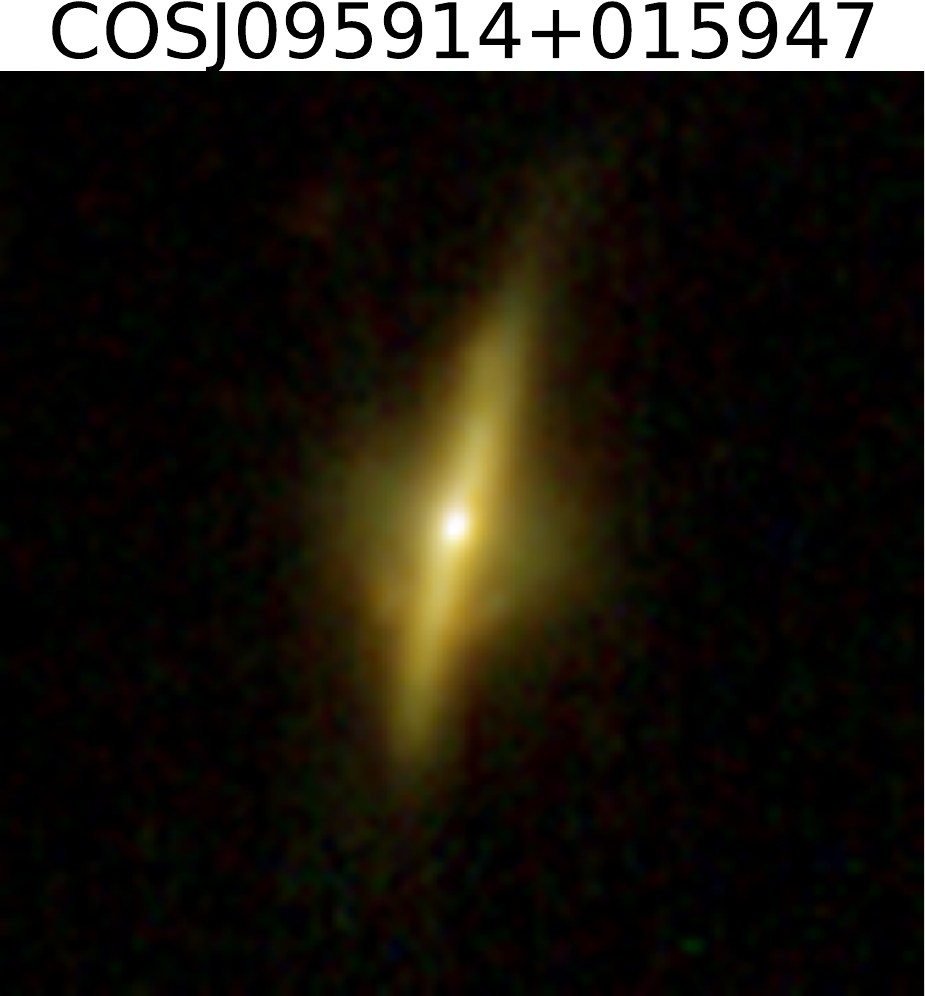}
\includegraphics[width=0.12\textwidth]{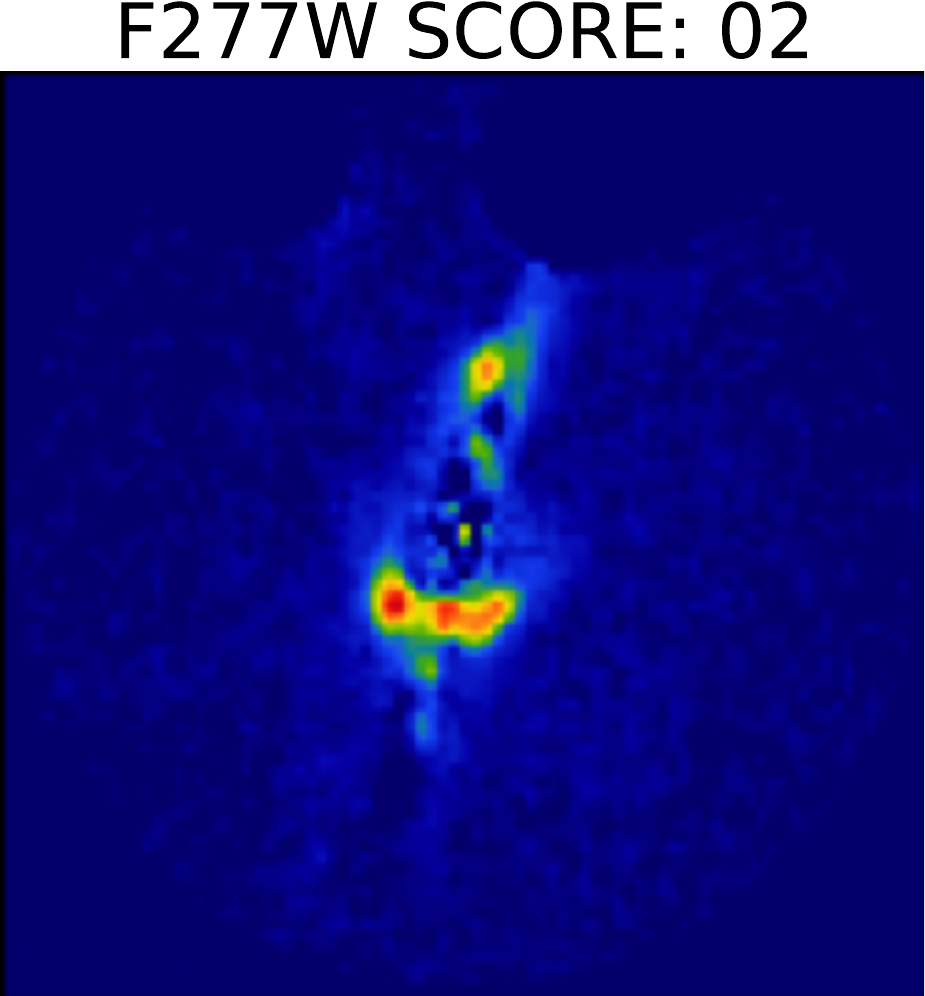}
\includegraphics[width=0.12\textwidth]{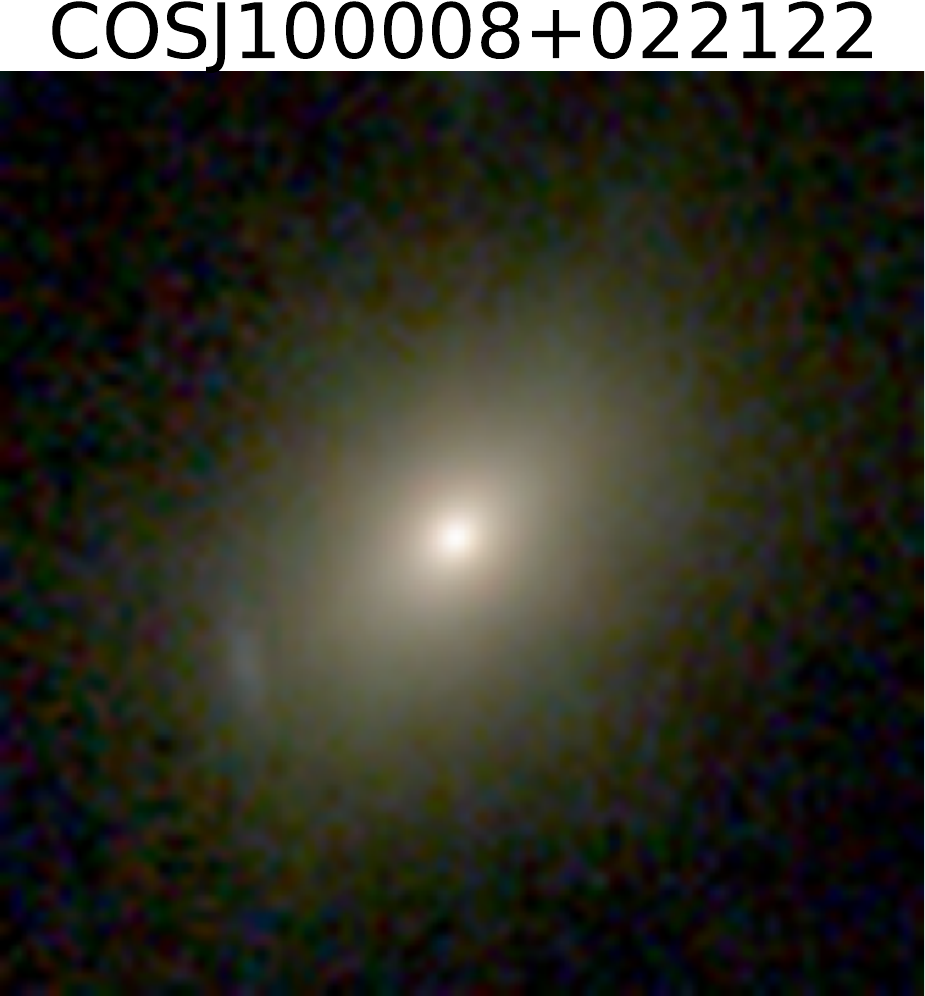}
\includegraphics[width=0.12\textwidth]{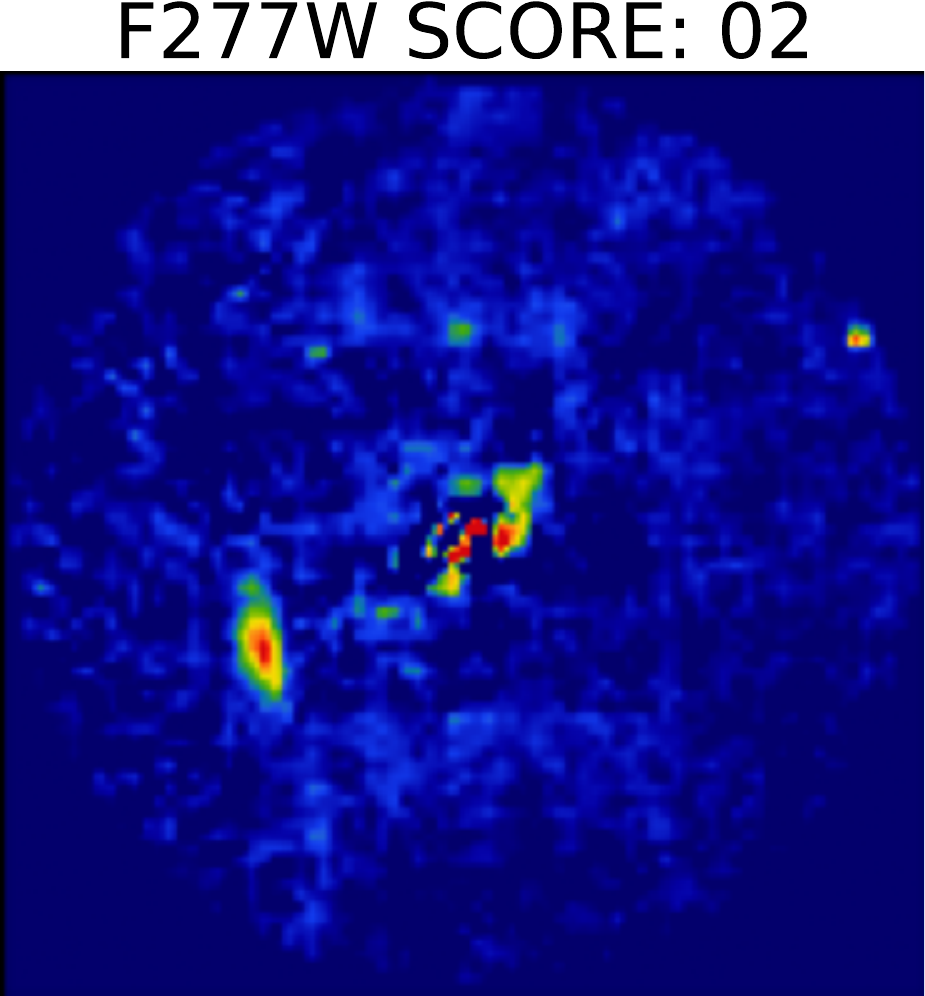}
\includegraphics[width=0.12\textwidth]{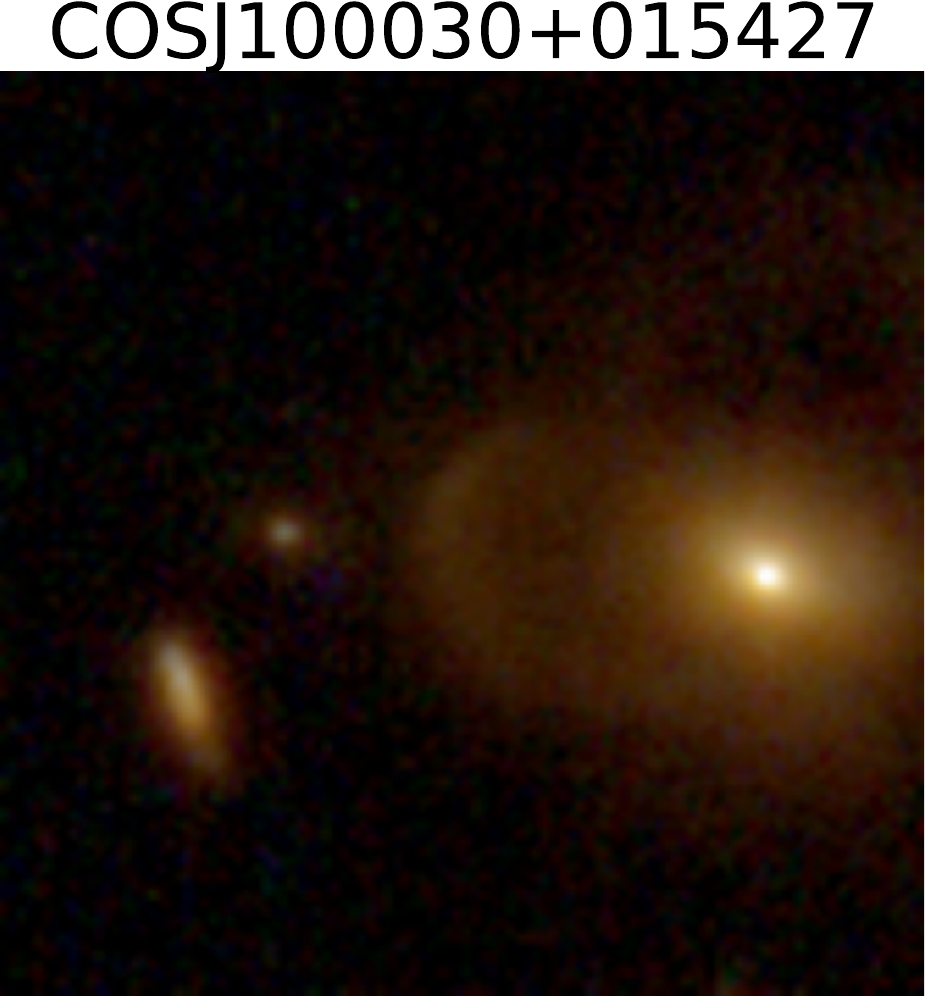}
\includegraphics[width=0.12\textwidth]{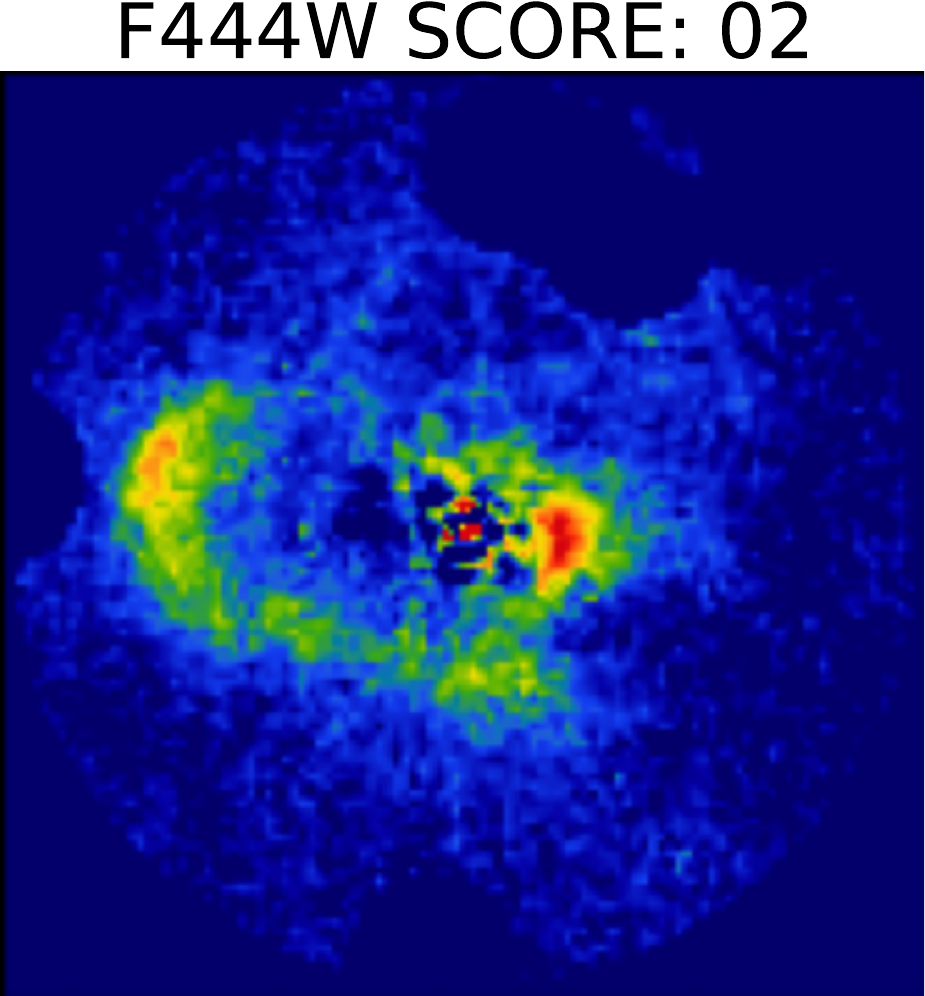}
\includegraphics[width=0.12\textwidth]{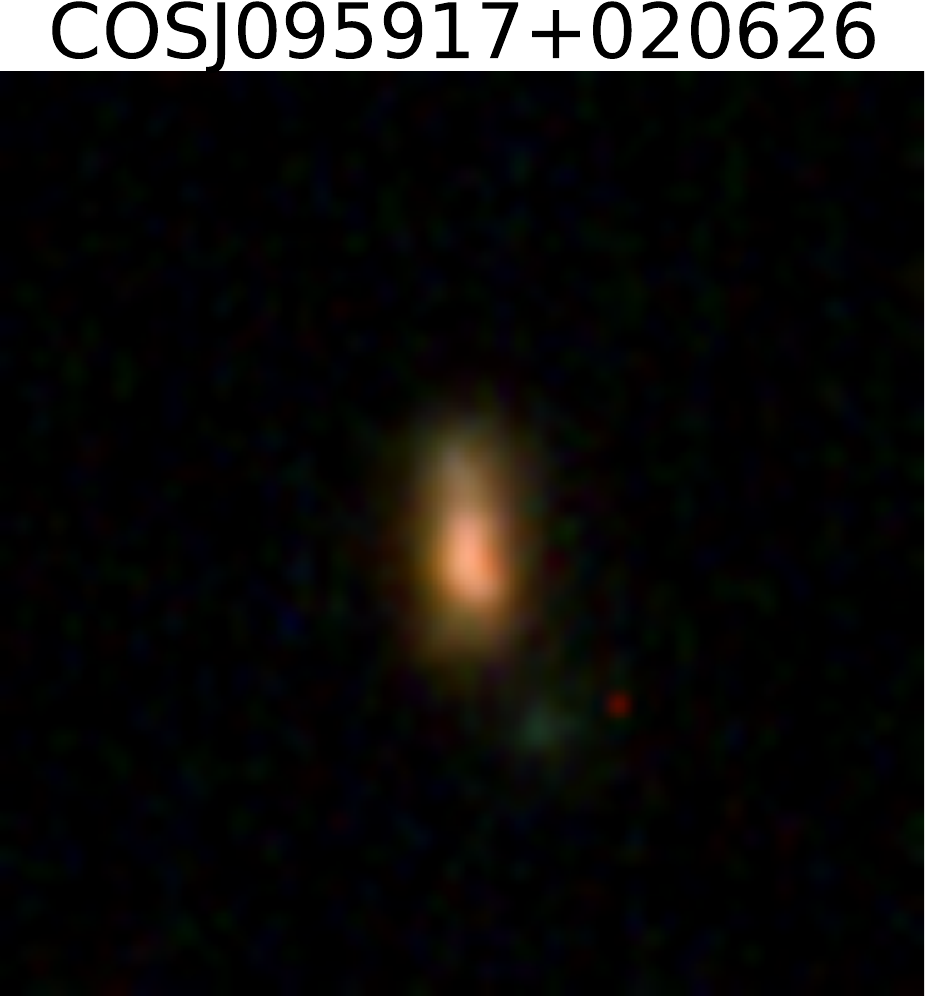}
\includegraphics[width=0.12\textwidth]{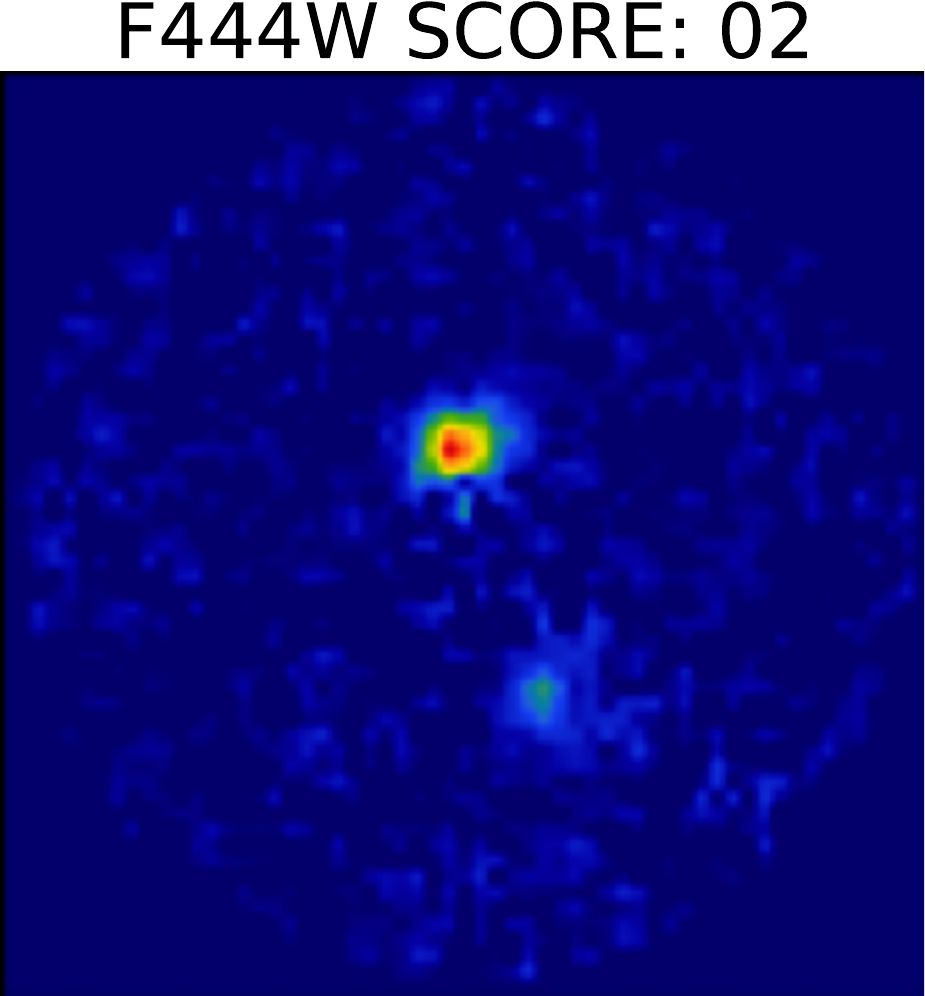}
\includegraphics[width=0.12\textwidth]{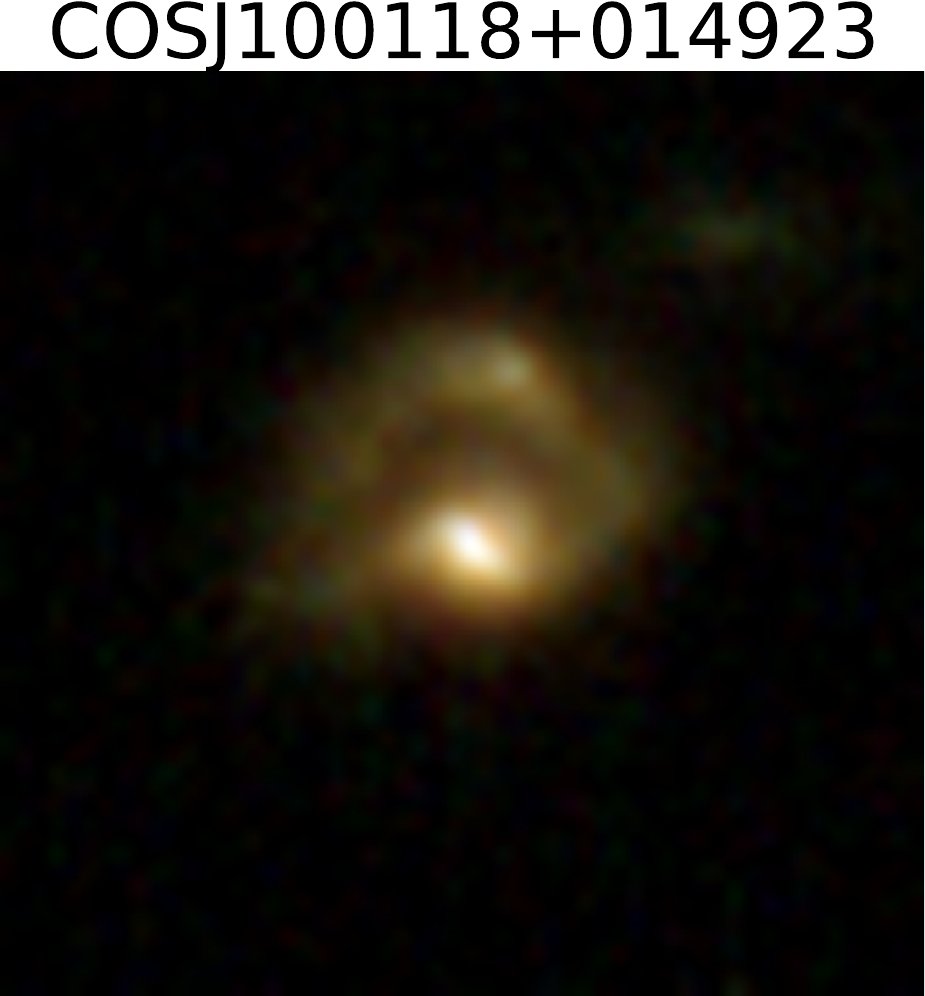}
\includegraphics[width=0.12\textwidth]{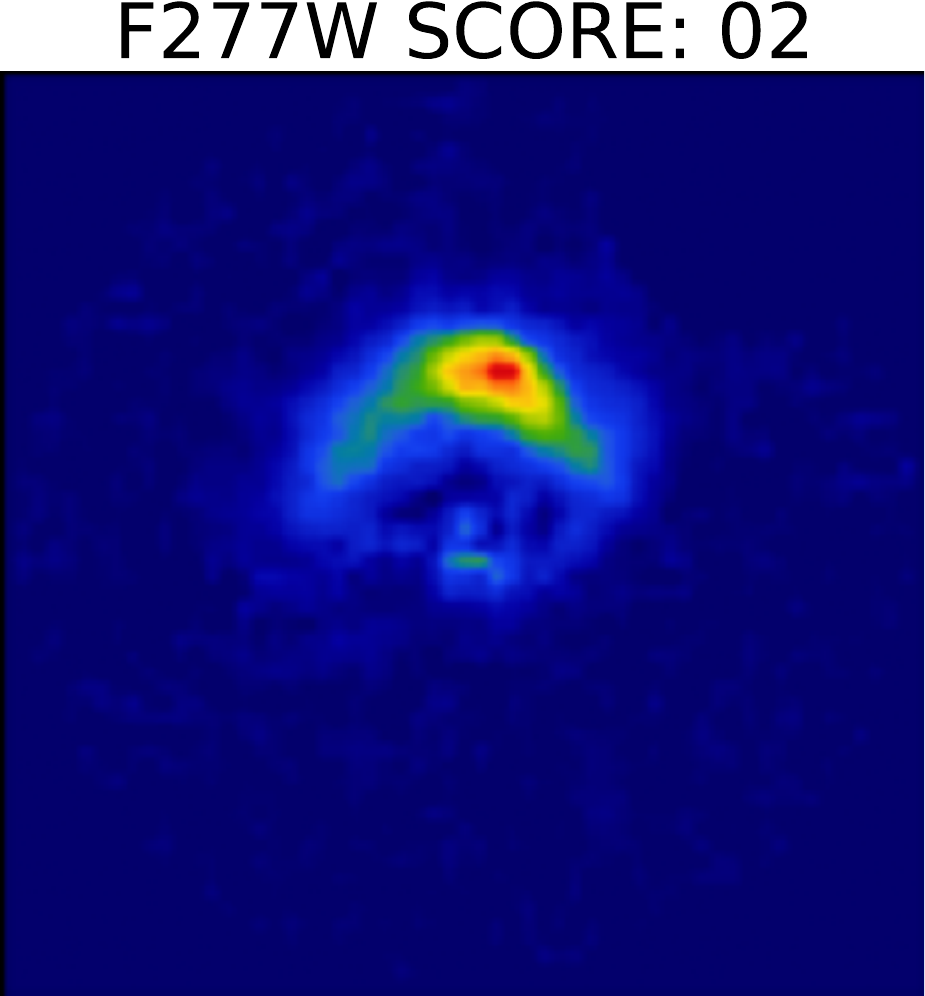}
\includegraphics[width=0.12\textwidth]{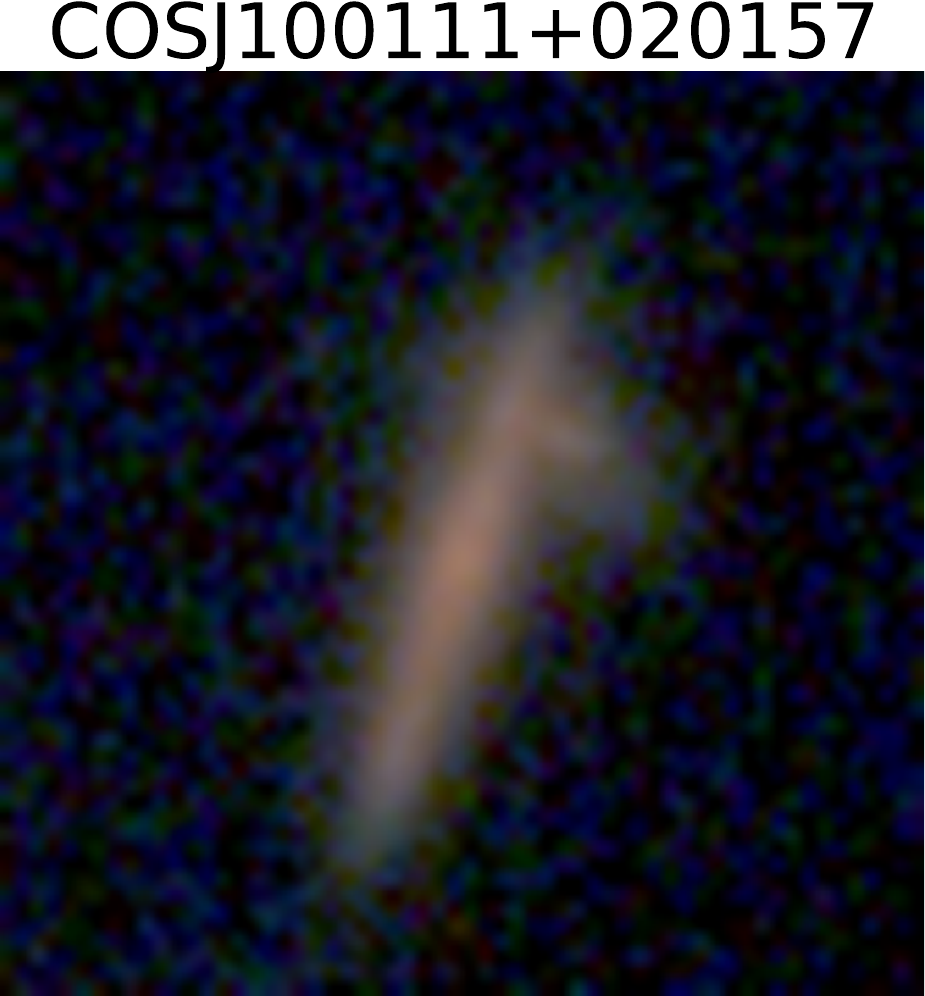}
\includegraphics[width=0.12\textwidth]{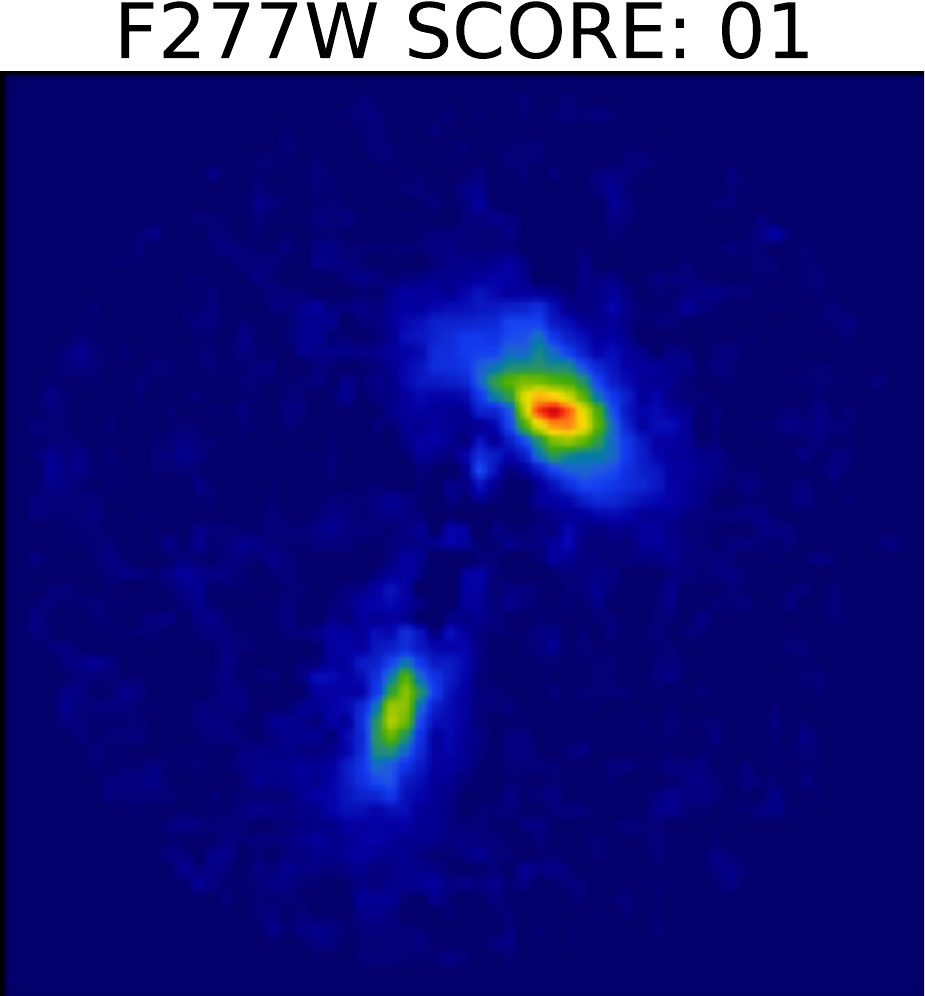}
\includegraphics[width=0.12\textwidth]{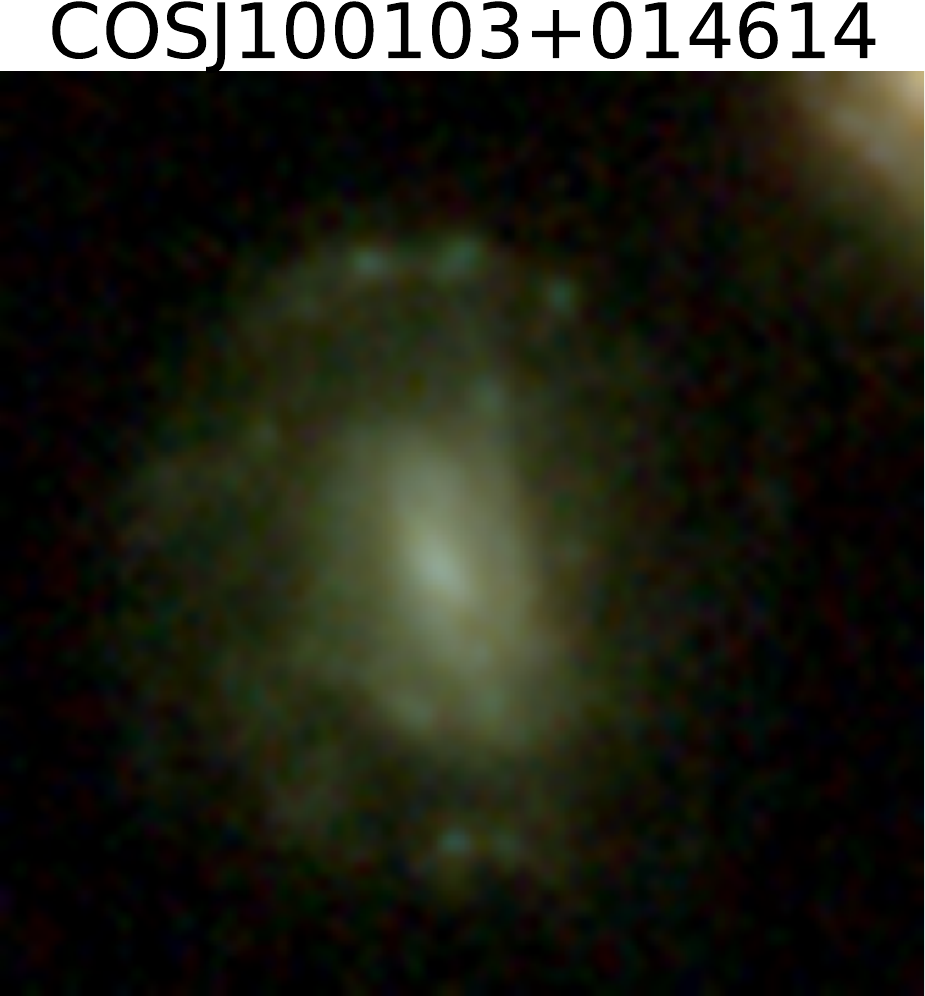}
\includegraphics[width=0.12\textwidth]{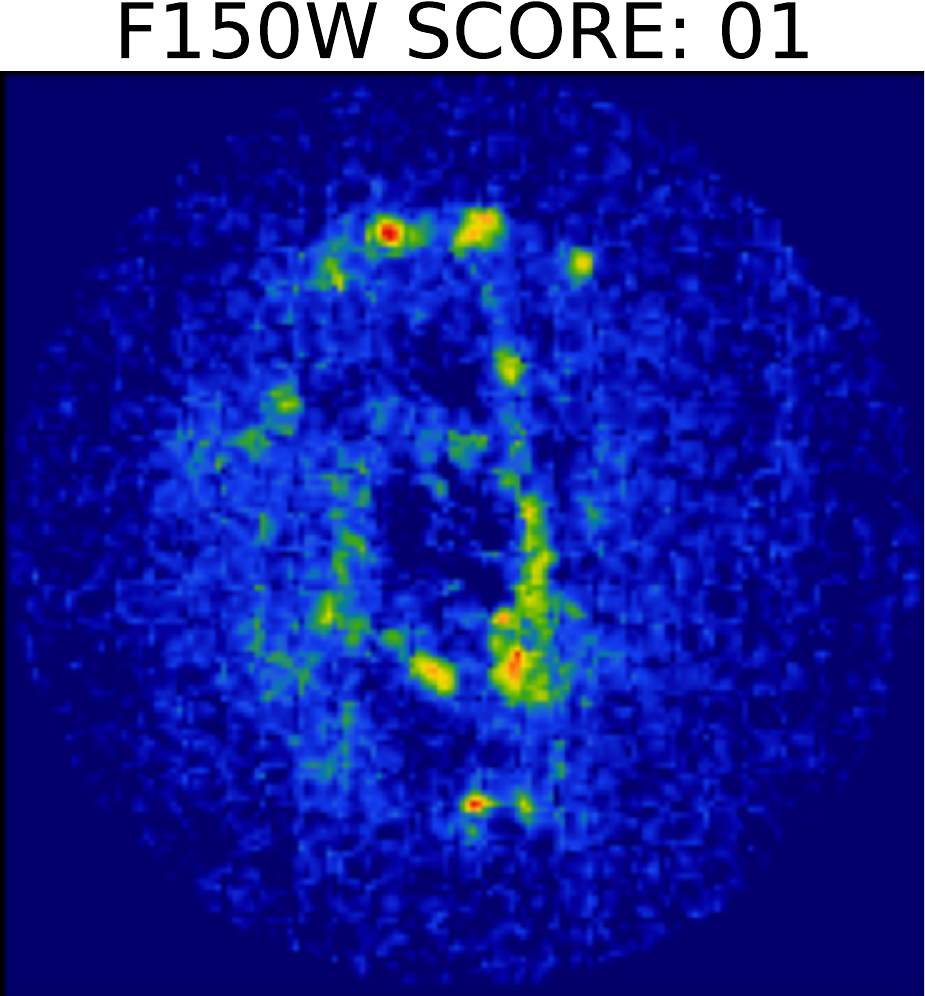}
\includegraphics[width=0.12\textwidth]{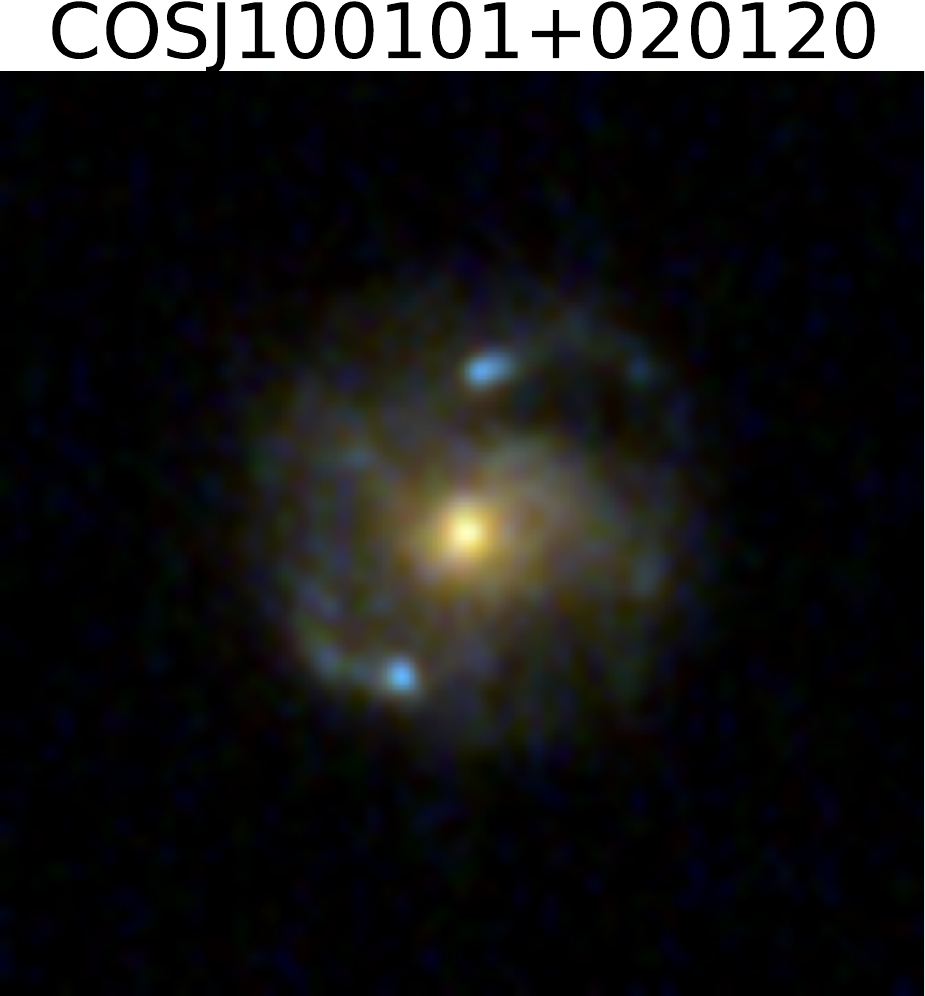}
\includegraphics[width=0.12\textwidth]{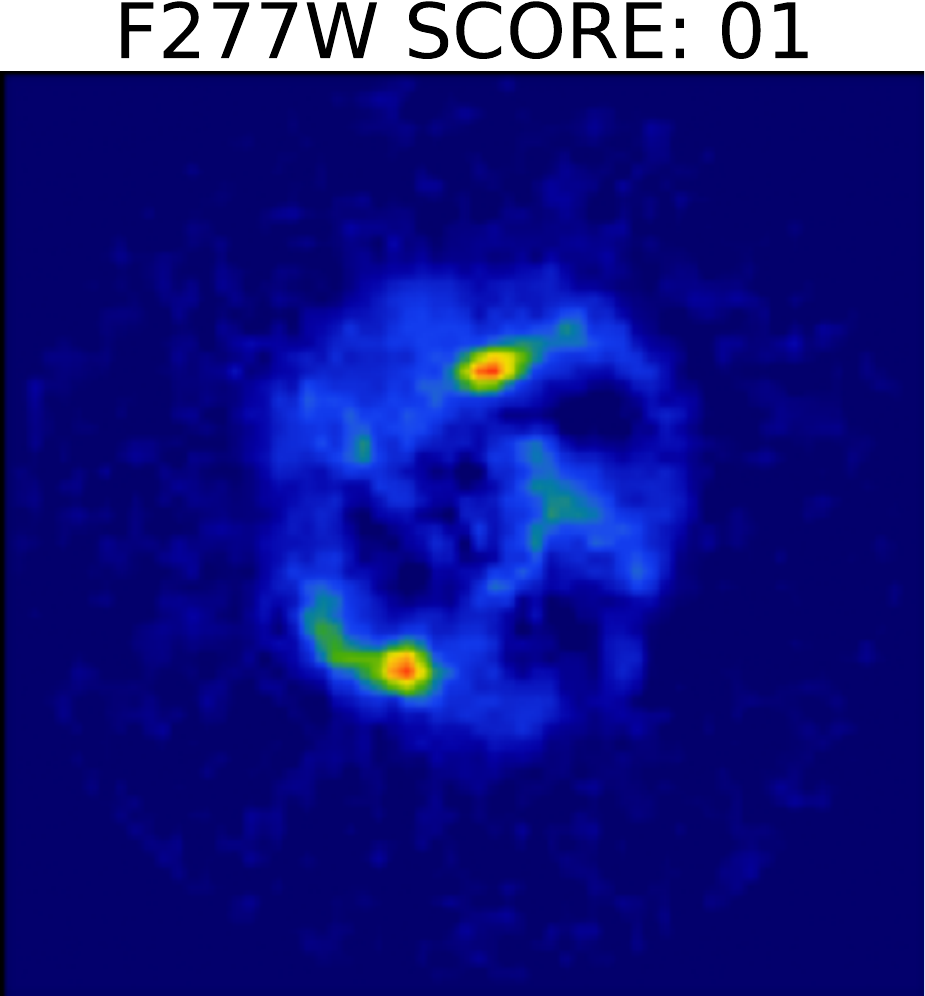}
\includegraphics[width=0.12\textwidth]{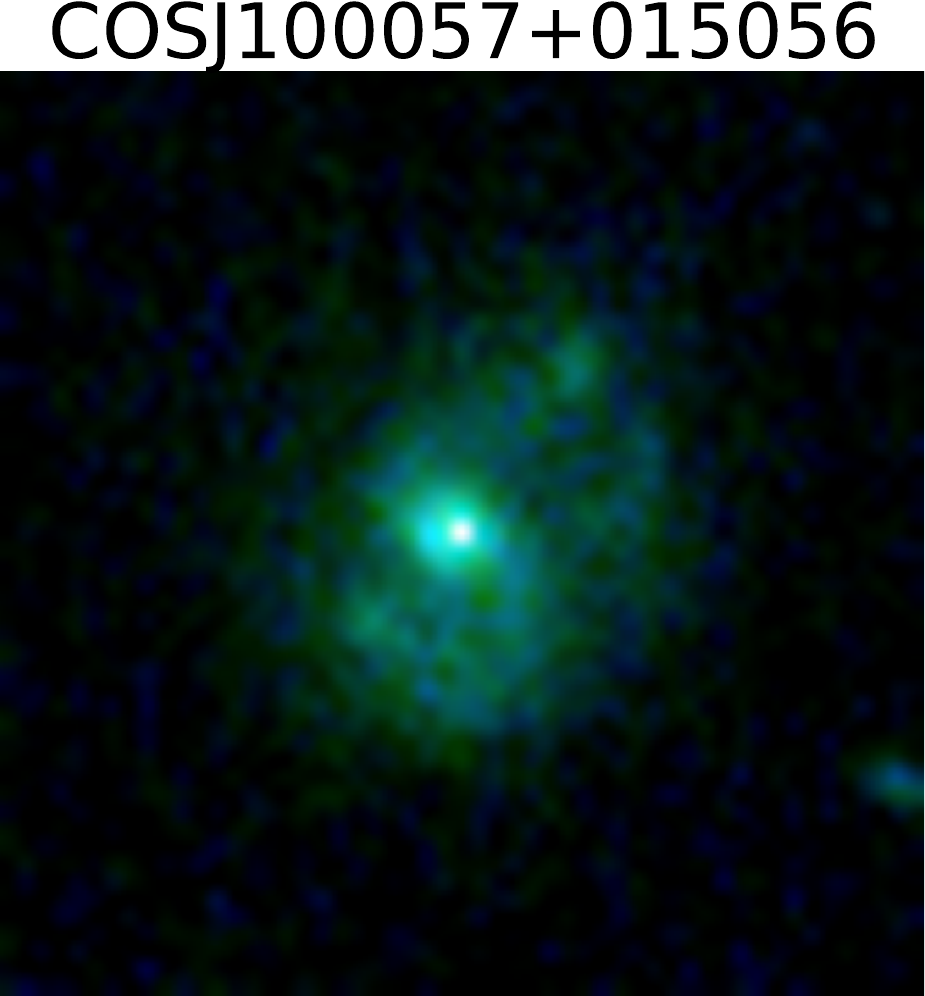}
\includegraphics[width=0.12\textwidth]{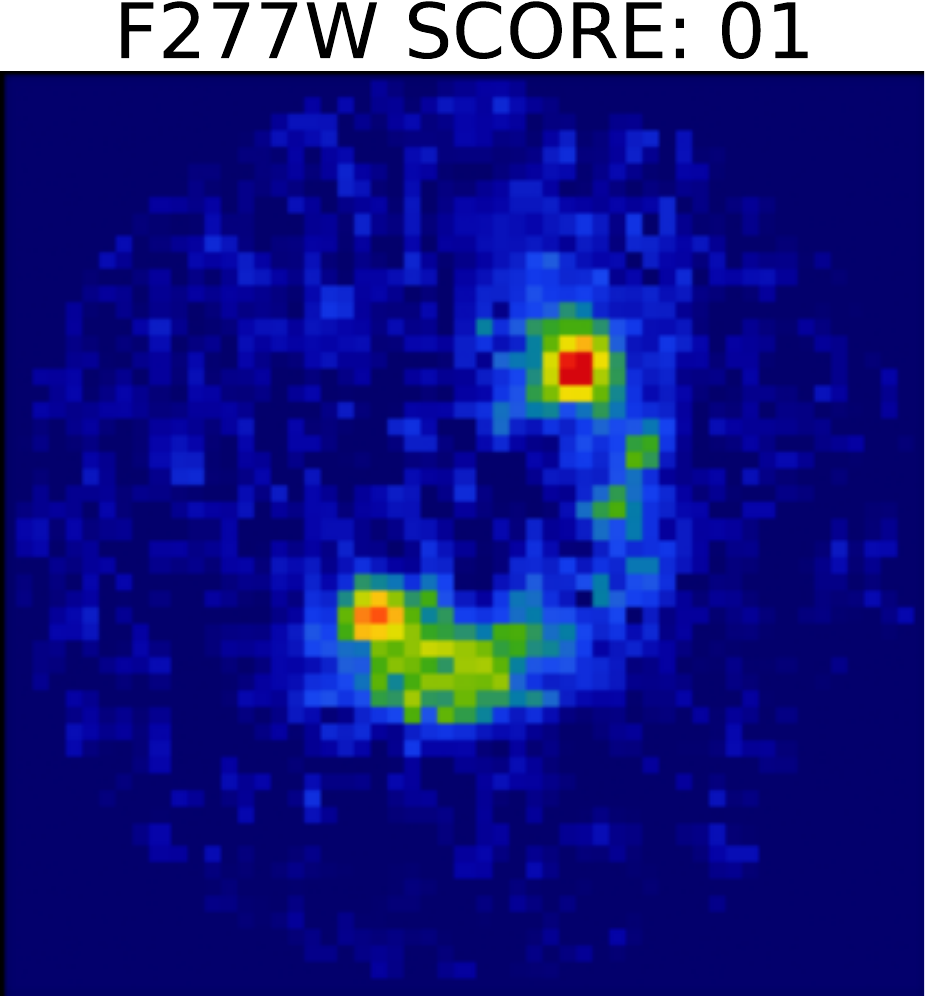}
\includegraphics[width=0.12\textwidth]{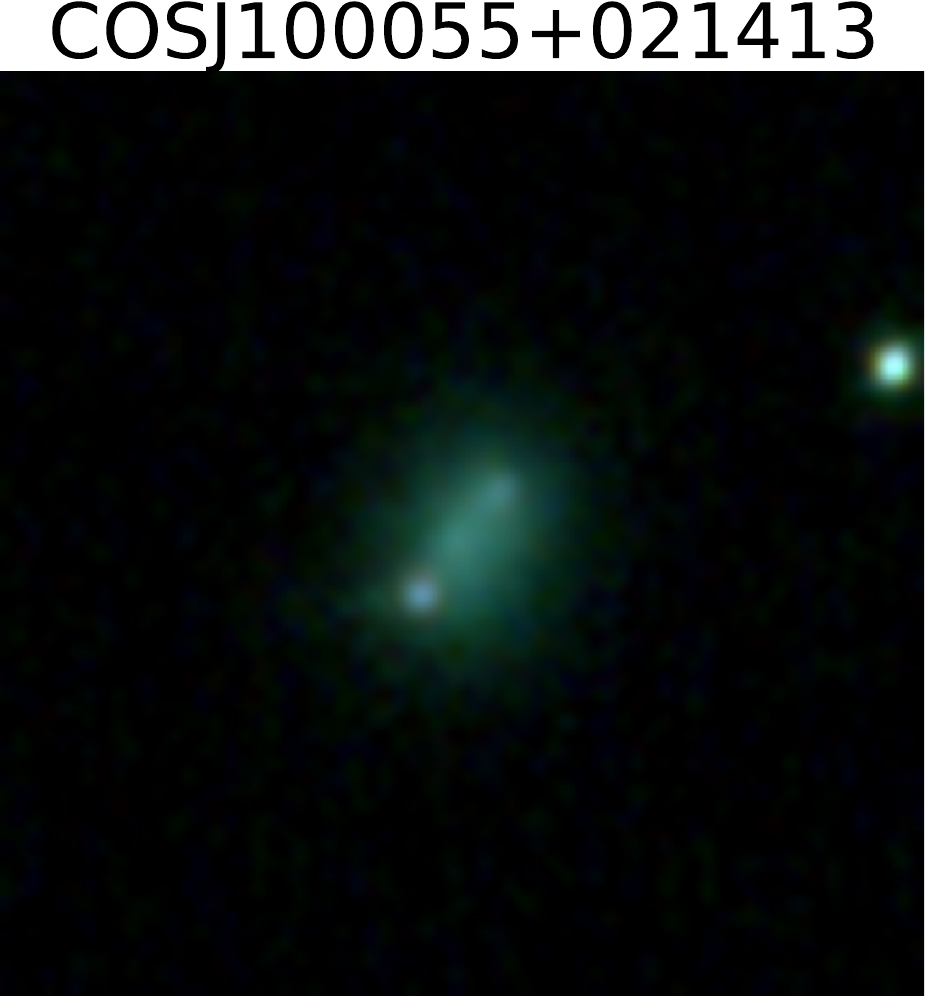}
\includegraphics[width=0.12\textwidth]{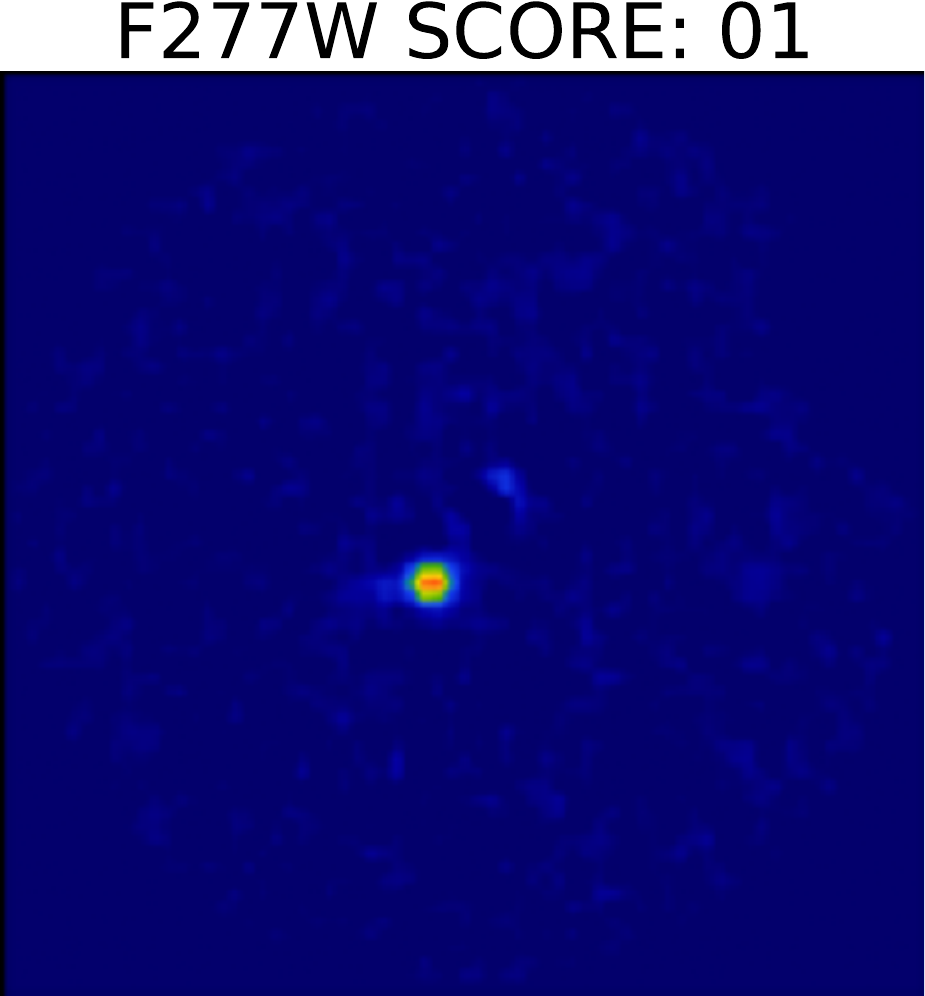}
\includegraphics[width=0.12\textwidth]{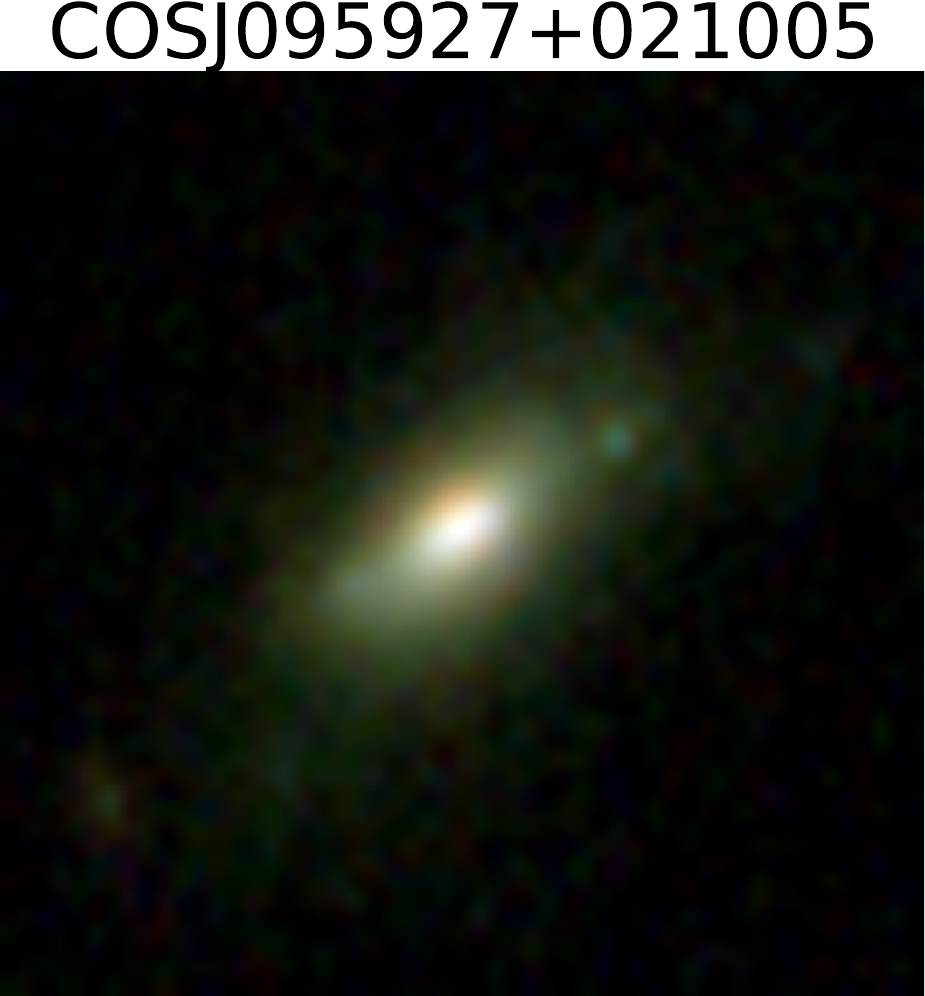}
\includegraphics[width=0.12\textwidth]{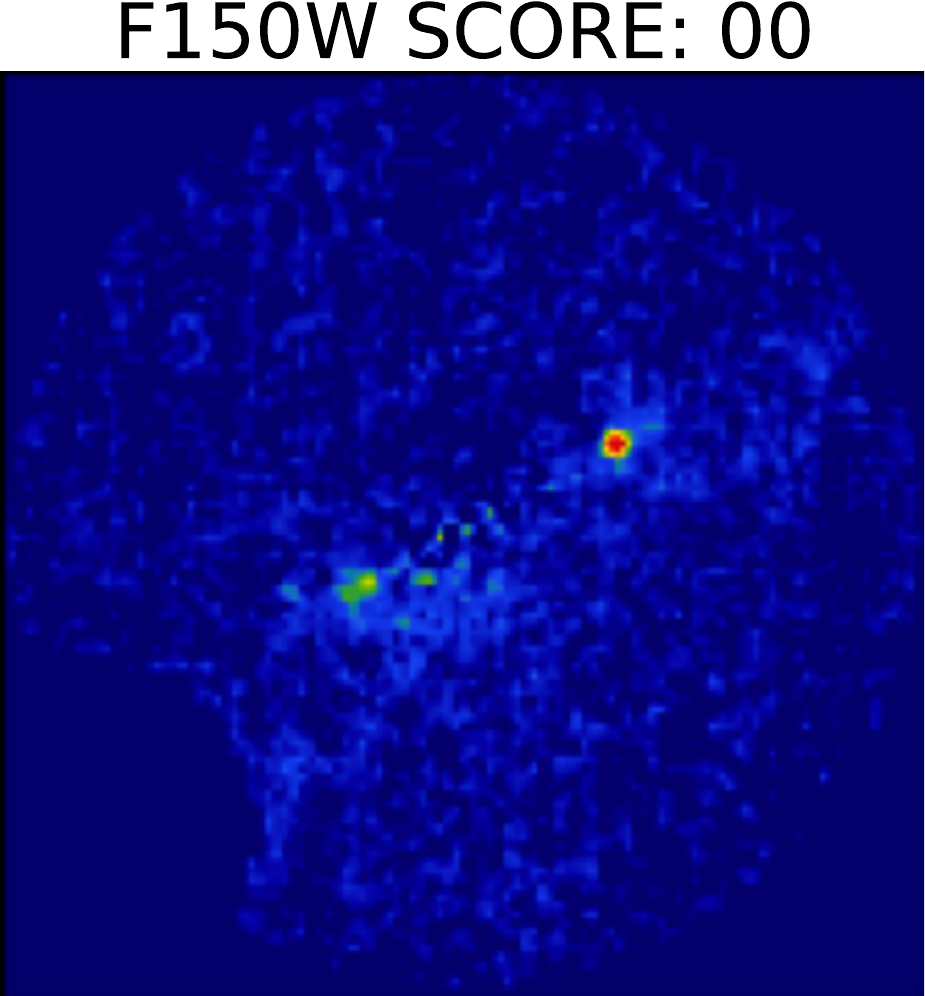}
\caption{
Identical to \cref{figure:CutoutA}, showing 40 candidates which scored below 5 in the second visual inspection round that have been chosen by JWN as those which show the clearest evidence for being strong lenses. The top left system, COSJ095939+02343, which scored 4, is confirmed as a lens by \citet{Guzzo2007}.
}
\label{figure:CutoutB}
\vspace{-30pt} 
\end{figure*}

Figure \ref{figure:CutoutB} shows forty candidates which scored four and below, which have subjectively been chosen by JWN as examples which show the most convincing evidence of being strong lenses. The aim of this figure is twofold: (i) to allow readers to assess the quality of the best candidates with low scores and; (ii) to provide visual insight into the regime where distinguishing genuine strong lenses from false positives is most challenging. All 40 candidates are successfully fitted with physically plausible lens models whereby the SIE plus shear focuses the candidate lensed source emission into consistent regions of the source plane. However, assuming that some candidates are false positives (most forecasts for COSMOS-Web predict $\sim 100$ lenses, \citealt{Holloway2023, Ferrami:2024obm}, H25), this means that successfully fitting a lens model is not sufficient to confirm a candidate is a strong lens. The top left candidate COSJ095939+02343, which scored 4, is confirmed as a lens by \citet{Guzzo2007}.

Candidate scoring is an inherently subjective process, and for candidates with scores of 8 and below it is common for some inspectors to score an object A whilst others gave it U or X. Different readers will view \cref{figure:CutoutA} and \cref{figure:CutoutB} and make different judgments regarding which they think are genuine lenses. Below a score of 10, it is therefore difficult to make definitive statements about whether any given candidate is definitively a strong lens or a false positive, and therefore how many genuine strong lenses are in the COWLS sample. Therefore, in the following sections which quantify the sample characteristics, all plots are shown for all $419$ candidates as a function of score.

\subsection{The COWLS Sample Properties}

\subsubsection{Lens Redshifts}

\begin{figure*}
\centering
\includegraphics[width=0.32\textwidth]{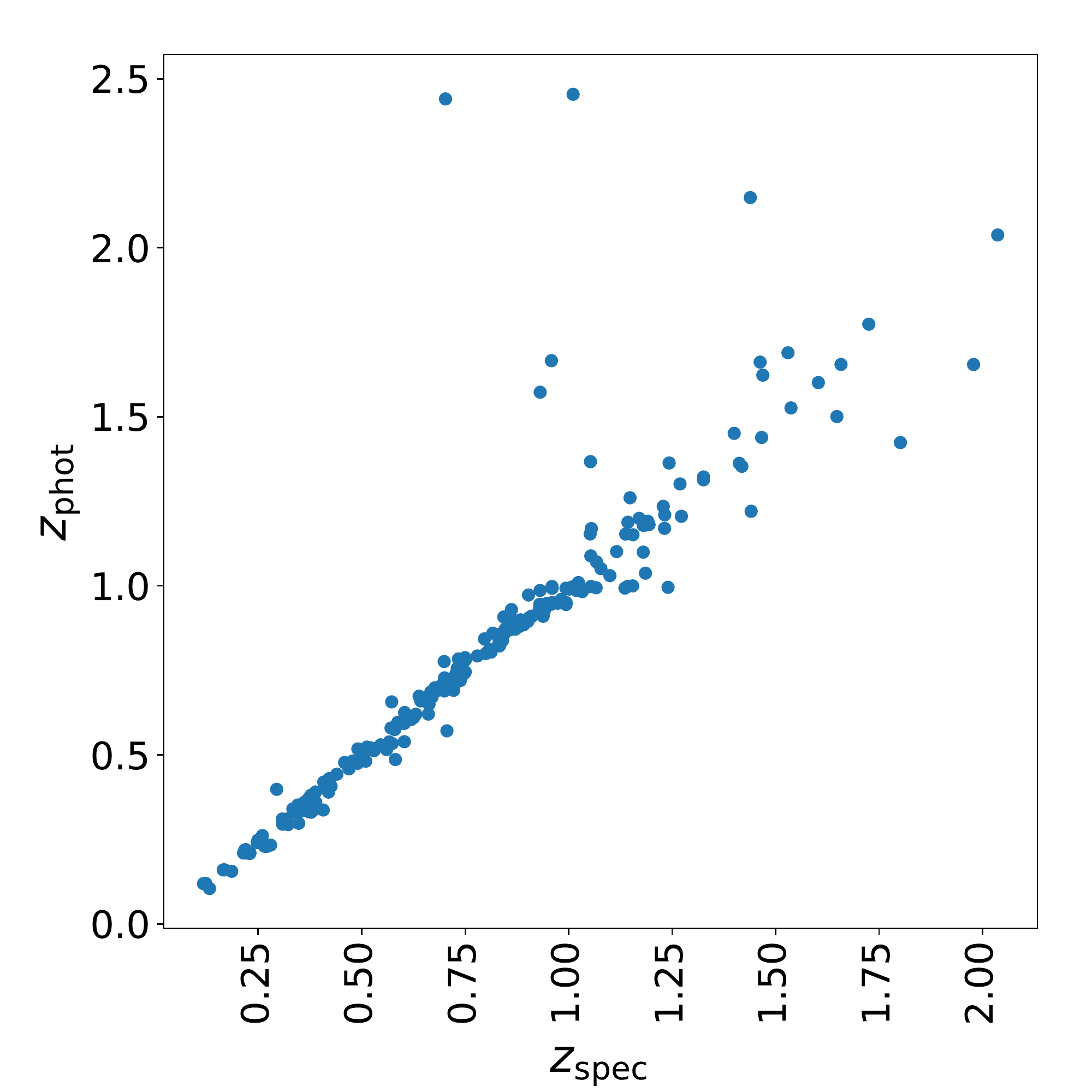}
\includegraphics[width=0.32\textwidth]{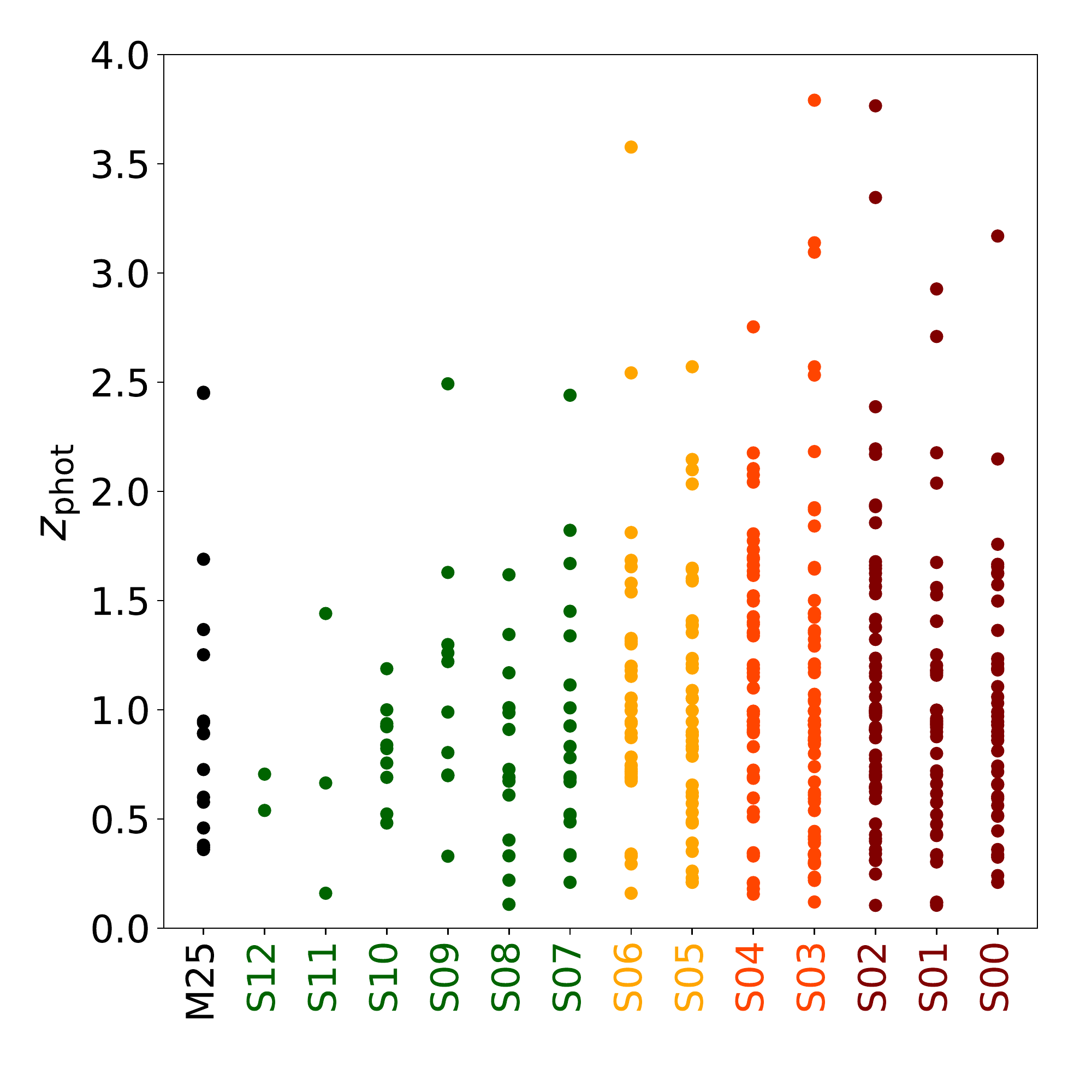}
\includegraphics[width=0.32\textwidth]{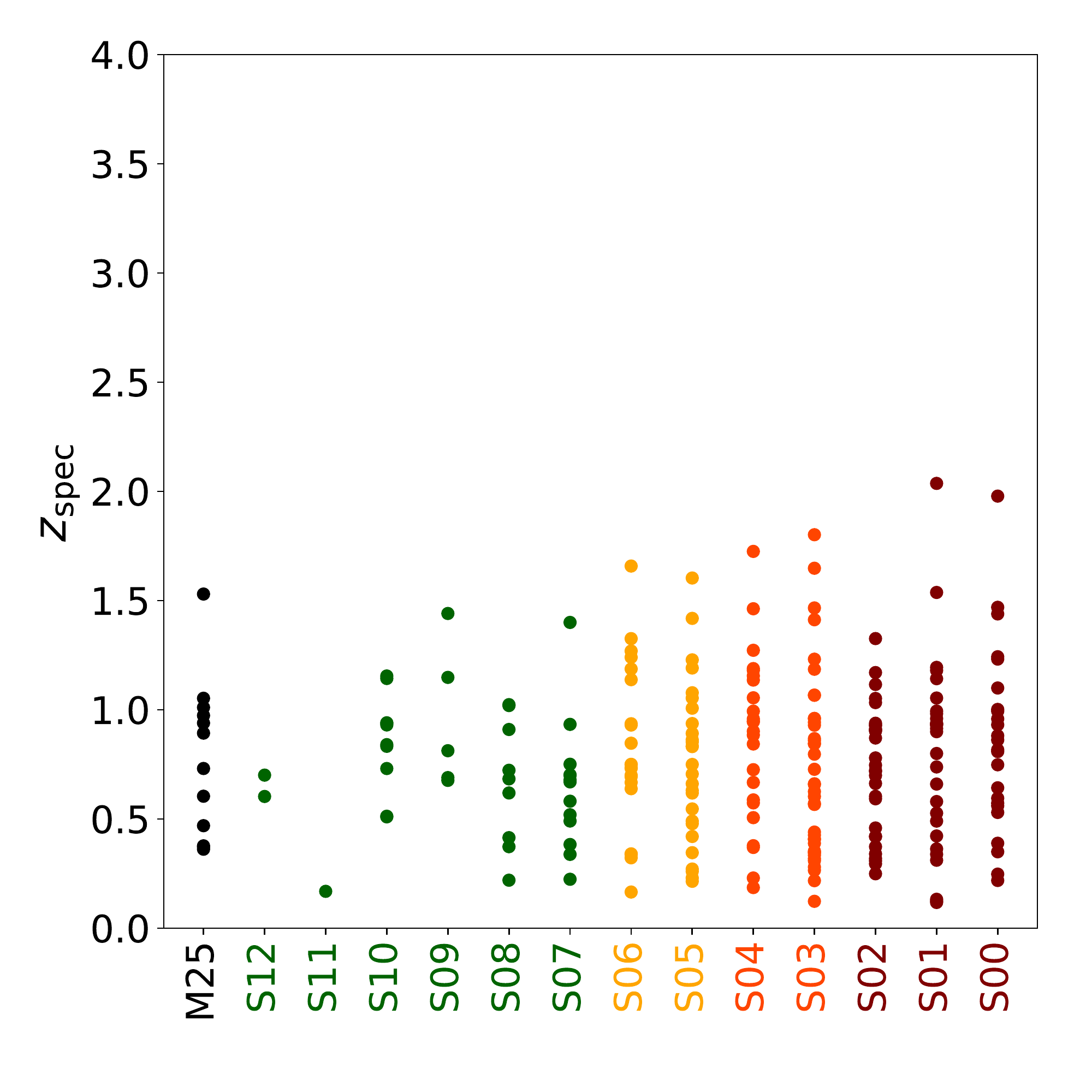}
\caption{
Spectroscopic and photometric redshift measurements for candidate lens galaxy emission in the COWLS sample are shown here. The full source of redshift information is described in \cref{Redshift}, which includes DESI DR1 spectroscopic data and photometric redshift estimates from COSMOS archive ground-based and \textit{HST} samples. The left panel compares spectroscopic and photometric redshift estimates for 243 candidates where both measurements are available. The near one-to-one correspondence reaffirms that the majority of photometric redshift measurements are reliable. The central panel shows photometric redshift estimates for 409 candidates where available, plotted as a function of their second-round visual inspection score. The right panel shows spectroscopic redshifts for 245 candidates with available measurements. Scores are colour-coded: high-ranked systems in green, mid-ranked in orange, and low-ranked in red (see \cref{visual:second_round}). Spectroscopic data is unavailable above redshifts of $z \sim 2$, while photometric estimates extend to redshifts around $z \sim 4$. There is no correlation between candidate score and redshift.}
\label{figure:Redshift}
\end{figure*}

Photometric and spectroscopic redshift estimates for candidate lens galaxies are available in the COSMOS archive (see \cref{Redshift}), and these are shown for all candidates from the second round of visual inspection in \cref{figure:Redshift}. The left panel compares photometric and spectroscopic estimates for 243 candidates where both measurements are available. The near one-to-one correspondence between the two confirms the majority of photometric redshift estimates are accurate, particularly up to $z \sim 2$, where spectroscopic data is accessible. Beyond this redshift, spectroscopic catalogues (e.g., DESI, see \cref{Redshift}) lack data. Overall, the sample has reliable redshift information for most of its candidate lenses.

The central and right panels show the relationship between photometric and spectroscopic redshifts and second-round scores. Based on photometric redshifts, about half of the sample's lens galaxies are above $z = 1$, with some candidates exceeding $z = 2$. There is no observable trend between candidate score and redshift estimate, indicating that high-ranking candidates are not more likely to be more distant lens galaxies.

\subsubsection{Weak Lensing Mass Map}

\begin{figure*}
\centering
\includegraphics[width=0.99\textwidth]{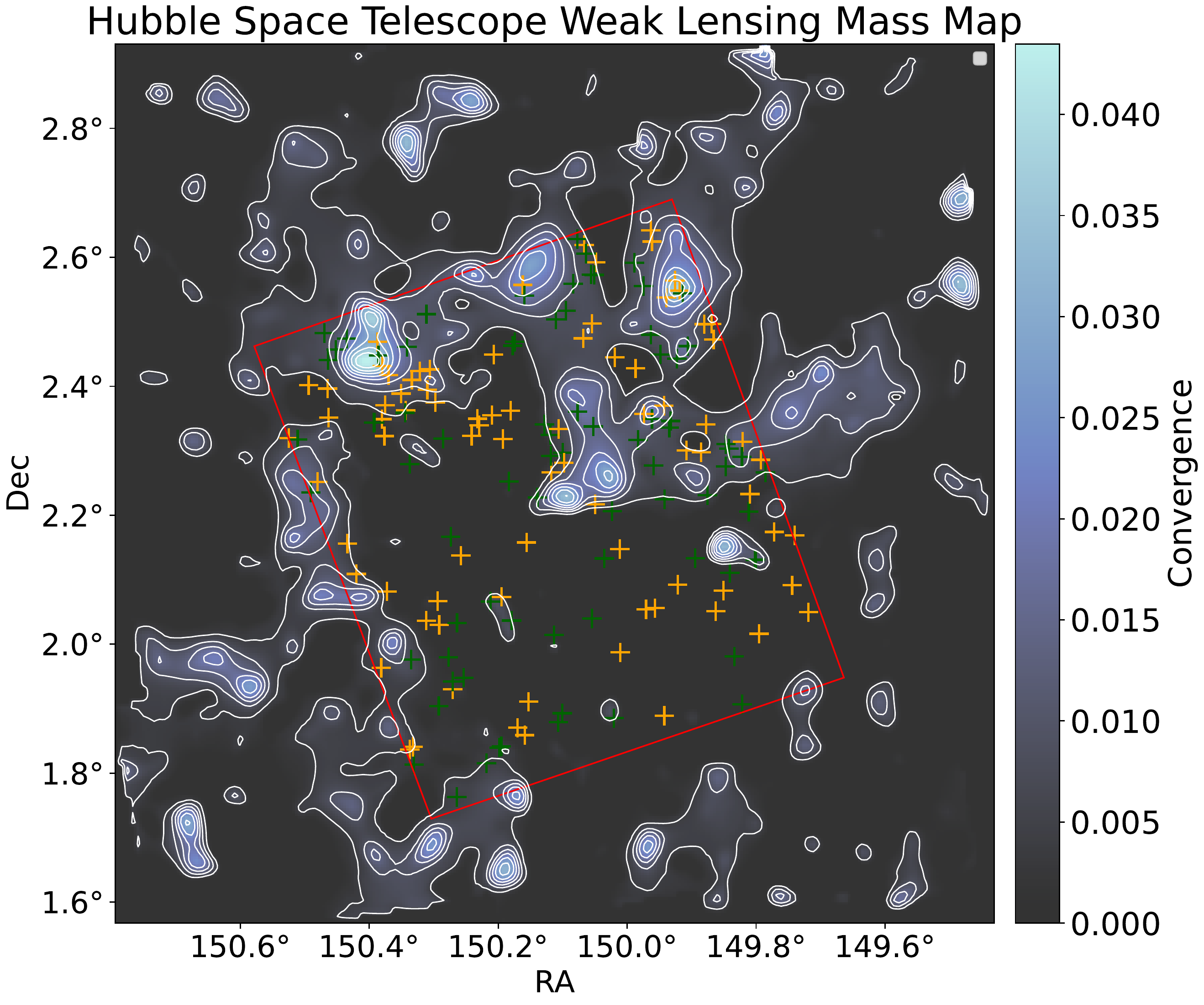}
\caption{
The weak lensing convergence estimated via the \textit{Hubble} Space Telescope weak lensing mass map analysis of \citet{Massey2007}. High-ranked strong lens candidates are marked as green (M25 spectacular lenses and scores of 7 and above) and yellow (scores of 5 and 6) crosses. The red rectangle represents the coverage of the COSMOS-Web JWST imaging, and strong lenses appear only within the red rectangle since they are exclusively identified through COSMOS-Web JWST data. The distribution of candidates shows correlation with regions of higher weak lensing convergence, but there are also many candidates in lower density regions. 
}
\label{figure:Weak}
\end{figure*}


Figure \ref{figure:Weak} shows the distribution of strong lens candidates scoring five and above, plotted over the
\textit{HST} weak lensing convergence map inferred by \citet{Massey2007}. Higher convergences indicate denser regions of dark matter, with two high-density peaks visible to the north of the COSMOS-Web field, along with filamentary structures connecting them. The red rectangle outlines the $0.54$\,deg$^2$ COSMOS-Web footprint, where \textit{JWST} imaging is available, while weak lensing analysis outside this rectangle uses \textit{HST} imaging. Our strong lens search was conducted using only the \textit{JWST} data, so all strong lens candidates (marked with crosses) fall within this red rectangle.

There is a noticeable clustering of strong lens candidates over the higher-density peaks of dark matter inferred from weak lensing, especially at the top-left and top-right corners of the red rectangle. However, many candidates are also found in lower-density regions. A detailed statistical comparison to determine if lens candidates are more likely to be found in denser regions of dark matter large-scale structure would require a comparison to the full sample of inspected candidates, which is beyond the scope of this work. Nevertheless, this figure highlights that all lenses have high-quality weak lensing data available, which can help inform their local environment, constrain their mass distribution on larger scales, and enable unique studies of the strong lens candidates in the COWLS sample.

\subsubsection{Lens Model Properties}

\begin{figure*}
\centering
\includegraphics[width=0.32\textwidth]{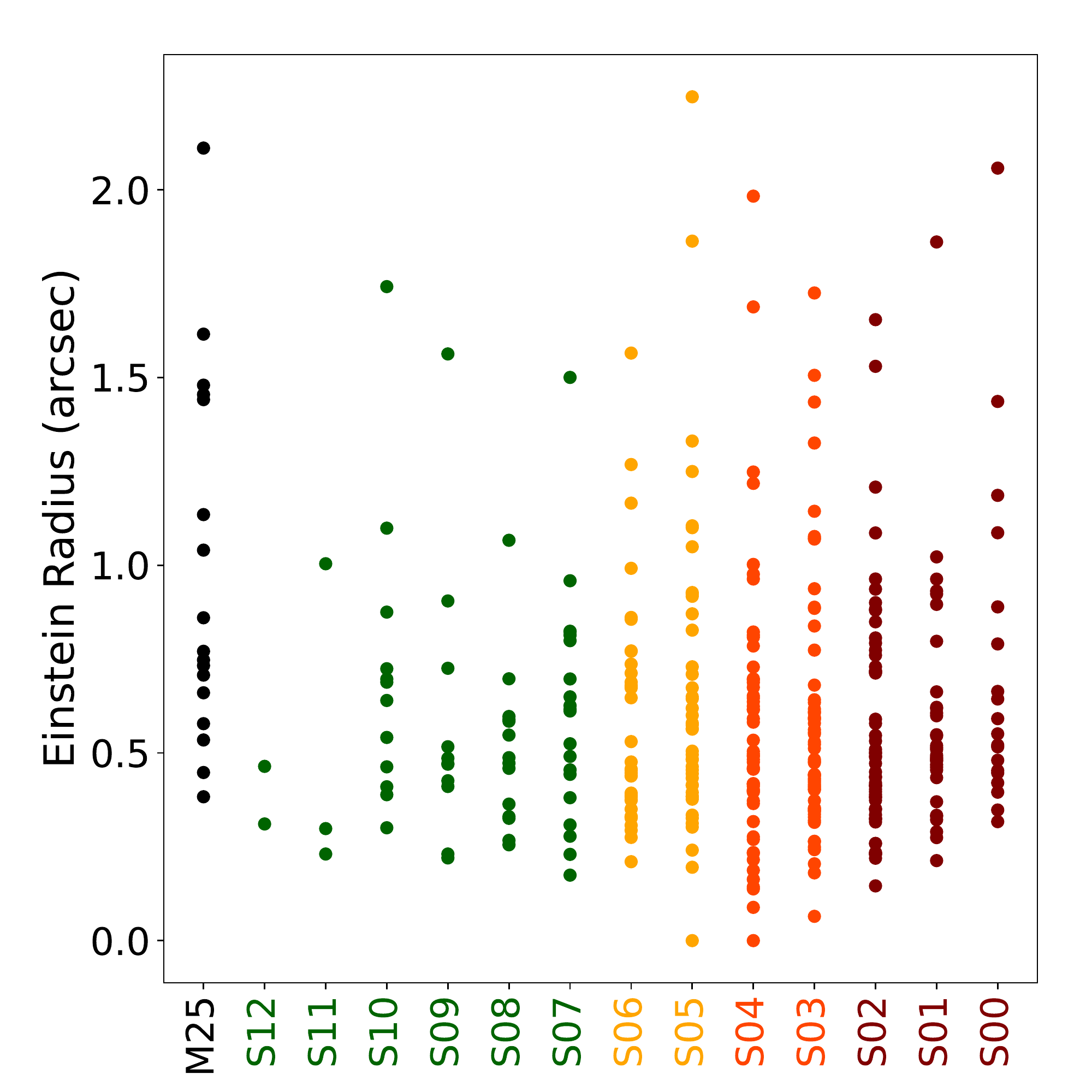}
\includegraphics[width=0.32\textwidth]{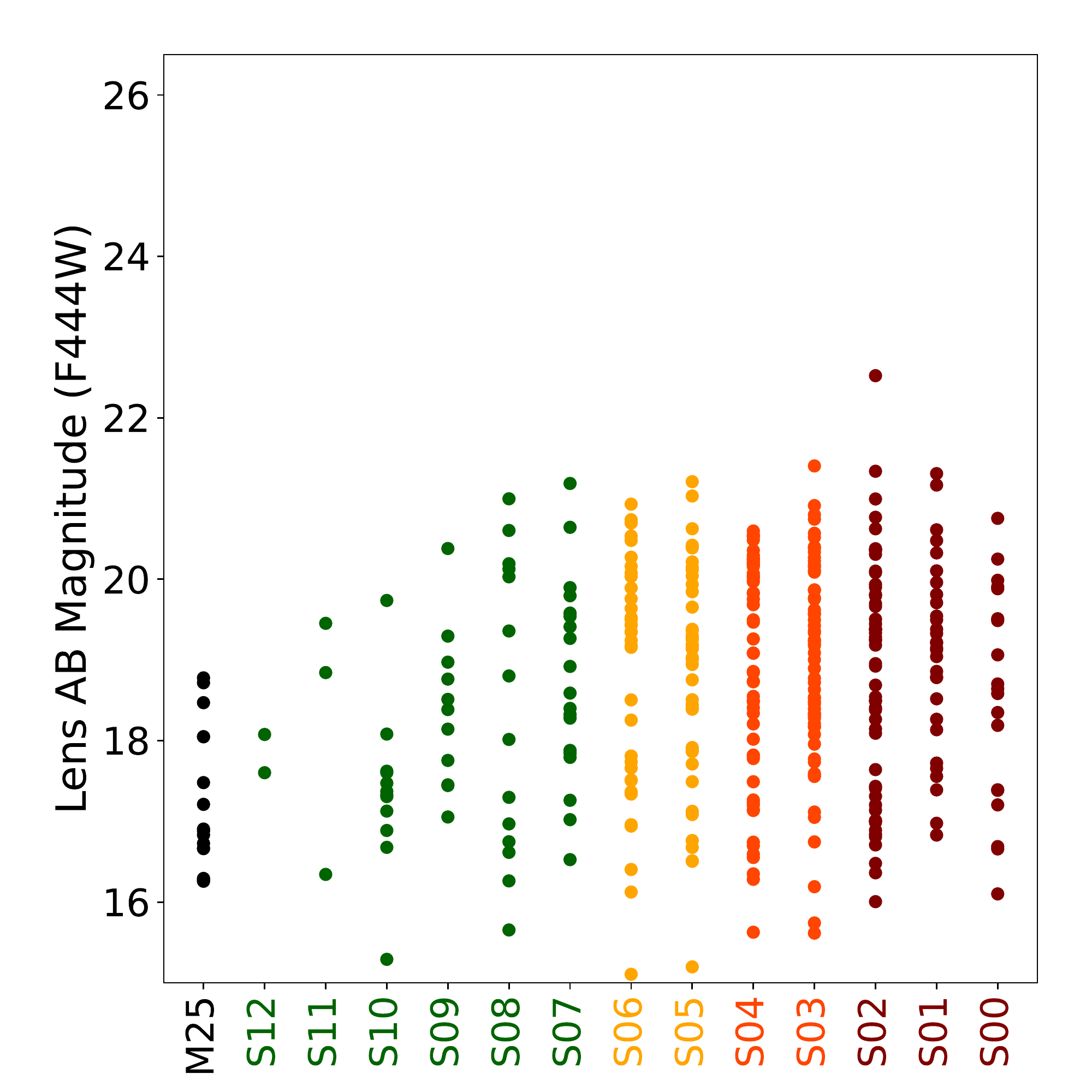}
\includegraphics[width=0.32\textwidth]{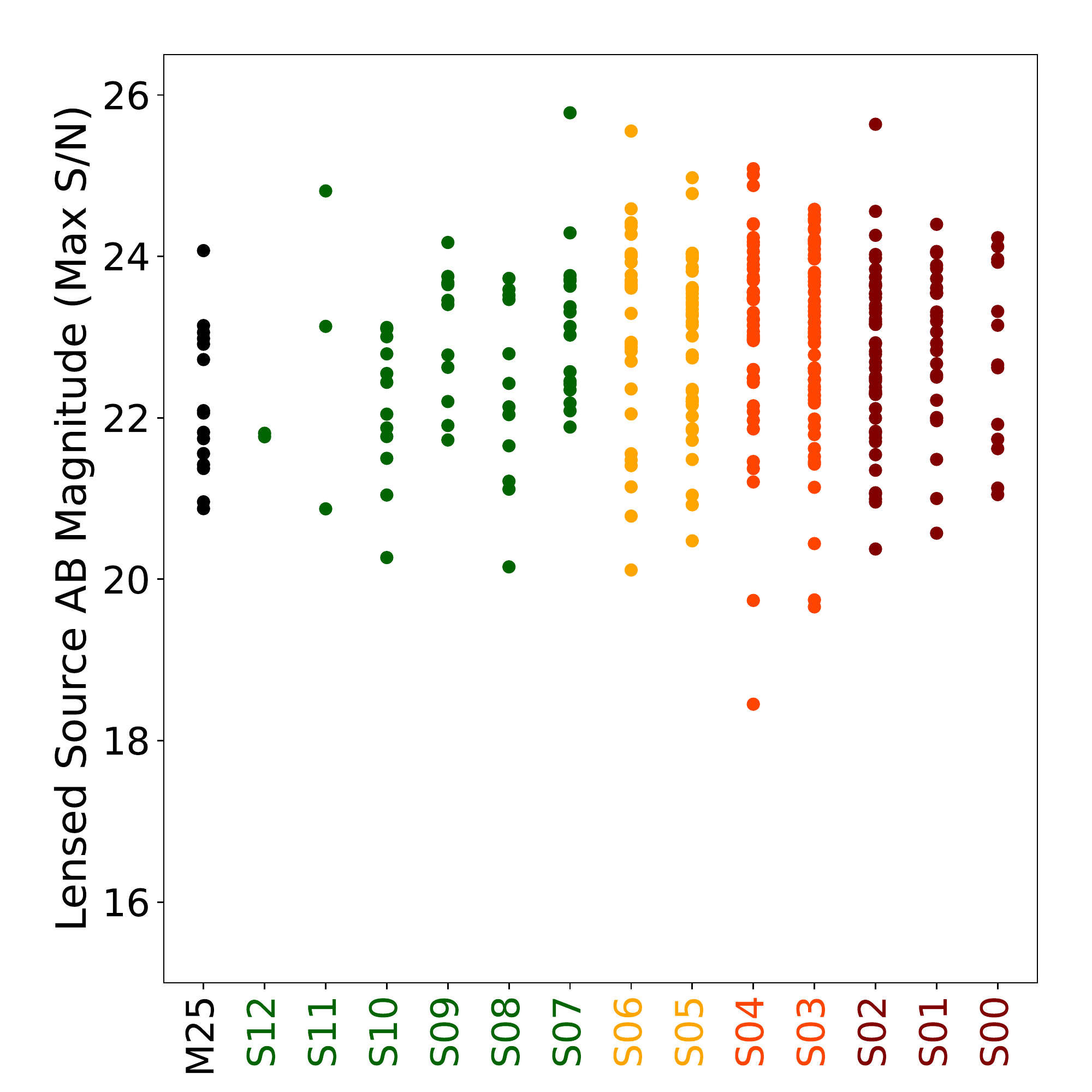}
\includegraphics[width=0.32\textwidth]{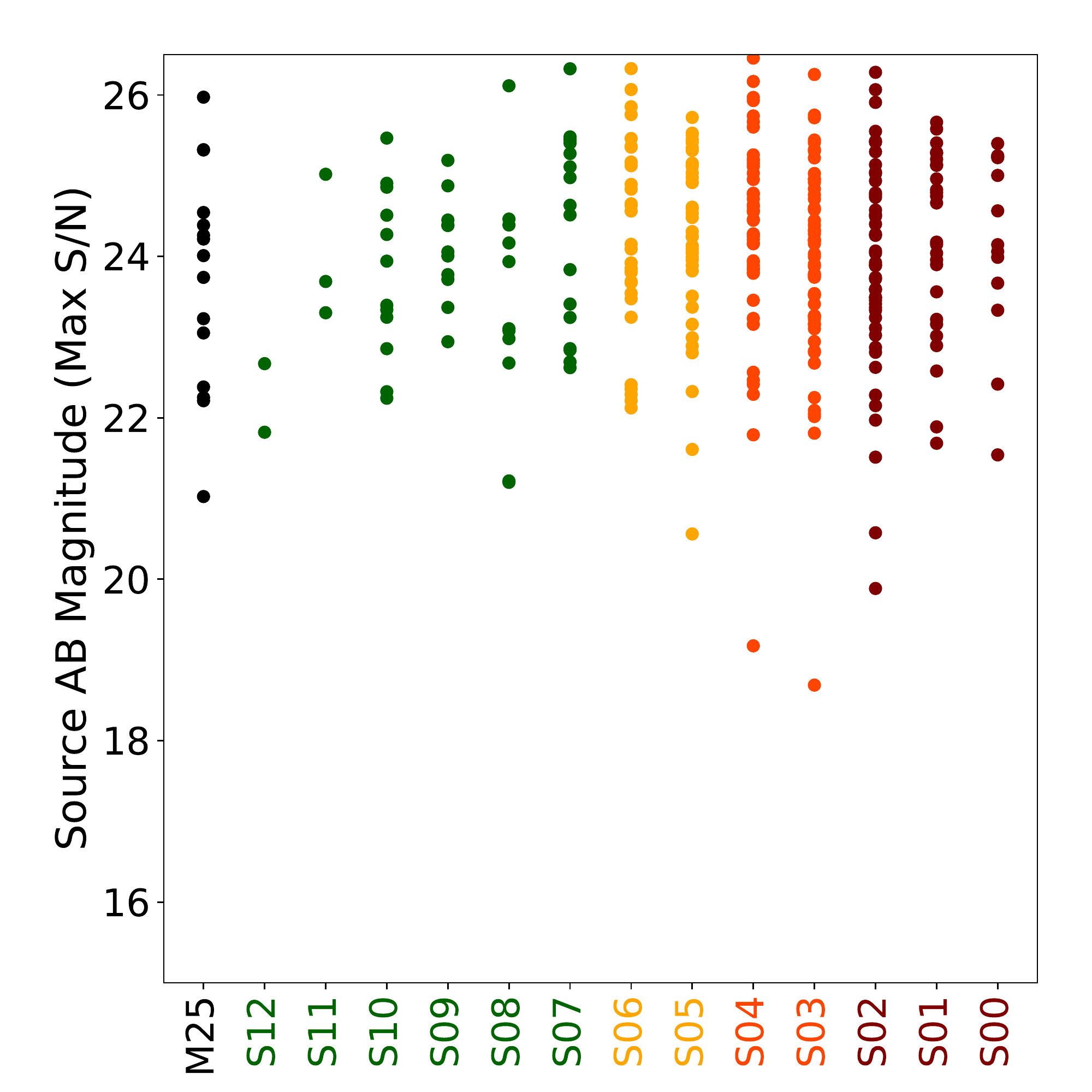}
\includegraphics[width=0.32\textwidth]{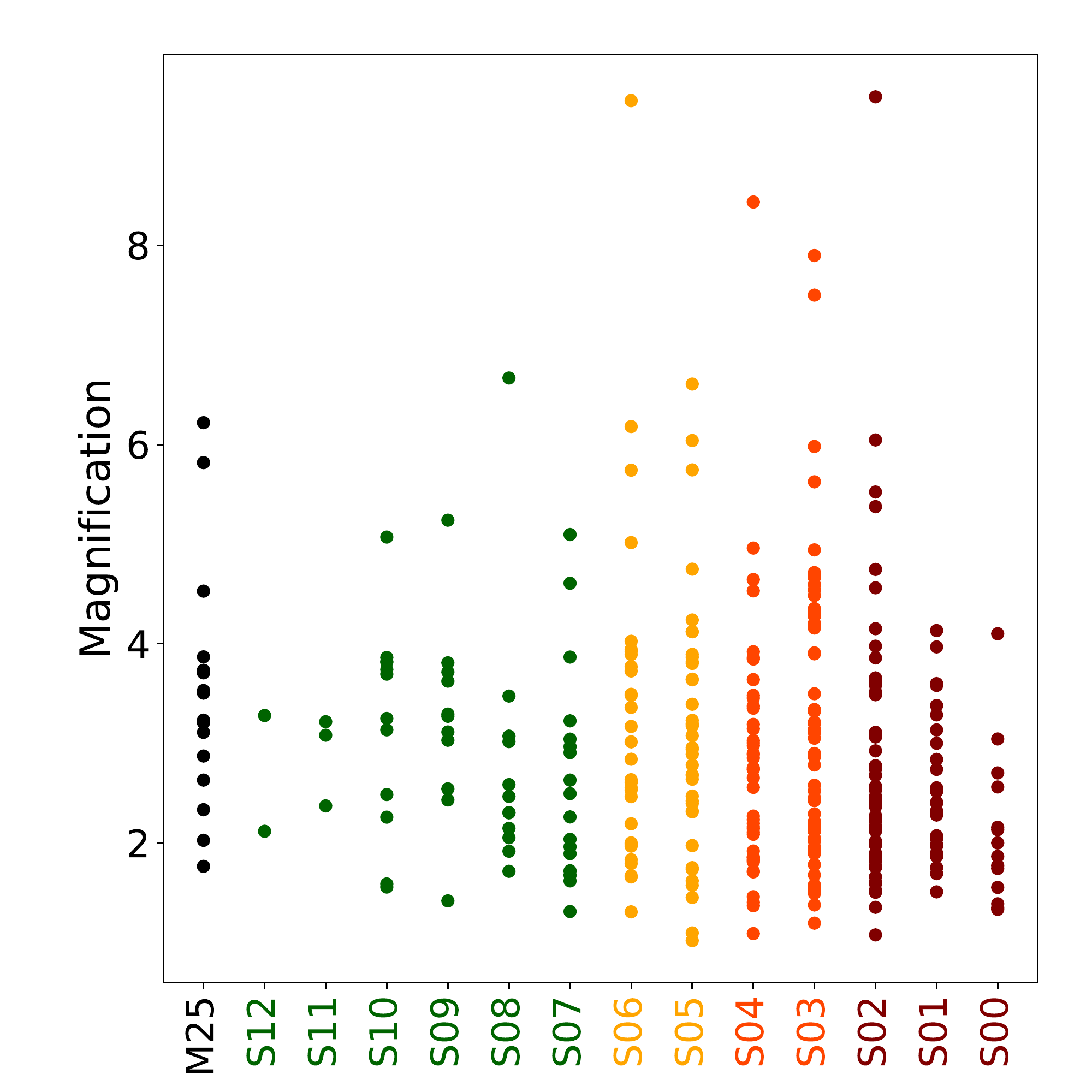}
\includegraphics[width=0.32\textwidth]{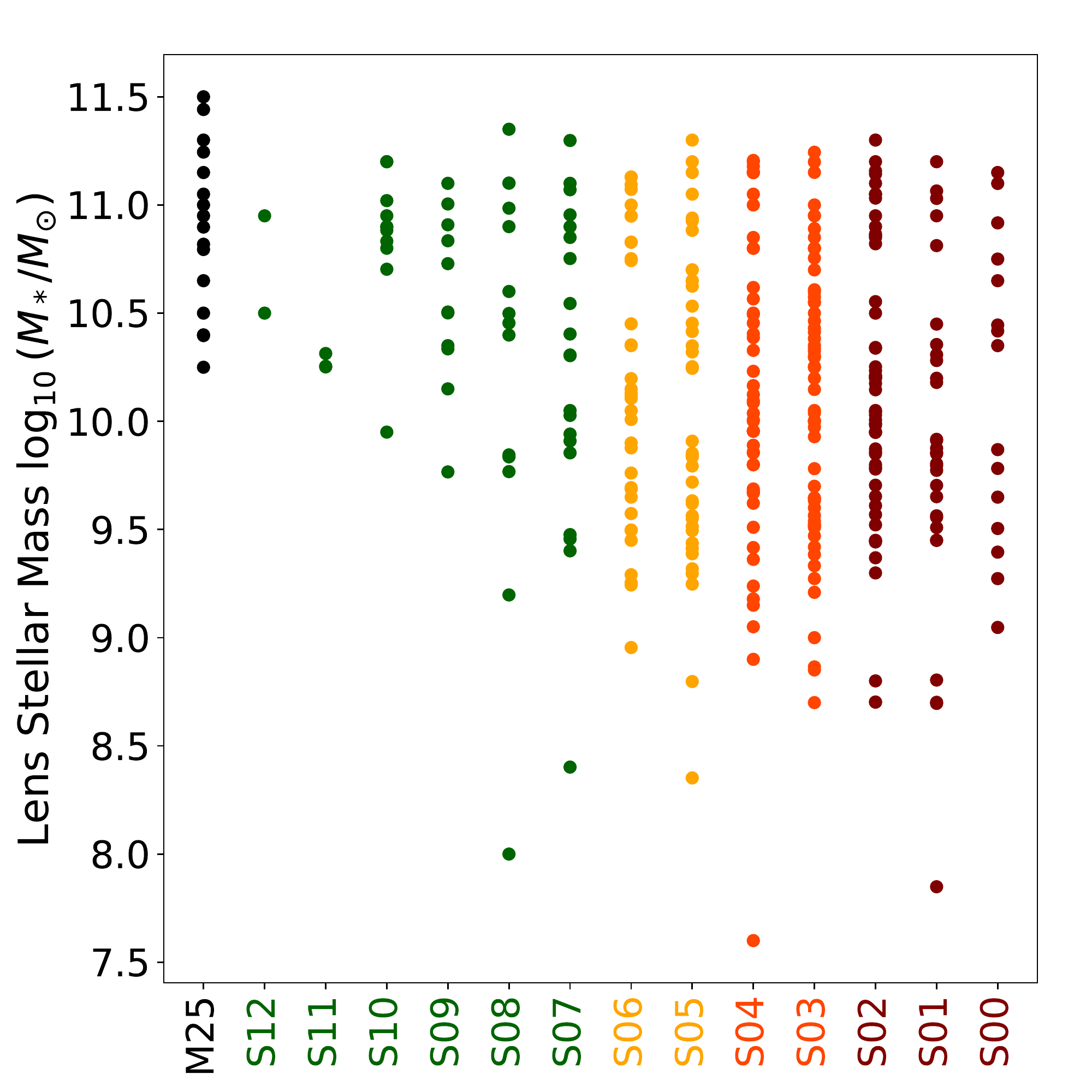}
\caption{
The Einstein radius (arcsec), lens galaxy AB magnitudes (F444W waveband), lensed source galaxy image-plane AB magnitudes (highest S/N waveband), source galaxy source-plane AB magnitudes (highest S/N waveband), magnification ratios and stellar masses as $\log_{10} (M_*/M_{\odot})$ for all candidates in the second round of visual inspection. The $x$-axis of each plot shows each candidate's score in the second round of visual inspection. Scores are coloured such that high ranked systems are green, mid ranked orange and low ranked red (see \cref{visual:second_round}). There is no correlation between candidate score and any quantity, except stellar mass, where higher scoring candidates have higher stellar masses.
}
\label{figure:Sample}
\end{figure*}

Figure \ref{figure:Sample} presents the Einstein radii (arcsec), AB magnitudes of lens galaxies (F444W waveband), lensed source galaxy magnitudes in both the image and source planes (highest S/N waveband), magnification ratios and stellar masses (collected from the COSMOS-Web catalogue) for all candidates from the second round of visual inspection. Most candidates, regardless of their score, have Einstein radii below $1.0\arcsec$, lens galaxy magnitudes ranging from 16 to 22, and source magnitudes between 20 and 25. These values differ from existing strong lens samples, such as SLACS \citep{Bolton2008a, Shu2017}, where Einstein radii are typically above $0.8\arcsec$ and lens magnitudes range from 15 to 17. The smaller Einstein radii and fainter lens galaxies are consistent with higher-redshift lens galaxies, as discussed in the redshift analysis. SLACS source magnitudes peak around 24 for the F814W \textit{HST} filter, so the candidate sources in our sample have similar magnitudes but observed in \textit{JWST}’s longer waveband filters \citep{Newton2011}. SLACS magnifications are higher, with a median of $\mu = 8.8$ and peak values above $40$ \citep{Newton2011}; their higher magnifications could partly explain why SLACS sources have similar magnitudes despite being selected based on SDSS spectroscopy. The stellar masses of SLACS lenses are nearly all above $\log_{10} (M_*/M_{\odot}) = 11$ for a Chabrier initial mass function, indicating that the majority of the COWLS sample consists of lower-mass lenses.

No correlation is observed between candidate score and the quantities shown in \cref{figure:Sample}, except for stellar mass, where the highest ranked candidates tend to correspond to the highest stellar mass lens galaxies. The lack of correlation between Einstein Radius and score further suggests that lens models fitted to false positives can yield physically plausible values, given that the H25 forecast suggests around $\sim 100$ of the $419$ are genuine lenses. While lens modelling provides crucial information for inspectors to better assess whether a system is a lens, quantitative cuts in modelling results alone therefore cannot be broadly applied across the sample to eliminate false positives.

\subsubsection{Source-Lens Image Separation}\label{SourceLensSeparation}

\begin{figure*}
\centering
\includegraphics[width=0.24\textwidth]{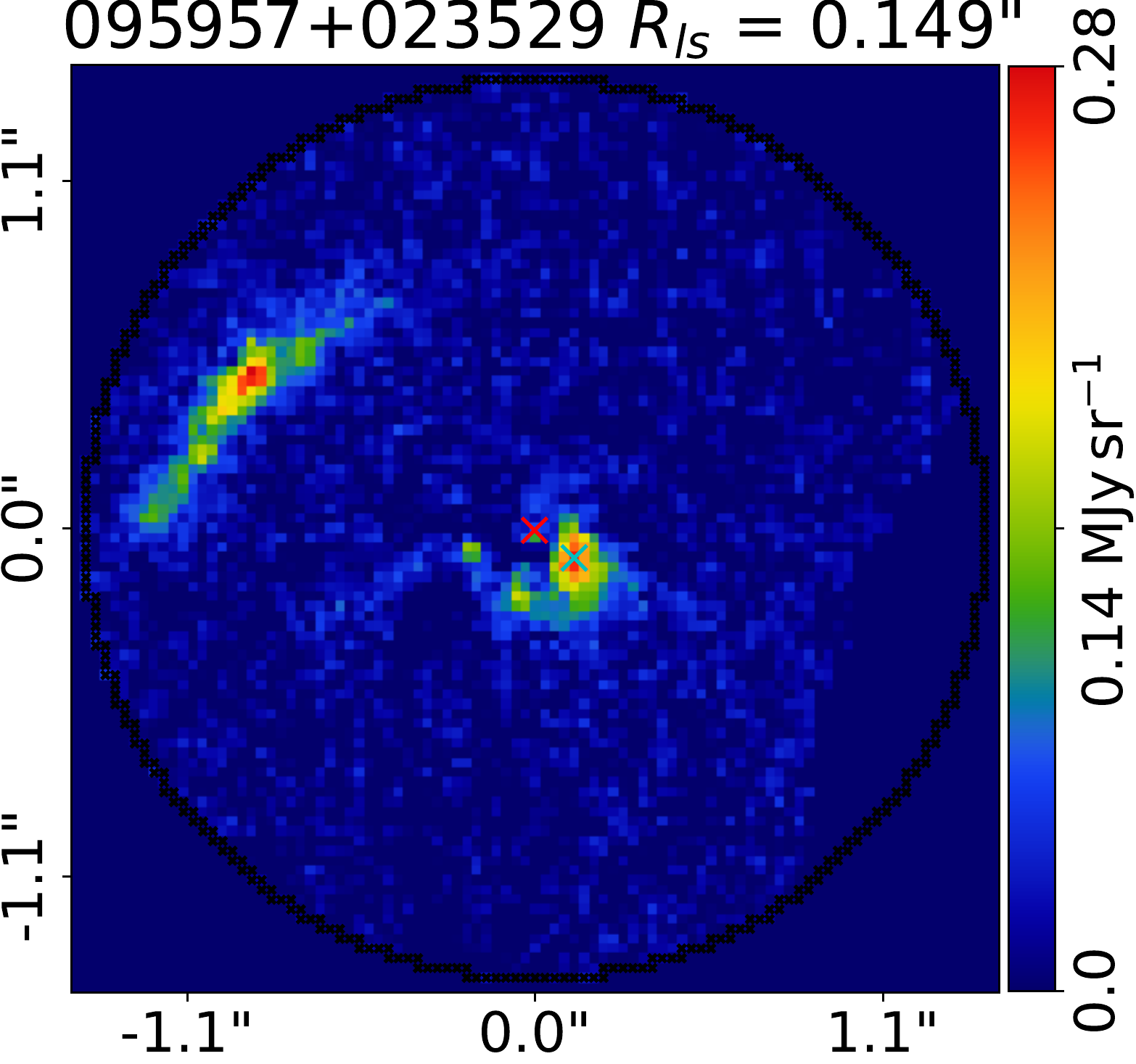}
\includegraphics[width=0.24\textwidth]{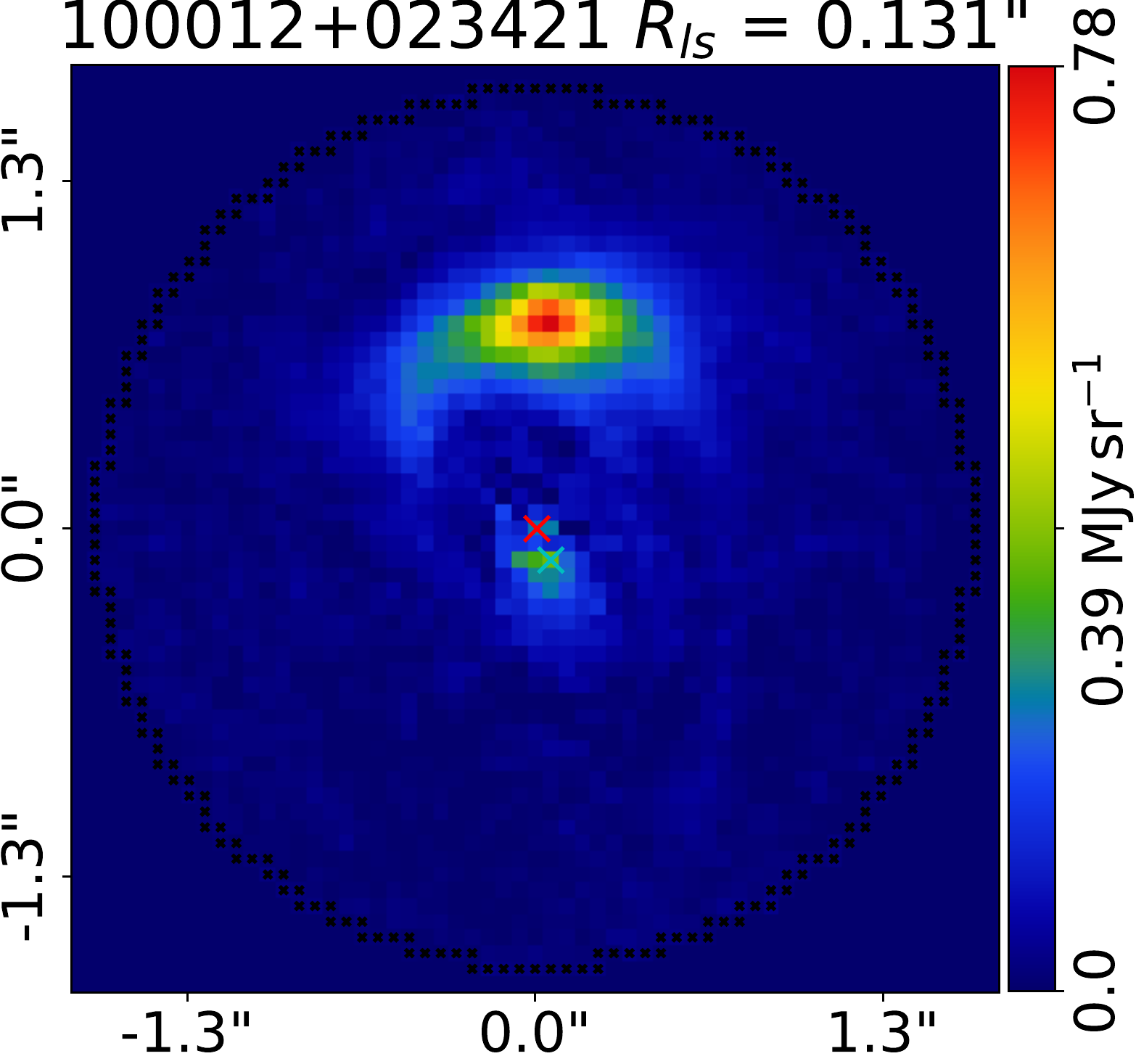}
\includegraphics[width=0.24\textwidth]{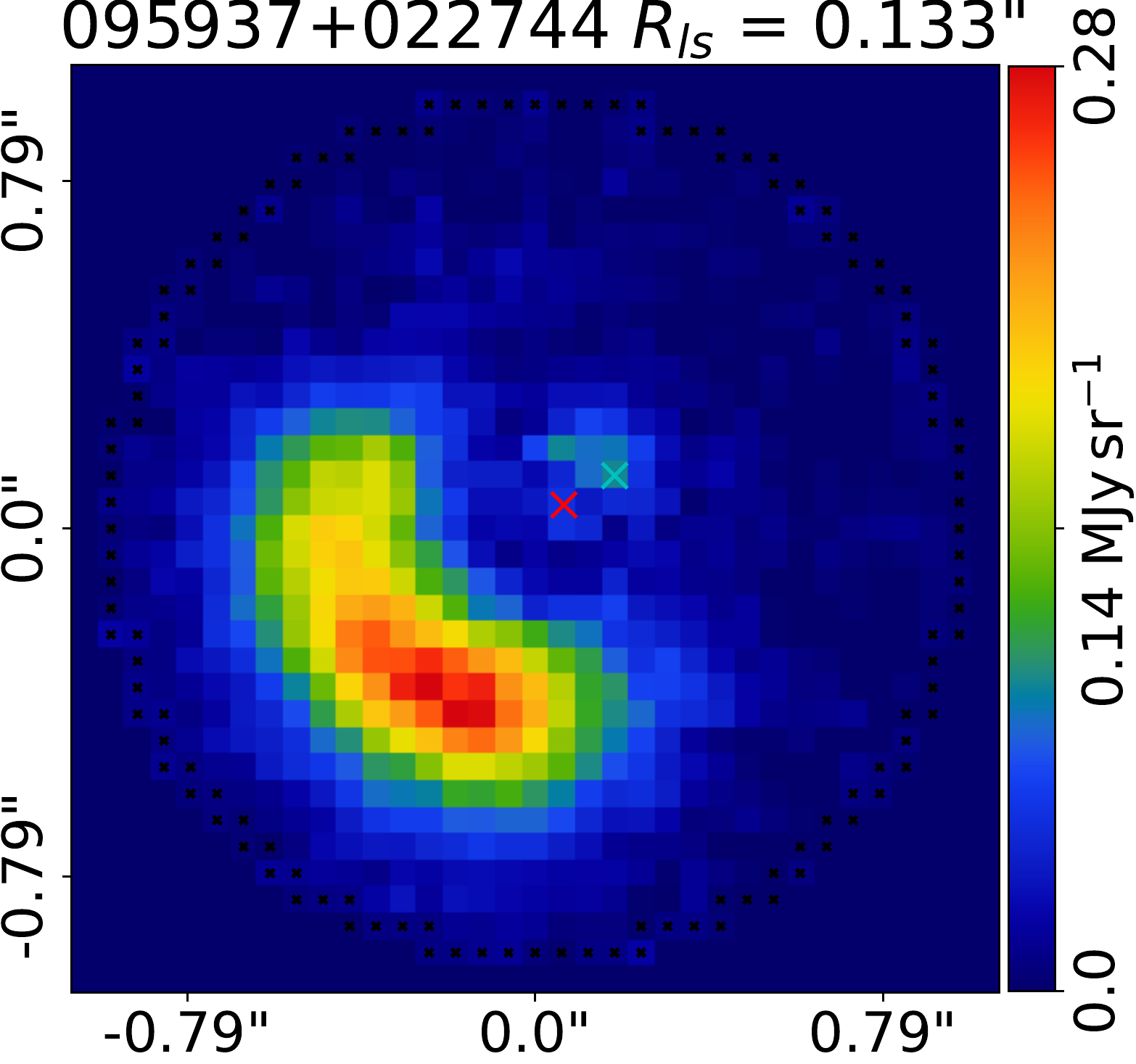}
\includegraphics[width=0.24\textwidth]{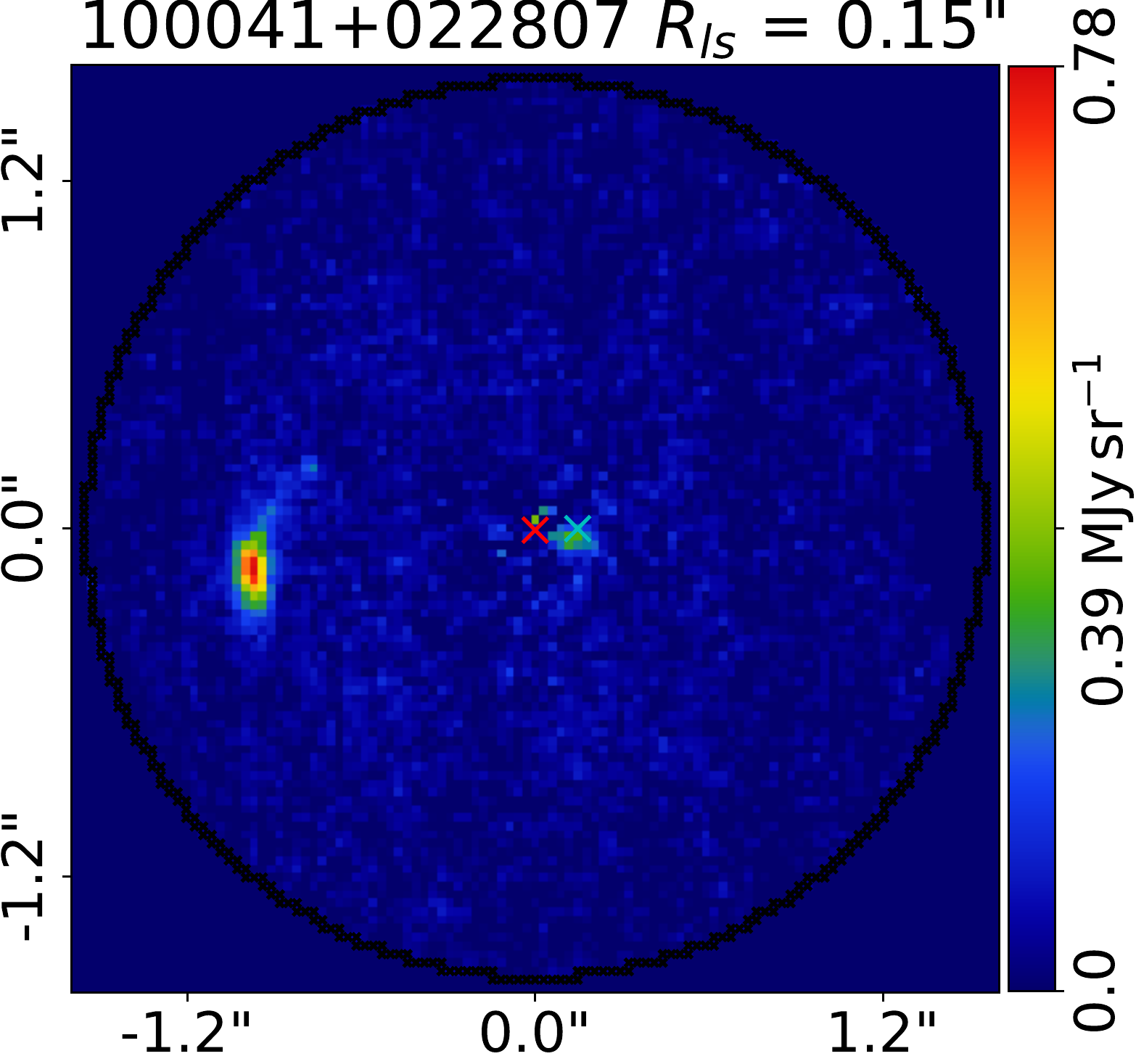}
\caption{
Lens subtracted images of four candidates which scored 7 or above where $R_{ls}$, the arcsecond radial distance between the lens galaxy centre (based on the MGE lens light model fit) and closest lensed source image, is below $0.15\arcsec$. The lens galaxy centres are shown with red crosses, the lensed source emission centres as cyan crosses.}
\label{figure:SourceLens1}
\end{figure*}

\begin{figure}
\centering
\includegraphics[width=0.48\textwidth]{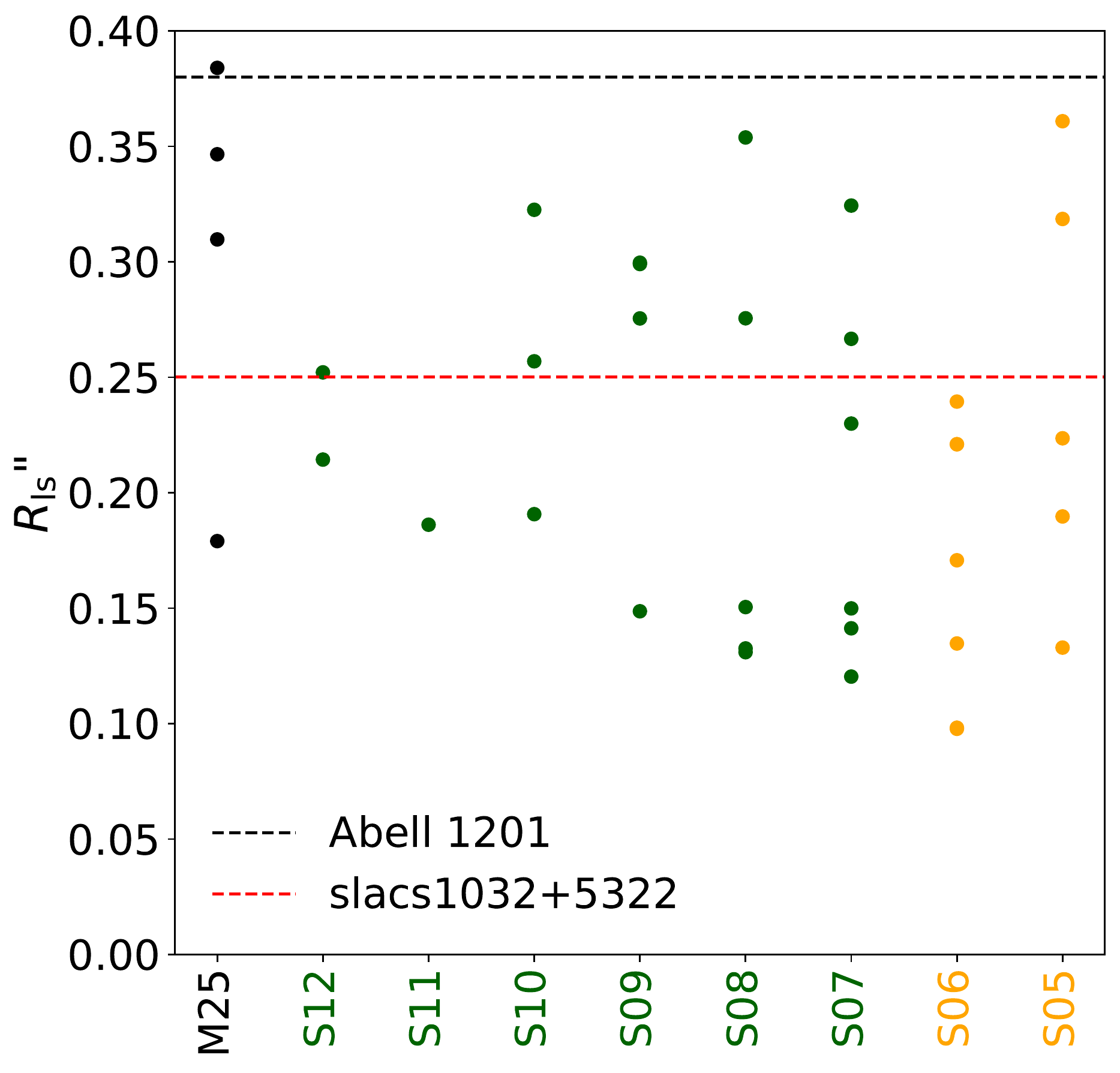}
\caption{
$R_{ls}$, the arcsecond radial distance between the lens galaxy centre (based on the MGE lens light model fit) and closest lensed source image, for $37$ candidates in the second round of visual inspection which scored 5 and above and had an $R_{ls}$ below 0.4. The $x$-axis of each plot shows each candidate's score in the second round of visual inspection. The black dashed line shows $R_{ls} = 0.38\arcsec$ for the strong lens Abell 1201 \citep{Nightingale2023}, where the separation was sufficiently small that the lens galaxy's super massive black hole was detected and measured to have a mass of $M_{\rm BH} = 3.27 \pm 2.12  \times 10^{10}$\,M$_{\rm \odot}$. The red dashed line shows this separation for the SLACS strong lens SDSSJ1032+5322, which of the sample presented in \citep{Etherington2022} has the lowest $R_{ls} = 0.38\arcsec$ value. The COWLS lenses have some of the lowest $R_{ls}$ values of any known strong lens.}
\label{figure:SourceLens2}
\end{figure}

The locations of the lensed source images in a number of candidates are close to the centre of their respective lens galaxies. To quantify this, we define $R_{ls}$ as the arcsecond radial distance between the lens galaxy centre (based on the MGE lens light model fit) and the closest lensed source image. \cref{figure:SourceLens1} shows four example candidates which scored 7 or above, where $R_{ls} < 0.15\arcsec$ and where in all cases the emission near the lens centre corresponds to the counter image of a larger arc located farther from the candidate lens on the opposite side. These candidate sources are therefore near the radial caustic in the source plane, which leads to these configurations.

A total of 37 candidates scored 5 or above and have $R_{ls} \leq 0.4\arcsec$. \cref{figure:SourceLens2} shows $R_{ls}$ versus score for these candidates, confirming that 15 have $R_{ls} < 0.2\arcsec$. The black dashed line shows the value $R_{ls} = 0.38\arcsec$, for the strong lens Abell 1201 \citep{Nightingale2023}, where the separation was small enough for the lens galaxy's supermassive black hole to be detected via lens modelling, with a mass measurement of $M_{\rm BH} = 3.27 \pm 2.12 \times 10^{10}\,M_\odot$. The red dashed line indicates this separation for the SLACS strong lens SDSSJ1032+5322, which, of the sample presented in \cite{Etherington2022}, has the smallest value of $R_{ls} = 0.25\arcsec$. The COWLS sample therefore contains candidates where the source emission passes closer to the lens centre than in other surveys, which in \cref{Discussion} we argue is a selection effect.

\subsubsection{Singly Imaged Arcs}

\begin{figure*}
\centering
\includegraphics[width=0.16\textwidth]{cutout/rgb/S09_AAABBS_COSJ100103+020159.pdf}
\includegraphics[width=0.16\textwidth]{cutout/lens_sub/S09_AAABBS_COSJ100103+020159_F150W.pdf}
\includegraphics[width=0.16\textwidth]{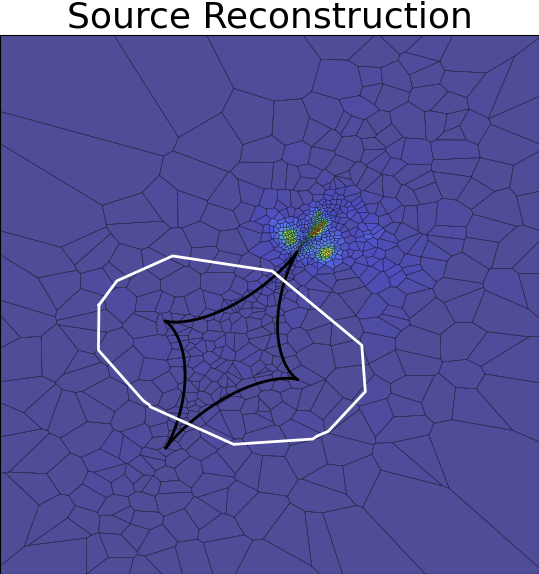}
\includegraphics[width=0.16\textwidth]{cutout/rgb/S08_AAABBU_COSJ095939+023239.pdf}
\includegraphics[width=0.16\textwidth]{cutout/lens_sub/S08_AAABBU_COSJ095939+023239_F444W.pdf}
\includegraphics[width=0.16\textwidth]{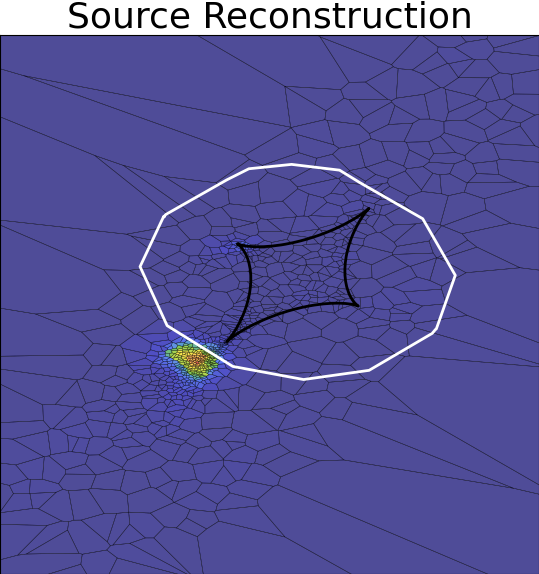}
\includegraphics[width=0.16\textwidth]{cutout/rgb/S07_AAASXX_COSJ095944+022008.pdf}
\includegraphics[width=0.16\textwidth]{cutout/lens_sub/S07_AAASXX_COSJ095944+022008_F277W.pdf}
\includegraphics[width=0.16\textwidth]{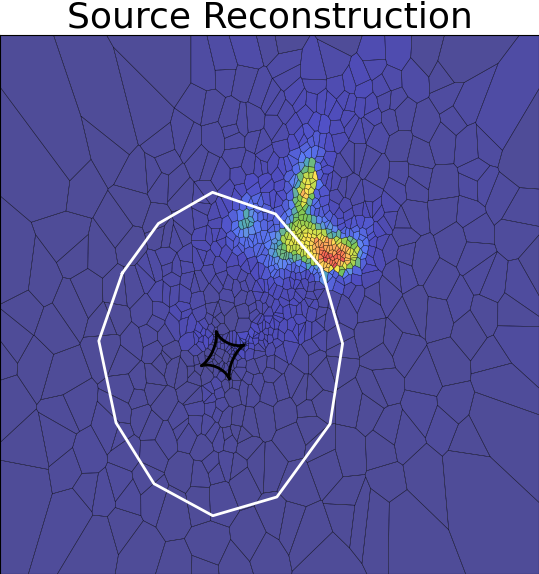}
\includegraphics[width=0.16\textwidth]{cutout/rgb/S05_ABBBUX_COSJ100155+021506.pdf}
\includegraphics[width=0.16\textwidth]{cutout/lens_sub/S05_ABBBUX_COSJ100155+021506_F277W.pdf}
\includegraphics[width=0.16\textwidth]{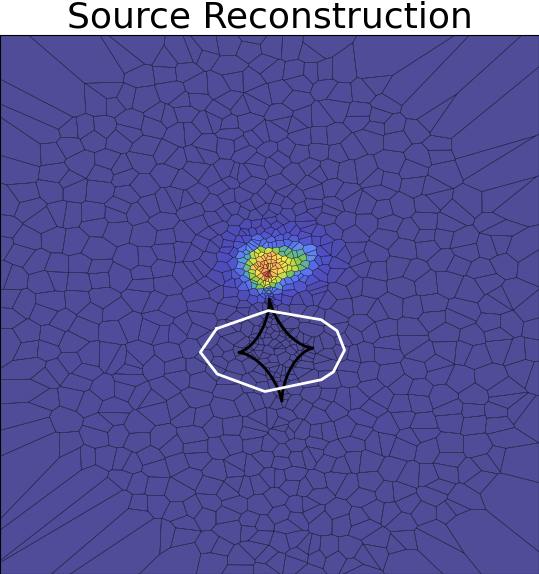}
\caption{
Four example candidates from the second round of visual inspection show evidence for singly imaged, almost-strong lensing. These objects had at least one inspector input `S: A singly imaged strong lens feature/arc (e.g., without an observed counter image)', scored above 5, and after inspection of the source reconstruction, it was confirmed that the majority of the emission lies outside both caustics. Each candidate is displayed three times: the left image shows the RGB cut-out, the next image the foreground lens-subtracted image, and the right image displays the source-plane reconstruction, using the single wavelength image that most clearly shows the candidate lensed source emission. Inspectors had access to more information when grading these lenses, including data from all four wavelengths, which for these systems further supports their singly imaged nature. There are at least 8 singly imaged candidates in total. Readers can access the complete set of visualisations at the following URL:~\github{https://github.com/Jammy2211/COWLS_COSMOS_Web_Lens_Survey}.
}
\label{figure:CutoutS}
\end{figure*}

Figure \ref{figure:CutoutS} shows four example candidates from the second round of visual inspection, which display evidence for singly imaged almost-strong lensing. These objects had at least one inspector input `S: a singly imaged strong lens feature/arc (e.g., without an observed counter image)', and inspection of the source reconstruction by JWN confirmed that the majority of the emission is located outside both caustics. The candidate lensed source arcs exhibit tangential shearing consistent with strong gravitational lensing and colours distinct from the lens galaxy. There are at least 8 candidates showing this behaviour. Singly imaged systems of this nature have been discovered and studied in only a few previous studies \citep{Shu2015, Smith2018}. Some systems may be genuine multiply imaged strong lenses, where the counter image is not observed due to it being too faint, blending with the foreground lens, or obscured by dust absorption.

\subsection{Lens Modelling}\label{VisualLensModel}

We now illustrate how lens modelling helped visual inspection.

\subsubsection{Extra Information}

\begin{figure*}
\centering
\includegraphics[width=0.128\textwidth]{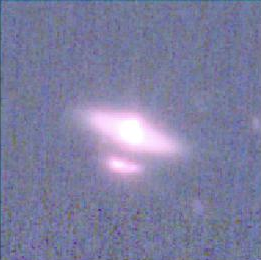}
\includegraphics[width=0.138\textwidth]{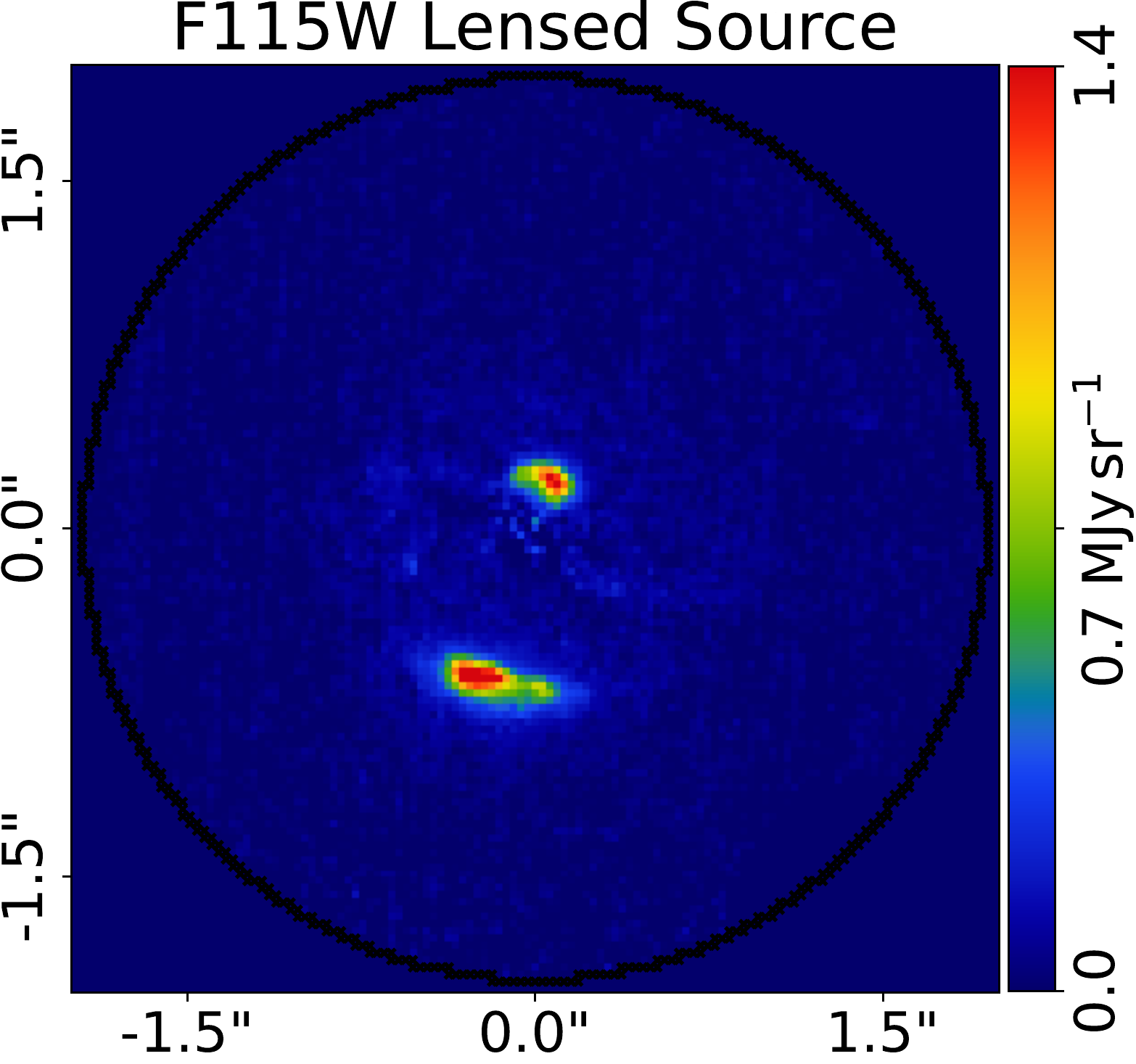}
\includegraphics[width=0.138\textwidth]{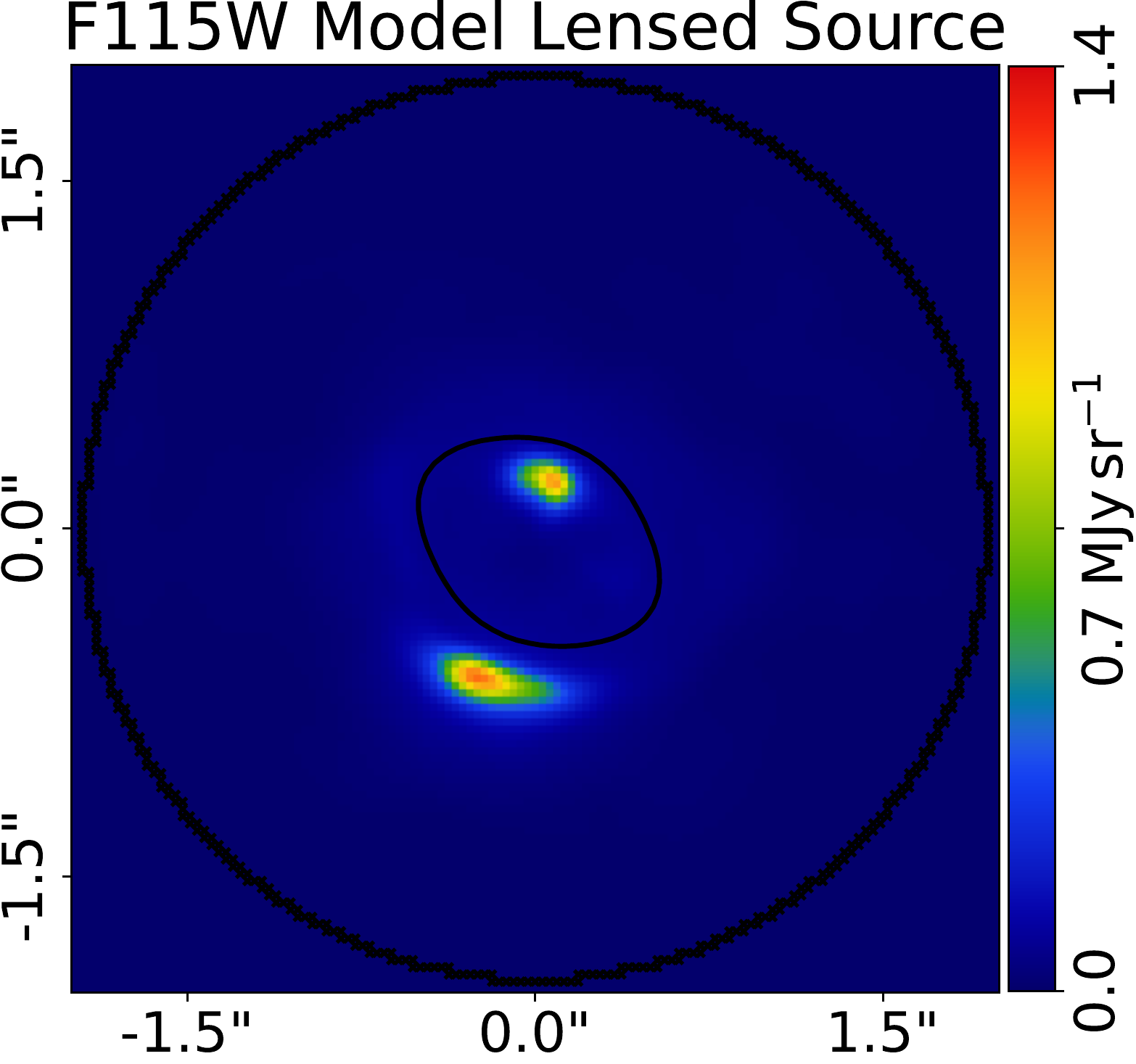}
\includegraphics[width=0.138\textwidth]{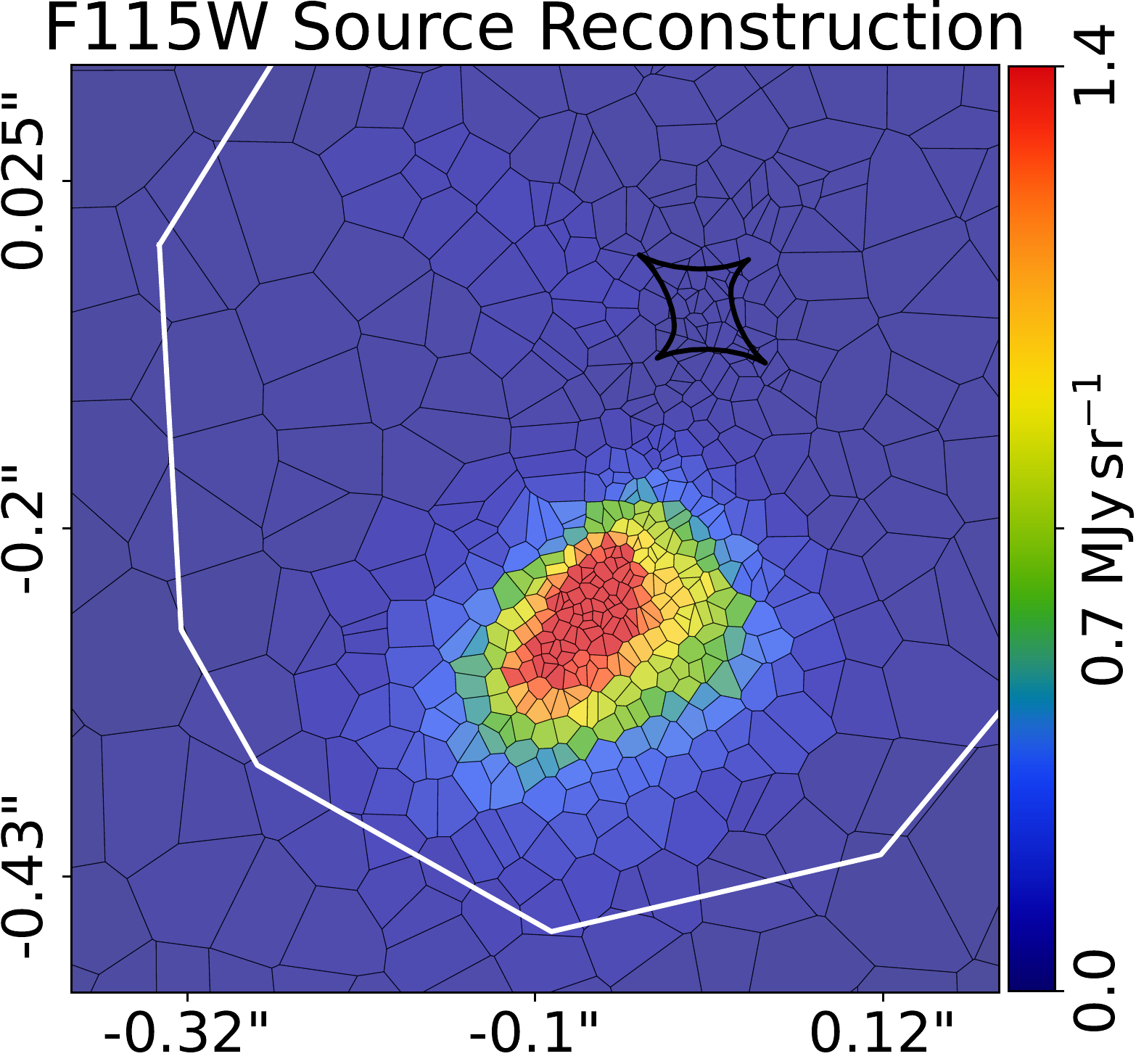}
\includegraphics[width=0.138\textwidth]{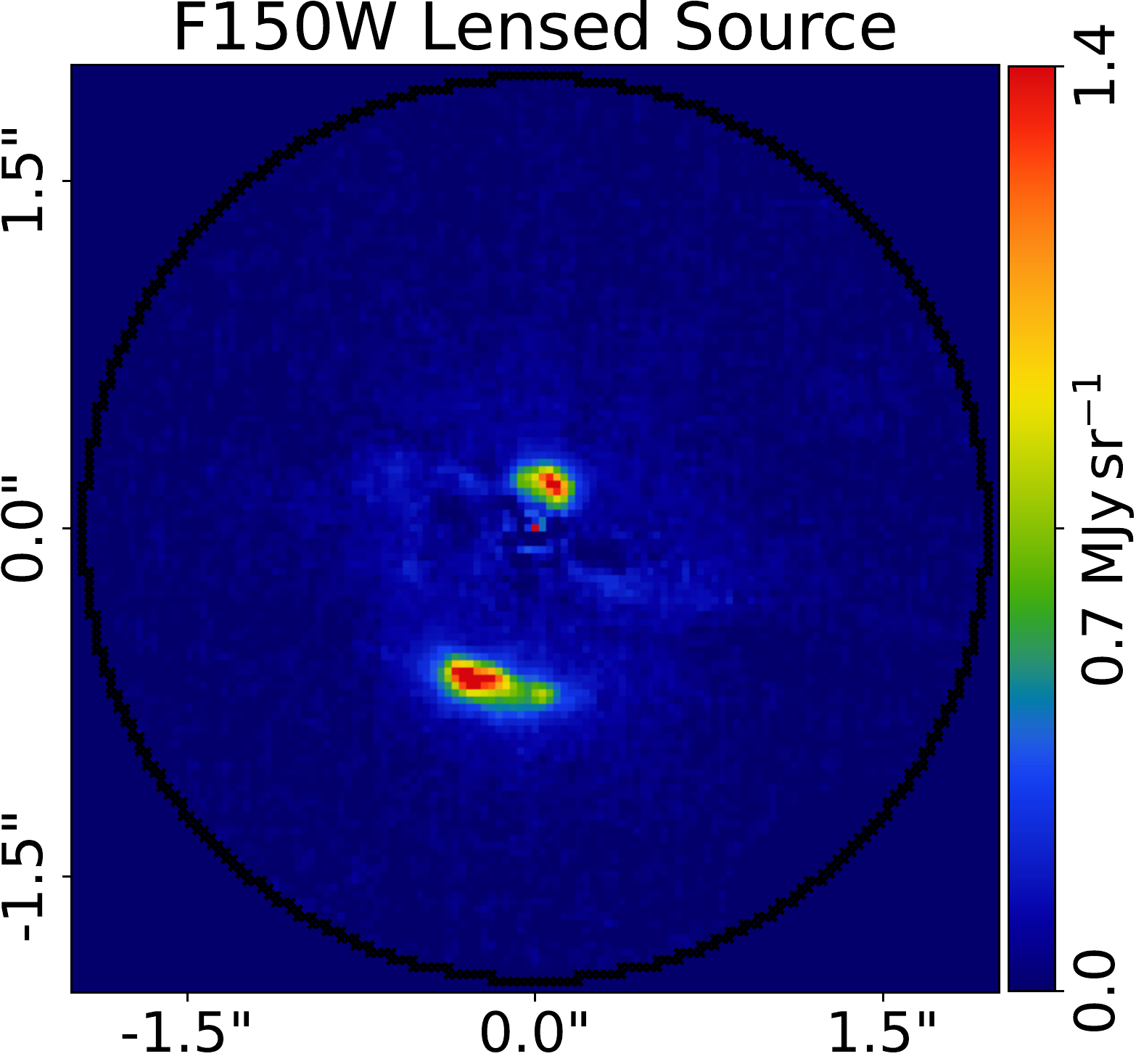}
\includegraphics[width=0.138\textwidth]{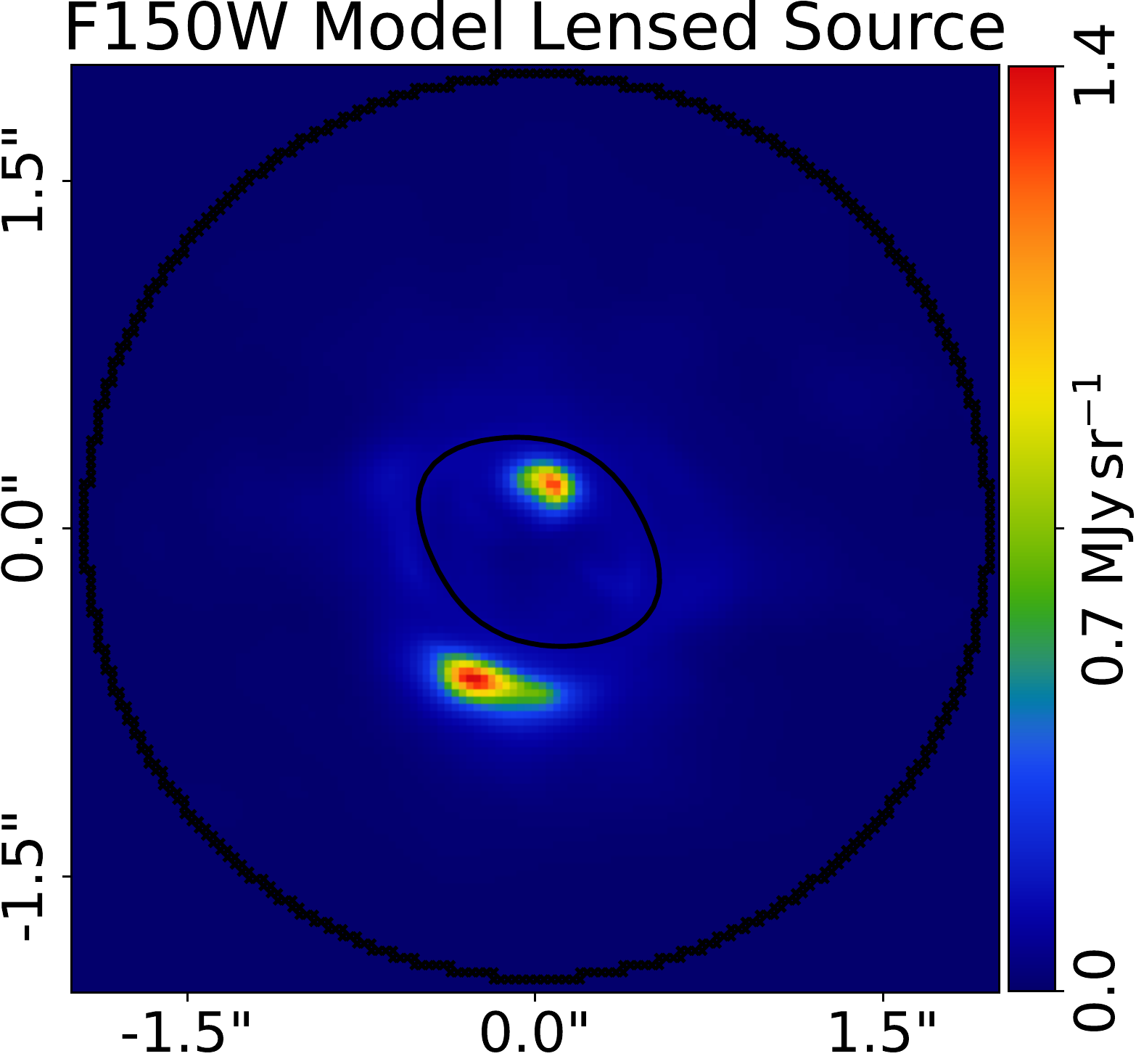}
\includegraphics[width=0.138\textwidth]{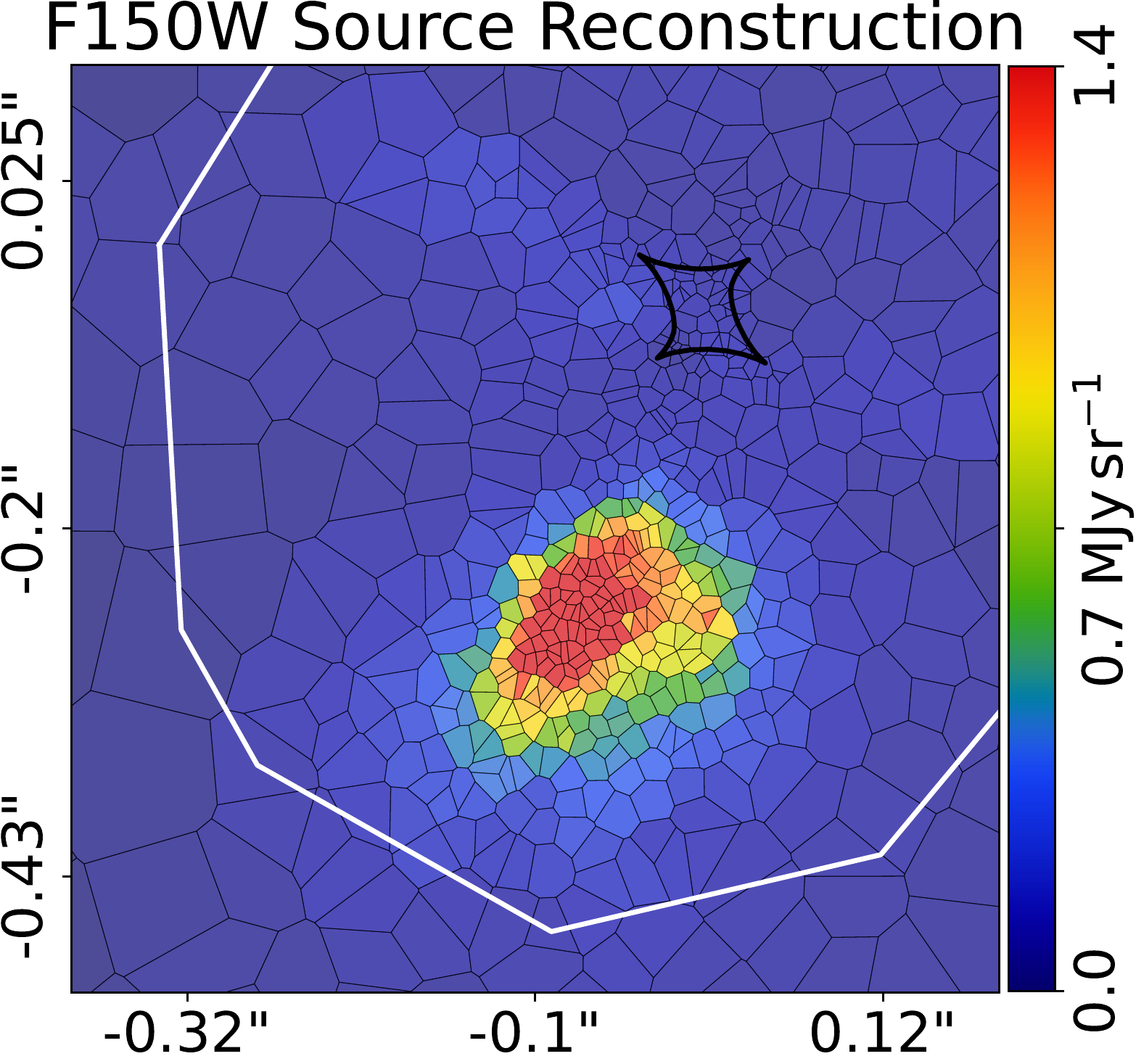}
\includegraphics[width=0.128\textwidth]{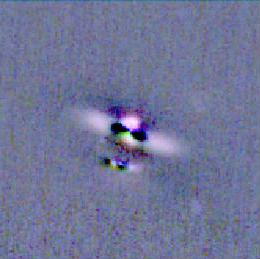}
\includegraphics[width=0.138\textwidth]{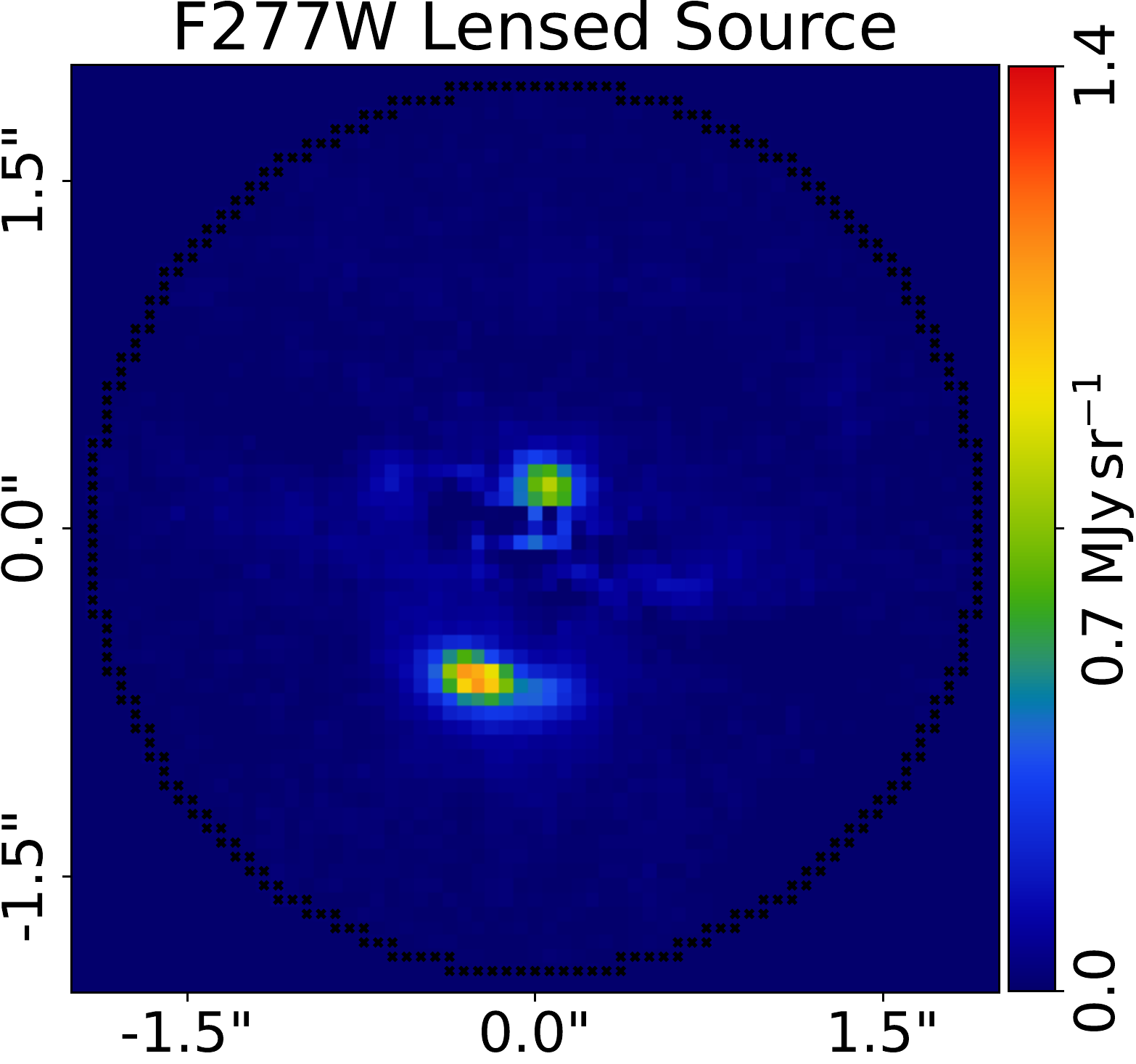}
\includegraphics[width=0.138\textwidth]{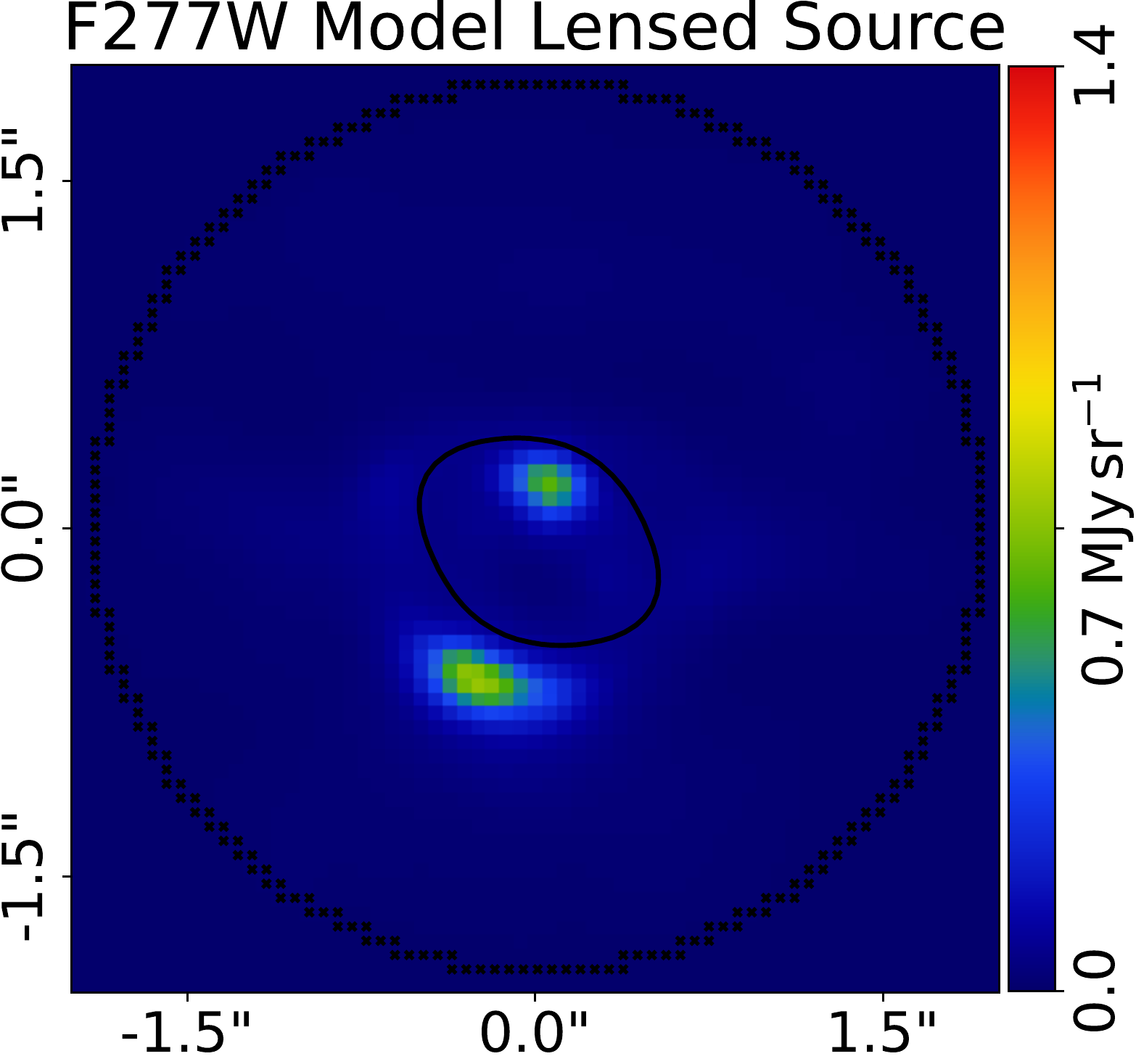}
\includegraphics[width=0.138\textwidth]{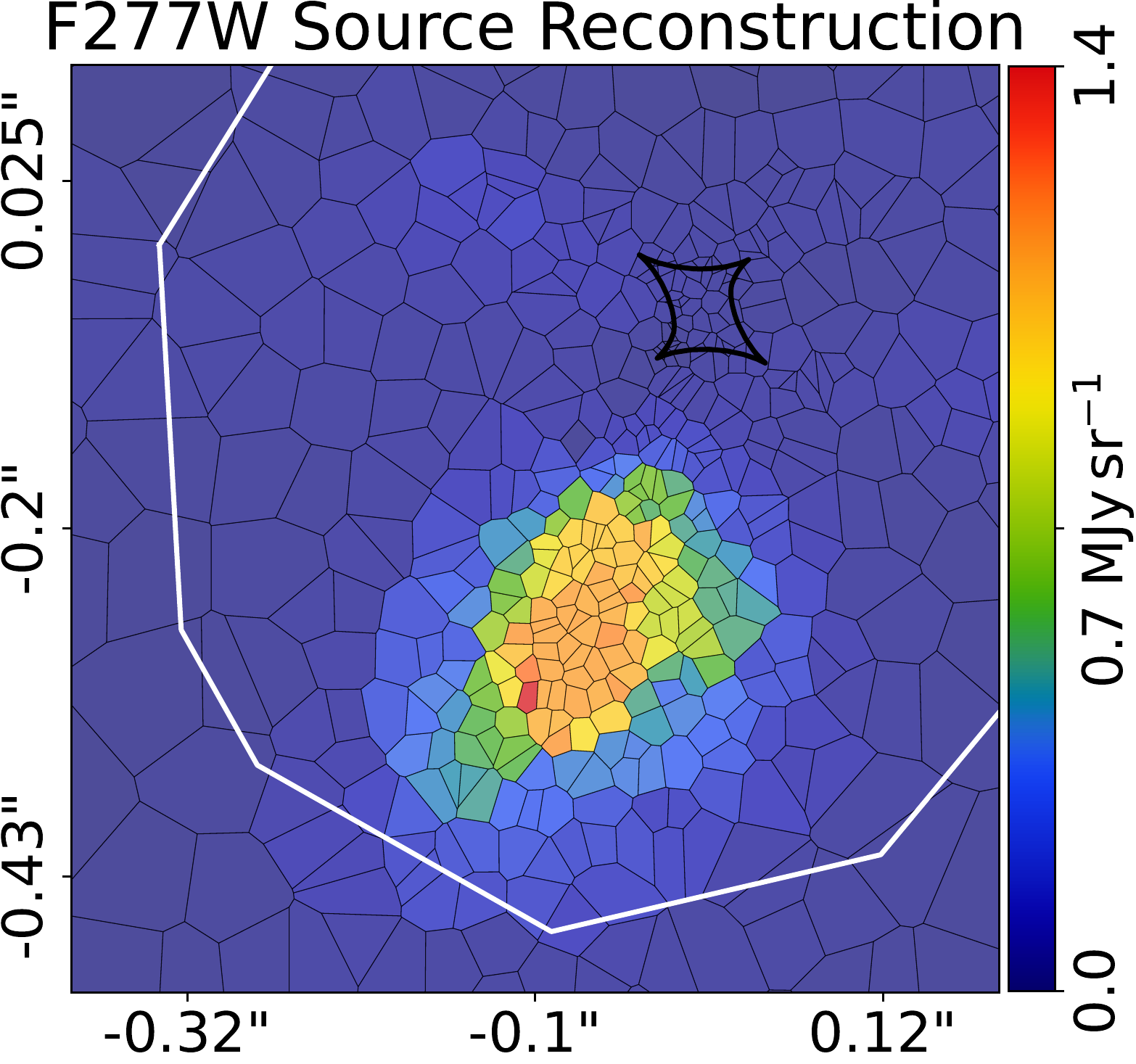}
\includegraphics[width=0.138\textwidth]{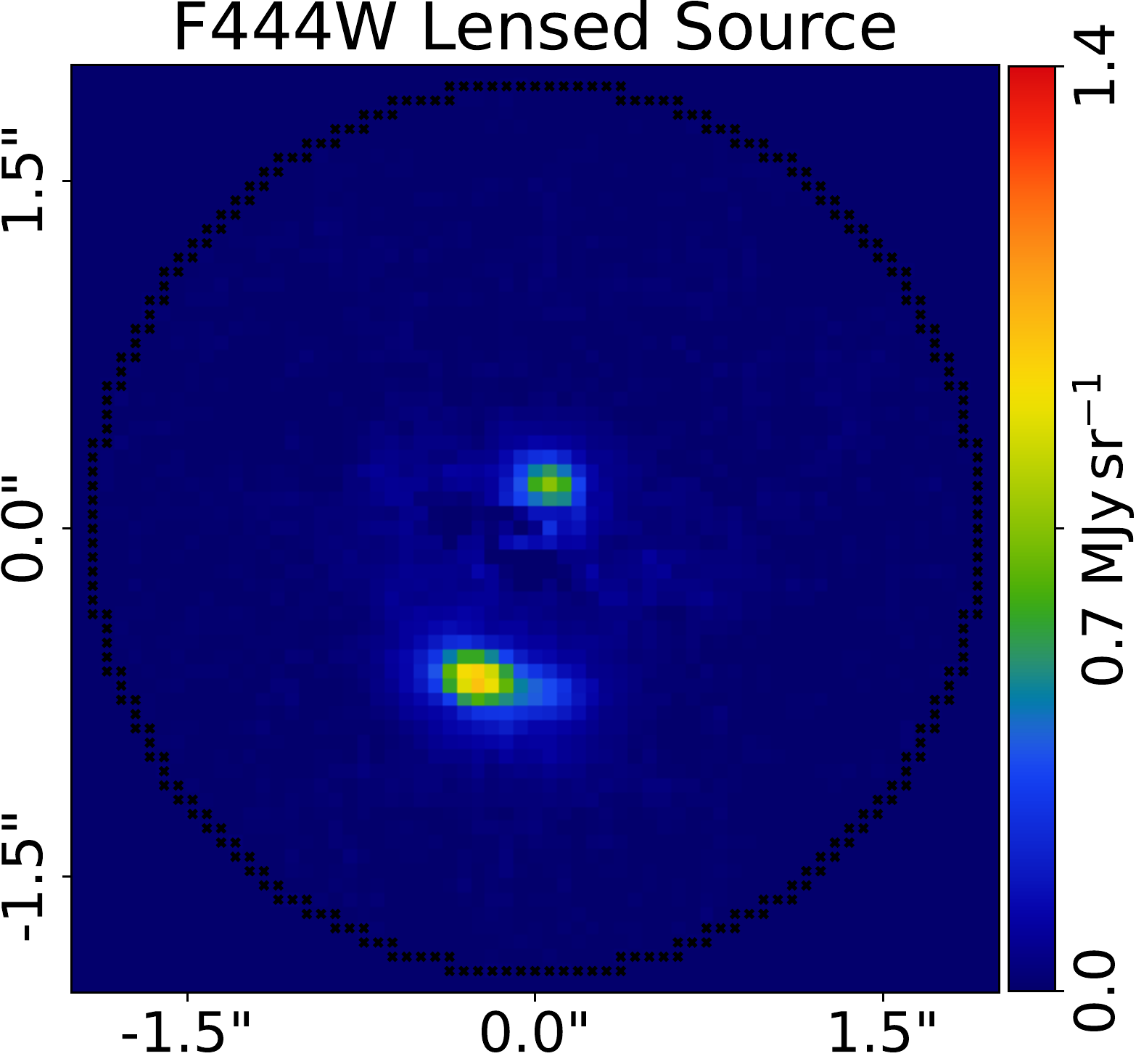}
\includegraphics[width=0.138\textwidth]{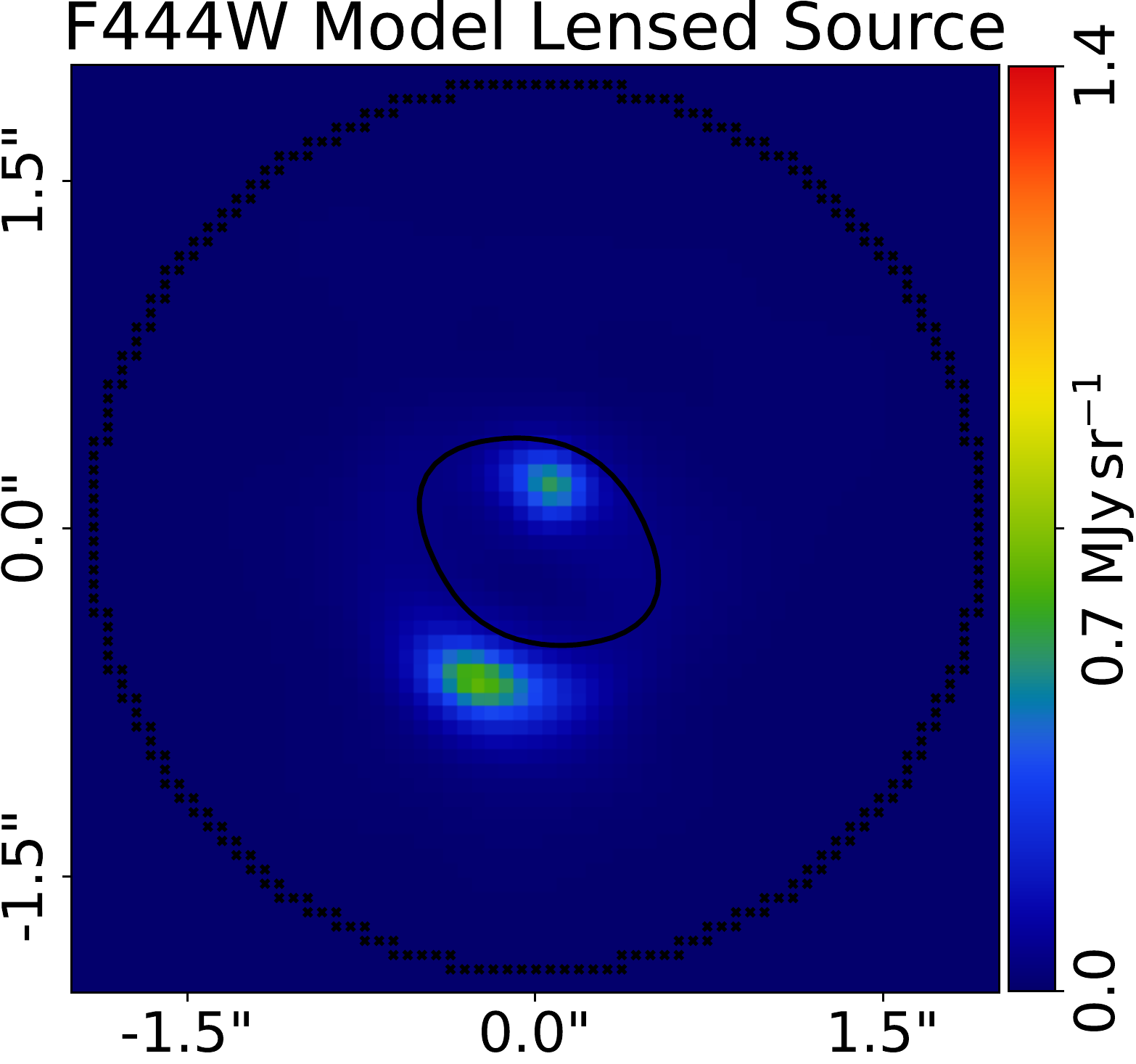}
\includegraphics[width=0.138\textwidth]{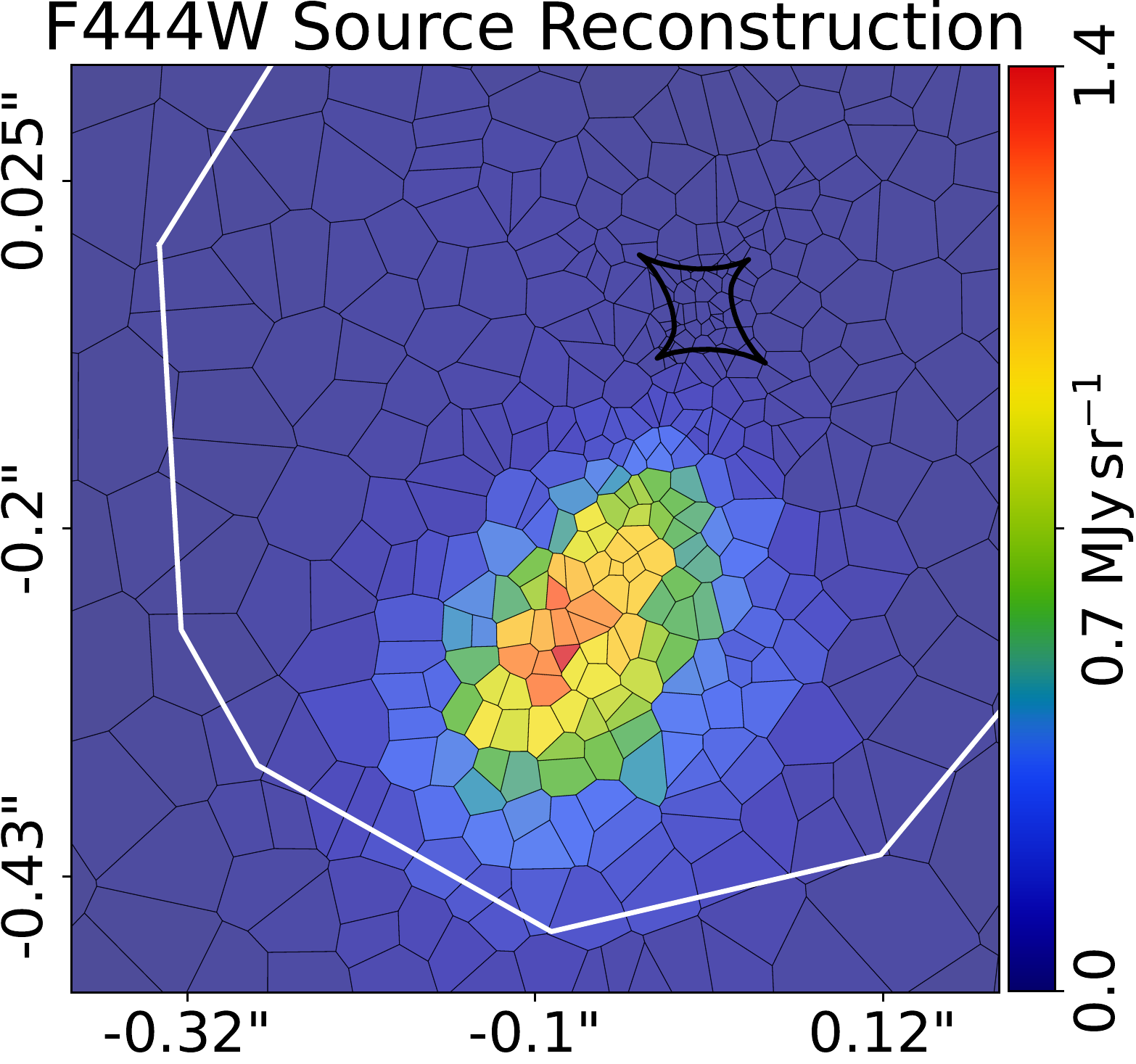}
\caption{
How lens modelling enhances inspector rankings during the second round of visual inspection. The first column shows the postage stamp cut-out images used in the initial round of inspection, including one with foreground emission and another with a single Sérsic fit foreground subtraction. The following six columns present additional information provided to inspectors during the second round, which for each of the four wavebands fitted (F115W, F150W, F277W, F444W) consists of: (i) a cleaner foreground-subtracted image using a Multi-Gaussian expansion (MGE) rather than a single Sérsic fit; (ii) the model lensed source in the image plane, highlighting features that the lens model can accurately reproduce; and (iii) source-plane reconstructions, demonstrating whether the candidate has a plausible source configuration. The black and white curves represent tangential and radial critical curves and caustics, providing further insights to assess the likelihood of the candidate being a strong lens. The example lens shown was an edge case in the first round of visual inspection, meaning only one inspector thought it might be a lens, but in the second round of visual inspection scored the maximum score of all A's. It is therefore the clearest example of how lens modelling provided information which changed the inspectors' opinion on if a candidate is a lens.
}
\label{figure:LensModelInfo}
\end{figure*}

For many images inspected in the first round of visual inspection, the emission from the candidate source galaxy is visible but is close to and blended with the much brighter emission from the candidate lens galaxy. This made it challenging and time-consuming to get a clear view of the source emission, even after single S\'ersic subtractions and careful adjustments to visualisation RGB colour maps and scalings. \cref{figure:LensModelInfo} illustrates this for the highest ranked candidate in the second round of visual inspection, which was an edge case after the first round. The cut-outs shown in the left column were used in the first round of visual inspection. These images do not clearly show the candidate's lensed source emission and have artefacts from the single S\'ersic subtraction, further complicating detailed inspection for lensing features, explaining why this high-confidence candidate was ranked poorly in the first round of visual inspection.

The remaining columns in \cref{figure:LensModelInfo} show some of the additional images provided by lens modelling in the second round of visual inspection. In each waveband, lens modelling effectively deblends the lens and source, producing clean images of the lensed source galaxy, which makes it easier to judge whether the system is a lens. The lensed source model image, source-plane reconstruction, and critical curves and caustics, all shown in \cref{figure:LensModelInfo}, further assist the inspector by confirming that the system fits a lens model. While we cannot quantify how many high-ranked candidates relied on this additional information, the large number of edge cases that were ranked highly in the second round of visual inspection (38.4\% of candidates scoring 5 and above) suggests that it is crucial for many candidates.

\subsubsection{Revealing Counter Images}

\begin{figure*}
\centering
\includegraphics[width=0.128\textwidth]{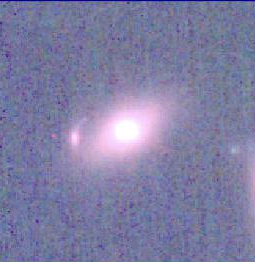}
\includegraphics[width=0.138\textwidth]{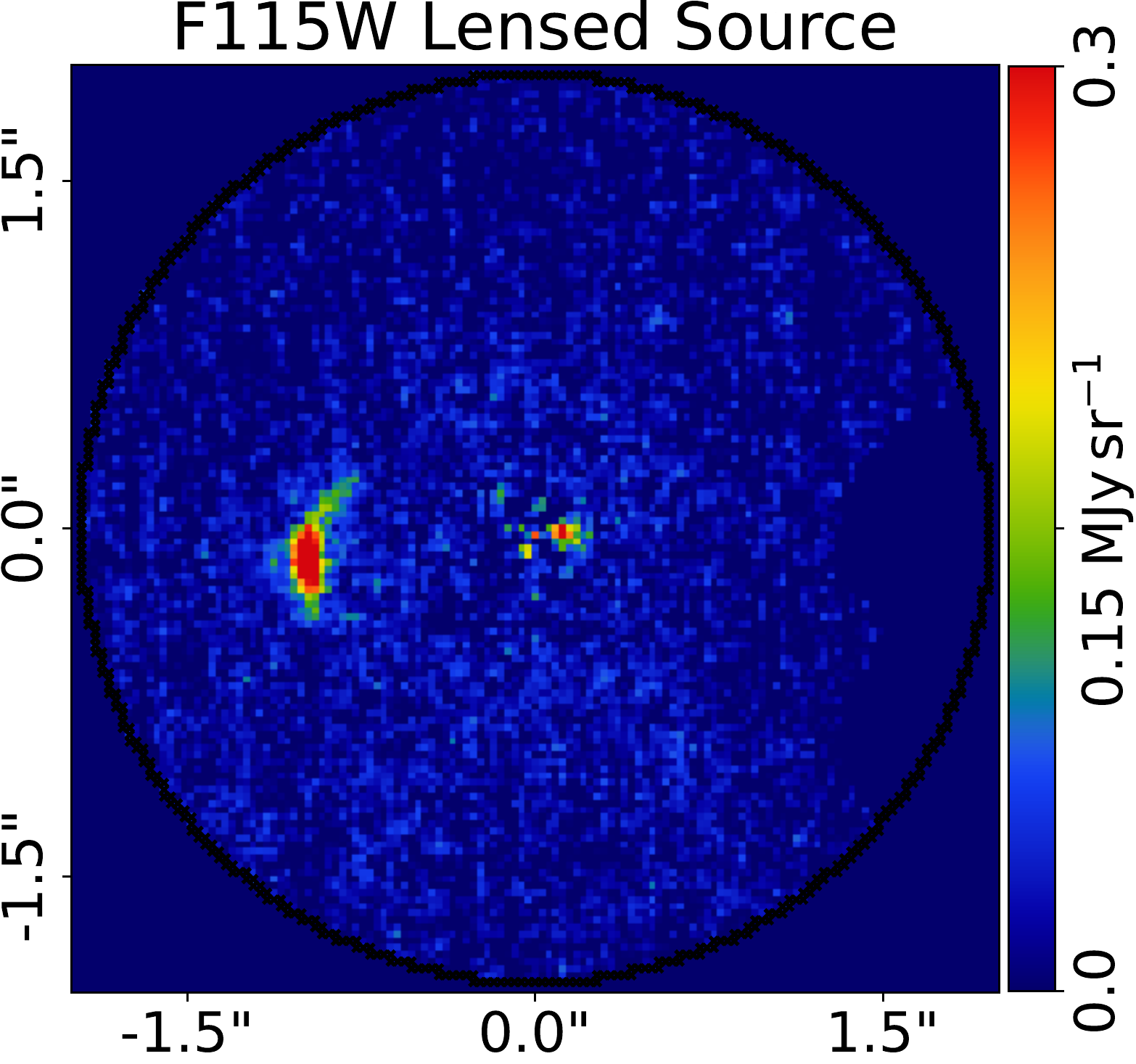}
\includegraphics[width=0.138\textwidth]{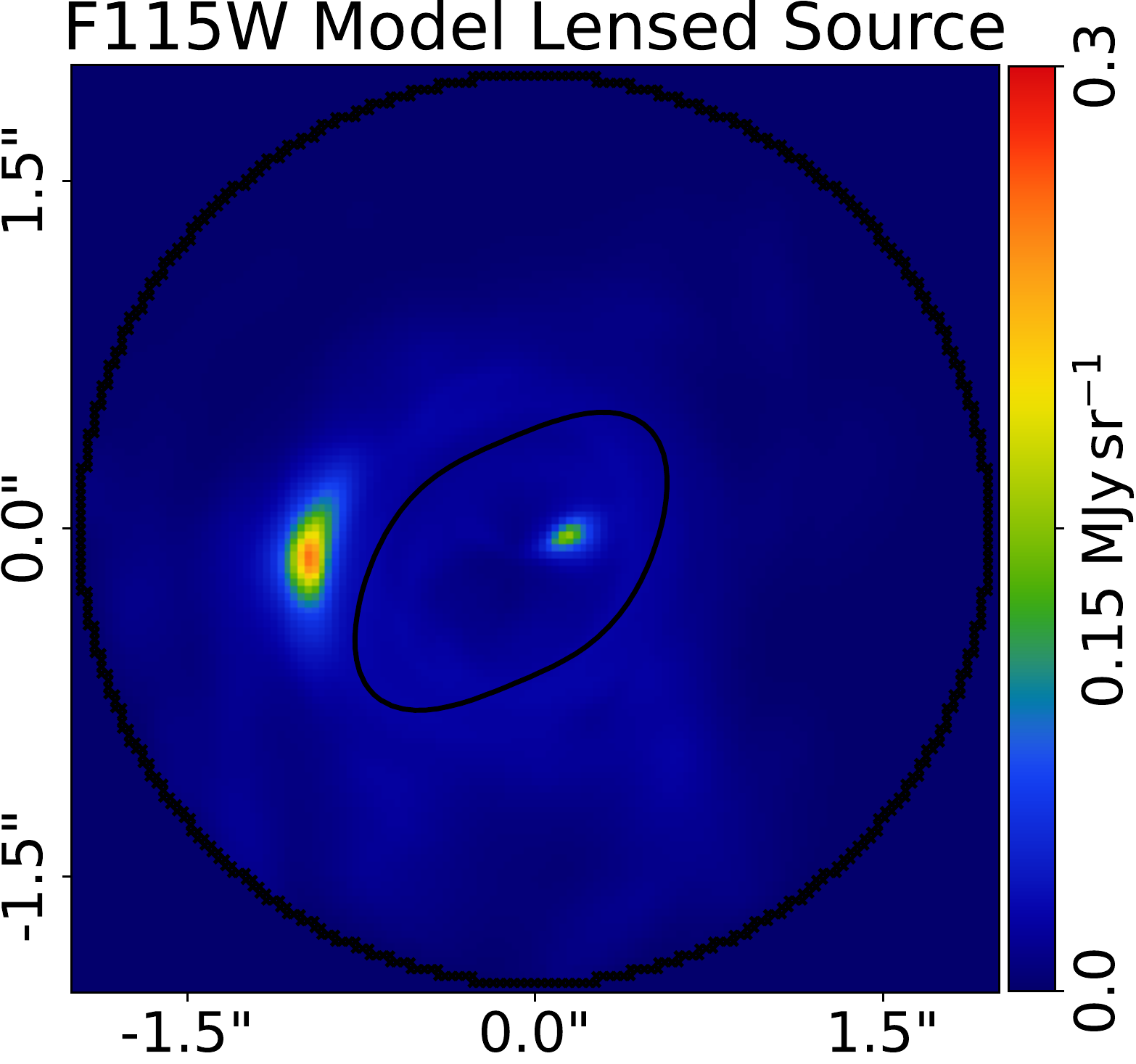}
\includegraphics[width=0.138\textwidth]{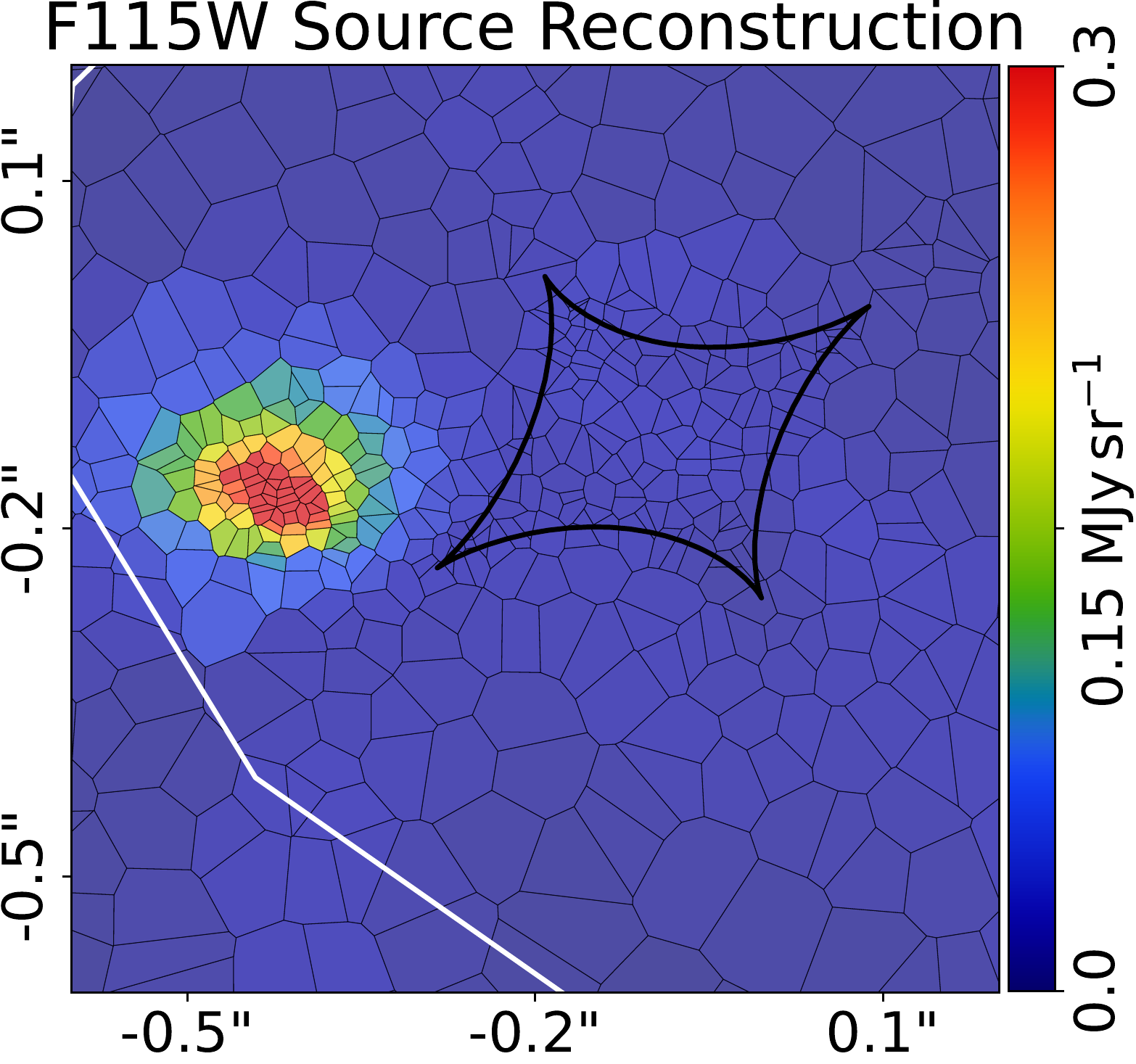}
\includegraphics[width=0.138\textwidth]{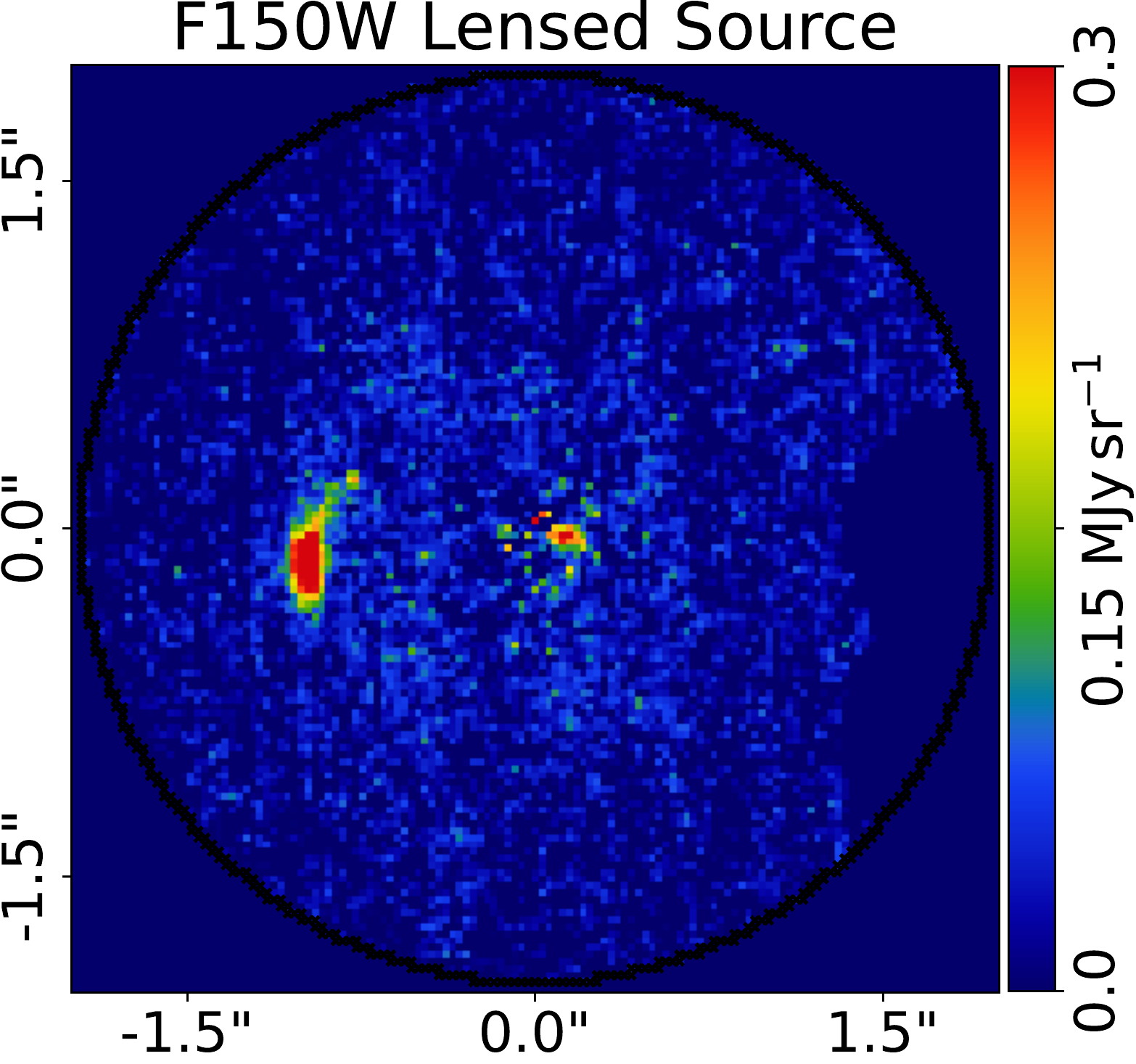}
\includegraphics[width=0.138\textwidth]{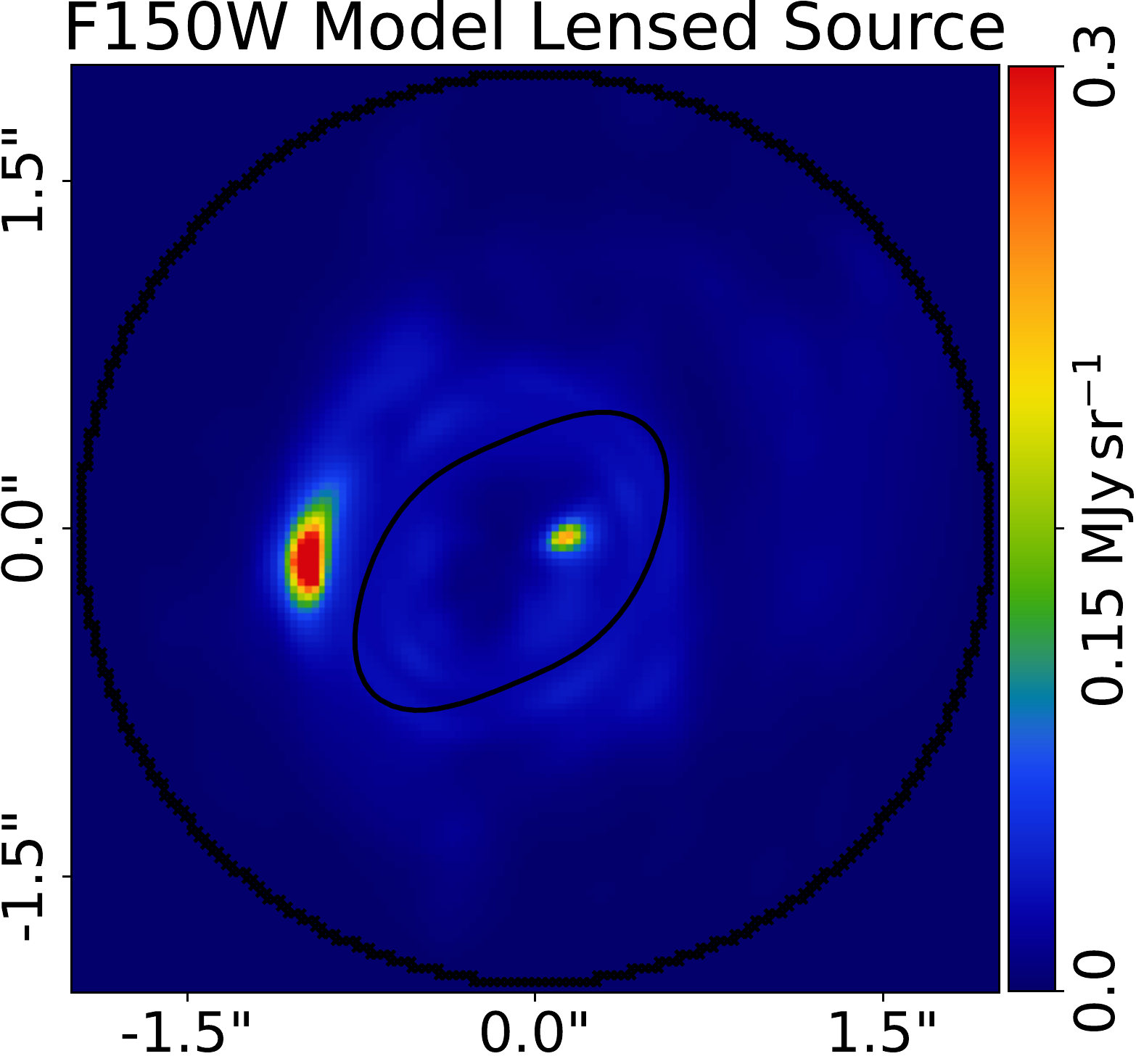}
\includegraphics[width=0.138\textwidth]{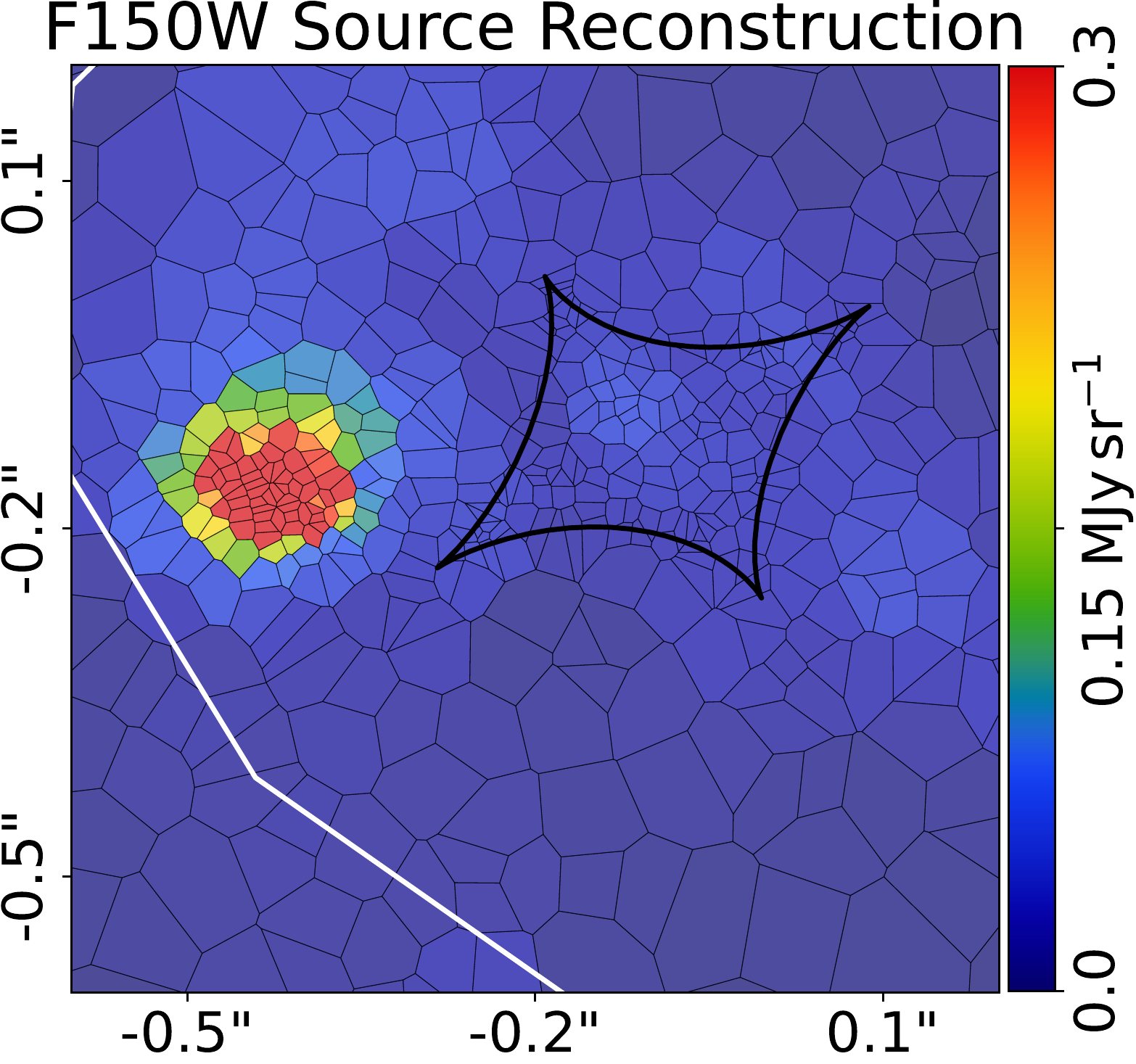}
\includegraphics[width=0.128\textwidth]{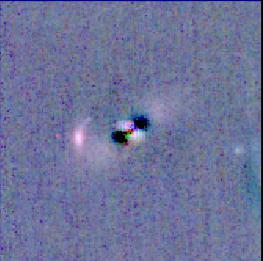}
\includegraphics[width=0.138\textwidth]{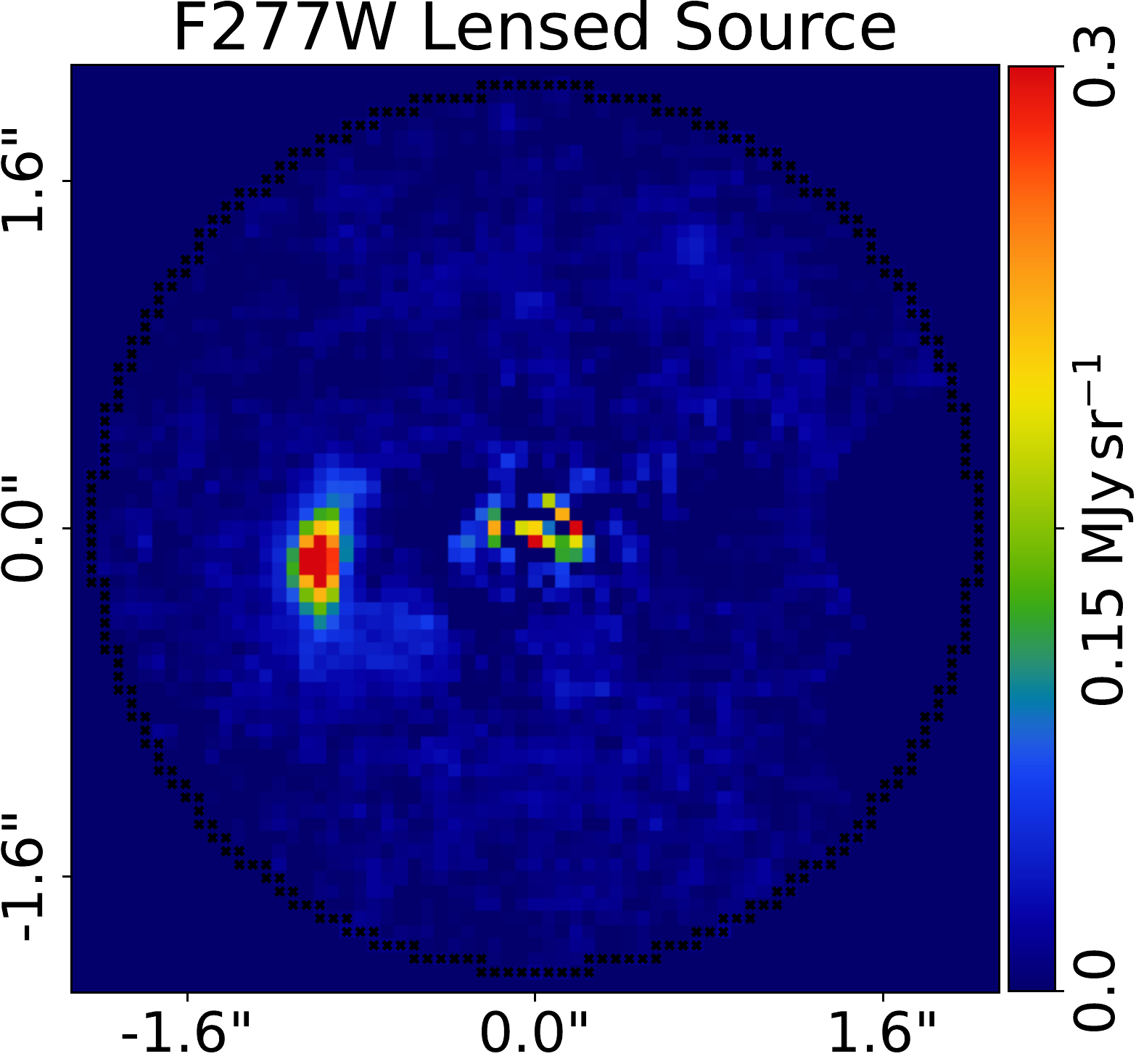}
\includegraphics[width=0.138\textwidth]{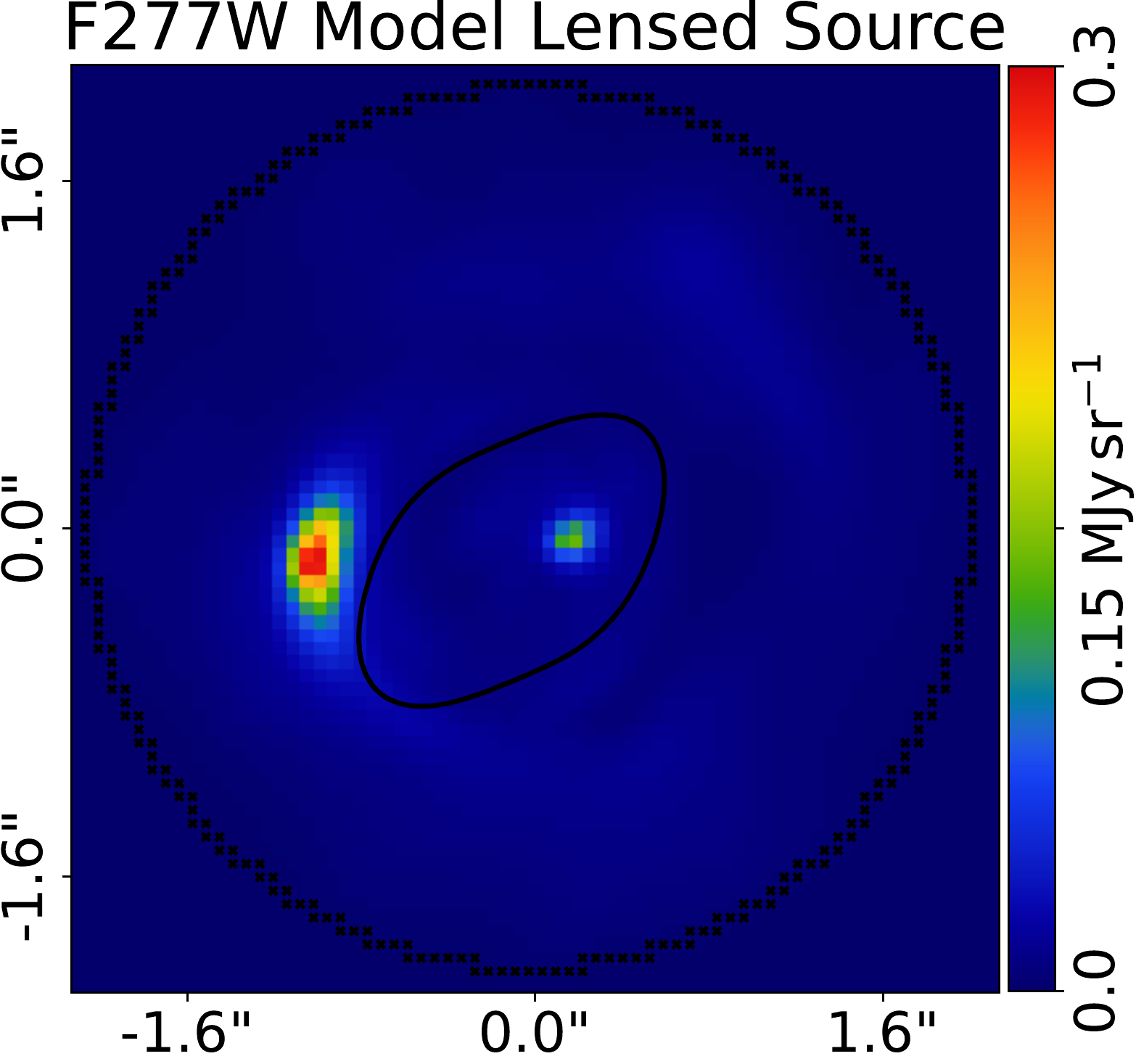}
\includegraphics[width=0.138\textwidth]{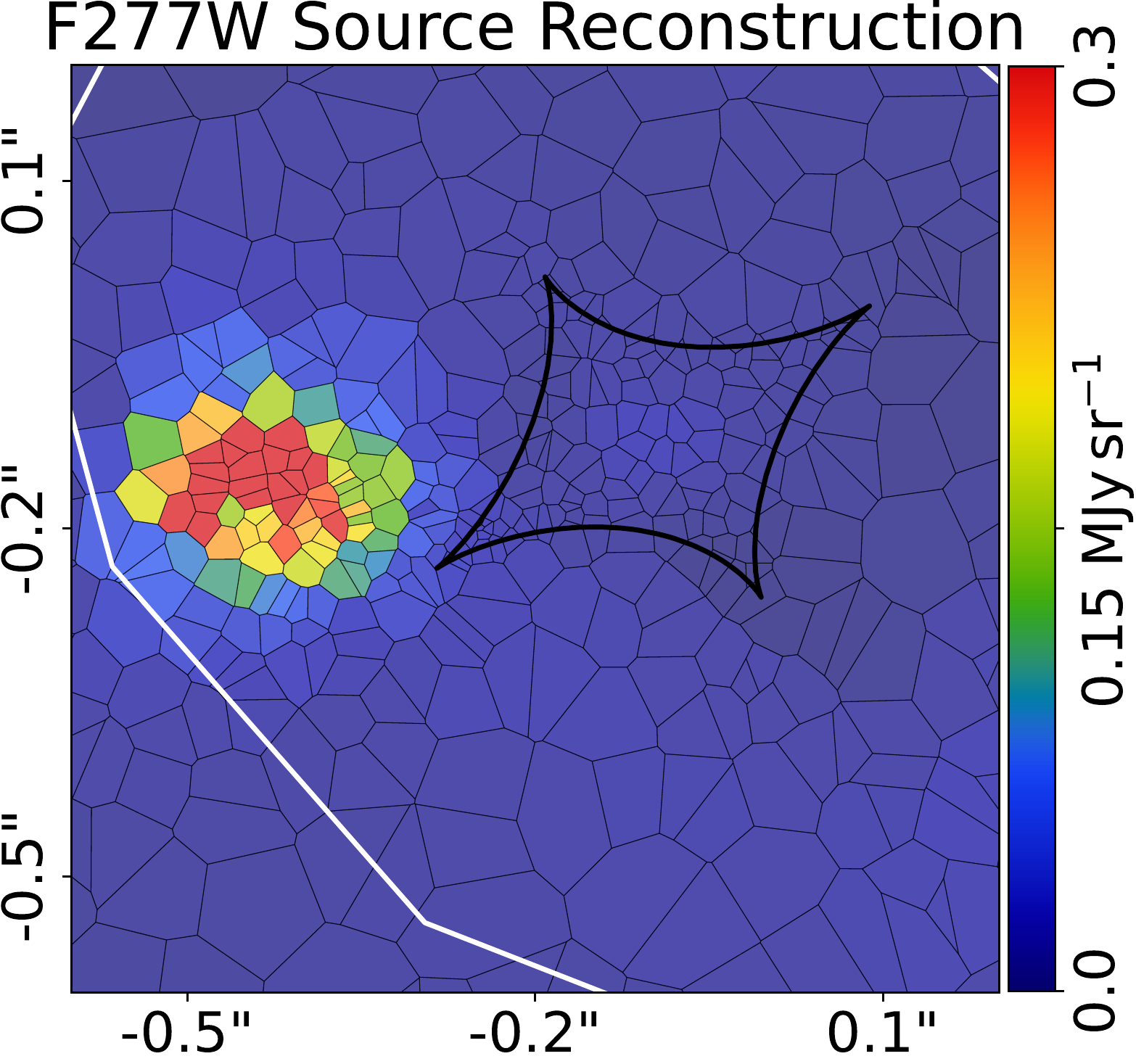}
\includegraphics[width=0.138\textwidth]{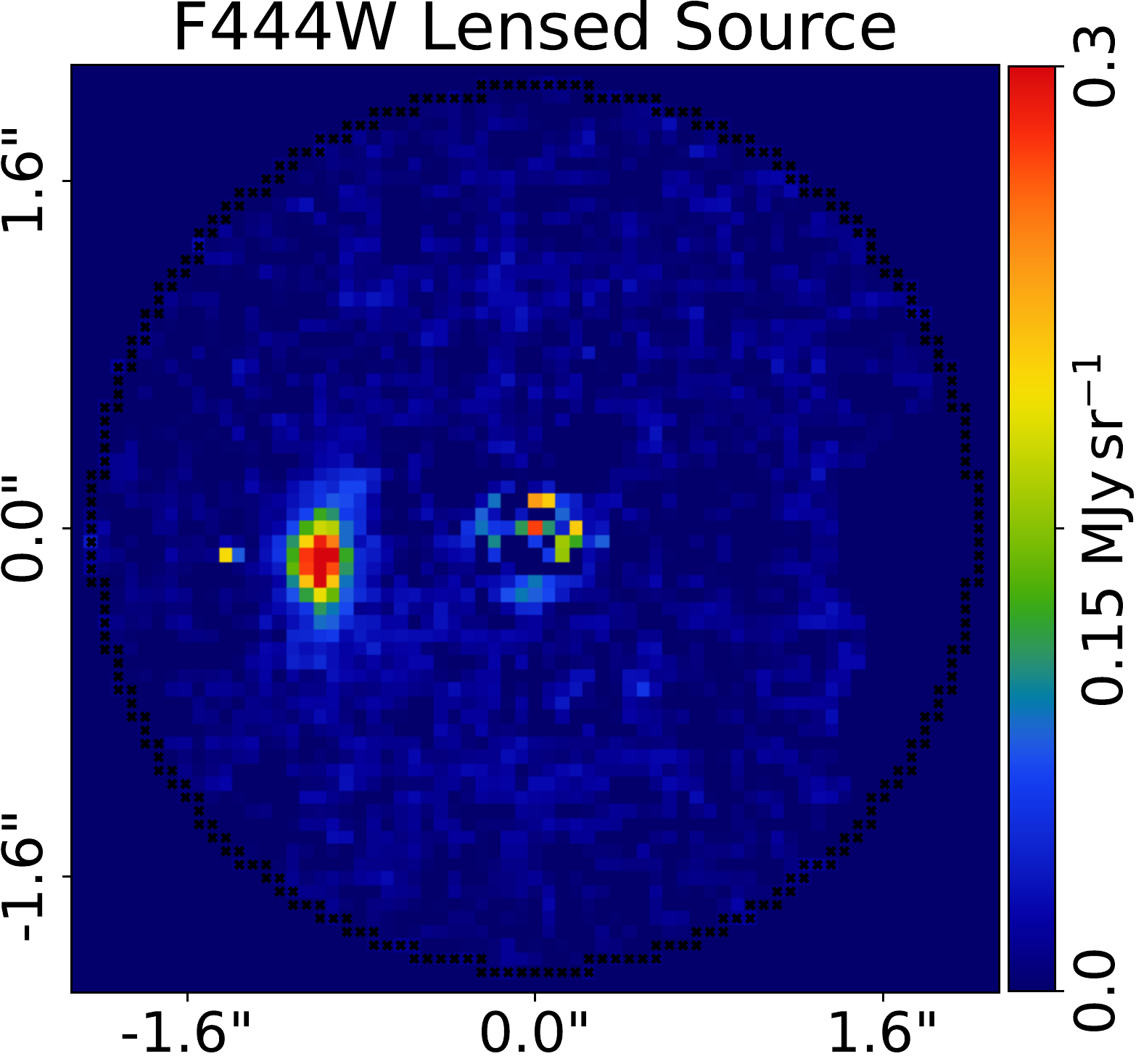}
\includegraphics[width=0.138\textwidth]{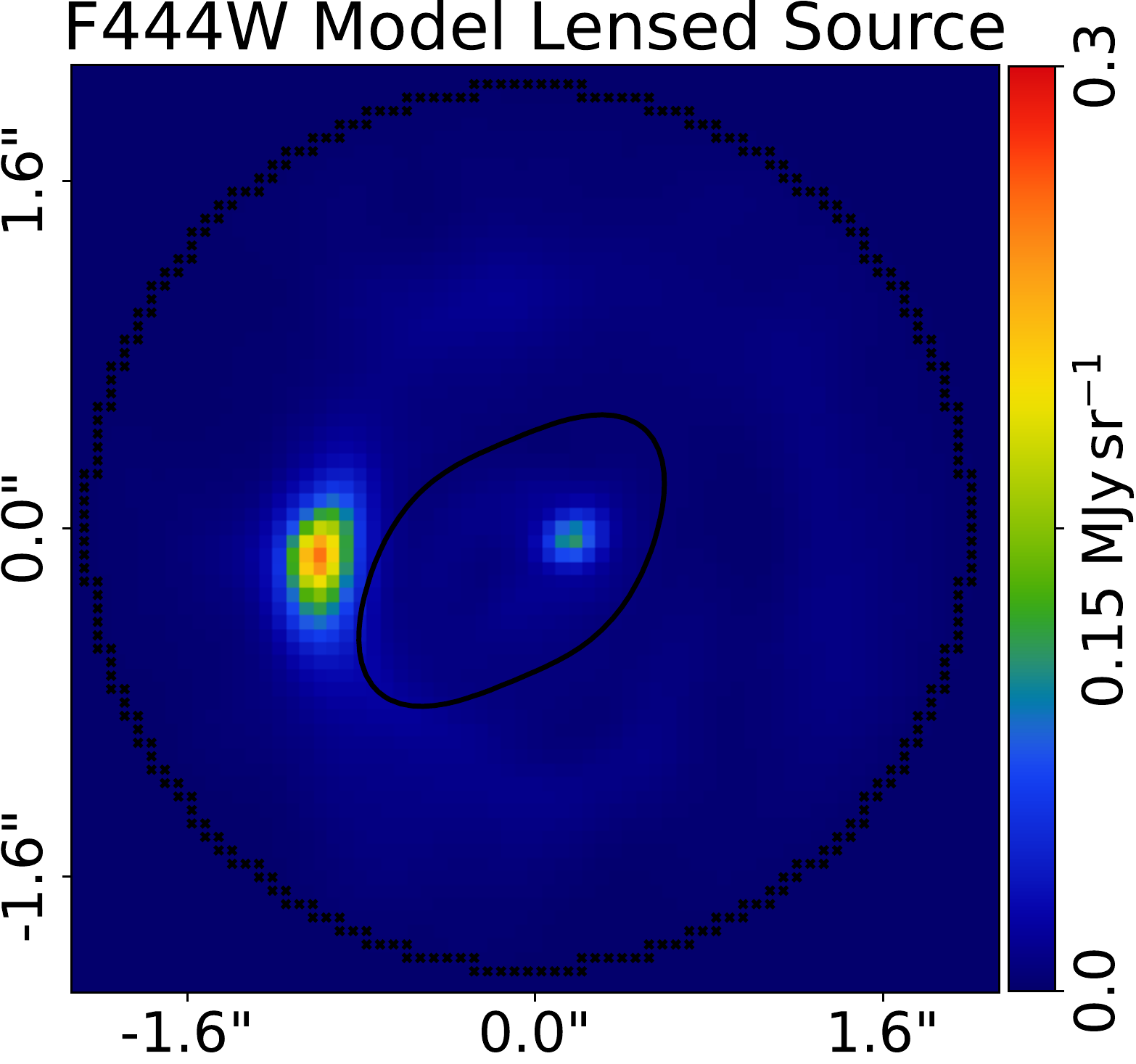}
\includegraphics[width=0.138\textwidth]{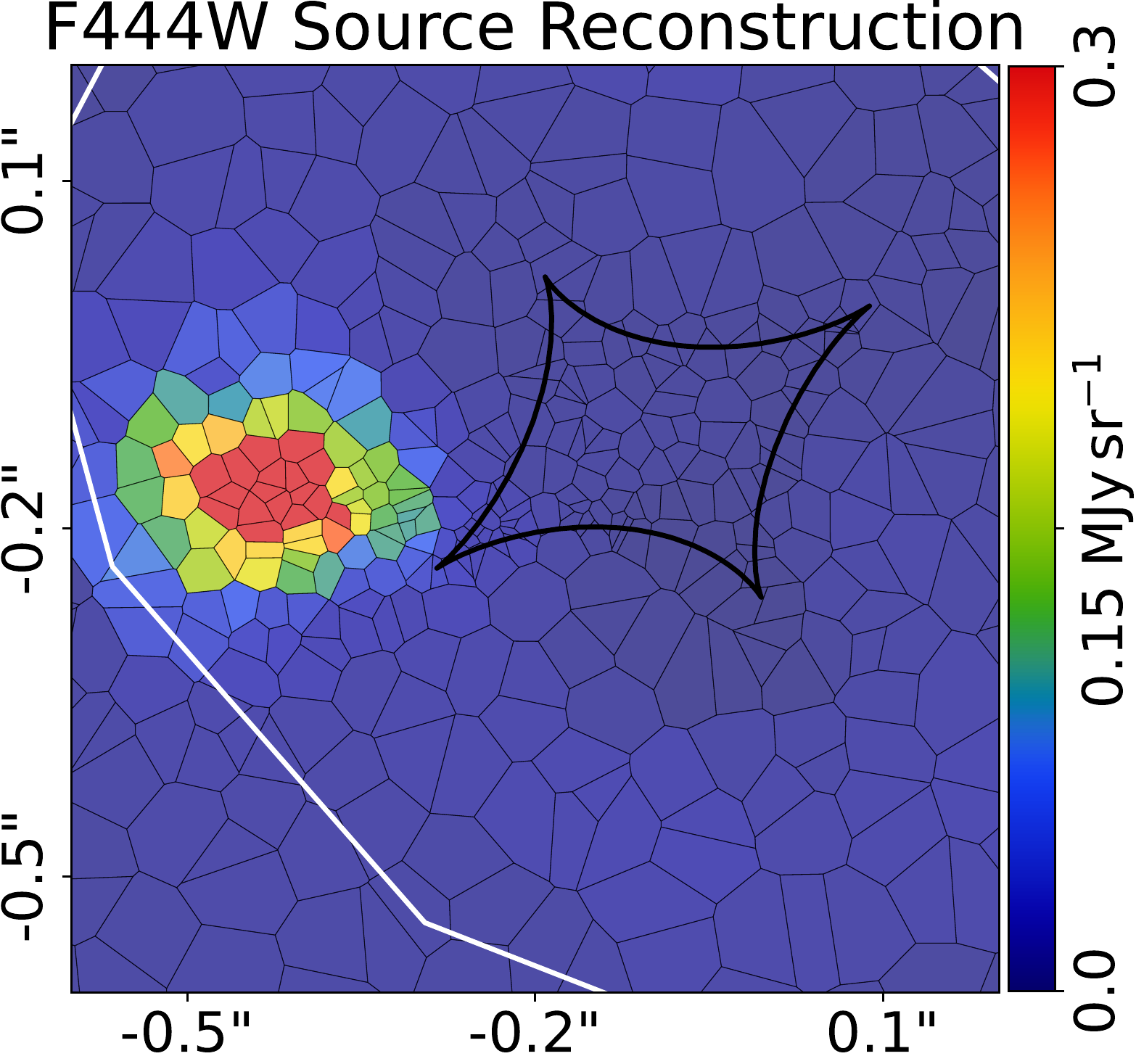}
\caption{
Illustration of how lens modelling reveals counter images of candidates in the second round of visual inspection. The first column shows the postage stamp cut-out images used in the first round of inspection, including cut-outs with foreground emission and where a single Sérsic fit subtracts it. An arc to the left, indicative of a candidate lensed source, is present, but the corresponding counter image cannot be seen in either image, including after the single Sérsic subtraction. The remaining six columns present for each of the four wavebands fitted (F115W, F150W, F277W, F444W): (i) a cleaner foreground-subtracted image using a Multi-Gaussian expansion; (ii) the model lensed source in the image plane; and (iii) source-plane reconstructions. The black and white curves represent tangential and radial critical curves and caustics. The lens model reveals faint emission near the candidate lens's centre indicative of the source's counter image, with this feature visible in all four wavebands. 
}
\label{figure:LensModelCounter}
\end{figure*}

The left column of \cref{figure:LensModelCounter} shows cut-outs from the first round of visual inspection of a lens where there is emission near the candidate lens galaxy that could be a lensed source. However, its counter image is too faint and outshone by the bright lens emission, making them impossible to visually separate, even after the S\'ersic subtraction. The remaining columns of \cref{figure:LensModelCounter} display the results of lens modelling, which effectively deblends the candidate lens and source emission, revealing additional emission indicative of the counter image of the lensed source. The emission appears in multiple wavebands, confirming that it is not an artefact of the modelling process. For this system, carefully tuned RGB multi-colour images did not make the counter image visible. The ability of lens modelling to reveal counter images in this case required detailed deblending, made possible only by lens modelling and its comprehensive treatment of each waveband’s point spread function.

Lens modelling can therefore reveal a lensed source's faint counter image, turning candidates that initially show no signs of multiple imaging—therefore likely to be ranked low—into candidates showing multiple images and thus becoming highly ranked. In total, approximately 30 candidates scoring 5 and above had counter images that were not visible in RGB images and only became visible after lens modelling, highlighting this as a crucial way lens modelling enables lens discovery. A large number of lenses with low $R_{ls}$ values discussed in \cref{SourceLensSeparation} were only identified because of this capability. Many previous lens searches do not use lens modelling to deblend candidate lens and source emission, which is why in \cref{Discussion} we argue that our approach overcomes a selection effect that leads other surveys to miss these types of systems. 

\subsubsection{Compact Multiple Image Pairing}

\begin{figure*}
\centering
\includegraphics[width=0.138\textwidth]{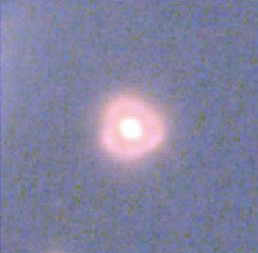}
\includegraphics[width=0.138\textwidth]{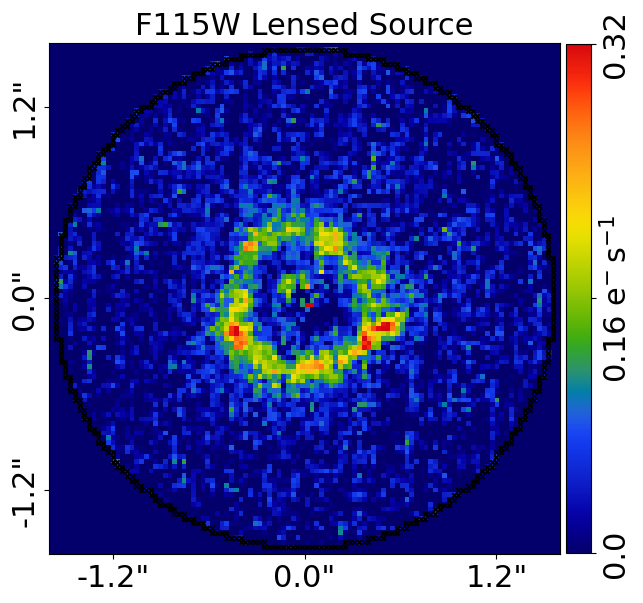}
\includegraphics[width=0.138\textwidth]{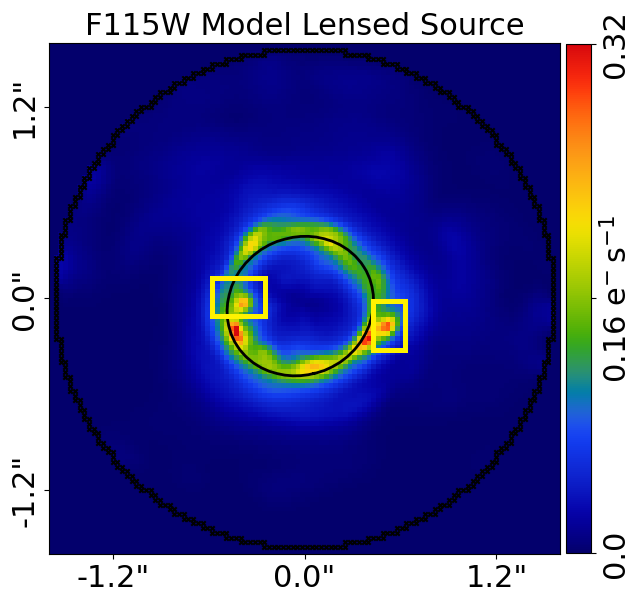}
\includegraphics[width=0.138\textwidth]{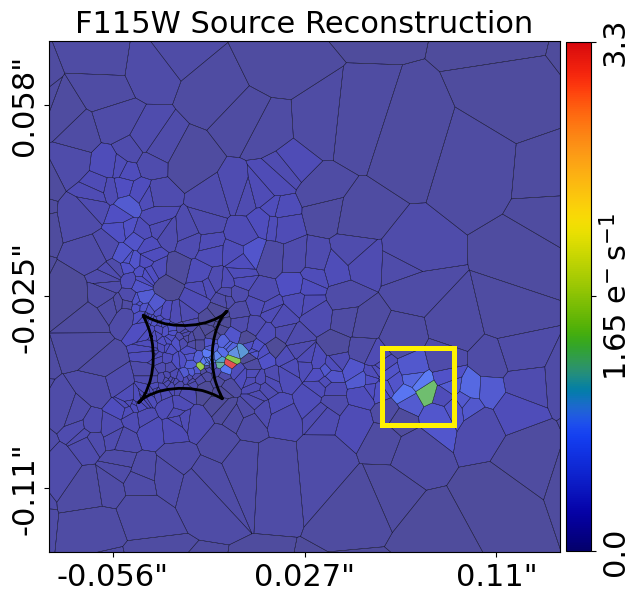}
\includegraphics[width=0.138\textwidth]{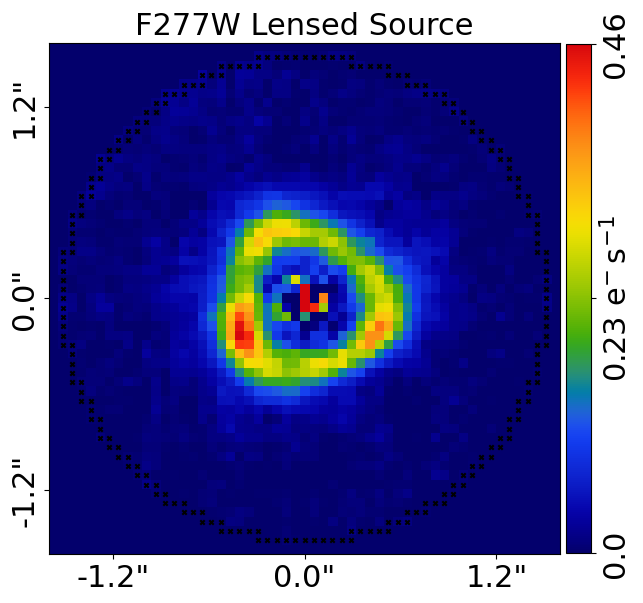}
\includegraphics[width=0.138\textwidth]{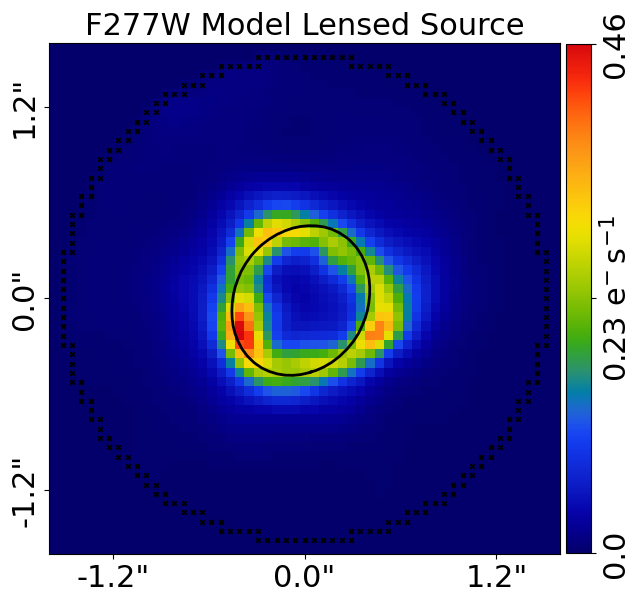}
\includegraphics[width=0.138\textwidth]{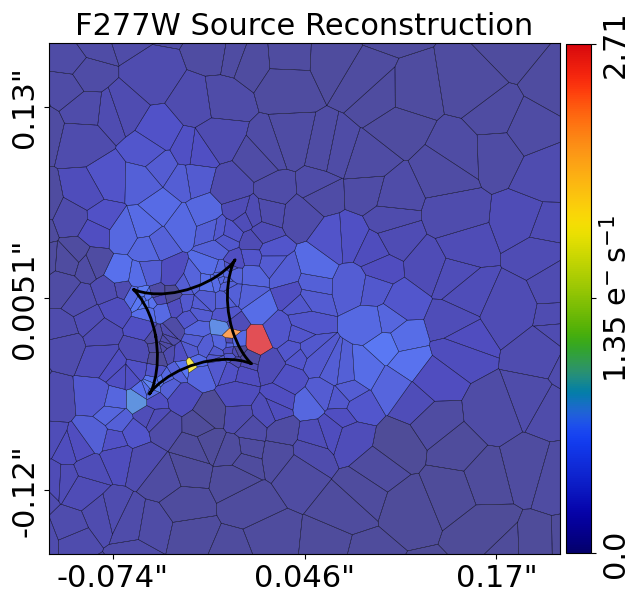}
\caption{
Illustration of how lens modelling enables inspectors to pair compact multiple images between the image and source planes. The first column displays the postage stamp cut-out images from the initial inspection round. The subsequent six columns, for the F115W and F277W wavebands, include: (i) a cleaner foreground-subtracted image using a Multi-Gaussian expansion (MGE); (ii) the model lensed source in the image plane; and (iii) source-plane reconstructions. Black and white curves represent tangential and radial critical curves and caustics. In the F115W images, two yellow boxes overlay the model source, indicating two multiple images in the image-plane that map to a single emission clump in the source plane. Due to lower spatial resolution, this clump is not resolved in the F277W images. These examples show how detailed inspection in high-resolution filters like F115W can reveal image-plane pairings (or even groupings of four images), which are strong indicators of genuine lens systems. This method effectively distinguishes true lenses from false positives, such as spiral arms, which are unlikely to exhibit image pairings consistent with a strong lens model.
}
\label{figure:LensModelMulti}
\end{figure*}

Figure \ref{figure:LensModelMulti} illustrates a candidate where multiply imaged compact and clumpy emissions appears in the candidate lensed source, \textit{in addition to more extended arc/ring like features}. Furthermore, the SIE plus shear mass model can produce an image-plane to source-plane mapping which simultaneously fits the extended ring-like emission and the compact features, with the latter shown in \cref{figure:LensModelMulti} using yellow boxes. These image-plane pairings (or even groupings of four images) are strong indicators of genuine lens systems and allow inspectors to effectively distinguish true lenses from false positives, like spiral arms, ring galaxies or nearby galaxies that happen to be aligned, which are unlikely to exhibit image pairings consistent with a strong lens model. This method of identifying lenses is particularly effective in the F115W filter, which has the highest spatial resolution, allowing these clumps to be resolved, and operates at the shortest wavelength, where galaxies generally appear more clumpy. In the second round of visual inspection, inspectors were provided with figures with polygons overlaid showing mappings like this. In total, four candidates ranked 5 and above showed evidence of this compact multiple image pairing.

\subsubsection{Other Remarks}

After the second round of visual inspection, the inspectors met and discussed other aspects of how lens modelling information impacted their candidate ranking, with a summary as follows:

\begin{itemize}
    \item The projected lens light image and SIE mass model convergence should show a degree of agreement for genuine lenses. Inspectors noted this information, but often found it hard to guide their judgement as even in genuine strong lenses the two can appear somewhat different (e.g. due to dark matter).
    \item The MGE mass model often gave poor fits, even for high confidence lenses, because the lens galaxy was often too faint to constrain the stellar mass distribution and due to the omission of dark matter in the mass model. Reviewers tended to therefore not use this information and relied mostly on the SIE plus shear mass models.
    \item ``Successful'' lens models were fitted to the majority of candidates, in the sense that they focused the candidate source emission into a single region of the source-plane and produced plausible critical curves and caustics. A lens model being successful was therefore not used by itself to rank candidates highly, it was the detailed inspection of the model thereafter that was key.
\end{itemize}

\subsection{Previous Searches}

The $1.64$\,deg$^2$ HST COSMOS field, which encompasses the $0.54$\,deg$^2$ COSMOS-Web field, has been extensively searched for strong lenses, yielding 64 candidates in \citet{Faure2008}, 112 in \citet{Jackson2008}, and 92 in \citet{Pourrahmani2018}, alongside individual discoveries \citep{Guzzo2007, More2012a, VanDerWel2013, Jin2018, Pearson2024, Mercier2024, VanDokkum2024}. Of the 17 ‘spectacular’ lenses in M25, nine were previously identified, while only 8 of the 419 second-round visual inspection candidates have been reported in prior studies. These include high-scoring lenses such as COSJ095930+021351 and COSJ100050+020357 (score of 10) from \citet{Faure2008}, COSJ100103+020159 (score of 9) from \citet{Jackson2008}, COSJ100028+021600 \citep{Pourrahmani2018}, COSJ095940+023253 \citep{Faure2008}, and COSJ100023+021652 \citep{Jackson2008}, all scoring 6, as well as COSJ095943+022829 and COSJ095939+023043 from \citet{Pourrahmani2018}, both scoring 4.

The majority of COWLS candidates are absent from prior \textit{HST}-based searches, suggesting most sources were too faint for detection with \textit{HST} and instead benefit from \textit{JWST}’s unprecedented depth. Additionally, some source galaxies are extremely red, appearing only in F444W and/or F277W bands, making them ‘\textit{HST}-dark’. \textit{HST} F814W imaging is available for most candidates, and visual inspection confirms that lensed source emission is rarely visible. A quantitative analysis, including lens modelling, is beyond this work’s scope but would further characterise these newly uncovered systems.

Previous searches also include 14 candidates from \citet{Faure2008}, 18 from \citet{Jackson2008}, and 18 from \citet{Pourrahmani2018} that did not advance to our second inspection round. We generated \textit{JWST} images and performed lens modelling for all of these, which are now available in the COWLS public data release. JWN individually inspected these candidates, finding most to be false positives due to: (i) spiral/barred galaxies that, at \textit{HST} depth, mimic strong lenses but reveal their structure in \textit{JWST} imaging, and (ii) candidate systems with a nearby line-of-sight galaxy but lacking counter-images in lens modelling. However, two candidates from \citet{Pourrahmani2018} would have been graded A by JWN, and therefore have a reasonable probability of being genuine strong lenses the COWLS search missed.


\section{Discussion}\label{Discussion}

\subsection{The COSMOS-Web Lens Survey}

This work presents the COSMOS-Web Lens Survey (COWLS), with the first public data available at this URL:~\github{https://github.com/Jammy2211/COWLS_COSMOS_Web_Lens_Survey}. It includes over 100 highly ranked strong lens candidates, including the spectacular lenses discussed in M25. Selected via \textit{JWST}, the sample exhibits unique characteristics compared to other lens surveys, which drive a variety of science cases which will only be possible with this sample for years to come:

\begin{itemize}

    \item  \textbf{Highest Redshift Source Galaxies:} While source redshifts remain unmeasured, the high lens redshifts and results of COWLS paper III imply the COWLS sample contains some of the most distant galaxy-scale sources which extend beyond $z > 6$ and into the epoch of reionisation. Sources are imaged in remarkable detail with \textit{JWST}’s deep multi-band data, for example at $z = 6$ the F115W filter will achieve effective $\sim 100$ pc resolution. Being galaxy-scale lenses, this allows precise reconstructions of their unlensed structure, enabling unprecedented studies of high-redshift galaxy morphology, as done for sources up to $z \sim 3$ \citep{Swinbank2015}. 

    \item \textbf{Highest Redshift Lens Galaxies:} Spectroscopic and photometric data indicate that half the lenses lie at $z > 1$, with some pushing beyond $z > 2$. This regime is crucial for studying galaxy density profile evolution \citep{Shajib2021, Etherington2023a, Tan2024}, dark matter substructure \citep{Vegetti2014, Nightingale2024} and leveraging large lens samples for cosmology \citep{Li2024, Geng2025}.  
    
    \item \textbf{Strong Lenses in a Single Cosmic Volume:} All lenses reside within a contiguously imaged $0.54$\,deg$^2$ region, enabling the combination of strong and weak lensing to measure cosmic shear with unparalleled precision \citep{Birrer2017, Fleury2021, Hogg2022, Hogg:2025wac}, and the study of lens environments within their large-scale structure \citep{Peng2010}.  

    \item  \textbf{Lensed Sources Near Lens Centres:} In a subset of COWLS candidates, lensed emission passes within $0.25\arcsec$ of the lens galaxy centre, closer than most previously known lenses. This may allow detection of the influence of the lens’s supermassive black hole on the lensing signal \citep{Nightingale2023} and provide insights into dust absorption effects \citep{Kreckel2013, Barone2024}.

\end{itemize}

The survey's top priority is confirming which candidates are genuine strong lenses, best achieved by measuring spectroscopic redshifts of the source galaxies. Many lensed sources are faint, visible only due to \textit{JWST}’s deep imaging, making NIRSpec follow-up or observations with eight-metre-class ground-based telescopes necessary -- though feasible for only a subset of the lenses. Candidates benefit from extensive multi-wavelength COSMOS data, including MIRI F770W, \textit{HST} F814W, and numerous ground-based exposures, which enabled the precise photometric lens redshifts presented here. Whilst faint lensed sources may not have wide wavelength coverage, photometric redshift analysis could still help determine whether the candidate lensed emission originates at a different redshift than the candidate lens galaxy, given that the lens redshift is known. 

The impact of source redshift availability varies by science case. For instance, studying the density slope evolution of lens galaxies requires only lens redshifts \citep{Etherington2023a}, while cosmology combining strong and weak lensing is not critically dependent on a complete set of source redshifts \citep{Hogg2022}. Dark matter subhalos and SMBHs near the lens galaxy can be detected without a source redshift, but converting their mass to physical units requires one. Similarly, source galaxy studies can identify key targets using multi-wavelength reconstructions and photometric data, but require redshifts to know the physical scales and time in the Universe's formation of the galaxy.

Another priority is studying the source galaxy population and identifying the most intriguing objects for detailed individual study, leveraging lensing magnification to extract more information than would otherwise be possible for galaxies in the distant Universe. To support this, the public data release includes unlensed source reconstructions for all candidates across four NIRCam bands and additional detected wavebands. Preliminary analysis (see M25) confirms a fraction of \textit{HST}-dark red galaxies \citep{Gonzalez2023, Barrufet2023}, indicative of intense star formation beyond $z > 4$. The sample also contains compact, red lensed sources, potentially belonging to the ``red nugget'' population -- progenitors of massive ellipticals \citep{Barro2013, Dekel2014, Oldham2016} -- or ``little red dots'' at even higher redshifts \citep{Matthee2024}.

\subsection{Undiscovered Lenses Observed With The \textit{James Webb} Space Telescope}

We identified numerous high-quality strong lens candidates within just $0.54$\,deg$^2$ of \textit{JWST} data. Given that \textit{JWST} has conducted science observations for over three years, its archive now contains extensive high-quality imaging across much larger areas. Yet, only one galaxy-scale \textit{JWST} strong lens discovery has been reported to date, which is included in our survey \citep{Mercier2024, VanDokkum2024}.  

This suggests that \textit{JWST} has likely already observed hundreds more undiscovered strong lenses. Some may have been spotted by scientists during data analysis but not formally reported. As demonstrated by COWLS, these lenses form a unique sample, distinct from existing catalogs and future wide-field surveys like \textit{Euclid} \citep{Collett2014}, with many likely probing the highest lens and source redshifts. We encourage further searches of the \textit{JWST} archive, using machine learning or citizen science approaches, as successfully done with \textit{HST} \citep{Garvin2022}.  

M25 focuses on the 17 most “Spectacular” lenses identified by COWLS, which are immediately recognizable by eye as strong lenses in \textit{JWST} data. M25 also estimates how many similar high-quality lens candidates are likely already detected but not yet reported. 

\subsection{Lens Modelling for Lens Finding}

A significant fraction of high-ranked lens candidates would not have been identified without lens modelling. For example, $38.4\%$ of candidates that scored highly in the second round of visual inspection (with lens modelling information) were edge cases, meaning only one inspector flagged them as potential lenses in the first round (without this information). The improved visualisation and detailed insights provided by lens models allowed inspectors to rank these candidates more accurately, even those that were previously overlooked. We also documented that 30 candidates exhibited faint counter-images near the centre of the lens galaxy, which were not visible by eye (even in manually tuned RGB images) but became apparent after lens modelling. This is because it cleanly deblends the lens and source emission, fully utilising the PSF in each waveband. 

Our results suggest that ongoing lens-finding efforts in wide-field surveys, such as \textit{Euclid} -- expected to discover over 100,000 strong lenses \citep{Collett:2015roa, Holloway2023, Ferrami:2024obm} -- will significantly benefit from incorporating lens modelling into their processes. 
Lens-finding studies using ground-based data and modeling (e.g., \citealt{Sonnenfeld2018b, Sonnenfeld2020, Rojas2022}) have highlighted several advantages, such as improved arc visibility through foreground lens subtraction and the ability to reject false positives by testing whether a candidate fits a lens model. Our study identifies two additional benefits: (i) deblending faint counter-images near the lens galaxy center, and (ii) matching compact source features to assess the plausibility of a lens candidate. These improvements are likely enabled by a combination of more advanced modeling techniques and the higher resolution of space-based imaging. The modeling approach used here was recently applied to Euclid Q1 data \citep{EuclidCollaboration2025, EuclidCollaboration2025a, EuclidCollaboration2025b}, where both methods of lens identification were again found to be effective.

However, lens modelling does not definitively determine whether a candidate is a lens. We found that most candidates could be fitted with a physically plausible lens model, even when other evidence (such as colours) suggested they were unlikely to be strong lenses. Measured lens model parameters (e.g., Einstein radius, magnification) showed no correlation with candidate rankings, indicating that lens models fitted to both high- and low-ranked candidates were statistically indistinguishable. Additionally, fit quality metrics -- such as reduced chi-squared values -- did not effectively differentiate between high- and low-ranked candidates.

Thus, our study highlights that lens modelling should serve as a tool to provide visual inspectors with enhanced information for ranking candidates, rather than relying solely on the successful fitting of a lens model as a diagnostic. This approach helps to identify and properly rank the most promising lens candidates.

\subsubsection{Comparison to Euclid Lens Search Experiment}

\citet{Acevedo2024} used lens modeling of 16 candidates from high-resolution \textit{Euclid} early release observations to identify lenses based on three criteria: (1) whether an SIE critical curve can enclose or exclude the correct number of bright components in the image plane; (2) whether the critical curve, as an isodensity contour, aligns with the lens galaxy light profile; and (3) whether the reconstructed source surface brightness distribution resembles a compact, focused object within a caustic.  

In our modeling of $419$ candidates, we found that defining strict criteria or metrics for identifying lenses is unreliable. Many objects that meet such criteria score low and are likely not lenses based on other factors, such as colors. While lens modeling significantly aids candidate assessment, we believe its usefulness is more subjective and does not adhere to rigid rules. Notably, lens modeling excels at revealing counter-images and multiple images, providing crucial information beyond straightforward criteria.

\subsection{Selection Effects}

The COWLS sample highlights a selection effect that previous lens samples built by inspecting wide-field ground-based surveys are subject to, for example, SL2S and DES \citep{Gavazzi2012a, Gavazzi2014, Sonnenfeld2013b, Jacobs2019, Tran2022}. The separation between their lens centres and the closest lensed source image is constrained by the resolution of the instrument, which for ground-based surveys is around $0.27\arcsec$. The higher resolution of the NIRCam allowed us to find lenses where this separation was as low as $0.1\arcsec$. This required that lens modelling provided a clean deblending of the lens and source light. We therefore recommend that future lens surveys based on wide-field imaging, most notably \textit{Euclid} with a resolution of $\sim 0.15 \arcsec$, incorporate lens modelling to ensure they recover these low-separation strong lenses.

This also raises the question of what selection effects COWLS itself is subject to, given that the selection function of strong lenses is now well-characterised \citep{Sonnenfeld2023, Sonnenfeld2024}. COWLS Paper III investigates this by forecasting lens properties such as Einstein radii, redshifts, and magnitudes, comparing these to the values measured in this paper. This analysis confirms that typical lensing selection biases are at play, for example, the tendency for lens galaxies to be more massive than galaxies in photometric catalogues. 

Future scientific studies will benefit from a more complete characterisation of our selection function, noting that achieving this for a complete sample enables an unbiased comparison to the general galaxy population \citep{Sonnenfeld2022}. This would require a probabilistic assessment of how inspectors identified lenses and which ones were overlooked. However, owing to its relatively small size and the high value of its \textit{JWST} data, COWLS could serve as an ideal test case for performing such an analysis on a smaller scale before applying it to much larger lens samples.
 


\section{Summary}\label{Summary}

The \textit{James Webb} Space Telescope (\textit{JWST}) provides a powerful new avenue for discovering strong gravitational lenses, thanks to its high-resolution imaging capabilities and the unprecedented depth offered by its 6.5 metre mirror. To capitalise on this, we performed a systematic search for strong lenses in COSMOS-Web \citep{Casey2023}, \textit{JWST}’s largest contiguous survey, covering $0.54$ deg$^2$ in four NIRCam bands (F115W, F150W, F277W, and F444W). We present the COSMOS-Web Lens Survey (COWLS), a sample of over 100 strong lens candidates. This includes 17 `spectacular' lenses reported in COWLS Paper II by \citet{Mahler2025} and is consistent with the lens forecasts presented in COWLS Paper III by \citet{Hogg2025b}. A key aspect of our approach was the incorporation of traditional lens modelling, using the open-source software {\tt PyAutoLens} \citep{Nightingale2015, Nightingale2018, pyautolens}, which enabled us to identify many lenses that would otherwise have been missed. 

The COWLS sample stands out due to several unique features that enhance its scientific potential. Firstly, the sample contains some of the most distant source galaxies of galaxy-scale strong lenses ever found, with redshifts anticipated to extend  beyond $z > 6$ and into the epoch of reionisation. Being galaxy-scale lenses, this allows precise reconstructions of their unlensed structure, providing an unprecedented opportunity to study high-redshift galaxy morphology. Secondly, the lens galaxies are at higher redshifts than previous samples, with approximately half at $z > 1$, and some reaching beyond $z > 2$. This redshift range is crucial for understanding the evolution of galaxy density profiles and strong lensing studies of cosmology. All COWLS lenses reside within a contiguously imaged $0.54$ deg$^2$ region, enabling the combination of strong and weak lensing to measure cosmic shear with unparalleled precision and to investigate lens environments within their large-scale structure. Finally, a subset of COWLS candidates features lensed emission passing within $0.25\arcsec$ of the lens galaxy centre, offering the potential to detect the influence of the supermassive black hole on the lensing signal \citep{Nightingale2023}, while also providing new insights into dust absorption effects.

\textit{JWST} has been operating for over three years, and the COWLS sample represents just the tip of the iceberg in terms of the strong lenses it has observed, many of which remain undiscovered. Future COWLS studies will demonstrate the potential of these lenses for probing the high-redshift Universe, highlighting the need for dedicated efforts to uncover and exploit the vast reservoir of strong lenses hidden within the \textit{JWST} archive.

\section*{Software Citations}

This work uses the following software packages:

\begin{itemize}

\item
\href{https://github.com/rhayes777/PyAutoFit}{{PyAutoFit}}
\citep{pyautofit}

\item
\href{https://github.com/Jammy2211/PyAutoGalaxy}{{PyAutoGalaxy}}
\citep{pyautogalaxy}

\item
\href{https://github.com/Jammy2211/PyAutoLens}{{PyAutoLens}}
\citep{Nightingale2015, Nightingale2018, Nightingale2021}

\item
\href{https://github.com/astropy/astropy}{{Astropy}}
\citep{astropy1, astropy2}

\item
\href{https://github.com/dfm/corner.py}{{Corner.py}}
\citep{corner}

\item
\href{https://github.com/joshspeagle/dynesty}{{Dynesty}}
\citep{Speagle2020}

\item
\href{https://github.com/matplotlib/matplotlib}{{Matplotlib}}
\citep{matplotlib}

\item
\href{numba` https://github.com/numba/numba}{{Numba}}
\citep{numba}

\item
\href{https://github.com/numpy/numpy}{{NumPy}}
\citep{numpy}

\item
\href{https://www.python.org/}{{Python}}
\citep{python}

\item
\href{https://github.com/scikit-image/scikit-image}{{Scikit-image}}
\citep{scikit-image}

\item
\href{https://github.com/scikit-learn/scikit-learn}{{Scikit-learn}}
\citep{scikit-learn}

\item
\href{https://github.com/scipy/scipy}{{Scipy}}
\citep{scipy}

\item
\href{https://www.sqlite.org/index.html}{{SQLite}}
\citep{sqlite}

\end{itemize}

\section*{Acknowledgements}


JWN is supported by an STFC/UKRI Ernest Rutherford Fellowship, Project Reference: ST/X003086/1. AA and QH acknowledge support from the European Research Council (ERC) through Advanced Investigator grant DMIDAS (GA 786910). GM, MvW-K and RM were supported in Durham by STFC via grant ST/X001075/1, and the UK Space Agency via grant ST/X001997/1.

This work used both the Cambridge Service for Data Driven Discovery (CSD3) and the DiRAC Data-Centric system, project code dp195, which are operated by the University of Cambridge and Durham University on behalf of the STFC DiRAC HPC Facility (www.dirac.ac.uk). These were funded by BIS capital grant ST/K00042X/1, STFC capital grants ST/P002307/1, ST/R002452/1, ST/H008519/1, ST/K00087X/1, STFC Operations grants ST/K003267/1, ST/K003267/1, and Durham University. DiRAC is part of the UK National E-Infrastructure.

This work was made possible by utilising the CANDIDE cluster at the Institut d’Astrophysique de Paris. The cluster was funded through grants from the PNCG, CNES, DIM-ACAV, the Euclid Consortium, and the Danish National Research Foundation Cosmic Dawn Center (DNRF140). It is maintained by Stephane Rouberol. The French contingent of the COSMOS team is partly supported by the Centre National d’Etudes Spatiales (CNES). OI acknowledges the funding of the French Agence Nationale de la Recherche for the project iMAGE (grant ANR-22-CE31-0007). 
DS acknowledge the Jet Propulsion Laboratory, California Institute of Technology, under a contract with the National Aeronautics and Space Administration (80NM0018D0004).

This work is based on observations made with the NASA/ESA/CSA James Webb Space Telescope. The data were obtained from the Mikulski Archive for Space Telescopes at the Space Telescope Science Institute, which is operated by the Association of Universities for Research in Astronomy, Inc., under NASA contract NAS 5-03127 for JWST. These observations are associated with program 1727.

\section*{Data Availability}

Analysis results are publicly available at \github{https://github.com/Jammy2211/COWLS_COSMOS_Web_Lens_Survey}.


\bibliographystyle{mnras}
\bibliography{main}

\begin{thebibliography}{}
\makeatletter
\relax
\def\mn@urlcharsother{\let\do\@makeother \do\$\do\&\do\#\do\^\do\_\do\%\do\~}
\def\mn@doi{\begingroup\mn@urlcharsother \@ifnextchar [ {\mn@doi@}
  {\mn@doi@[]}}
\def\mn@doi@[#1]#2{\def\@tempa{#1}\ifx\@tempa\@empty \href
  {http://dx.doi.org/#2} {doi:#2}\else \href {http://dx.doi.org/#2} {#1}\fi
  \endgroup}
\def\mn@eprint#1#2{\mn@eprint@#1:#2::\@nil}
\def\mn@eprint@arXiv#1{\href {http://arxiv.org/abs/#1} {{\tt arXiv:#1}}}
\def\mn@eprint@dblp#1{\href {http://dblp.uni-trier.de/rec/bibtex/#1.xml}
  {dblp:#1}}
\def\mn@eprint@#1:#2:#3:#4\@nil{\def\@tempa {#1}\def\@tempb {#2}\def\@tempc
  {#3}\ifx \@tempc \@empty \let \@tempc \@tempb \let \@tempb \@tempa \fi \ifx
  \@tempb \@empty \def\@tempb {arXiv}\fi \@ifundefined
  {mn@eprint@\@tempb}{\@tempb:\@tempc}{\expandafter \expandafter \csname
  mn@eprint@\@tempb\endcsname \expandafter{\@tempc}}}

\bibitem[\protect\citeauthoryear{{Acevedo Barroso} et~al.,}{{Acevedo Barroso}
  et~al.}{2025}]{Acevedo2024}
{Acevedo Barroso} J.~A.,  et~al., 2025, \mn@doi [\aap]
  {10.1051/0004-6361/202451868}, \href
  {https://ui.adsabs.harvard.edu/abs/2025A&A...697A..14A} {697, A14}

\bibitem[\protect\citeauthoryear{Amvrosiadis et~al.,}{Amvrosiadis
  et~al.}{2025}]{Amvrosiadis2025}
Amvrosiadis A.,  et~al., 2025, \mn@doi [MNRAS] {10.1093/mnras/staf048}, 537,
  1163

\bibitem[\protect\citeauthoryear{Aretxaga et~al.,}{Aretxaga
  et~al.}{2011}]{Aretxaga2011}
Aretxaga I.,  et~al., 2011, \mn@doi [MNRAS] {10.1111/j.1365-2966.2011.18989.x},
  415, 3831

\bibitem[\protect\citeauthoryear{{Arnouts}, {Cristiani}, {Moscardini},
  {Matarrese}, {Lucchin}, {Fontana}  \& {Giallongo}}{{Arnouts}
  et~al.}{1999}]{Arnouts1999}
{Arnouts} S.,  {Cristiani} S.,  {Moscardini} L.,  {Matarrese} S.,  {Lucchin}
  F.,  {Fontana} A.,   {Giallongo} E.,  1999, \mn@doi [\mnras]
  {10.1046/j.1365-8711.1999.02978.x}, \href
  {https://ui.adsabs.harvard.edu/abs/1999MNRAS.310..540A} {310, 540}

\bibitem[\protect\citeauthoryear{{Astropy Collaboration} et~al.,}{{Astropy
  Collaboration} et~al.}{2013}]{astropy1}
{Astropy Collaboration} et~al., 2013, \mn@doi [A\&A]
  {10.1051/0004-6361/201322068}, \href
  {http://adsabs.harvard.edu/abs/2013A%26A...558A..33A} {558, A33}

\bibitem[\protect\citeauthoryear{Auger, Treu, Bolton, Gavazzi, Koopmans,
  Marshall, Moustakas  \& Burles}{Auger et~al.}{2010}]{Auger2009}
Auger M.~W.,  Treu T.,  Bolton A.~S.,  Gavazzi R.,  Koopmans L.~V.,  Marshall
  P.~J.,  Moustakas L.~A.,   Burles S.,  2010, \mn@doi [ApJ]
  {10.1088/0004-637X/724/1/511}, 724, 511

\bibitem[\protect\citeauthoryear{{Barbary}}{{Barbary}}{2016}]{K.Barbary2016}
{Barbary} K.,  2016, \mn@doi [J. Open Source Softw.] {10.21105/joss.00058},
  \href {https://ui.adsabs.harvard.edu/abs/2016JOSS....1...58B} {1, 58}

\bibitem[\protect\citeauthoryear{Barone et~al.,}{Barone
  et~al.}{2024}]{Barone2024}
Barone T.~M.,  et~al., 2024, \mn@doi [Nature Communications Physics]
  {10.1038/s42005-024-01778-4}, 7, 1

\bibitem[\protect\citeauthoryear{Barro et~al.,}{Barro et~al.}{2013}]{Barro2013}
Barro G.,  et~al., 2013, \mn@doi [ApJ] {10.1088/0004-637X/765/2/104}, 765

\bibitem[\protect\citeauthoryear{Barrufet et~al.,}{Barrufet
  et~al.}{2023}]{Barrufet2023}
Barrufet L.,  et~al., 2023, \mn@doi [MNRAS] {10.1093/mnras/stad947}, 522, 449

\bibitem[\protect\citeauthoryear{{Bertin}}{{Bertin}}{2011}]{2011ASPC..442..435B}
{Bertin} E.,  2011, in {Evans} I.~N.,  {Accomazzi} A.,  {Mink} D.~J.,   {Rots}
  A.~H.,  eds,  Astronomical Society of the Pacific Conference Series Vol. 442,
  Astronomical Data Analysis Software and Systems XX. p.~435

\bibitem[\protect\citeauthoryear{{Bertin}, {Mellier}, {Radovich}, {Missonnier},
  {Didelon}  \& {Morin}}{{Bertin} et~al.}{2002}]{2002ASPC..281..228B}
{Bertin} E.,  {Mellier} Y.,  {Radovich} M.,  {Missonnier} G.,  {Didelon} P.,
  {Morin} B.,  2002, in {Bohlender} D.~A.,  {Durand} D.,   {Handley} T.~H.,
  eds,  Astronomical Society of the Pacific Conference Series Vol. 281,
  Astronomical Data Analysis Software and Systems XI. p.~228

\bibitem[\protect\citeauthoryear{{Bertin}, {Schefer}, {Apostolakos},
  {{\'A}lvarez-Ayll{\'o}n}, {Dubath}  \& {K{\"u}mmel}}{{Bertin}
  et~al.}{2020}]{Bertin2020}
{Bertin} E.,  {Schefer} M.,  {Apostolakos} N.,  {{\'A}lvarez-Ayll{\'o}n} A.,
  {Dubath} P.,   {K{\"u}mmel} M.,  2020, in {Pizzo} R.,  {Deul} E.~R.,  {Mol}
  J.~D.,  {de Plaa} J.,   {Verkouter} H.,  eds,  Astronomical Society of the
  Pacific Conference Series Vol. 527, Astronomical Data Analysis Software and
  Systems XXIX. p.~461

\bibitem[\protect\citeauthoryear{Birrer, Welschen, Amara  \& Refregier}{Birrer
  et~al.}{2017}]{Birrer2017}
Birrer S.,  Welschen C.,  Amara A.,   Refregier A.,  2017, \mn@doi [Journal of
  Cosmology and Astroparticle Physics] {10.1088/1475-7516/2017/04/049}, 04, 049

\bibitem[\protect\citeauthoryear{Birrer, Refregier  \& Amara}{Birrer
  et~al.}{2018}]{Birrer:2017sge}
Birrer S.,  Refregier A.,   Amara A.,  2018, \mn@doi [ApJ]
  {10.3847/2041-8213/aaa1de}, 852, L14

\bibitem[\protect\citeauthoryear{Birrer et~al.}{Birrer
  et~al.}{2020}]{Birrer2020}
Birrer S.,  et~al., 2020, \mn@doi [A\&A] {10.1051/0004-6361/202038861}, 643,
  A165

\bibitem[\protect\citeauthoryear{{Bolton}, {Burles}, {Koopmans}, {Treu}  \&
  {Moustakas}}{{Bolton} et~al.}{2006}]{Bolton2006}
{Bolton} A.~S.,  {Burles} S.,  {Koopmans} L. V.~E.,  {Treu} T.,   {Moustakas}
  L.~A.,  2006, \mn@doi [ApJ] {10.1086/498884}, \href
  {https://ui.adsabs.harvard.edu/abs/2006ApJ...638..703B} {638, 703}

\bibitem[\protect\citeauthoryear{Bolton, Burles, Koopmans, Treu, Gavazzi,
  Moustakas, Wayth  \& Schlegel}{Bolton et~al.}{2008}]{Bolton2008a}
Bolton A.~S.,  Burles S.,  Koopmans L. V.~E.,  Treu T.,  Gavazzi R.,  Moustakas
  L.~A.,  Wayth R.,   Schlegel D.~J.,  2008, \mn@doi [ApJ] {10.1086/589327},
  682, 964

\bibitem[\protect\citeauthoryear{Bolton et~al.,}{Bolton
  et~al.}{2012}]{Bolton2012}
Bolton A.~S.,  et~al., 2012, \mn@doi [ApJ] {10.1088/0004-637X/757/1/82}, 757,
  82

\bibitem[\protect\citeauthoryear{Bro \& {De Jong}}{Bro \& {De
  Jong}}{1997}]{Bro1997}
Bro R.,  {De Jong} S.,  1997, \mn@doi [Journal of Chemometrics]
  {10.1002/(SICI)1099-128X(199709/10)11:5<393::AID-CEM483>3.0.CO;2-L}, 11, 393

\bibitem[\protect\citeauthoryear{Cao et~al.,}{Cao et~al.}{2022}]{Cao2021}
Cao X.,  et~al., 2022, \mn@doi [Research in Astronomy and Astrophysics]
  {10.1088/1674-4527/ac3f2b}, 22, 30 pp

\bibitem[\protect\citeauthoryear{Cappellari}{Cappellari}{2002}]{Cappellari2002}
Cappellari M.,  2002, \mn@doi [MNRAS] {10.1046/j.1365-8711.2002.05412.x}, 333,
  400

\bibitem[\protect\citeauthoryear{{Casey} et~al.,}{{Casey}
  et~al.}{2024}]{Casey2023}
{Casey} C.~M.,  et~al., 2024, \mn@doi [ApJ] {10.3847/1538-4357/ad2075}, \href
  {https://ui.adsabs.harvard.edu/abs/2024ApJ...965...98C} {965, 98}

\bibitem[\protect\citeauthoryear{Collett}{Collett}{2015}]{Collett:2015roa}
Collett T.~E.,  2015, \mn@doi [ApJ] {10.1088/0004-637X/811/1/20}, 811, 20

\bibitem[\protect\citeauthoryear{Collett \& Auger}{Collett \&
  Auger}{2014}]{Collett2014}
Collett T.~E.,  Auger M.~W.,  2014, \mn@doi [MNRAS] {10.1093/mnras/stu1190},
  443, 969

\bibitem[\protect\citeauthoryear{Collett, Oldham, Smith, Auger, Westfall,
  Bacon, Nichol  \& Masters}{Collett et~al.}{2018}]{Collett2018}
Collett T.~E.,  Oldham L.~J.,  Smith R.~J.,  Auger M.~W.,  Westfall K.~B.,
  Bacon D.,  Nichol R.~C.,   Masters K.~L.,  2018, Science, 360, 1342

\bibitem[\protect\citeauthoryear{Dekel \& Burkert}{Dekel \&
  Burkert}{2014}]{Dekel2014}
Dekel A.,  Burkert A.,  2014, \mn@doi [MNRAS] {10.1093/mnras/stt2331}, 438,
  1870

\bibitem[\protect\citeauthoryear{Despali, Vegetti, White, Giocoli  \& van~den
  Bosch}{Despali et~al.}{2018}]{Despali2018}
Despali G.,  Vegetti S.,  White S. D.~M.,  Giocoli C.,   van~den Bosch F.~C.,
  2018, \mn@doi [MNRAS] {10.1093/mnras/sty159}, 475, 5424

\bibitem[\protect\citeauthoryear{Despali, Vegetti, White, Powell, Stacey,
  Fassnacht, Rizzo  \& Enzi}{Despali et~al.}{2022}]{Despali2022}
Despali G.,  Vegetti S.,  White S.~D.,  Powell D.~M.,  Stacey H.~R.,  Fassnacht
  C.~D.,  Rizzo F.,   Enzi W.,  2022, \mn@doi [MNRAS] {10.1093/mnras/stab3537},
  510, 2480

\bibitem[\protect\citeauthoryear{{Dey} et~al.,}{{Dey} et~al.}{2019}]{Dey2019}
{Dey} A.,  et~al., 2019, \mn@doi [\aj] {10.3847/1538-3881/ab089d}, \href
  {https://ui.adsabs.harvard.edu/abs/2019AJ....157..168D} {157, 168}

\bibitem[\protect\citeauthoryear{{Drlica-Wagner} et~al.,}{{Drlica-Wagner}
  et~al.}{2018}]{Drlica-Wagner2018}
{Drlica-Wagner} A.,  et~al., 2018, \mn@doi [\apjs] {10.3847/1538-4365/aab4f5},
  \href {https://ui.adsabs.harvard.edu/abs/2018ApJS..235...33D} {235, 33}

\bibitem[\protect\citeauthoryear{Duboscq, Hogg, Fleury  \& Larena}{Duboscq
  et~al.}{2024}]{Duboscq:2024asf}
Duboscq T.,  Hogg N.~B.,  Fleury P.,   Larena J.,  2024, \mn@doi [Journal of
  Cosmology and Astroparticle Physics] {10.1088/1475-7516/2024/08/021}, 08, 021

\bibitem[\protect\citeauthoryear{Etherington et~al.,}{Etherington
  et~al.}{2022}]{Etherington2022}
Etherington A.,  et~al., 2022, \mn@doi [MNRAS] {10.1093/mnras/stac2639}, 517,
  3275

\bibitem[\protect\citeauthoryear{Etherington et~al.,}{Etherington
  et~al.}{2023}]{Etherington2023}
Etherington A.,  et~al., 2023, MNRAS, 521, 6005

\bibitem[\protect\citeauthoryear{Etherington et~al.}{Etherington
  et~al.}{2024}]{Etherington2023a}
Etherington A.,  et~al., 2024, \mn@doi [MNRAS] {10.1093/mnras/stae1375}, 531,
  3684

\bibitem[\protect\citeauthoryear{{Euclid Collaboration} et~al.,}{{Euclid
  Collaboration} et~al.}{2025a}]{EuclidCollaboration2025}
{Euclid Collaboration} et~al., 2025a, https://arxiv.org/abs/2503.15324

\bibitem[\protect\citeauthoryear{{Euclid Collaboration} et~al.,}{{Euclid
  Collaboration} et~al.}{2025c}]{EuclidCollaboration2025b}
{Euclid Collaboration} et~al., 2025c, https://arxiv.org/pdf/2503.15325

\bibitem[\protect\citeauthoryear{{Euclid Collaboration} et~al.,}{{Euclid
  Collaboration} et~al.}{2025b}]{EuclidCollaboration2025a}
{Euclid Collaboration} et~al., 2025b, https://arxiv.org/pdf/2503.15326

\bibitem[\protect\citeauthoryear{Faure et~al.}{Faure et~al.}{2008}]{Faure2008}
Faure C.,  et~al., 2008, \mn@doi [ApJ Supplement] {10.1086/526426}, 176, 19

\bibitem[\protect\citeauthoryear{Ferrami \& Wyithe}{Ferrami \&
  Wyithe}{2024}]{Ferrami:2024obm}
Ferrami G.,  Wyithe J. S.~B.,  2024, \mn@doi [MNRAS] {10.1093/mnras/stae1607},
  532, 1832

\bibitem[\protect\citeauthoryear{Fleury, Larena  \& Uzan}{Fleury
  et~al.}{2021}]{Fleury2021}
Fleury P.,  Larena J.,   Uzan J.-P.,  2021, \mn@doi [Journal of Cosmology and
  Astroparticle Physics] {10.1088/1475-7516/2021/08/024}, 08, 024

\bibitem[\protect\citeauthoryear{Foreman-Mackey}{Foreman-Mackey}{2016}]{corner}
Foreman-Mackey D.,  2016, \mn@doi [J. Open Source Softw.]
  {10.21105/joss.00024}, 1, 24

\bibitem[\protect\citeauthoryear{Garvin, Kruk, Cornen, Bhatawdekar,
  Ca{\~{n}}ameras  \& Mer{\'{i}}n}{Garvin et~al.}{2022}]{Garvin2022}
Garvin E.~O.,  Kruk S.,  Cornen C.,  Bhatawdekar R.,  Ca{\~{n}}ameras R.,
  Mer{\'{i}}n B.,  2022, \mn@doi [A\&A] {10.1051/0004-6361/202243745}, 667

\bibitem[\protect\citeauthoryear{Gavazzi, Treu, Rhodes, Koopmans, Bolton,
  Burles, Massey  \& Moustakas}{Gavazzi et~al.}{2007}]{Gavazzi2007}
Gavazzi R.,  Treu T.,  Rhodes J.~D.,  Koopmans L. V.~E.,  Bolton A.~S.,  Burles
  S.,  Massey R.~J.,   Moustakas L.~A.,  2007, \mn@doi [ApJ] {10.1086/519237},
  667, 176

\bibitem[\protect\citeauthoryear{Gavazzi, Treu, Marshall, Brault  \&
  Ruff}{Gavazzi et~al.}{2012}]{Gavazzi2012a}
Gavazzi R.,  Treu T.,  Marshall P.~J.,  Brault F.,   Ruff A.,  2012, \mn@doi
  [ApJ] {10.1088/0004-637X/761/2/170}, 761

\bibitem[\protect\citeauthoryear{Gavazzi, Marshall, Treu  \&
  Sonnenfeld}{Gavazzi et~al.}{2014}]{Gavazzi2014}
Gavazzi R.,  Marshall P.~J.,  Treu T.,   Sonnenfeld A.,  2014, \mn@doi [ApJ]
  {10.1088/0004-637X/785/2/144}, 785, 144

\bibitem[\protect\citeauthoryear{Geng, Grespan, Thuruthipilly, Harikumar, Pollo
   \& Biesiada}{Geng et~al.}{2025}]{Geng2025}
Geng S.,  Grespan M.,  Thuruthipilly H.,  Harikumar S.,  Pollo A.,   Biesiada
  M.,  2025, arXiv:2501.02577

\bibitem[\protect\citeauthoryear{Guzzo et~al.,}{Guzzo et~al.}{2007}]{Guzzo2007}
Guzzo L.,  et~al., 2007, \mn@doi [ApJSS] {10.1086/516588}, 172, 254

\bibitem[\protect\citeauthoryear{Harrington et~al.,}{Harrington
  et~al.}{2016}]{Harrington2016}
Harrington K.~C.,  et~al., 2016, \mn@doi [MNRAS] {10.1093/mnras/stw614}, 458,
  4383

\bibitem[\protect\citeauthoryear{He et~al.,}{He et~al.}{2022a}]{He2022a}
He Q.,  et~al., 2022a, \mn@doi [MNRAS] {10.1093/mnras/stac191}, 511, 3046

\bibitem[\protect\citeauthoryear{He et~al.,}{He et~al.}{2022b}]{He2022b}
He Q.,  et~al., 2022b, \mn@doi [MNRAS] {10.1093/mnras/stac759}, 512, 5862

\bibitem[\protect\citeauthoryear{He et~al.,}{He et~al.}{2023}]{He2023}
He Q.,  et~al., 2023, \mn@doi [MNRAS] {10.1093/mnras/stac2779}, 518, 220

\bibitem[\protect\citeauthoryear{He et~al.}{He et~al.}{2024}]{He2024}
He Q.,  et~al., 2024, \mn@doi [MNRAS] {10.1093/mnras/stae1577}, 532, 2441

\bibitem[\protect\citeauthoryear{Hipp}{Hipp}{2020}]{sqlite}
Hipp R.~D.,  2020, {SQLite}, \url {https://www.sqlite.org/index.html}

\bibitem[\protect\citeauthoryear{Hogg}{Hogg}{2024}]{Hogg:2023khs}
Hogg N.~B.,  2024, \mn@doi [MNRAS: Letters] {10.1093/mnrasl/slae005}, 528, L95

\bibitem[\protect\citeauthoryear{Hogg, Fleury, Larena  \& Martinelli}{Hogg
  et~al.}{2023}]{Hogg2022}
Hogg N.~B.,  Fleury P.,  Larena J.,   Martinelli M.,  2023, \mn@doi [MNRAS]
  {10.1093/mnras/stad512}, 520, 5982

\bibitem[\protect\citeauthoryear{Hogg et~al.}{Hogg et~al.}{2025a}]{Hogg2025b}
Hogg N.~B.,  et~al., 2025a, {The COSMOS-Web Lens Survey (COWLS) III: forecasts
  versus data} (\mn@eprint {arXiv} {2503.08785})

\bibitem[\protect\citeauthoryear{Hogg, Shajib, Johnson  \& Larena}{Hogg
  et~al.}{2025b}]{Hogg:2025wac}
Hogg N.~B.,  Shajib A.~J.,  Johnson D.,   Larena J.,  2025b, {Line-of-sight
  shear in SLACS strong lenses} (\mn@eprint {arXiv} {2501.16292})

\bibitem[\protect\citeauthoryear{Holloway, Verma, Marshall, More  \&
  Tecza}{Holloway et~al.}{2023}]{Holloway2023}
Holloway P.,  Verma A.,  Marshall P.~J.,  More A.,   Tecza M.,  2023, \mn@doi
  [MNRAS] {10.1093/mnras/stad2371}, 525, 2341

\bibitem[\protect\citeauthoryear{Hunter}{Hunter}{2007}]{matplotlib}
Hunter J.~D.,  2007, \mn@doi [Comput Sci Eng] {10.1109/MCSE.2007.55}, 9, 90

\bibitem[\protect\citeauthoryear{{Ilbert} et~al.,}{{Ilbert}
  et~al.}{2006}]{Ilbert2006}
{Ilbert} O.,  et~al., 2006, \mn@doi [\aap] {10.1051/0004-6361:20065138}, \href
  {https://ui.adsabs.harvard.edu/abs/2006A&A...457..841I} {457, 841}

\bibitem[\protect\citeauthoryear{Jackson}{Jackson}{2008}]{Jackson2008}
Jackson N.,  2008, \mn@doi [MNRAS] {10.1111/j.1365-2966.2008.13629.x}, 389,
  1311

\bibitem[\protect\citeauthoryear{Jacobs et~al.,}{Jacobs
  et~al.}{2019}]{Jacobs2019}
Jacobs C.,  et~al., 2019, \mn@doi [MNRAS] {10.1093/mnras/stz272}, 484, 5330

\bibitem[\protect\citeauthoryear{{Jin} et~al.,}{{Jin} et~al.}{2018}]{Jin2018}
{Jin} S.,  et~al., 2018, \mn@doi [ApJ] {10.3847/1538-4357/aad4af}, \href
  {https://ui.adsabs.harvard.edu/abs/2018ApJ...864...56J} {864, 56}

\bibitem[\protect\citeauthoryear{{Jin} et~al.,}{{Jin} et~al.}{2024}]{Jin24}
{Jin} S.,  et~al., 2024, \mn@doi [\aap] {10.1051/0004-6361/202451445}, \href
  {https://ui.adsabs.harvard.edu/abs/2024A&A...690L..16J} {690, L16}

\bibitem[\protect\citeauthoryear{Koekemoer et~al.,}{Koekemoer
  et~al.}{2011}]{Grogin2011}
Koekemoer A.~M.,  et~al., 2011, \mn@doi [ApJSS] {10.1088/0067-0049/197/2/36},
  197, 3535

\bibitem[\protect\citeauthoryear{Koopmans et~al.,}{Koopmans
  et~al.}{2009}]{Koopmans2009}
Koopmans L.~V.,  et~al., 2009, \mn@doi [ApJ] {10.1088/0004-637X/703/1/L51},
  703, L51

\bibitem[\protect\citeauthoryear{{Kreckel} et~al.,}{{Kreckel}
  et~al.}{2013}]{Kreckel2013}
{Kreckel} K.,  et~al., 2013, \mn@doi [\apj] {10.1088/0004-637X/771/1/62}, \href
  {https://ui.adsabs.harvard.edu/abs/2013ApJ...771...62K} {771, 62}

\bibitem[\protect\citeauthoryear{{K{\"u}mmel}, {Bertin}, {Schefer},
  {Apostolakos}, {{\'A}lvarez-Ayll{\'o}n}  \& {Dubath}}{{K{\"u}mmel}
  et~al.}{2020}]{Kummel2020}
{K{\"u}mmel} M.,  {Bertin} E.,  {Schefer} M.,  {Apostolakos} N.,
  {{\'A}lvarez-Ayll{\'o}n} A.,   {Dubath} P.,  2020, in {Pizzo} R.,  {Deul}
  E.~R.,  {Mol} J.~D.,  {de Plaa} J.,   {Verkouter} H.,  eds,  Astronomical
  Society of the Pacific Conference Series Vol. 527, Astronomical Data Analysis
  Software and Systems XXIX. p.~29

\bibitem[\protect\citeauthoryear{Lam, Pitrou  \& Seibert}{Lam
  et~al.}{2015}]{numba}
Lam S.~K.,  Pitrou A.,   Seibert S.,  2015, \mn@doi [Proceedings of the Second
  Workshop on the LLVM Compiler Infrastructure in HPC - LLVM '15]
  {10.1145/2833157.2833162}, pp~1--6

\bibitem[\protect\citeauthoryear{{Lange}}{{Lange}}{2023}]{nautilus}
{Lange} J.~U.,  2023, \mn@doi [\mnras] {10.1093/mnras/stad2441}, \href
  {https://ui.adsabs.harvard.edu/abs/2023MNRAS.525.3181L} {525, 3181}

\bibitem[\protect\citeauthoryear{{Li}, {Collett}, {Krawczyk}  \& {Enzi}}{{Li}
  et~al.}{2024}]{Li2024}
{Li} T.,  {Collett} T.~E.,  {Krawczyk} C.~M.,   {Enzi} W.,  2024, \mn@doi
  [MNRAS] {10.1093/mnras/stad3514}, \href
  {https://ui.adsabs.harvard.edu/abs/2024MNRAS.527.5311L} {527, 5311}

\bibitem[\protect\citeauthoryear{Liu et~al.,}{Liu et~al.}{2024}]{Liu2024}
Liu D.,  et~al., 2024, \mn@doi [Nature Astronomy] {10.1038/s41550-024-02296-7},
  8

\bibitem[\protect\citeauthoryear{Mahler et~al.}{Mahler
  et~al.}{2025}]{Mahler2025}
Mahler G.,  et~al., 2025, {The COSMOS-Web Lens Survey (COWLS) II: depth,
  resolution, and NIR coverage from JWST reveal 17 spectacular lenses}
  (\mn@eprint {arXiv} {2503.08782})

\bibitem[\protect\citeauthoryear{Maresca, Dye  \& Li}{Maresca
  et~al.}{2021}]{Maresca2021}
Maresca J.,  Dye S.,   Li N.,  2021, \mn@doi [MNRAS] {10.1093/mnras/stab387},
  503, 2229

\bibitem[\protect\citeauthoryear{Massey et~al.,}{Massey
  et~al.}{2007}]{Massey2007}
Massey R.,  et~al., 2007, \mn@doi [ApJS] {10.1086/516599}, 172, 239

\bibitem[\protect\citeauthoryear{Matthee et~al.,}{Matthee
  et~al.}{2024}]{Matthee2024}
Matthee J.,  et~al., 2024, \mn@doi [ApJ] {10.3847/1538-4357/ad2345}, 963, 129

\bibitem[\protect\citeauthoryear{Melo-Carneiro, Furlanetto  \&
  Chies-Santos}{Melo-Carneiro et~al.}{2023}]{Melo2023}
Melo-Carneiro C.~R.,  Furlanetto C.,   Chies-Santos A.~L.,  2023, \mn@doi
  [MNRAS] {10.1093/mnras/stad162}, 520, 1613

\bibitem[\protect\citeauthoryear{{Mercier} et~al.,}{{Mercier}
  et~al.}{2024}]{Mercier2024}
{Mercier} W.,  et~al., 2024, \mn@doi [A\&A] {10.1051/0004-6361/202348095},
  \href {https://ui.adsabs.harvard.edu/abs/2024A&A...687A..61M} {687, A61}

\bibitem[\protect\citeauthoryear{More, Cabanac, More, Alard, Limousin, Kneib,
  Gavazzi  \& Motta}{More et~al.}{2012}]{More2012a}
More A.,  Cabanac R.,  More S.,  Alard C.,  Limousin M.,  Kneib J.~P.,  Gavazzi
  R.,   Motta V.,  2012, \mn@doi [ApJ] {10.1088/0004-637X/749/1/38}, 749

\bibitem[\protect\citeauthoryear{{Nagam} et~al.,}{{Nagam}
  et~al.}{2025}]{Nagam2025}
{Nagam} B.~C.,  et~al., 2025, \mn@doi [arXiv e-prints]
  {10.48550/arXiv.2502.09802}, \href
  {https://ui.adsabs.harvard.edu/abs/2025arXiv250209802N} {p. arXiv:2502.09802}

\bibitem[\protect\citeauthoryear{{Navarro}, {Frenk}  \& {White}}{{Navarro}
  et~al.}{1996}]{Navarro1996}
{Navarro} J.~F.,  {Frenk} C.~S.,   {White} S. D.~M.,  1996, \mn@doi [ApJ]
  {10.1086/177173}, \href
  {https://ui.adsabs.harvard.edu/abs/1996ApJ...462..563N} {462, 563}

\bibitem[\protect\citeauthoryear{Negrello et~al.,}{Negrello
  et~al.}{2014}]{Negrello2014}
Negrello M.,  et~al., 2014, \mn@doi [MNRAS] {10.1093/mnras/stu413}, 440, 1999

\bibitem[\protect\citeauthoryear{Newton, Marshall, Treu, Auger, Gavazzi,
  Bolton, Koopmans  \& Moustakas}{Newton et~al.}{2011}]{Newton2011}
Newton E.~R.,  Marshall P.~J.,  Treu T.,  Auger M.~W.,  Gavazzi R.,  Bolton
  A.~S.,  Koopmans L.~V.,   Moustakas L.~A.,  2011, \mn@doi [ApJ]
  {10.1088/0004-637X/734/2/104}, 734, 104

\bibitem[\protect\citeauthoryear{{Nightingale} \& {Dye}}{{Nightingale} \&
  {Dye}}{2015}]{Nightingale2015}
{Nightingale} J.~W.,  {Dye} S.,  2015, \mn@doi [MNRAS] {10.1093/mnras/stv1455},
  \href {https://ui.adsabs.harvard.edu/abs/2015MNRAS.452.2940N} {452, 2940}

\bibitem[\protect\citeauthoryear{Nightingale, Dye  \& Massey}{Nightingale
  et~al.}{2018}]{Nightingale2018}
Nightingale J.,  Dye S.,   Massey R.,  2018, \mn@doi [MNRAS]
  {10.1093/mnras/sty1264}, 478, 4738

\bibitem[\protect\citeauthoryear{Nightingale, Hayes  \& Griffiths}{Nightingale
  et~al.}{2021a}]{pyautofit}
Nightingale J.~W.,  Hayes R.~G.,   Griffiths M.,  2021a, \mn@doi [J. Open
  Source Softw.] {10.21105/joss.02550}, 6, 2550

\bibitem[\protect\citeauthoryear{Nightingale et~al.,}{Nightingale
  et~al.}{2021b}]{pyautolens}
Nightingale J.~W.,  et~al., 2021b, \mn@doi [J. Open Source Softw.]
  {10.21105/joss.02825}, 6, 2825

\bibitem[\protect\citeauthoryear{{Nightingale} et~al.}{{Nightingale}
  et~al.}{2021c}]{Nightingale2021}
{Nightingale} J.,  et~al., 2021c, \mn@doi [J. Open Source Softw.]
  {10.21105/joss.02825}, \href
  {https://ui.adsabs.harvard.edu/abs/2021JOSS....6.2825N} {6, 2825}

\bibitem[\protect\citeauthoryear{Nightingale et~al.,}{Nightingale
  et~al.}{2023a}]{pyautogalaxy}
Nightingale J.~W.,  et~al., 2023a, \mn@doi [J. Open Source Softw.]
  {10.21105/joss.04475}, 8, 4475

\bibitem[\protect\citeauthoryear{Nightingale et~al.,}{Nightingale
  et~al.}{2023b}]{Nightingale2023}
Nightingale J.~W.,  et~al., 2023b, MNRAS, 521, 3298

\bibitem[\protect\citeauthoryear{Nightingale et~al.,}{Nightingale
  et~al.}{2024}]{Nightingale2024}
Nightingale J.~W.,  et~al., 2024, \mn@doi [MNRAS] {10.1093/mnras/stad3694},
  527, 10480

\bibitem[\protect\citeauthoryear{{O'Riordan} et~al.,}{{O'Riordan}
  et~al.}{2025}]{ORiordan2025}
{O'Riordan} C.~M.,  et~al., 2025, \mn@doi [\aap] {10.1051/0004-6361/202453014},
  \href {https://ui.adsabs.harvard.edu/abs/2025A&A...694A.145O} {694, A145}

\bibitem[\protect\citeauthoryear{Oldham et~al.,}{Oldham
  et~al.}{2017}]{Oldham2016}
Oldham L.,  et~al., 2017, \mn@doi [MNRAS] {10.1093/mnras/stw2832}, 465, 3185

\bibitem[\protect\citeauthoryear{{Pearce-Casey} et~al.,}{{Pearce-Casey}
  et~al.}{2025}]{Pearce-Casey2024}
{Pearce-Casey} R.,  et~al., 2025, \mn@doi [\aap] {10.1051/0004-6361/202453152},
  \href {https://ui.adsabs.harvard.edu/abs/2025A&A...696A.214P} {696, A214}

\bibitem[\protect\citeauthoryear{Pearson et~al.}{Pearson
  et~al.}{2023}]{Pearson2024}
Pearson J.,  et~al., 2023, \mn@doi [MNRAS] {10.1093/mnras/stad3916}, 527, 12044

\bibitem[\protect\citeauthoryear{Pedregosa et~al.,}{Pedregosa
  et~al.}{2011}]{scikit-learn}
Pedregosa F.,  et~al., 2011, Journal of Machine Learning Research, 12, 2825

\bibitem[\protect\citeauthoryear{Peng, Ho, Impey  \& Rix}{Peng
  et~al.}{2010}]{Peng2010}
Peng C.~Y.,  Ho L.~C.,  Impey C.~D.,   Rix H.~W.,  2010, \mn@doi [Astronomical
  Journal] {10.1088/0004-6256/139/6/2097}, 139, 2097

\bibitem[\protect\citeauthoryear{P{\'{e}}rez-Gonz{\'{a}}lez
  et~al.,}{P{\'{e}}rez-Gonz{\'{a}}lez et~al.}{2023}]{Gonzalez2023}
P{\'{e}}rez-Gonz{\'{a}}lez P.~G.,  et~al., 2023, \mn@doi [ApJL]
  {10.3847/2041-8213/acb3a5}, 946, L16

\bibitem[\protect\citeauthoryear{{Pourrahmani}, {Nayyeri}  \&
  {Cooray}}{{Pourrahmani} et~al.}{2018}]{Pourrahmani2018}
{Pourrahmani} M.,  {Nayyeri} H.,   {Cooray} A.,  2018, \mn@doi [ApJ]
  {10.3847/1538-4357/aaae6a}, \href
  {https://ui.adsabs.harvard.edu/abs/2018ApJ...856...68P} {856, 68}

\bibitem[\protect\citeauthoryear{{Price-Whelan} et~al.,}{{Price-Whelan}
  et~al.}{2018}]{astropy2}
{Price-Whelan} A.~M.,  et~al., 2018, \mn@doi [AJ] {10.3847/1538-3881/aabc4f},
  \href {https://ui.adsabs.harvard.edu/#abs/2018AJ....156..123T} {156, 123}

\bibitem[\protect\citeauthoryear{Ritondale, Vegetti, Despali, Auger, Koopmans
  \& McKean}{Ritondale et~al.}{2019}]{Ritondale2019a}
Ritondale E.,  Vegetti S.,  Despali G.,  Auger M.~W.,  Koopmans L.~V.,   McKean
  J.~P.,  2019, \mn@doi [MNRAS] {10.1093/mnras/stz464}, 485, 2179

\bibitem[\protect\citeauthoryear{Rizzo, Vegetti, Powell, Fraternali, McKean,
  Stacey  \& White}{Rizzo et~al.}{2020}]{Rizzo2020}
Rizzo F.,  Vegetti S.,  Powell D.,  Fraternali F.,  McKean J.~P.,  Stacey
  H.~R.,   White S.~D.,  2020, \mn@doi [Nature] {10.1038/s41586-020-2572-6},
  584, 201

\bibitem[\protect\citeauthoryear{Rizzo, Vegetti, Fraternali, Stacey  \&
  Powell}{Rizzo et~al.}{2021}]{Rizzo2021}
Rizzo F.,  Vegetti S.,  Fraternali F.,  Stacey H.~R.,   Powell D.,  2021,
  \mn@doi [MNRAS] {10.1093/mnras/stab2295}, 507, 3952

\bibitem[\protect\citeauthoryear{Rojas et~al.,}{Rojas et~al.}{2022}]{Rojas2022}
Rojas K.,  et~al., 2022, \mn@doi [A\&A] {10.1051/0004-6361/202142119}, 668, 1

\bibitem[\protect\citeauthoryear{Shajib}{Shajib}{2019}]{Shajib2019}
Shajib A.~J.,  2019, \mn@doi [MNRAS] {10.1093/mnras/stz1796}, 14, 1387

\bibitem[\protect\citeauthoryear{{Shajib}, {Treu}, {Birrer}  \&
  {Sonnenfeld}}{{Shajib} et~al.}{2021}]{Shajib2021}
{Shajib} A.~J.,  {Treu} T.,  {Birrer} S.,   {Sonnenfeld} A.,  2021, \mn@doi
  [MNRAS] {10.1093/mnras/stab536}, \href
  {https://ui.adsabs.harvard.edu/abs/2021MNRAS.503.2380S} {503, 2380}

\bibitem[\protect\citeauthoryear{Shu et~al.,}{Shu et~al.}{2015}]{Shu2015}
Shu Y.,  et~al., 2015, \mn@doi [ApJ] {10.1088/0004-637X/803/2/71}, 803, 1

\bibitem[\protect\citeauthoryear{Shu et~al.,}{Shu et~al.}{2016}]{Shu2016}
Shu Y.,  et~al., 2016, \mn@doi [ApJ] {10.3847/0004-637x/824/2/86}, 824, 86

\bibitem[\protect\citeauthoryear{Shu et~al.,}{Shu et~al.}{2017}]{Shu2017}
Shu Y.,  et~al., 2017, \mn@doi [ApJ] {10.3847/1538-4357/aa9794}, 851, 48

\bibitem[\protect\citeauthoryear{{Shuntov} et~al.,}{{Shuntov}
  et~al.}{2025a}]{Shuntov2024}
{Shuntov} M.,  et~al., 2025a, \mn@doi [\aap] {10.1051/0004-6361/202452570},
  \href {https://ui.adsabs.harvard.edu/abs/2025A&A...695A..20S} {695, A20}

\bibitem[\protect\citeauthoryear{{Shuntov} et~al.,}{{Shuntov}
  et~al.}{2025b}]{Shuntov2025}
{Shuntov} M.,  et~al., 2025b, \mn@doi [\aap] {10.1051/0004-6361/202554273},
  \href {https://ui.adsabs.harvard.edu/abs/2025A&A...696L..14S} {696, L14}

\bibitem[\protect\citeauthoryear{Sibson}{Sibson}{1981}]{Sibson1981}
Sibson R.,  1981, Interpreting Multivariate Data.
John Wiley and Sons, New York

\bibitem[\protect\citeauthoryear{Smith, Lucey  \& Collier}{Smith
  et~al.}{2018}]{Smith2018}
Smith R.~J.,  Lucey J.~R.,   Collier W.~P.,  2018, \mn@doi [MNRAS]
  {10.1093/MNRAS/STY2328}, 481, 2115

\bibitem[\protect\citeauthoryear{{Sonnenfeld}}{{Sonnenfeld}}{2022}]{Sonnenfeld2022}
{Sonnenfeld} A.,  2022, \mn@doi [A\&A] {10.1051/0004-6361/202142301}, \href
  {https://ui.adsabs.harvard.edu/abs/2022A&A...659A.132S} {659, A132}

\bibitem[\protect\citeauthoryear{{Sonnenfeld}}{{Sonnenfeld}}{2024}]{Sonnenfeld2024}
{Sonnenfeld} A.,  2024, \mn@doi [A\&A] {10.1051/0004-6361/202451341}, \href
  {https://ui.adsabs.harvard.edu/abs/2024A&A...690A.325S} {690, A325}

\bibitem[\protect\citeauthoryear{Sonnenfeld, Gavazzi, Suyu, Treu  \&
  Marshall}{Sonnenfeld et~al.}{2013}]{Sonnenfeld2013b}
Sonnenfeld A.,  Gavazzi R.,  Suyu S.~H.,  Treu T.,   Marshall P.~J.,  2013,
  \mn@doi [ApJ] {10.1088/0004-637X/777/2/97}, 777, 97

\bibitem[\protect\citeauthoryear{Sonnenfeld et~al.,}{Sonnenfeld
  et~al.}{2018}]{Sonnenfeld2018b}
Sonnenfeld A.,  et~al., 2018, \mn@doi [PASJ] {10.1093/pasj/psx062}, 70, 1

\bibitem[\protect\citeauthoryear{{Sonnenfeld} et~al.,}{{Sonnenfeld}
  et~al.}{2020}]{Sonnenfeld2020}
{Sonnenfeld} A.,  et~al., 2020, \mn@doi [A\&A] {10.1051/0004-6361/202038067},
  \href {https://ui.adsabs.harvard.edu/abs/2020A&A...642A.148S} {642, A148}

\bibitem[\protect\citeauthoryear{Sonnenfeld, Li, Despali, Gavazzi, Shajib  \&
  Taylor}{Sonnenfeld et~al.}{2023}]{Sonnenfeld2023}
Sonnenfeld A.,  Li S.~S.,  Despali G.,  Gavazzi R.,  Shajib A.~J.,   Taylor
  E.~N.,  2023, \mn@doi [A\&A] {10.1051/0004-6361/202346026}, 678, 1

\bibitem[\protect\citeauthoryear{Speagle}{Speagle}{2020a}]{dynesty}
Speagle J.~S.,  2020a, \mn@doi [MNRAS] {10.1093/mnras/staa278}, 493, 3132

\bibitem[\protect\citeauthoryear{Speagle}{Speagle}{2020b}]{Speagle2020}
Speagle J.~S.,  2020b, \mn@doi [MNRAS] {10.1093/mnras/staa278}, 493, 3132

\bibitem[\protect\citeauthoryear{{Swinbank} et~al.,}{{Swinbank}
  et~al.}{2015}]{Swinbank2015}
{Swinbank} A.~M.,  et~al., 2015, \mn@doi [ApJ: Letters]
  {10.1088/2041-8205/806/1/L17}, \href
  {https://ui.adsabs.harvard.edu/abs/2015ApJ...806L..17S} {806, L17}

\bibitem[\protect\citeauthoryear{Szalay, Connolly  \& Szokoly}{Szalay
  et~al.}{1999}]{szalay_simultaneous_1999}
Szalay A.~S.,  Connolly A.~J.,   Szokoly G.~P.,  1999, \mn@doi [AJ]
  {10.1086/300689}, 117, 68

\bibitem[\protect\citeauthoryear{{Tan} et~al.,}{{Tan} et~al.}{2024}]{Tan2024}
{Tan} C.~Y.,  et~al., 2024, \mn@doi [MNRAS] {10.1093/mnras/stae884}, \href
  {https://ui.adsabs.harvard.edu/abs/2024MNRAS.530.1474T} {530, 1474}

\bibitem[\protect\citeauthoryear{{Tessore} \& {Metcalf}}{{Tessore} \&
  {Metcalf}}{2015}]{Tessore2015}
{Tessore} N.,  {Metcalf} R.~B.,  2015, \mn@doi [A\&A]
  {10.1051/0004-6361/201526773}, \href
  {https://ui.adsabs.harvard.edu/abs/2015A&A...580A..79T} {580, A79}

\bibitem[\protect\citeauthoryear{Tran et~al.,}{Tran et~al.}{2022}]{Tran2022}
Tran K.-V.~H.,  et~al., 2022, \mn@doi [AJ] {10.3847/1538-3881/ac7da2}, 164, 148

\bibitem[\protect\citeauthoryear{Van~Rossum \& Drake}{Van~Rossum \&
  Drake}{2009}]{python}
Van~Rossum G.,  Drake F.~L.,  2009, Python 3 Reference Manual.
CreateSpace, Scotts Valley, CA

\bibitem[\protect\citeauthoryear{Van~der Walt, Sch{\"o}nberger, Nunez-Iglesias,
  Boulogne, Warner, Yager, Gouillart  \& Yu}{Van~der Walt
  et~al.}{2014}]{scikit-image}
Van~der Walt S.,  Sch{\"o}nberger J.~L.,  Nunez-Iglesias J.,  Boulogne F.,
  Warner J.~D.,  Yager N.,  Gouillart E.,   Yu T.,  2014, PeerJ, 2, e453

\bibitem[\protect\citeauthoryear{Vegetti, Koopmans, Auger, Treu  \&
  Bolton}{Vegetti et~al.}{2014}]{Vegetti2014}
Vegetti S.,  Koopmans L. V.~E.,  Auger M.~W.,  Treu T.,   Bolton A.~S.,  2014,
  \mn@doi [MNRAS] {10.1093/mnras/stu943}, 442, 2017

\bibitem[\protect\citeauthoryear{Vieira et~al.,}{Vieira
  et~al.}{2010}]{Vieira2010}
Vieira J.~D.,  et~al., 2010, \mn@doi [ApJ] {10.1088/0004-637X/719/1/763}, 719,
  763

\bibitem[\protect\citeauthoryear{Vieira et~al.,}{Vieira
  et~al.}{2013}]{Vieira2013}
Vieira J.~D.,  et~al., 2013, \mn@doi [Nature] {10.1038/nature12001}, 495, 344

\bibitem[\protect\citeauthoryear{{Virtanen} et~al.,}{{Virtanen}
  et~al.}{2020}]{scipy}
{Virtanen} P.,  et~al., 2020, \mn@doi [Nature Methods]
  {10.1038/s41592-019-0686-2}, \href {https://rdcu.be/b08Wh} {17, 261}

\bibitem[\protect\citeauthoryear{Warren \& Dye}{Warren \&
  Dye}{2003}]{Warren2003}
Warren S.,  Dye S.,  2003, \mn@doi [ApJ] {10.1086/375132}, 590, 673

\bibitem[\protect\citeauthoryear{van Dokkum, Brammer, Wang, Leja  \&
  Conroy}{van Dokkum et~al.}{2024}]{VanDokkum2024}
van Dokkum P.,  Brammer G.,  Wang B.,  Leja J.,   Conroy C.,  2024, \mn@doi
  [Nature Astronomy] {10.1038/s41550-023-02103-9}, 8, 119

\bibitem[\protect\citeauthoryear{{van der Walt}, {Colbert}  \&
  {Varoquaux}}{{van der Walt} et~al.}{2011}]{numpy}
{van der Walt} S.,  {Colbert} S.~C.,   {Varoquaux} G.,  2011, \mn@doi [Comput
  Sci Eng] {10.1109/MCSE.2011.37}, 13, 22

\bibitem[\protect\citeauthoryear{van~der Wel et~al.}{van~der Wel
  et~al.}{2013}]{VanDerWel2013}
van~der Wel A.,  et~al., 2013, \mn@doi [ApJ Letters]
  {10.1088/2041-8205/777/1/L17}, 777, L17

\makeatother
\end{thebibliography}

\noindent\rule{\columnwidth}{0.4pt}
$^{1}$\Newcastle\\
$^{2}$\Liege\\
$^{3}$\DurhamCEA\\
$^{4}$\DurhamICC\\
$^{5}$\Northeastern\\
$^{6}$\LUPM\\
$^{7}$\Aalto\\
$^{8}$\Helsinki\\
$^{9}$\LAM\\
$^{10}$\JPL\\
$^{11}$\PMO\\
$^{12}$\DAWN\\
$^{13}$\NBI\\
$^{14}$\UTAustin\\
$^{15}$\IAP\\
$^{16}$\UCSB\\
$^{17}$\Rochester\\
$^{18}$\STScI\\
$^{19}$\UCSC\\
$^{20}$\Hawaii\\
$^{21}$\Caltech\\
$^{22}$\DTU\\

\appendix
\section{Visual Inspection Details}\label{AppVisual}

\begin{table}
\begin{adjustbox}{max width=\textwidth}
\normalsize
\begin{tabular}{ l | l | l | l} 
\multicolumn{1}{p{1.2cm}|}{Inspector} 
& \multicolumn{1}{p{1.1cm}|}{Yes} 
& \multicolumn{1}{p{1.1cm}|}{Maybe} 
& \multicolumn{1}{p{1.1cm}}{Unique} \\  \hline
& & & \\[-6pt]
1 & 19 (0.05\%) & 615 (1.44\%) & 345 (0.81\%)\\[2pt]
2 & 106 (0.25\%) & 868 (2.03\%) & 518 (1.21\%) \\[2pt]
3 & 23 (0.05\%)  & 247 (0.57\%) & 99 (0.22\%) \\[2pt]
4 & 1082 (2.50\%) & 2500 (5.86\%) & 2950 (6.91\%) \\[2pt]
5 & 45 (0.17\%) & 188 (0.69\%) & 92 (0.34\%) \\[2pt]
\end{tabular}
\end{adjustbox}
\caption{
The table presents the number and percentage of objects each inspector assigned the inputs `Y - Yes, this is a lens' and `M - Maybe this is a lens' during the first round of visual inspection. Inspectors 1-4 inspected all 42,660 objects, whereas inspector 5 only inspected the January 2024 data comprising 27,125 objects, which is reflected in the percentages. The final column provides the number of candidates an inspector flagged as Yes' or Maybe` that are unique because no other inspector flagged them.
}
\label{table:visual_1_inspect}
\end{table}

\cref{table:visual_1_inspect} shows the distribution of `Yes' and `Maybe' inputs from the 5 inspectors. Inspector 4 was notably optimistic, classifying over 8\% of objects as `Yes' or `Maybe'. There is also significant variation among the remaining inspectors. The final column highlights the number of unique objects each inspector flagged as `Yes' or `Maybe' that no other inspector did. Inspectors 1 to 5 had 345, 518, 99, 2950 and 92 unique candidates each, illustrating that it is common for inspectors to disagree.

\cref{table:visual_1} displays the distribution of categories for the three dataset groups: January 2023, April 2023, and January 2024. Notably, the `At Least $50\%$ Maybe' category, which dictates if a candidates make it to the second round of visual inspection, contains over double the percentage of candidates in April 2024 compared to January 2024. This is because, for April 2023 (and January 2023), there were only 4 inspectors, including the optimistic inspector, making it more likely a candidate reached this category as only one of the other 3 inspectors had to agree with the optimistic inspector. With 5 inspectors, 2 inspectors had to agree, making it less likely for candidates to make it into this category.

The optimistic inspector influenced the number of candidates advancing to the second round of visual inspection. This effect primarily comes from lenses making it into the `At Least 50\% Maybe' category. However, they also contributed a number of the edge cases put forward after reinspection by JWN. Out of the 29 edge cases that became highly ranked, 14 qualified as edge cases due to the optimistic inspector. 


\section{Catalogue}\label{AppCatalogue}

\begin{table*}
\begin{adjustbox}{max width=\textwidth}
\normalsize
\begin{tabular}{ l | l | l | l | l | l | l | l | l | l | l | l | l | l | l | l | l | l | l | l} 
\multicolumn{1}{p{1.8cm}|}{Lens Name} 
& \multicolumn{1}{p{0.7cm}|}{Score} 
& \multicolumn{1}{p{1.1cm}|}{RA [deg]} 
& \multicolumn{1}{p{1.1cm}|}{Dec [deg]}
& \multicolumn{1}{p{0.85cm}|}{$z_\text{spec}$} 
& \multicolumn{1}{p{0.85cm}|}{$z_\text{phot}$} 
& \multicolumn{1}{p{1.2cm}|}{$\log_{{10}}$ $(M_*/M_{{\odot}})$} 
& \multicolumn{1}{p{0.85cm}|}{$R_{\rm Ein}$  (\arcsec)}
& \multicolumn{1}{p{0.85cm}|}{Lens $m_{\mathrm{F115W}}$}
& \multicolumn{1}{p{0.85cm}|}{Lens $m_{\mathrm{F150W}}$}
& \multicolumn{1}{p{0.85cm}|}{Lens $m_{\mathrm{F277W}}$}
& \multicolumn{1}{p{0.85cm}|}{Lens $m_{\mathrm{F444W}}$}
& \multicolumn{1}{p{0.85cm}|}{Source $m_{\mathrm{F115W}}$}
& \multicolumn{1}{p{0.85cm}|}{Source $m_{\mathrm{F150W}}$}
& \multicolumn{1}{p{0.85cm}|}{Source $m_{\mathrm{F277W}}$}
& \multicolumn{1}{p{0.85cm}|}{Source $m_{\mathrm{F444W}}$}
& \multicolumn{1}{p{0.85cm}|}{$\mu_{\mathrm{F115W}}$}
& \multicolumn{1}{p{0.85cm}|}{$\mu_{\mathrm{F150W}}$}
& \multicolumn{1}{p{0.85cm}|}{$\mu_{\mathrm{F277W}}$}
& \multicolumn{1}{p{0.85cm}}{$\mu_{\mathrm{F444W}}$}
\\   \hline
& & & & & & & & & & & & & & & & & & & \\ [-6pt]
COSJ100121+022740 & M25 & 150.341363 & 2.461297 & 0.371 & 0.371 & 10.819 & 0.860 & 18.05 & 17.68 & 15.89 & 16.73 & 25.32 & 24.61 & 22.97 & 22.25 & 3.11 & 3.39 & 4.49 & 5.25 \\ [2pt]
COSJ100119+014849 & M25 & 150.330526 & 1.813616 &  &  &  & 1.135 & 22.68 & 21.74 & 19.17 & 18.72 & 23.23 & 22.03 & 19.47 & 19.40 & 1.77 & 1.85 & 1.87 & 1.86 \\ [2pt]
COSJ100047+015023 & M25 & 150.198584 & 1.839785 & 0.893 & 0.893 & 11.441 & 1.616 & 18.77 & 18.28 & 16.08 & 16.29 & 24.22 & 24.20 & 20.92 & 21.78 & 3.87 & 2.86 & 3.74 & 4.21 \\ [2pt]
COSJ100028+021928 & M25 & 150.119067 & 2.324503 & 0.604 & 0.600 & 11.000 & 0.660 & 19.24 & 18.68 & 16.28 & 16.88 & 22.21 & 21.87 & 20.13 & 20.37 & 3.53 & 3.29 & 3.75 & 3.87 \\ [2pt]
COSJ100025+015245 & M25 & 150.106708 & 1.879287 &  & 2.449 & 10.396 & 0.535 & 22.67 & 22.18 & 19.23 & 18.78 & 24.26 & 23.25 & 21.15 & 20.22 & 2.63 & 3.12 & 4.49 & 4.52 \\ [2pt]
COSJ100024+021749 & M25 & 150.100089 & 2.297133 & 0.362 & 0.360 & 10.500 & 0.733 & 18.63 & 18.25 & 16.53 & 17.21 & 24.39 & 23.99 & 21.98 & 21.03 & 3.71 & 3.34 & 3.37 & 4.44 \\ [2pt]
COSJ100024+015334 & M25 & 150.100467 & 1.893029 & & 2.454 & 10.250 & 0.771 & 22.68 & 21.56 & 18.64 & 18.05 & 24.54 & 23.63 & 22.03 & 20.57 & 6.22 & 5.39 & 7.11 & 6.98 \\ [2pt]
COSJ100018+022138 & M25 & 150.076935 & 2.360757 & 1.530 & 1.689 & 10.794 & 0.383 & 22.23 & 21.56 & 18.89 & 18.47 & 22.38 & 22.59 & 21.31 & 21.48 & 3.50 & 5.11 & 5.69 & 4.36 \\ [2pt]
COSJ100013+023424 & M25 & 150.055775 & 2.573347 &  & 0.890 & 10.950 & 1.455 & 19.93 & 19.42 & 17.02 & 17.48 & 45.35 & 26.86 & 22.19 & 21.60 & 3.73 & 3.91 & 4.41 & 4.80 \\ [2pt]
COSJ100012+022015 & M25 & 150.052588 & 2.337706 & 0.377 & 0.380 & 11.050 & 0.748 & 17.61 & 17.36 & 15.54 & 16.29 & 23.05 & 22.99 & 22.33 & 23.13 & 2.33 & 2.97 & 2.62 & 2.79 \\ [2pt]
COSJ095955+021900 & M25 & 149.983162 & 2.316881 &  & 0.577 & 10.400 & 2.112 &  & 18.66 & 15.98 & 16.27 &  & 21.82 & 22.16 & 22.43 &  & 3.64 & 3.18 & 3.43 \\ [2pt]
COSJ095953+023319 & M25 & 149.974678 & 2.555446 & 0.731 & 0.727 & 11.500 & 1.441 & 18.65 & 17.87 & 15.62 & 16.26 & 22.25 & 22.86 & 20.77 & 19.89 & 3.23 & 4.29 & 4.60 & 4.85 \\ [2pt]
COSJ095950+022057 & M25 & 149.961430 & 2.349412 & 0.939 & 0.940 & 11.150 & 1.040 & 19.38 & 18.67 & 16.47 & 16.66 & 25.32 & 25.21 & 24.04 & 24.12 & 5.82 & 5.85 & 7.68 & 8.25 \\ [2pt]
COSJ095921+020638 & M25 & 149.840689 & 2.110653 & 0.469 & 0.459 & 10.650 & 0.707 & 19.09 & 18.33 & 16.21 & 16.90 & 21.02 & 22.30 & 21.81 & 23.04 & 2.87 & 3.25 & 3.54 & 3.94 \\ [2pt]
COSJ095920+015851 & M25 & 149.833887 & 1.980927 & 0.974 & 0.949 & 11.244 & 0.448 & 19.54 & 18.95 & 16.57 & 16.67 & 23.74 & 22.83 & 20.78 & 21.22 & 2.03 & 2.39 & 2.17 & 2.21 \\ [2pt]
COSJ095917+015424 & M25 & 149.821462 & 1.906873 &  & 1.252 & 10.898 & 0.578 & 21.11 & 20.54 & 18.02 & 18.05 & 24.01 & 24.47 & 22.77 & 20.61 & 3.21 & 3.45 & 5.04 & 5.10 \\ [2pt]
COSJ095914+021219 & M25 & 149.811422 & 2.205440 & 1.053 & 1.367 & 11.300 & 1.480 & 20.15 & 19.44 & 16.81 & 16.83 & 25.97 & 25.04 & 23.90 & 24.67 & 4.53 & 4.59 & 4.17 & 3.41 \\ [2pt]
COSJ100110+015415 & S12 & 150.292189 & 1.904211 & 0.603 & 0.539 & 10.500 & 0.464 & 19.82 & 19.40 & 17.47 & 18.08 & 21.82 & 21.87 & 21.06 & 21.49 & 3.28 & 3.33 & 2.82 & 3.37 \\ [2pt]
COSJ095908+021559 & S12 & 149.785542 & 2.266506 & 0.701 & 0.705 & 10.950 & 0.311 & 19.54 & 19.14 & 16.95 & 17.60 & 22.67 & 22.62 & 21.52 & 21.72 & 2.12 & 1.98 & 2.84 & 3.02 \\ [2pt]
COSJ100152+022856 & S11 & 150.469730 & 2.482339 &  & 0.665 & 10.252 & 0.231 & 20.60 & 20.27 & 18.21 & 18.84 & 23.30 & 23.29 & 22.89 & 23.53 & 2.37 & 2.81 & 3.47 & 5.66 \\ [2pt]
COSJ100052+014856 & S11 & 150.218253 & 1.815644 & 0.169 & 0.160 & 10.314 & 1.004 & 17.30 & 16.97 & 15.69 & 16.34 & 23.69 & 23.62 & 22.03 & 22.70 & 3.08 & 3.13 & 2.96 & 2.95 \\ [2pt]
COSJ095941+022634 & S11 & 149.922755 & 2.442972 &  & 1.441 & 10.255 & 0.298 & 22.85 & 22.25 & 19.75 & 19.45 & 25.02 & 25.11 & 23.95 & 23.59 & 3.22 & 3.01 & 3.41 & 4.41 \\ [2pt]
COSJ095930+021351 & S10 & 149.875355 & 2.231100 &  &  &  & 1.742 & 16.63 & 16.37 & 14.56 & 15.29 & 22.32 & 22.11 & 19.90 & 19.66 & 3.82 & 3.85 & 4.22 & 4.20 \\ [2pt]
COSJ100050+020357 & S10 & 150.211908 & 2.066090 & 0.941 & 0.922 & 10.800 & 0.389 & 20.27 & 19.51 & 16.94 & 17.13 & 24.27 & 24.38 & 23.74 & 23.63 & 1.59 & 1.94 & 2.09 & 2.11 \\ [2pt]
COSJ100020+023332 & S10 & 150.083433 & 2.559067 & 0.510 & 0.481 & 10.703 & 0.724 & 19.28 & 18.84 & 16.92 & 17.60 & 24.86 & 23.39 & 22.35 & 22.28 & 3.70 & 3.46 & 3.33 & 4.09 \\ [2pt]
COSJ095922+021812 & S10 & 149.842223 & 2.303377 & 0.513 & 0.523 & 11.020 & 0.697 & 18.38 & 17.86 & 15.97 & 16.68 & 24.51 & 23.97 & 22.32 & 22.08 & 2.26 & 2.72 & 2.44 & 2.81 \\ [2pt]
COSJ100157+021407 & S10 & 150.490523 & 2.235357 & 0.833 & 0.823 & 10.833 & 0.640 & 19.89 & 19.43 & 17.16 & 17.62 & 24.91 & 24.06 & 22.76 & 22.98 & 5.07 & 4.08 & 2.96 & 3.53 \\ [2pt]
COSJ100104+015631 & S10 & 150.270455 & 1.942178 & 0.931 & 0.925 & 9.950 & 0.410 & 21.71 & 21.59 & 19.54 & 19.74 & 23.94 & 24.24 & 22.81 & 23.08 & 3.14 & 3.63 & 4.23 & 4.27 \\ [2pt]
COSJ100027+020051 & S10 & 150.113222 & 2.014218 & 0.840 & 0.839 & 10.899 & 0.875 & 19.77 & 19.26 & 16.99 & 17.37 & 25.47 & 25.89 & 23.19 & 23.96 & 2.49 & 2.40 & 2.48 & 2.34 \\ [2pt]
COSJ100005+021223 & S10 & 150.023177 & 2.206483 & 0.938 & 0.937 & 10.950 & 0.541 & 19.94 & 19.43 & 17.10 & 17.33 & 22.86 & 22.94 & 21.09 & 21.58 & 3.82 & 3.93 & 5.22 & 5.28 \\ [2pt]
COSJ095943+022046 & S10 & 149.932435 & 2.346200 &  & 0.691 & 11.200 & 0.688 & 18.65 & 18.15 & 16.08 & 16.89 & 23.39 & 23.00 & 20.96 & 22.37 & 3.74 & 4.02 & 4.36 & 4.85 \\ [2pt]
COSJ095923+021836 & S10 & 149.846863 & 2.310166 & 0.731 & 0.756 & 11.200 & 0.301 & 19.48 & 18.91 & 16.70 & 17.31 & 22.24 & 21.68 & 19.46 & 20.11 & 1.56 & 1.65 & 1.86 & 1.86 \\ [2pt]
COSJ095912+020751 & S10 & 149.800636 & 2.131076 & 1.143 & 1.188 & 10.899 & 0.463 & 21.33 & 20.79 & 18.07 & 18.08 & 23.25 & 22.88 & 21.34 & 21.43 & 3.86 & 4.41 & 5.89 & 7.09 \\ [2pt]
COSJ100028+021731 & S10 & 150.118246 & 2.292093 & 1.155 & 1.000 & 10.883 & 1.099 & 20.48 & 19.90 & 17.53 & 17.47 & 23.34 & 22.67 & 19.73 & 18.75 & 3.25 & 3.35 & 3.44 & 3.29 \\ [2pt]
COSJ100015+023621 & S09 & 150.063861 & 2.605837 & 0.690 & 0.702 & 10.729 & 1.563 & 19.65 & 19.18 & 17.07 & 17.76 & 24.45 & 24.37 & 21.99 & 21.51 & 3.27 & 3.84 & 3.00 & 3.25 \\ [2pt]
COSJ095957+023529 & S09 & 149.988569 & 2.591581 &  & 1.629 & 10.335 & 0.486 & 22.99 & 22.49 & 19.52 & 18.97 & 24.39 & 23.79 & 22.77 & 22.46 & 2.54 & 2.32 & 2.39 & 2.42 \\ [2pt]
COSJ095912+020750 & S09 & 149.801596 & 2.130823 &  & 0.330 & 10.350 & 0.905 & 18.68 & 18.26 & 16.40 & 17.05 & 24.00 & 23.33 & 21.83 & 21.69 & 3.72 & 3.66 & 3.77 & 4.17 \\ [2pt]
COSJ100122+022132 & S09 & 150.344407 & 2.358976 &  & 2.493 & 10.908 & 0.426 & 23.44 & 21.78 & 19.05 & 18.51 & 23.77 & 23.82 & 23.27 & 23.51 & 3.81 & 2.89 & 3.17 & 3.06 \\ [2pt]
COSJ100103+020159 & S09 & 150.263879 & 2.033091 & 1.441 & 1.221 & 10.835 & 0.726 & 20.98 & 20.47 & 18.21 & 18.14 & 23.37 & 22.67 & 22.16 & 21.36 & 3.11 & 3.37 & 3.15 & 3.17 \\ [2pt]
COSJ100133+022036 & S09 & 150.391497 & 2.343529 &  & 1.299 & 10.501 & 0.411 & 23.11 & 22.27 & 19.23 & 18.76 & 24.87 & 23.94 & 23.15 & 21.87 & 3.29 & 2.92 & 3.04 & 3.34 \\ [2pt]
COSJ100144+022826 & S09 & 150.434774 & 2.474108 & 1.148 & 1.261 & 11.005 & 0.220 & 20.92 & 20.30 & 17.62 & 17.45 & 25.19 & 24.41 & 21.29 & 20.85 & 1.42 & 1.57 & 1.83 & 1.92 \\ [2pt]
COSJ100018+023741 & S09 & 150.077182 & 2.628220 &  & 0.699 & 11.100 & 0.470 & 19.45 & 19.04 & 16.83 & 17.44 & 23.71 & 22.98 & 21.72 & 22.17 & 3.03 & 2.89 & 3.18 & 3.44 \\ [2pt]
COSJ100013+020223 & S09 & 150.054630 & 2.039860 & 0.677 & 0.699 & 9.766 & 0.231 & 21.78 & 21.66 & 19.74 & 20.38 & 24.06 & 24.27 & 22.34 & 22.89 & 2.43 & 2.97 & 4.48 & 4.24 \\ [2pt]
COSJ095917+021725 & S09 & 149.821914 & 2.290477 &  & 0.990 & 10.150 & 0.471 & 21.68 & 21.26 & 19.13 & 19.30 & 24.38 & 24.45 & 24.19 & 25.53 & 3.62 & 3.33 & 4.67 & 4.68 \\ [2pt]
COSJ100046+015029 & S09 & 150.195146 & 1.841506 & 0.813 & 0.804 & 10.505 & 0.517 & 20.87 & 20.29 & 18.04 & 18.39 & 22.94 & 23.28 & 21.50 & 22.33 & 5.24 & 5.58 & 7.62 & 7.60 \\ [2pt]
COSJ100106+015848 & S08 & 150.277169 & 1.980003 & 0.910 & 0.910 & 9.768 & 0.255 & 22.28 & 22.14 & 19.81 & 20.13 & 24.46 & 24.44 & 23.59 & 24.00 & 2.15 & 2.04 & 2.71 & 2.85 \\ [2pt]
COSJ100100+015653 & S08 & 150.253332 & 1.948107 & 1.024 & 1.010 & 11.101 & 0.589 &  &  & 16.62 & 16.75 &  &  & 21.47 & 19.73 &  &  & 2.58 & 2.66 \\ [2pt]
COSJ100132+022652 & S08 & 150.385677 & 2.447834 & 0.684 & 0.691 & 10.985 & 0.548 & 19.70 & 19.19 & 16.66 & 17.30 & 24.17 & 23.40 & 20.89 & 22.09 & 2.31 & 2.50 & 2.50 & 2.58 \\ [2pt]
COSJ100012+023421 & S08 & 150.050936 & 2.572700 & 0.619 & 0.610 & 11.101 & 0.459 & 18.87 & 18.36 & 16.27 & 16.97 & 23.10 & 22.44 & 20.82 & 21.22 & 2.05 & 2.36 & 2.73 & 2.60 \\ [2pt]
COSJ100042+020212 & S08 & 150.179078 & 2.036798 & 0.220 & 0.220 & 10.499 & 0.487 & 17.66 & 17.26 & 16.05 & 16.62 & 23.08 & 22.21 & 21.48 & 21.64 & 2.30 & 2.64 & 3.12 & 3.43 \\ [2pt]
COSJ100033+021339 & S08 & 150.139142 & 2.227557 &  & 1.619 & 10.454 & 0.585 & 22.97 & 22.32 & 19.65 & 19.36 & 26.11 & 24.30 & 23.63 & 23.53 & 6.67 & 5.57 & 5.68 & 6.24 \\ [2pt]
COSJ095946+021328 & S08 & 149.941930 & 2.224700 &  & 1.169 & 9.836 & 0.364 & 22.01 & 21.88 & 19.88 & 20.03 &  & 25.62 & 24.97 & 25.49 &  & 2.41 & 2.88 & 2.64 \\ [2pt]
COSJ100108+021907 & S08 & 150.285654 & 2.318835 &  & 1.345 & 9.845 & 0.267 & 22.58 & 22.15 & 20.23 & 20.19 & 24.39 & 24.08 & 23.86 & 23.00 & 1.92 & 2.15 & 1.97 & 2.02 \\ [2pt]
COSJ095939+023239 & S08 & 149.913891 & 2.544270 &  & 0.673 & 10.600 & 1.067 & 19.98 & 19.57 & 17.41 & 18.02 & 26.61 & 25.99 & 22.48 & 21.68 & 3.07 & 5.30 & 4.03 & 4.77 \\ [2pt]
COSJ095937+022744 & S08 & 149.905670 & 2.462283 & 0.415 & 0.404 & 9.198 & 0.330 & 21.94 & 21.75 & 20.04 & 20.60 & 23.94 & 23.78 & 22.73 & 23.94 & 3.02 & 3.05 & 3.33 & 3.08 \\ [2pt]
COSJ095923+021632 & S08 & 149.847381 & 2.275630 &  & 0.109 & 8.000 & 0.473 & 23.11 & 23.00 & 21.18 & 21.00 & 21.22 & 20.99 & 20.91 & 21.81 & 2.47 & 2.54 & 2.83 & 2.88 \\ [2pt]
COSJ100120+021646 & S08 & 150.336999 & 2.279514 & 0.373 & 0.332 & 10.900 & 0.597 & 17.18 & 16.74 & 14.92 & 15.65 & 22.98 & 23.27 & 20.68 & 20.50 & 1.72 & 1.87 & 2.27 & 2.26 \\ [2pt]
COSJ100151+022625 & S08 & 150.463597 & 2.440480 & 1.019 & 0.986 & 11.350 & 0.698 & 18.92 & 18.16 & 16.09 & 16.26 & 21.20 & 19.02 & 22.17 & 22.56 & 2.59 & 3.15 & 2.52 & 2.01 \\ [2pt]
COSJ095947+022657 & S08 & 149.948402 & 2.449396 & 0.723 & 0.727 & 10.399 & 0.326 & 20.81 & 20.50 & 18.34 & 18.80 & 22.68 & 22.54 & 20.01 & 21.36 & 3.48 & 3.59 & 4.39 & 4.49 \\ [2pt]
COSJ100114+023042 & S07 & 150.311267 & 2.511921 & 0.520 & 0.521 & 10.955 & 0.959 & 18.19 & 17.78 & 15.78 & 16.53 & 24.98 & 24.74 & 23.42 & 24.60 & 3.23 & 3.68 & 3.84 & 4.07 \\ [2pt]
COSJ095950+021636 & S07 & 149.959149 & 2.276917 & 0.933 & 0.926 & 11.100 & 1.501 & 19.42 & 19.01 & 16.77 & 17.02 & 25.45 & 25.78 & 23.43 & 23.59 & 5.10 & 6.39 & 7.69 & 6.76 \\ [2pt]
COSJ100103+014547 & S07 & 150.264049 & 1.763196 & 0.678 & 0.693 & 10.404 & 0.278 & 20.83 & 20.34 & 18.06 & 18.92 & 25.11 & 24.42 & 22.83 & 23.12 & 1.31 & 1.70 & 2.47 & 2.65 \\ [2pt]
COSJ100148+022724 & S07 & 150.451587 & 2.456779 & 0.224 & 0.210 & 9.456 & 0.381 & 19.37 & 19.15 & 17.79 & 18.32 & 22.69 & 21.97 & 21.30 & 20.34 & 1.68 & 1.67 & 1.66 & 1.68 \\ [2pt]
COSJ100041+022807 & S07 & 150.174477 & 2.468759 & 0.670 & 0.671 & 10.850 & 0.626 & 19.78 & 19.35 & 17.06 & 17.88 & 23.84 & 23.65 & 22.83 & 23.01 & 2.91 & 3.44 & 2.71 & 2.56 \\ [2pt]
COSJ100004+015307 & S07 & 150.020616 & 1.885320 & 0.338 & 0.331 & 11.070 & 0.814 &  & 16.95 &  &  &  & 21.21 &  &  &  & 2.58 &  &  \\ [2pt]
COSJ100105+020959 & S07 & 150.273591 & 2.166530 & 0.751 & 0.781 & 9.402 & 0.491 & 21.43 & 21.30 & 19.37 & 19.80 & 23.24 & 23.46 & 22.89 & 22.44 & 3.04 & 3.05 & 3.02 & 2.94 \\ [2pt]
COSJ100038+023226 & S07 & 150.159410 & 2.540809 & 0.491 & 0.518 & 9.941 & 0.174 & 21.04 & 20.87 & 18.97 & 19.58 & 22.86 & 22.70 & 22.38 & 22.20 & 1.62 & 1.57 & 1.73 & 1.61 \\ [2pt]
COSJ100202+021902 & S07 & 150.511003 & 2.317466 &  & 1.114 & 10.900 & 0.455 & 20.51 & 20.08 & 17.82 & 17.79 & 25.40 & 24.22 & 23.05 & 23.06 & 2.26 & 2.42 & 2.78 & 2.80 \\ [2pt]
COSJ100030+022026 & S07 & 150.128364 & 2.340789 &  & 1.009 & 10.307 & 0.823 & 21.08 & 20.61 & 18.47 & 18.59 & 23.41 & 23.96 & 22.82 & 23.92 & 1.89 & 2.29 & 2.64 & 2.14 \\ [2pt]
COSJ095951+022847 & S07 & 149.963086 & 2.479945 & 0.383 & 0.336 & 8.402 & 0.824 & 22.21 & 21.96 & 20.55 & 21.19 & 26.32 & 25.39 & 24.82 & 24.94 & 4.61 & 5.41 & 5.54 & 5.41 \\ [2pt]
COSJ095944+022008 & S07 & 149.934511 & 2.335714 & 0.700 & 0.689 & 9.854 & 0.799 & 20.90 & 20.68 & 18.80 & 19.41 & 22.84 & 22.88 & 21.16 & 21.49 & 2.50 & 2.60 & 2.45 & 2.42 \\ [2pt]
COSJ100134+022037 & S07 & 150.393082 & 2.343643 &  & 1.822 & 11.298 & 0.443 & 22.10 & 20.67 & 17.67 & 17.26 & 25.27 & 24.67 & 24.28 & 22.48 & 2.63 & 2.92 & 2.81 & 3.30 \\ [2pt]
COSJ100042+022746 & S07 & 150.177370 & 2.463002 &  & 1.670 & 10.304 & 0.612 & 22.47 & 24.15 & 19.61 & 19.27 & 26.74 &  & 23.65 & 23.70 & 1.96 &  & 2.42 & 2.47 \\ [2pt]
COSJ100120+015834 & S07 & 150.334823 & 1.976380 & 0.702 & 2.440 & 10.050 & 0.230 & 21.16 & 21.37 & 19.28 & 19.54 & 24.63 & 24.37 & 21.89 & 20.72 & 1.72 & 1.75 & 1.93 & 2.06 \\ [2pt]
COSJ100044+021508 & S07 & 150.183583 & 2.252395 & 0.581 & 0.486 & 10.027 & 0.525 & 20.78 & 20.17 & 17.81 & 18.28 & 22.62 & 21.44 & 18.72 & 17.91 & 2.97 & 3.05 & 2.88 & 2.84 \\ [2pt]
\end{tabular}
\end{adjustbox}
\caption{
The full catalogue of COWLS lens candidates. The columns are, from left to right: (i) the lens name; (ii) its score after the second round of visual inspection; (iii) right ascension; (iv) declination; (v) spectroscopic redshift, when available (see \cref{Redshift}); (vi) photometric redshift, when available (see \cref{Redshift}); (vii) stellar mass estimate, when available, in units of solar masses; (viii) Einstein radius in arcseconds; (ix)–(xii) lens galaxy magnitude estimates in the F115W, F150W, F277W, and F444W wavebands, respectively; (xiii)–(xvi) source galaxy magnitude estimates in the same wavebands; (xvii)–(xx) the estimated magnification of the candidate lens in the F115W, F150W, F277W, and F444W wavebands, respectively. Blank entries mean that information is not available; in the case of redshifts and stellar masses it means the necessary data does not exist in the COSMOS archive (see \cref{Redshift}), whereas when $R_{\rm Ein}$, magnitudes and magnifications are blank it means the lens model did not converge for that dataset.
}
\label{table:catalogue}
\vspace{-18pt} 
\end{table*}

\begin{table*}\ContinuedFloat
\begin{adjustbox}{max width=\textwidth}
\normalsize
\begin{tabular}{ l | l | l | l | l | l | l | l | l | l | l | l | l | l | l | l | l | l | l | l} 
\multicolumn{1}{p{1.8cm}|}{Lens Name} 
& \multicolumn{1}{p{0.7cm}|}{Score} 
& \multicolumn{1}{p{1.1cm}|}{RA [deg]} 
& \multicolumn{1}{p{1.1cm}|}{Dec [deg]}
& \multicolumn{1}{p{0.85cm}|}{$z_\text{spec}$} 
& \multicolumn{1}{p{0.85cm}|}{$z_\text{phot}$} 
& \multicolumn{1}{p{1.2cm}|}{$\log_{{10}}$ $(M_*/M_{{\odot}})$} 
& \multicolumn{1}{p{0.85cm}|}{$R_{\rm Ein}$  (\arcsec)}
& \multicolumn{1}{p{0.85cm}|}{Lens $m_{\mathrm{F115W}}$}
& \multicolumn{1}{p{0.85cm}|}{Lens $m_{\mathrm{F150W}}$}
& \multicolumn{1}{p{0.85cm}|}{Lens $m_{\mathrm{F277W}}$}
& \multicolumn{1}{p{0.85cm}|}{Lens $m_{\mathrm{F444W}}$}
& \multicolumn{1}{p{0.85cm}|}{Source $m_{\mathrm{F115W}}$}
& \multicolumn{1}{p{0.85cm}|}{Source $m_{\mathrm{F150W}}$}
& \multicolumn{1}{p{0.85cm}|}{Source $m_{\mathrm{F277W}}$}
& \multicolumn{1}{p{0.85cm}|}{Source $m_{\mathrm{F444W}}$}
& \multicolumn{1}{p{0.85cm}|}{$\mu_{\mathrm{F115W}}$}
& \multicolumn{1}{p{0.85cm}|}{$\mu_{\mathrm{F150W}}$}
& \multicolumn{1}{p{0.85cm}|}{$\mu_{\mathrm{F277W}}$}
& \multicolumn{1}{p{0.85cm}}{$\mu_{\mathrm{F444W}}$}
\\   \hline
& & & & & & & & & & & & & & & & & & & \\ [-6pt]
COSJ100022+023102 & S07 & 150.094884 & 2.517344 &  & 1.339 & 10.545 & 0.616 & 21.56 & 21.13 & 18.62 & 18.40 & 25.48 & 24.26 & 23.00 & 22.38 & 3.87 & 4.12 & 3.65 & 3.59 \\ [2pt]
COSJ100008+020758 & S07 & 150.035355 & 2.133043 &  & 0.832 & 9.476 & 0.697 &  & 22.06 & 19.98 & 20.64 &  & 25.88 & 24.58 & 24.60 &  & 5.13 & 6.30 & 6.05 \\ [2pt]
COSJ095934+020759 & S07 & 149.894333 & 2.133300 &  & 0.926 & 9.909 & 0.308 & 22.45 & 21.86 & 19.67 & 19.90 & 24.51 & 24.23 & 22.87 & 23.19 & 2.04 & 1.78 & 2.66 & 2.73 \\ [2pt]
COSJ100026+023013 & S07 & 150.110616 & 2.503795 & 1.400 & 1.451 & 10.752 & 0.649 & 21.23 & 20.59 & 17.90 & 17.84 &  & 25.69 & 23.83 & 21.99 &  & 2.33 & 2.84 & 2.89 \\ [2pt]
COSJ095940+023253 & S06 & 149.918705 & 2.548235 &  &  &  & 1.166 &  & 17.58 & 15.66 & 16.40 &  & 22.36 & 21.67 & 21.64 &  & 2.33 & 3.55 & 4.07 \\ [2pt]
COSJ100015+023708 & S06 & 150.066509 & 2.619129 & 0.936 & 0.937 & 9.574 & 0.439 & 21.86 & 21.75 & 19.58 & 19.64 & 23.69 & 23.46 & 22.24 & 22.80 & 2.64 & 2.60 & 2.42 & 2.71 \\ [2pt]
COSJ100050+022118 & S06 & 150.209564 & 2.355255 & 0.166 & 0.160 & 10.350 & 0.393 &  &  & 14.68 & 15.11 &  &  & 19.17 & 19.18 &  &  & 2.13 & 1.87 \\ [2pt]
COSJ100130+021921 & S06 & 150.376429 & 2.322744 &  & 0.716 & 9.688 & 0.307 & 21.61 & 21.39 & 19.68 & 19.18 & 23.82 & 24.29 & 20.29 & 21.64 & 3.02 & 2.82 & 3.25 & 3.60 \\ [2pt]
COSJ100151+022347 & S06 & 150.464650 & 2.396479 &  & 3.577 & 11.072 & 0.387 & 25.31 & 24.60 & 20.80 & 19.89 &  &  & 23.88 & 23.44 &  &  & 5.13 & 3.95 \\ [2pt]
COSJ095926+020304 & S06 & 149.862499 & 2.051265 & 1.326 & 1.314 & 10.197 & 0.331 & 21.99 & 21.65 & 19.62 & 19.53 & 23.80 & 23.32 & 21.79 & 21.33 & 2.53 & 2.83 & 3.04 & 3.14 \\ [2pt]
COSJ100120+022435 & S06 & 150.333943 & 2.409961 &  & 2.543 & 10.828 & 0.856 &  & 22.73 & 19.63 & 19.16 &  & 26.07 & 24.57 & 24.58 &  & 2.64 & 2.85 & 2.85 \\ [2pt]
COSJ100013+022949 & S06 & 150.054307 & 2.497079 &  & 1.812 & 10.106 & 0.329 & 23.20 & 22.70 & 20.40 & 20.08 & 26.07 & 24.92 & 23.38 & 23.06 & 2.61 & 2.78 & 5.94 & 5.08 \\ [2pt]
COSJ100002+015914 & S06 & 150.010652 & 1.987338 & 1.187 & 1.180 & 9.694 & 0.713 & 22.22 & 21.77 & 19.72 & 19.50 & 23.92 & 23.28 & 22.77 & 22.49 & 5.74 & 5.35 & 5.56 & 5.25 \\ [2pt]
COSJ095932+021750 & S06 & 149.885141 & 2.297500 &  & 0.893 & 10.743 & 0.861 & 20.07 & 19.69 & 17.47 & 17.81 &  & 26.33 & 25.06 & 24.49 &  & 2.99 & 4.76 & 4.39 \\ [2pt]
COSJ100023+021652 & S06 & 150.097987 & 2.281363 & 0.750 & 0.746 & 11.129 & 0.992 & 19.32 & 18.74 & 16.34 & 16.96 & 23.67 & 23.32 & 22.12 & 21.88 & 1.97 & 1.70 & 1.86 & 1.92 \\ [2pt]
COSJ100114+020211 & S06 & 150.311349 & 2.036662 & 0.339 & 0.331 & 10.149 & 0.327 & 19.15 & 18.83 & 17.05 & 17.74 & 22.41 & 21.96 & 20.54 & 20.72 & 1.67 & 1.85 & 1.96 & 2.20 \\ [2pt]
COSJ100046+021906 & S06 & 150.192514 & 2.318347 &  & 1.580 & 10.129 & 0.374 & 21.91 & 21.71 & 19.66 & 19.43 & 25.46 & 23.48 & 22.66 & 22.25 & 1.79 & 2.25 & 2.61 & 2.52 \\ [2pt]
COSJ100040+015213 & S06 & 150.169786 & 1.870477 & 0.930 & 0.945 & 10.950 & 0.680 & 20.41 & 19.82 & 17.29 & 17.50 & 23.25 & 22.21 & 21.52 & 21.40 & 5.02 & 4.49 & 4.49 & 4.00 \\ [2pt]
COSJ100140+020632 & S06 & 150.419986 & 2.108970 & 0.322 & 0.294 & 10.828 & 0.673 & 17.49 & 17.25 & 15.41 & 16.12 & 22.21 & 21.50 & 20.12 & 20.44 & 2.56 & 2.95 & 3.62 & 3.50 \\ [2pt]
COSJ100122+022144 & S06 & 150.343542 & 2.362431 &  & 1.684 & 10.009 & 0.329 & 22.87 & 22.29 & 19.79 & 19.35 & 23.85 & 23.38 & 22.98 & 22.55 & 1.99 & 2.16 & 2.63 & 2.90 \\ [2pt]
COSJ100025+022001 & S06 & 150.106193 & 2.333773 &  & 0.690 & 9.900 & 0.677 & 21.07 & 20.85 & 18.84 & 19.24 & 24.09 & 23.86 & 23.83 & 23.74 & 3.48 & 3.54 & 2.84 & 2.91 \\ [2pt]
COSJ095956+022539 & S06 & 149.987013 & 2.427728 & 0.694 & 0.707 & 10.049 & 0.350 & 21.62 & 21.06 & 18.89 & 19.52 & 24.89 & 24.57 & 23.34 & 23.70 & 2.19 & 1.81 & 4.53 & 4.55 \\ [2pt]
COSJ095942+023350 & S06 & 149.925885 & 2.564035 & 0.340 & 0.330 & 9.406 &  &  &  &  &  &  &  &  &  &  &  &  &  \\ [2pt]
COSJ095916+021850 & S06 & 149.820733 & 2.313949 & 1.138 & 1.153 & 9.496 & 0.275 & 22.66 & 22.42 & 20.58 & 20.70 & 23.54 & 23.49 & 23.89 & 24.17 & 2.84 & 2.45 & 3.56 & 3.51 \\ [2pt]
COSJ100151+022105 & S06 & 150.462928 & 2.351429 & 0.734 & 0.784 & 8.955 & 0.737 & 22.05 & 21.91 & 20.06 & 20.27 & 24.56 & 23.75 & 22.93 & 24.05 & 3.36 & 2.83 & 3.76 & 3.40 \\ [2pt]
COSJ100132+022807 & S06 & 150.387190 & 2.468770 & 0.340 & 0.340 & 10.353 & 0.374 & 18.57 & 18.19 & 16.59 & 17.37 & 22.12 & 21.62 & 20.57 & 23.08 & 1.83 & 1.91 & 1.87 & 1.86 \\ [2pt]
COSJ100119+015027 & S06 & 150.331903 & 1.840885 & 1.658 & 1.655 & 11.129 & 0.530 & 21.43 & 20.79 & 18.09 & 17.66 &  & 25.99 & 23.62 & 23.05 &  & 3.27 & 4.13 & 4.83 \\ [2pt]
COSJ095946+015320 & S06 & 149.942291 & 1.889142 &  & 1.054 & 10.117 & 0.210 & 22.21 & 21.82 & 19.37 & 19.35 & 25.12 & 24.20 & 22.70 & 23.66 & 1.31 & 1.71 & 1.63 & 2.21 \\ [2pt]
COSJ095945+023214 & S06 & 149.939255 & 2.537480 &  & 0.693 & 9.450 & 0.689 & 22.56 & 22.19 & 20.26 & 20.93 & 24.15 & 24.30 & 23.86 & 23.98 & 3.49 & 2.87 & 3.60 & 3.36 \\ [2pt]
COSJ095905+021026 & S06 & 149.771861 & 2.174120 & 1.240 & 0.996 & 9.244 & 0.647 & 22.20 & 21.76 & 19.58 & 19.43 & 25.76 & 24.99 & 24.43 & 23.87 & 3.92 & 2.31 & 5.70 & 3.95 \\ [2pt]
COSJ100205+021910 & S06 & 150.524561 & 2.319479 &  & 1.019 & 9.761 & 0.294 & 21.99 & 21.76 & 19.87 & 20.03 & 24.65 & 24.71 & 23.65 & 24.28 & 2.00 & 1.81 & 2.01 & 2.09 \\ [2pt]
COSJ100101+020814 & S06 & 150.257957 & 2.137432 & 0.639 & 0.674 & 10.752 & 0.772 & 19.35 & 18.94 & 16.89 & 17.51 & 26.33 & 25.35 & 24.70 & 24.17 & 2.47 & 3.02 & 3.89 & 3.37 \\ [2pt]
COSJ100016+022827 & S06 & 150.068306 & 2.474389 & 0.700 & 0.728 & 11.000 & 1.268 & 19.36 & 18.96 & 16.77 & 17.34 & 25.35 & 25.11 & 23.54 & 24.32 & 3.77 & 3.11 & 4.67 & 5.88 \\ [2pt]
COSJ095953+022126 & S06 & 149.973833 & 2.357376 &  & 1.540 & 10.948 & 1.565 & 21.62 & 20.87 & 18.51 & 18.25 & 25.85 & 24.26 & 21.92 & 22.11 & 9.45 & 8.49 & 9.84 & 9.09 \\ [2pt]
COSJ095941+020532 & S06 & 149.921702 & 2.092473 & 0.733 & 0.718 & 9.291 & 0.383 & 22.30 & 22.09 & 20.21 & 20.74 & 25.36 & 24.56 & 24.28 & 24.97 & 3.17 & 2.70 & 4.30 & 4.84 \\ [2pt]
COSJ095930+022026 & S06 & 149.877791 & 2.340688 & 0.847 & 0.873 & 9.877 & 0.454 & 21.55 & 21.37 & 19.37 & 19.76 & 23.47 & 23.70 & 22.95 & 23.86 & 3.94 & 6.03 & 6.51 & 6.21 \\ [2pt]
COSJ100158+022406 & S06 & 150.493900 & 2.401722 &  & 1.327 & 9.498 & 0.476 & 22.79 & 22.56 & 20.62 & 20.54 & 24.83 & 25.08 & 23.80 & 23.65 & 4.03 & 4.22 & 5.84 & 5.36 \\ [2pt]
COSJ100055+022059 & S06 & 150.232111 & 2.349762 & 1.269 & 1.302 & 9.254 & 0.457 & 22.51 & 22.31 & 20.55 & 20.48 & 24.63 & 23.65 & 24.12 & 25.32 & 1.66 & 1.68 & 1.72 & 1.63 \\ [2pt]
COSJ100037+020928 & S06 & 150.156104 & 2.157867 & 0.667 & 0.685 & 10.450 & 0.439 & 20.66 & 20.29 & 18.11 & 18.50 & 22.36 & 22.31 & 20.46 & 21.70 & 3.73 & 3.49 & 4.18 & 4.65 \\ [2pt]
COSJ100028+021600 & S06 & 150.117709 & 2.266758 & 0.750 & 0.744 & 11.094 & 0.771 & 19.13 & 18.58 & 16.44 & 16.94 & 22.29 & 22.29 & 20.20 & 20.57 & 6.18 & 6.64 & 6.91 & 7.08 \\ [2pt]
COSJ095852+020258 & S06 & 149.718750 & 2.049643 &  & 1.200 & 9.650 & 0.447 & 21.85 & 21.82 & 20.05 & 20.16 & 25.17 & 24.23 & 22.39 & 22.96 & 3.89 & 3.99 & 4.34 & 3.51 \\ [2pt]
COSJ095950+023727 & S05 & 149.961402 & 2.624443 & 0.750 & 0.788 & 9.296 & 0.395 & 21.99 & 21.93 & 20.13 & 20.63 & 27.14 & 24.89 & 24.06 & 24.19 & 2.69 & 1.96 & 2.71 & 2.42 \\ [2pt]
COSJ095928+022948 & S05 & 149.869224 & 2.496899 & 0.832 & 0.832 & 11.050 & 0.674 & 19.00 & 18.58 & 16.39 & 16.76 & 23.89 & 23.93 & 22.87 & 23.60 & 1.75 & 1.76 & 1.76 & 1.73 \\ [2pt]
COSJ100129+020455 & S05 & 150.372603 & 2.081955 & 1.191 & 1.191 & 10.415 & 0.505 & 21.90 & 21.58 & 19.05 & 18.94 & 25.34 & 24.65 & 23.56 & 23.87 & 2.47 & 2.83 & 3.90 & 5.54 \\ [2pt]
COSJ100104+015548 & S05 & 150.270676 & 1.930063 & 0.833 & 0.857 & 9.495 & 0.574 & 21.74 & 21.68 & 19.85 & 20.15 & 24.61 & 24.62 & 24.13 & 24.16 & 3.82 & 4.06 & 4.88 & 4.96 \\ [2pt]
COSJ095931+022946 & S05 & 149.880427 & 2.496315 &  & 1.590 & 10.700 & 0.495 & 22.11 & 21.30 & 18.51 & 18.43 & 25.45 & 24.67 & 24.67 & 24.40 & 2.31 & 2.19 & 2.17 & 2.51 \\ [2pt]
COSJ100131+022553 & S05 & 150.379659 & 2.431396 & 0.853 & 0.898 & 10.700 & 0.445 & 20.59 & 20.13 & 17.61 & 17.88 & 24.05 & 23.30 & 21.37 & 21.54 & 3.17 & 2.77 & 2.78 & 2.54 \\ [2pt]
COSJ100131+022056 & S05 & 150.380090 & 2.348997 & 0.863 & 0.883 & 9.514 & 0.482 & 22.51 & 22.36 & 20.02 & 20.42 & 25.31 & 25.02 &  &  & 3.89 & 3.25 &  &  \\ [2pt]
COSJ100131+015748 & S05 & 150.381308 & 1.963502 &  & 2.034 & 10.349 & 1.105 & 22.31 & 21.89 & 19.62 & 19.27 & 25.04 & 24.20 & 22.67 & 23.12 & 6.04 & 4.03 & 7.23 & 6.74 \\ [2pt]
COSJ100130+022214 & S05 & 150.375498 & 2.370832 & 0.215 & 0.210 & 10.625 & 0.000 & 17.50 & 17.12 & 15.84 & 16.51 & 22.33 & 22.61 & 21.75 & 21.24 & 1.10 & 1.10 & 1.14 & 1.27 \\ [2pt]
COSJ100004+022641 & S05 & 150.019115 & 2.444799 & 0.271 & 0.230 & 9.248 & 0.650 & 20.27 & 19.97 & 18.28 & 19.03 & 25.12 & 24.68 & 24.01 & 23.85 & 2.64 & 2.46 & 2.28 & 2.57 \\ [2pt]
COSJ095949+020321 & S05 & 149.956660 & 2.055960 & 0.661 & 0.656 & 9.388 & 0.644 & 21.61 & 21.49 & 19.68 & 20.11 & 24.91 & 25.16 & 23.37 & 24.09 & 2.89 & 3.78 & 4.78 & 4.54 \\ [2pt]
COSJ095858+020529 & S05 & 149.744014 & 2.091401 &  & 0.489 & 10.533 & 1.250 & 19.71 & 19.09 & 17.03 & 17.49 & 22.89 & 24.05 & 22.25 & 21.93 & 1.02 & 1.86 & 1.93 & 2.25 \\ [2pt]
COSJ100155+021506 & S05 & 150.480051 & 2.251751 &  & 1.389 & 9.621 & 0.385 & 23.20 & 22.44 & 20.04 & 19.94 & 23.37 & 22.69 & 22.16 & 22.22 & 2.42 & 2.49 & 2.33 & 2.48 \\ [2pt]
COSJ100114+022340 & S05 & 150.309886 & 2.394468 & 0.420 & 0.390 & 9.438 & 0.414 & 20.71 & 20.46 & 18.81 & 19.19 & 24.31 & 23.94 & 23.26 & 23.70 & 2.40 & 1.82 & 2.41 & 2.00 \\ [2pt]
COSJ100109+020148 & S05 & 150.291484 & 2.030071 &  & 1.053 & 9.836 & 0.710 & 21.87 & 21.65 & 19.79 & 19.84 & 25.41 & 24.90 & 24.48 & 24.88 & 3.19 & 6.28 & 6.68 & 5.98 \\ [2pt]
COSJ100046+020424 & S05 & 150.194840 & 2.073440 & 0.937 & 0.945 & 11.150 & 1.100 & 19.05 & 18.73 & 16.46 & 16.68 & 24.59 & 24.70 & 21.80 & 22.33 & 4.12 & 4.54 & 4.84 & 5.97 \\ [2pt]
COSJ100038+015132 & S05 & 150.158923 & 1.858997 & 0.546 & 0.530 & 9.908 & 0.414 & 20.84 & 20.63 & 18.61 & 19.36 & 22.81 & 22.58 & 20.96 & 22.17 & 3.23 & 3.49 & 4.59 & 4.76 \\ [2pt]
COSJ095927+022821 & S05 & 149.865918 & 2.472688 &  & 2.571 & 10.321 & 0.730 & 23.85 & 23.10 & 20.65 & 20.40 & 25.15 & 23.77 & 23.17 & 23.19 & 3.20 & 2.81 & 4.69 & 4.96 \\ [2pt]
COSJ095924+020459 & S05 & 149.850672 & 2.083105 &  & 1.386 & 11.300 & 0.927 & 20.31 & 19.71 & 17.31 & 17.08 & 24.54 & 24.19 & 22.64 & 21.75 & 4.24 & 5.35 & 6.86 & 6.68 \\ [2pt]
COSJ100117+022524 & S05 & 150.321635 & 2.423451 & 0.619 & 0.604 & 9.412 & 0.334 & 21.84 & 21.64 & 19.73 & 20.39 & 24.05 & 23.87 & 23.21 & 23.50 & 2.68 & 3.90 & 5.07 & 5.62 \\ [2pt]
COSJ100111+022230 & S05 & 150.297521 & 2.375047 &  &  &  & 0.563 &  & 19.74 & 17.48 & 17.86 &  & 23.22 & 21.62 & 22.79 &  & 1.75 & 1.81 & 1.95 \\ [2pt]
COSJ100110+020400 & S05 & 150.293704 & 2.066721 & 0.631 & 0.620 & 9.318 & 0.312 & 21.86 & 21.80 & 19.77 & 20.21 & 24.92 & 25.04 & 24.53 &  & 1.74 & 2.03 & 3.02 &  \\ [2pt]
COSJ100055+022022 & S05 & 150.229649 & 2.339559 &  &  &  & 0.383 &  & 20.34 & 18.21 & 18.51 &  & 23.93 & 23.75 & 23.91 &  & 1.83 & 2.15 & 2.31 \\ [2pt]
COSJ100038+023326 & S05 & 150.161899 & 2.557231 &  & 1.643 & 9.719 & 0.923 &  &  &  & 19.65 &  &  &  & 25.26 &  &  &  & 6.10 \\ [2pt]
COSJ100002+020851 & S05 & 150.011238 & 2.147548 & 1.007 & 0.996 & 9.843 & 0.302 & 22.10 & 21.81 & 19.36 & 19.00 & 24.24 & 23.83 & 23.71 & 23.60 & 2.32 & 2.62 & 2.43 & 3.21 \\ [2pt]
COSJ095946+022212 & S05 & 149.942909 & 2.370067 & 1.054 & 1.089 & 10.883 & 0.195 & 20.77 & 20.22 & 17.77 & 17.71 & 23.16 & 22.59 & 21.28 & 21.20 & 1.58 & 1.62 & 2.02 & 2.30 \\ [2pt]
COSJ095914+021358 & S05 & 149.809763 & 2.232818 & 1.078 & 1.051 & 10.650 & 0.464 & 21.21 & 20.72 & 18.31 & 18.39 & 25.72 & 25.33 & 23.47 & 23.55 & 3.64 & 2.28 & 4.72 & 3.64 \\ [2pt]
COSJ095910+021709 & S05 & 149.792802 & 2.285942 & 0.479 & 0.482 & 10.939 & 0.917 & 18.37 & 17.93 & 16.01 & 16.51 & 21.61 & 19.84 & 19.66 & 20.69 & 5.75 & 5.31 & 5.64 & 5.84 \\ [2pt]
COSJ095857+021007 & S05 & 149.740337 & 2.168803 & 0.491 & 0.482 & 8.797 & 0.241 & 22.20 & 22.16 & 20.34 & 21.03 & 24.98 & 24.15 & 24.24 &  & 1.62 & 1.65 & 2.73 &  \\ [2pt]
COSJ100011+023531 & S05 & 150.048603 & 2.592090 & 0.706 & 0.571 & 9.842 & 2.248 & 20.75 & 20.12 & 18.00 & 18.39 & 20.56 & 20.98 & 19.23 & 20.74 & 6.61 & 4.50 & 0.97 & 2.87 \\ [2pt]
COSJ100011+021301 & S05 & 150.049707 & 2.217117 &  &  &  & 1.864 &  &  &  & 15.20 &  &  &  & 18.32 &  &  &  & 4.00 \\ [2pt]
COSJ095952+020314 & S05 & 149.970607 & 2.054110 & 0.892 & 0.885 & 10.453 & 1.049 & 21.32 & 20.71 & 18.14 & 18.75 & 25.51 & 24.90 & 23.88 & 23.99 & 2.78 & 4.10 & 4.19 & 4.00 \\ [2pt]
COSJ100128+022502 & S05 & 150.370052 & 2.417308 & 0.230 & 0.210 & 9.633 & 1.331 & 19.02 & 18.84 & 17.42 & 17.91 & 24.48 & 24.05 & 23.48 & 23.30 & 3.86 & 4.66 & 3.71 & 3.33 \\ [2pt]
COSJ100124+022319 & S05 & 150.350407 & 2.388722 & 0.260 & 0.261 & 9.550 & 0.377 & 19.88 & 19.62 & 17.92 & 18.45 & 23.82 & 22.72 & 22.21 & 20.96 & 1.97 & 1.89 & 2.03 & 1.98 \\ [2pt]
COSJ100120+015011 & S05 & 150.337031 & 1.836610 &  & 2.099 & 10.926 & 0.827 & 22.67 & 22.05 & 19.12 & 19.01 & 24.13 & 24.99 & 23.43 & 23.54 & 4.75 & 4.61 & 5.65 & 4.38 \\ [2pt]
COSJ100043+022143 & S05 & 150.180878 & 2.362144 & 1.229 & 1.235 & 10.250 & 0.484 & 21.94 & 21.66 & 19.38 & 19.30 & 22.99 & 22.77 & 21.12 & 21.06 & 3.39 & 3.26 & 3.52 & 3.31 \\ [2pt]
COSJ095937+021802 & S05 & 149.908134 & 2.300610 & 1.603 & 1.601 & 10.245 & 0.564 & 22.65 & 22.06 & 19.63 & 19.18 & 23.99 & 23.38 & 21.81 & 21.91 & 3.80 & 3.82 & 5.41 & 5.22 \\ [2pt]
COSJ095910+020058 & S05 & 149.795472 & 2.016143 &  & 1.209 & 10.253 & 0.619 & 21.94 & 21.52 & 19.18 & 19.15 & 25.53 & 25.08 & 23.09 & 24.44 & 3.08 & 2.87 & 2.49 & 2.69 \\ [2pt]
COSJ100144+020921 & S05 & 150.433737 & 2.156104 & 0.346 & 0.352 & 9.564 & 0.871 & 20.68 & 20.32 & 18.67 & 19.38 &  & 25.40 & 23.90 & 24.40 &  & 2.74 & 3.22 & 2.72 \\ [2pt]
\end{tabular}
\end{adjustbox}
\caption{
Table \ref{table:catalogue} continued.
}
\label{table:catalogue2}
\end{table*}

\begin{table*}\ContinuedFloat
\begin{adjustbox}{max width=\textwidth}
\normalsize
\begin{tabular}{ l | l | l | l | l | l | l | l | l | l | l | l | l | l | l | l | l | l | l | l} 
\multicolumn{1}{p{1.8cm}|}{Lens Name} 
& \multicolumn{1}{p{0.7cm}|}{Score} 
& \multicolumn{1}{p{1.1cm}|}{RA [deg]} 
& \multicolumn{1}{p{1.1cm}|}{Dec [deg]}
& \multicolumn{1}{p{0.85cm}|}{$z_\text{spec}$} 
& \multicolumn{1}{p{0.85cm}|}{$z_\text{phot}$} 
& \multicolumn{1}{p{1.2cm}|}{$\log_{{10}}$ $(M_*/M_{{\odot}})$} 
& \multicolumn{1}{p{0.85cm}|}{$R_{\rm Ein}$  (\arcsec)}
& \multicolumn{1}{p{0.85cm}|}{Lens $m_{\mathrm{F115W}}$}
& \multicolumn{1}{p{0.85cm}|}{Lens $m_{\mathrm{F150W}}$}
& \multicolumn{1}{p{0.85cm}|}{Lens $m_{\mathrm{F277W}}$}
& \multicolumn{1}{p{0.85cm}|}{Lens $m_{\mathrm{F444W}}$}
& \multicolumn{1}{p{0.85cm}|}{Source $m_{\mathrm{F115W}}$}
& \multicolumn{1}{p{0.85cm}|}{Source $m_{\mathrm{F150W}}$}
& \multicolumn{1}{p{0.85cm}|}{Source $m_{\mathrm{F277W}}$}
& \multicolumn{1}{p{0.85cm}|}{Source $m_{\mathrm{F444W}}$}
& \multicolumn{1}{p{0.85cm}|}{$\mu_{\mathrm{F115W}}$}
& \multicolumn{1}{p{0.85cm}|}{$\mu_{\mathrm{F150W}}$}
& \multicolumn{1}{p{0.85cm}|}{$\mu_{\mathrm{F277W}}$}
& \multicolumn{1}{p{0.85cm}}{$\mu_{\mathrm{F444W}}$}
\\   \hline
& & & & & & & & & & & & & & & & & & & \\ [-6pt]
COSJ100057+021924 & S05 & 150.241171 & 2.323353 &  & 0.820 & 8.352 & 0.455 & 23.50 & 23.39 & 21.15 & 21.21 & 24.09 & 24.03 & 23.57 & 23.68 & 3.64 & 3.03 & 3.35 & 3.82 \\ [2pt]
COSJ100049+022656 & S05 & 150.207062 & 2.449048 & 1.419 & 1.354 & 11.200 & 0.326 & 21.93 & 20.63 & 17.62 & 17.12 & 23.95 & 23.45 & 21.02 & 20.34 & 1.45 & 1.54 & 1.97 & 2.43 \\ [2pt]
COSJ100113+022534 & S05 & 150.306232 & 2.426242 &  & 2.146 & 10.650 & 0.600 & 22.47 & 21.80 & 19.49 & 19.14 & 25.13 & 24.01 & 24.92 & 24.93 & 2.94 & 3.39 & 3.70 & 3.98 \\ [2pt]
COSJ100036+015439 & S05 & 150.152807 & 1.910859 &  & 1.408 & 9.794 & 0.433 & 22.31 & 22.18 & 20.09 & 20.04 & 24.25 & 23.88 & 24.10 & 23.42 & 4.12 & 4.01 & 4.19 & 3.84 \\ [2pt]
COSJ095951+023830 & S05 & 149.963237 & 2.641757 &  & 1.648 & 9.853 & 0.580 & 21.55 & 21.50 & 19.50 & 19.25 & 23.51 & 23.48 & 23.06 & 23.37 & 2.96 & 2.74 & 2.75 & 3.09 \\ [2pt]
COSJ100043+022356 & S04 & 150.180125 & 2.398930 &  &  &  & 0.591 &  & 23.45 & 20.62 & 20.23 &  & 24.73 & 22.41 & 22.24 &  & 1.71 & 1.75 & 1.81 \\ [2pt]
COSJ100033+022752 & S04 & 150.141599 & 2.464573 & 0.506 & 0.509 & 10.500 & 0.976 & 19.28 & 18.90 & 17.01 & 17.78 & 24.28 & 23.91 & 22.95 & 23.60 & 3.03 & 2.46 & 2.52 & 2.83 \\ [2pt]
COSJ100001+020942 & S04 & 150.006812 & 2.161927 &  & 1.188 & 11.050 & 0.616 & 21.23 & 20.54 & 17.97 & 17.82 & 25.93 & 24.54 & 22.52 & 22.71 & 2.74 & 2.06 & 2.53 & 2.42 \\ [2pt]
COSJ100015+022850 & S04 & 150.062843 & 2.480703 & 0.994 & 0.993 & 10.165 & 0.822 & 21.50 & 21.01 & 18.11 & 18.34 & 24.64 & 23.13 & 23.65 & 24.08 & 4.96 & 5.04 & 4.57 & 4.41 \\ [2pt]
COSJ100027+015014 & S04 & 150.113602 & 1.837452 & 0.574 & 0.534 & 9.150 & 0.493 & 21.73 & 21.54 & 19.79 & 20.53 & 23.85 & 23.34 & 23.18 & 23.35 & 2.75 & 3.16 & 3.30 & 3.95 \\ [2pt]
COSJ095924+020807 & S04 & 149.851192 & 2.135431 & 1.136 & 0.994 & 9.858 & 0.143 & 23.50 & 22.59 & 19.17 & 19.09 & 23.46 & 23.10 & 22.46 & 23.03 & 1.40 & 1.31 & 1.61 & 1.31 \\ [2pt]
COSJ095955+015522 & S04 & 149.980299 & 1.922941 & 1.156 & 1.151 & 9.510 & 0.372 & 22.22 & 22.01 & 19.66 & 20.25 & 24.16 & 23.97 & 24.28 & 23.86 & 2.23 & 2.25 & 3.28 & 2.63 \\ [2pt]
COSJ095953+022000 & S04 & 149.974165 & 2.333568 &  & 0.831 & 9.678 & 1.002 & 22.62 & 22.02 & 19.71 & 19.97 & 24.16 & 23.61 & 22.76 & 22.72 & 3.64 & 3.00 & 4.94 & 5.11 \\ [2pt]
COSJ100113+020313 & S04 & 150.306889 & 2.053826 &  & 1.690 & 10.001 & 0.233 & 22.08 & 21.41 & 18.26 & 17.82 & 24.27 & 23.27 & 22.14 & 23.28 & 1.71 & 1.85 & 2.49 & 2.42 \\ [2pt]
COSJ100038+022455 & S04 & 150.158351 & 2.415390 &  & 0.980 & 10.404 & 0.963 & 20.86 & 20.29 & 17.61 & 17.20 & 25.11 & 25.31 & 23.01 & 23.31 & 2.11 & 1.95 & 2.10 & 2.24 \\ [2pt]
COSJ100021+022333 & S04 & 150.089337 & 2.392717 &  &  &  & 0.695 &  & 22.12 & 17.87 &  &  & 24.42 &  &  &  & 1.28 &  &  \\ [2pt]
COSJ095953+022159 & S04 & 149.971582 & 2.366628 &  & 0.206 & 9.362 & 0.814 & 19.81 & 19.55 & 18.05 & 18.55 & 24.55 & 24.09 & 22.64 & 23.01 & 4.53 & 4.51 & 5.05 & 5.25 \\ [2pt]
COSJ100017+020443 & S04 & 150.073298 & 2.078722 & 0.948 & 0.948 & 9.417 & 0.412 & 22.50 & 22.27 & 20.44 & 20.60 & 25.67 & 25.55 & 24.50 &  & 2.14 & 2.04 & 4.51 &  \\ [2pt]
COSJ095933+022242 & S04 & 149.887670 & 2.378363 &  & 0.693 & 9.799 & 1.688 & 21.60 & 21.23 & 19.19 & 19.83 & 26.46 & 24.72 & 25.06 & 24.98 & 8.44 & 8.44 & 9.79 & 11.45 \\ [2pt]
COSJ095917+021629 & S04 & 149.823854 & 2.274896 & 1.180 & 1.100 & 9.854 & 0.636 & 22.79 & 22.15 & 20.14 & 20.16 & 25.03 & 24.98 & 23.58 & 23.46 & 3.45 & 4.93 & 3.60 & 3.49 \\ [2pt]
COSJ100020+023047 & S04 & 150.085188 & 2.513140 &  & 1.401 & 9.956 & 0.396 & 22.09 & 21.81 & 19.91 & 19.83 & 25.26 & 24.54 & 24.53 & 23.70 & 2.90 & 2.58 & 2.75 & 2.63 \\ [2pt]
COSJ095954+022203 & S04 & 149.976598 & 2.367605 & 0.587 & 0.596 & 9.239 & 0.317 & 21.31 & 21.25 & 19.48 & 20.03 & 23.80 & 23.32 & 22.88 & 22.85 & 2.27 & 2.47 & 2.45 & 2.60 \\ [2pt]
COSJ095950+023226 & S04 & 149.961913 & 2.540768 &  & 1.698 & 10.800 & 0.089 & 21.74 & 21.00 & 18.73 & 18.41 & 22.56 & 22.62 & 22.32 & 22.25 & 1.09 & 1.16 & 1.20 & 1.20 \\ [2pt]
COSJ100110+023301 & S04 & 150.292935 & 2.550323 &  & 1.356 & 10.006 & 0.491 & 21.92 & 21.69 & 19.80 & 19.68 &  & 25.39 & 23.95 & 24.58 &  & 3.28 & 6.03 & 2.84 \\ [2pt]
COSJ095951+020423 & S04 & 149.964287 & 2.073058 &  &  &  & 0.400 &  & 20.19 & 18.19 & 18.73 &  & 23.47 & 21.32 & 21.05 &  & 1.89 & 2.12 & 2.14 \\ [2pt]
COSJ100135+022033 & S04 & 150.397352 & 2.342531 &  & 1.806 & 10.800 & 1.248 & 22.70 & 21.71 & 19.16 & 18.74 & 24.25 & 23.34 & 21.91 & 22.84 & 2.74 & 2.97 & 2.79 & 2.94 \\ [2pt]
COSJ100022+021312 & S04 & 150.095620 & 2.220134 & 0.186 & 0.156 & 10.095 & 0.187 & 17.07 & 16.83 & 15.55 & 15.63 & 19.17 & 18.72 & 17.87 & 18.24 & 1.72 & 1.68 & 1.85 & 1.89 \\ [2pt]
COSJ100132+020533 & S04 & 150.385462 & 2.092557 & 0.902 & 0.895 & 11.000 & 0.534 & 19.10 & 18.74 & 16.48 & 16.74 & 24.95 & 24.38 & 21.53 & 21.92 & 1.81 & 1.69 & 1.72 & 1.80 \\ [2pt]
COSJ100028+015805 & S04 & 150.118316 & 1.968167 & 1.463 & 1.662 & 11.178 & 0.675 & 21.25 & 20.05 & 17.55 & 17.23 & 23.23 & 22.51 & 21.77 & 21.65 & 2.20 & 2.45 & 2.34 & 2.36 \\ [2pt]
COSJ100020+021308 & S04 & 150.085037 & 2.219133 &  & 0.929 & 9.563 &  &  &  &  &  &  &  &  &  &  &  &  &  \\ [2pt]
COSJ095856+015821 & S04 & 149.735333 & 1.972660 & 1.055 & 1.169 & 9.179 & 0.785 & 22.34 & 22.01 & 19.90 & 20.04 & 24.77 & 24.39 & 23.75 & 24.83 & 4.64 & 4.44 & 4.86 & 4.38 \\ [2pt]
COSJ100138+023003 & S04 & 150.412229 & 2.500934 &  & 2.176 & 9.687 & 0.498 & 23.24 & 23.04 & 20.77 & 20.18 &  & 26.89 & 23.72 & 23.05 &  & 2.46 & 2.63 & 2.56 \\ [2pt]
COSJ100121+020944 & S04 & 150.338211 & 2.162236 &  &  &  & 0.402 &  & 18.63 & 16.73 & 17.14 &  & 21.22 & 18.76 & 18.94 &  & 2.73 & 3.07 & 3.18 \\ [2pt]
COSJ100101+021108 & S04 & 150.254849 & 2.185816 &  &  &  & 0.698 &  & 19.90 & 17.36 & 17.13 &  & 21.82 & 20.02 & 20.94 &  & 1.96 & 1.87 & 1.97 \\ [2pt]
COSJ100002+022710 & S04 & 150.010592 & 2.452880 & 0.726 & 0.724 & 10.328 & 1.218 & 20.50 & 20.06 & 18.00 & 18.50 & 25.60 & 25.02 & 23.22 & 24.50 & 3.86 & 4.01 & 5.23 & 5.75 \\ [2pt]
COSJ095939+023043 & S04 & 149.913198 & 2.512220 &  &  &  & 1.983 &  & 17.90 & 15.78 & 16.35 &  & 24.40 & 21.81 & 23.84 &  & 5.26 & 4.17 & 3.88 \\ [2pt]
COSJ095926+021023 & S04 & 149.861822 & 2.173086 & 0.884 & 0.899 & 11.206 & 0.163 & 19.09 & 18.80 & 16.18 & 16.28 & 22.47 & 22.33 & 19.91 & 20.58 & 1.37 & 1.26 & 1.32 & 1.36 \\ [2pt]
COSJ100206+022349 & S04 & 150.528728 & 2.397060 &  & 1.635 & 10.387 & 0.688 & 24.04 & 22.72 & 19.91 & 19.26 & 25.97 & 25.93 & 25.18 &  & 2.89 & 3.87 & 3.68 &  \\ [2pt]
COSJ100143+021020 & S04 & 150.433136 & 2.172319 &  & 1.733 & 9.954 & 0.365 & 23.65 & 23.00 & 20.60 & 20.49 & 24.62 & 24.25 & 23.20 & 23.35 & 2.16 & 2.24 & 2.24 & 2.25 \\ [2pt]
COSJ100123+015608 & S04 & 150.345953 & 1.935636 &  & 1.522 & 9.799 & 0.476 & 22.18 & 22.00 & 20.29 & 20.20 & 23.93 & 23.58 & 23.73 & 23.65 & 3.48 & 3.28 & 3.46 & 3.42 \\ [2pt]
COSJ100106+015025 & S04 & 150.278472 & 1.840333 &  & 0.345 & 10.494 & 0.582 & 18.58 & 18.31 & 16.67 & 17.27 & 22.42 & 22.29 & 20.90 & 22.42 & 2.56 & 2.56 & 2.63 & 2.55 \\ [2pt]
COSJ100101+021819 & S04 & 150.256601 & 2.305431 &  & 2.075 & 10.037 & 0.418 & 22.75 & 22.46 & 20.38 & 20.07 & 25.15 & 24.64 & 23.91 & 24.57 & 1.83 & 2.39 & 2.95 & 3.11 \\ [2pt]
COSJ100057+020245 & S04 & 150.240982 & 2.045926 & 1.273 & 1.206 & 9.801 & 0.417 & 22.51 & 22.26 & 20.23 & 20.21 & 24.46 & 24.31 & 23.99 & 23.90 & 3.92 & 3.74 & 4.71 & 5.04 \\ [2pt]
COSJ100054+014936 & S04 & 150.227839 & 1.826850 &  & 0.180 & 7.600 & 0.138 & 21.68 & 21.18 & 19.59 & 18.86 & 25.74 & 24.87 & 22.86 & 23.36 & 1.46 & 1.41 & 1.69 & 1.50 \\ [2pt]
COSJ100036+020354 & S04 & 150.150024 & 2.065182 & 1.189 & 1.190 & 9.671 & 0.457 & 22.30 & 22.15 & 20.27 & 20.26 & 24.45 & 24.44 & 23.98 & 23.74 & 2.65 & 3.15 & 3.96 & 4.07 \\ [2pt]
COSJ100025+023225 & S04 & 150.106859 & 2.540447 & 0.667 & 0.686 & 10.566 & 0.624 & 20.46 & 19.97 & 17.63 & 18.21 & 23.95 & 23.58 & 23.19 & 23.31 & 3.35 & 2.88 & 3.42 & 2.28 \\ [2pt]
COSJ095959+023537 & S04 & 149.999957 & 2.593876 &  & 1.498 & 10.124 & 0.418 & 22.25 & 21.80 & 19.35 & 19.08 & 24.71 & 24.20 & 23.55 & 23.19 & 3.85 & 3.91 & 5.19 & 5.02 \\ [2pt]
COSJ095953+022016 & S04 & 149.973134 & 2.338000 & 0.945 & 0.946 & 11.152 & 0.461 & 19.30 & 18.76 & 16.45 & 16.59 & 23.84 & 23.68 & 22.20 & 21.97 & 1.86 & 1.89 & 1.83 & 1.81 \\ [2pt]
COSJ095943+022829 & S04 & 149.929907 & 2.474917 &  &  &  &  &  &  &  &  &  &  &  &  &  &  &  &  \\ [2pt]
COSJ095918+021939 & S04 & 149.827582 & 2.327748 &  & 2.753 & 11.150 & 0.504 & 22.62 & 21.58 & 19.14 & 18.48 & 24.21 & 23.74 & 18.54 & 21.45 & 1.82 & 2.62 & 2.14 & 2.54 \\ [2pt]
COSJ095847+015837 & S04 & 149.696298 & 1.977037 &  & 1.426 & 9.955 & 0.215 & 22.47 & 22.14 & 20.12 & 20.01 & 24.78 & 23.96 & 23.59 & 23.75 & 1.86 & 1.85 & 1.93 & 2.06 \\ [2pt]
COSJ100116+022354 & S04 & 150.320014 & 2.398582 & 1.725 & 1.774 & 11.200 & 0.729 & 20.50 & 19.91 & 17.68 & 17.49 &  & 25.39 & 24.50 & 24.04 &  & 2.92 & 4.82 & 3.95 \\ [2pt]
COSJ100011+023501 & S04 & 150.049305 & 2.583731 &  & 2.042 & 10.086 & 0.235 & 23.61 & 22.89 & 20.31 & 20.30 &  & 24.46 &  & 24.65 &  & 1.62 &  & 1.82 \\ [2pt]
COSJ100120+015538 & S04 & 150.335202 & 1.927250 &  & 1.339 & 9.668 & 0.270 & 22.29 & 22.41 & 20.53 & 20.55 & 24.57 & 24.35 & 23.61 & 23.84 & 1.92 & 1.90 & 2.62 & 2.57 \\ [2pt]
COSJ095853+020707 & S04 & 149.721324 & 2.118717 & 0.843 & 0.908 & 10.454 & 0.000 & 22.20 & 21.73 & 18.48 & 18.84 & 22.29 & 21.16 & 19.22 & 20.13 & 1.85 & 2.08 & 2.31 & 2.33 \\ [2pt]
COSJ100121+022213 & S04 & 150.338729 & 2.370360 &  & 1.615 & 9.889 & 0.277 & 23.16 & 22.71 & 20.66 & 20.53 & 25.18 & 24.08 & 23.82 & 23.73 & 3.01 & 3.29 & 3.89 & 3.97 \\ [2pt]
COSJ100038+014850 & S04 & 150.159129 & 1.814150 &  & 1.391 & 10.619 & 0.398 & 21.60 & 20.92 & 18.39 & 18.02 & 23.89 & 23.56 & 22.30 & 22.56 & 3.19 & 3.52 & 3.92 & 4.13 \\ [2pt]
COSJ095951+021800 & S04 & 149.963021 & 2.300163 & 0.370 & 0.341 & 9.051 & 0.457 & 21.15 & 20.97 & 19.17 & 19.47 & 23.79 & 23.32 & 22.88 & 23.88 & 2.09 & 2.13 & 2.41 & 2.72 \\ [2pt]
COSJ100124+022027 & S04 & 150.351516 & 2.341058 &  & 2.104 & 10.231 & 0.616 & 23.21 & 22.86 & 20.61 & 20.36 &  & 25.96 & 25.02 & 24.91 &  & 6.62 & 11.30 & 8.37 \\ [2pt]
COSJ100013+022023 & S04 & 150.054497 & 2.339809 & 0.378 & 0.331 & 10.850 & 0.808 & 18.32 & 17.87 & 15.99 & 16.70 & 21.79 & 21.20 & 19.11 & 18.74 & 3.38 & 3.80 & 3.39 & 3.84 \\ [2pt]
COSJ100112+023057 & S04 & 150.302111 & 2.515906 & 0.230 & 0.209 & 8.900 & 0.652 & 20.39 & 20.26 & 18.93 & 19.50 & 25.20 & 24.70 & 24.80 & 24.33 & 3.15 & 2.71 & 2.95 & 2.29 \\ [2pt]
COSJ100053+020042 & S04 & 150.222520 & 2.011675 & 0.958 & 0.946 & 11.149 & 0.646 & 19.57 & 18.98 & 16.48 & 16.55 & 23.16 & 23.03 & 20.90 & 21.27 & 2.98 & 3.08 & 2.97 & 2.92 \\ [2pt]
COSJ100128+021853 & S04 & 150.370272 & 2.314920 &  & 1.350 & 9.621 & 0.482 & 22.55 & 22.14 & 20.07 & 19.75 & 26.17 & 25.19 & 23.43 & 23.82 & 2.85 & 2.33 & 3.89 & 2.92 \\ [2pt]
COSJ100130+021936 & S03 & 150.378633 & 2.326751 & 0.661 & 0.621 & 10.550 & 0.642 & 19.75 & 19.18 & 16.72 & 17.12 & 23.75 & 23.65 & 22.25 & 22.62 & 1.58 & 2.37 & 2.12 & 1.90 \\ [2pt]
COSJ100054+015225 & S03 & 150.227194 & 1.873627 & 0.440 & 0.443 & 10.249 & 0.264 & 20.15 & 19.67 & 17.73 & 18.39 & 23.52 & 23.08 & 21.87 & 22.19 & 1.93 & 1.72 & 2.43 & 2.40 \\ [2pt]
COSJ100000+021201 & S03 & 150.002442 & 2.200468 & 0.797 & 0.843 & 10.574 & 1.144 & 20.77 & 20.26 & 17.73 & 18.17 & 24.71 & 24.41 & 22.60 & 23.40 & 2.46 & 2.64 & 2.58 & 2.58 \\ [2pt]
COSJ095938+022520 & S03 & 149.909732 & 2.422317 & 0.334 & 0.341 & 8.851 & 0.885 & 21.79 & 21.67 & 19.98 & 20.74 & 25.41 & 25.42 &  &  & 4.16 & 3.19 &  &  \\ [2pt]
COSJ100103+021610 & S03 & 150.265336 & 2.269616 & 0.408 & 0.420 & 10.003 & 1.435 & 20.26 & 19.77 & 17.71 & 18.18 & 23.25 & 22.98 & 21.22 & 20.88 & 2.90 & 3.02 & 3.07 & 3.02 \\ [2pt]
COSJ095942+023423 & S03 & 149.928386 & 2.573265 &  & 1.652 & 10.755 & 0.352 & 22.18 & 21.47 & 18.76 & 18.34 & 24.58 & 24.24 & 23.62 & 23.29 & 2.05 & 1.98 & 2.28 & 2.19 \\ [2pt]
COSJ100031+022615 & S03 & 150.130084 & 2.437658 & 0.265 & 0.230 & 10.608 & 0.562 & 17.26 & 17.02 & 15.54 & 16.19 & 22.25 & 22.31 & 21.17 & 21.21 & 2.58 & 2.36 & 2.29 & 2.78 \\ [2pt]
COSJ100039+020305 & S03 & 150.165332 & 2.051400 &  & 1.292 & 10.337 & 0.441 & 22.34 & 21.77 & 19.54 & 19.34 & 23.88 & 23.45 & 22.97 & 22.64 & 4.66 & 4.55 & 5.61 & 4.99 \\ [2pt]
COSJ100000+022042 & S03 & 150.000622 & 2.345062 & 0.218 & 0.219 & 8.865 & 0.681 & 21.43 & 21.37 & 19.77 & 20.23 & 24.92 & 24.57 & 23.59 & 24.60 & 3.90 & 3.62 & 3.56 & 3.39 \\ [2pt]
COSJ095955+022901 & S03 & 149.981820 & 2.483833 & 0.407 & 0.337 & 9.471 & 0.065 & 21.29 & 21.00 & 18.88 & 19.36 & 23.23 & 22.80 & 20.93 & 23.19 & 1.20 & 1.25 & 1.23 & 1.20 \\ [2pt]
COSJ095955+015520 & S03 & 149.980525 & 1.922315 &  & 1.194 & 9.781 & 0.423 & 22.74 & 22.31 & 19.99 & 19.87 & 25.22 & 23.37 & 23.85 & 23.44 & 2.52 & 2.21 & 2.24 & 2.26 \\ [2pt]
COSJ095944+021432 & S03 & 149.934447 & 2.242458 &  & 0.799 & 10.383 & 0.442 & 20.76 & 20.30 & 17.94 & 18.28 & 23.10 & 23.09 & 20.63 & 20.48 & 3.34 & 3.11 & 3.48 & 3.09 \\ [2pt]
COSJ095941+021442 & S03 & 149.922904 & 2.245162 &  & 1.644 & 10.500 & 0.634 & 22.25 & 21.86 & 19.23 & 19.18 & 23.90 & 23.83 & 23.39 & 22.18 & 3.50 & 3.33 & 3.67 & 2.92 \\ [2pt]
COSJ100119+014658 & S03 & 150.329637 & 1.782938 & 1.066 & 0.995 & 9.647 & 0.593 & 22.36 & 22.03 & 20.03 & 20.15 & 24.76 & 24.32 & 25.11 & 24.70 & 2.90 & 4.30 & 3.88 & 4.38 \\ [2pt]
COSJ100115+020015 & S03 & 150.312657 & 2.004201 & 0.319 & 0.305 & 9.700 & 0.579 & 20.27 & 19.81 & 18.19 & 18.77 & 24.32 & 24.37 & 22.75 & 24.56 & 2.87 & 2.48 & 2.82 & 2.71 \\ [2pt]
COSJ100106+014743 & S03 & 150.278152 & 1.795430 & 0.426 & 0.407 & 9.333 & 0.408 & 20.95 & 20.71 & 18.88 & 19.42 & 25.03 & 24.59 & 24.46 & 24.61 & 2.17 & 2.56 & 3.46 & 3.85 \\ [2pt]
COSJ100028+020024 & S03 & 150.119278 & 2.006790 & 1.466 & 1.439 & 10.038 & 0.348 & 22.36 & 22.07 & 20.04 & 19.21 & 23.74 & 23.65 & 23.38 & 22.77 & 1.91 & 2.25 & 2.22 & 2.05 \\ [2pt]
COSJ100002+015753 & S03 & 150.012474 & 1.964970 &  & 0.995 & 10.412 & 0.442 & 22.32 & 21.55 & 18.75 & 18.48 & 24.19 & 23.23 & 20.93 & 20.73 & 3.21 & 3.30 & 3.49 & 3.83 \\ [2pt]
COSJ095952+021514 & S03 & 149.969886 & 2.254161 & 0.846 & 0.860 & 9.634 & 0.555 & 21.93 & 21.67 & 19.79 & 20.18 & 22.83 & 23.37 & 22.15 & 23.48 & 5.63 & 5.81 & 6.25 & 7.51 \\ [2pt]
\end{tabular}
\end{adjustbox}
\caption{
Table \ref{table:catalogue} continued.
}
\label{table:catalogue3}
\vspace{-12pt} 
\end{table*}

\begin{table*}\ContinuedFloat
\begin{adjustbox}{max width=\textwidth}
\normalsize
\begin{tabular}{ l | l | l | l | l | l | l | l | l | l | l | l | l | l | l | l | l | l | l | l} 
\multicolumn{1}{p{1.8cm}|}{Lens Name} 
& \multicolumn{1}{p{0.7cm}|}{Score} 
& \multicolumn{1}{p{1.1cm}|}{RA [deg]} 
& \multicolumn{1}{p{1.1cm}|}{Dec [deg]}
& \multicolumn{1}{p{0.85cm}|}{$z_\text{spec}$} 
& \multicolumn{1}{p{0.85cm}|}{$z_\text{phot}$} 
& \multicolumn{1}{p{1.2cm}|}{$\log_{{10}}$ $(M_*/M_{{\odot}})$} 
& \multicolumn{1}{p{0.85cm}|}{$R_{\rm Ein}$  (\arcsec)}
& \multicolumn{1}{p{0.85cm}|}{Lens $m_{\mathrm{F115W}}$}
& \multicolumn{1}{p{0.85cm}|}{Lens $m_{\mathrm{F150W}}$}
& \multicolumn{1}{p{0.85cm}|}{Lens $m_{\mathrm{F277W}}$}
& \multicolumn{1}{p{0.85cm}|}{Lens $m_{\mathrm{F444W}}$}
& \multicolumn{1}{p{0.85cm}|}{Source $m_{\mathrm{F115W}}$}
& \multicolumn{1}{p{0.85cm}|}{Source $m_{\mathrm{F150W}}$}
& \multicolumn{1}{p{0.85cm}|}{Source $m_{\mathrm{F277W}}$}
& \multicolumn{1}{p{0.85cm}|}{Source $m_{\mathrm{F444W}}$}
& \multicolumn{1}{p{0.85cm}|}{$\mu_{\mathrm{F115W}}$}
& \multicolumn{1}{p{0.85cm}|}{$\mu_{\mathrm{F150W}}$}
& \multicolumn{1}{p{0.85cm}|}{$\mu_{\mathrm{F277W}}$}
& \multicolumn{1}{p{0.85cm}}{$\mu_{\mathrm{F444W}}$}
\\   \hline
& & & & & & & & & & & & & & & & & & & \\ [-6pt]
COSJ095943+023800 & S03 & 149.929479 & 2.633600 &  & 1.210 & 10.950 & 0.529 & 21.68 & 20.58 & 17.78 & 17.74 & 24.16 & 24.19 & 22.41 & 22.62 & 2.22 & 2.02 & 2.19 & 2.37 \\ [2pt]
COSJ095939+023707 & S03 & 149.913357 & 2.618778 &  & 0.950 & 9.273 & 0.417 & 22.62 & 22.75 & 20.76 & 20.80 &  & 25.70 & 25.98 &  &  & 4.21 & 7.93 &  \\ [2pt]
COSJ100107+022736 & S03 & 150.282213 & 2.460207 & 0.123 & 0.120 & 9.538 & 1.077 & 17.46 & 17.27 & 16.25 & 16.75 & 18.69 & 18.64 & 18.30 & 18.39 & 4.94 & 3.45 & 3.58 & 3.89 \\ [2pt]
COSJ100018+022034 & S03 & 150.078137 & 2.342983 &  & 1.841 & 10.550 & 0.594 & 22.71 & 21.90 & 19.45 & 19.00 & 23.16 & 22.81 & 22.97 & 23.97 & 1.95 & 2.13 & 3.08 & 2.96 \\ [2pt]
COSJ095936+015703 & S03 & 149.903097 & 1.951014 & 1.412 & 1.363 & 11.000 & 0.353 & 21.72 & 20.78 & 18.06 & 17.77 & 23.53 & 22.38 & 20.85 & 20.32 & 1.38 & 1.61 & 1.65 & 1.72 \\ [2pt]
COSJ100057+021443 & S03 & 150.239695 & 2.245484 &  & 0.930 & 10.800 & 0.329 & 18.30 & 18.25 & 17.39 & 17.56 & 21.81 & 22.38 & 21.02 & 22.83 & 1.78 & 1.92 & 1.85 & 1.48 \\ [2pt]
COSJ100109+020536 & S03 & 150.290905 & 2.093475 & 0.280 & 0.233 & 9.000 & 0.512 & 20.27 & 20.13 & 18.58 & 19.24 & 22.94 & 22.99 & 22.37 & 22.41 & 3.11 & 3.58 & 3.44 & 4.15 \\ [2pt]
COSJ095908+021435 & S03 & 149.785937 & 2.243176 & 0.348 & 0.298 & 8.700 & 0.551 & 21.52 & 21.31 & 19.92 & 20.38 & 24.96 & 24.86 & 24.41 &  & 2.78 & 2.43 & 2.32 &  \\ [2pt]
COSJ100115+023131 & S03 & 150.315840 & 2.525394 & 0.571 & 0.580 & 9.210 & 0.477 & 21.62 & 21.50 & 19.66 & 20.27 & 24.20 & 24.04 & 22.88 & 23.96 & 2.43 & 2.52 & 2.38 & 2.25 \\ [2pt]
COSJ100104+015032 & S03 & 150.270193 & 1.842269 & 0.603 & 0.593 & 10.000 & 0.607 & 20.69 & 20.15 & 17.63 & 18.08 & 23.26 & 23.16 & 22.68 & 22.89 & 3.05 & 3.12 & 3.08 & 3.19 \\ [2pt]
COSJ100047+015313 & S03 & 150.199438 & 1.887132 & 0.961 & 0.950 & 9.644 & 0.439 & 22.09 & 21.87 & 19.95 & 20.09 & 23.54 & 23.42 & 23.35 & 24.22 & 1.68 & 1.59 & 1.58 & 1.51 \\ [2pt]
COSJ100045+020601 & S03 & 150.188045 & 2.100533 & 0.930 & 0.932 & 9.516 & 0.589 & 22.31 & 22.15 & 20.25 & 20.34 & 26.25 & 26.34 & 24.35 & 25.33 & 5.98 & 4.13 & 6.33 & 5.53 \\ [2pt]
COSJ100044+022404 & S03 & 150.183750 & 2.401388 &  & 3.791 & 10.147 & 0.205 & 23.26 & 23.05 & 20.66 & 20.91 & 23.27 & 23.69 & 23.77 & 22.93 & 1.96 & 2.11 & 2.13 & 1.91 \\ [2pt]
COSJ100037+021117 & S03 & 150.155729 & 2.188254 & 1.232 & 1.170 & 9.973 & 0.523 & 22.18 & 21.88 & 19.82 & 19.62 & 24.03 & 23.86 & 22.90 & 23.55 & 4.60 & 5.75 & 7.66 & 6.69 \\ [2pt]
COSJ100034+015047 & S03 & 150.142051 & 1.846442 & 1.648 & 1.501 & 9.385 & 0.242 & 23.72 & 23.56 & 22.24 & 21.40 & 24.60 & 23.30 & 23.11 & 23.99 & 2.02 & 1.97 & 2.44 & 2.35 \\ [2pt]
COSJ095954+021900 & S03 & 149.977319 & 2.316730 &  & 3.138 & 10.598 & 0.404 & 24.65 & 24.48 & 21.00 & 20.40 & 24.45 & 23.36 & 22.27 & 22.82 & 1.54 & 1.67 & 1.88 & 1.91 \\ [2pt]
COSJ095946+015749 & S03 & 149.943094 & 1.963865 &  & 1.925 & 10.850 & 0.838 & 21.44 & 20.68 & 18.55 & 18.31 & 24.39 & 24.14 & 23.66 & 23.16 & 3.91 & 2.32 & 3.32 & 3.87 \\ [2pt]
COSJ095945+023622 & S03 & 149.937984 & 2.606272 & 0.345 & 0.340 & 11.199 & 0.402 & 16.87 & 16.50 & 14.93 & 15.74 & 22.05 & 21.43 & 21.58 & 22.86 & 1.49 & 1.66 & 1.74 & 1.52 \\ [2pt]
COSJ095942+015628 & S03 & 149.927184 & 1.941219 & 1.802 & 1.424 & 9.523 & 0.181 & 22.97 & 22.52 & 19.93 & 19.56 & 23.15 & 22.89 & 21.81 & 21.71 & 2.29 & 2.00 & 2.05 & 1.98 \\ [2pt]
COSJ095941+023306 & S03 & 149.924449 & 2.551688 & 0.844 & 0.861 & 9.419 & 0.432 & 22.18 & 22.01 & 20.23 & 20.52 & 25.72 & 25.08 & 24.19 & 24.35 & 1.89 & 1.72 & 2.18 & 2.27 \\ [2pt]
COSJ095924+021345 & S03 & 149.850428 & 2.229186 & 1.186 & 1.038 & 9.564 & 0.373 & 22.85 & 22.59 & 20.50 & 20.57 & 24.00 & 23.70 & 24.22 & 23.58 & 4.71 & 3.94 & 5.47 & 5.20 \\ [2pt]
COSJ095920+022515 & S03 & 149.836266 & 2.420999 & 0.728 & 0.740 & 10.500 &  &  &  &  &  &  &  &  &  &  &  &  &  \\ [2pt]
COSJ095912+020656 & S03 & 149.801539 & 2.115563 & 0.353 & 0.334 & 9.600 & 0.774 & 20.13 & 19.92 & 18.06 & 18.21 & 23.78 & 23.40 & 22.87 & 23.51 & 3.32 & 4.08 & 3.99 & 3.50 \\ [2pt]
COSJ100128+022544 & S03 & 150.369087 & 2.429054 &  & 2.533 & 10.302 & 0.412 & 23.44 & 23.52 & 20.48 & 19.58 &  & 26.59 & 24.15 & 23.22 &  & 3.78 & 4.77 & 3.71 \\ [2pt]
COSJ100128+021951 & S03 & 150.367951 & 2.331101 &  &  &  & 0.344 &  & 20.78 & 18.62 & 18.72 &  & 24.36 & 22.03 & 22.62 &  & 3.19 & 4.18 & 3.19 \\ [2pt]
COSJ100124+021640 & S03 & 150.352479 & 2.277845 & 0.625 & 0.610 & 10.296 & 0.320 & 20.35 & 19.98 & 17.86 & 18.63 & 23.41 & 23.00 & 22.89 & 22.70 & 2.18 & 2.15 & 1.86 & 1.85 \\ [2pt]
COSJ100050+014728 & S03 & 150.210379 & 1.791277 & 0.309 & 0.295 & 11.244 & 1.506 & 16.84 & 16.48 & 14.92 & 15.62 & 22.81 & 22.25 & 20.70 & 21.24 & 4.21 & 4.26 & 4.41 & 4.49 \\ [2pt]
COSJ100045+022738 & S03 & 150.189209 & 2.460605 &  & 2.182 & 11.151 & 0.938 & 22.43 & 21.21 & 18.46 & 17.96 & 25.44 & 25.90 & 23.77 & 24.16 & 1.56 & 2.12 & 3.24 & 2.77 \\ [2pt]
COSJ100033+015709 & S03 & 150.137699 & 1.952645 &  & 1.353 & 10.000 & 0.315 & 22.57 & 22.32 & 20.19 & 20.11 & 24.21 & 23.66 & 23.40 & 22.96 & 4.28 & 3.43 & 4.98 & 5.32 \\ [2pt]
COSJ100002+020932 & S03 & 150.011953 & 2.158962 & 0.389 & 0.391 & 10.464 & 0.888 & 19.00 & 18.64 & 17.00 & 17.59 & 24.28 & 23.99 & 22.17 & 23.77 & 3.20 & 3.17 & 3.17 & 3.02 \\ [2pt]
COSJ095959+023649 & S03 & 149.999152 & 2.613663 & 0.658 & 0.669 & 9.510 & 0.610 & 21.47 & 21.29 & 19.37 & 19.86 & 25.32 & 25.32 & 24.21 & 24.46 & 4.32 & 3.80 & 5.20 & 4.95 \\ [2pt]
COSJ095948+022017 & S03 & 149.950381 & 2.338282 &  & 1.322 & 10.700 & 0.483 & 21.57 & 21.17 & 18.74 & 18.53 & 24.96 & 24.13 & 23.45 & 22.99 & 3.11 & 3.33 & 3.32 & 3.30 \\ [2pt]
COSJ095942+022136 & S03 & 149.926915 & 2.360025 &  & 1.443 & 10.800 & 0.474 & 21.62 & 21.02 & 18.62 & 18.46 & 24.84 & 24.21 & 21.64 & 20.94 & 3.14 & 3.28 & 3.04 & 2.50 \\ [2pt]
COSJ095941+020259 & S03 & 149.924961 & 2.049744 & 1.067 & 1.071 & 10.324 & 0.249 & 21.49 & 20.91 & 18.84 & 18.89 & 22.02 & 22.32 & 21.88 & 21.31 & 2.14 & 1.97 & 2.16 & 2.07 \\ [2pt]
COSJ095938+022213 & S03 & 149.911917 & 2.370337 & 0.868 & 0.897 & 10.891 & 1.326 & 20.03 & 19.49 & 16.89 & 17.05 & 22.09 & 21.38 & 20.06 & 21.37 & 4.49 & 4.40 & 3.92 & 3.95 \\ [2pt]
COSJ100145+022949 & S03 & 150.441643 & 2.496974 &  & 3.095 & 10.800 & 1.726 & 23.94 & 22.76 & 19.85 & 19.49 & 27.01 & 25.88 & 24.59 & 24.89 & 4.54 & 5.04 & 5.18 & 5.38 \\ [2pt]
COSJ100112+022310 & S03 & 150.300059 & 2.386126 &  & 2.570 & 10.350 & 0.348 & 24.73 & 23.42 & 20.10 & 19.09 & 25.75 & 25.75 & 22.78 & 22.06 & 2.12 & 1.55 & 2.60 & 2.67 \\ [2pt]
COSJ100110+022157 & S03 & 150.292790 & 2.365948 &  & 1.916 & 10.199 & 0.336 & 22.80 & 22.29 & 20.11 & 19.75 & 24.32 & 24.24 & 23.81 & 23.71 & 2.88 & 3.19 & 3.06 & 3.26 \\ [2pt]
COSJ100028+021710 & S03 & 150.117388 & 2.286311 & 0.943 & 0.932 & 10.050 & 0.428 & 21.30 & 21.02 & 18.93 & 19.25 & 23.77 & 23.67 & 22.30 & 22.77 & 4.35 & 3.30 & 3.73 & 3.63 \\ [2pt]
COSJ100027+023137 & S03 & 150.116203 & 2.527036 & 0.859 & 0.871 & 10.431 & 0.617 &  & 20.36 & 18.23 & 18.54 &  & 25.84 & 24.61 & 24.34 &  & 2.87 & 2.91 & 2.93 \\ [2pt]
COSJ095943+022719 & S03 & 149.931072 & 2.455459 & 0.567 & 0.539 & 10.253 & 0.265 & 20.53 & 20.04 & 17.75 & 18.33 & 22.68 & 22.42 & 22.40 & 22.33 & 1.92 & 1.72 & 2.41 & 2.49 \\ [2pt]
COSJ095952+022216 & S03 & 149.967376 & 2.371269 &  & 1.045 & 9.929 & 1.072 & 22.19 & 21.56 & 19.64 & 19.77 & 26.67 & 26.47 & 24.25 & 24.69 & 7.90 & 5.97 & 7.90 & 8.32 \\ [2pt]
COSJ095908+020157 & S03 & 149.783335 & 2.032604 & 0.960 & 0.993 & 10.950 & 1.070 & 20.22 & 19.73 & 17.38 & 17.57 & 25.31 & 24.85 & 22.92 & 23.18 & 7.50 & 6.20 & 6.01 & 5.83 \\ [2pt]
COSJ100031+023300 & S02 & 150.130217 & 2.550243 & 0.249 & 0.248 & 10.865 & 0.936 & 17.03 & 16.73 & 15.32 & 16.01 & 24.57 & 23.89 & 22.16 & 22.74 & 1.76 & 1.85 & 2.08 & 1.74 \\ [2pt]
COSJ100021+023600 & S02 & 150.090661 & 2.600006 & 1.033 & 0.983 & 11.032 & 0.900 & 19.62 & 19.12 & 16.66 & 16.81 & 20.58 & 23.65 & 22.43 & 22.26 & 2.68 & 2.51 & 2.84 & 2.35 \\ [2pt]
COSJ095950+021936 & S02 & 149.959276 & 2.326929 &  &  &  & 0.718 &  & 22.06 & 19.65 & 19.81 &  & 24.96 & 24.12 & 23.81 &  & 5.13 & 3.27 & 3.21 \\ [2pt]
COSJ100138+023041 & S02 & 150.409660 & 2.511620 & 0.871 & 0.872 & 11.100 & 0.880 & 19.70 & 19.21 & 16.66 & 16.89 & 24.93 & 24.50 & 22.81 & 23.31 & 1.90 & 1.74 & 1.56 & 1.78 \\ [2pt]
COSJ100058+015400 & S02 & 150.245599 & 1.900155 &  &  &  & 0.231 & 23.63 & 22.27 & 19.95 & 19.25 & 23.37 & 22.70 & 21.56 & 21.13 & 2.27 & 2.06 & 2.19 & 1.74 \\ [2pt]
COSJ100038+023304 & S02 & 150.161421 & 2.551388 & 0.592 & 0.593 & 10.043 & 0.233 & 21.04 & 20.65 & 18.51 & 18.92 & 24.40 & 23.11 & 22.52 & 23.44 & 2.17 & 1.86 & 1.80 & 1.82 \\ [2pt]
COSJ100026+020345 & S02 & 150.109390 & 2.062568 & 0.319 & 0.310 & 9.450 & 0.230 & 19.83 & 19.63 & 17.93 & 18.26 & 21.51 & 21.61 & 20.36 & 20.97 & 2.17 & 2.13 & 2.31 & 2.37 \\ [2pt]
COSJ100146+022109 & S02 & 150.445770 & 2.352582 &  & 0.995 & 10.341 & 1.208 & 21.30 & 21.01 & 18.50 & 18.41 & 36.54 & 20.62 & 21.80 & 22.24 & 3.98 & 0.67 & 0.82 & 0.61 \\ [2pt]
COSJ100058+022721 & S02 & 150.244793 & 2.455928 & 0.910 & 0.909 & 8.919 &  &  &  &  &  &  &  &  &  &  &  &  &  \\ [2pt]
COSJ100046+014924 & S02 & 150.191795 & 1.823552 & 0.420 & 0.410 & 9.299 & 0.374 & 21.07 & 20.76 & 18.99 & 19.39 & 23.59 & 24.31 & 22.75 & 23.28 & 1.59 & 2.23 & 2.14 & 3.30 \\ [2pt]
COSJ100041+020145 & S02 & 150.172118 & 2.029267 &  & 1.414 & 9.653 & 0.730 & 22.48 & 22.32 & 20.46 & 20.36 & 26.07 & 25.04 & 24.28 & 24.58 & 2.12 & 2.46 & 2.98 & 2.83 \\ [2pt]
COSJ100011+023713 & S02 & 150.046815 & 2.620396 & 0.718 & 0.717 & 11.200 & 0.509 & 19.28 & 18.78 & 16.48 & 17.01 & 23.50 & 22.92 & 21.64 & 22.29 & 2.78 & 2.86 & 3.18 & 3.11 \\ [2pt]
COSJ100005+023753 & S02 & 150.022061 & 2.631444 & 0.340 & 0.341 & 10.553 & 0.774 & 18.21 & 17.85 & 16.17 & 16.84 & 25.55 & 25.12 & 23.45 & 22.43 & 1.85 & 1.94 & 2.08 & 1.99 \\ [2pt]
COSJ100000+022327 & S02 & 150.004003 & 2.390916 &  & 1.168 & 11.050 & 0.492 & 20.58 & 19.96 & 17.62 & 17.43 & 23.89 & 23.66 & 22.18 & 21.78 & 2.37 & 2.24 & 2.47 & 2.28 \\ [2pt]
COSJ095959+020634 & S02 & 149.999999 & 2.109498 &  & 0.993 & 10.209 & 0.259 & 21.19 & 21.05 & 18.72 & 18.69 & 42.19 & 24.44 & 24.49 & 24.50 & 1.98 & 1.72 & 2.17 & 1.70 \\ [2pt]
COSJ095949+023301 & S02 & 149.956201 & 2.550537 & 1.326 & 1.322 & 10.029 & 0.532 & 21.65 & 21.41 & 19.56 & 19.50 & 23.88 & 23.50 & 23.29 & 23.52 & 3.86 & 2.97 & 6.24 & 5.76 \\ [2pt]
COSJ095945+021006 & S02 & 149.940865 & 2.168468 & 0.722 & 0.691 & 10.199 & 0.335 & 21.29 & 20.88 & 18.73 & 19.38 & 22.28 & 21.88 & 20.00 & 20.52 & 2.46 & 2.48 & 2.55 & 2.71 \\ [2pt]
COSJ095924+022130 & S02 & 149.853978 & 2.358575 & 0.604 & 0.625 & 10.900 & 0.883 & 18.17 & 17.82 & 15.79 & 16.48 & 23.24 & 22.88 & 20.70 & 20.49 & 2.43 & 2.42 & 2.64 & 2.96 \\ [2pt]
COSJ095913+015448 & S02 & 149.806378 & 1.913426 &  & 1.236 & 9.918 &  &  &  &  &  &  &  &  &  &  &  &  &  \\ [2pt]
COSJ095903+021139 & S02 & 149.764070 & 2.194361 &  & 2.170 & 10.252 & 0.326 & 22.76 & 22.61 & 19.57 & 19.18 & 23.72 & 23.27 & 21.97 & 22.20 & 1.77 & 1.97 & 4.18 & 4.15 \\ [2pt]
COSJ100138+022126 & S02 & 150.411476 & 2.357414 & 0.373 & 0.360 & 10.500 & 0.760 & 18.47 & 18.08 & 16.22 & 16.71 & 21.97 & 21.67 & 19.41 & 18.82 & 3.52 & 3.31 & 3.27 & 3.73 \\ [2pt]
COSJ100122+015013 & S02 & 150.341779 & 1.837170 &  & 1.596 & 10.211 & 0.316 &  & 22.95 & 19.72 & 19.25 &  & 23.70 & 22.05 & 22.62 &  & 3.12 & 2.62 & 2.89 \\ [2pt]
COSJ100116+020041 & S02 & 150.320143 & 2.011404 & 0.308 & 0.310 & 11.000 &  &  &  &  &  &  &  &  &  &  &  &  &  \\ [2pt]
COSJ100111+022233 & S02 & 150.296551 & 2.375987 &  & 0.699 & 9.950 & 0.420 &  &  & 18.59 &  &  &  &  &  &  &  &  &  \\ [2pt]
COSJ100046+022321 & S02 & 150.192875 & 2.389317 &  & 1.678 & 9.463 &  &  &  &  &  &  &  &  &  &  &  &  &  \\ [2pt]
COSJ100035+015102 & S02 & 150.148007 & 1.850826 & 0.927 & 0.920 & 9.872 & 0.579 & 21.61 & 21.34 & 19.32 & 19.33 & 24.76 & 24.59 & 23.34 & 23.69 & 4.15 & 5.03 & 7.29 & 4.95 \\ [2pt]
COSJ100010+023634 & S02 & 150.042969 & 2.609662 & 1.116 & 1.101 & 9.369 & 0.436 & 22.45 &  & 20.51 &  & 24.28 &  & 24.76 &  & 2.53 &  & 2.16 &  \\ [2pt]
COSJ095956+020100 & S02 & 149.984463 & 2.016805 &  & 1.647 & 10.850 & 0.412 & 22.64 & 21.79 & 19.17 & 18.50 & 24.26 & 23.98 & 23.28 & 23.59 & 2.45 & 3.01 & 3.67 & 3.18 \\ [2pt]
COSJ095951+022522 & S02 & 149.963750 & 2.422816 & 0.905 & 0.906 & 11.049 & 0.220 & 20.30 & 19.71 & 17.21 & 17.41 & 23.11 & 23.43 & 22.97 & 23.54 & 1.75 & 1.59 & 2.11 & 1.99 \\ [2pt]
COSJ095947+015615 & S02 & 149.948352 & 1.937556 &  & 1.009 & 10.821 & 0.389 & 21.61 & 20.82 & 18.22 & 18.09 & 23.41 & 22.53 & 20.01 & 20.37 & 2.57 & 2.81 & 3.48 & 3.91 \\ [2pt]
COSJ095938+021318 & S02 & 149.912366 & 2.221786 &  & 0.998 & 9.991 & 0.350 & 22.01 & 21.58 & 19.28 & 19.28 & 24.73 & 24.17 & 24.40 & 24.92 & 1.61 & 1.97 & 2.14 & 2.87 \\ [2pt]
COSJ095937+020742 & S02 & 149.907207 & 2.128461 &  & 1.856 & 9.949 & 0.472 & 22.67 & 22.42 & 20.47 & 19.93 & 25.04 & 25.11 & 24.02 & 24.30 & 3.11 & 2.98 & 5.30 & 3.86 \\ [2pt]
COSJ095936+021947 & S02 & 149.903881 & 2.329788 &  & 2.195 & 10.006 & 0.713 & 23.73 & 22.70 & 20.48 & 20.08 & 26.28 & 25.53 & 24.04 & 24.26 & 2.47 & 3.22 & 3.12 & 4.00 \\ [2pt]
COSJ095914+015947 & S02 & 149.809838 & 1.996563 & 0.459 & 0.478 & 10.146 & 0.590 & 20.40 & 19.86 & 17.73 & 18.14 & 25.14 & 24.07 & 22.34 & 22.86 & 3.06 & 2.62 & 2.34 & 3.26 \\ [2pt]
COSJ100101+020230 & S02 & 150.255135 & 2.041694 & 1.052 & 1.154 & 9.850 & 0.435 & 21.71 & 21.54 & 19.35 & 19.45 & 24.50 & 24.31 & 24.37 & 24.06 & 1.66 & 1.61 & 1.87 & 2.08 \\ [2pt]
COSJ100031+015335 & S02 & 150.129925 & 1.893308 & 0.911 & 0.910 & 9.983 & 0.547 & 21.49 & 21.63 & 19.47 & 19.79 & 24.52 & 23.72 & 22.81 & 23.24 & 3.49 & 4.18 & 4.78 & 4.58 \\ [2pt]
COSJ095959+023441 & S02 & 149.997208 & 2.578056 &  & 3.765 & 11.157 & 0.351 & 25.04 & 24.37 & 19.49 & 18.95 &  &  & 23.21 & 22.06 &  &  & 5.17 & 4.53 \\ [2pt]
COSJ095956+022902 & S02 & 149.984395 & 2.484000 & 0.780 & 0.793 & 8.703 & 0.146 & 22.47 & 22.83 & 21.07 & 21.34 & 23.59 & 23.57 & 23.40 & 23.77 & 1.50 & 1.62 & 1.75 & 1.69 \\ [2pt]
COSJ095945+022340 & S02 & 149.939514 & 2.394477 & 0.938 & 0.910 & 11.300 & 0.793 & 19.30 & 18.60 & 16.14 & 16.36 & 23.92 & 23.37 & 22.31 & 22.57 & 2.22 & 2.29 & 2.41 & 2.35 \\ [2pt]
COSJ095931+020601 & S02 & 149.882791 & 2.100279 & 0.699 & 0.776 & 9.569 & 0.323 & 21.85 & 21.77 & 19.69 & 19.90 & 24.07 & 23.79 & 22.87 & 24.26 & 2.74 & 2.69 & 2.58 & 1.86 \\ [2pt]
\end{tabular}
\end{adjustbox}
\caption{
Table \ref{table:catalogue} continued.
}
\label{table:catalogue4}
\vspace{-20pt} 
\end{table*}

\begin{table*}\ContinuedFloat
\begin{adjustbox}{max width=\textwidth}
\normalsize
\begin{tabular}{ l | l | l | l | l | l | l | l | l | l | l | l | l | l | l | l | l | l | l | l} 
\multicolumn{1}{p{1.8cm}|}{Lens Name} 
& \multicolumn{1}{p{0.7cm}|}{Score} 
& \multicolumn{1}{p{1.1cm}|}{RA [deg]} 
& \multicolumn{1}{p{1.1cm}|}{Dec [deg]}
& \multicolumn{1}{p{0.85cm}|}{$z_\text{spec}$} 
& \multicolumn{1}{p{0.85cm}|}{$z_\text{phot}$} 
& \multicolumn{1}{p{1.2cm}|}{$\log_{{10}}$ $(M_*/M_{{\odot}})$} 
& \multicolumn{1}{p{0.85cm}|}{$R_{\rm Ein}$  (\arcsec)}
& \multicolumn{1}{p{0.85cm}|}{Lens $m_{\mathrm{F115W}}$}
& \multicolumn{1}{p{0.85cm}|}{Lens $m_{\mathrm{F150W}}$}
& \multicolumn{1}{p{0.85cm}|}{Lens $m_{\mathrm{F277W}}$}
& \multicolumn{1}{p{0.85cm}|}{Lens $m_{\mathrm{F444W}}$}
& \multicolumn{1}{p{0.85cm}|}{Source $m_{\mathrm{F115W}}$}
& \multicolumn{1}{p{0.85cm}|}{Source $m_{\mathrm{F150W}}$}
& \multicolumn{1}{p{0.85cm}|}{Source $m_{\mathrm{F277W}}$}
& \multicolumn{1}{p{0.85cm}|}{Source $m_{\mathrm{F444W}}$}
& \multicolumn{1}{p{0.85cm}|}{$\mu_{\mathrm{F115W}}$}
& \multicolumn{1}{p{0.85cm}|}{$\mu_{\mathrm{F150W}}$}
& \multicolumn{1}{p{0.85cm}|}{$\mu_{\mathrm{F277W}}$}
& \multicolumn{1}{p{0.85cm}}{$\mu_{\mathrm{F444W}}$}
\\   \hline
& & & & & & & & & & & & & & & & & & & \\ [-6pt]
COSJ100022+014924 & S02 & 150.092249 & 1.823386 &  & 1.625 & 9.860 & 0.383 & 23.53 & 23.26 & 20.76 & 20.99 & 23.33 & 23.26 & 22.44 & 22.12 & 3.66 & 3.85 & 4.38 & 4.56 \\ [2pt]
COSJ100008+022122 & S02 & 150.035617 & 2.356388 & 0.746 & 0.740 & 10.950 & 0.714 & 19.25 & 18.79 & 16.52 & 17.14 & 24.79 & 24.88 & 23.79 & 24.39 & 2.41 & 2.28 & 1.94 & 2.16 \\ [2pt]
COSJ095910+020937 & S02 & 149.794945 & 2.160474 & 0.663 & 0.650 & 9.448 & 0.259 & 21.25 & 21.02 & 19.14 & 19.70 & 22.63 & 22.58 & 21.83 & 22.61 & 2.02 & 1.78 & 2.12 & 1.98 \\ [2pt]
COSJ100134+022028 & S02 & 150.393924 & 2.341301 &  & 0.104 & 8.800 & 1.654 & 19.87 & 19.73 & 18.63 & 19.37 & 45.29 & 25.48 &  &  & 9.49 & 11.06 &  &  \\ [2pt]
COSJ100103+015816 & S02 & 150.263846 & 1.971350 & 0.931 & 0.987 & 11.143 & 0.718 & 20.11 & 19.56 & 17.08 & 17.20 & 22.15 & 23.34 & 19.87 & 20.21 & 4.56 & 4.40 & 4.03 & 5.23 \\ [2pt]
COSJ100044+020652 & S02 & 150.184108 & 2.114691 &  & 1.938 & 9.950 & 0.489 &  & 21.05 & 19.48 & 17.31 &  & 21.15 & 22.03 & 22.50 &  & 1.47 & 1.05 & 1.65 \\ [2pt]
COSJ100030+015427 & S02 & 150.127549 & 1.907704 &  &  &  & 1.086 & 20.22 & 19.66 & 17.23 & 17.01 & 25.41 & 24.48 & 22.28 & 22.11 & 1.08 & 3.05 & 1.87 & 2.61 \\ [2pt]
COSJ100149+022734 & S02 & 150.455510 & 2.459588 &  &  &  &  &  &  &  &  &  &  &  &  &  &  &  &  \\ [2pt]
COSJ100113+020321 & S02 & 150.304564 & 2.055838 & 0.295 & 0.398 & 8.702 & 0.963 & 21.92 & 21.73 & 20.10 & 20.77 & 22.87 & 23.05 & 22.10 & 22.19 & 6.05 & 5.84 & 5.37 & 5.17 \\ [2pt]
COSJ100051+023219 & S02 & 150.213542 & 2.538649 &  & 0.427 & 9.946 &  &  &  &  &  &  &  &  &  &  &  &  &  \\ [2pt]
COSJ100023+020958 & S02 & 150.096811 & 2.166291 &  & 1.060 & 9.705 & 0.325 & 22.51 & 24.91 & 19.62 & 22.52 & 25.03 &  & 24.04 & 21.88 & 1.53 &  & 2.64 & 3.83 \\ [2pt]
COSJ095926+022727 & S02 & 149.859248 & 2.457768 & 0.420 &  &  &  &  &  &  &  &  &  &  &  &  &  &  &  \\ [2pt]
COSJ100141+022738 & S02 & 150.423524 & 2.460609 &  & 3.346 & 11.150 &  &  &  &  &  &  &  &  &  &  &  &  &  \\ [2pt]
COSJ100015+023341 & S02 & 150.066424 & 2.561556 &  & 0.704 & 10.338 & 1.530 & 20.45 & 20.02 & 17.44 & 17.64 & 25.91 & 25.57 & 24.07 & 42.29 & 2.93 & 2.12 & 2.15 & 1.64 \\ [2pt]
COSJ095926+022040 & S02 & 149.862057 & 2.344538 &  & 1.564 & 10.858 & 0.381 & 22.29 & 21.54 & 18.82 & 18.38 & 23.46 & 23.28 & 22.24 & 21.97 & 1.60 & 1.60 & 1.64 & 1.70 \\ [2pt]
COSJ095917+020626 & S02 & 149.824573 & 2.107423 &  & 1.661 & 9.800 & 0.412 & 22.30 & 22.10 & 20.25 & 20.10 & 25.43 & 25.11 & 24.73 & 23.76 & 1.82 & 2.58 & 2.53 & 2.27 \\ [2pt]
COSJ100138+022225 & S02 & 150.410304 & 2.373836 &  & 1.930 & 9.780 & 0.414 & 22.86 & 22.49 & 20.70 & 20.63 & 23.74 & 23.35 & 23.59 & 23.32 & 5.38 & 4.89 & 6.91 & 7.05 \\ [2pt]
COSJ100118+014923 & S02 & 150.328835 & 1.823299 & 1.171 & 1.200 & 9.788 & 0.396 & 21.59 & 21.38 & 19.55 & 19.66 & 23.02 & 22.94 & 22.29 & 22.26 & 3.58 & 3.26 & 3.87 & 3.95 \\ [2pt]
COSJ100116+022409 & S02 & 150.319391 & 2.402554 &  & 2.388 & 10.234 & 0.235 & 23.40 & 23.58 & 20.72 & 20.38 & 25.30 & 23.69 & 22.96 & 23.15 & 1.35 & 1.71 & 1.82 & 1.70 \\ [2pt]
COSJ100101+023258 & S02 & 150.254932 & 2.549520 &  & 1.531 & 9.442 & 0.499 &  &  &  & 20.31 &  &  &  & 24.16 &  &  &  & 2.56 \\ [2pt]
COSJ100039+020800 & S02 & 150.165592 & 2.133493 &  & 0.995 & 10.201 & 0.806 & 21.20 & 20.41 & 17.62 & 16.99 & 23.49 & 23.05 & 20.98 & 20.25 & 5.52 & 5.33 & 5.08 & 4.51 \\ [2pt]
COSJ100025+021538 & S02 & 150.104349 & 2.260572 &  & 0.641 & 10.050 & 0.403 & 20.21 & 19.84 & 17.69 & 18.54 & 19.89 & 20.14 & 17.39 & 18.57 & 4.75 & 4.49 & 3.92 & 3.72 \\ [2pt]
COSJ095959+022254 & S02 & 149.996687 & 2.381836 &  &  &  &  &  &  &  &  &  &  &  &  &  &  &  &  \\ [2pt]
COSJ095956+022028 & S02 & 149.987281 & 2.341249 & 0.903 & 0.973 & 10.175 & 0.449 & 21.58 & 21.22 & 18.50 & 18.49 & 22.81 & 22.66 & 20.36 & 20.92 & 3.08 & 3.18 & 3.15 & 3.16 \\ [2pt]
COSJ095929+020320 & S02 & 149.873681 & 2.055789 &  & 0.644 & 9.611 & 0.849 &  &  & 20.32 &  &  &  & 24.33 &  &  &  & 5.95 &  \\ [2pt]
COSJ095912+020033 & S02 & 149.802096 & 2.009175 &  & 1.379 & 9.522 & 0.502 & 22.71 & 22.27 & 20.44 & 20.36 & 24.04 & 23.90 & 23.45 & 23.16 & 3.63 & 2.71 & 3.60 & 3.66 \\ [2pt]
COSJ100031+022639 & S01 & 150.129890 & 2.444361 & 1.537 & 1.526 & 11.063 &  &  &  &  &  &  &  &  &  &  &  &  &  \\ [2pt]
COSJ100034+021133 & S01 & 150.143320 & 2.192625 & 0.800 & 0.800 & 11.200 & 0.606 & 19.50 & 18.81 & 16.53 & 16.98 & 24.04 & 23.16 & 22.30 & 22.79 & 1.90 & 2.05 & 2.22 & 2.19 \\ [2pt]
COSJ100001+021455 & S01 & 150.005960 & 2.248797 & 0.930 & 0.936 & 9.912 & 0.663 & 21.79 & 21.68 & 19.44 & 19.81 & 25.13 & 24.85 & 23.55 & 23.88 & 3.38 & 3.82 & 2.79 & 3.78 \\ [2pt]
COSJ095944+020652 & S01 & 149.933337 & 2.114585 &  &  &  & 0.923 &  & 19.60 & 16.96 & 17.39 &  & 22.17 & 19.84 & 21.98 &  & 2.11 & 2.39 & 2.26 \\ [2pt]
COSJ100130+021315 & S01 & 150.375893 & 2.221091 & 0.915 &  &  & 0.963 &  &  & 18.81 & 29.32 &  &  & 22.69 &  &  &  & 2.96 &  \\ [2pt]
COSJ100116+020221 & S01 & 150.319173 & 2.039403 & 0.994 & 0.951 & 9.876 & 0.548 & 22.07 & 21.08 & 18.46 & 18.27 & 23.22 & 22.60 & 21.66 & 21.41 & 2.55 & 2.48 & 3.05 & 2.50 \\ [2pt]
COSJ100115+022423 & S01 & 150.315195 & 2.406501 & 0.124 & 0.117 & 9.001 &  &  &  &  &  &  &  &  &  &  &  &  &  \\ [2pt]
COSJ100043+015638 & S01 & 150.183249 & 1.944001 & 1.194 & 1.181 & 10.032 &  &  &  &  &  &  &  &  &  &  &  &  &  \\ [2pt]
COSJ095929+020338 & S01 & 149.873864 & 2.060783 & 1.180 & 1.179 & 11.300 &  &  &  &  &  &  &  &  &  &  &  &  &  \\ [2pt]
COSJ100143+020435 & S01 & 150.430436 & 2.076459 & 0.311 & 0.303 & 9.200 &  &  &  &  &  &  &  &  &  &  &  &  &  \\ [2pt]
COSJ100111+020157 & S01 & 150.297628 & 2.032618 &  & 0.877 & 9.804 & 0.619 & 22.04 & 21.70 & 19.30 & 19.33 & 24.82 & 24.54 & 22.76 & 23.38 & 2.40 & 3.46 & 3.15 & 3.39 \\ [2pt]
COSJ100103+015121 & S01 & 150.266640 & 1.856077 & 1.143 & 0.998 & 10.355 & 1.022 & 21.68 & 21.09 & 18.82 & 18.78 & 24.75 & 24.42 & 22.52 & 22.57 & 3.58 & 3.79 & 0.72 & 1.61 \\ [2pt]
COSJ100103+014614 & S01 & 150.263836 & 1.770571 & 0.527 & 0.519 & 9.704 & 0.798 & 20.69 & 20.57 & 18.53 & 18.86 & 23.90 & 23.39 & 22.46 & 23.78 & 3.60 & 3.64 & 4.09 & 4.19 \\ [2pt]
COSJ100059+020519 & S01 & 150.247628 & 2.088699 &  & 1.158 & 10.449 & 1.861 &  &  & 18.75 &  &  &  & 37.29 &  &  &  & 21.05 &  \\ [2pt]
COSJ100021+014905 & S01 & 150.089860 & 1.818217 & 0.422 & 0.430 & 9.450 & 0.462 &  & 20.87 & 19.04 &  &  & 24.24 & 23.70 &  &  & 1.83 & 1.65 &  \\ [2pt]
COSJ095954+020539 & S01 & 149.977835 & 2.094293 & 0.900 & 0.899 & 10.308 & 0.546 & 21.11 & 20.63 & 18.42 & 18.78 & 23.16 & 22.95 & 21.47 & 22.16 & 4.13 & 4.38 & 4.21 & 4.51 \\ [2pt]
COSJ095948+022859 & S01 & 149.951191 & 2.483094 & 0.738 & 0.720 & 11.065 & 0.481 & 19.01 & 18.52 & 16.16 & 16.83 & 25.20 & 24.42 & 23.01 & 24.58 & 1.69 & 1.87 & 1.65 & 1.53 \\ [2pt]
COSJ095937+023754 & S01 & 149.906127 & 2.631909 &  & 1.560 & 9.628 &  &  &  &  &  &  &  &  &  &  &  &  &  \\ [2pt]
COSJ095927+020315 & S01 & 149.863022 & 2.054210 &  & 2.927 & 10.778 &  &  &  &  &  &  &  &  &  &  &  &  &  \\ [2pt]
COSJ100152+022504 & S01 & 150.470603 & 2.417932 &  & 2.709 & 11.030 & 0.479 & 24.78 & 23.39 & 19.86 & 19.22 &  & 25.61 & 23.98 & 23.25 &  & 3.13 & 2.97 & 2.66 \\ [2pt]
COSJ100134+022039 & S01 & 150.392104 & 2.344240 &  & 0.424 & 9.039 &  &  &  &  &  &  &  &  &  &  &  &  &  \\ [2pt]
COSJ100109+015549 & S01 & 150.290601 & 1.930532 &  & 0.616 & 10.199 & 0.489 & 21.03 & 20.57 & 18.49 & 19.13 & 24.18 & 23.97 & 22.78 & 23.25 & 2.28 & 2.01 & 2.27 & 2.36 \\ [2pt]
COSJ100103+015747 & S01 & 150.265759 & 1.963065 & 0.932 & 0.920 & 11.450 &  &  &  &  &  &  &  &  &  &  &  &  &  \\ [2pt]
COSJ100101+020120 & S01 & 150.257378 & 2.022389 &  & 1.203 & 9.557 & 0.622 & 22.04 & 21.77 & 19.66 & 19.54 & 24.15 & 23.20 & 23.10 & 22.85 & 3.00 & 3.91 & 3.84 & 3.55 \\ [2pt]
COSJ100031+020046 & S01 & 150.130091 & 2.012904 & 0.982 & 0.960 & 8.696 & 0.213 & 22.78 & 22.86 & 20.72 & 21.31 & 23.56 & 23.77 & 23.70 & 24.56 & 1.86 & 2.01 & 2.37 & 2.03 \\ [2pt]
COSJ100017+023020 & S01 & 150.072815 & 2.505736 & 0.960 & 0.998 & 10.950 &  &  &  &  &  &  &  &  &  &  &  &  &  \\ [2pt]
COSJ100016+022131 & S01 & 150.067221 & 2.358656 & 0.660 & 0.660 & 9.450 & 0.434 & 21.01 & 20.96 & 19.18 & 19.71 & 23.96 & 23.77 & 23.31 & 24.34 & 2.53 & 2.46 & 2.61 & 2.94 \\ [2pt]
COSJ100010+022601 & S01 & 150.042498 & 2.433817 & 0.338 & 0.332 & 10.180 & 0.514 & 19.36 & 18.88 & 16.94 & 17.56 & 21.68 & 21.36 & 19.49 & 19.90 & 3.29 & 3.40 & 3.01 & 3.04 \\ [2pt]
COSJ100007+015240 & S01 & 150.032008 & 1.877780 &  &  &  & 0.370 & 22.75 & 22.53 & 20.57 & 20.61 & 25.13 & 24.68 & 24.20 & 24.91 & 1.97 & 1.67 & 2.70 & 3.63 \\ [2pt]
COSJ100002+021136 & S01 & 150.009540 & 2.193369 &  & 2.177 & 10.281 & 0.520 & 22.40 & 21.98 & 19.86 & 19.50 & 25.41 & 23.08 & 22.58 & 22.37 & 1.75 & 2.24 & 2.06 & 2.34 \\ [2pt]
COSJ095956+023353 & S01 & 149.983780 & 2.564968 &  & 0.703 & 9.805 &  &  &  &  &  &  &  &  &  &  &  &  &  \\ [2pt]
COSJ095948+015722 & S01 & 149.950702 & 1.956244 &  & 1.407 & 9.652 & 0.334 & 22.71 & 22.42 & 19.84 & 19.21 & 25.58 & 23.78 & 21.89 & 21.16 & 2.74 & 3.73 & 3.58 & 3.79 \\ [2pt]
COSJ095939+020309 & S01 & 149.916030 & 2.052506 & 0.363 & 0.337 & 8.804 & 0.509 & 21.46 & 21.32 & 19.87 & 20.48 & 24.66 & 24.50 & 24.52 & 24.56 & 2.41 & 2.38 & 2.82 & 2.38 \\ [2pt]
COSJ095929+020914 & S01 & 149.873815 & 2.153938 & 0.937 & 0.930 & 9.850 & 0.331 & 21.60 & 21.15 & 18.94 & 19.15 & 22.58 & 22.93 & 21.81 & 22.02 & 2.07 & 1.89 & 2.48 & 2.36 \\ [2pt]
COSJ095929+015348 & S01 & 149.871128 & 1.896779 & 0.132 & 0.105 & 8.701 & 0.455 & 18.99 & 19.04 & 17.85 & 18.52 & 23.01 & 22.75 & 21.70 & 22.00 & 2.04 & 2.16 & 2.07 & 2.62 \\ [2pt]
COSJ095921+020218 & S01 & 149.839544 & 2.038512 &  & 1.405 & 10.950 & 0.485 & 20.69 & 20.26 & 17.81 & 17.72 & 25.66 & 24.93 & 23.24 & 23.59 & 1.98 & 2.18 & 2.47 & 2.46 \\ [2pt]
COSJ095916+022021 & S01 & 149.819959 & 2.339357 & 0.490 & 0.475 & 11.141 &  &  &  &  &  &  &  &  &  &  &  &  &  \\ [2pt]
COSJ095913+021702 & S01 & 149.804343 & 2.283997 & 2.037 & 2.038 & 9.563 & 0.274 & 22.92 & 22.88 & 20.93 & 20.33 & 22.89 & 22.77 & 22.03 & 22.45 & 3.14 & 3.12 & 3.18 & 3.14 \\ [2pt]
COSJ100057+015056 & S01 & 150.241491 & 1.849084 & 1.054 & 0.998 & 9.856 & 0.469 & 22.13 & 21.90 & 19.82 & 19.96 & 25.29 & 24.93 & 23.49 & 23.68 & 2.32 & 2.90 & 3.10 & 3.13 \\ [2pt]
COSJ100055+021413 & S01 & 150.230044 & 2.237013 & 0.118 & 0.119 & 7.849 & 0.896 & 21.53 & 21.44 & 20.51 & 21.17 & 24.79 & 24.46 & 23.69 & 24.14 & 2.84 & 3.21 & 2.67 & 2.83 \\ [2pt]
COSJ100045+015244 & S01 & 150.191537 & 1.879114 &  & 1.169 & 9.796 & 0.494 & 22.18 & 21.98 & 20.06 & 20.10 & 25.28 & 25.44 & 24.46 & 24.25 & 3.97 & 4.86 & 8.68 & 6.63 \\ [2pt]
COSJ095857+020404 & S01 & 149.739971 & 2.067781 &  & 1.675 & 9.917 & 0.599 &  & 23.17 & 20.11 & 19.04 &  & 24.10 & 21.14 & 20.64 &  & 2.99 & 3.07 & 3.24 \\ [2pt]
COSJ095857+020404 & S01 & 149.739971 & 2.067781 &  &  & 9.917 &  &  &  &  &  &  &  &  &  &  &  &  &  \\ [2pt]
COSJ100127+022709 & S01 & 150.363955 & 2.452706 &  & 1.252 & 9.509 & 0.932 & 22.34 & 22.49 & 19.71 & 19.38 &  & 26.57 & 25.20 & 45.66 &  & 2.25 & 1.79 & 1.75 \\ [2pt]
COSJ100036+015220 & S01 & 150.152066 & 1.872323 & 0.580 & 0.575 & 9.773 & 0.290 & 19.48 & 19.29 & 17.57 & 18.14 & 21.89 & 21.23 & 20.93 & 21.44 & 2.52 & 2.70 & 3.00 & 3.45 \\ [2pt]
COSJ095956+021920 & S01 & 149.987497 & 2.322245 & 0.937 & 0.944 & 10.812 & 0.322 & 20.34 & 19.86 & 17.39 & 17.66 & 24.96 & 24.81 & 23.97 & 24.24 & 1.51 & 1.46 & 1.93 & 1.75 \\ [2pt]
COSJ100051+015724 & S00 & 150.215892 & 1.956802 & 0.573 & 0.657 & 10.100 &  &  &  &  &  &  &  &  &  &  &  &  &  \\ [2pt]
COSJ100048+015412 & S00 & 150.203284 & 1.903476 &  & 1.183 & 9.973 &  &  &  &  &  &  &  &  &  &  &  &  &  \\ [2pt]
COSJ100045+014712 & S00 & 150.189165 & 1.786751 & 0.809 & 0.812 & 10.913 &  &  &  &  &  &  &  &  &  &  &  &  &  \\ [2pt]
COSJ100021+014915 & S00 & 150.090770 & 1.821096 & 0.994 & 0.945 & 10.418 & 0.453 & 21.29 & 21.00 & 18.27 & 18.19 & 21.54 & 20.41 & 18.56 & 18.05 & 3.04 & 3.16 & 3.69 & 3.82 \\ [2pt]
COSJ095959+020105 & S00 & 149.999986 & 2.018308 &  & 1.234 & 9.774 &  &  &  &  &  &  &  &  &  &  &  &  &  \\ [2pt]
COSJ095846+020304 & S00 & 149.692205 & 2.051371 & 1.002 & 0.991 & 11.000 &  &  &  &  &  &  &  &  &  &  &  &  &  \\ [2pt]
COSJ100125+020326 & S00 & 150.358234 & 2.057231 & 0.817 & 0.860 & 9.812 &  &  &  &  &  &  &  &  &  &  &  &  &  \\ [2pt]
COSJ100105+015027 & S00 & 150.273782 & 1.840914 & 0.248 & 0.241 & 11.251 &  &  &  &  &  &  &  &  &  &  &  &  &  \\ [2pt]
COSJ100054+014602 & S00 & 150.228301 & 1.767308 & 0.350 & 0.336 & 10.450 &  &  &  &  &  &  &  &  &  &  &  &  &  \\ [2pt]
\end{tabular}
\end{adjustbox}
\caption{
Table \ref{table:catalogue} continued.
}
\label{table:catalogue5}
\vspace{-4pt} 
\end{table*}

\begin{table*}\ContinuedFloat
\begin{adjustbox}{max width=\textwidth}
\normalsize
\begin{tabular}{ l | l | l | l | l | l | l | l | l | l | l | l | l | l | l | l | l | l | l | l} 
\multicolumn{1}{p{1.8cm}|}{Lens Name} 
& \multicolumn{1}{p{0.7cm}|}{Score} 
& \multicolumn{1}{p{1.1cm}|}{RA [deg]} 
& \multicolumn{1}{p{1.1cm}|}{Dec [deg]}
& \multicolumn{1}{p{0.85cm}|}{$z_\text{spec}$} 
& \multicolumn{1}{p{0.85cm}|}{$z_\text{phot}$} 
& \multicolumn{1}{p{1.2cm}|}{$\log_{{10}}$ $(M_*/M_{{\odot}})$} 
& \multicolumn{1}{p{0.85cm}|}{$R_{\rm Ein}$  (\arcsec)}
& \multicolumn{1}{p{0.85cm}|}{Lens $m_{\mathrm{F115W}}$}
& \multicolumn{1}{p{0.85cm}|}{Lens $m_{\mathrm{F150W}}$}
& \multicolumn{1}{p{0.85cm}|}{Lens $m_{\mathrm{F277W}}$}
& \multicolumn{1}{p{0.85cm}|}{Lens $m_{\mathrm{F444W}}$}
& \multicolumn{1}{p{0.85cm}|}{Source $m_{\mathrm{F115W}}$}
& \multicolumn{1}{p{0.85cm}|}{Source $m_{\mathrm{F150W}}$}
& \multicolumn{1}{p{0.85cm}|}{Source $m_{\mathrm{F277W}}$}
& \multicolumn{1}{p{0.85cm}|}{Source $m_{\mathrm{F444W}}$}
& \multicolumn{1}{p{0.85cm}|}{$\mu_{\mathrm{F115W}}$}
& \multicolumn{1}{p{0.85cm}|}{$\mu_{\mathrm{F150W}}$}
& \multicolumn{1}{p{0.85cm}|}{$\mu_{\mathrm{F277W}}$}
& \multicolumn{1}{p{0.85cm}}{$\mu_{\mathrm{F444W}}$}
\\   \hline
& & & & & & & & & & & & & & & & & & & \\ [-6pt]
COSJ100049+014721 & S00 & 150.206861 & 1.789354 &  & 0.325 & 8.550 &  &  &  &  &  &  &  &  &  &  &  &  &  \\ [2pt]
COSJ100033+015039 & S00 & 150.138106 & 1.844238 &  & 0.561 & 8.903 &  &  &  &  &  &  &  &  &  &  &  &  &  \\ [2pt]
COSJ100031+020113 & S00 & 150.130831 & 2.020307 & 1.439 & 2.148 & 11.150 &  &  &  &  &  &  &  &  &  &  &  &  &  \\ [2pt]
COSJ100030+014754 & S00 & 150.126138 & 1.798424 &  & 1.190 & 10.466 &  &  &  &  &  &  &  &  &  &  &  &  &  \\ [2pt]
COSJ100016+020956 & S00 & 150.068073 & 2.165623 & 0.931 & 1.573 & 11.352 &  &  &  &  &  &  &  &  &  &  &  &  &  \\ [2pt]
COSJ100013+023205 & S00 & 150.058265 & 2.534838 & 0.643 & 0.660 & 10.650 & 0.664 & 19.50 & 19.12 & 16.75 & 17.39 & 25.24 & 24.46 & 23.16 & 24.82 & 1.87 & 2.37 & 2.19 & 1.58 \\ [2pt]
COSJ100010+023549 & S00 & 150.045173 & 2.597030 &  & 0.900 & 8.627 &  &  &  &  &  &  &  &  &  &  &  &  &  \\ [2pt]
COSJ100010+021022 & S00 & 150.043171 & 2.172803 & 1.232 & 1.210 & 9.500 &  &  &  &  &  &  &  &  &  &  &  &  &  \\ [2pt]
COSJ100141+022159 & S00 & 150.422924 & 2.366500 &  &  &  & 1.087 &  & 18.75 & 17.11 & 17.38 &  & 23.26 & 21.63 & 21.49 &  & 5.45 & 4.38 & 4.48 \\ [2pt]
COSJ100109+015436 & S00 & 150.288717 & 1.910055 & 0.219 & 0.210 & 10.039 &  &  &  &  &  &  &  &  &  &  &  &  &  \\ [2pt]
COSJ100108+014922 & S00 & 150.283473 & 1.822985 &  & 1.626 & 9.782 & 0.591 & 21.89 & 21.78 & 20.00 & 19.88 & 25.00 & 24.72 & 23.37 & 22.75 & 1.33 & 3.52 & 2.61 & 2.51 \\ [2pt]
COSJ100103+015207 & S00 & 150.265278 & 1.868658 & 0.530 & 0.513 & 11.100 & 0.790 & 18.68 & 18.13 & 16.04 & 16.69 & 23.99 & 23.67 & 22.89 & 23.98 & 1.74 & 1.74 & 1.65 & 1.65 \\ [2pt]
COSJ100050+015421 & S00 & 150.210807 & 1.906014 &  & 0.445 & 11.044 &  &  &  &  &  &  &  &  &  &  &  &  &  \\ [2pt]
COSJ100049+021945 & S00 & 150.205720 & 2.329233 &  & 1.498 & 10.100 &  &  &  &  &  &  &  &  &  &  &  &  &  \\ [2pt]
COSJ100049+014829 & S00 & 150.207031 & 1.808253 & 1.243 & 1.363 & 9.709 &  &  &  &  &  &  &  &  &  &  &  &  &  \\ [2pt]
COSJ100012+023351 & S00 & 150.050830 & 2.564307 &  & 3.169 & 10.488 &  &  &  &  &  &  &  &  &  &  &  &  &  \\ [2pt]
COSJ100005+021159 & S00 & 150.022019 & 2.199805 & 0.959 & 1.666 & 10.369 &  &  &  &  &  &  &  &  &  &  &  &  &  \\ [2pt]
COSJ095951+023722 & S00 & 149.965639 & 2.623026 &  & 0.716 & 10.350 & 1.186 & 20.39 & 20.07 & 18.08 & 18.58 & 24.15 & 23.91 & 23.54 & 23.82 & 2.13 & 2.19 & 2.04 & 2.60 \\ [2pt]
COSJ095950+020013 & S00 & 149.961835 & 2.003707 & 0.389 & 0.360 & 10.750 & 0.317 & 17.69 & 17.27 & 15.50 & 16.10 & 23.67 & 23.14 & 21.66 & 22.28 & 1.39 & 1.44 & 1.72 & 1.78 \\ [2pt]
COSJ095941+023210 & S00 & 149.922996 & 2.536189 & 0.881 & 0.880 & 9.395 & 0.447 & 22.47 & 22.35 & 20.28 & 20.75 & 25.40 & 24.75 & 24.20 & 24.93 & 2.00 & 2.17 & 1.92 & 2.07 \\ [2pt]
COSJ095920+020558 & S00 & 149.834487 & 2.099582 &  & 0.970 & 10.299 &  &  &  &  &  &  &  &  &  &  &  &  &  \\ [2pt]
COSJ095855+015828 & S00 & 149.731997 & 1.974491 & 0.748 & 0.743 & 11.150 & 1.437 & 18.70 & 18.31 & 15.93 & 16.66 & 25.25 & 24.81 & 22.75 & 23.86 & 4.10 & 4.13 & 4.16 & 4.17 \\ [2pt]
COSJ100030+021410 & S00 & 150.125654 & 2.236134 & 0.560 & 0.516 & 9.951 &  &  &  &  &  &  &  &  &  &  &  &  &  \\ [2pt]
COSJ100123+022005 & S00 & 150.346287 & 2.334865 &  &  &  & 0.348 & 24.30 & 23.12 & 19.02 & 19.06 &  &  & 22.43 & 21.23 &  &  & 2.54 & 2.23 \\ [2pt]
COSJ100120+022426 & S00 & 150.336462 & 2.407291 &  & 0.603 & 9.505 & 0.889 & 21.61 & 21.37 & 19.31 & 19.99 & 43.86 & 25.85 &  &  & 1.77 & 3.06 &  &  \\ [2pt]
COSJ100103+020433 & S00 & 150.266103 & 2.076059 & 1.100 & 1.030 & 9.047 & 0.420 & 21.81 & 21.74 & 20.21 & 20.25 & 22.42 & 21.80 & 21.71 & 22.72 & 2.16 & 3.61 & 3.17 & 2.99 \\ [2pt]
COSJ095953+022200 & S00 & 149.971875 & 2.366768 &  &  &  & 0.551 & 20.02 & 19.74 & 18.03 & 18.64 & 24.06 & 23.35 & 22.81 & 23.04 & 2.70 & 2.82 & 2.70 & 2.58 \\ [2pt]
COSJ095953+015437 & S00 & 149.973412 & 1.910466 & 1.469 & 1.623 & 10.900 &  &  &  &  &  &  &  &  &  &  &  &  &  \\ [2pt]
COSJ095952+020313 & S00 & 149.970691 & 2.053875 &  & 1.059 & 9.650 & 2.058 &  & 20.99 & 19.65 & 19.91 &  &  & 24.33 & 24.71 &  &  & 3.24 & 4.31 \\ [2pt]
COSJ095927+021005 & S00 & 149.865199 & 2.168084 & 0.595 & 0.594 & 9.273 & 0.516 & 21.08 & 21.06 & 19.06 & 19.49 & 24.56 & 24.60 & 23.60 & 25.01 & 1.55 & 1.71 & 2.38 & 1.75 \\ [2pt]
COSJ100116+022235 & S00 & 150.317122 & 2.376585 &  & 1.758 & 10.917 & 0.395 & 22.37 & 21.20 & 18.78 & 18.35 &  &  & 25.17 & 24.18 &  &  & 2.48 & 1.71 \\ [2pt]
COSJ100030+014853 & S00 & 150.128490 & 1.814993 & 1.978 & 1.655 & 9.732 &  &  &  &  &  &  &  &  &  &  &  &  &  \\ [2pt]
COSJ095941+021432 & S00 & 149.924781 & 2.242476 &  & 1.106 & 9.869 & 0.521 & 21.91 & 21.59 & 19.37 & 19.51 & 25.22 & 25.07 & 24.50 & 23.77 & 1.35 & 1.54 & 2.43 & 2.30 \\ [2pt]
COSJ095940+023802 & S00 & 149.917297 & 2.633927 & 0.862 & 0.930 & 10.445 & 0.481 & 20.95 &  & 18.51 & 18.70 & 23.33 &  & 21.61 & 21.99 & 2.56 &  & 2.83 & 2.76 \\ [2pt]
COSJ100004+023543 & S00 & 150.019141 & 2.595436 &  &  &  & 0.644 &  & 18.25 & 16.53 & 17.20 &  & 21.85 & 20.21 & 21.25 &  & 2.41 & 2.56 & 2.69 \\ [2pt]
\end{tabular}
\end{adjustbox}
\caption{
Table \ref{table:catalogue} continued.
}
\label{table:catalogue6}
\end{table*}

The full catalogue of candidate lenses, visual inspection scores and measured properties is given in \cref{table:catalogue}.

\label{lastpage}

\end{document}